\documentclass[fleqn,twoside,openany]{book}
\usepackage{espcrc2}

\DeclareMathAlphabet   {\mathsc}{OT1}{cmr}{m}{sc}
\usepackage{amssymb,amsmath,graphics,graphicx,epsfig}
\usepackage{epigraph}
\usepackage{latexsym}
\usepackage{psfrag}
\usepackage{amsfonts}
\usepackage{multicol}
\usepackage{ifthen}
\usepackage{bm}
\usepackage[figuresright]{rotating}

\begin{document}

%%%%%%%%%%%%%%%%%%%%%%%%%%%%%%%%%%%%%%%%%%%%%%%%%%%%%%%%%%%%%%%%%%%%%%%%%%%%%%%%%%%%%%%%%%%%%%
%%%%%%%%%%%%%%%%%%%%%%%%%%%%%%%%%%%%%%%%%%%%%%%%%%%%%%%%%%%%%%%%%%%%%%%%%%%%%%%%%%%%%%%%%%%%%%

\title{\Large \textbf{The Hunt for New Physics at the Large Hadron
Collider}}

%%%%%%%%%%%%%%%%%%%%%%%%%%%%%%%%%%%%%%%%%%%%%%%%%%%%%
% AUTHORS
%%%%%%%%%%%%%%%%%%%%%%%%%%%%%%%%%%%%%%%%%%%%%%%%%%%%%
\author{Principal Conveners: Pran Nath$^{a}$\\
 ~~~~~~~~~~~~~~~~~~~~~~~~~~~~Brent Nelson$^{a}$\\~\\
 Conveners for New Physics Sections:  Hooman Davoudiasl$^{b}$ (Extra Dimensions)\\
      ~~~~~~~~~~~~~~~~~~~~~~~~~~~~~~~~~~~~~~~~~~~~~~~~~Bhaskar Dutta$^{c}$ (Dark Matter)\\
       ~~~~~~~~~~~~~~~~~~~~~~~~~~~~~~~~~~~~~~~~~~~~~~~~~Daniel Feldman$^{d}$ and Zuowei Liu$^{e}$ (Hidden Sectors)\\
      ~~~~~~~~~~~~~~~~~~~~~~~~~~~~~~~~~~~~~~~~~~~~~~~~~Tao Han$^{f}$ (Top)\\
        ~~~~~~~~~~~~~~~~~~~~~~~~~~~~~~~~~~~~~~~~~~~~~~~~~Paul Langacker$^{g}$ (Z Prime)\\
        ~~~~~~~~~~~~~~~~~~~~~~~~~~~~~~~~~~~~~~~~~~~~~~~~~Rabi Mohapatra$^{h}$ and Jose Valle$^{i}$ (Neutrinos)\\
     ~~~~~~~~~~~~~~~~~~~~~~~~~~~~~~~~~~~~~~~~~~~~~~~~ Pran Nath$^{a}$ (SUSY)\\
~~~~~~~~~~~~~~~~~~~~~~~~~~~~~~~~~~~~~~~~~~~~~~~~ Brent Nelson$^{a}$ (Strings)\\
  ~~~~~~~~~~~~~~~~~~~~~~~~~~~~~~~~~~~~~~~~~~~~~~~~~Apostolos Pilaftsis$^{j}$ (CP violation)\\
  ~~~~~~~~~~~~~~~~~~~~~~~~~~~~~~~~~~~~~~~~~~~~~~~~ Dirk Zerwas$^{k}$ (Higgs)\\~\\
 \small Shehu AbdusSalam$^{l,bb}$, Claire Adam-Bourdarios$^{k}$,
 J.A.~Aguilar-Saavedra$^{m}$, Benjamin Allanach$^{l}$,
 B.~Altunkaynak$^{a}$, Luis~A. Anchordoqui$^{n}$, Howard Baer$^{o}$,
 Borut Bajc$^{p}$, O.~Buchmueller$^{q}$, M.~Carena$^{r,s}$,
 R.~Cavanaugh$^{t,u}$, S.~Chang$^{v}$, Kiwoon Choi$^{w}$,
 C.~Cs\'aki$^{x}$, S.~Dawson$^{b}$, F.~de~Campos$^{y}$,
 A.~De~Roeck$^{q,z}$, M.~D\"uhrssen$^{aa}$, O.J.P.~\'Eboli$^{ab}$,
 J.R.~Ellis$^{q}$, H.~Fl\"acher$^{q}$, H.~Goldberg$^{a}$,
 W.~Grimus$^{ac}$, U.~Haisch$^{ad}$, S.~Heinemeyer$^{ae}$,
 M.~Hirsch$^{i}$, M.~Holmes$^{a}$, Tarek Ibrahim$^{af}$,
 G.~Isidori$^{ag}$, Gordon Kane$^{d}$, K.~Kong$^{ah}$, Remi
 Lafaye$^{ai}$, G.~Landsberg$^{aj}$, L.~Lavoura$^{ak}$, Jae Sik
 Lee$^{al}$, Seung~J.~Lee$^{am}$, M.~Lisanti$^{ah}$,
 Dieter~L\"ust$^{an,ao}$, M.B.~Magro$^{ap}$, R.~Mahbubani$^{t}$,
 M.~Malinsky$^{aq}$, Fabio Maltoni$^{ar}$, S.~Morisi$^{i}$,
 M.M.~M\"uhlleitner$^{as}$, B.~Mukhopadhyaya$^{at}$,
 M.~Neubert$^{ad}$, K.A.~Olive$^{au}$, Gilad Perez$^{am}$, Pavel Fileviez
 P{\'e}rez$^{f}$, T.~Plehn$^{av}$, E.~Pont\'{o}n$^{aw}$, Werner
 Porod$^{ax}$, F.~Quevedo$^{l}$, M.~Rauch$^{as}$,
 D.~Restrepo$^{ay}$, T.G.~Rizzo$^{ah}$, J.~C.~Rom\~ao$^{ak}$,
 F.J.~Ronga$^{az}$, J.~Santiago$^{m}$, J.~Schechter$^{bb}$,
 G.~Senjanovi\'c$^{bc}$, J.~Shao$^{bb}$, M.~Spira$^{bd}$,
 S.~Stieberger$^{an}$, Zack Sullivan$^{be}$, Tim M.P.~Tait$^{bf}$, Xerxes
 Tata$^{f,bg}$, T.R.~Taylor$^{a}$, M.~Toharia$^{h}$,
 J.~Wacker$^{ah}$, C.E.M.~Wagner$^{s,bh,bi}$, Lian-Tao Wang$^{bj}$,
 G.~Weiglein$^{bk}$, D.~Zeppenfeld$^{as}$, K.~Zurek$^{d}$}

\maketitle

%Here is the list of addresses:

(a) Department of Physics, Northeastern University, Boston, MA
02115, USA

(b) Department of Physics, Brookhaven National Laboratory, Upton, NY
11973, USA

(c) Department of Physics, Texas A\&M University, College Station,
TX 77843-4242, USA

(d) Michigan Center for Theoretical Physics, Randall Lab.,
University of Michigan, Ann Arbor, MI 48109

(e) C.N. Yang Institute for Theoretical Physics, Stony Brook
University, Stony Brook, NY 11794, USA

(f) Department of Physics, University of Wisconsin, Madison, WI
53706, USA

(g) Institute for Advanced Study ,Princeton, NJ 08540

(h) Maryland Center for Fundamental Physics and Department of
Physics, University of Maryland, College Park, MD, 20742

(i) AHEP Group, Instituto de F\'{\i}sica Corpuscular --
C.S.I.C./Universitat de Val{\`e}ncia, Campus de Paterna, Aptdo
22085, E--46071 Val{\`e}ncia, Spain

(j) School of Physics and Astronomy, University of Manchester,
Manchester M13 9PL, United Kingdom

(k) LAL, Universit\'e Paris-Sud, IN2P3/CNRS, Orsay, France

(l) Department of Applied Mathematics and Theoretical Physics,
Wilberforce Road, Cambridge, CB3 0WA, United Kingdom

(m) Departamento de F\'{\i}sica Te\'orica y del Cosmos and CAFPE,
Universidad de Granada, E-18071 Granada, Spain

(n) Department of Physics, University of Wisconsin-Milwaukee,
Milwaukee, WI 53201, USA

(o) Dept. of Physics and Astronomy, University of Oklahoma, Norman,
OK, 73019, USA

(p) J. Stefan Institute, 1000 Ljubljana, Slovenia

(q) CERN, CH-1211 Gen\`eve 23, Switzerland

(r) Theoretical Physics Department, Fermilab, Batavia, IL 60510, USA

(s) EFI and Physics Department, University of Chicago 5640 S. Ellis
Ave., Chicago, IL 60637, USA

(t) Fermi National Accelerator Laboratory, P.O. Box 500, Batavia,
Illinois 60510, USA

(u) Physics Department, University of Illinois at Chicago, Chicago,
Illinois 60607-7059, USA

(v) Physics Department, University of California Davis, Davis, CA
95616

(w) Physics Department, KAIST, Daejeon, 305-701, Korea

(x) Institute for High Energy Phenomenology, Laboratory of
Elementary Particle Physics, Cornell University, Ithaca, NY 14853,
USA

(y) Departamento de F\'{\i}sica e Qu\'{\i}mica, Universidade
Estadual Paulista, Guaratinguet\'a -- SP, Brazil

(z) Antwerp University, B-2610 Wilrijk, Belgium

(aa) Physikalisches Institut, Universit\"at Freiburg, Germany

(ab) Instituto de F\'{\i}sica, Universidade de S\~ao Paulo, S\~ao
Paulo -- SP, Brazil

(ac) University of Vienna, Faculty of Physics, Boltzmanngasse 5,
A--1090 Vienna, Austria

(ad) Institut f\"ur Physik (THEP), Johannes Gutenberg-Universit\"at,
D-55099 Mainz, Germany

(ae) Instituto de F\'{\i}sica de Cantabria (CSIC-UC), E--39005
Santander, Spain

(af) Department of Physics, Faculty of Science, University of
Alexandria, Alexandria, Egypt

(ag) INFN, Laboratori Nazionali di Frascati, Via E. Fermi 40,
I--00044 Frascati, Italy

(ah) SLAC National Accelerator Laboratory 2575 Sand Hill Rd., Menlo
Park, CA, 94025, USA

(ai) LAPP, Universit\'e de Savoie, IN2P3/CNRS, Annecy, France

(aj) Department of Physics, Brown University, 182 Hope St,
Providence, RI 02912, USA

(ak) Technical University of Lisbon, Centre for Theoretical Particle
Physics, 1049-001 Lisbon, Portugal

(al) Physics Division, National Center for Theoretical Sciences,
Hsinchu, Taiwan

(am) Department of Particle Physics, Weizmann Institute of Science,
Rehovot 76100, Israel

(an) Max--Planck--Institut f\"ur Physik,
Werner--Heisenberg--Institut, 80805 M\"unchen, Germany

(ao) Arnold Sommerfeld Center for Theoretical Physics,
Ludwig-Maximilians-Universit\"at M\"unchen, 80333 M\"unchen, Germany

(ap) Centro Universit\'ario Funda\c{c}\~ao Santo Andr\'e, Santo
Andr\'e -- SP, Brazil

(aq) Theoretical Particle Physics Group, Department of Theoretical
Physics, Royal Institute of Technology (KTH), Roslagstullsbacken 21,
SE-106 91 Stockholm, Sweden

(ar) Center for Particle Physics and Phenomenology, Universit\'e
Catholique de Louvain Chemin du Cyclotron 2, B-1348,
Louvain-la-Neuve, Belgium

(as) Institut f\"ur Theoretische Physik, Universit\"at Karlsruhe,
KIT, D--76128 Karlsruhe, Germany

(at) Regional Centre for Accelerator-based Particle Physics,
Harish-Chandra Research Institute, Chhatnag Road, Jhunsi, Allahabad
- 211 019, India

(au) School of Physics and Astronomy, University of Minnesota,
Minneapolis, Minnesota 55455, USA

(av) Institut f\"ur Theoretische Physik, Universit\"at Heidelberg,
Germany

(aw) Department of Physics, Columbia University, New York, NY 10027,
USA

(ax) Institut f\"ur Theoretische Physik und Astronomie,
Universit\"at W\"urzburg, D-97074 W\"urzburg, Germany

(ay) Instituto de F\'{\i}sica, Universidad de Antioquia, A.A 1226,
Medellin, Colombia

(az) Institute for Particle Physics, ETH Z\"urich, CH-8093 Z\"urich,
Switzerland

(ba) CAFPE and Departamento de F\'{\i}sica Te\'orica y del Cosmos,
Universidad de Granada, E-18071 Granada, Spain

(bb) Department of Physics, Syracuse University, Syracuse, NY
13244-1130, USA

(bc) International Centre for Theoretical Physics, 34100 Trieste,
Italy

(bd) Paul Scherrer Institut, CH--5232 Villigen PSI, Switzerland

(be) Department of Biological, Chemical, and Physical Sciences,
Illinois Institute of Technology, 3101 S.\ Dearborn St., Chicago, IL
60616-3793, USA

(bf) Department of Physics and Astronomy, University of California,
Irvine, CA 92697, USA

(bg) Dept. of Physics and Astronomy, University of Hawaii, Honolulu,
HI , USA

(bh) KICP, University of Chicago 5640 S. Ellis Ave., Chicago, IL
60637, USA

(bi) HEP Division, Argonne National Laboratory 9700 S. Cass Ave.,
Argonne, IL 60439, USA

(bj) Department of Physics, Princeton University, Princeton, NJ.
08544, USA

(bk) IPPP, University of Durham, Durham DH1 3LE, United Kingdom

\onecolumn
\chapter*{The Hunt for New Physics at the Large Hadron Collider}

\begin{center} \begin{large} Abstract \end{large} \end{center}

%\begin{abstract}
%abs
The Large Hadron Collider presents an unprecedented opportunity to
probe  the realm of new physics in the TeV region and shed light on
some of the core unresolved issues of particle physics. These
include the nature of electroweak symmetry breaking,  the origin of
mass,  the possible constituent of cold dark matter, new sources of
CP violation needed to explain the baryon excess in the universe,
the possible existence of extra gauge groups and extra matter, and
importantly the path  Nature chooses to resolve the hierarchy
problem - is it supersymmetry or extra dimensions. Many models of
new physics beyond the standard model contain a hidden sector which
can be probed at the LHC. Additionally, the LHC will be a top
factory and accurate measurements of the properties of the top and
its rare decays will  provide  a window to  new physics. Further,
the LHC could shed light on the origin of neutralino masses if the
new physics associated with their generation lies in the TeV region.
Finally, the LHC is also a laboratory to test the hypothesis of TeV
scale strings and D brane models. An overview of these possibilities
is presented in the spirit that it will serve as a companion  to the
Technical Design Reports (TDRs) by the particle detector groups
ATLAS and CMS  to facilitate the test of the new theoretical ideas
at the LHC.  Which of these ideas stands the test of the LHC data
will govern the course of particle  physics in the subsequent
decades.
\\

%\end{abstract}

%%%%%%%%%%%%%%%%%%%%%%%%%%%%%%%%%%%%%%%%%%%%%%%
\tableofcontents
%%%%%%%%%%%%%%%%%%%%%%%%%%%%%%%%%%%%%%%%%%%%%%%

\twocolumn
%%%%%%%%%%%%%%%%%%%%%%%%%%%%%%%%%%%%%%%%%%%%%%%%%%%%%%%%%%%%%%%%%%%%%%%%%%%%%%%%%%%%%%%%%%%%%%
%%%%%%%%%%%%%%%%%%%%%%%%%%%%%%%%%%%%%%%%%%%%%%%%%%%%%%%%%%%%%%%%%%%%%%%%%%%%%%%%%%%%%%%%%%%%%%
\chapter{Introduction}
\epigraph{\Large {\em Pran Nath}}{}
The Large Hadron Collider (LHC) when fully operational will have an
 optimal center of mass   energy in proton -proton collisions of $\sqrt s= 14$ TeV
and a design luminosity of $10^{34} cm^{-2}s^{-1}$. The main
experiments at the LHC are: ALICE, ATLAS, CMS, LHCb, and TOTEM. Of
these ALICE is devote to the study of heavy ion collisions, LHCb to
the study of B physics, and TOTEM to the study of  total cross
section, elastic scattering and diffraction dissociation at the LHC.
Thus ATLAS\footnote{A Torroidal LHC ApparatuS.} and
CMS\footnote{Compact Muon Solenoid.} are the primary detectors
dedicated to  the discovery of new physics.  It is expected that
initially LHC will run at $\sqrt s=7$ TeV to collect data for
calibration, later ramping the CM energy to $\sqrt  s =10$ TeV,
and then to $\sqrt s=14$ TeV.\\

The particle physics capabilities of the ATLAS and CMS  detectors
are described  in
 their technical design reports (TDRs)~\cite{:2008zzm,CMS}
 which give an overview of their performance as the LHC begins its operation.
The purpose of the present document is to present a broad overview
of the new physics possibilities that the LHC is likely to see. Of
course, irrespective of the particular nature of new physics the end
product at the LHC would be an excess of  observed leptons, photons,
jets and missing energy in some combination. It is then necessary to
devise ways in which one may connect the observed deviations from
the Standard Model prediction to the underlying new physics.  \\

Thus the underlying theme of this report is to provide an overview
for experimentalists of the testable new physics  at the LHC.  The
main topics covered in the report are the following.
\begin{enumerate}
 \item
 Hunt for supersymmetry
 \item
 Hunt for the Higgs boson
 \item
 CP violation at the LHC
 \item
  LHC and dark matter
 \item
 Top quark physics at the LHC
 \item
$Z'$ physics at the LHC
 \item
 Visible signatures from the hidden sector at the LHC
 \item
 Probing the origin of neutrino mass at the LHC
 \item
 Hunt for extra dimensions at the LHC

 \item
 Hunt for strings  at the LHC

 \end{enumerate}

We discuss below each of these topics briefly.

\section{Hunt for supersymmetry}
Supersymmetry provides a technically natural solution to the so
called gauge hierarchy problem that arises in the non-supersymmetric
unified  theories with various mass scales.  Gauging of
supersymmetry necessarily requires gravity and the gauged
supersymmetry known as supergravity can be coupled to matter and to
Yang Mills gauge fields providing a framework for model building.
The effective potential in supergravity coupled with chiral matter
and gauge fields is not positive definite allowing for the
possibility of fine tuning the vacuum energy to be small.  Various
mechanisms exist for the spontaneous breaking of supersymmetry. They
include gravity mediation, gauge mediation and anomaly
mediation and other possible schemes which combine them.\\

With R parity  the lightest supersymmetric particle (LSP) is
absolutely stable, and thus production and decays of supersymmetric
particles with R parity necessarily involve at least a pair of LSPs.
Two of the leading candidates for the LSP are the neutralino and the
gravitino both of which are charge neutral and thus also candidates
for dark matter. The production and decay of sparticles  will thus
contain an even number of LSPs and result in significant missing
energy. There are many possible signatures available for the
discover of supersymmetry at the LHC. The details of the SUSY
signatures depend on the specific scenario of SUSY breaking, i.e.,
gravity mediation, gauge mediation or anomaly mediation or
combinations thereof. They are briefly discussed in this report.

\section{Hunt for the Higgs boson}
In the SM there is just one Higgs  doublet and thus after
spontaneous breaking of the electroweak symmetry, where $W^{\pm}$
and $Z^0$ become massive, there is only one residual neutral Higgs
boson left. However, in the MSSM one has two Higgs doublets, and
after spontaneous breaking one is left with four residual Higgs
bosons. Of these three are neutral with two CP even Higgs $h^0,H^0$,
one CP odd Higgs $A^0$, and a charged Higgs $H^{\pm}$. Within  the
MSSM  framework, for a broad class of soft breaking  with scale
O(TeV). The mass of the lightest Higgs boson is limited from above
by about 150 GeV. The Higgs boson will certainly be probed at the
LHC and the Higgs phenomenology  explored in considerable detail.

\section{CP violation at the LHC}
The Standard Model of particle interactions has two sources of CP
violation, one that enters in the Cabibbo-Kobayashi-Maskawa (CKM)
matrix and the other  that enters in the strong interaction
dynamics. These phases are constrained by the neutron electric
dipole moment (edm). However, it is known that the CP violation in
the standard model is not sufficient to generate the desired baryon
asymmetry in the universe and new sources of CP violation are
needed.  Such new sources  can arise in new physics models. Thus,
for example, softly broken  supersymmetric theories contain a large
number of new sources of CP violation which can be large and still
consistent with the experimental constraints on the edms of the
electron and of the neutron as well as with the edms of Mercury and of Thallium.\\

Large CP phases affect the Higgs sector of MSSM leading to a mixing
between the CP even and the CP odd neutral Higgs bosons. Such
mixings can lead to interesting signatures which can be observed at
the LHC. A test of new sources of CP violation can be done in
several other processes such as  in sparticle productions and decays
and in signatures including counting signatures and kinematical
signatures  as in missing energy, and tranverse  momenta of leptons
and jets. Thus the LHC is an excellent laboratory for the discovery
of new sources of CP violation some of which
may  enter in the analyses of baryogenesis.  \\

\section{LHC and dark matter}
Current estimates indicate that as much as 96\% of the physical
universe consists of objects other than the normal (atomic) form of
matter while  the remainder is constituted of  either dark energy
($\sim 73\%$) or cold dark matter ($\sim
23\%$)~\cite{Komatsu:2008hk}.
  Most  main stream approaches to physics beyond the standard model contain possible candidates for
cold dark matter. Thus, e.g., in supergravity based models with  R
parity, the LSP  is often a neutralino and thus a candidate for cold
dark matter. Similarly, in extra dimension models the lightest
Kaluza -Klein particle (LKP) could be a possible dark matter
candidate. These massive dark particles would carry  a lot of
missing energy and can be probed at the LHC. Thus the production of
dark  particles can be detected and even their masses and their
interactions measured with a significant degree of accuracy. For
instance, for the neutralino LSP theoretical estimates show that
purely from the LHC measurements with about 30 fb$^{-1}$ of LHC data
one can make predictions on the relic density with the same degree
of uncertainty  as the Wilkinson Microwave Anisotropy Probe (WMAP).
Thus from the LHC data alone one would be able to shed light on one
of the great mysteries, i.e., the composition of cold dark matter in
the Universe.

\section{Top  physics at the LHC}
LHC would also be  a  top  factory. Thus the LHC data will provide
an accurate determination of the top
 mass, its couplings and its spin correlations. Additionally the phenomenology of
 the top    can provide  a window to new physics
via study of its rare decays  and via  modifications of its
couplings from new physics at the loop level, or  from a study of
top events in associated production.

\section{$Z'$ physics at the LHC}
Another area of considerable interest is the study of additional
$Z'$ bosons. Such bosons occur in a variety of extensions of the
Standard Model, including grand unified models, strings and branes,
extra dimension models and models utilizing alternative schemes of
symmetry breaking. If such bosons exist with masses in the TeV
region they can be explored at the LHC.

\section{Visible signatures from the hidden sector at the LHC}
In a broad class of particles physics models,  including models
based on strings and branes, one has a new sector of physics, often
labeled the hidden sector (HS), which is typically a gauge singlet
under the standard model gauge group. However, communication with
the hidden sector may occur in a variety of ways including fields
which connect with the visible and the hidden sector, e.g.,  via
kinetic mixing, mass mixing
 or via higher dimensional  operators. In this circumstance signatures exist which
 can be explored at the LHC.
 Some hidden sector models also produce a $Z'$ boson  which, however,
 can be very narrow with width which could be just a fraction of a GeV. The possible
 observation of  such a narrow resonance would  be a clear indication of  a hidden
 sector and possibly of an underlying string framework.

\section{Probing the origin of neutrino mass at the LHC}
 A very interesting possibility not fully appreciated is that the LHC may also
 be helpful in shedding light on the origin of neutrino mass for which
 evidence now exists  via neutrino oscillations in
 solar and atmospheric neutrino data along  with data
from reactors and accelerators. However, the origin of neutrino mass
which is much smaller than the masses of the other elementary
particles, such as of  the electron or of  the muon,  remains a
mystery.
 If the new physics that generates such a mass lies in the TeV region,
 it could be explored at the LHC.

\section{Hunt for extra dimensions at the LHC}
Models with a large extra dimension offer an alternative to
supersymmetry for the solution to the hierarchy problem.  These
models produce a rich array of signatures which can be tested at the
LHC. They include signatures for  black holes in models with weak
scale quantum gravity, and of  Kaluza Klein excitations in models
with compactification radii of size 1/TeV with signatures detectable
at the LHC in dilepton signals in Drell-Yan processes as well as in
jet production.

\section{Hunt for strings  at the LHC}
String theory offers the possibility of unifying all the forces of
nature including gravity. Considerable progress has occurred over
the past two and a half decades in decoding the implications of this
theory at low energies. Although there is no single model yet that
can be labeled unique, there are many possibilities some of which
are discussed in this report. These relate to models based on
heterotic strings, on D branes, as well as on M theory. Recently
several works have presented model independent predictions for TeV
scale strings.      The signatures from these various possibilities
are discussed and one finds
these models  testable at the LHC.\\

 Each of the  main sections  in this report
 was organized by a convener (or conveners) who was (were)  responsible for
 synthesizing several individual contributions to that section and providing
 a summary and a brief abstract.
  {\em
 The document contains many diverse ideas and approaches
 which are often diametrically opposite: such  is the case regarding solution to
 the hierarchy problem, i.e., supersymmetry vs large extra dimensions.
 Further, even within a section, different authors  present their individual,
 often competitive approaches.  Thus the list of names on the face page simply implies
 that
 the authors contributed to one or more sections, but there is no implication that they endorse either
 the write up of the other sections, or for that matter the write ups  of other authors even within
 the same section. }

%%%%%%%%%%%%%%%%%%%%%%%%%%%%%%%%%%%%%%%%%%%%%%%%%%%%%%%%%%%%%%%%%%%%%%%%%%%%%%%%%%%%%%%%%%%%%%
%%%%%%%%%%%%%%%%%%%%%%%%%%%%%%%%%%%%%%%%%%%%%%%%%%%%%%%%%%%%%%%%%%%%%%%%%%%%%%%%%%%%%%%%%%%%%%
\chapter{Hunt for Supersymmetry at the LHC}
\setlength{\epigraphrule}{1pt}
\epigraphhead[20]{\epigraph{\large {\em S.S.~AbdusSalam, Claire
Adam-Bourdarios, B.C.~Allanach, Howard Baer, Kiwoon Choi, Remi
Lafaye, Pran Nath, Tilman Plehn, F.~Quevedo, Michael Rauch, Xerxes
Tata, Dirk Zerwas}}{\large Pran Nath (Convener)}}
%
%%%%%%%%%% espcrc2.tex %%%%%%%%%%
%
% $Id: espcrc2.tex 1.2 2000/07/24 09:12:51 spepping Exp spepping $
%
%\documentclass[fleqn,twoside]{article}
%\usepackage{espcrc2}

% change this to the following line for use with LaTeX2.09
% \documentstyle[twoside,fleqn,espcrc2]{article}

% if you want to include PostScript figures
%\usepackage{graphicx}
% if you have landscape tables
%\usepackage[figuresright]{rotating}

% put your own definitions here:
%   \newcommand{\cZ}{\cal{Z}}
%   \newtheorem{def}{Definition}[section]
%   ...
\newcommand{\beqn}{\begin{eqnarray}}
\newcommand{\eeqn}{\end{eqnarray}}
\newcommand{\be}{\begin{equation}}
\newcommand{\ee}{\end{equation}}
%%%
%   ...
%   ...
\newcommand{\bea}{\begin{eqnarray}}
\newcommand{\eea}{\end{eqnarray}}

\def \cha{\widetilde{\chi}^{\pm}_1}
\def \chb{\widetilde{\chi}^{\pm}_2}

\def \na{\widetilde{\chi}^{0}_1}
\def \nb{\widetilde{\chi}^{0}_2}
\def \nc{\widetilde{\chi}^{0}_3}
\def \nd{\widetilde{\chi}^{0}_4}

\def \g{\widetilde{g}}
\def \ql{\widetilde{q}_L}
\def \qr{\widetilde{q}_R}

\def \dl{\widetilde{d}_L}
\def \dr{\widetilde{d}_R}
\def \ul{\widetilde{u}_L}
\def \ur{\widetilde{u}_R}

\def \ccl{\widetilde{c}_L}
\def \ccr{\widetilde{c}_R}
\def \ssl{\widetilde{s}_L}
\def \ssr{\widetilde{s}_R}

\def \ta{\widetilde{t}_1}
\def \tb{\widetilde{t}_2}
\def \ba{\widetilde{b}_1}
\def \bb{\widetilde{b}_2}

\def \sta{\widetilde{\tau}_1}
\def \stb{\widetilde{\tau}_2}

\def \smr{\widetilde{\mu}_R}
\def \ser{\widetilde{e}_R}
\def \sml{\widetilde{\mu}_L}
\def \sel{\widetilde{e}_L}

\def \slr{\widetilde{l}_R}
\def \sll{\widetilde{l}_L}

\def \snl{\widetilde{\nu}_{\tau}}
\def \snm{\widetilde{\nu}_{\mu}}
\def \sne{\widetilde{\nu}_{e}}

\def \hc{H^{\pm}}

\def \lra{\longrightarrow}

% -------- end Macro definition
%%%%%%%%%%%%%%%%%%%%%%%%%%%%%%%%%%%%%%%%%%%%%%%%%%%%%%%%%%%%%%%%%%%%%%%%
% -------- begin Macro definition
\newcommand{\GeV}      {~\mathrm{GeV}}
\newcommand{\pb}      {~\mathrm{pb}}
\def\nj{$n_{jet}$}
\def\njs{$n_{jet}^*$}
\def\co{coannihilation~}
\def\pts{$P_T^{\rm * miss}$}
\def\pt{\not\!\!{P_T}}
\def\no{\nonumber\\}
\def \chan{\widetilde{\chi}}
\def \cha{\widetilde{\chi}^{\pm}_1}
\def \chb{\widetilde{\chi}^{\pm}_2}
\def \nb{\widetilde{\chi}^{0}_2}
\def \nc{\widetilde{\chi}^{0}_3}
\def \nd{\widetilde{\chi}^{0}_4}
\def \g{\tilde{g}}
\def \ql{\widetilde{q}_L}
\def \qr{\widetilde{q}_R}
\def \dl{\widetilde{d}_L}
\def \dr{\widetilde{d}_R}
\def \ul{\widetilde{u}_L}
\def \ur{\widetilde{u}_R}
\def \ccl{\widetilde{c}_L}
\def \ccr{\widetilde{c}_R}
\def \ssl{\widetilde{s}_L}
\def \ssr{\widetilde{s}_R}
\def \ta{\widetilde{t}_1}
\def \tb{\widetilde{t}_2}
\def \ba{\widetilde{b}_1}
\def \bb{\widetilde{b}_2}
\def \sta{\widetilde{\tau}_1}
\def \stb{\widetilde{\tau}_2}
\def \smr{\widetilde{\mu}_R}
\def \ser{\widetilde{e}_R}
\def \sml{\widetilde{\mu}_L}
\def \sel{\widetilde{e}_L}
\def \slr{\widetilde{l}_R}
\def \sll{\widetilde{l}_L}
\def \snl{\widetilde{\nu}_{\tau}}
\def \snm{\widetilde{\nu}_{\mu}}
\def \sne{\widetilde{\nu}_{e}}
\def \hc{H^{\pm}}
\def \lra{\longrightarrow}
\def \ETmiss{${\not\!\!{E_T}}~$}
\def \missET{${\not\!\!{E_T}}$}
\def \mh{m_{1/2}}
\def\.4{\vspace{-.5cm}}

%%%%%%%%%%%%%%%%%%%%%%%%%%%%%%%%%%%%%%%%%%%%%%%%%%%%%%%%%%%%%%%%%%%%%%%
\def \non{nonuniversalities~}
\def\a{GNLSP$_{\rm A}~$}
\def\b{GNLSP$_{\rm B}~$}
\def\c{GNLSP$_{\rm C}~$}

%%%%%
%%%%
\def\eslt{E_T^{\rm miss}}
\def\emiss{\not\!\!{E}}
\def\to{\rightarrow}
\def\Phat{\hat{\Phi}}
\def\bi{\begin{itemize}}
 \def\ei{\end{itemize}}
\def\te{\tilde e}
\def\c1p{C1^\prime}
\def\ta{\tilde a}
\def\tG{\tilde G}
\def\th{\tilde h}
\def\tH{\tilde H}
\def\tl{\tilde l}
\def\tu{\tilde u}
\def\tc{\tilde c}
\def\ta{\tilde a}
\def\ts{\tilde s}
\def\tb{\tilde b}
\def\tf{\tilde f}
\def\td{\tilde d}
\def\tQ{\tilde Q}
\def\tL{\tilde L}
\def\tH{\tilde H}
\def\tst{\tilde t}
\def\ttau{\tilde \tau}
\def\tmu{\tilde \mu}
\def\tg{\tilde g}
\def\tnu{\tilde\nu}
\def\tell{\tilde\ell}
\def\tq{\tilde q}
\def\tB{\widetilde B}
\def\tw{\tilde\chi^\pm}
\def\twp{\tilde\chi^+}
\def\twm{\tilde\chi^-}
\def\twpm{\tilde\chi^\pm}
\def\tz{\tilde\chi^0}
\def\alt{\stackrel{<}{\sim}}
\def\agt{\stackrel{>}{\sim}}
\def\be{\begin{equation}}
\def\ee{\end{equation}}
\def\bea{\begin{eqnarray}}
\def\eea{\end{eqnarray}}
\def\CM{\cal M}
\newcommand\annp[3]{{\it Annals\ Phys.\ }{\bf #1} (#2) #3}
\newcommand\sjp[3]{{\it Sov.\ J.\ Nucl.\ }{\bf #1} (#2) #3}
\newcommand\prd[3]{{\it Phys.\ Rev.\ }{\bf D #1} (#2) #3}
\newcommand\prep[3]{{\it Phys.\ Rept.\ }{\bf #1} (#2) #3}
\newcommand\prl[3]{{\it Phys.\ Rev.\ Lett.\ }{\bf #1} (#2) #3}
\newcommand\plb[3]{{\it Phys.\ Lett.\ }{\bf B #1} (#2) #3}
\newcommand\jhep[3]{{\it J. High Energy Phys.\ }{\bf #1} (#2) #3}
\newcommand\app[3]{{\it Astropart.\ Phys.\ }{\bf #1} (#2) #3}
\newcommand\apj[3]{{\it Astrophys.\ J. }{\bf #1} (#2) #3}
\newcommand\ijmpd[3]{{\it Int.\ J.\ Mod.\ Phys.\ }{\bf D #1} (#2) #3}
\newcommand\npb[3]{{\it Nucl.\ Phys.\ }{\bf B #1} (#2) #3}
\newcommand\epjc[3]{{\it Eur.\ Phys.\ J. }{\bf C #1} (#2) #3}
\newcommand\ptp[3]{{\it Prog.\ Theor.\ Phys.\ }{\bf #1} (#2) #3}
\newcommand\zpc[3]{{\it Z.\ Physik }{\bf C #1} (#2) #3}
\newcommand\cpc[3]{{\it Comput.\ Phys.\ Commun.}{\bf #1} (#2) #3}
\newcommand\mpla[3]{{\it Mod.\ Phys.\ Lett.}{\bf A #1} (#2) #3}
\newcommand\arnps[3]{{\it Ann.\ Rev.\ Nucl.\ Part.\ Sci.}{\bf  #1} (#2) #3}
\newcommand\njp[3]{{\it New\ Jou.\ Phys.}{\bf  #1} (#2) #3}
\newcommand\ppnp[3]{{\it Prog.\ Part.\ Nucl.\ Phys.}{\bf  #1} (#2) #3}
\newcommand\jphg[3]{{\it J. Phys.\ }{\bf G #1} (#2) #3}

Supersymmetry is one of the leading candidates for discovery at the
LHC. However, the fact that SUSY partners degenerate with known
particles have not been observed requires that supersymmetry must be
softly broken in a phenomenologically consistent manner. Many
schemes accomplish this prominent among them are the SUGRA grand
unified models  with gravity mediated breaking,  models based  on
gauge and anomaly mediation and a variety of models  using
admixtures of the above. In this section we give a brief discussion
of some of these topics. We list  signatures for weak scale
supersymmetry (SUSY) which may be expected at the LHC. From each
signature, we provide a description of why the signature might
occur, and possible SUSY models which give rise to each specific
SUSY signature channel. If new physics is to be discovered at the
LHC, the next step would be to reconstruct the underlying theory,
and this endeavor should not be biased by any assumption on
high-scale models. SFitter and its weighted Markov chain technique
is a tool of choice to perform such a task. Using the example of the
TeV-scale MSSM Lagrangian we illustrate in detail how it will be
possible to analyze
 this high dimensional physics parameter spaces and extrapolate parameters to the high scale, to test unification.
Next in a bottom-up approach, we present global fit results of a
phenomenological parametrization of the weak-scale minimal
supersymmetric standard model (MSSM) with 25 relevant parameters
known as the phenomenological MSSM. Finally, we discuss the recently
proposed $M_{T2}$-kink method to measure the sparticle  masses  in
hadron collider events with missing energy. Here
 a new kinematic variable, the
$M_{T2}$-Assisted-On-Shell (MAOS) momentum, is introduced which can
be useful for spin measurement of new particles produced at the LHC.

\section{Hunt for SUSY}
\noindent
{\it Pran Nath}\\

Supersymmetry initially postulated in two~\cite{pr} and then
extended to four dimensions~\cite{golf,wz}
 possesses the remarkable  property of the so called non renormalization theorem~\cite{grs}.
Models based on supersymmetry provide a technically natural
solution~\cite{georgi} to the so called gauge hierarchy problem that
arises in the non-supersymmetric unified  theories
 with various mass scales.
 The main problem
in building models based on supersymmetry centers around the issue
of how to break supersymmetry.  One could add to the Lagrangian
arbitrary amounts of soft breaking~\cite{grisaru}. However, the
number of such possibilities is enormous. Thus it is desirable to
generate  a spontaneous breaking of supersymmetry, which however,
turns out to be  difficult  to achieve in a phenomenologically
viable manner within global supersymmetry. Gauging of supersymmetry
necessarily brings  in gravity~\cite{Nath:1975nj},
  leading to a natural fusion of supersymmetry
and gravity in supergravity~\cite{Freedman:1976xh}. To build models
based on supergravity one needs to couple an arbitrary number of
chiral fields and  gauge fields  in the adjoint representation of
the gauge group~\cite{Cham1982,Cremmer:1982wb,Nath:1983fp}. Such
constructions depend on three  arbitrary functions, the
superpotential $W(\phi_i)$ which a holomorphic function of the
chiral fields $\phi_i$,  a K\"ahler potential for the chiral scalar
fields $K(\phi_i, \phi_i^{\dagger})$, and the gauge kinetic energy
function. One remarkable result  of this construction which may be
appropriately called  applied supergravity is that the scalar
potential is not positive definite. This allows  one to fine tune
the vacuum energy to an arbitrary small value after the breaking of
supersymmetry and thus allows one to build phenomenologically viable
models based on supersymmetry. Thus the first viable models were
build incorporating these features using what is now called gravity
mediation~\cite{Cham1982,Barb1982,Hall:1983iz,Nath:1983aw}.
 In gravity  mediation supersymmetry
is broken in the hidden sector and communicated via gravity to the
visible sector by gravity- generated soft masses. Supergravity grand
unified models~\cite{Cham1982,Hall:1983iz,Nath:1983aw} also exhibit
the further remarkable phenomenon that the soft parameters are
independent of the grand unification scale. Supersymmetry breaking
in this class of models is governed by the ratio
$m_s=m^2/M_{Planck}$ where $m$ is a mass scale that enters the
hidden sector and $M_{Planck} =(8\pi G_N)^{-1/2}=2.4\times 10^{18}
GeV$ and thus $m=10^{10-11}$ GeV corresponds to a soft mass of
$m_s\sim 10^{3}$ GeV. Such size  scales could arise in supergravity,
e.g., via   gaugino condensation~\cite{nilles1}. However, the actual
implementation of such a mechanism is rather intricate since it is
non-perturbative. One important modification in gaugino condensation
is that the soft masses will be typically of size $m_s \sim <\lambda
\lambda>/M_{Pl}^2$ and thus $m_s \sim 10^{3}$ requires a
condensation scale of around $10^{13}$ GeV.

In minimal supergravity the parameters at the  GUT scale consist of
$m_0, m_{1/2}, A_0, B_0$ and $\mu_0$ where $m_0$ is the universal
scalar mass, $m_{1/2}$ is the universal gaugino mass, $A_0$ is the
universal trilinear coupling, $B_0$ is the universal bilinear
coupling and $\mu_0$ is the Higgs mixing parameter of the two Higgs
doublets, $H_2$ and $H_1$ which give masses to the up quark and to
the down quark and the lepton. The  parameter $\mu_0$  arises in
supergravity in a natural way and is  typically of the size of soft
breaking~\cite{Giudice:1988yz}.
 There exist now several other mechanisms for the breaking of supersymmetry  the chief among these are gauge mediation~\cite{Dine:1995ag,Giudice:1998bp}
and anomaly mediation~\cite{Randall:1998uk,Giudice:1998xp}. Several
other mediation mechanisms have also been discussed in the
literature. The phenomenology of the supersymmetric models with soft
breaking have been discussed extensively in the literature and some
recent reviews can be found
in~\cite{Nilles:1983ge,Haber:1984rc},\cite{Martin:1997ns},\cite{Nath:2003zs,Baer:2006rs,Drees:2004jm}.
\begin{table}
%[h]
    \begin{center}
\begin{tabular}{|l||l|c|}
\hline\hline mSP&     Mass Pattern & $\mu$
\\\hline\hline
mSP1    &   $\na$   $<$ $\cha$  $<$ $\nb$   $<$ $\nc$   &
$\mu_{\pm}$    \cr mSP2    &   $\na$   $<$ $\cha$  $<$ $\nb$   $<$
$A/H$  & $\mu_{\pm}$    \cr mSP3    &   $\na$   $<$ $\cha$  $<$
$\nb$ $<$ $\sta$    & $\mu_{\pm}$    \cr mSP4    &   $\na$   $<$
$\cha$ $<$ $\nb$   $<$ $\g$    & $\mu_{\pm}$    \cr \hline mSP5    &
$\na$ $<$ $\sta$  $<$ $\slr$  $<$ $\snl$      & $\mu_{\pm}$    \cr
mSP6 &   $\na$   $<$ $\sta$  $<$ $\cha$  $<$ $\nb$     & $\mu_{\pm}$
\cr mSP7    &   $\na$   $<$ $\sta$  $<$ $\slr$  $<$ $\cha$  &
$\mu_{\pm}$    \cr mSP8    &   $\na$ $<$ $\sta$  $<$ $A\sim H$
& $\mu_{\pm}$    \cr mSP9    &   $\na$   $<$ $\sta$  $<$ $\slr$ $<$
$A/H$    & $\mu_{\pm}$    \cr mSP10   &   $\na$   $<$ $\sta$ $<$
$\tilde t_1$ $<$ $\slr$     & $\mu_{+}$    \cr
 \hline
mSP11   &   $\na$ $<$ $\tilde t_1$ $<$ $\cha$  $<$ $\nb$      &
$\mu_{\pm}$    \cr mSP12 &   $\na$ $<$$\tilde t_1$$<$ $\sta$ $<$
$\cha$   & $\mu_{\pm}$    \cr mSP13   & $\na$   $<$ $\tilde t_1$ $<$
$\sta$  $<$ $\slr$      & $\mu_{\pm}$    \cr \hline mSP14   &
$\na$   $<$  $A\sim H$ $<$ $\hc$       & $\mu_{+}$    \cr mSP15   &
$\na$   $<$ $ A\sim H$ $<$ $\cha$    & $\mu_{+}$    \cr mSP16   &
$\na$   $<$ $A\sim H$ $<$ $\sta$         & $\mu_{+}$    \cr
\hline\hline
 \end{tabular}
\caption{\small The Sparticle Landscape of mass hierarchies in
mSUGRA.
 In patterns mSP14,15,16  the LSP
$\tilde \chi_1^0$  and the Higgs bosons $(A,H)$ can switch  their
order. (From  Refs.(1,3) of~\cite{Feldman:2007zn}.) }
\label{msptable}
\end{center}
 \end{table}

\begin{figure*}[ht]
\centering
\includegraphics[width=6.0cm,height=5.0 cm]{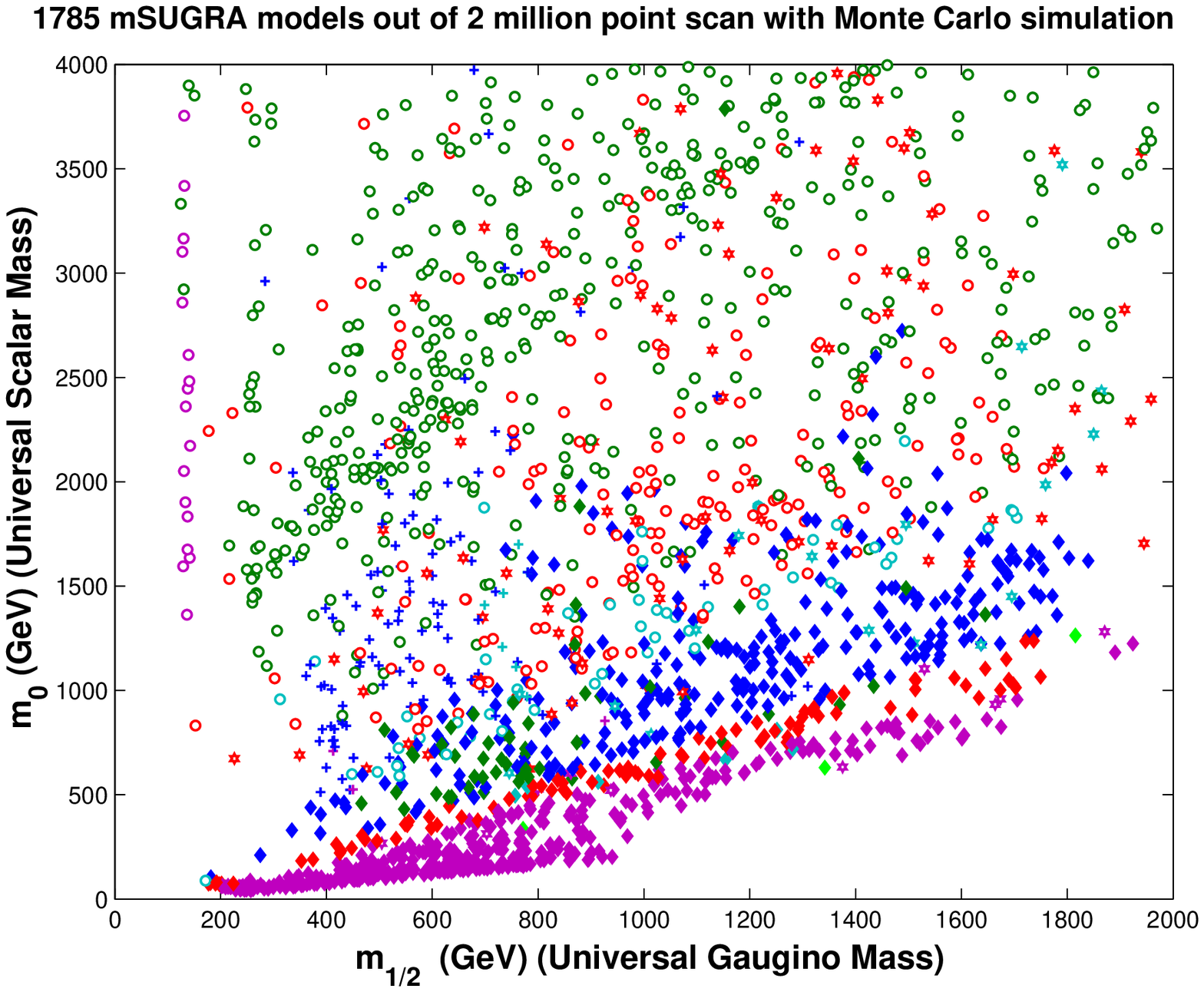}
\includegraphics[width=6.0cm,height=5.0 cm]{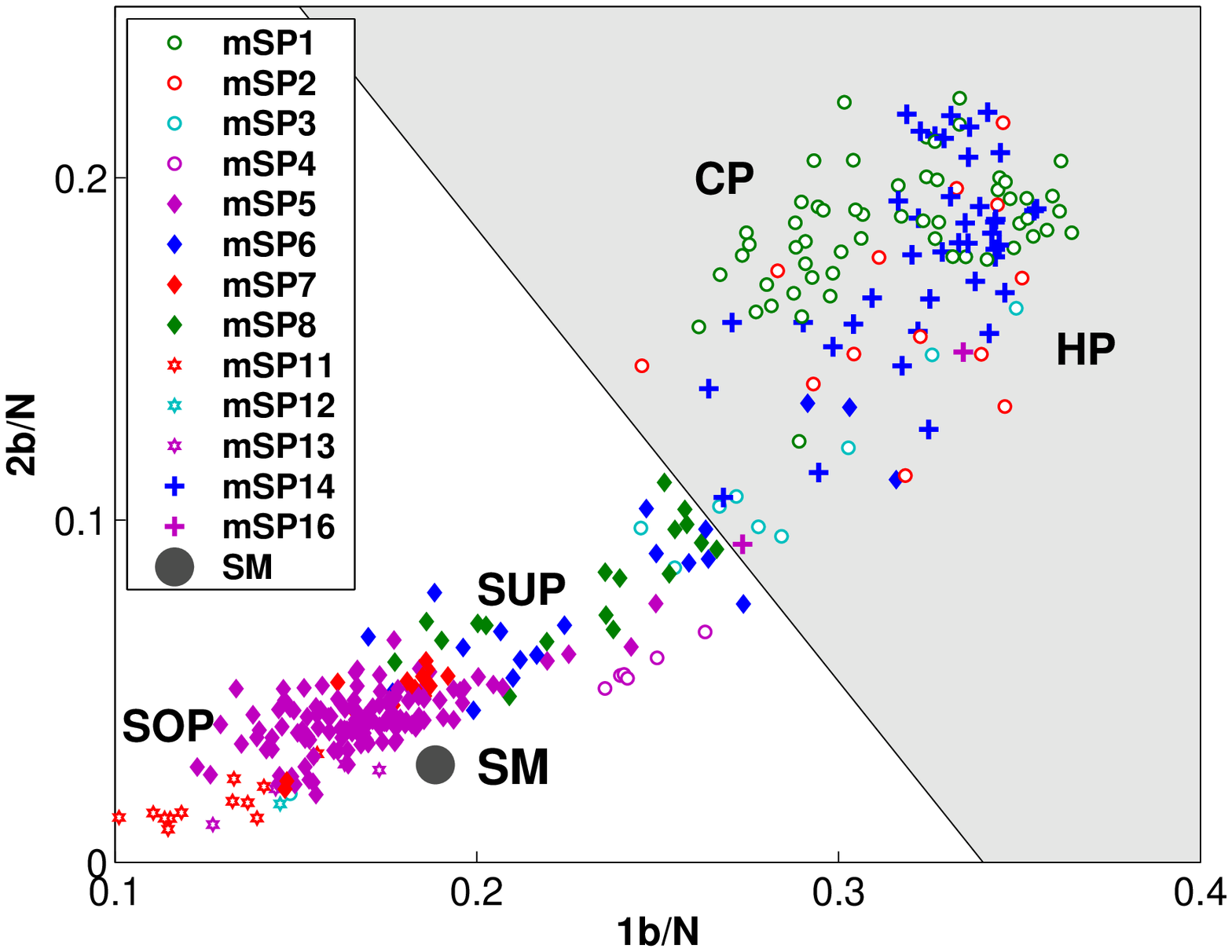}
\caption{\small Left panel: The allowed parameter space in the
$m_0-m_{\frac{1}{2}}$ plane in the mSUGRA model when all relevant
constraints are imposed. Left panel: Simulation with 10fb$^{-1}$  of the fraction
$2b/N$ vs the fraction $1b/N$ which exhibits a  wide
dispersion among patterns and separates the signal from the
background. From Refs.(1,2,3) of~\cite{Feldman:2007zn}. . }
\label{bigpic1}
\end{figure*}

One feature which is generic to a variety of schemes is the breaking
of the electroweak symmetry by radiative effects~\cite{inoue}. We
focus here on the radiative breaking in the context of supergravity
models but it can be appropriately adopted for other breaking
schemes as well. In SUGRA models one evolves the physical quantities
such as gauge couplings, Yukawa couplings and  sparticle masses from
the GUT scale to low energies by renormalization
group~\cite{marie,martin}.
 The renormalization group effects
then trigger electroweak symmetry breaking reducing $SU(2)\times
U(1)_Y$ to $U(1)_{em}$. For the case of the minimal supergravity
unified model the magnitude of $\mu$ at the electroweak scale can be
determined by using one of the radiative electroweak symmetry
breaking conditions, while the parameter $B_0$ can be eliminated in
favor of $\tan\beta =<H_2>/<H_1>$. Thus after the breaking of the
electroweak symmetry the parameter space of the minimal supergravity
model,  mSUGRA ,  consists of 4 parameters and the sign of $\mu$,
i.e., the parameters~\cite{earlypheno} \beqn m_0, m_{\frac{1}{2}},
A_0, \tan\beta, {\rm sign}(\mu). \label{mp} \eeqn One important
consequence of the supergravity unification is that it leads to a
unification of gauge couplings\cite{drw}  consistent with the LEP
data~\cite{couplings}. Further, the sparticle spectrum can be
computed by the renormalization group
evolution~\cite{Arnowitt:1992aq,rr,bbo,kkrw,Baer:1994zt}
 in terms of the parameters of Eq.(\ref{mp}).

%\section{Tables}
\begin{table}
%[h]
    \begin{center}
\begin{tabular}{|l||l|c|}
\hline\hline mSP&     Mass Pattern & $\mu$
\\\hline\hline
mSP17   &   $\na$   $<$ $\sta$ $<$ $\nb$ $<$ $\cha$     & $\mu_{-}$
\cr mSP18   &  $\na$   $<$ $\sta$  $<$ $\slr$  $<$$\tilde t_1$&
$\mu_{-}$    \cr mSP19   &  $\na$ $<$ $\sta$ $<$ $\tilde t_1$   $<$
$\cha$    & $\mu_{-}$    \cr
 \hline
mSP20  & $\na$ $<$ $\tilde t_1$   $<$ $\nb$   $<$ $\cha$    &
$\mu_{-}$    \cr mSP21   & $\na$   $<$$\tilde t_1$$<$ $\sta$  $<$
$\nb$      & $\mu_{-}$    \cr \hline mSP22   & $\na$   $<$ $\nb$
$<$ $\cha$  $<$ $\g$   & $\mu_{-}$    \cr \hline\hline
 \end{tabular}
\caption{\small The Sparticle Landscape of mass hierarchies in
mSUGRA.
 In patterns mSP14,15,16  the LSP
$\tilde \chi_1^0$  and the Higgs bosons $(A,H)$ can switch  their
order. (From  Refs.(1,3) of~\cite{Feldman:2007zn}.) }
\label{msptable}
\end{center}
 \end{table}

\begin{table}
%[h]
    \begin{center}
\begin{tabular}{|l|l|c|c|}
\hline\hline NUSP &   Mass Pattern  &     Model\\\hline\hline NUSP1
&   $\na$   $<$ $\cha$  $<$ $\nb$   $<$$\tilde t_1$ & NU3,NUG  \cr
NUSP2   &   $\na$   $<$ $\cha$  $<$ $A\sim H$               & NU3
\cr NUSP3   &   $\na$   $<$ $\cha$  $<$ $\sta$  $<$ $\nb$       &
NUG \cr NUSP4   &   $\na$   $<$ $\cha$  $<$ $\sta$  $<$ $\slr$     &
NUG \cr \hline NUSP5   &   $\na$   $<$ $\sta$  $<$ $\snl$  $<$
$\stb$       & NU3     \cr NUSP6  &   $\na$   $<$ $\sta$ $<$ $\snl$
$<$ $\cha$        & NU3  \cr NUSP7   &   $\na$ $<$ $\sta$
$<$$\tilde t_1$$<$ $A/H$         & NUG \cr NUSP8   &   $\na$   $<$
$\sta$  $<$ $\slr$  $<$ $\snm$      & NUG   \cr NUSP9   &   $\na$
$<$ $\sta$  $<$ $\cha$  $<$ $\slr$       & NUG  \cr \hline NUSP10  &
$\na$   $<$ $\tilde t_1$   $<$ $\g$    $<$ $\cha$         & NUG  \cr
NUSP11  &   $\na$   $<$ $\tilde t_1$   $<$ $A\sim H$
& NUG  \cr \hline NUSP12  &   $\na$   $<$ $A\sim H$   $<$ $\g$
& NUG  \cr \hline NUSP13  &   $\na$   $<$ $\g$    $<$ $\cha$ $<$
$\nb$         & NUG  \cr NUSP14  &   $\na$   $<$ $\g$    $<$$\tilde
t_1$ $<$ $\cha$            & NUG  \cr NUSP15  &   $\na$   $<$ $\g$
$<$ $A\sim H$               & NUG  \cr \hline\hline
\end{tabular}
%}
\end{center}
\label{patternstable2} \caption{ New sparticle mass hierarchies
above and beyond those in the minimal framework in NUSUGRA where NUG
corresponds to non-universalities in the gaugino sector and NU3
corresponds to non-universalities in the third generation sector.
(From  Refs.(2,3) of~\cite{Feldman:2007zn}.)}
 \end{table}

We note that the nature of physics at the Planck scale is not fully
known and thus deformations from universality should be considered.
This is what is done in non-universal supergravity models, where one
considers modifications of universality consistent with flavor
changing neutral currents. The above possibility allows the
following sets of allowed non-universalities: (i) non-universalities
in the Higgs sector (NUH), non-universalities in the third
generation sector (N3q),
 and (iii)  nonuniversalities in the gaugino sector (NUG).

\subsection{Hyperbolic Branch / Focus Point (HB/FP)} The radiative breaking of the electroweak symmetry
exhibits two important branches. One of the these is the
conventional branch where the soft parameters lie on the surface of
an ellipsoid. For a given amount of fine tuning the soft parameters
can move around on the ellipsoid surface  but  cannot get very large
for fixed radii. However, there is another branch the Hyperbolic
Branch (HB) where for certain regions of the parameter space the
ellipsoid turns into a hyperboloid (see the first paper
of~\cite{Chan:1997bi}). On the hyperbolic branch the scalar masses
can get very large (5-10 TeV or even larger)
 consistent with a small fine tuning and other experimental constraints. This region is also often
 called the Focus Point region (see the second paper of~\cite{Chan:1997bi}).

After the breaking of the electroweak symmetry one generates  the
masses of all the 32 sparticles in terms of a small number of  soft
parameters, and because of this  small number many sum rules on
sparticle masses result~\cite{ramond}. The allowed parameter space
is limited by a variety of constraints such as color and charge
conservation as well as  experimental lower limits on sparticle
masses from LEP and from the Tevatron. Further, there are
constraints arising from the Brookhaven experiment on
$g_{\mu}-2$\cite{bnl}, and from the flavor changing processes $b\to
s \gamma$, and $B^0_{s}\to \mu^+\mu^-$. Regarding $g_{\mu}-2$, it in
known that the supersymmetric contribution to the anomalous magnetic
moment of the muon can be as large or larger than the standard model
electroweak contribution~\cite{yuan}.   The most recent analysis of
$a_{\mu}=(g_{\mu}-2)/2$ gives for $\delta a_{\mu}=
a_{\mu}^{exp}-a_{\mu}^{SM}$ the value~\cite{Davier:2009zi}
 \beqn
 \delta a_{\mu}=(24.6\pm 8.0)\times 10^{-10}
 \label{bnl2009}
 \eeqn
  which is a $3.1\sigma$ deviation from the Standard Model.  This result is
  similar to the Brookhaven 2001 result which was about $2.6\sigma$ deviation
  and led to the prediction that there should be upper limits on the sparticle
  masses\cite{fm}. Thus if the result of Eq.(\ref{bnl2009}) holds  up, it would imply
  that sparticles  must be observed at the LHC.

  Regarding the FCNC decay $b\to s\gamma$ it arises only at loop
  level and the supersymmetric contributions are typically comparable to the
  Standard Model contributions.
   Consequently the difference between the experimental
  value and the Standard Model value acts as a strong constraint on new physics
  (For theoretical analyses of this decay in supersymmetry see~\cite{bert}).
  The most recent evaluations of the Standard Model result including the next to next
  leading order contributions to this process  give at $O(\alpha^2_s)$~\cite{Misiak:2006zs}
  the result
  \beqn
  {\mathcal
{{\mathcal BR}}}(b\rightarrow s\gamma) =(3.15\pm 0.23) \times
10^{-4}. \label{bsgsm} \eeqn The above is to be compared with the
experimental central value given by The Heavy Flavor Averaging Group
(HFAG)~\cite{Barberio:2008fa} along with the BABAR, Belle and CLEO
experimental results: ${\mathcal Br}(B \to X_s \gamma) =(352\pm
23\pm 9) \times 10^{-6}$. The difference between the experiment and
the Standard Model result acts as a strong constraint on new
physics. The current discrepancy between theory and experiment
requires about a $1.5\sigma$ correction from supersymmetry which
points to the possibility of relatively light charged  Higgs,
charginos, and stops~\cite{Chen:2009cw}.

Additionally, if one assumes that R parity is conserved, which is
what is assumed in a large class of models discussed in the
literature, then this results in the lightest sparticle (LSP) being
absolutely stable.  If the LSP is neutral it is a possible candidate
for dark matter. In SUGRA models over most of the allowed parameter
space the neutralino turns out to be the LSP~\cite{Arnowitt:1992aq}
and thus a candidate for dark matter~\cite{hg}\cite{goodman}. More recently other
possibilities have also been considered as discussed in the section
below.

As mentioned above there are 32 sparticles in the MSSM which after
breaking of supersymmetry and after electroweak supersymmetry
breaking acquire masses. These masses arrange themselves in a
hierarchical pattern and as many as $10^{25-28}$ possibilities  may
arise (depending on additional constraints imposed) leading to a
vast landscape of sparticle mass hierarchies. It is interesting to
ask how this  landscape shrinks within a specific model of soft
breaking. The result for the general case of 32 sparticle tower is
currently unknown although partial results were given for the mSUGRA
case in Ref.~\cite{Feldman:2008hs} However,  if one limits oneself
to the first four lightest sparticles aside from the LSP and the
lightest Higgs boson, then there are  only 22 such possibilities in
mSUGRA  for both signs of $\mu$ which are labeled  as the minimal
supergravity patterns mSP1-mSP22~\cite{Feldman:2008hs}.
 These are
exhibited in Table~\ref{msptable}. Here mSP1-mSP4 are the ones where
the next to the lightest particle (NLSP) is the chargino and they
can be labeled  Chargino Patterns (CP), mSP5-mSP10, mSP17-mSP19 are
the ones where NLSP is the stau and  they can be labeled  Stau
Patterns (SUP), mSP11-mSP13, mSP20-mSP21 are the ones where NLSP is
the stop and hence they can be labeled  Stop Patterns (SOP),
mSP14-mSP16 are the ones where NLSP is either the CP odd  Higgs  $A$
or the heavy CP even Higgs $H^0$ and they can be labeled  Higgs
Patterns (SUP), and finally we have mSP22 where the second
neutralino is the LSP and it can be labeled a Neutralino Pattern
(NEP). The $\mu$ sign for which these patterns can be realized is
listed in the last column of Table~\ref{msptable}. As may  be seen
from this table most of the patterns appear for both signs of $\mu$
while a small number appears only for one sign of $\mu$.

As mentioned already the nature of Planck scale physics is not fully
understood and  thus it is useful to consider inclusion of
non-universalities in the
analysis~\cite{Ellis:1985jn,Nath97qm,nonuni2,Baer:2005bu,Anderson:1999uia,choinilles,aghkn}.
A similar analysis but including non-universalities is given in
Table \ref{patternstable2} where the last column indicates the type
of non-universality~\cite{Feldman:2008hs} Here the patterns
corresponding to the lightest four particles  are labeled as
non-universal SUGRA models NUSP and they range  from NUSP1-NUSP15.
One interesting new feature is that the gluino can be an NLSP.

   Signatures of supersymmetry at colliders have been discussed in many works.
   Some early work on signatures and search for supersymmetry can be found in~\cite{earlypheno,bht,Nath:1987sw} and
     an early review on the search for supersymmetric particles in hadron-hadron collisions
is~\cite{Dawson:1983fw} and a more recent review is given
in~\cite{Feng:2009te}.
 Many interesting questions arise regarding such searches,
e.g., how one distinguishes SUSY from extra
dimensions~\cite{Datta:2005vx,Smillie:2005ar}, how one can
extrapolate back from the LHC data to hopefully a unique point  in
the parameter space of a new physics
model~\cite{ArkaniHamed:2005px,Altunkaynak:2008ry,Balazs:2009it},
how well one can measure sparticle
masses~\cite{Arnowitt:2008bz,Arnowitt:2007nt,Nojiri:2008hy} and what
one  may  learn from the early runs at the
LHC~\cite{Hubisz:2008gg,Baer:2008kc,Baer:2008ey,early2,Edsjo:2009rr}.

An illustration of how an appropriate combination of signatures  can
discriminate among models is given in Fig.(\ref{bigpic1}). The left
panel of Fig.(\ref{bigpic1}) exhibits the allowed parameter space of
the mSUGRA model (used here as an illustration)  in the $m_0-
m_{\frac{1}{2}}$ plane under the constraints of radiative breaking,
naturalness assumptions, and under WMAP and other experimental
constraints.  Using this parameter space the  right panel of
Fig.(\ref{bigpic1}) exhibits the discrimination of the Chargino,
Higgs, Stau, and Stop Patterns in the signature space of the fraction $2b/N$ 
 vs the fraction $1b/N$ with 10 fb$^{-1}$ of LHC data
at $\sqrt s=14$ TeV. The analysis shows the sparticle patterns can
be easily discriminated from the Standard Model background and
further they can also be discriminated from each other in most
cases. Of course a full discrimination among models  would require a
combination of many signatures. A more complete list of such
signatures can be found in~\cite{ArkaniHamed:2005px,Feldman:2008hs}
and a more detailed discussion of sparticle signatures is given
below.

\section{A Brief Catalogue of SUSY Signatures at the LHC}
\noindent
{\it Howard Baer and Xerxes Tata}\\

We list  signatures for weak scale supersymmetry (SUSY) which may be
expected at the LHC. From each signature, we provide a description
of why the signature might occur, and possible SUSY models
which give rise to each specific SUSY signature channel.\\

Particle physics models that include weak scale supersymmetry
(supersymmetric matter  at the weak scale: $M_{\rm weak}\sim 250$
GeV) are highly motivated by both theory and
experiment~\cite{Baer:2006rs,Drees:2004jm}. A generic prediction of
such models is the existence of new matter states -- the
superpartners of ordinary matter -- with the same gauge quantum
numbers as ordinary matter, but spins differing by 1/2, and masses
in the $10^2-10^4$ GeV range. The CERN Large Hadron Collider (LHC)
is a proton-proton collider which is expected to begin operating in
November, 2009, with the start-up energy of $\sqrt{s}\simeq 7$ TeV
increasing to 8-10~TeV in 2010, with the ultimate goal of running at
its design energy of 14~TeV. With such high energies, production
cross sections for TeV-scale new matter states with SM gauge
interactions, such as the SUSY superpartners or heavy Higgs bosons,
should be at an observable level.

In SUSY models with a conserved $R$-parity, heavy sparticles
produced at LHC decay to lighter sparticles plus Standard Model (SM)
particles until this cascade terminates in the lightest SUSY
particle (LSP) which is stable. This is also the case in $R$-parity
violating models if these $R$-violating couplings are small compared
with gauge couplings, except that then the would-be-stable LSP also
decays into SM particles.  By including sparticle production
reactions, sparticle decay channels, initial and final state QCD
radiation, hadronization, and beam remnant modeling, one can predict
using event generator programs the sorts of collider events expected
from SUSY, along with various SM background processes.

How superpartners acquire SUSY-breaking masses and couplings is
unknown, and a generic parametrization requires 178
parameters~\cite{Baer:2006rs}, making phenomenology intractable.
Various economic models, with mass patterns and corresponding
characteristic collider signatures have been constructed. Here, we
catalogue a wide variety of LHC SUSY signatures together with
associated SM background sources, and list the SUSY models from
which they might arise.

SUSY models divide into  three main classes characterized by the
SUSY breaking mediation mechanism:
\begin{itemize}
\item Models with gravity-mediated SUSY breaking (SUGRA), where
supergravity is broken by a $vev$ $F\sim 10^{11}$~GeV in a ``hidden
sector'' resulting in a massive gravitino. The gravitino mass sets
the overall mass scale for the superpartners, and is expected to be
at or around the
TeV-scale~\cite{Cremmer:1982wb,Barb1982,Hall:1983iz,Ohta:1982wn}.
  Three well-motivated LSP
candidates include: 1. the lightest neutralino $\tz_1$ (a WIMP dark
matter candidate) 2. the gravitino itself~\cite{gravlsp} (although
constraints from gravitino overproduction and Big Bang
Nucleosynthesis must be respected) and 3. if the Peccei-Quinn
solution to the strong $CP$ problem is invoked, the
axino~\cite{rtw,ckkr} (here, dark matter might then consist of an
axion/axino admixture~\cite{bbs}). Active sneutrinos are disfavored,
while gauge singlet sneutrinos are another possibility~\cite{moroi}.

\item Gauge-mediated SUSY breaking models~\cite{Dine:1995ag,Giudice:1998bp}
(GMSB) contain a hidden sector which interacts with a messenger
sector, and where the messenger sector experiences SM gauge forces.
If messengers are relatively light, the SUSY breaking scale can be
low, and the gravitino mass ($\sim F/M_P$) can be of order eV-GeV,
in which case it is the LSP.

\item Anomaly-mediated SUSY breaking models~\cite{Randall:1998uk,Giudice:1998xp}
(AMSB) include a hidden sector  geometrically separated from the
visible sector in extra dimensions, suppressing the tree level
contribution to SM superpartner masses and the loop level SUSY
breaking Weyl anomaly contribution dominates. The gravitino is
expected to be 1-2 orders of magnitude heavier than the TeV scale.
AMSB needs to be augmented by an additional source of SUSY breaking
to avoid a tachyonic slepton. A wino-like neutralino is usually the
LSP.
\end{itemize}
Combinations of SUSY-breaking mediation mechanisms~\cite{comb} that
can lead to very interesting phenomenology~\cite{phen} are also
possible.

\subsection{Catalogue of SUSY signatures}
%{bt}
%The classic collider signature for SUSY particle production with
%$R$-parity conserving decays is the multi-jet $+$ large $\eslt$
%channel. In SUGRA, GMSB and AMSB models, one expects at the LHC large
Unless gluinos and squarks are very heavy, one expects copious
gluino and/or squark production at the LHC \cite{hl}.
%$pp\to \tg\tg,\ \tg\tq,\ \tq\tq +X$ production cross sections\cite{hl},
%where $X$ stands for assorted hadronic debris, $\tg$ is the gluino and
%$\tq$ stands for any of the possible squark or anti-squark types.
%Gluinos and squarks then cascade decay\cite{cascade}.
Gluinos can decay either via two-body modes $\tg\to q\tq$ or three
body modes $\tg\to q\bar{q}\tz_i$ or $\tg\to q\bar{q}'\tw_j$.
Squarks almost always decay via the two body modes: $\tq\to q\tg,\
q\tz_i$ or, for left-squarks, also via $q'\tw_j$. In special cases,
loop-level decays of sparticles may also be
important~\cite{loop,z2z1g}. Gluino/squark production generically
leads to multi-jet plus multilepton (from decays of daughter $\tw_i$
and $\tz_j$) with, in $R$-parity conserving models, also large
$\eslt$ from the undetected LSPs and sometimes also from neutrinos.
Recently, correlations between sparticle mass patterns and ensuing
signatures have been examined~\cite{Feldman:2007zn,Feldman:2008hs}.
% \cite{fln}.

%For a recent study of the correlation between sparticle mass patterns
%and the ensuing signatures, see Ref.\cite

%A recent catalogue of SUSY signatures appears in Ref. \cite{fln}.

\subsection{Events with missing $E_T$}

\subsubsection{Jets $+\eslt$ with charged lepton veto}
This is the classic SUSY signature in all $R$-parity conserving
models.  The dominant background comes from QCD multi-jet
production, where $\eslt$ arises from missed jets, or hadronic
energy mis-measurement. This background is detector-dependent.
Important physics backgrounds come from $Z+jets$ production where
$Z\to\nu\bar{\nu}$, $W+jets$ production where $W\to\ell\nu_\ell$
($\ell= e,\mu, \tau$), and the lepton is mis-measured, soft or
non-isolated~\cite{bcpt1} and $t\bar{t}$ production where again the
leptons from the decay are mismeasured or soft or not isolated.
Numerous other SM $2\to n$ hard scattering backgrounds exist,
usually at lower rates.  The hard $\eslt$ and $E_T(jet)$ spectrum
coming from the heavy SUY particles usually allows for signal to be
distinguished from BG in that signal has a much harder distribution
in $\eslt$, $E_T(jets)$, $H_T\equiv\sum E_T(jets)$ or $M_{\rm
eff}\equiv \eslt +H_T$.
%It is usually
%difficult to distinguish mass edges from squark decays, since the LSP is
%assumed to give rise to missing energy. In the case of $\tg\to
%q\bar{q}\tz_i$, then a kinematic mass edge is expected at
%$m(q\bar{q})=m_{\tg}-m_{\tz_i}$. However, picking out the appropriate
%di-jet pair can be problematic in most SUSY cases.

\subsubsection{$1\ell +$jets$+\eslt$}

In most models, cascade decays of gluinos and squarks to $W$s or
$\tw_j$s, with $W\to\ell\nu_\ell$ or $\tw_j\to \ell\nu_\ell\tz_1$
($\ell=e,\mu$), occur without a big rate suppression because the
lepton can come from any one of many decay chains.
%then an addition hard isolated lepton can populate the
%events.
Requiring a hard isolated lepton gets rid of much of QCD BG, but
leaves BG from processes such as $W+jets$ and $t\bar{t}$
production~\cite{bcpt2}.

\subsubsection{Opposite sign (OS) dilepton $+$jets$+\eslt$}

Gluino and squark cascade decays readily lead to a pair of hard
isolated different flavor $e^\pm\mu^\mp$ where mostly each lepton
originates in a chargino, or the same flavor $e^+ e^-$ or
$\mu^+\mu^-$ where the leptons come from either a single neutralino
or a pair of charginos in the decay cascade, along with jets and
$\eslt$.  The neutralino contribution is statistically isolated in
the flavor-subtracted $e^+e^-+\mu^+\mu^--e^+\mu^--e^-\mu^+$ cross
section. Then, the dilepton invariant mass is kinematically bounded
by $m_{\tz_2}-m_{\tz_1}$ for decays from $\tz_2$ (though for small
values of $|\mu|$ contributions from $\tz_3$ are also identifiable).
If neutralinos decay to real sleptons the mass edge occurs instead
at
$$m_{\rm max}^{\ell\ell}
=m_{\tz_2}\sqrt{1-\frac{m_{\tell}^2}{m_{\tz_2}^2}}\sqrt{1-\frac{m_{\tz_1}^2}{m_{\tell}^2}}\leq
m_{\tz_2}-m_{\tz_1}.$$
%In the case where a $\tz_2$ (or sometimes $\tz_3$) is produced
%in a cascade decay, followed by $\tz_2\to\ell\bar{\ell}\tz_1$, then the
%two isolated leptons have an invariant mass bounded by
%$m_{\tz_2}-m_{\tz_1}$ in the case of neutralino 3-body
%decays\cite{mlledge}, or a related formula in the case of a sequence of
%2-body decays\cite{frank}: $\tz_2\to\ell\tell$ followed by
%$\tell\to\ell\tz_1$.
The dilepton mass edge~\cite{mlledge} is a smoking gun for SUSY
cascade decays and often serves as a starting point for the
reconstruction of decay chains~\cite{frank}, assuming the neutralino
leptonic branching fraction is large.
%If 2-body ``spoiler'' modes such as $\tz_2\to\tz_1 Z$ or $\tz_1
%h$ are kinematically allowed, then these instead will dominate the decay
%branching fractions.
SM backgounds to the OS dilepton signal from neutralinos mainly come
from  $Z+jets$ (followed by $Z\to\tau\bar{\tau}$), $t\bar{t}$ and
$W^+W^-$ pair production.

\subsubsection{Same sign (SS) dilepton $+$jets$+\eslt$}

Majorana gluinos are equally likely to decay into positive/negative
charginos via $\tg\to q\bar{q}'\tw_j$ so that gluino pair production
followed by the cascade decay $\tg\to\tw_j\to \ell^\pm$  of both
gluinos leads to SS, isolated dilepton plus jets plus $\eslt$
events~\cite{ss,btw}. This signature also arises from $\tg\tq_L$ and
$\tq_L\tq_L$ production followed by cascade decays. In fact, since
LHC is a $pp$ collider, then a charge asymmetry in $++$ vs. $--$
events is expected if $\tg\tq$ or $\tq\tq$ production is dominant,
while no charge asymmetry is expected from $\tg\tg$
production~\cite{btw,bcpt2}.  SM  BGs come from $WZ$ production
(where one lepton from a $Z$ decay is lost, $W^\pm W^\pm$
production, or $2\to 3$ processes such as $Wt\bar{t}$ production and
are much smaller than in the OS dilepton channel.

\subsubsection{$3\ell +$jets$+\eslt$}

Gluino and squark cascade decays also lead to three-isolated lepton
plus jet events, albeit with a lower rate.  These events are
nonetheless important because as isolated lepton multiplicity
increases, SM backgrounds usually drop much more rapidly than SUSY
signal. This makes it possible to use the trilepton signal to pick
out SUSY signals from SM backgrounds in early stages of LHC running
when reliable $\eslt$ measurements are not
possible~\cite{Baer:2008kc,Baer:2008ey,early2}.  SM BGs include
$t\bar{t}$ production, where one of the $b$ semi-leptonic decays
yields a hard, isolated lepton, together with other $2\to 3$
processes.

\subsubsection{$\ge 4\ell +$jets$+\eslt$}
Multi-jet $+\eslt$ events with $\ge 4$ isolated leptons are
ubiquitous in GMSB models where the selectron/smuon/stau are
together the next-to-lightest SUSY particle (NLSP) produced as the
penultimate step in the SUSY decay cascade. The NLSP then decays via
$\tell\to \ell\tG$ into the gravitino LSP so that every SUSY event
has at least two leptons (and frequently more). The SM background to
$\ge 4$~lepton events (where the leptons do not reconstruct the $Z$
mass is very small, and in this case LHC experiments can probe
gluino masses up to 3~TeV with just 10~fb$^{-1}$ of integrated
luminosity~\cite{yili} to be compared with a reach of $\alt 2$~TeV
in the corresponding case where the LSP escapes the detector
undetected.

%events\cite{bcpt2}.
%As one goes to higher lepton multiplcities, signal rates drop, although SM BG drops more quickly.

\subsubsection{$b$- and $\tau$-jets in SUSY events}

Gluino and squark cascade decays are often expected to be rich in
$b$-jets, so these can be used to reduce SM backgrounds.  There are
several reasons~\cite{ltanb}: 1)~large top and botom Yukawa
couplings-- especially at large $\tan\beta$-- enhance decays to
third generation quarks, especially if $\tz_1$ has significant
higgsino content as favored by the measured density of cold dark
matter, 2)~in many models third generation squarks are  lighter than
other
squarks, %leading to enhanced production cross sections and enhanced
resulting in an enhancement of sparticle decays to $b$-quarks, and
3)~real or virtual Higgs bosons, produced in cascade decays,
dominantly decay to $t$- and $b$-quarks. $b$-jet tagging thus allows
an increased SUSY reach at the LHC, in models with first generation
squarks substantially heavier than gluinos~\cite{btag}.
%\subsubsection{Presence of $\tau$-jets}

An enhanced multiplicity of $\tau$ leptons, identified by their
decays to 1 or 3 charged particles, is expected in SUSY cascade
decay events at large $\tan\beta$, for much the same reasons as high
$b$ multiplicities are expected~\cite{tau}.

\subsubsection{Leptonic $Z$ bosons in SUSY events}

In the case where either $\tz_i\to \tz_j Z$, or $\tw_2\to \tw_1 Z$
have significant branching fractions (frequently so for the former
if $\tz_2 \to h\tz_1$ is suppressed), then cascade decay events
containing real $Z\to\ell\bar{\ell}$ events are expected at high
rates~\cite{susyz} compared to SM BGs from $Z+jets$, $WZ$ or $ZZ$
production, especially if high $E_T$ jets and $\eslt$ are also
required in the signal.

\subsubsection{Higgs bosons in SUSY events}

It is entirely possible that the the lightest Higgs scalar $h$ will
be discovered first in the $\tg\to\tz_2\to h\tz_1$ SUSY cascade
rather than via usual SM search strategies that limit the search to
its rare decays. The reason is that with hard jet and $\eslt$ cuts
it is possible to search for $h$ via a mass bump in its dominant
$h\to b\bar{b}$~\cite{susyh} (and also the $h\to \tau\tau$) decay
mode without being overwhelmed by QCD backgrounds.

% the  , since
%the gluino and squark production reactions are distinctive and can occur
%at large rates.
%Frequently the light Higgs boson $h$ arises from cascade decays
%containing {\it e.g.} $\tz_2\to\tz_1 h$ decays\cite{susyh}. In many
%cases, the $h\to b\bar{b}$ or $\tau\bar{\tau}$ mass bump can be
%reconstructed. In fact,

%\subsubsection{Presence of heavy Higgs $A,\ H,\ H^\pm$}

Heavy Higgs bosons $A,\ H$ and $H^\pm$ can sometimes also be
produced in SUSY cascade decay events~\cite{bisset}.  It may be
possible to reconstruct mass bumps such as $H,A\to b\bar{b}$. Also,
heavy Higgs decay to SUSY particles is sometimes possible, such as
$H\to\tz_2\tz_2\to 4\ell+\eslt$ if $\tz_2\to \ell\bar{\ell}\tz_1$.
This would impact upon searches for heavy Higgs bosons via their
decays to SM particles.

\subsection{Jet-free  multilepton$+\eslt$ events}

\subsubsection{OS-dilepton $+\eslt$}

Same-flavor OS dileptons $+\eslt$ events (clean, or jet free) can
arise from slepton pair production~\cite{slepton}, {\it e.g.}
$pp\to\tell_R^+\tell_R^-$ followed by $\tell_R\to \ell\tz_1$.
Variables such as $\Delta\phi (\ell^+\ell^- )$ or $M_{T2}$ can be
used to see slepton signals above SM BGs such as $W^+W^-$ production
for $m_{\tell}\alt 350$ GeV. Determination of slepton spin also
appears to be possible~\cite{barr}
%OS dileptons can also arise from $pp\to
%\tw_i^+\tw_i^-$ production. If $\tw_i\to\ell\nu_\ell\tz_1$ decay occurs,
%frequently the leptons are too soft to easily be extracted from SM BG.

\subsubsection{Clean trilepton $+\eslt$}

Electroweak production of charginos and neutralinos via
$pp\to\tw_i\tz_j+X$, followed by $\tw_i\to\ell'\nu_{\ell'}\tz_1$ and
$\tz_2\to \ell\bar{\ell}\tz_1$ decay~\cite{clean3l} yields clean
trilepton events for which SM backgrounds are very small. The signal
is largest and readily
observable %above SM BGs if the production and decay rates are large
over background when $\tw_1$ and $\tz_2$ are wino-like and the
$\tz_2$ spoiler decay modes are kinematically closed.  The OS
dilepton mass edge from the $\tz_2$ decay should again be visible,
corroborating its SUSY origin.

\subsection{Signals with isolated photons}

In GMSB models with a gravitino LSP and $\tz_1$ the next-to-lightest
SUSY particle (NLSP),  $\tz_1\to \tG\gamma$ is often the dominant
decay mode of $\tz_1$. Then, gluino and squark production followed
by their  cascade decays will always yield at least two $\tz_1$s,
both of which decay to hard, isolated photons.  Thus, GMSB models
with a small number of messenger fields are expected to yield large
rates for multi-jet$+$ multi-lepton $+\eslt +2\gamma $
events~\cite{gmsbgamma}.

Hard isolated photons can also arise in SUGRA-type models, where the
branching fraction for the loop decay $\tz_2\to\tz_1\gamma$ is
significant~\cite{z2z1g}.  These loop decays are enhanced if the
$\tz_2$-$\tz_1$ mass gap is small as in small $|\mu|$ models or in
models with $|M_1|\simeq |M_2|$ {\em at the weak
scale}~\cite{nonugam}, where the 3-body decays of $\tz_2$ are
strongly suppressed by phase space.

In some cases, if $h$ production is large in cascade decay events,
then the decay $h\to \gamma\gamma$ can be reconstructed in the SUSY
event sample~\cite{susyhgg}.

\subsection{Signals from long-lived charged sparticles}

\subsubsection{Highly ionizing tracks (HITs)}

In the simplest GMSB models with large enough number of messenger
fields, the slepton (usually the lighter stau $\ttau_1$) is the
NLSP. The NLSP then decays via $\ttau_1\to\tG\tau$ will take place,
but with a rate suppressed by its tiny coupling to the goldstino
component of $\tG$.  In such a case, the relatively slow-moving
heavy $\ttau$  is long-lived, and leaves a highly ionizing track as
it traverses the detector.
%Such tracks can be picked out from $e$ or $\mu$ tracks, and can be
%distinctive.
These tracks may terminate, or leave a kink, depending on where the
delayed NLSP decay occurs. A determination of the NLSP lifetime, and
hence the fundamental SUSY breaking scale, is possible if the NLSP
decay length is between 0.5~m to 1~km~\cite{HITs}.  The wino-like
chargino of AMSB models has a decay length of order {\it
centimeters} and so leaves a short stubby track potentially with
kinks from its pion daughter.

%In AMSB with a wino-like chargino and neutralino, with $m_{\tw_1}\simeq
%m_{\tz_1}$, then $\tw_1$ can be long enough lived to leave a HIT of
%length of order {\it centimeters}.

%\subsubsection{Length of track as measure of sparticle lifetime}
%
%In the case where the HIT is due to a long lived quasi-stable particle, the termination of the
%HIT, or its termination in a kink, will allow a measure of the quasi-stable particle's lifetime,
%due to a measure of its track length. Such a measurement is aided by the HIT bend in a magnetic
%field, which allows a detemination of its velocity if the particle mass is known\cite{track}.

\subsubsection{Trapping sleptons}

If the gravitino is heavy enough, the charged slepton NLSP of GMSB
models may live days or months or even longer.  In this case, it is
possible to capture these sleptons produced in collider experiments
in, for instance, a water tank surrounding the detector. The water
can be siphoned off, and the slepton decay properties can then be
well-measured: {\it e.g.} its lifetime, and mass (based on energy
release from an at-rest slepton decay)~\cite{slep_trap}.

An intriguing variant of this idea is to trigger on events with
$\eslt > 100$~GeV and high jet activity that contain an isolated
track from a slow-moving stau (or any charged massive particle, the
CHAMP) stopped in the calorimeter, and at this stage dump the beams
(or at least change their orbit) so there are no collisions (in at
least the triggered detector) for about an hour, during which the
focus is on the detection of the decay products of CHAMPs trapped in
the calorimeter~\cite{asai} if the lifetime is $\alt$~1~hour.
Longer-lived CHAMPs can be studied during collider shut-downs. It is
claimed that, with an integrated luminosity of 100~fb$^{-1}$, stau
lifetimes ranging from $10^{-1}-10^{10}$~s will be measureable at
ATLAS, and that this idea may be extendable to other quasi-stable
CHAMPs.

%\subsubsection{Trapping stops}

\subsection{Events with displaced vertices}

%\subsubsection{Non-pointing $\gamma$s or $\ell$s}

In the case of GMSB models with a long-lived neutralino LSP decaying
via $\tz_1\to \tG\gamma$, or $Z\tG$ or $h\tG$, the decay vertex will
be dispaced from the primary interaction point and the EM shower
induced by the $\gamma$ or the decay products of the $Z$ or $h$ will
likely not point back to the interaction point. The same is true for
a neutralino LSP decaying via tiny $R$-parity violating couplings.
%and is another signature of a long
%lived particle decaying.  In the case of GMSB where $\tell\to\tG\ell$,
%then the track from the $\ell$ will not point back to the interaction
%region.
The case of the photon decay of $\tz_1$ has been studied in
detail~\cite{kkno} and it was shown that for an NLSP decay length of
10~cm-20~m, the secondary vertex could be well-determined from
events where the photon converts to an electron-positron pair so
that reconstruction of the entire SUSY event is possible. It is
claimed that this reconstruction is also possible using events where
the photon does not convert, since the degradation in the precision
is compensated by the much larger number of events. The NLSP
lifetime is determined to within a few percent. This is an important
measurement as it determines the fundamental scale of SUSY breaking.

%\subsubsection{Non-pointing jets}
%
%In models such as GMSB, or $R$-violating models with a tiny $R$-violating coupling, then
%delayed decays such as $\tz_1\to\tG h$ (followed by $h\to b\bar{b}$) or $\tz_1\to bcs$ can occur.
%In cases such as these, the hadronic shower will not point back to the interaction region,
%and it will be a signal for the presence of a particle with a delayed decay.

\subsection{Events containing intermittent tracks}

Scenarios with stable \cite{bcg} or long-lived \cite{splitsusy}
gluinos or squarks (usually $\tst_1$) have been considered. Once
produced at colliders, the squark or gluino quickly hadronizes by
picking up an antiquark or a gluon/$q\bar{q}$, respectively, and
traverses the detector as an $R$-hadron that may be electrically
charged or neutral. This $R$-hadron interacts with nuclei in the
detector material via pion exchanges, and so may move between its
charged and neutral states, thereby manifesting itself as an
intermittent track in a collider event: a track that suddenly
appears, disappears and reappears along its path \cite{drees,bcg}.
%
%along its path of travel, and can get stripped of its quark
%content, and re-hadronize. If it moves between charged and charge
%neutral gluino hadron states, then collider events will contain
%intermittent tracks: ones that appear and dis-appear along the path of
%propagation.
%
%Models where the $\tg$ is the LSP and $R$-parity is
%conserved yield such stable gluinos, while models of split
%supersymmetry \cite{splitsusy}, wherein squarks can be in the $10^{8}-10^{13}$ GeV range,
%thus suppressing gluino 3-body decays, yield quasi-stable gluinos.
In the quasi-stable particle case, the intermittent track might
terminate in a burst of hadronic showers which of course would not
point back to the interaction region.

\subsection{Inclusive multilepton events without $\eslt$} In
$R$-parity violating models where the LSP decays into SM particles,
neutrinos are the only physics source of $\eslt$ and the classic
$\eslt$ signature is greatly reduced (though even in the worst-case
scenario where the LSP decays hadronically, the 10~fb$^{-1}$ reach
extends to 1~TeV in $m_{\tg}$ \cite{rviol}. In the favorable case
that the neutralino LSP decays purely leptonically via $\tz_1\to
\ell\bar{\ell}\nu$, SUSY events will be awash in multileptons and
the reach will be greatly increased even without $\eslt$. There are
no reach calculations available for the LHC, but even the Tevatron
is sensitive to mSUGRA parameter values that give $m_{\tg}=800$~GeV
\cite{tev}. Event shapes in the OS dilepton channel (especially
dilepton mass distributions) \cite{deba} and the rate for SS
dilepton production \cite{bartl} at the LHC are sensitive to
$R$-parity violating interactions,.

\subsection{Resonance sparticle production} In $R$-parity violating
scenarios with $\hat{L}\hat{Q} {\hat D}^c$-type couplings, it is
possible to resonantly produce sleptons and sneutrinos at the LHC
\cite{hall}. The phenomenology is very sensitive to details of the
model, and potentially to interesting multilepton signals. Even
assuming just a single $R$-parity violating coupling, the
phenomenology depends on the scale at which this single coupling is
assumed to be present, since renormalization effects induce small
(but phenomenologically significant) values for other $R$-violating
couplings at the weak scale. For a recent analysis, see
Ref.~\cite{herbi}, and references therein.

\subsection{Rapity gap events from SUSY} Very recently \cite{rapgap}
it has been pointed out that production of squark pairs by
$t$-channel exchanges of colour singlet -inos would lead to events
with large ``rapidity gaps'', {\it i.e.} little energy deposition
between squark decay products. If this observation survives scrutiny
and such events turn out to be observable, they could be used to
separate electroweak squark pair production from the much larger QCD
squark pair production, and provide a new, potentially interesting
ways to separate SUSY contributions at the LHC.

\subsection{Final Remarks}

We have listed a number of signals via which SUSY may be discovered
at the LHC.  While some of the catalogued signals are quite generic,
and so present in wide classes of models, others occur only in
specific scenarios, or only for special regions of model parameter
space. Seeing a signal in several channels will corroborate that the
origin of the new physics is supersymmetry, while their relative
rates (together with measurements of masses, branching ratios, {\it
etc.}) will serve to zero in on the underlying framework.
Observation of special signatures will be particularly useful as
these occur only in specific models.

\section{LHC Measurements}
%{alprz}
\noindent
{\it Claire Adam-Bourdarios,  Remi Lafaye, Tilman Plehn, Michael Rauch, and Dirk Zerwas}\\

If new physics is to be discovered at the LHC, the next step would
be to reconstruct the underlying theory, and this endeavor should
not be biased by any assumption on high-scale models. SFitter
\cite{SFitter} and its weighted Markov chain technique is a tool of
choice to perform such a task.\footnote{Fittino~\cite{Fittino06}
follows a very similar logic to SFitter, including a scan of the
high dimensional MSSM parameter space.} Using the example of the
TeV-scale MSSM Lagrangian we illustrate in detail how it will be
possible to analyze this high dimensional
physics parameter spaces and extrapolate parameters to the high scale, to test unification.\\

The analysis  critically depends on detailed experimental
simulations of measurements and errors at the LHC. Therefore the
well-understood parameter point SPS1a~\cite{Allanach01} is used.

The parameter point SPS1a is characterized by moderately heavy
squarks and gluinos, which leads to long cascades including
neutralinos and sleptons. Gauginos are lighter than Higgsinos, and
the mass of the lightest Higgs boson is close to the mass limit
determined at LEP. At the LHC, the mass measurements are obtained
from measurements of kinematical endpoints and mass differences. The
particle mass measurements used by SFitter \cite{SFitter} are taken
from Ref. \cite{Weiglein06}, while the central values are calculated
by SuSpect~\cite{Kneur02}.

In order to obtain reliable error estimates for the fundamental
parameters, a proper treatment of experimental and theory errors
depending on their origin is mandatory. The CKMfitter
prescription~\cite{Hocker01} is largely followed. The complete set
of errors includes statistical experimental errors, systematic
experimental errors, and theory errors.

The statistical experimental errors are treated as uncorrelated
among the measured observables, in contrast to the systematic
experimental errors, essentially due to the uncertainty in the
lepton and jet energy scales, expected to be 0.1\% and 1\%,
respectively, at the LHC. These energy-scale errors are each taken
to be 99\% correlated. Theory errors are propagated from the masses
to the measurements and are not taken to be gaussian but flat
box-shaped. Thus, the probability assigned to any measurement does
not depend on its actual value, as long as it is within the interval
covered by the theory error. Outside this interval, normal
statistical and systematic errors treatment is used.

\subsection{mSUGRA}

mSUGRA is an example of a model with few parameters, most of which
are defined at the grand unification scale (GUT).

SFitter approaches the problem of the high dimensional parameter
space, producing first a set of Markov chains over the entire
parameter space. Then, Minuit resolves the local maxima in the
likelihood map. Once the global best fitting parameter point is
identified, the errors on all parameters are determined using
smeared sets of pseudo measurements and flat theory
errors~\cite{SFitter}.

The precision obtained with LHC alone is at the level of percent for
the determination of the parameters. It is improved by the ILC by
about an order of magnitude. Including the theoretical errors has an
impact on the precision at both machines, the errors are larger by a
factor of three to four. Thus the precision of the parameter
determination at the LHC is limited by the precision of the
theoretical predictions.

\begin{figure*}[htb]
\includegraphics[width=\columnwidth]{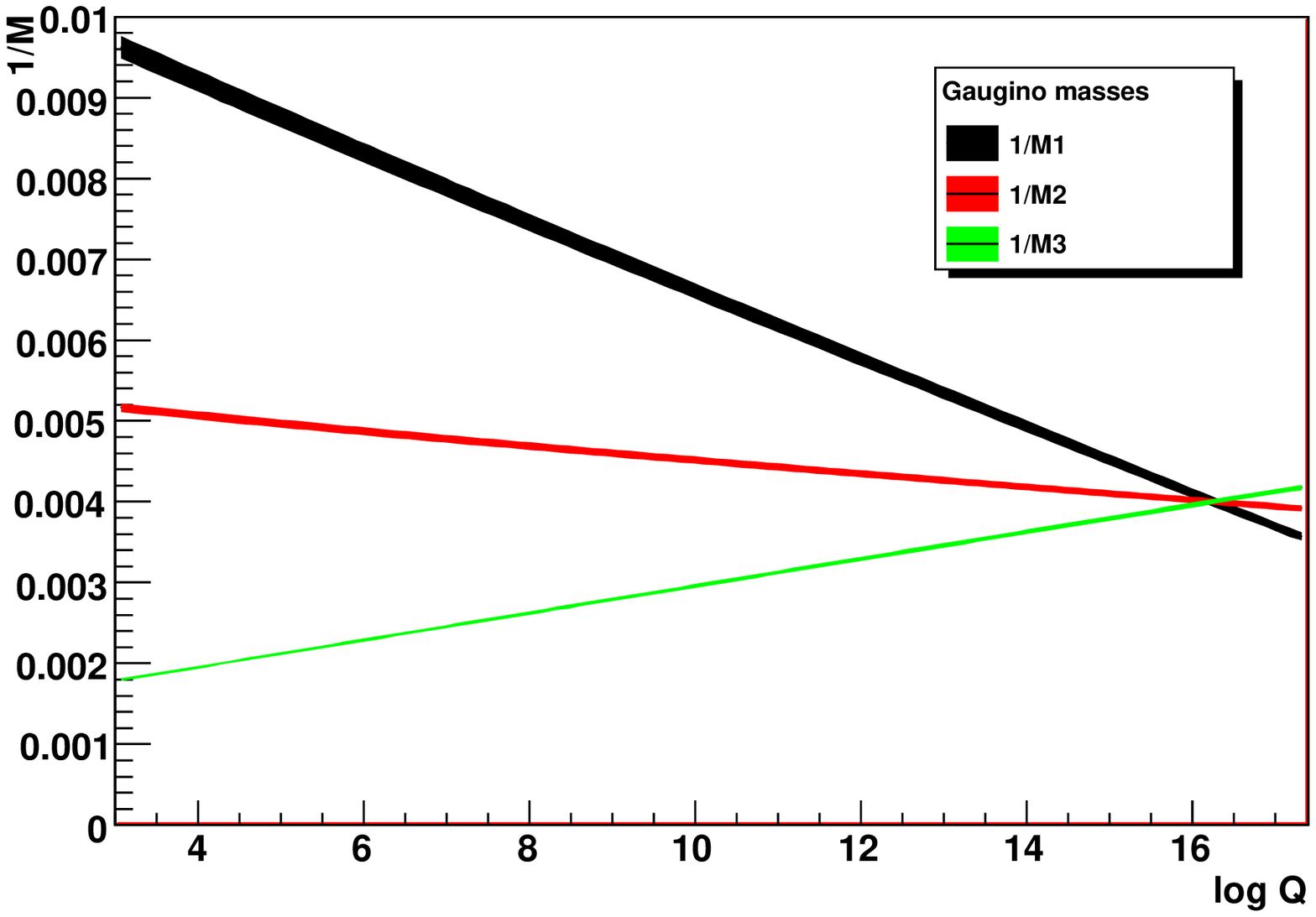}
\hspace{0.5cm}
\includegraphics[width=\columnwidth]{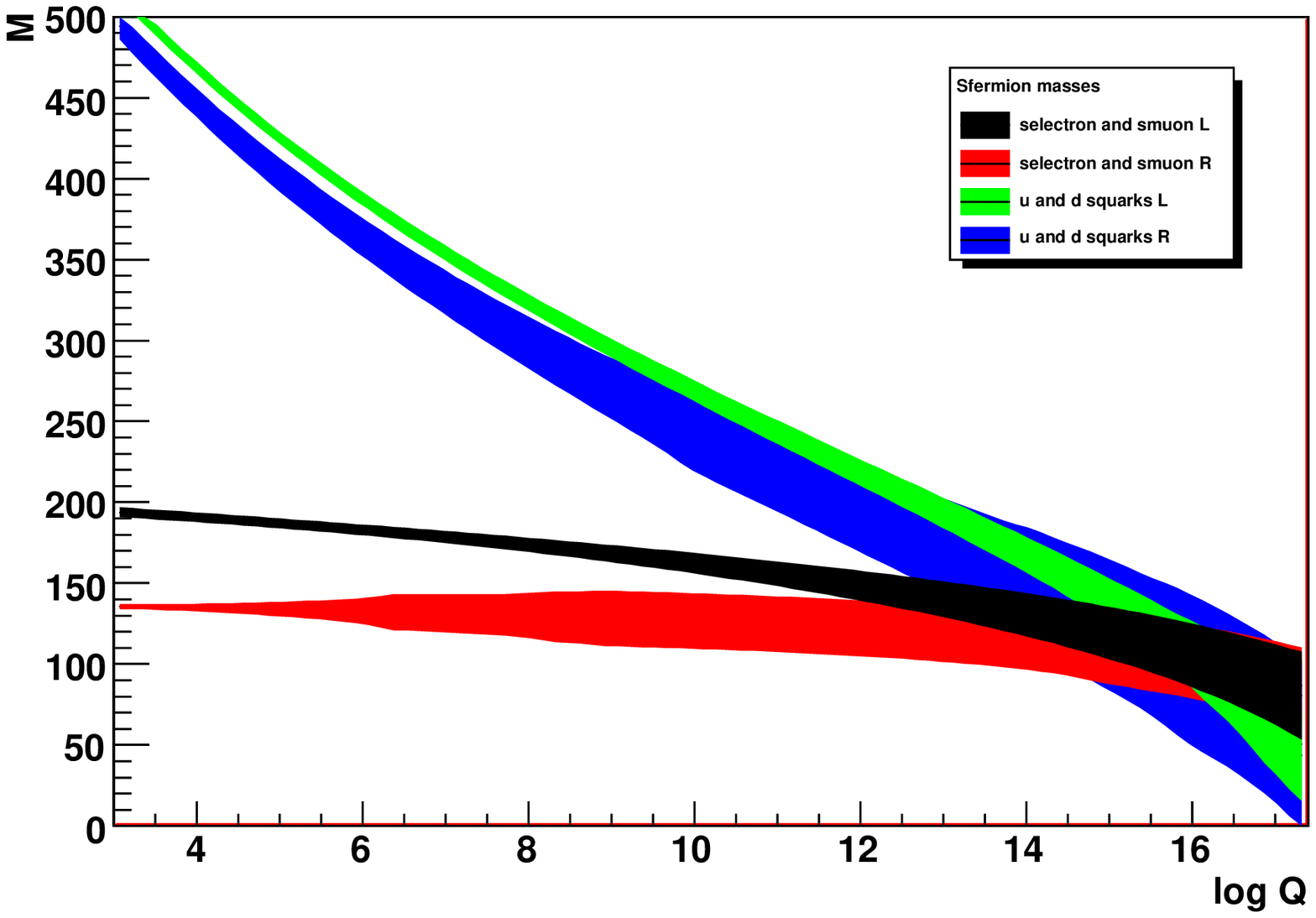}
\caption{Extrapolation of the inverse of the gaugino mass parameters
(left) and the first and second generation scalar mass parameters
(right) to the GUT scale, for one of 8 the degenerate solutions at
the LHC.} \label{fig:SPS1aExtrapol}
\end{figure*}

\subsection{MSSM}

The complete parameter space of the MSSM can have more than 100
parameters. However, at experiments like the LHC some new physics
parameters can be fixed, because no information on them is expected.
Properly including phenomenological constraints and $m_t$ leads to
an effective 19-dimensional parameter space.

LHC provides 22 measurements, counting the measurements involving
$m_{\tilde{l}}$ separately for electrons and muons. Using these
naively it should be possible to completely constrain a
19-dimensional parameter space. However, the situation is more
complicated. These 22 measurements are constructed from only 15
underlying masses. The additional measurements will resolve
ambiguities and improve errors, but they will not constrain any
additional parameters.

SFitter approaches the problem of the higher dimensional MSSM
parameter space by an iterative procedure. With LHC measurements
only, eight solutions are found. The errors obtained for one of the
minima, the one closest to the SPS1a point, are shown in
Table~\ref{table:1} : while many parameters are well determined,
some model parameters turn out to be not well constrained. Some of
them, namely the trilinear mixing terms $A_i$, are fixed in the fit
because their impact is close to zero. Others, like the heavier
stau-mass and stop-mass parameters or the pseudoscalar Higgs mass,
turn out to be unconstrained because they do not appear directly in
any of the measurements.

\begin{table}[htb]
\caption{Results for the general MSSM parameter determination in
SPS1a assuming flat theory errors.} \label{table:1}
\newcommand{\m}{\hphantom{$-$}}
\newcommand{\cc}[1]{\multicolumn{1}{c}{#1}}
\newcommand{\ccthree}[1]{\multicolumn{3}{c}{#1}}
\begin{tabular}{@{}lr|lr}
\hline
$\tan\beta$ & $10.0\pm 4.5$ & $M_1$ & $102.1\pm 7.8$\\
$M_2$ & $193.3\pm 7.8$ & $M_3$ & $577.2\pm 14.5$\\
$M_{\tilde{\tau}_L}$ & $227.8\pm O(10^3)$ & $M_{\tilde{\tau}_R}$ & $164.1\pm O(10^3)$\\
$M_{\tilde{\mu}_L}$ & $193.2\pm 8.8$ & $M_{\tilde{\mu}_R}$ & $135.0\pm 8.3$\\
$M_{\tilde{e}_L}$ & $193.3\pm 8.8$ & $M_{\tilde{e}_R}$ & $135.0\pm 8.3$\\
$M_{\tilde{q}3_L}$ & $481.4\pm 22.0$ & $M_{\tilde{t}_R}$ & $415.8\pm O(10^2)$\\
$M_{\tilde{b}_R}$ & $501.7\pm 17.9$ & $M_{\tilde{q}_L}$ & $524.6\pm 14.5$\\
$M_{\tilde{q}_R}$ & $507.3\pm 17.5$ & $A_\tau$ &  fixed 0\\
$A_t$ &  $-509.1\pm 86.7$ & $A_b$ &  fixed 0\\
$A_{l1,2}$ & fixed 0 & $A_{u1,2}$ & fixed 0\\
$A_{d1,2}$ & fixed 0 & $m_A$ & $406.3\pm O(10^3)$\\
$\mu$ & $350.5\pm 14.5$ & $m_t$ & $171.4\pm 1.0$\\
\hline
\end{tabular}\\[2pt]
All values are given in GeV.
\end{table}

Moreover, there is no good direct measurement of $\tan\beta$.
Looking at the neutralino and sfermion mixing matrices, any effect
in changing $\tan\beta$ can always be accommodated by a
corresponding change in another parameter. Here, information from
flavour physics or the anomalous magnetic moment of the muon can
help~\cite{Houches07}.

\subsection{Extrapolation to High Scale}

The MSSM is defined at the electroweak scale. The definition of its
parameters does not depend on the model of supersymmetry breaking.
Thus, having determined the parameters at the weak scale, the
extrapolation of the parameters to the GUT scale can be performed,
using RGE's : instead of assuming unification at the GUT scale as in
mSUGRA, it can now be tested.

Figure \ref{fig:SPS1aExtrapol} shows that, for the point chosen in
Table~\ref{table:1}, unification is observed. However, this is only
true for one of the 8 solutions : with LHC only, requiring
unification can be used to reduce the number of degenerate
solutions, but cannot be proven.
%
%\begin{figure*}[htb]
%\includegraphics[width=\columnwidth]{Tab6_9_susy09_M123.eps}
%\hspace{0.5cm}
%\includegraphics[width=\columnwidth]{Tab6_9_susy09_Sleptons.eps}
%\caption{Extrapolation of the inverse of the gaugino mass parameters (left) and the first and second generation scalar mass parameters (right) to the GUT scale, for one of 8 the degenerate solutions at the LHC.}
%\label{fig:SPS1aExtrapol}
%\end{figure*}

%\section{CONCLUSION}

In conclusion a sophisticated tool such as SFitter is necessary to
determine the underlying parameters of supersymmetry from the
correlated measurments of the LHC. mSUGRA can be determined at
the percent level, and most of the MSSM parameters are measured
precisely. The MSSM parameters can be extrapolated to the GUT scale
to test grand unification for possibly degenerate solutions. \\

%%%%%%%%%%%%%%%%%%%%%%%%%%%%%%%%%%%%%%%%%%%%%%%%%%%%%
\section{Fits to the Phenomenological MSSM}
\noindent
{\it Sehu S AbdusSalam, Benjamin Allanach, Fernando Quevedo}\\

With the LHC experiments about to start we are at a new and exciting
scientific era where enquiries about the fundamentals of nature will
be governed by experiments and hence complement the past decades of
theoretically motivated efforts. A great deal of research directed
towards understanding the origin of SUSY breaking from ultra-violet
theories lead to a plethora of models but no single one clearly
favoured over the others. Many of them, however, fall into some
subset of the phenomenologically viable MSSM parameter space.

For a phenomenological MSSM (pMSSM) fit, the source of SUSY breaking
is completely decoupled from the problem such that all relevant
parameters are varied simultaneously at the weak energy scale. This
is a general approach to SUSY breaking phenomenology and has been
studied in~\cite{AbdusSalam:2009qd}. Here we present a summary of
the pMSSM global fit construction and results.

Only real soft SUSY breaking terms were considered, with all
off-diagonal elements in the sfermions mass terms and trilinear
couplings set to zero, and the first and second generation soft
terms equalised. The effects of the trilinear coupling terms $A_t$,
$A_b$, $A_\tau$ on SUSY effects are not negligible. So also $A_e =
A_\mu$, which is relevant for the computation of the anomalous
magnetic moment of the muon, $(g-2)_\mu$. %\cite{Martin:2001st}.
pMSSM Higgs-sector parameters are specified by $m^2_{H_1}$,
$m^2_{H_2}$, $\tan \beta$ and $sign(\mu)$. Important SM parameters
we varied are $m_Z$, $m_t$, $m_b(m_b)$, $\alpha_{em}(m_Z)$ and
$\alpha_{s}(m_Z)$.

The above parameters form a vector, $\underline m$, such that the
combined prior PDF for the model (H = pMSSM) is $p(\underline m | H)
= p(m_1| H) \, p(m_2| H) \, \ldots \, p(m_{25}| H)$. Two widely
different, the linear and log, priors were used to check the
robustness of the fits.

The constraints employed include the $W$-boson mass, $m_W$, the
effective leptonic weak mixing angle, $\sin^2 \theta^{lep}_{eff}$,
the total $Z$-boson decay width, $\Gamma_Z$, $(g - 2)_\mu$, Z-pole
asymmetry parameters and the mass of the lightest MSSM Higgs boson,
$m_h$; branching ratios $BR(B \rightarrow X_s\gamma)$, $BR(B_s
\rightarrow \mu^+ \mu^-)$, $BR(B_{u^-} \rightarrow \tau^- \nu)$,
$BR(B_u \rightarrow K^* \gamma)$ and the $B_s$ mass-mixing parameter
$\Delta M_{B_s}$; and the dark matter relic density from WMAP5
results. These make the dataset $\underline d$.

The set of pMSSM predictions, $\underline O$, for the above
observables were obtained from the 25 input parameters, sampled
using \texttt{MultiNest}~\cite{Feroz:2007kg}, in the SLHA
format~\cite{hep-ph/0311123} via
\texttt{SOFTSUSY2.0.17}~\cite{Allanach:2001kg} for producing the
MSSM spectrum; \texttt{micrOMEGAs2.1}~\cite{Belanger:2004yn} for
computing neutralino dark matter relic density, the branching ratio
$BR(B_s \rightarrow \mu^+ \mu^-)$ and $(g-2)_\mu$;
\texttt{SuperIso2.0}~\cite{arXiv:0808.3144} for predicting the
Isospin asymmetry in the decays $B \rightarrow K^* \gamma$ and $BR(b
\rightarrow s \gamma)$ with all NLO SUSY QCD and NNLO SM QCD
contributions included; and \texttt{susyPOPE}~\cite{arXiv:0710.2972}
for computing $W$-boson mass $m_W$, the effective leptonic mixing
angle variable $\sin^2 \theta^{lep}_{eff}$, the total $Z$-boson
decay width, $\Gamma_Z$ and other Z-pole asymmetry parameters from
$e^+e^- \rightarrow f \bar{f}$ processes.
\begin{figure}[!ht]
\begin{center}
  \unitlength=1.0in
\begin{picture}(3.2,1.6)(0,0.55)
%\put(0.35,2.){\includegraphics[angle=0, width=2.5in]{ewwgtypeN.eps} }
\put(0.35,0){\includegraphics[angle=0,
width=3.0in]{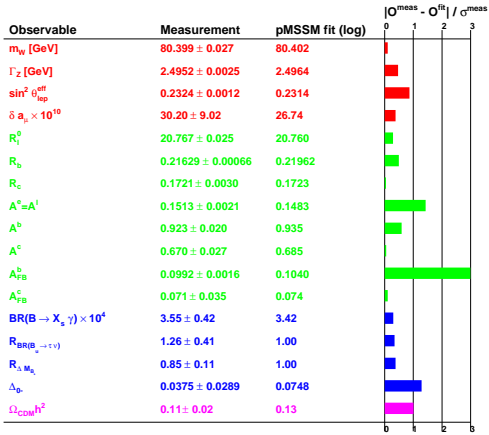}}
\end{picture}
\end{center}
\caption{Statistical pull of various observables at the
  best-fit point.}
 \label{f.ewwg1}
\end{figure}

The combined likelihood from the different predictions $O_i$ of the
observables $d_i$,  together with the prior PDF via the Bayes'
theorem, Eq.~\ref{eq:bayes}, is used to compute the posterior PDF of
the parameters, observables and predictions for the MSSM SUSY mass
spectrum. \be p(\underline d|\underline m,H) = \prod_{i=1}^{20}
\frac{1}{\sqrt{2\pi \sigma_i^2}}\exp\left[- \frac{(O_i -
    \mu_i)^2}{2\sigma_i^2}\right]
    \label{eq:bayes}
    \ee
where $\mu_i$ and $\sigma_i$ are respectively the experimental
central values and errors.

We found that the sparticle mass spectrum for the log prior
assumption can have slepton, squarks and neutralino LSP masses as
low as about 251 GeV, 383 GeV and 243 GeV respectively. The masses
are much heavier for the linear prior assumption as expected. The
Higgs boson mass turned out to be approximately prior independent
with an almost equal mass range of about $117$ GeV to $129$ GeV for
both prior assumptions.
\begin{figure*}[ht]
  \begin{tabular}{llll}
(a) & (b) & (c) & (d) \\
   \begin{minipage}[t]{3.5cm}
     \begin{center}
     \includegraphics[width=.9\textwidth]{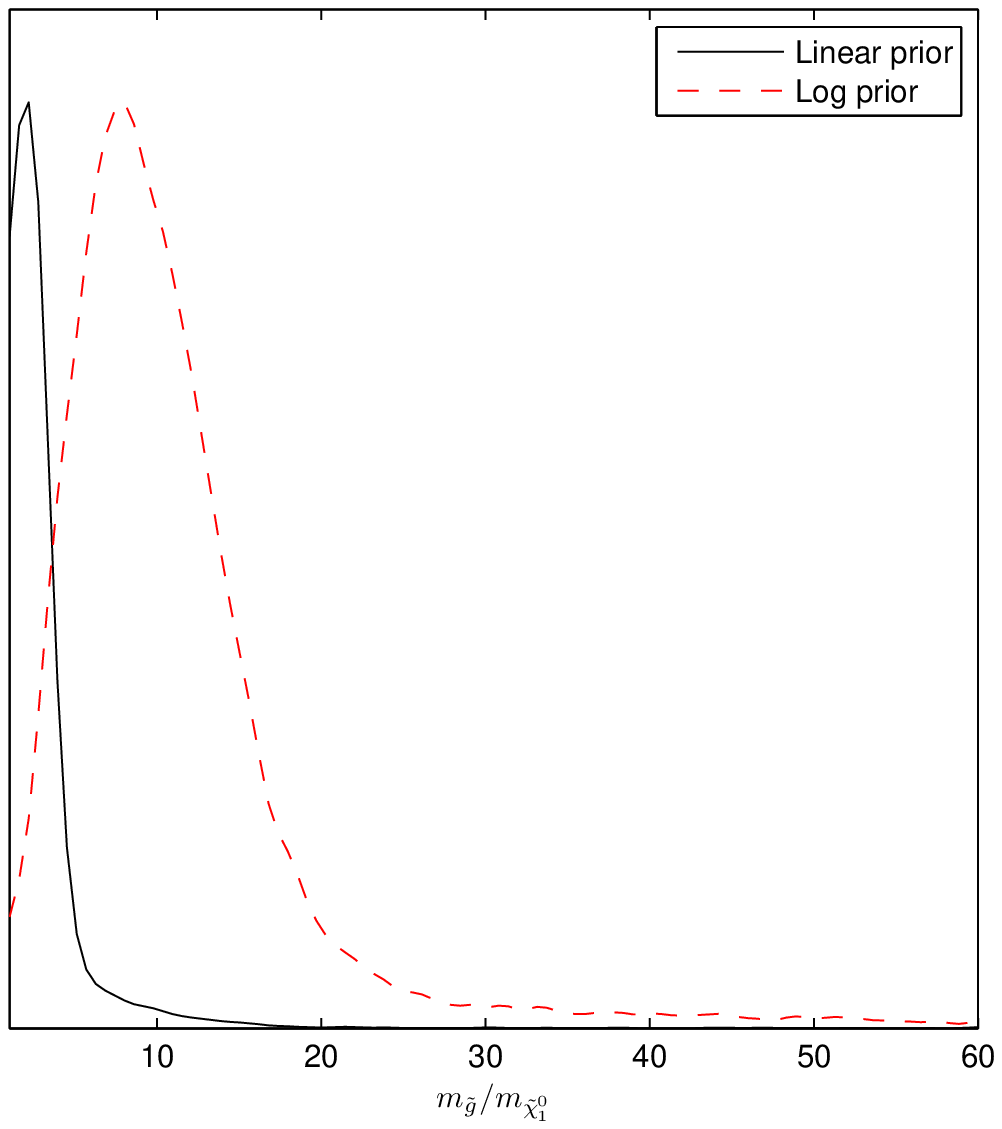}
     \end{center}
   \end{minipage}
&
  \begin{minipage}[t]{3.5cm}
    \begin{center}
    \includegraphics[angle=0,
    width=.9\textwidth]{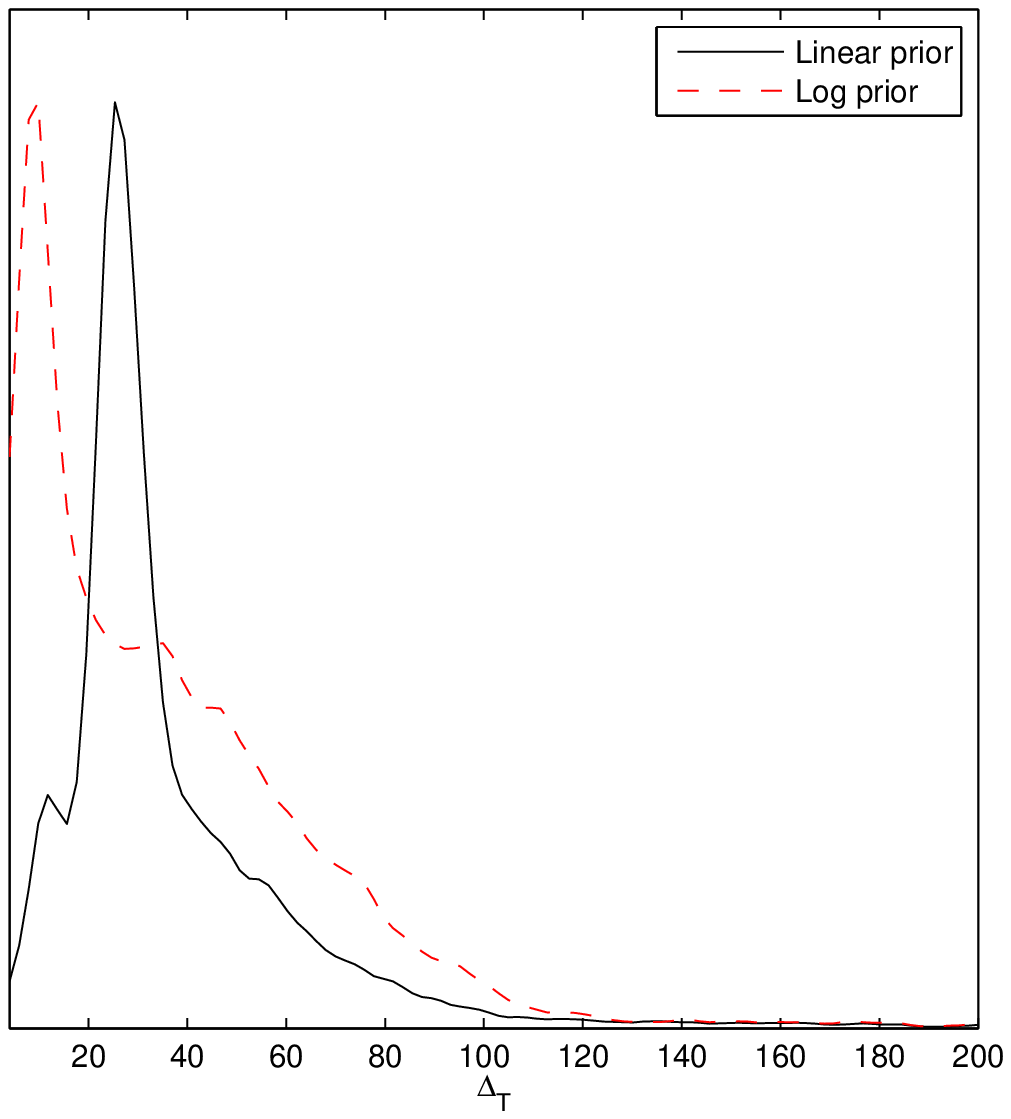}
     \end{center}
   \end{minipage}
&
   \begin{minipage}[t]{3.5cm}
     \begin{center}
     \includegraphics[width=.9\textwidth]{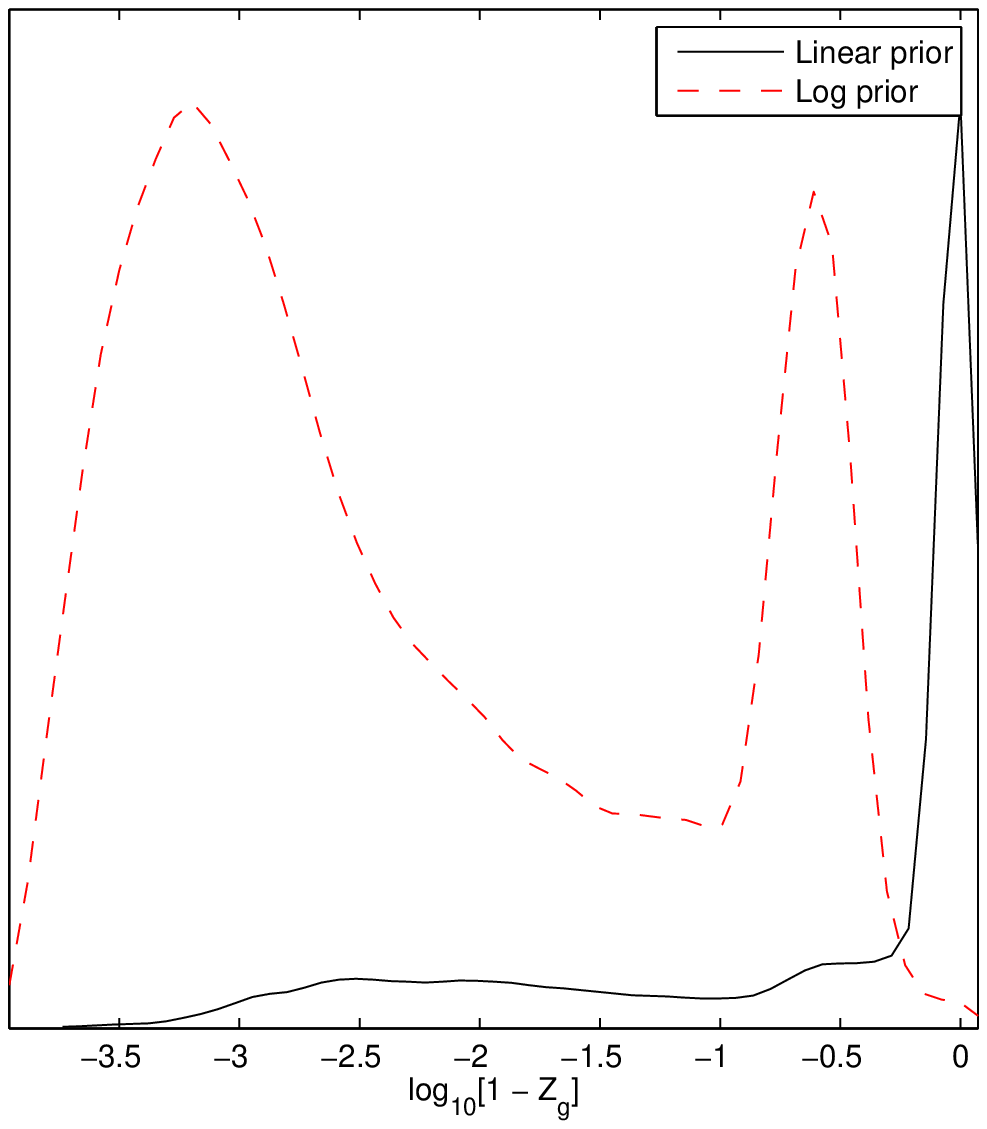}
     \end{center}
   \end{minipage}
&
   \begin{minipage}[t]{4.2cm}
     \begin{center}
     \includegraphics[width=1.\textwidth]{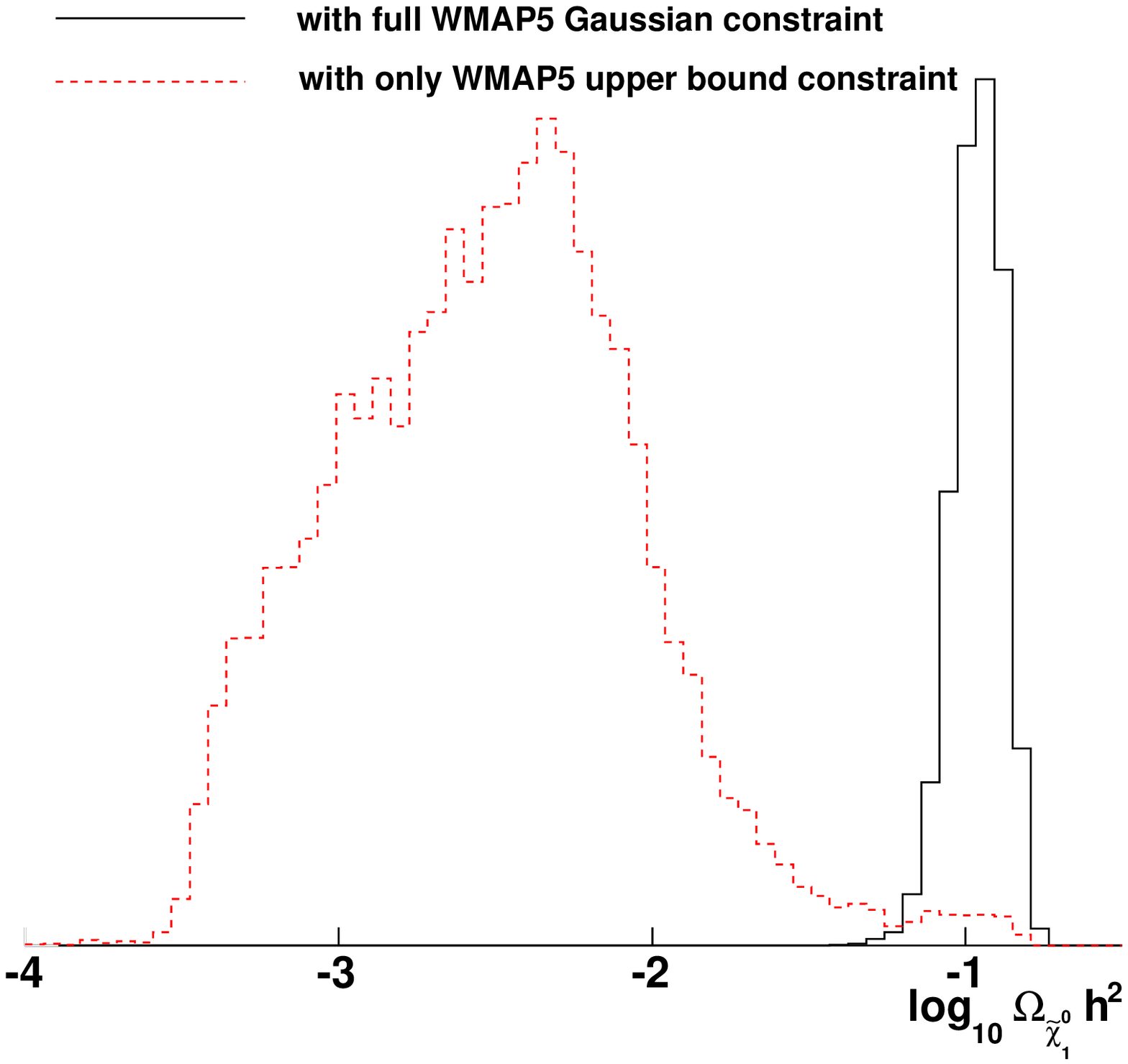}
     \end{center}
   \end{minipage}
   \end{tabular}
  \caption{{\bf (a)}: pMSSM
    $m_{\tilde{g}}$-$m_{\tilde{\chi}^0_1}$ mass ratio PDFs. {\bf (b)}:
    Fine-tuning PDFs in the
    pMSSM. {\bf (c)} pMSSM neutralino gaugino-Higgsino admixture
    fractions, $Z_g = |N_{11}|^2 + |N_{22}|^2$, PDFs. {\bf (d)}: Neutralino
    relic density assuming WMAP5 as a Gaussian likelihood constraint
    or as an upper bound. This plots is for a 2 TeV range pMSSM
    with settings with linear priors, as in
    Ref.~\protect\cite{AbdusSalam:2008uv}.}
    \label{f.gs}
\end{figure*}

The statistical pulls of the various observables are shown in
Fig.~\ref{f.ewwg1} at the best-fit point. We see from the figure
that, like the Standard Model, the forward-backward asymmetry in
$e^+e^-$ $\rightarrow$  $b\bar b$
%$e^+e^- \rightarrow b$
provides the greatest discrepancy, being at odds with data at the
$3\sigma$
level. Notably, an extra-SM component of the $(g-2)_\mu$ %anomalous
                %magnetic moment of the muon, $\delta
                %a_\mu$
and the relic density of cold dark matter, $\Omega_{CDM} h^2$  are
well fit. Both quantities are ill-fitting in the Standard Model.

We now summarise the marginalised posterior PDFs for various pMSSM
quantities, but the full set of plots can be found in
Ref.~\cite{AbdusSalam:2009qd}. Some differences between the
posterior PDFs for the two prior cases can be observed. They are
mostly due to the fact that the sparticle masses are larger in the
linear prior measure, leading to a suppression of SUSY effects in
the loop calculations of most observables. As such, only the EW
physics observables show significant difference between the two
prior cases while the other observables (except for, notably, the DM
constraint) are relatively weaker in constraining the parameter
space.
%The observable that most discriminates between some two best-fit points
%is $(g-2)_\mu$, which is better fit for the log prior point since
%it receives large corrections from lighter slepton and gaugino masses and
%hence fits the apparent non-SM contribution implied by data.
Linear pMSSM global fit results show a mild preference for $\mu >
0$, depending on the $(g-2)_\mu$ constraint. However, there is no
conclusive evidence  for one particular  $sign(\mu)$ over the other.
%However based on the Jeffrey's
%scale, Tab.~\ref{tab:Jeffreys}, there is no conclusive evidence for
%one particular $sign(\mu)$ over the other.

The gluino-neutralino mass ratio quantifies the amount of visible
energy that could be seen in sparticle production from LHC
collisions and therefore could be used to discriminate between
different models. mSUGRA (AMSB) with predominantly bino (wino) LSP
predicts $m_{\tilde{g}}/m_{\tilde{\chi}^0_1} \approx$ 6(and
9)~\cite{nilles}. The mirage mediation~\cite{Lebedev:2005ge} and the
LARGE volume~\cite{Conlon:2006wz,Conlon:2007xv} scenarios have the
characteristic ratio less than 6 and between 3 to 4 respectively.
However by construction, the pMSSM is a more generic approach for
MSSM phenomenology. We show the pMSSM posterior PDFs for the
gluino-neutralino mass ratio in Fig.~\ref{f.gs}(a). The linear prior
predicts a compact $m_{\tilde{g}}/m_{\tilde{\chi}^0_1} \approx 2.5$
while for the log prior a much broader distribution mostly around
10.

The amount of fine-tuning in the pMSSM is quantified by considering
the sensitivity of $m_Z$ to parameter variations $ \Delta(\xi) =
\left| \frac{\partial \log m_Z^2}{\partial \log \xi}\right|$, where
$\xi=m_{H_1}^2, m_{H_2}^2, m_3^2$ and $\mu$ are the relevant
parameters. The over-all measure of fine-tuning,  $\Delta_T$, is \be
\Delta^2_T =
\Delta(\mu)^2+\Delta(m_3^2)^2+\Delta(m_{H_1}^2)^2+\Delta(m_{H_2}^2)^2.\ee
Values of $\Delta_T$ far greater than unity indicate large
fine-tuning. As shown in Fig.~\ref{f.gs}(b) fine-tuning typically
mild for either prior measure. The log prior has a lower $\Delta_T$
because SUSY breaking terms are much reduced compared to the linear
prior case.

\begin{figure}[!ht]
  \begin{tabular}{ll}
    \includegraphics[angle=0,
      width=.23\textwidth]{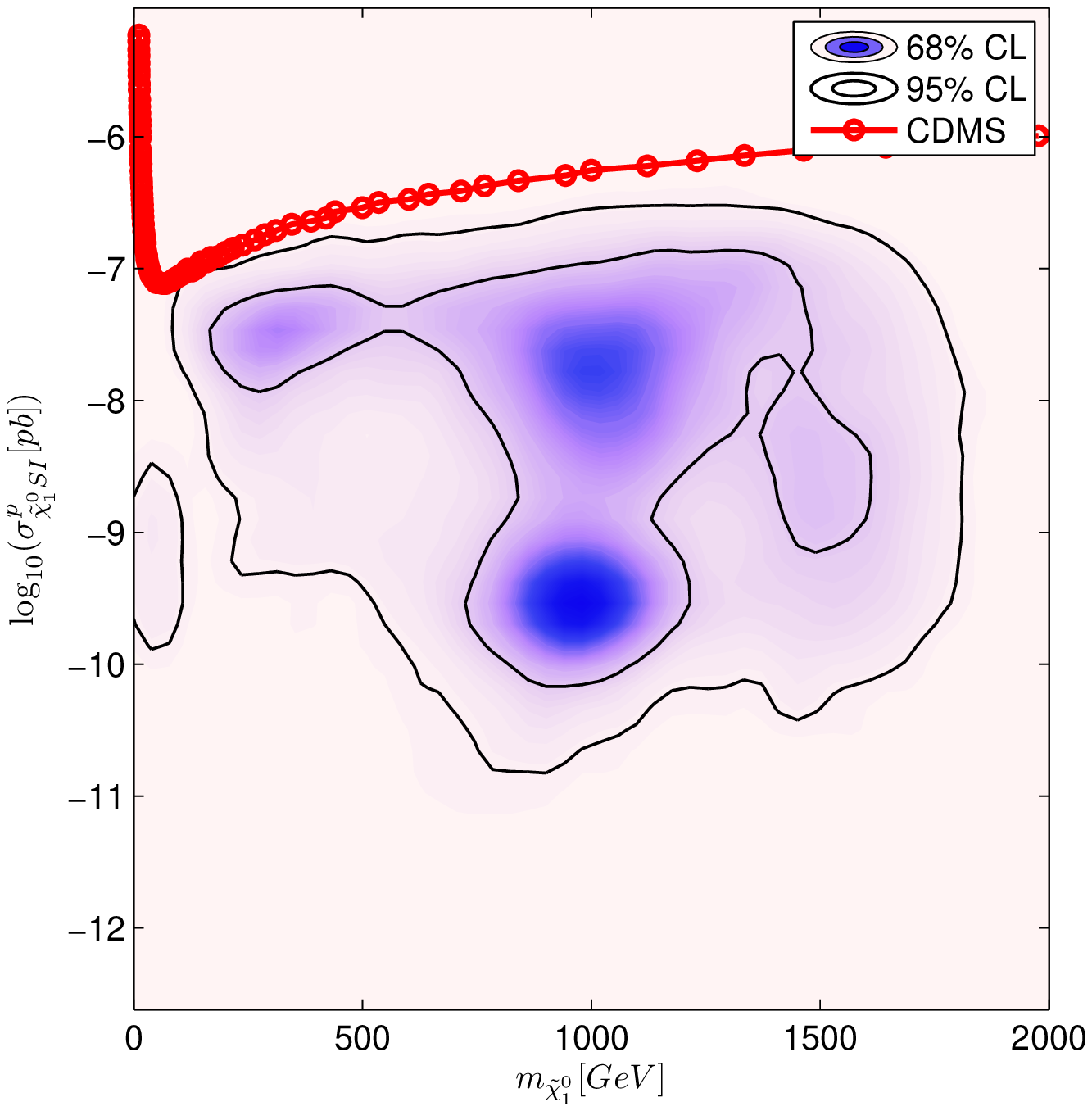}
    \includegraphics[angle=0,
      width=.23\textwidth]{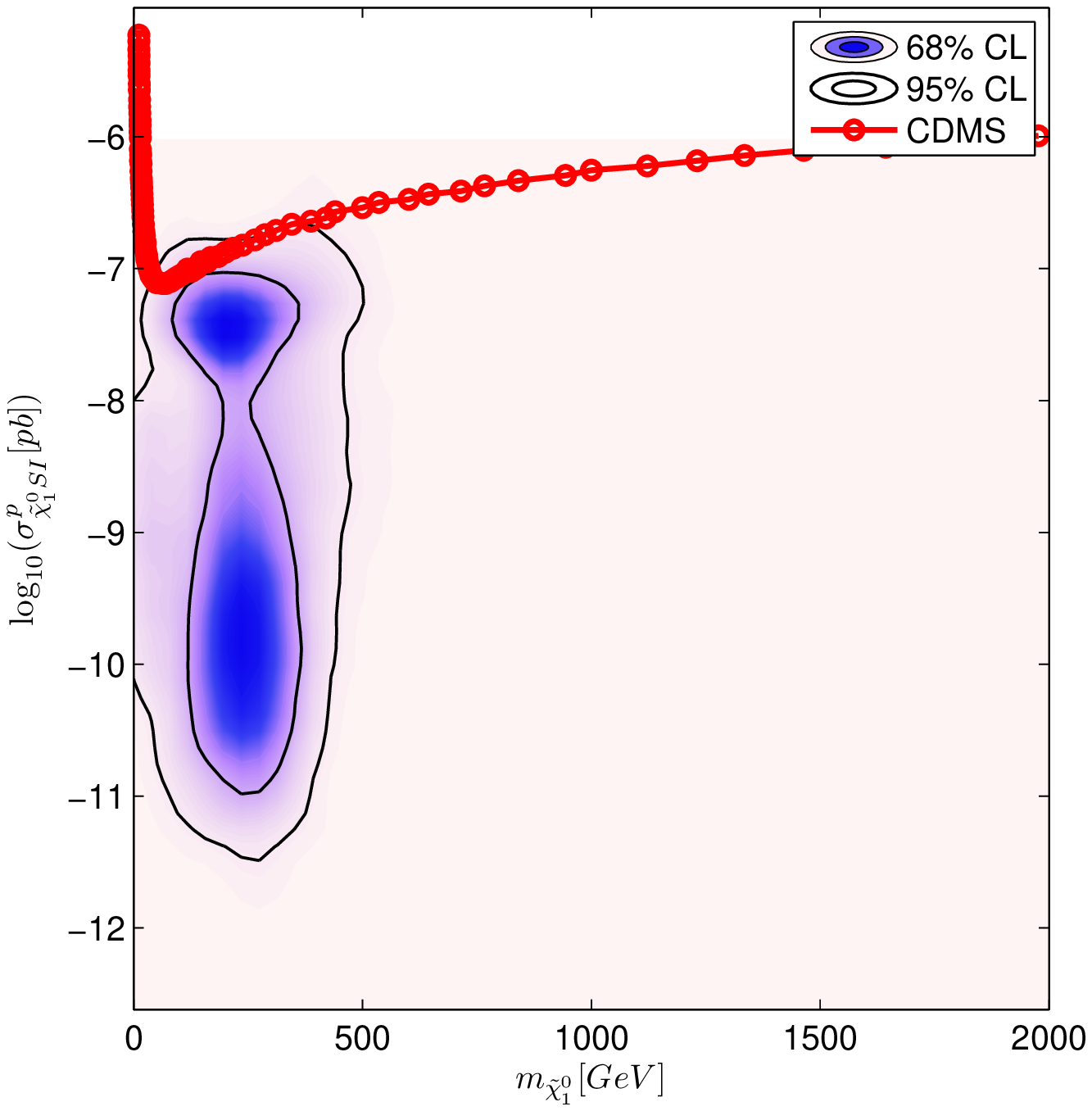}
  \end{tabular}
 \caption{Posterior PDFs of the
 neutralino-proton spin-independent scattering cross-section for the pMSSM
 with linear (left) and log (right) prior measures. The CDMS
 90\% confidence level upper bound is also shown.} %, assuming a local DM density
% of 0.3 GeV/cm$^{3}$.}
\label{fig.ddetect}
\end{figure}
The LSP neutralino mass eigenstate is a mixture of bino
($\tilde{b}$), wino($\tilde{w}$) and Higgsino ($\tilde{H_{1,2}^0}$)
gauge fields with real coefficients \be \label{neut1}
\tilde{\chi}_1^0 = N_{11}\tilde{b} + N_{12}\tilde{w}^3 +
   N_{13}\tilde{H_1^0} + N_{14}\tilde{H_2^0}. \ee
Gaugino/Higgsino admixture PDFs of the LSP are shown in
Fig.~\ref{f.gs}(c). The LSP is almost purely Higgsino with a
neutralino-chargino dominant co-annihilation channel in the linear
prior case. For log priors, it is an admixture of mostly gaugino and
to a much lesser extent Higgsino with a neutralino-slepton dominant
co-annihilation channel. Thus, current data do not unambiguously
constrain the LSP.

An independent sampling with the WMAP relic density constraint used
only as an upper bound (i.e.\ allowing for non LSP DM components)
favours very low DM relic densities, typically in the range
$\Omega_{CDM} h^2=10^{-2}-10^{-3}$, compared to the case of purely
LSP DM assumption, as shown in Fig.~\ref{f.gs}(d). Thus once one
allows an additional component of DM to the LSP, the model prefers
the additional component to dominate the relic density.

The pMSSM global fit results are also consistent with DM direct
detection bounds although current data are insufficient to constrain
the direct detection cross-sections. The constraint from the
cryogenic cold dark matter search (CDMS) experiments on the pMSSM is
shown in Fig.~\ref{fig.ddetect}.

The strong prior dependence of the fits is a measure of the
insufficient information from experimental data to derive robust
results about the preferred SUSY parameter space. In any case this
type of study would be most relevant and needed in the near future
once more information about possible SUSY extension of the SM is
obtained at LHC experiments.

\section{Mass and Spin Measurement with the Transverse Mass Variable $M_{T2}$}

\noindent
{\it Kiwoon Choi}\\
%Abstact: We discuss the recently proposed $M_{T2}$-kink method to measure the
%sparticle  masses  in hadron collider events with missing energy. We
%also introduce a new kinematic variable, the
%$M_{T2}$-Assisted-On-Shell (MAOS) momentum, which can be useful for
%spin measurement of new particles produced at the LHC.\\
\subsection{Introduction}
$R$-parity conserving  supersymmetry (SUSY) predicts a clear
signature at the LHC: excessive multi-jet (possibly with isolated
leptons) events with a large missing transverse momentum, which are
due to pair-produced squarks or gluinos subsequently decaying to the
invisible lightest supersymmetric particle (LSP) through
model-dependent decay chains. It is highly challenging to determine
the masses and spins of sparticles in such events because of missing
kinematic information. In recent years, several methods to measure
the unknown masses or spins in hadron collider events with missing
energy have been proposed \cite{review}. Here we briefly discuss the
methods that rely on the kinematic variable $M_{T2}$
\cite{lester,mt2kink1,mt2kink2,maos,higgsmaos}.

\subsection{$M_{T2}$ Kink}
A typical SUSY event at the LHC takes the form: \bea \label{susy} &&
Y(p+k)
+\bar{Y}(q+l)+U(u)\nonumber \\
&\rightarrow& V(p)\chi(k)+V(q)\chi(l)+U(u),\eea where $Y\hskip
-0.05cm+\bar{Y}$ denote pair-produced (mother) sparticles each of
which decays to a set of visible SM particles ($V$) and the
invisible LSP ($\chi$), and $U$ stands for visible particles not
coming from $Y+\bar{Y}$.  The event variable $M_{T2}$ \cite{lester}
is defined as
\begin{equation}
M_{T2}= \min_{\mathbf{k}_{T}+\mathbf{l}_{T}=\mathbf{p}_T^{\rm miss}}
\Big[ \mathrm{max}\Big\{M_{T}(Y), M_{T}(\bar{Y})\Big\}\Big],
\label{mt2_def}
\end{equation}
where the transverse mass is given by
$M_T^2(Y)=p^2+\tilde{m}_\chi^2+2E_T(p)E_T(k)-2{\bf p}_T\cdot {\bf
k}_T$ for generic trial LSP mass $\tilde{m}_\chi$ and trial
transverse momentum ${\bf k}_T$, $E_T^2(p)={p^2+|{\bf p}_T|^2}$,
$E_T^2(k)={\tilde{m}_\chi^2+|{\bf k}_T|^2}$,
%for the transverse momenta ${\bf p}_T$ and ${\bf k}_T$,
and the missing transverse momentum is given by  $\mathbf{p}_T^{\rm
miss}=-(\bf{p}_T+\bf{q}_T+\bf{u}_T)$.  For each event, the
corresponding  $M_{T2}(\tilde{m}_\chi)$ is a monotonically
increasing function of $\tilde{m}_\chi$, and its value at
$\tilde{m}_\chi=m_\chi$ is bounded as \bea
M_{T2}(\tilde{m}_\chi=m_\chi)\leq m_Y,\eea where $m_Y$ and $m_\chi$
are the true masses of $Y$ and $\chi$, respectively. Then,
generically there can be multiple events that saturate the above
upper bound at $\tilde{m}_\chi=m_\chi$, but have different values of
$(dM_{T2}/d\tilde{m}_\chi)_{\tilde{m}_\chi=m_\chi}$. This simple
observation implies that the endpoint values of $M_{T2}$ generically
exhibit  a kink \cite{mt2kink1,mt2kink2} at
$(M_{T2},\tilde{m}_\chi)=(m_Y,m_\chi)$.
%\cite{mt2kink1,mt2kink2}.

\begin{figure}[ht!]
\begin{center}
\includegraphics[width=13pc]{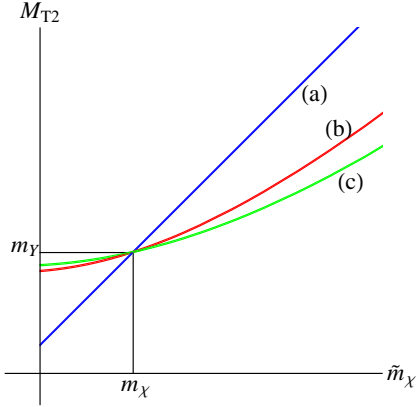}
\end{center}
\vskip -1.3cm \caption{$M_{T2}(\tilde{m}_\chi)$ showing a kink at
$\tilde{m}_\chi=m_\chi$.} \label{kink_origin}
\end{figure}

In Fig.~\ref{kink_origin}, we depict $M_{T2}(\tilde{m}_\chi)$ for
some events with $p^2=q^2$, ${\bf p}_T={\bf q}_T$ and
$M_{T2}(\tilde{m}_\chi=m_\chi)=m_Y$, when $m_Y/m_\chi=6$. The curve
(a) stands for an event with ${\bf u}_T=0$, $p^2=(m_Y-m_\chi)^2$,
(b) is for ${\bf u}_T=p^2=0$, and (c) is for $|{\bf u}_T|=m_Y$,
$p^2=0$. They show that the kink can be sharp enough if $V(p)$ is a
multi-particle state having a wide range of $p^2$ and/or $|{\bf
u}_T|$ is large enough to be of ${\cal O}(m_Y)$. There are some
cases known to give a visible kink \cite{mt2kink1,mt2kink2}, e.g.
(i) $Y=\tilde{g}\rightarrow q\bar{q}\chi$ in heavy sfermion
scenario, for which $0\leq {p^2}\leq (m_{\tilde{g}}-m_\chi)^2$, and
(ii) $Y=\chi_2\rightarrow \ell\bar{\ell}\chi$ for which a large
${\bf u}_T$ is provided by the gluino/squark decay producing
$\chi_2$. It remains to be seen if
%Although kink is quite generic, it requires a detailed
%study to see if
the $M_{T2}$-kink method can be applied to a wider class of SUSY
events.
%implemented in each case.

%\bea M_{T2}^{\rm
%max}(\tilde{m_\chi})= {\rm max}_{\mbox{all events}}\Big[
%M_{T2}(p,q;\tilde{m}_\chi)\Big] \eea
%\[ M_{T2}^{\rm max}(\tilde{m}_\chi)=\max_{p,q,u}
%\Big[\,M_{T2}(p^2,{\bf p}_T,q^2,{\bf q}_T,{\bf
%u}_T;\tilde{m}_\chi)\,\Big],
%\]

\subsection{MAOS Momentum}

\begin{figure}[ht!]
\begin{center}
\includegraphics[width=15pc]{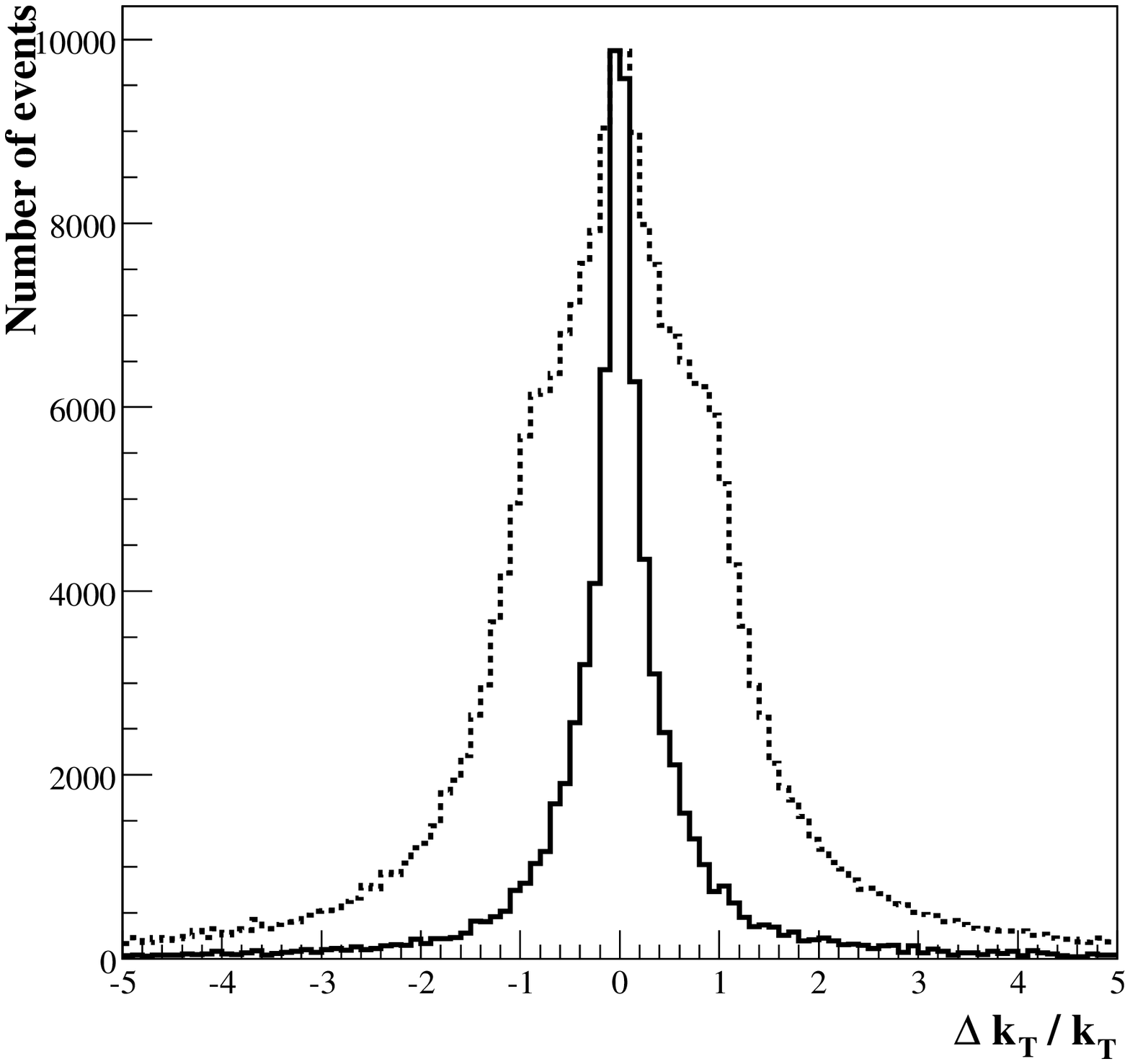}
\end{center}
\vskip -1.3cm \caption{Distribution of $\Delta{\bf k}_T/{\bf
k}_T^{\rm true}$.} \label{susy_dalitz}
\end{figure}

The $M_{T2}$-Assisted-On-Shell (MAOS) momentum is an event variable
designed to systematically  approximate the invisible LSP momentum
in the SUSY event (\ref{susy}) \cite{maos}. The transverse
components, ${\bf k}_T^{\rm maos}$ and ${\bf l}_T^{\rm maos}$,
correspond to the trial LSP transverse momenta which determine
$M_{T2}$, while the longitudinal and energy components are
determined by the on-shell conditions:
 \bea
\label{maos1} && \hskip -0.65cm k^2_{\rm maos}=l^2_{\rm
maos}=m_\chi^2,\nonumber
\\
&&\hskip -0.65cm (p+k_{\rm maos})^2=(q+l_{\rm maos})^2=m_Y^2.
  \eea
An interesting feature of the MAOS momentum is that
%For the endpoint
%events of balanced $M_{T2}$ \cite{lester,mt2kink1}, i.e. the events
%with $M_{T2}=M_T(Y)=M_T(\bar{Y})=m_Y$, one finds $k^{\rm maos}_\mu =
%k^{\rm true}_\mu$ and $l^{\rm maos}_\mu=l^{\rm true}_\mu.$  On the
%other hand,  for the endpoint events of unbalanced $M_{T2}$ having
%$M_T(\bar{Y})<M_T(Y)=M_{T2}=m_Y$, only $k_\mu^{\rm maos}=k_\mu^{\rm
%true}$.
it corresponds to the true LSP momentum for the endpoint events of
$M_{T2}$. This indicates that the MAOS momentum might provide a
reasonable approximation to the true LSP momentum even for generic
non-endpoint events, and the approximation can be systematically
improved by selecting an event subset near the $M_{T2}$ endpoint.
This can be confirmed by Fig. (\ref{susy_dalitz})
 showing the distribution of \bea
{\Delta {\bf k}_T}/{{\bf k}^{\rm true}_T} \equiv ({{\bf k}^{\rm
maos}_T-{\bf k}^{\rm true}_T})/{{\bf k}_T^{\rm true}}\eea for the
gluino pair decays: $\tilde{g}+\tilde{g}\rightarrow q\bar{q}\chi+
q\bar{q}\chi$ in the focus point scenario of mSUGRA. Here the dotted
line is the distribution
%of $\Delta{\bf k}_T/{\bf k}_T^{\rm true}$
over the full event set, while the solid line is the distribution
over the 10\% subset near the $M_{T2}$ endpoint.

In certain cases, the MAOS momenta can be useful for spin
measurement \cite{maos}. For instance, for the 3-body decay
$\tilde{g}\rightarrow q(p_1)\bar{q}(p_2)\chi(k)$, the gluino spin
can be easily read off from  $d\Gamma/dsdt_{\rm maos}$, where
$s=(p_1+p_2)^2$ and $t_{\rm maos}=(p_1+k_{\rm maos})^2$ or
$(p_2+k_{\rm maos})^2$. One can also determine the slepton spin with
the production angle distribution obtained from $p+k_{\rm maos}$ for
$q\bar{q}\rightarrow Z^0/\gamma\rightarrow
\tilde{\ell}+\tilde{\ell}^* \rightarrow \ell(p)\chi(k)
+\bar\ell(q)\chi(l)$. MAOS momentum can be used also to determine
the Higgs boson mass  in $H\rightarrow WW\rightarrow
\ell\nu\bar{\ell}\nu$ \cite{higgsmaos}.

%%%%%%%%%%%%%%%%%%%%%%%%%%%%%%%%%%%%%%%%%%%%%%%%%%%%%%%%%%%%%%%%%%%%%%%%%%%%%%%%%%%%%%%%%%%%%%
%%%%%%%%%%%%%%%%%%%%%%%%%%%%%%%%%%%%%%%%%%%%%%%%%%%%%%%%%%%%%%%%%%%%%%%%%%%%%%%%%%%%%%%%%%%%%%
\chapter{Higgs Physics}
\epigraphhead[20]{\epigraph{\large {\em O.~Buchmueller,
R.~Cavanaugh, S.~Chang, S.~Dawson, A.~De Roeck, M.~D\"uhrssen,
J.R.~Ellis, D.~Feldman, H.~Fl\"acher, T.~Han, S.~Heinemeyer,
G.~Isidori, R.~Lafaye, M. Lisanti, Z.~Liu, M.M.~M\"uhlleitner,
P.~Nath, K.A.~Olive, T.~Plehn, M.~Rauch, F.J.~Ronga, M.~Spira,
J.~Wacker, G.~Weiglein, D.~Zeppenfeld, D.~Zerwas}}{\large Dirk
Zerwas (Convener)}}
%
%%%%%%%%%% espcrc2.tex %%%%%%%%%%
%
% $Id: espcrc2.tex 1.2 2000/07/24 09:12:51 spepping Exp spepping $
%
%\documentclass[fleqn,twoside]{article}
%\usepackage{amsmath,amssymb,espcrc2}

% change this to the following line for use with LaTeX2.09
% \documentstyle[twoside,fleqn,espcrc2]{article}

% if you want to include PostScript figures
%\usepackage{psfrag}
%\usepackage{graphicx}
%\usepackage{epsfig}
% if you have landscape tables
%\usepackage[figuresright]{rotating}

\newcommand{\gsimST}{\raisebox{-0.13cm}{~\shortstack{$>$ \\[-0.07cm] $\sim$}}~}
\newcommand{\lsimST}{\raisebox{-0.13cm}{~\shortstack{$<$ \\[-0.07cm] $\sim$}}~}
\newcommand{\ifb}{{\ensuremath\rm fb^{-1}}}
\newcommand{\ie}{{\sl i.e.\ }}
\newcommand{\eg}{{\sl e.g. }}
\newcommand{\tevTSG}{{\ensuremath\rm TeV}}
\newcommand{\gevTSG}{{\ensuremath\rm GeV}}
\newcommand{\mevTSG}{{\ensuremath\rm MeV}}
\def\chaPN{\widetilde{\chi}^{\pm}_1}
\def\naPN{\chi}
\def\staPN{\widetilde{\tau}_1}
\def\hcPN{H^{\pm}}

% From Lisanti
\newcommand{\OO}{\mathcal{O}}
\newcommand{\LL}{\mathcal{L}}
\newcommand{\DO}{{\mbox{DO\hspace{-0.105in}$\not$\hspace{0.11in}}}}
\newcommand{\MET}{\mbox{$E_T\hspace{-0.21in}\not\hspace{0.15in}$}}

\hyphenation{author another created financial paper re-commend-ed Post-Script}

{
\newcommand{\cp}{{\cal CP}}
\newcommand{\lsim}
{\;\raisebox{-.3em}{$\stackrel{\displaystyle <}{\sim}$}\;}
\newcommand{\gsim}
{\;\raisebox{-.3em}{$\stackrel{\displaystyle >}{\sim}$}\;}
\newcommand{\gmt}{$(g-2)_\mu$}
\newcommand{\br}{{\rm BR}}
\newcommand{\bsg}{BR($b \to s \gamma$)}
\newcommand{\btn}{BR($B_u \to \tau \nu_\tau$)}
\newcommand{\bmm}{BR($B_s \to \mu^+\mu^-$)}
\newcommand{\ssi}{\sigma^{\rm SI}_p}
\newcommand{\Och}{\ensuremath{\Omega_\chi h^2}}
\newcommand{\sweff}{\sin^2\theta_{\mathrm{eff}}}
\newcommand{\MW}{M_W}
\newcommand{\MZ}{M_Z}
\newcommand{\Mh}{M_h}
\newcommand{\MA}{M_A}
\newcommand{\MH}{M_H}
\newcommand{\MHp}{M_{H^\pm}}
\newcommand{\mt}{m_t}
\newcommand{\mgl}{m_{\tilde g}}
\renewcommand{\cha}[1]{\tilde \chi^\pm_{#1}}
\newcommand{\mcha}[1]{m_{\tilde \chi^\pm_{#1}}}
\newcommand{\neu}[1]{\tilde \chi^0_{#1}}
\newcommand{\mneu}[1]{m_{\tilde \chi^0_{#1}}}
\newcommand{\mste}{m_{\tilde t_1}}
\newcommand{\mstaue}{m_{\staue}}
\newcommand{\staue}{\tilde \tau_1}
\renewcommand{\tb}{\tan\beta}
\newcommand{\ecm}{\sqrt{s}}
\newcommand{\tev}{\,\, \mathrm{TeV}}
\newcommand{\gev}{\,\, \mathrm{GeV}}
\newcommand{\mev}{\,\, \mathrm{MeV}}

\maketitle
%%%%%%%%%%%%%%%%%%%%%%%%%%%%%%%%%%%%%%%%%%%%%%%%%%%%%%%%%%%%%%%%%%%%%%%%%%%%%%
%%%%%%%%%%%%%%%%%%%%%%%%%%%%%%%%%%%%%%%%%%%%%%%%%%%%%%%%%%%%%%%%%%%%%%%%%%%%%%

\section{Predictions for SUSY Higgses at the LHC}
{\it O.~Buchmueller, R.~Cavanaugh, A.~De Roeck, J.R.~Ellis, H.~Fl\"acher, S.~Heinemeyer, G.~Isidori, K.A.~Olive, F.J.~Ronga and  G.~Weiglein}\medskip

\label{sec:HiggsPred}
%\section{INTRODUCTION}

One of the main goals of the LHC is the identification of the mechanism
of electroweak symmetry breaking. The most frequently investigated
models are the Higgs mechanism within the Standard
Model (SM) and within the Minimal Supersymmetric Standard Model
(MSSM)~\cite{mssm}. Contrary to the case of the SM, in the MSSM
two Higgs doublets are required.
This results in five physical Higgs bosons instead of the single Higgs
boson in the SM. These are the light and heavy $\cp$-even Higgs bosons, $h$
and $H$, the $\cp$-odd Higgs boson, $A$, and the charged Higgs bosons,
$H^\pm$.
The Higgs sector of the MSSM can be specified at lowest
order in terms of the gauge couplings, the ratio of the two Higgs vacuum
expectation values, $\tb \equiv v_2/v_1$, and the mass of the $\cp$-odd
Higgs boson, $\MA$.
Consequently, the masses of the $\cp$-even neutral and the charged Higgs
bosons are dependent quantities that can be
predicted in terms of the Higgs-sector parameters.
Higgs phenomenology
in the MSSM is strongly affected by higher-order corrections, in
particular from the sector of the third generation quarks and squarks,
so that the dependencies on various other MSSM parameters can be
important, see e.g.\ \cite{PomssmRep,habilSH,mhiggsAWB} for reviews.
The mass of the lightest Higgs boson is bounded from above by
$\Mh \lsim 135 \gev$~\cite{mhiggsAEC}. For $\MA \gsim 150 \gev$
the other Higgs bosons are close to each other,
$\MA \approx \MH \approx \MHp$.

Predictions for the MSSM Higgs bosons masses, which are needed to evaluate the
LHC discovery potential, are bedeviled by the large dimensionality of
the MSSM. For this reason, simplifying assumptions that may be more or
less well motivated are often made, so as to reduce the parameter space to
a manageable dimensionality. We focus here on the
framework of the constrained MSSM (CMSSM), in which the soft
supersymmetry-breaking scalar and gaugino masses are each assumed to be
equal at some GUT input scale. In this case, the new independent MSSM
parameters are just four in number: the universal gaugino mass $m_{1/2}$,
the scalar mass $m_0$, the trilinear soft supersymmetry-breaking parameter
$A_0$, and the ratio $\tb$ of Higgs vacuum expectation values. The
pseudoscalar Higgs mass $\MA$ and the magnitude of the Higgs mixing
parameter $\mu$ can be determined by using the electroweak vacuum
conditions, leaving the sign of $\mu$ as a residual ambiguity.
An extension of the CMSSM is obtained in the NUHM1
in which the soft SUSY-breaking
contributions to the Higgs masses are allowed a different but common
value with respect to the scalar fermion mass parameter
$m_0$. Effectively this yields either $\MA$ or $\mu$ as an additional
free parameter at the electroweak (EW) scale.

%%%%%%%%%%%%%%%%%%%%%%%%%%%%%%%%%%%%%%%%%%%%%%%%%%%%%%%%%%%%%%%%%%%%%%%%%%%%%%
%%%%%%%%%%%%%%%%%%%%%%%%%%%%%%%%%%%%%%%%%%%%%%%%%%%%%%%%%%%%%%%%%%%%%%%%%%%%%%

\subsection{Frequentist Fit}

We will review the results for the predictions of Higgs boson masses and
other properties of the Higgs sector in the CMSSM and NUHM1, based on a
frequentist approach~\cite{Master3}.
In our frequentist analysis we use the Markov chain Monte Carlo (MCMC)
technique to sample efficiently the CMSSM and NUHM1 parameter spaces,
and we generate sufficiently many chains to sample these parameter
spaces completely.

Our treatments of the experimental constraints from electroweak
precision observables, $B$-physics observables and cosmological data
are explained in detail in~\cite{Master3,Master1,Master2}.
We define a global $\chi^2$ likelihood function, which combines all
theoretical predictions with experimental constraints:
\begin{eqnarray}
\chi^2 &=& \sum^N_i \frac{(C_i - P_i)^2}{\sigma(C_i)^2 +
\sigma(P_i)^2} \nonumber \\
%\nonumber \\[.2em]
%&+ {\chi^2(\Mh) + \chi^2(\br(B_s \to \mu\mu))}
%\nonumber \\[.2em]
& &+ \sum^M_i \frac{(f^{\rm obs}_{{\rm SM}_i}
              - f^{\rm fit}_{{\rm SM}_i})^2}{\sigma(f_{{\rm SM}_i})^2}
\label{eqn:chi2}
\end{eqnarray}
Here $N$ is the number of observables studied, $C_i$ represents an
experimentally measured value (constraint) and each $P_i$ defines a
prediction for the corresponding constraint that depends on the
supersymmetric parameters. The constraints comprise a variety of
electroweak precision observables (e.g., $\MW$, $A_\ell({\rm SLD})$
and $A_{\rm fb}(b)$(LEP), $(g-2)_\mu$, $\Mh$, \ldots), flavour
related observables (e.g., \bsg, \bmm, \ldots) and the relic
abundance of cold dark matter (CDM), \Och, see~\cite{Master3} for
details. The experimental uncertainty, $\sigma(C_i)$, of each
measurement is taken to be both statistically and systematically
independent of the corresponding theoretical uncertainty,
$\sigma(P_i)$, in its prediction.
%We denote by
%$\chi^2(\Mh)$ and $\chi^2(\br(B_s \to \mu\mu))$ the $\chi^2$
%contributions from the two measurements for which only one-sided
%bounds are available so far.
%We stress that, as in~\cite{Master1,Master2},
The three standard model parameters
$f_{\rm SM} = \{\Delta\alpha_{\rm had}, \mt, \MZ \}$ are included as fit
parameters and allowed to vary with their current experimental
resolutions $\sigma(f_{\rm SM})$. We do not
include $\alpha_s$, which would have only a minor impact on
the analysis.

%%%%%%%%%%%%%%%%%%%%%% F I G U R E %%%%%%%%%%%%%%%%%%%%%%%%%%%%%%%%%%%
\begin{figure*}[htb]
\vspace{-2em}
%%%%%%%%%%%%%%%%%%%%%%%%%%%%%%%
\resizebox{8cm}{!}{\includegraphics{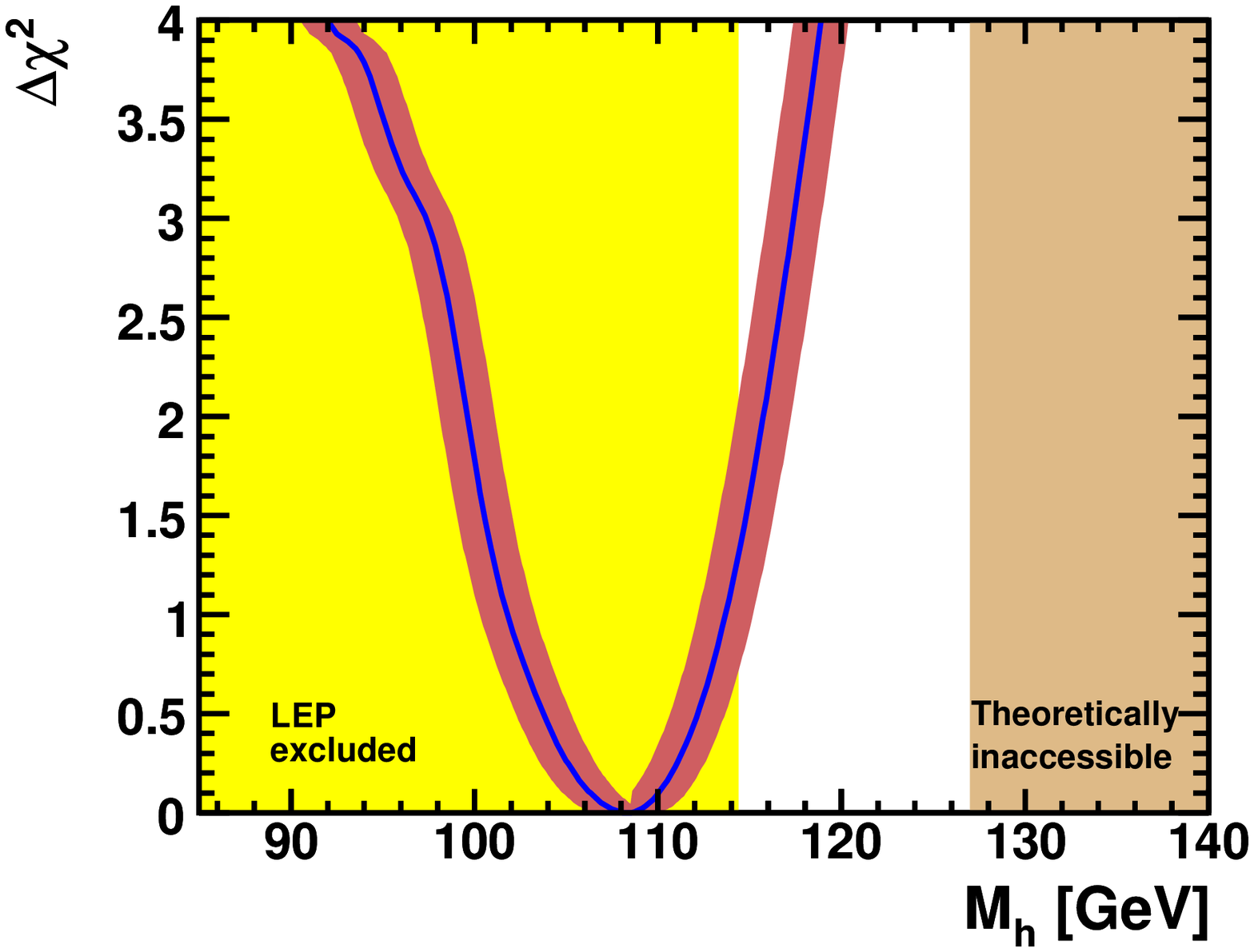}}
\resizebox{8cm}{!}{\includegraphics{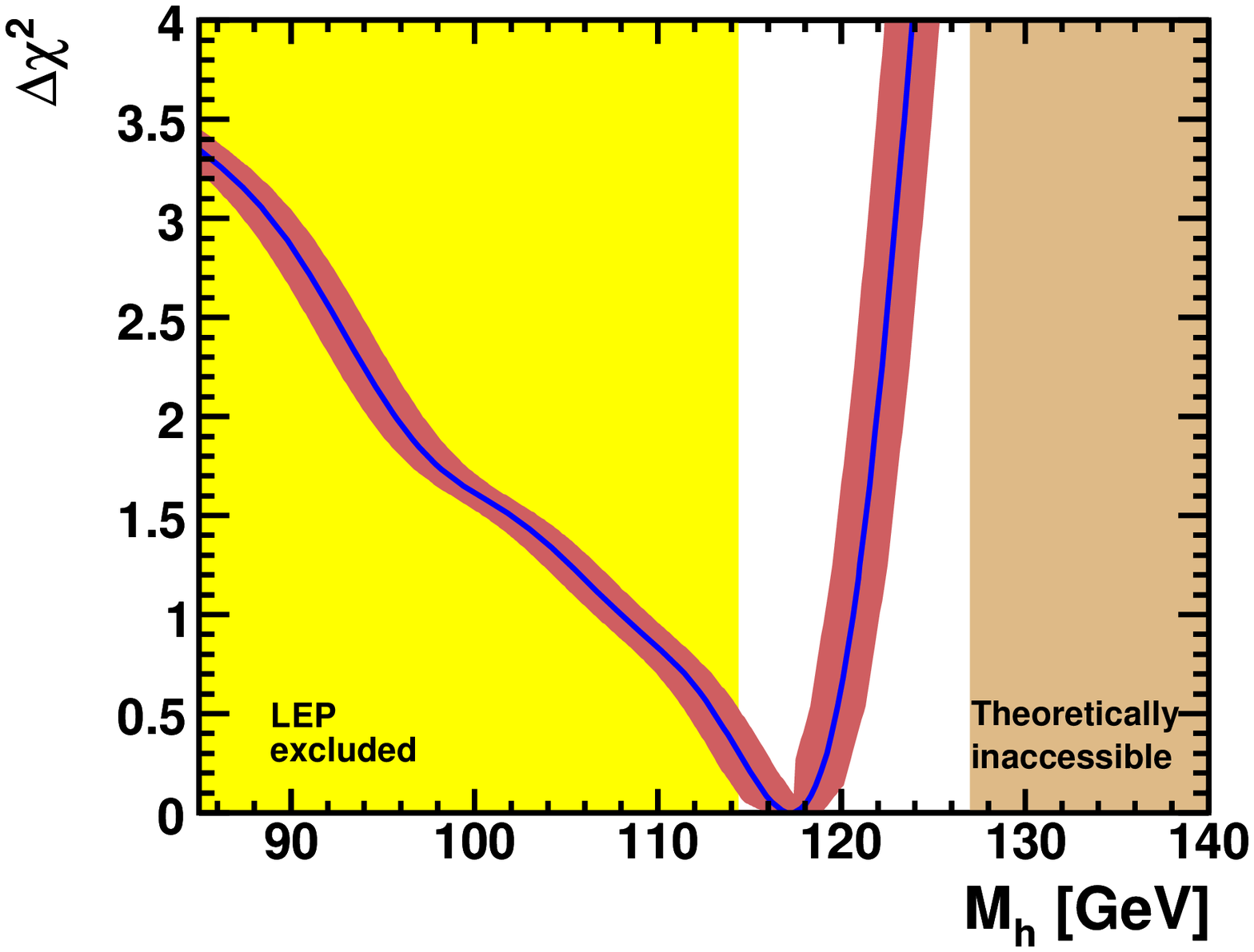}}
%%%%%%%%%%%%%%%%%%%%%%%%%%%%%%%
\vspace{-1cm}
\caption{The $\chi^2$ functions for $\Mh$ in the CMSSM (left) and
  the NUHM1 (right),
including the theoretical uncertainties (red bands). Also shown is the mass
range excluded for a SM-like Higgs boson (yellow shading),
and the ranges theoretically inaccessible in the supersymmetric models studied.
}
\label{fig:redband}
\end{figure*}
%%%%%%%%%%%%%%%%%%%%%% F I G U R E %%%%%%%%%%%%%%%%%%%%%%%%%%%%%%%%%%%

The numerical evaluation of the frequentist likelihood function
using these constraints has been performed with the
{\tt MasterCode}~\cite{Master3,Master1,Master2},
which includes several up-to-date codes for the calculations/evaluations
in the various sectors, see~\cite{Master3} for a complete list of codes
and references.

%%%%%%%%%%%%%%%%%%%%%%%%%%%%%%%%%%%%%%%%%%%%%%%%%%%%%%%%%%%%%%%%%%%%%%%%%%%%%%
%%%%%%%%%%%%%%%%%%%%%%%%%%%%%%%%%%%%%%%%%%%%%%%%%%%%%%%%%%%%%%%%%%%%%%%%%%%%%%

\subsection{Results for $\mathbf{\Mh}$}

%%%%%%%%%%%%%%%%%%%%%% F I G U R E %%%%%%%%%%%%%%%%%%%%%%%%%%%%%%%%%%%
\begin{figure*}[htb]
%%%%%%%%%%%%%%%%%%%%%%%%%%%%%%%
\resizebox{8cm}{!}{\includegraphics{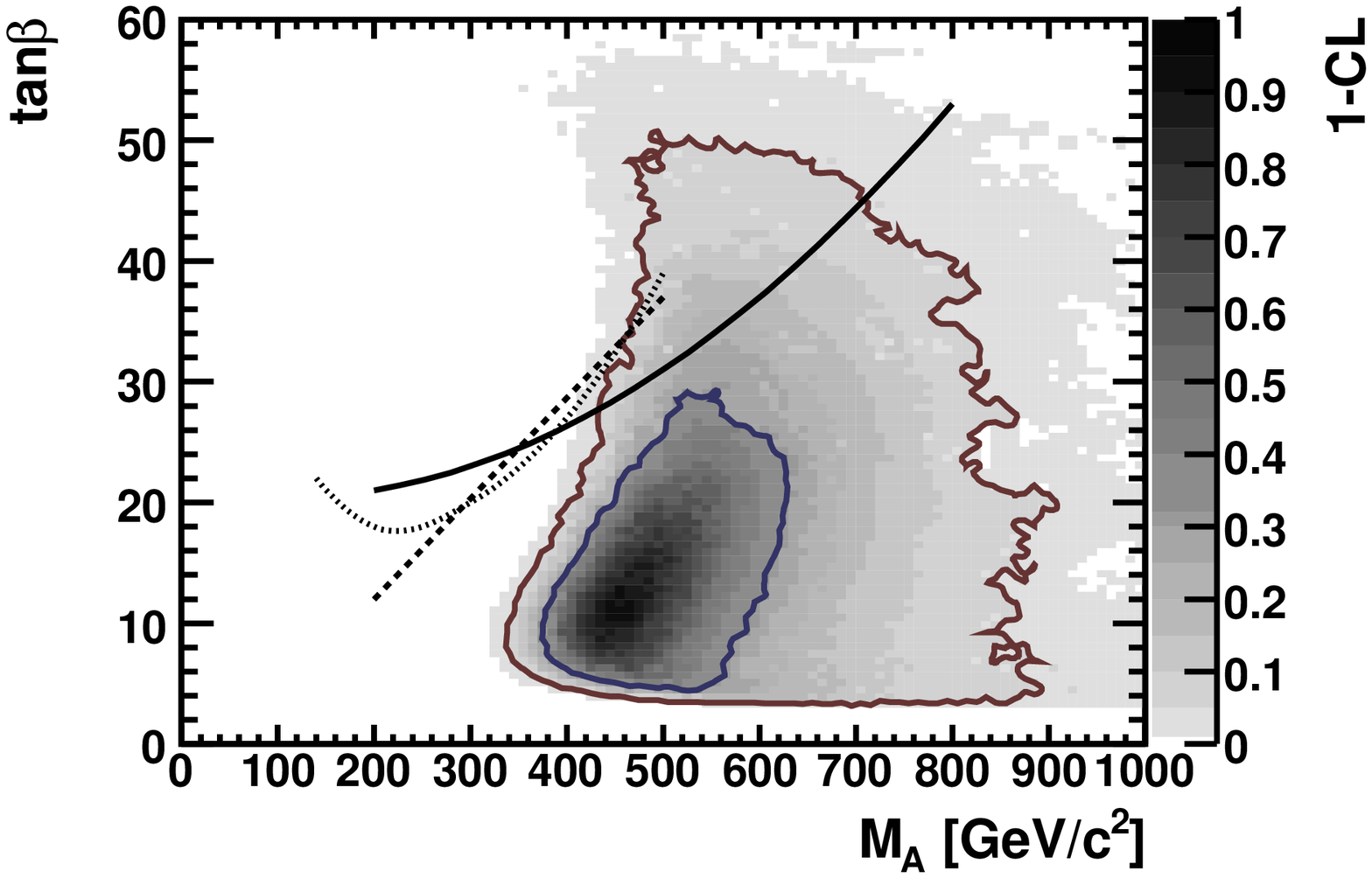}}
\resizebox{8cm}{!}{\includegraphics{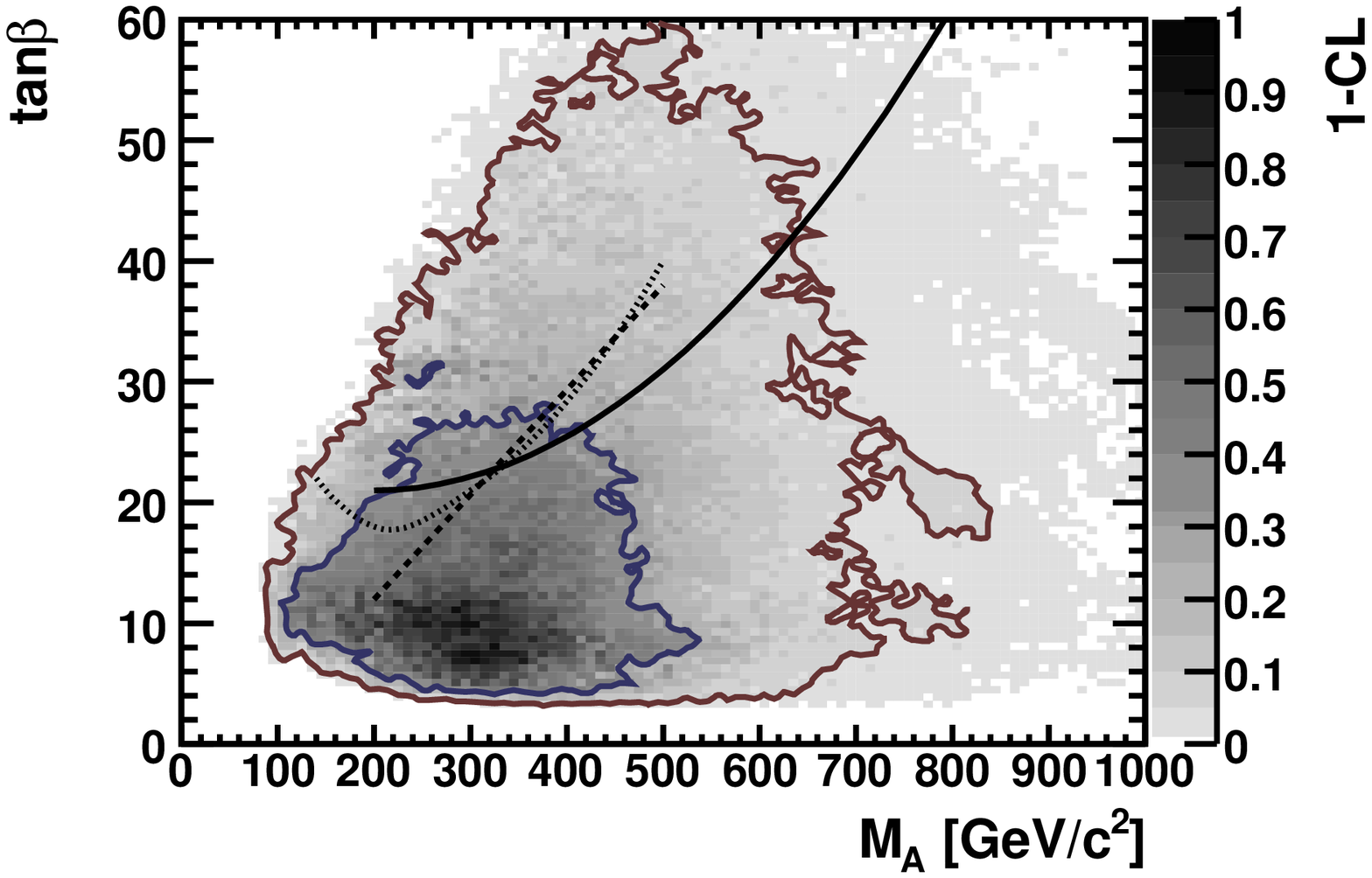}}
%%%%%%%%%%%%%%%%%%%%%%%%%%%%%%%
\vspace{-4em}
\caption{The correlations between $\MA$ and $\tb$
in the CMSSM (left panel) and in the NUHM1 (right panel).
Also shown are the 5-$\sigma$ discovery contours
for observing the heavy MSSM Higgs bosons $H, A$
in the three decay channels
$H,A \to \tau^+\tau^- \to {\rm jets}$ (solid line),
$\rm{jet}+\mu$ (dashed line), $\rm{jet}+e$ (dotted line)
at the LHC. The discovery contours have been obtained using an
analysis that assumed 30 or 60~fb$^{-1}$
collected with the CMS detector~\cite{Ball:2007zza,cmsHiggs}.}
\label{fig:mAtb}
%\vspace{3em}
\end{figure*}
%%%%%%%%%%%%%%%%%%%%%% F I G U R E %%%%%%%%%%%%%%%%%%%%%%%%%%%%%%%%%%%

We start the discussion of predictions by showing
the likelihood functions for $\Mh$ within the CMSSM and NUHM1
frameworks obtained when dropping the
contribution to $\chi^2$ from the direct Higgs searches at LEP,
shown in the left and right panels of Fig.~\ref{fig:redband}, respectively.
%The left plot updates that for the CMSSM given in~\cite{Master1}.
%
It is well
known that the central value of the Higgs mass in a SM
fit to the precision electroweak data lies below
100~GeV~\cite{lepewwg},
but the theoretical and experimental uncertainties
in the SM fit are such that there is no significant
discrepancy with the direct lower limit of
114.4~GeV~\cite{Barate:2003sz,Schael:2006cr} derived
from searches at LEP. In the case of the CMSSM and NUHM1,
one may predict $\Mh$ on the basis of the underlying model
parameters, with a one-$\sigma$ uncertainty of 1.5~GeV~\cite{mhiggsAEC},
shown as a red band in Fig.~\ref{fig:redband}. Also shown in
Fig.~\ref{fig:redband} are the LEP exclusion (yellow shading)
and the ranges that are theoretically inaccessible in the
supersymmetric models studied (beige shading).
The LEP exclusion is directly applicable to the CMSSM,
since the $h$ couplings are essentially indistinguishable from
those of the SM Higgs boson~\cite{Ellis:2001qv,Ambrosanio:2001xb}.
The NUHM1 case is more involved, see Ref.~\cite{Master3} for
details.
%this is not
%necessarily the case in the NUHM1, as discussed earlier in this
%paper.

In the case of the CMSSM, we see in the left panel of
Fig.~\ref{fig:redband} that the minimum of the $\chi^2$
function occurs below the formal LEP lower limit. However, as in the
case of the SM, this discrepancy is not significant, and
a global fit including the LEP constraint has acceptable $\chi^2$.
In the case of the NUHM1, shown in the right panel of
Fig.~\ref{fig:redband}, we see that the minimum of the $\chi^2$
function occurs {\it above} the formal LEP lower limit. Thus,
within the NUHM1 the combination of all other experimental
contraints {\em naturally} evade the LEP Higgs constraints, and no
tension between $\Mh$ and the experimental bounds exist.

%%%%%%%%%%%%%%%%%%%%%%%%%%%%%%%%%%%%%%%%%%%%%%%%%%%%%%%%%%%%%%%%%%%%%%%%%%%%%%
%%%%%%%%%%%%%%%%%%%%%%%%%%%%%%%%%%%%%%%%%%%%%%%%%%%%%%%%%%%%%%%%%%%%%%%%%%%%%%

\subsection{Results for the Heavy Higgs Bosons}

Fig.~\ref{fig:mAtb} displays the favoured regions in the $(\MA, \tb)$ planes
for the CMSSM and NUHM1. We see that there is little
correlation between the two parameters in either the CMSSM or the NUHM1, though the
preferred range of $m_A$ is somewhat smaller in the latter model. Superposed
on the likelihood contours are the LHC reaches in various channels, based
on the production and decay modes discussed later. The contours shown
in Fig.~\ref{fig:mAtb} are based on
the analysis in~\cite{cmsHiggs}, which assumed 30
or 60~fb$^{-1}$ collected with the CMS detector,
evaluating radiative corrections using the soft
SUSY-breaking parameters of the best-fit points in the CMSSM and the
NUHM1, respectively.
We show in Fig.~\ref{fig:mAtb} the 5-$\sigma$
discovery contours for the three decay channels
$H,A \to \tau^+\tau^- \to {\rm jets}$ (solid lines), $\rm{jet}+\mu$ (dashed
lines) and $\rm{jet}+e$ (dotted lines).
The parameter regions above and to the left of the curves are within reach
of the LHC with about 30~fb$^{-1}$ of integrated luminosity.
We see that the highest-likelihood regions lie beyond this reach.

}

%%%%%%%%%%%%%%%%%%%%%%%%%%%%HIGGSPRODUCTION%%%%%%%%%%%%%%%%%%%%%%%%%%%%%%%%%%%

\section{Higgs Boson Production at the LHC}

{\it M. Spira and D. Zeppenfeld}\medskip

The production of the Higgs boson at the LHC will be discussed
in the following sections. First the status of the
calculations of the Standard Model production will be presented,
followed by the status of the production
in the MSSM.

\subsection{Standard Model}
\begin{figure}[htb]
\begin{picture}(130,250)(30,0)
%%%%%%%%%\put(-20,-150){\includegraphics[scale=0.6]{dia1.ps}}
\put(-20,-180){\includegraphics[scale=0.6]{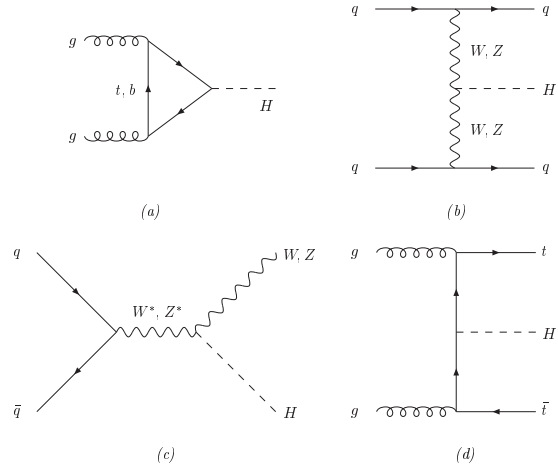}}
\end{picture} \\[-3.5cm]
\caption{\label{fg:lodiapro} Typical diagrams for all relevant Higgs
boson production mechanisms at leading order: {\it (a)} gluon fusion,
{\it (b)} vector boson fusion, {\it (c)} Higgs-strahlung, {\it (d)}
Higgs bremsstrahlung off top quarks.}
\vspace*{-0.5cm}
\end{figure}

%        ==============
The dominant Higgs production mechanism at the LHC will be the
gluon-fusion process $gg\to H$ \cite{glufus}.  This process is mediated
by top and bottom quark loops (see Fig.~\ref{fg:lodiapro}a). Due to the
large size of the top Yukawa couplings and the gluon densities gluon
fusion comprises the dominant Higgs boson production mechanism for the
whole Higgs mass range of interest.

The QCD corrections to the top and bottom quark loops are well known
including the full Higgs and quark mass dependences
\cite{gghnlo}. They increase the total cross section by $50-100\%$. The
limit of very heavy top quarks provides an approximation within $\sim
10\%$ for all Higgs masses \cite{gghnlo,limit,higgsrev,Djouadi:2005gi,mhiggsAWB}. In this limit
the NLO QCD corrections have been calculated before
\cite{gghnlo,htoggdecay1,gghnlolim} and more recently the NNLO QCD corrections
\cite{gghnnlo} with the latter increasing the total cross section
further by $\sim 20\%$. A full massive NNLO calculation is only partly
available \cite{gghnnlovirt}, so that the NNLO results can only be
trusted for small and intermediate Higgs masses. The approximate NNLO
results have been improved by a soft-gluon resummation at the
next-to-next-to-leading log (NNLL) level, which yields another increase
of the total cross section by $\sim 10\%$ \cite{gghresum}. Electroweak
corrections have been computed, too, and turn out to be small
\cite{gghelw1,gghelw2,gghqcdelw}. The theoretical uncertainties of the total
cross section can be estimated as $\sim 15\%$ at NNLO due to the
residual scale dependence, the uncertainties of the parton densities and
due to neglected quark mass effects.

At LO the Higgs boson does not acquire any transverse momentum in the
gluon fusion process, so that Higgs bosons with non-vanishing transverse
momentum can only be produced in the gluon fusion process, if an
additional gluon is radiated.  This contribution is part of the real NLO
corrections to the total gluon fusion cross section. The LO $p_T$
distribution of the Higgs boson is known including the full quark mass
dependence \cite{Ellis:1987xu,higgsptlo}. The NLO corrections, however, are only
known in the heavy quark limit, so that they can only be trusted for
small and moderate Higgs masses and $p_T$ \cite{higgsptnlo}. In this
limit a NLL soft gluon resummation has been performed \cite{higgsptnll},
which has recently been extended to the NNLL level \cite{higgsptnnll}
thus yielding a reliable description of the small $p_T$ range. It should
be noted that these results are only reliable, if the top quark loops
provide the dominant contribution and $p_T$ is not too large.  In the
regions where the NLO and resummed results are valid the theoretical
uncertainties have been reduced to $\sim 20\%$.
Higgs production cross sections in association with two jets, via
gluon fusion, have been calculated for full top and bottom quark
mass dependence at LO only \cite{DelDuca:2001fn}. Comparison with the
large top mass limit \cite{dawson} shows that the latter is again reliable
for not too large Higgs masses and jet $p_T$ (roughly below $m_t$).
Recently, also the NLO QCD corrections to the $Hjj$ cross section have
been calculated \cite{HjjNLO}, in the $m_t\to \infty$ limit.
They lead to a modest cross section increase of about 20 to 30\%
compared to the LO results.

For large Higgs masses the $W$ and $Z$ boson-fusion processes (see
Fig.~\ref{fg:lodiapro}b) $qq\to qq+W^*W^*/Z^*Z^*\to qqH$ become
competitive \cite{vvh}. These processes are relevant in the intermediate
Higgs mass range, too, since the additional forward jets offer the
opportunity to reduce the background processes significantly.  The NLO
QCD corrections turn out to be ${\cal O}(10\%)$ for the total cross
section \cite{vvhqcd,higgsrev,Djouadi:2005gi}. Quite recently the full NLO QCD and
electroweak corrections to the differential cross sections have been
computed, resulting in modifications of the relevant distributions
by up to $\sim 20\%$ \cite{vvhqcddist,vvhelw}. The residual
uncertainties are of ${\cal O}(5\%)$.

In the intermediate mass range $M_H\lsimST 2M_Z$ Higgs-strahlung off $W,Z$
gauge bosons (see Fig.~\ref{fg:lodiapro}c) $q\bar q\to Z^*/W^* \to H+
Z/W$ provides alternative signatures for the Higgs boson search
\cite{vhv}.  The NLO QCD corrections increase the total cross section by
${\cal O}(30\%)$ \cite{vhvqcd,higgsrev,Djouadi:2005gi}.  Recently this calculation has
been extended up to NNLO \cite{vhvnnlo}.  The NNLO corrections are
small. Moreover, the full electroweak corrections have been obtained in
Ref.~\cite{vhvelw} resulting in a decrease of the total cross sections
by $5-10\%$. The total theoretical uncertainty is of ${\cal O}(5\%)$.

Higgs radiation off top quarks (see Fig.~\ref{fg:lodiapro}d) $q\bar
q/gg\to Ht\bar t$ plays a role for smaller Higgs masses below $\sim 150$
GeV.  The LO cross section has been computed a long time ago \cite{htt}.
The full NLO QCD corrections have been calculated resulting in a
moderate increase of the total cross section by $\sim 20\%$ at the LHC
\cite{httqcd}. These results confirm former estimates based on an
effective Higgs approximation \cite{httapprox}.  The effects on the
relevant parts of final state particle distribution shapes are of
moderate size, i.e.  ${\cal O}(10\%)$, so that former experimental
analyses are not expected to alter much due to these results.

\subsection{Minimal Supersymmetric Extension}
%        ================================
The dominant neutral MSSM Higgs production mechanisms for small and
moderate values of $\tan\beta$ are the gluon fusion processes $gg \to
h,H,A$, which are mediated by top and bottom loops as in the SM case,
but in addition by stop and sbottom loops for the scalar Higgs bosons
$h,H$, if the squark masses are below about 400 GeV \cite{gghnlosq1,gghnlosq2}. The
NLO QCD corrections to the quark loops are known in the heavy quark
limit as well as including the full quark mass dependence
\cite{gghnlo,htoggdecay1,gghnlolim}. They increase the cross sections by up to about
100\% for smaller $\tan\beta$ and up to about 40\% for large
$\tan\beta$, where the bottom loop contributions become dominant due to
the strongly enhanced bottom Yukawa couplings.  The limit of heavy
quarks is only applicable for $\tan\beta\lsimST 5$ within about 20--25\%,
if the full mass dependence of the LO terms is taken into account
\cite{limit,higgsrev,Djouadi:2005gi}. Thus the available NNLO QCD corrections in the
heavy quark limit \cite{gghnnlo} can only be used for small and moderate
$\tan\beta$, while for large $\tan\beta$ one has to rely on the fully
massive NLO results \cite{gghnlo}. Recently the QCD corrections to the
squark loops \cite{gghnlosq1,gghnlosq2} and the full SUSY--QCD corrections have
been calculated \cite{gghnlosqcd1,gghnlosqcd2,gghnlosqcd3}.  The pure QCD corrections are of
about the same size as those to the quark loops thus rendering the total
$K$ factor of similar size as for the quark loops alone with a maximal
deviation of about 10\% \cite{gghnlosq1,gghnlosq2}. The pure SUSY--QCD corrections
are small \cite{gghnlosqcd1,gghnlosqcd2,gghnlosqcd3}. The NNLL resummation of the SM Higgs cross
section \cite{gghresum} can also be applied to the MSSM Higgs cross
sections in the regions, where the heavy quark and squark limits are
valid. The same is also true for the NLO QCD corrections to the $p_T$
distributions \cite{higgsptnlo} and the NNLL resummation of soft gluon
effects \cite{higgsptnnll}, i.e. for small values of $\tan\beta, M_H$
and $p_T$ only. However, for large values of $\tan\beta$ the $p_T$
distributions are only known at LO, since the bottom loops are dominant
and the heavy top limit is not valid. An important consequence is that
the $p_T$ distributions of the neutral Higgs bosons are softer than for
small values of $\tan\beta$ \cite{higgsptbottom}.

The vector-boson fusion processes $qq\to qq+W^*W^*/Z^*Z^*\to qq + h/H$
\cite{vvh} play an important role for the light scalar Higgs boson $h$
close to its upper mass bound, where it becomes SM-like, and for the
heavy scalar Higgs particle $H$ at its lower mass bound
\cite{Plehn:1999nw}. In the other regions the cross sections are suppressed
by the additional SUSY-factors of the Higgs couplings. The NLO QCD
corrections to the total cross section and the distributions can be
taken from the SM Higgs case and are of the same size
\cite{vvhqcd,vvhqcddist}. The SUSY--QCD corrections mediated by virtual
gluino and squark exchange at the vertices turned out to be small
\cite{susyqcd}.

Higgs-strahlung off $W,Z$ gauge bosons $q\bar q\to Z^*/W^* \to h/H+ Z/W$
\cite{vhv} does not play a major role for the neutral MSSM Higgs bosons
at the LHC.  The NLO \cite{vhvqcd} and NNLO \cite{vhvnnlo} QCD
corrections are the same as in the SM case, and the SUSY--QCD
corrections are small \cite{susyqcd}.

Higgs radiation off top quarks $q\bar q/gg\to h/H/A + t\bar t$
\cite{htt} plays a role at the LHC for the light scalar Higgs particle
only. The NLO QCD corrections are the same as for the SM Higgs boson
with modified top and bottom Yukawa couplings and are thus of moderate
size \cite{httqcd}. The SUSY--QCD corrections have been computed
recently \cite{httsqcd}. They are of similar size as the pure QCD
corrections.

For large values of $\tan\beta$ Higgs radiation off bottom quarks
\cite{htt} $q\bar q/gg\to h/H/A + b\bar b$ constitutes the dominant
Higgs production process. The NLO QCD corrections can be taken from the
analagous calculation involving top quarks. However, they turn out to be
large \cite{hbbqcd}. The main reason is that the integration over the
transverse momenta of the final state bottom quarks generates large
logarithmic contributions. The resummation of the latter requires the
introduction of bottom quark densities in the proton, since the large
logarithms can be resummed by the DGLAP-evolution of these densities.
This leads to an approximate approach starting from the processes $b\bar
b\to h/H/A$ at LO \cite{bb2h}, where the transverse momenta of the
incoming bottom quarks, their masses and their off-shellness are
neglected. The NLO \cite{bb2hnlo} and NNLO \cite{bb2hnnlo} QCD
corrections to this bottom-initiated process are known and of moderate
size, if the running bottom Yukawa coupling at the scale of the Higgs
mass is introduced. The SUSY--QCD \cite{bb2hsqcd1,bb2hsqcd2,bb2hsqcd3} and SUSY-electroweak
corrections \cite{bb2helw} can be well approximated by the corresponding
universal $\Delta_b$ terms of the bottom Yukawa couplings.  The fully
exclusive $gg\to h/H/A + b\bar b$ process, calculated with four active
parton flavors in a fixed flavour number scheme (FFNS), and this
improved resummed result, calculated with 5 active parton flavours in
the variable flavour number scheme (VFNS), will converge against the
same value at higher perturbative orders \cite{hbbcomp}. If only one of
the final state bottom jets accompanying the Higgs particle is tagged,
the LO bottom-initiated process is $gb\to b+h/H/A$, the NLO QCD
corrections of which have been calculated \cite{bg2hbnlo}. They reach
${\cal O}(40-50\%)$.  The situation concerning the comparison with the
FFNS at NLO is analogous to the total cross section \cite{hbbcomp}.  If
both bottom jets accompanying the Higgs boson in the final state are
tagged, one has to rely on the fully exclusive calculation for $gg\to
b\bar b + h/H/A$.

\begin{figure}[hbt]
\begin{picture}(130,290)(30,0)
\put(-20,-150){\includegraphics[scale=0.6]{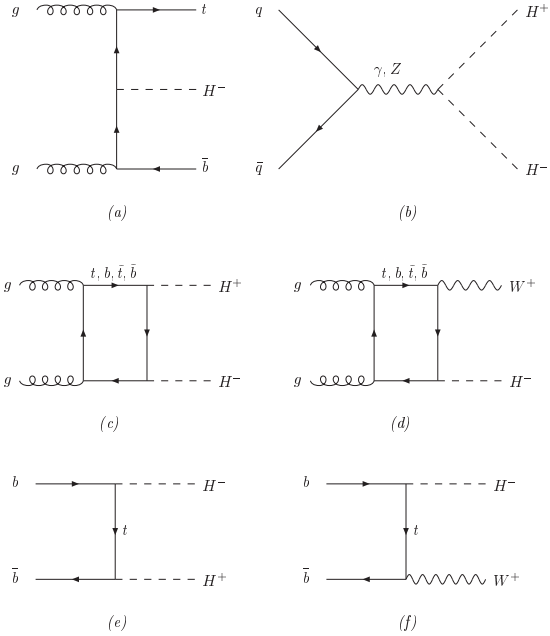}}
\end{picture} \\[-2.7cm]
\caption{\label{fg:hcpro} Typical diagrams for charged Higgs boson
production mechanisms at leading order:
{\it (a)} $gg\to H^-t\bar b$, {\it (b)} $q\bar q\to H^+H^-$,
{\it (c)} $gg\to H^+H^-$, {\it (d)} $gg\to W^+H^-$, {\it (e)} $b\bar
b\to H^+H^-$, {\it (f)} $b\bar b\to W^+H^-$.}
\vspace*{-0.7cm}
\end{figure}

The dominant charged Higgs production process is the associated
production with heavy quarks (see Fig.~\ref{fg:hcpro}a) $q\bar q,gg\to
H^-t\bar b, H^+\bar tb$ \cite{hctb}.  The NLO QCD and SUSY--QCD
corrections have very recently been computed \cite{hctbnlo,hctbnlo2}.
They are of significant size due to the large logarithms arising from
the transverse-momentum integration of the bottom quark in the final
state and the large SUSY--QCD corrections to the bottom Yukawa coupling.
The large logarithms can be resummed by the introduction of bottom quark
densities in the proton in complete analogy to the neutral Higgs case.
In this approach the LO process is $gb\to H^-t$ and its charge
conjugate. The NLO SUSY--QCD corrections have been derived in
\cite{hctnlo} and found to be of significant size. This process,
however, relies on the same approximations as all bottom-initiated
processes. A quantitative comparison of the processes $gb\to H^-t$ and
$gg\to H^- t\bar b$ at NLO shows significant differences, i.e.~poor
agreement for the relevant scale choices \cite{hctbnlo}.

The second important charged Higgs production process is charged Higgs
pair production in a Drell--Yan type process (see Fig.~\ref{fg:hcpro}b)
$q\bar q\to H^+H^-$ which is mediated by $s$-channel photon and
$Z$-boson exchange. The NLO QCD corrections are of moderate size as in
the case of the neutral Higgs-strahlung process discussed before. The
genuine SUSY--QCD corrections, mediated by virtual gluino and squark
exchange in the initial state, are small \cite{susyqcd}.

Charged Higgs pairs can also be produced from $gg$ intital states by the
loop-mediated process (see Fig.~\ref{fg:hcpro}c) $gg\to H^+H^-$
\cite{gg2hc,bb2hc} where the dominant contributions emerge from top and
bottom quark loops as well as stop and sbottom loops, if the squark
masses are light enough. The NLO corrections to this process are
unknown. This cross section is of similar size as the bottom-initiated
process (see Fig.~\ref{fg:hcpro}e) $b\bar b\to H^+H^-$ \cite{bb2hc}
which relies on the approximations required by the introduction of the
bottom densities as discussed before and is known at NLO
\cite{bb2hcnlo}. The SUSY--QCD corrections are of significant size. The
pure QCD corrections and the genuine SUSY--QCD corrections can be of
opposite sign.

Finally, charged Higgs bosons can be produced in association with a $W$
boson (see Fig.~\ref{fg:hcpro}d) $gg\to H^\pm W^\mp$
\cite{gg2hcw1,gg2hcw2} which is generated by top-bottom quark loops and
stop-sbottom loops, if the squark masses are small enough. This process
is known at LO only.  The same final state also arises from the process
(see Fig.~\ref{fg:hcpro}f) $b\bar b\to H^\pm W^\mp$
\cite{gg2hcw1,bb2hcw} which is based on the approximations of the VFNS.
The QCD corrections have been calculated and turn out to be of moderate
size \cite{bb2hcwnlo}.

%%%%%%%%%%%%%%%%%%%%%%%%%%%%%%%HIGGSDECAY%%%%%%%%%%%%%%%%%%%%%%%%%%%%%%%%%%%%

\section{Higgs decays}

{\it M.M.~M\"uhlleitner}\medskip

In this section, we discuss
the decay modes of the Higgs boson and the status of the theoretical
calculations. We begin with a description of the the Standard Model Higgs decays
followed by MSSM Higgs decays.

\subsection{Standard Model Higgs decays}

The profile of the Standard Model (SM) Higgs boson is uniquely
determined once its mass $M_H$ is fixed. The scale of the Higgs
couplings to the fermions and massive gauge bosons is set by the mass
of these particles. The trilinear and quartic Higgs self couplings are
uniquely determined by the Higgs boson mass.

The Higgs branching ratios and total width are determined by these
parameters. A measurement of the decay properties will therefore serve as
a first test of the Higgs mechanism, a consequence of which is that the
Higgs boson couplings to the particles grow with the particle masses.

A Higgs boson in the intermediate mass range, ${\cal O}(M_Z) \le M_H \le
{\cal O}(2M_Z)$, dominantly decays into a $b\bar{b}$ pair and a
pair of massive gauge bosons, one or two of them being virtual. Above
the gauge boson threshold, it almost exclusively decays into $WW,ZZ$,
with a small admixture of top decays near the $t\bar{t}$
threshold. Below $\sim 140$ GeV, the decays into $\tau^+\tau^-$, $c\bar{c}$
and $gg$ are important besides the dominant $b\bar{b}$ decay. The
$\gamma\gamma$ decay, though being very small, provides a clean
2-body signature for the Higgs production in this mass range.

\subsubsection{Higgs decays into fermions}
The decays into fermions are suppressed near threshold by a cubic factor
in  the velocity. For asymptotic energies there is only a linear
dependence on the Higgs boson mass. The QCD corrections to the Higgs
decays into quarks are known to three-loop order \cite{hffqcd} and the
electroweak (EW) corrections up to next-to-leading order (NLO)
\cite{hffew}, the latter being also valid for leptonic decay
modes. Whereas the effect of the EW radiative corrections in the
branching ratios is negligible, the QCD corrections can be large.
The bulk of the corrections can be absorbed into
the scale dependent quark mass, evaluated at the Higgs mass. The residual QCD
corrections modify the widths only slightly. Whereas the precise
value of the running quark mass at the Higgs boson scale represents a
significant source of uncertainty in the decays to the $c$ quark pair,
the $b\bar{b}$ and $\tau^+\tau^-$ predictions can be obtained
with accuracies comparable to the experimental uncertainties. Due to the
smallness of the effective $c$-quark mass, the colour factor 3 in the
ratio between charm and $\tau$ decays is overcompensated.

\subsubsection{Higgs decays into $WW$ and $ZZ$ pairs}
Above the $WW$ and $ZZ$ thresholds, the Higgs decays almost exclusively
into these gauge boson pairs \cite{hvvdecay} except for the mass range
above the $t\bar{t}$ threshold. For large Higgs masses, the vector
bosons are longitudinally polarized and characterized by wave functions
linear in the energy. The widths therefore grow with the third power of
the Higgs boson mass. Below the decay threshold into two real bosons,
the Higgs can decay into a pair of real and virtual vector bosons
\cite{hvvoffdecay}. Decays into $W^{(*)}W^{(*)}$ pairs become comparable
to the $b\bar{b}$ mode at $M_H \sim 140$~GeV.

For $M_H \gsimST 140$~GeV, the $Z^{(*)}Z^{(*)}$ channel becomes
relevant. Above the  threshold, the $4$-lepton channel $H\to ZZ \to
4l^\pm$ provides a very
clean signature for the Higgs boson search. If the on-shell $ZZ$ decay
is still closed kinematically, the $WW$ decay channel is very useful,
despite the escaping neutrinos in the leptonic $W$ decays. The QCD and
electroweak radiative corrections to the decays $H\to WW/ZZ \to 4f$ have
been evaluated in \cite{hvv4l}. The EW corrections amount to a few percent
and increase with growing Higgs mass. The QCD corrections for
quark final states are of
${\cal O} (\alpha_s/\pi)$. The distributions, important for the
reconstruction of the Higgs mass and the suppression of the background,
are in general distorted by the corrections.

\begin{figure}[t]
%\begin{center}
%\epsfig{figure=Higgs/wtotbr1.ps,bbllx=100,bblly=350,bburx=500,bbury=500,
%width=4cm,angle=-90,clip=}
%\end{center}
\includegraphics[width=5.7cm,angle=270]{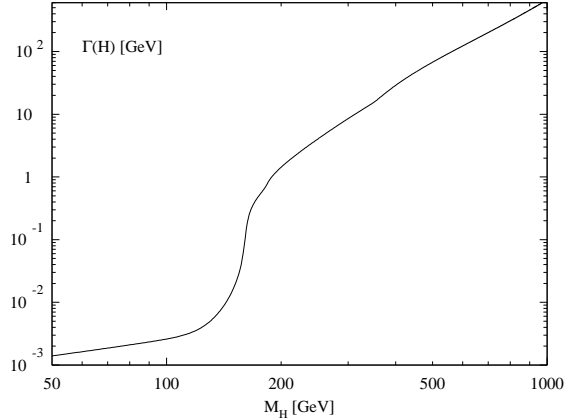}
\vspace{-0.5cm} \caption{Total decay width as a function of the
Higgs boson
  mass, taking into account all relevant higher order corrections and
  virtual decays. Code: HDECAY \cite{hdecay}.}
\label{fig:higgstot}
\end{figure}

\subsubsection{Higgs decays to $gg$ and $\gamma\gamma$ pairs}
The SM gluonic Higgs decays are mediated by $t$- and $b$-quark loops.
The photonic decay involves in addition $W$ boson loops.
%Being
%significant only far below the top and $W$ thresholds and with the SM
%b-quark loop contribution being negligeable, they can be
%described by approximate formulae \cite{htoggdecay,htoggdecay1} in the
%limit where the loop masses are taken to infinity.
The QCD corrections to the decay into gluon pairs include the $ggg$ and
$gq\bar{q}$ final states. They have been calculated in \cite{htoggdecay1,htoggdecay2}
and amount up to $\sim 70$\%. The NNLO \cite{nnlogg} and the NNNLO
\cite{nnnlogg} corrections have been evaluated in the heavy top mass
limit, increasing the reliability of the perturbative expansion of the
decay rate.
The NLO QCD \cite{qcdgg} and electroweak \cite{gghelw1,ewgg} corrections to the
photonic decay are known and small in the mass range relevant for
experiment. Despite being very suppressed, the photonic Higgs decays
provide an attractive resonance-type search channel at the LHC for the low mass
Higgs boson.

\subsubsection{Summary}
The total Higgs width, shown in Fig.\ref{fig:higgstot}, is obtained by
adding up all possible decay channels. For $M_H \lsimST 140$~GeV,
the Higgs width remains very small, $\Gamma (H) \le 10$~MeV. Once the real and
virtual gauge boson channels open up, it rapidly increases, reaching
$\sim 1$~GeV at the $ZZ$ threshold. In the intermediate mass range the total
width cannot be measured directly. It can be determined indirectly
by combining Higgs production and decay channels. Above $M_H
\approx 250$~GeV, the width becomes large enough to be resolved
experimentally.

\begin{figure}[t]
%\begin{center}
%\epsfig{figure=Higgs/wtotbr2.ps,bbllx=100,bblly=350,bburx=500,bbury=500,
%width=4cm,angle=-90,clip=}
%\end{center}
\includegraphics[width=5.7cm,angle=270]{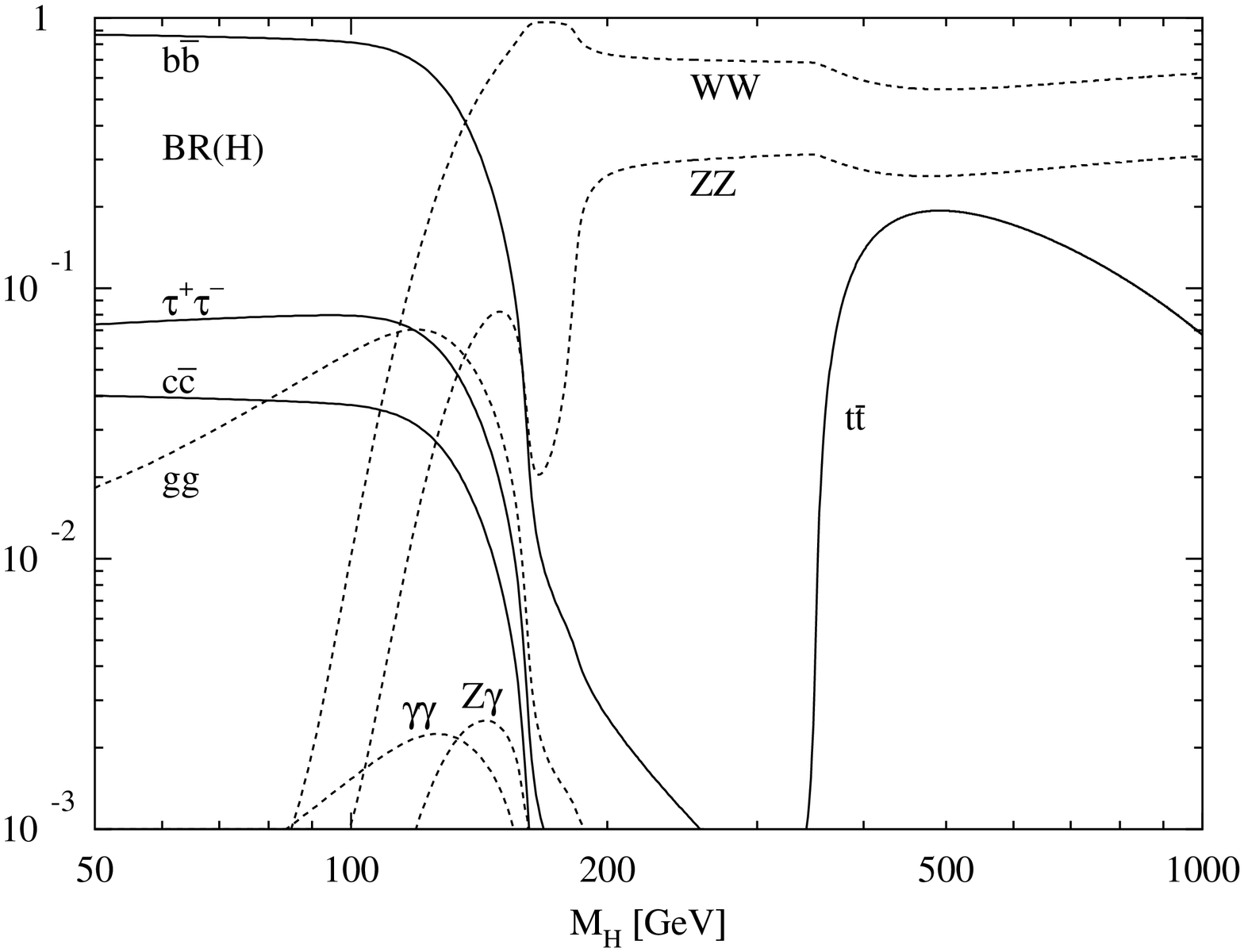}
\vspace{-0.5cm} \caption{Branching ratios of the dominant SM Higgs
decay modes.} \label{fig:higgsbran}
\end{figure}

\begin{figure*}[t]
%\begin{center}
\includegraphics[width=4.9cm,angle=270]{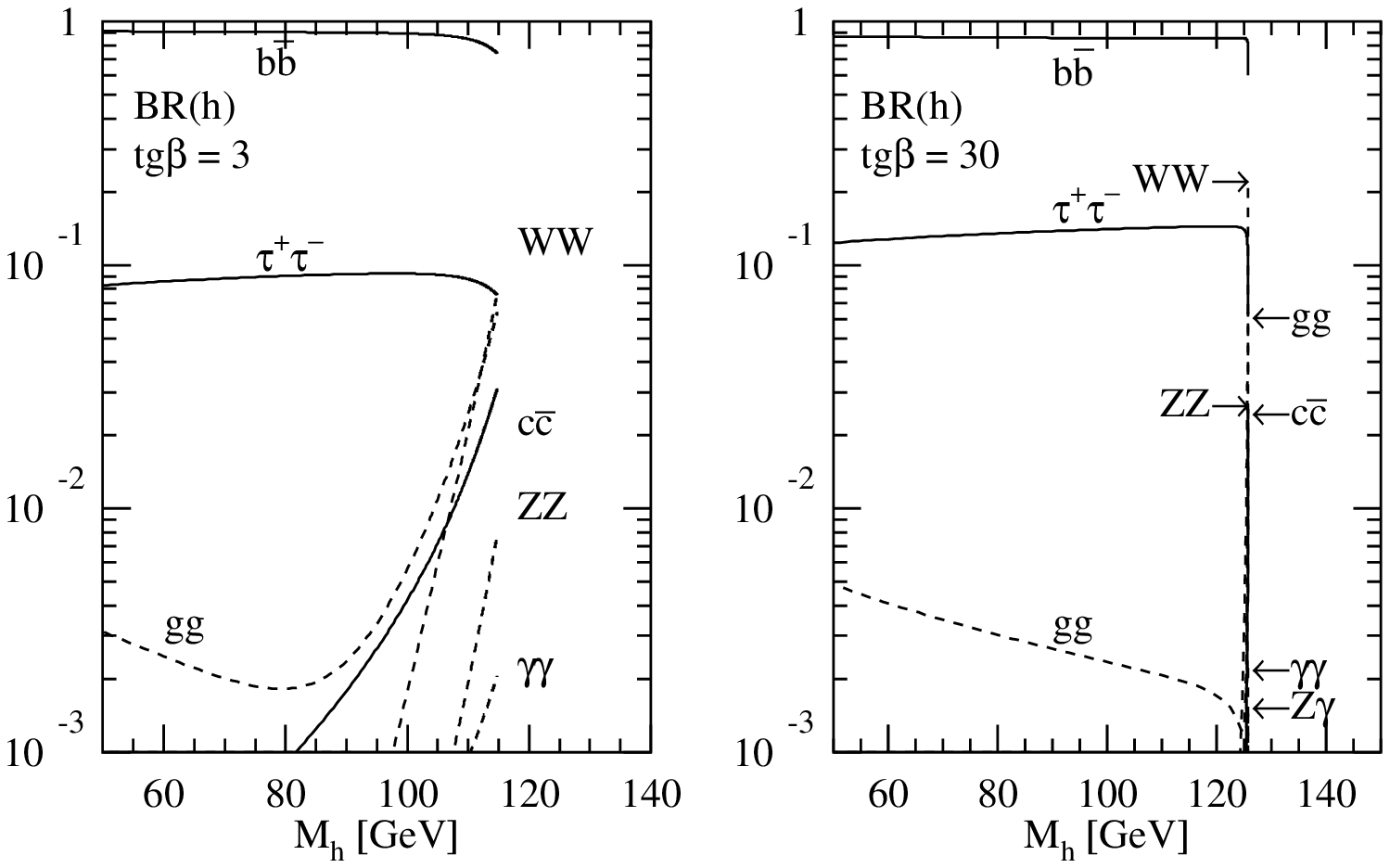}
\includegraphics[width=4.9cm,angle=270]{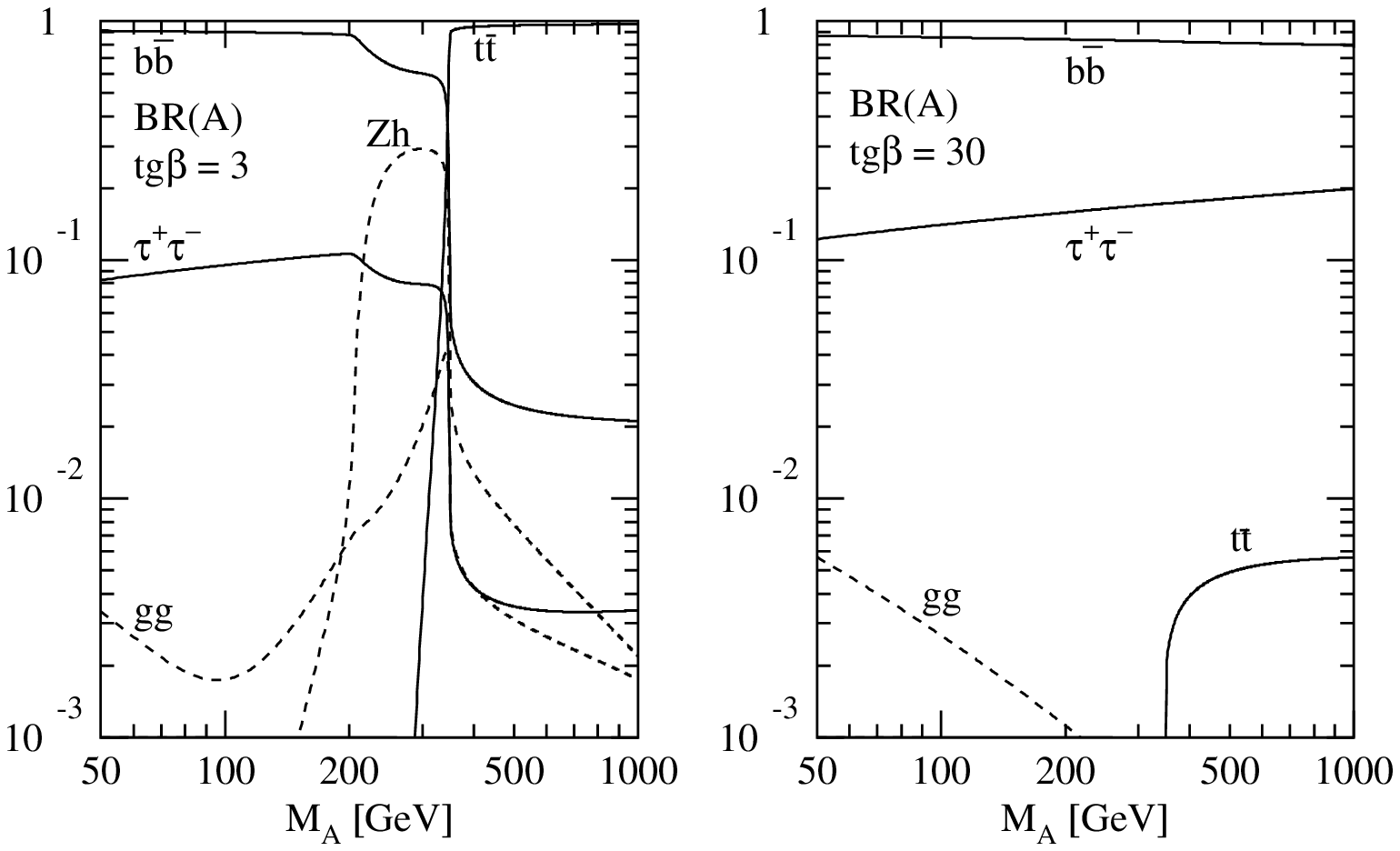}\\
\includegraphics[width=4.9cm,angle=270]{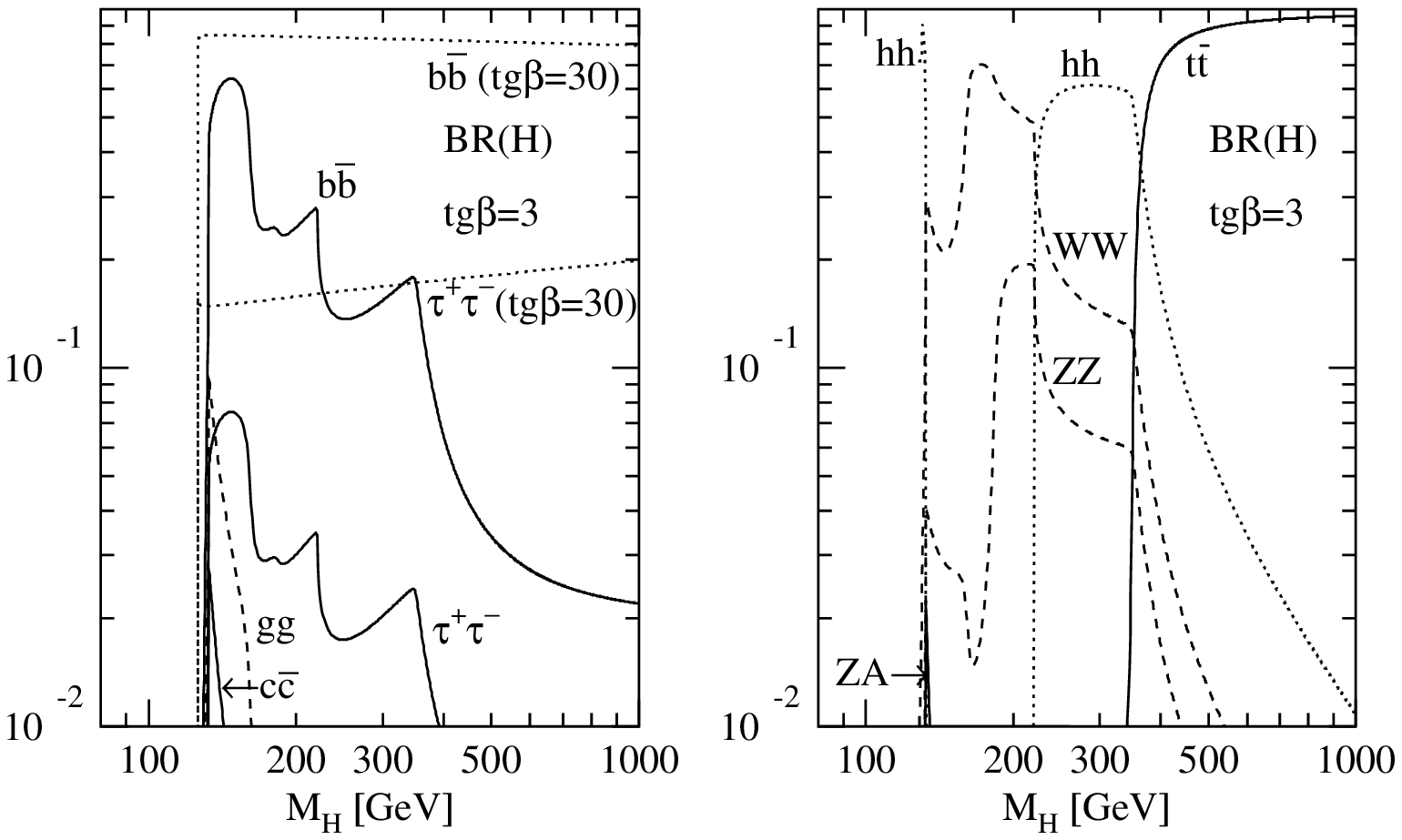}
\includegraphics[width=4.9cm,angle=270]{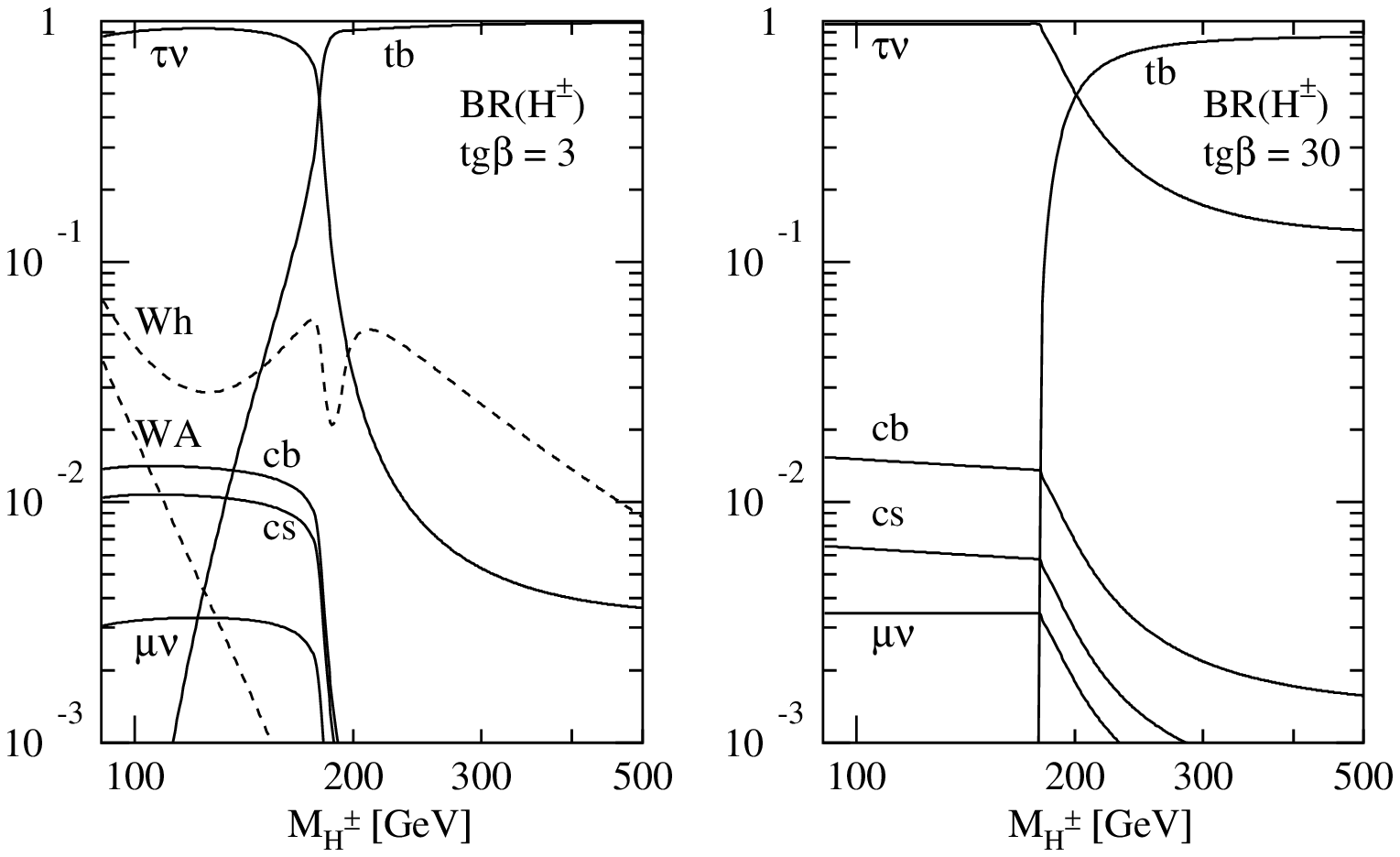}
%\epsfig{figure=Higgs/mssmhcbr.ps,bbllx=100,bblly=350,bburx=400,bbury=500,
%width=5cm,angle=-90,clip=}
%\end{center}
\vspace*{-0.2cm}
\caption[]{\label{fig:hmssmbran} Branching ratios of the MSSM Higgs
 bosons $h,H, A, H^\pm$ for non-SUSY decay modes as a function of the
 masses for two values of $\tan\beta=3, 30$ and vanishing mixing. The
 common squark mass has been chosen as $M_S=1$ TeV.}
\end{figure*}
The branching ratios of the main decay modes are shown in
Fig. \ref{fig:higgsbran}. For $M_H \lsimST 140$ GeV, various channels
will be accessible. The dominant decay mode $b\bar{b}$ with a branching
ratio of $\sim 85$\% is followed by the decay into $\tau^+\tau^-$ with a
ratio of $\sim 8$\%. The decays into $c\bar{c}$ and $gg$, reach the level of
several per-cent. The photonic branching ratio occurs at the permille
level. Above 140 GeV the decay into $W$ bosons becomes dominant. Once
the decay into real $W$'s is kinematically possible it overwhelms all
other decays. Far above the thresholds, the $ZZ$ and $WW$ decays are
given at a ratio of $1:2$, modified slightly by the top decays
just above the $t\bar{t}$ threshold. The Higgs particle gets very wide
asymptotically, since the decay widths into vector boson pairs grow as
$M_H^3$.

\subsection{MSSM Higgs boson decays}
Compared to the SM couplings, the MSSM Higgs couplings to
fermions and gauge bosons are modified by the mixing angle $\alpha$ in
the neutral CP-even Higgs sector and the ratio of the two vacuum
expectation values of the Higgs doublet $\tan\beta$. The couplings to
the massive gauge bosons are suppressed by these mixing angles compared
to the SM Higgs-gauge couplings. At tree-level they are are absent for
the pseudoscalar Higgs boson. The couplings to down-(up-)type quarks are
enhanced (suppressed) by $\tan\beta$. In the decoupling limit, where the
mass of the pseudoscalar is large, the $h$-couplings approach
the SM couplings, whereas the heavy Higgs $H$ decouples from the vector
bosons.
%\addtocounter{figure}{-1}
%\begin{figure}[t]
%\begin{center}
%\epsfig{figure=mssmabr.ps,bbllx=100,bblly=350,bburx=400,bbury=500,
%width=5cm,angle=-90,clip=} \\
%\epsfig{figure=mssmhcbr.ps,bbllx=100,bblly=350,bburx=400,bbury=500,
%width=5cm,angle=-90,clip=}
%\end{center}
%\vspace*{-0.2cm}
%\caption[]{Continued.}
%\end{figure}

\subsection{Higher order corrections}
The higher order corrections to the MSSM Higgs couplings also involve
contributions from supersymmetric (SUSY) particles running in the
loops. The (SUSY-)QCD \cite{bb2hsqcd1,bb2hsqcd2,hffqcd,susyqcd1} and
(SUSY-)electroweak \cite{hffew,susyqcd1,susyew} corrections
to the fermionic decay
modes are sizeable. Additional significant corrections arise from virtual
sbottom/stop and gluino/gaugino exchange in the $h,H,A\to b\bar{b}$ and
$H^\pm \to tb$ decays \cite{bb2hsqcd1,bb2hsqcd2,susyqcd1,susyew}. The dominant part
of the latter corrections can be absorbed in improved bottom Yukawa
couplings, so that these contributions can also be resummed up to all
orders and yield reliable perturbative results \cite{bb2hsqcd3,resum}. The
two-loop corrections to the improved couplings have been provided in
\cite{2loopb} reducing the residual theoretical error to the per-cent level.

The rare photonic decays are mediated by $W$, $t$ and $b$-loops as in
the SM, the $b$-contribution being important for large $\tan\beta$
values. In addition, contributions from charged Higgs bosons, charginos
and sfermions arise, if these virtual particles are light enough. The
QCD corrections amount to a few percent in the relevant mass regions
\cite{qcdgg}. The SUSY-QCD corrections are of similar
size~\cite{gghnlosqcd1,gghnlosqcd2,gghnlosq1,susyqcdgg}.

If decays into gauginos and sfermions are possible, they acquire
significant branching ratios and can even be the dominant decay modes
\cite{susyfinal}.

The self-couplings of the Higgs bosons induce heavy Higgs decays into two
lighter Higgs states, if kinematically possible. The measurement of the
Higgs self-couplings is a crucial ingredient for the reconstruction of
the Higgs potential and verification of the Higgs mechanism
\cite{higgsself}. The NLO SUSY correction to the self-couplings of the
lightest Higgs boson can almost completely be absorbed into the Higgs
boson mass \cite{nlohhh}.

\begin{figure}[t]
\vspace*{-0.2cm}
%\hspace*{2cm}
\includegraphics[width=4.8cm,angle=270]{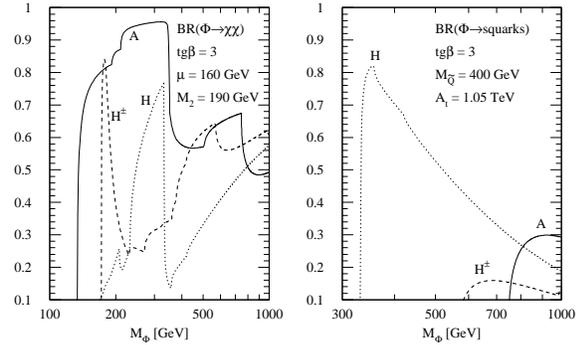}
\vspace*{-0.2cm} \caption[]{Branching
ratios of the MSSM Higgs boson $H,A,H^\pm$ decays
  into charginos/neutralinos and squarks as a function of their
  masses for $\tan\beta=3$. The mixing parameters have been chosen as
  $\mu=160$ GeV,
  $A_t=1.05$ TeV, $A_b=0$ and the squark masses of the first two generations as
  $M_{\widetilde{Q}}=400$ GeV. The gaugino mass parameter has been set to
  $M_2=190$ GeV.}
\label{fig:susydec}
\end{figure}

\subsection{Branching ratios and total widths}
The lightest \underline{neutral Higgs boson} $h$ mainly decays into
fermion pairs, since its mass is smaller than $\sim 140$~GeV,
c.f. Fig.\ref{fig:hmssmbran}. This is, in
general, also the dominant decay mode for the pseudoscalar $A$. For
large $\tan\beta$ and masses below $\sim 140$~GeV, the main decay modes of
the neutral Higgs bosons are into $b\bar{b}$ and $\tau^+\tau^-$ with
branching ratios of order $\sim 90$\% and 8\%, respectively. The decays
into $c\bar{c}$ and $gg$ are suppressed, especially for large
$\tan\beta$ values. Above the kinematic threshold, the decays $H,A\to
t\bar{t}$ open up. This mode remains suppressed for large values of
$\tan\beta$, however, and the neutral Higgs bosons decay almost
exclusively into $b\bar{b}$ and $\tau^+\tau^-$ pairs. Contrary to the
pseudoscalar $A$, the heavy CP-even Higgs boson $H$ can decay into
massive gauge bosons, if its mass is large enough. However, the partial
widths are in general strongly suppressed by cos/sin of the mixing
angles. As a result, the total widths of the SUSY Higgs bosons are much
smaller than in the SM.

The heavy $H$ can also decay into two lighter Higgs bosons. Furthermore,
Higgs cascade decays and decays into other SUSY particles are possible
and can even be dominant in regions of the MSSM parameter space
\cite{susyfinal}, c.f. Fig.\ref{fig:susydec}. Decays
of $h$ into the lightest neutralino are also important and exceed 50\%
in parts of the parameter space. These decays therefore strongly affect
the experimental search techniques.

The \underline{charged Higgs particles} decay into fermions and, if
kinematically possible, into the lightest neutral Higgs and a $W$
boson. Below the $tb$ and $Wh$ thresholds, they decay mostly into
$\tau\nu_\tau$ and $cs$ pairs, with the former being dominant for
$\tan\beta$ larger than unity. For large $H^\pm$ masses, the decay into
$tb$ becomes dominant. In some parts of the SUSY parameters space the
decays into SUSY particles make up more than 50\%.

The total widths are obtained by adding up the various decay modes. They
are quite narrow for all five Higgs bosons, of order 10 GeV even for
large masses.

%%%%%%%%%%%%%%%%%%%%%%%%%%%HIGGSSIGNATURES%%%%%%%%%%%%%%%%%%%%%%%%%%%%%%%%%%%%%

\section{Higgs Signatures}
{\it Sally Dawson and Tilman Plehn\footnote{The authors would like to
    thank Gavin Salam for many enlightening discussions and for
    providing a large fraction of the new physics results discussed
    below.}}\medskip

We consider discovery channels for the Higgs boson at the LHC and
emphasize the prospects for a measurement of the Higgs
boson mass. The importance of understanding theory uncertainties
for the interpretation of results is emphasized.

\subsection{Standard channels}

Higgs boson production at the LHC has been extensively studied,
and the most important channels for discovery in the Standard Model
are summarized below. An important development for example as compared
to the ATLAS TDR\cite{atlastdr} or earlier CMS studies\cite{Ball:2007zza}
 is that by now each Higgs mass
point is covered by at least two discovery channels of similar
strength~\cite{Djouadi:2005gi,reviews}. Based on electroweak precision data a lot of effort has been
invested in low-mass Higgs channels, but we note that the preference
of a light Higgs boson is strongly linked to the assumption that no
physics beyond the Standard Model impacts the electroweak
scale. Otherwise, a larger Higgs mass could be required to reach the
experimentally preferred ellipse in the $S$-$T$ plane. Neglecting
systematic uncertainties, to reach the quoted sensitivity for a given
channel at $\sqrt{s}=10~\tevTSG$ requires roughly twice as much
luminosity as at $\sqrt{s}=14~\tevTSG$.

\subsubsection{$H\rightarrow \gamma \gamma$}

A very light Higgs boson in the mass range, $120~\gevTSG< M_H <140~\gevTSG$,
can be searched for as a narrow
resonance in the $H\rightarrow \gamma\gamma$
channel,
thanks to the excellent electromagnetic
resolution of both ATLAS and CMS. Although
the rate is small,  this channel has the advantage
that the irreducible $\gamma
\gamma$ background can be measured from the sidebands. The dominant
reducible backgrounds are jet mis-identification and converted photons
from material in the detector.

Using a cut based analysis, ATLAS finds that with $10~\ifb$ the
significance is less than four above $M_H = 120~\gevTSG$.  For the same
mass region, CMS finds a $7-10\sigma$ significance with $30~\ifb$
using an optimized analysis.  Higgs plus jets signatures with a
slightly boosted Higgs boson offer improved signal to background
ratios, but the number of events is reduced in this formally
next-to-leading order QCD process.

\subsubsection{$H\rightarrow ZZ$}

The channel $H\rightarrow ZZ\rightarrow 4l$ (where $l=e,\mu$) has been
termed the {\sl golden channel}, because it produces a clear peak on
top of a smooth background which can be estimated from the sidebands.
The major backgrounds are $t {\overline t}$, $ZZ$, and $Z b {\overline
  b}$ and are significantly reduced by kinematic cuts.
Fig.~\ref{fg:zz} demonstrates the cleanliness of this signal.  Except
for the region near $M_H = 2M_W$, where the off-shell $H\rightarrow
ZZ$ branching ratio is suppressed, this is a discovery channel up to
$M_H \sim 500~\gevTSG$ with $30~\ifb$.  A Standard Model Higgs boson over
the entire mass range allowed by LEP2 can be excluded using only this channel
with $10~\ifb$.
%, but discovery
%is not possible in the $Z\rightarrow 4l$ channel with $1~\ifb$.

\begin{figure}[t]
\begin{center}
\includegraphics[scale=0.2]{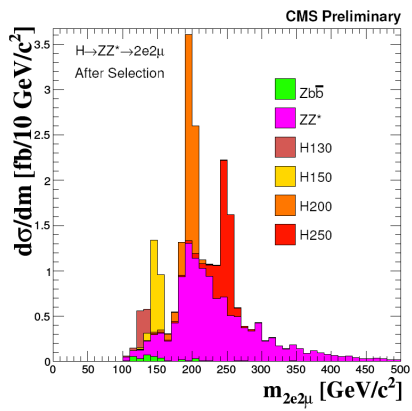}
\end{center}
\vspace*{-10mm}
\caption[]{ Invariant mass after cuts with $M_H=130, 150,200$~ and
  $250~\gevTSG$ for the $H\rightarrow ZZ\rightarrow e^+e^-\mu^+\mu^-$
  signal using the CMS detector.  From Ref.~\cite{cms_zz}. }
\label{fg:zz}
\end{figure}

\subsubsection{$H\rightarrow W^+W^-$}

In the mass region between $135~\gevTSG < M_H < 2M_Z$, the dominant Higgs
branching ratio is to $WW$. Off-shell effects play an important role
because the bottom Yukawa coupling is significantly smaller than the
weak gauge coupling, so the $WW$ decay can surpass the $bb$ decay
channel several $W$ widths below threshold.  The Higgs can be produced
in this region by both gluon fusion and weak boson fusion and
discovery is possible in the $WW\rightarrow l\nu l \nu$ channel, where
$l=e,\mu$.  There is no mass peak and in particular the gluon induced
process suffers from large backgrounds from $WW$, $Wt$ and $t
{\overline t}$ production.  The main background rejection cut on the
$W$ pair is the angular correlation of the two leptons coming from a
scalar resonance~\cite{dittmar_dreiner}, which also enhances the
relative impact for the loop-induced $gg \to WW$
background~\cite{gg_ww}.  In the weak boson fusion channel, cuts on
the forward jets and the QCD activity reduce the background to a level
of $S/B>1$~\cite{wbf_ww}. Both production channels can be extended to
Higgs masses around 120~GeV and allow both ATLAS and CMS to obtain a
$5\sigma$ discovery reach for $135~\gevTSG < M_H < 180~\gevTSG$ with
$10~\ifb$. Even with only $1~\ifb$ a Higgs boson in the region
$160~\gevTSG< M_H < 170~\gevTSG$ can be discovered, provided the missing
energy vector can be used for the transverse mass reconstruction.

\begin{figure}[t]
\begin{center}
\includegraphics[scale=0.4]{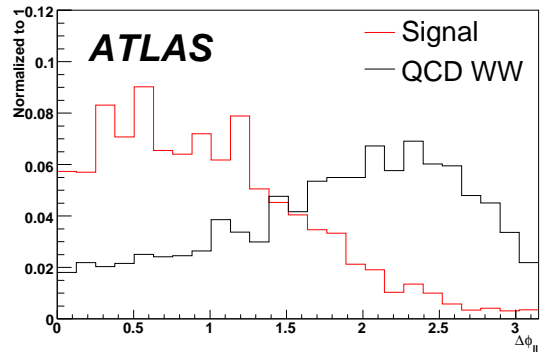}
\end{center}
\vspace*{-10mm}
\caption[]{Angle between the leptons from $H\rightarrow WW\rightarrow
  l\nu l\nu$ with $M_H=170~\gevTSG$ and those from the QCD background
  using the ATLAS detector.  From Ref.\cite{Aad:2009wy}. }
\label{fg:vbf_4l}
\end{figure}
% From ATLAS CSC notes

\subsubsection{$H\rightarrow \tau^+\tau^-$}

The vector boson fusion production of a Higgs boson, followed by the
decay of the Higgs to $\tau \tau$, with at least one $\tau$ decaying
leptonically, is a discovery channel with $30~\ifb$ only for a
very light Higgs
boson, $M_H < 125~\gevTSG$, with the significance
falling quickly with increasing Higgs mass owing to the sharp drop in
the fermionic branching ratios. For the mass reconstruction this
channel requires a sizeable transverse momentum of the Higgs, which
singles out the weak boson fusion production channel. This Higgs
signature is of particular interest in the MSSM, where we expect a
light  Higgs boson with a slightly enhanced branching ratio to
down-type fermions~\cite{Plehn:1999nw,Plehn:1999xi}. It implies that to discover at
least one (light) supersymmetric Higgs boson we can rely on Standard
Model search channels alone.

\subsubsection{What to expect}

A summary of the ATLAS and CMS results for Higgs production are shown
in Fig.~\ref{fg:finalsig}.  With the exception of
the $H\rightarrow \gamma\gamma$ channel, the two experiments have
similar sensitivities for Higgs discovery. Each Higgs mass value is
covered by at least two different analyses. Note that the $t \bar{t}H$
production mode with a decay $H \to b \bar{b}$ is not present any
longer, and that the subjet analysis for $WH/ZH$ with $H \to b\bar{b}$
discussed later in this contribution is not yet included.

\begin{figure*}[t]
\begin{center}
\includegraphics[scale=0.34]{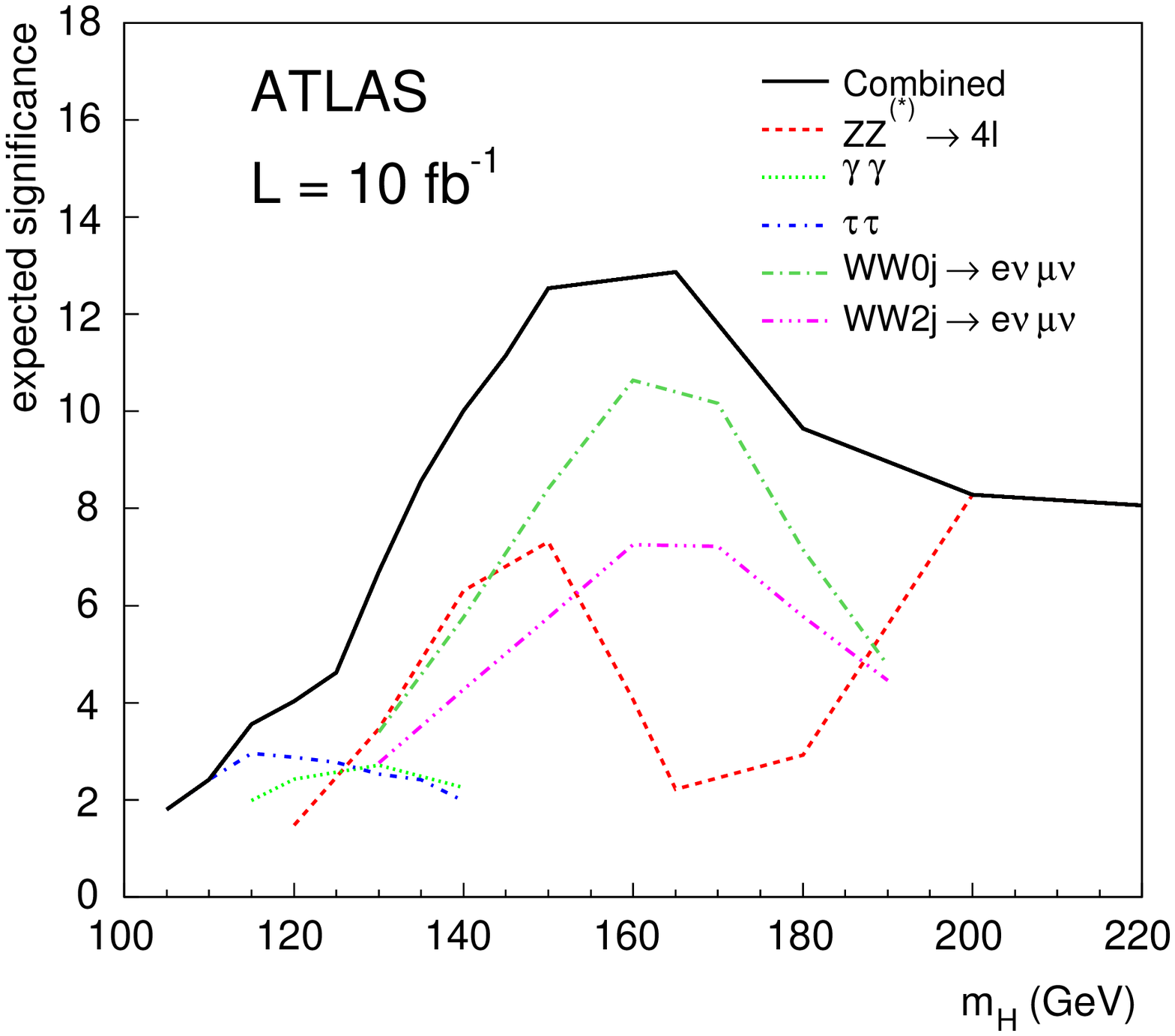}
\hspace*{15mm}
\raisebox{-1mm}{\includegraphics[scale=0.72]{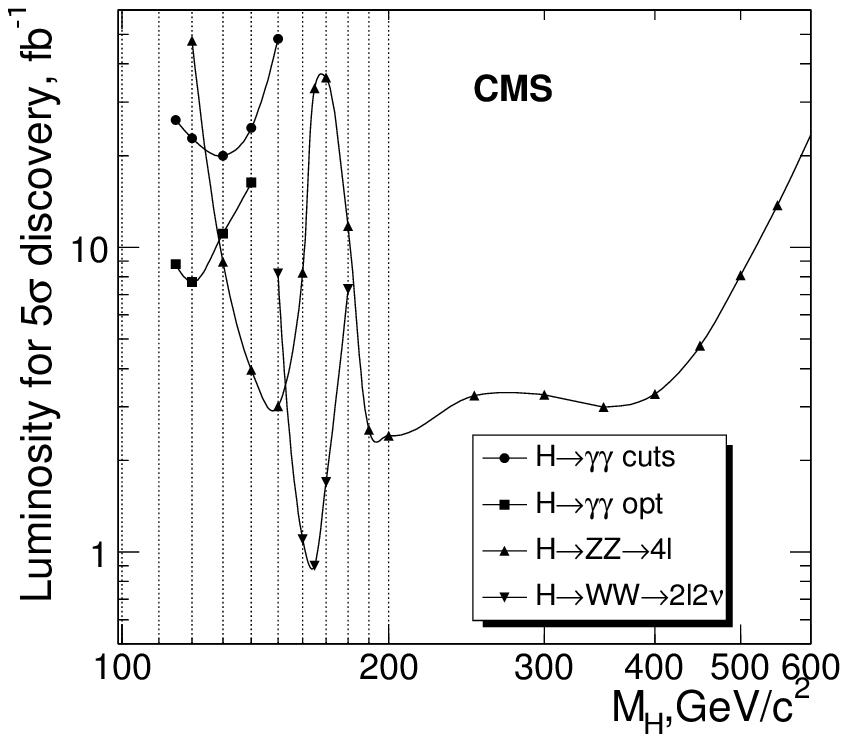}}
\end{center}
\vspace*{-10mm}
\caption[]{ Left: ATLAS significance for a Higgs discovery with
  $10~\ifb$ \cite{Aad:2009wy}.
Right: luminosity needed for a $5\sigma$ discovery using
  the CMS detector\cite{Ball:2007zza}.}
\label{fg:finalsig}
\end{figure*}

\subsubsection{What to wait for}

All Higgs production and decay channels currently explored at the LHC
involve either tree-level couplings to third-generation fermions or
weak gauge bosons or higher-dimensional couplings to gluons and photons
induced by those tree-level couplings. Making use of the strength of
the CMS detector the only second-generation Higgs coupling we might
observe at the LHC is the muon Yukawa coupling~\cite{wbf_mu}.

More importantly, the final proof that an observed Higgs scalar is
actually a result of the spontaneous breaking of electroweak symmetry
would be a measurement of the Higgs self coupling. For large enough
Higgs masses the appropriate channel would be $HH \to 4W$ leading to
like-sign dileptons~\cite{self_sd,self_tp}.  Because of the delicate
cancellations of the self-coupling and continuum contributions at
threshold we might be able to place a lower limit on the self coupling
from an upper limit on the pair production rate. The most dangerous
background is top pair production which we can hope to understand by
the time the LHC will have accumulated enough luminosity to probe this
channel.

\subsection{Mass measurements}

For an experimental confirmation of the Higgs mechanism of the
Standard Model (aside from the Higgs self coupling) we need to show
several features:

1-~The different Higgs signals actually arise from the same
fundamental scalar, \ie~ the masses of gauge bosons as well as up-type
and down-type fermions are all linked to the same Higgs vacuum expectation
value.

2-~The observed branching ratios correspond to the
theoretical predictions. The branching
ratios strongly vary with the Higgs mass, in
particular in the region ($M_H\sim 130-160~\gevTSG$)
where the off-shell $H \to WW$ decay slowly
starts competing with the decay to bottom quarks~\cite{hdecay,Djouadi:2006bz}. As
discussed elsewhere in this volume this means we have to include the
Higgs mass as one of the model parameters which we extract from the
Higgs sector~\cite{Lafaye:2009vr}.\medskip

The different production and decay channels discussed above yield very
different prospects for the measurement of the Higgs mass.
Close to perfect mass measurements can be expected from the decays $H
\to \gamma \gamma$ and $H \to ZZ \to  4\mu$, based on the energy
resolutions of ATLAS and CMS for photons and muons. In the Standard
Model, the width of a light Higgs boson is much smaller than the best
possible detector resolution, so we expect a
gaussian resonance peak over a smooth background. In this situation
the mass resolution is not limited by the experimental
resolution, because we can fit a
gaussian to the observed peak. In the signal-only limit the
resulting mass resolution is $\delta M_H/M_H = \Delta_{\rm
  res}/\sqrt{N_S}$, in terms of the number of signal events $N_S$. For the
$H
\to \gamma \gamma$ and $H \to 4\mu$
channels  this implies a measurement to
$\mathcal{O}(100~\mevTSG)$. Note, however, that this number does not take
into account systematic errors from the lepton and photon energy
scales, which would simply shift the mass peak by an unknown
factor. An expected scale uncertainty of the order of $0.1~\%$
again gives us an expected Higgs mass measurement to
$\mathcal{O}(100~\mevTSG)$.

For the weak-boson-fusion Higgs production and a decay to tau leptons we can
rely on the sizeable kinematic
boost of the Higgs boson as
well as its tau decay products. In the collinear approximation we can
then reconstruct the invariant mass of the $\tau \tau$
system~\cite{Ellis:1987xu}. The experimental resolution of the
reconstructed Higgs mass is dominated by the missing transverse energy
resolution. With an experimental resolution around $15~\gevTSG$, the
resulting Higgs mass measurement for $\mathcal{O}(15)$ events in
$30~\ifb$ is expected to be around $5~\gevTSG$.

The transverse mass in the decay $H \to WW$ can be defined in two
ways, depending on how we generalize the transverse mass formula for
$W$, which now involves $M_{\nu \nu}$. If we are interested in a
realistic (central) value for $M_{T,WW}$ we can set $M_{\nu \nu} =
M_{\ell \ell}$, simply based on the symmetry of the final state. This
definition leads to a sharp peak in the $M_{T,WW}$ distribution in,
for example, the vector boson fusion subprocess $qq \to qqH \to
qqWW$~\cite{mtww_dieter}. If instead we want to keep the original edge
shape $M_{T,X} < M_X$ we need to set $M_{\nu \nu}=0$, which allows for
the best Higgs mass measurement in this
channel~\cite{hww_cambridge}. A similar analysis using an
$M_{T,2}$-assisted momentum reconstruction indicates that we might be
able to measure the Higgs mass to a $1 - 2\%$ precision in this
channel~\cite{hww_korea}. Preliminary experimental studies show that
systematic errors and detector effects might decrease this accuracy to
$\mathcal{O}(5\%)$.

\subsection{Error estimates}

While the error estimate for a Higgs mass measurement is fairly
straightforward --- as  is usually the case for kinematic features
even at hadron colliders --- new experimental
analysis techniques seriously
challenge the estimate of theory uncertainties for rate
measurements. Because one of  the most interesting aspects of a Higgs sector
analysis is the measurement of the Higgs
coupling strengths to different gauge bosons and
fermions, we need to have a firm understanding of
the theory uncertainties. A good example is the Higgs
search in the $WW$ decay channel~\cite{bryan_tev,keung}.

The problem of simulating Higgs events with high precision is
perturbative QCD: computing the {\sl inclusive Higgs production} rate
at NNLO in $\alpha_s$ also predicts the Higgs distributions at the
same order.
The error estimates for the total cross
section for Higgs production via gluon fusion are well understood and
range around ${\cal{ O}} (5-10\%)$~\cite{gghqcdelw,gf_errors}.
At small transverse momenta, these predictions have to be
complemented with a collinear resummation to regularize the
small-$p_T$ regime. Such effects can be taken into account
in a Monte Carlo by
re-weighting the events in the Higgs phase space, based on a
perturbative series in collinear logarithsm. However, the
accuracy of the Higgs distributions when a finite transverse
momentum is included
is not matched by the accuracy of the  distributions of
the
recoiling jets.

At NLO the radiation of one additional jet from the initial state
contributes to the total rate and is needed to regularize the infrared
divergences from virtual gluon exchange. The kinematic distributions
of this jet are only included to leading order, even though the total
rate is known to NLO. This is why matching schemes like
MC\@@NLO~\cite{mcnlo} or POWHEG~\cite{powheg} work in spite of the
fact that we only know the parton shower with finite contributions at
the leading-order level. Computing the total rate and the Higgs
distributions to NNLO by counting powers of $\alpha_s$ includes NLO
corrections to the {\sl Higgs plus one jet} process, and with it NLO
kinematic distributions of this one jet. Strictly following the
definition of the parton distributions and the DGLAP equation, we would
expect this additional jet to be the leading jet, but given our
freedom in choosing the factorization scales this is not automatically
the case in practice. Moreover, none of the currently used schemes consistently
match these NLO jet distributions with a parton shower.

To regularize the two-loop virtual corrections in the total NNLO Higgs
production cross section we need to compute {\sl Higgs plus two jets}
at leading order. Hence, the distributions of this second jet as part
of the complete NNLO computation are known to the same accuracy as we
would obtain from  a simple tree-level $n$-jet merging scheme like
CKKW~\cite{ckkw} or MLM~\cite{mlm}.\medskip

For the Tevatron Higgs search in the inclusive $H \to WW$ channel the
contributions of the different topologies have been analyzed in
detail~\cite{bryan_tev}. The theory uncertainties estimated by a
simultaneous renormalization and factorization scale variation $\mu
\epsilon [m_H/2, 2 m_H]$ are weighted with the relative contributions
from the three dominant topologies included in the
resummation-improved NNLO prediction and give a signal uncertainty of
\begin{alignat}{5}
\hspace{-1.2cm}
\frac{\Delta N_S}{N_S} &=
60\%  \left({^{+5\%}_{-9\%}} \right)
+29\% \left({^{+24\%}_{-23\%}} \right)
+11\%  \left({^{+91\%}_{-44\%}} \right) \notag \\
&= \left({^{+20.0\%}_{-16.9\%}} \right)
\qquad \qquad {\rm [Tevatron]}
\end{alignat}
which is larger than expected in current Tevatron analyses. Note that
any error estimate based on a scale variation can only give us a lower
limit of the theory uncertainty, because it probes certain
higher-order contributions while neglecting others.

This illustrative example
indicates how, for example, the theory uncertainty of a neural net
analysis at the LHC would have to be analyzed. First, we
identify the regions of phase space contributing at a given rate to
the combined result. For each of these regions we quantify the theory
uncertainty, and combine them for a final number. While for
backgrounds this argument seems to call for theory-independent (or Monte-Carlo
independent) search channels, such a
thing does not exist for the measurement of the Higgs rates.
Unfortunately,
we
cannot make conclusive statements about a Higgs discovery without
estimating the Higgs couplings from rate measurements at hadron
colliders and hence the issue of theory uncertainties in Higgs
signals remains as a crucial issue.

\subsection{Subjet analyses for $H \to b\bar{b}$}

Until recently, there did not exist a Higgs discovery channel
involving the (dominant) decay to bottom jets. The key to such a
measurement is to focus on boosted Higgs bosons with two collimated
bottom jets which can in turn be analyzed as one fat Higgs
jet~\cite{Butterworth:2008iy}. The size of such a jet can be estimated by
\begin{equation}
 R_{bb} \sim \frac{1}{\sqrt{z(1-z)}} \; \frac{M_H}{p_T}
\end{equation}
where $z$ and $1-z$ are the momentum fractions of the two decay
jets. The cleanest Higgs production mode with a guaranteed trigger
even for a fully hadronic Higgs decay is the associated production
with a leptonic $W$ or $Z$ boson.  Applying a cut $p_{T,H} > 200~\gevTSG$
reduces the available rate in this process to around 5\% and suggests
a starting size of the fat Higgs jet of $R < 1.2$. This fat jet we
de-cluster and search for a signature of a heavy Higgs decay into two
light bottom jets. One measure of such a massive decay is a drop in
the jet mass at a given de-clustering step which we can supplement
with a balance criterion to separate symmetric Higgs decays from
asymmetric QCD jet radiation.

\begin{figure}[t]
\begin{center}
\includegraphics[scale=0.38]{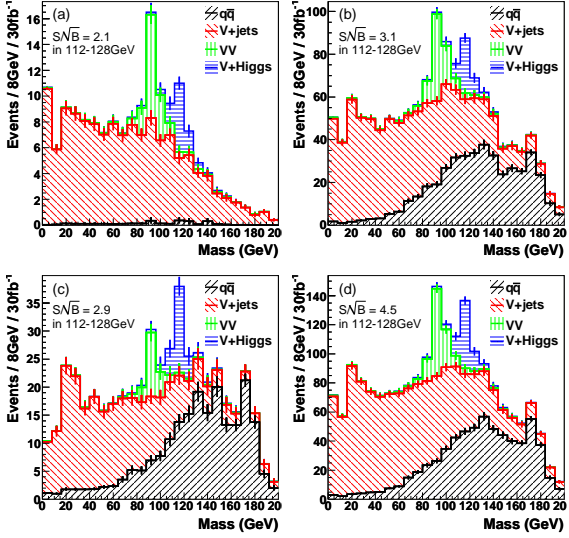}
\end{center}
\vspace*{-10mm}
\caption[]{ Signals and backgrounds for the subjet analysis in the
  three different channels --- (a) $Z_\ell H$; (b) $Z_{\rm inv} H$;
  (c) $W_\ell H$; and (d) all three channels combined. The nominal
  Higgs mass is 115~GeV. From Ref.~\cite{Butterworth:2008iy}. }
\label{fg:gavin}
\end{figure}

It turns out that for such a mass drop criterion and a not too large
Higgs boost the Cambridge-Aachen jet algorithm is suited best, in
particular better than the $k_T$ or anti-$k_T$ algorithms. The free
parameter in the asymmetry criterion for the two splitting products
$j_{1,2}$
\begin{equation}
\frac{\min (p_{T,j_1}^2, p_{T,j_2}^2)}{m_{\rm jet}^2}
\; ( \Delta R_{j_1,j_2} )^2 > y_{\rm cut} = 0.09
\end{equation}
is chosen to balance signal efficiency and background rejection for
QCD jets with a wide variety of topologies and is therefore process
dependent.

The typical energy scales of the Higgs constituent jets is not much
above the transverse mass scale of the underlying event at the
LHC. This means that
%unless we prune the jets before applying a fat
%jet analysis~\cite{pruning}
we need to remove softer jet radiation
from the reconstructed Higgs jet~\cite{Butterworth:2008iy,pruning}. At the same time, one radiated QCD
jet is likely to contribute to the jet mass reconstructing the Higgs
mass, so we cannot simply veto a third jet inside the Higgs jet. One
way to remove underlying event (or pileup) contamination is by
filtering the contents of the fat Higgs jet with a lower resolution
$R_{\rm filter} = \min(0.3, R_{bb}/2)$. At this finer resolution we
combine the three leading objects to form the Higgs resonance, which
sharpens the Higgs mass peak while at the same time preserving its
peak position at the nominal Higgs mass value, shown in
Fig.~\ref{fg:gavin}.

At the hadron level but without detector simulation the significance
of the combined $ZH$ and $WH$ search channels with $m_H = 115~\gevTSG$ is
$4.5\sigma$ for an integrated luminosity of $30~\ifb$ or $8.2\sigma$
for $100~\ifb$. These numbers are based on a $b$ tagging performance
of 60\% and a light-flavor mis-tagging probability of 2\%.\medskip

The result of the analysis Ref.~\cite{Butterworth:2008iy} has been checked
including full ATLAS detector simulation~\cite{atlsubjet}. The two
results are compatible and only differ slightly in two respects:
first, the $b$ tagging performance of constituents inside a filtered
fat is actually improved compared to regular jets, because the
filtered constituents are closer to the $B$ baryon's direction. On the
other hand, charm backgrounds from $t \to b \bar{c} s$ where the
$b\bar{c}$ system fakes the Higgs are dangerous and require a good
charm rejection in the $b$ tag. The resulting shifts in the final
significance due to these two effects largely balance each
other.\medskip

\begin{figure}[t]
\begin{center}
\includegraphics[scale=0.70]{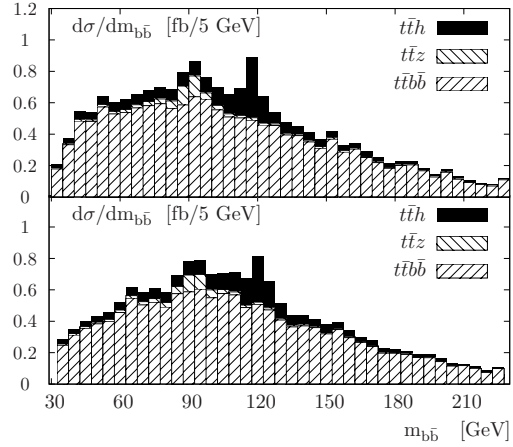}
\end{center}
\vspace*{-10mm}
\caption[]{ Signals and background for the subjet analysis of the $t
  \bar{t} H$ analysis assuming $m_H = 120~\gevTSG$. Shown are results
  without (upper) and with (lower) underlying event and
  filtering. From Ref.~\cite{tth}.}
\label{fg:tth}
\end{figure}

The same basic idea of boosted fat jets can be used to resurrect the
$t\bar{t}H$ analysis with $H \to b \bar{b}$. The two lethal problems
of the original analysis are the combinatorics of bottom jets in the
signal (and which bottoms to choose to reconstruct the Higgs mass) and
the lack of an effective cut against the $t\bar{t} b\bar{b}$ continuum
background. This background has recently been evaluated at
next-to-leading order~\cite{ttbb_nlo}, reducing the systematic
uncertainties on this Higgs search channel.

Semileptonic $t \bar{t} H$ production with a Higgs decay to bottoms is
well suited for a Higgs search involving two fat jets, one from the
Higgs and one from the hadronic top, while the leptonic top decay
ensures reliable triggering~\cite{tth}. To account for the limited
phase space, both fat jets are expanded to $R < 1.5$. The top tagger
again searches for mass drops inside the fat jet, but then requires
them to reconstruct the $W$ and top masses. This twofold mass
constraint reduces the mis-tagging probability to the 5\% level. The
Higgs tagger, in contrast, has to extract the correct Higgs mass peak,
so it cannot be biased by a given Higgs mass. Instead, it orders the
observed mass drops by the modified Jade distance
\begin{equation}
J = p_{T,j_1} p_{T,j_2} (\Delta R_{j_1,j_2} )^4  \; .
\end{equation}
The jet substructure analysis alone is sufficient to control the continuum $t
\bar{t} b\bar{b}$ background alone to $S/B \sim 1/2.5$, but rejecting the
different topologies of the mis-tagged $t \bar{t} jj$ background
requires three $b$ tags, for example two inside the Higgs jet and one
outside the top and Higgs constituents. The resulting significance at
the hadron level is $4.8\sigma$ with $S/B\sim 1/2$ assuming $100~\ifb$ for a Higgs mass
of 115~GeV. The reconstructed Higgs mass peak is shown in
Fig.~\ref{fg:tth}.

\subsection{Conclusions}
With $30~\ifb$ at an energy of $\sqrt{s}=14~\tevTSG$, we expect discovery
of a Higgs like signal in at least two channels at the LHC. New Higgs
search channels utilizing the decay $H \to b \bar{b}$ and jet
substructure analyses are expected to improve this situation
further~\cite{Butterworth:2008iy,tth}. The task remaining will be to verify that this particle is
the Higgs boson of the Standard Model, which necessitates the
measurement of Higgs couplings (See section~\ref{sec:HiggsCouplings} ). For
each of the signatures included this requires a solid understanding of
the experimental and theory uncertainties, with a focus on modern
analysis techniques.

%%%%%%%%%%%%%%%%%%%%%%%%%%%HIGGSNONSTANDARD%%%%%%%%%%%%%%%%%%%%%%%%%%%%%%%%%%%%%

\section{Alternative Higgs Scenarios}

\subsection{Nonstandard Higgs Models and Decays}

{\it S.~Chang and T.~Han}\medskip

In addition to the standard Higgs decay scenarios, nonstandard Higgs decay scenarios
can be envisaged. In the nonstandard scenarios
the Higgs dominantly decays into new light states, ultimately resulting in a cascade decay
into multiple Standard Model particles.

\subsubsection{Introduction}
%Electroweak symmetry breaking (EWSB) through the vacuum expectation value of a fundamental scalar, as in the Standard Model (SM), predicts the existence of a physical scalar called the Higgs boson.  Despite many searches, most notably those at LEP2, the Higgs boson remains to be undetected.  In the near future, the Tevatron and LHC experiments will be the primary hopes for its discovery.  Given the Higgs boson's prominent place in EWSB, these negative search results place interesting constraints on many theories.  In the Standard Model, there is a slight tension between the Higgs mass allowed by LEP2 direct search constraints ($m_H > 114.4$ GeV at 95\% C.L. \cite{Barate:2003sz}) and the precision electroweak data (which prefer a light Higgs mass of 87 GeV and requires it to be less than 157 GeV at 95\% C.L. \cite{lepewwg}).  Similarly, in many theories beyond the Standard Model, the Higgs mass direct search limit has ruled out much of the ``natural'' parameter space of the theory.  The best known example is the minimal supersymmetric Standard Model (MSSM), where the Higgs mass limit requires heavy scalar top quarks, which reintroduces a $\sim 5\%$ fine-tuning for proper EWSB (see ref.~\cite{Dermisek:2005ar}).

Given the Higgs boson's prominent place in electroweak symmetry
breaking (EWSB), the negative search results at LEP and the Tevatron
place interesting constraints on many theories. There is a slight
tension between the Higgs mass allowed by the LEP2 direct search
constraints ($m_H > 114.4$ GeV at 95\% C.L. \cite{Barate:2003sz})
and the precision electroweak data in the Standard Model preferring
a mass of 87 GeV~\cite{lepewwg} as discussed in
Section~\ref{sec:HiggsPred}. In many theories beyond the Standard
Model, the Higgs mass direct search limit has ruled out much of the
``natural'' parameter space of the theory.  The best known example
is the MSSM, where the Higgs mass limit requires heavy scalar top
quarks, which reintroduces a $\sim 5\%$ fine-tuning for proper EWSB
(see ref.~\cite{Dermisek:2005ar}).

In the interim between LEP2's shutdown and the LHC's startup, there
has been a lot of work in what this might imply for Higgs physics.
In particular, of recent interest is the notion that the Higgs is
actually lighter than the LEP2 limit.  Such a situation can be
consistent if the Higgs boson has new decays that dominate over the
standard decay modes.  Since for such Higgs masses, the Higgs decay
width is quite small, new decay modes can very easily be the
dominant modes.  This was realized as early as
ref.~\cite{Shrock:1982kd} and emphasized in supersymmetric models in
\cite{Gunion:1996fb}.  In summary, this nonstandard Higgs decay
scenario is a situation where the Higgs coupling to SM particles is
normal, but interactions with new light fields allow the Higgs to
decay into them.  These new particles themselves decay, producing a
Higgs ``cascade'' whose limits are typically weaker than the LEP2
limit.  Hence, the scenario alleviates the tensions between theory
and experiment.  More details on the motivations and implications
can be found in a recent review \cite{Chang:2008cw}.

\subsubsection{Nonstandard Higgs Decays}

The nonstandard Higgs scenario's most important phenomenological
consequence is the fact that standard searches may no longer be
sensitive to the new decays.\footnote{For an exception, see
\cite{Chang:2009zpa}.}  Since the standard decays are suppressed by
\begin{equation}
\epsilon = \frac{1}{(1+\Gamma_{new}/\Gamma_{SM}) },
\end{equation}
the significance of any standard Higgs search can be reduced by a
factor of $\epsilon$ or equivalently require a factor of
$1/\epsilon^2$ increase in luminosity relative to the standard
expectation to reach the same statistical significance. For the
nonstandard Higgs lighter than the LEP2 constraint, $\epsilon$ is
required to be less than about 20\%, thus discovery requires $> 25$
times more luminosity than before.  This makes LHC searches
extremely challenging, so it is important to see if the new dominant
decays can be searched for as well.

In this short note, we highlight some of the recent progress in this
regard, references to earlier work can be found in
\cite{Chang:2008cw}.  We focus on decays to a pair of a light scalar
$a$, which decays into a pair of SM fermions:
\begin{equation}
h \to aa,\quad a \to f\bar f.
\end{equation}
If it is heavy enough, $a$ dominantly decays into $b\bar{b}$, while
below the $b$ threshold, it will decay mostly into $\tau
\bar{\tau}$.  Thus, depending on the $a$ mass, the dominant Higgs
decays are $h \to 4b, 4\tau$.\footnote{In some models the scalar
dominantly decays to gluons, giving weaker Higgs mass limits, see
for e.g. \cite{Bellazzini:2009xt}.}  In particular, the $4b$ decay
was still strongly constrained at LEP2, but the $4\tau$ decay was
much more weakly constrained \cite{Schael:2006cr}.

\vspace{.1cm} {\noindent $\mathbf {h\to 4b}$ \bf :} Even though it
is strongly constrained, the Higgs decay into four $b$ quarks may
still occur for Higgs masses above 110 GeV.  This can be searched
for at ATLAS/CMS experiments by looking at $W h$ associated
production \cite{Carena:2007jk}.  Other interesting searches can be
done by focusing on displaced vertices. In Hidden Valley models, the
$a$ decays can be highly displaced \cite{Strassler:2006ri}, which
has been searched for by D0 \cite{Abazov:2009ik} with no significant
excess. Interestingly, at LHCb with its capabilities in displaced
vertices, a preliminary analysis suggests that it could also
discover this decay mode \cite{Kaplan:2009qt}.

\vspace{.1cm} {\noindent $\mathbf {h\to 4\tau}$ \bf :} Given the
weaker constraints on the $4\tau$ decay mode, it is important to
consider lower Higgs masses.  An interesting approach to searching
for this mode is to take a branching ratio hit by requiring one of
the $a$ bosons to decay into a pair of muons and the other into
hadronic tau modes.  The muons form a mass peak at $m_a$, helping
pick this out of background. A preliminary Tevatron/LHC analysis was
done in \cite{Lisanti:2009uy} and was performed at D0
\cite{Abazov:2009yi}, with no significant excess.

The light scalar $a$ can also be searched for in Upsilon decays
\cite{Hiller:2004ii}, $\Upsilon \to a \gamma$. Babar has recently
searched for this in the $a\to 2\tau, 2\mu$ modes
\cite{Aubert:2009cka,Aubert:2009cp} placing constraints that are
starting to limit the expected parameter space
\cite{Dermisek:2006py}.

As all of these studies illustrate,  there are many challenging and
exciting ways to probe the nonstandard Higgs scenario.  If this
scenario is realized, it will take new studies such as these, to
maintain the ability to discover the Higgs boson at future
colliders.  Thus, discovering what breaks electroweak symmetry could
hinge on the careful exploration of these new possibilities.

%%%%%%%%%%%%%%%%%%%%%%%%%%%Low Mass Tau Pairs%%%%%%%%%%%%%%%%%%%%%%%%%%%%%%%%%%%%%

\subsection{Discovering the Higgs with Low Mass Muon Pairs}

{\it M.~Lisanti and J.~Wacker}\medskip

A primary goal of current collider programs is to discover the Higgs
boson and the mechanism of electroweak symmetry breaking.  Direct
and indirect searches for the Standard Model (SM) Higgs have set
bounds on the allowed masses.  LEP has excluded a Higgs that decays
directly to $b\bar{b}$ or $\tau^+ \tau^-$ with mass $m_{h^0} \leq
114$ GeV \cite{Barate:2003sz}.  Combined Higgs searches from $\DO$
and CDF have recently excluded $163$ GeV $< m_{h^0} < 166$ GeV
\cite{Collaboration:2009je}.  While direct searches point to a heavy
Higgs, measurements of electroweak observables that depend
logarithmically on the Higgs mass impose upper limits:  the best fit
for a SM Higgs is 77 GeV, with a 95\% upper bound of $m_{h^0} \leq
167$ GeV \cite{LEP:2008}.

Alternate models of electroweak symmetry breaking that lead to
naturally light Higgses with non-standard decay modes are less
constrained than the Standard Model scenario (see
\cite{Chang:2008cw} for a review).  LEP's model-independent bound on
the Higgs mass is 82 GeV \cite{Abbiendi:2002qp} and its bound for a
Higgs boson that decays to four taus is 86 GeV
\cite{Abbiendi:2002in}.  These alternate Higgs models are motivated
by the desire to reduce fine-tuning and to satisfy the bounds from
indirect searches.  They often contain additional scalar fields and
more complicated scalar potentials with approximate global
symmetries.  If these symmetries are explicitly broken, they can
result in a light pseudo-Goldstone boson that has $\OO(1)$ coupling
to the Higgs, leading to a substantial branching fraction into new
light scalar states.

For specificity, consider a two Higgs doublet model with an
additional scalar field $S$.  All three fields acquire vacuum
expectation values: $v_u = v\sin \beta$, $v_d = v \cos \beta$, and
$\langle S \rangle$.\footnote{This example shares the essential
features of NMSSM-like theories.}  Assume that there is an
approximate symmetry that acts upon the Higgs doublets as $H_i
\rightarrow e^{i \theta_{q_i}} H_i$, with the singlet compensating
by $S \rightarrow e^{i \theta_{q_s}} S$.  When electroweak symmetry
is broken, $S$ acquires a vev, spontaneously breaking the global
symmetry.  The phase of $S$ becomes a pseudo-Goldstone boson, $a^0$,
that has small interactions with the Standard Model when $\langle S
\rangle \gg v$.  In such models, the dominant interaction between
the Higgs and the pseudoscalar arises from %%
\begin{equation}
\LL_{\text{int}} \sim  \frac{h^0 \partial_{\mu} a^0 \partial^{\mu}
a^0}{1 + \Big(\frac{\langle S \rangle}{\sin 2\beta}\Big)^2 }.
\end{equation}
The ratio $\langle S \rangle/ \sin2\beta$ parameterizes the strength
of the Higgs-pseudoscalar coupling.  In particular, the coupling
strength increases as  the value of $\langle S \rangle/ \sin2\beta$
decreases.  The light pseudoscalar $a^0$ also couples to the SM
fermions through the following interaction: %%
\begin{equation}
\LL_{\text{int}}=i g_{f} \frac{m_f}{v} \bar{f} \gamma_5 f a^0,
\end{equation}
where
\begin{eqnarray}
g_{f} \sim \frac{v \sin 2\beta}{\langle S \rangle} \begin{cases}
\cot\beta  & \mbox{      (up-type quarks)} \\
\tan\beta & \mbox{     (down-type quarks/leptons)}
 \end{cases}
\end{eqnarray}
Below the b-quark threshold, the pseudoscalar decays primarily to
tau leptons, rather than charm quarks.

In the presence of the new light pseudoscalar state, the primary
Higgs decay mode is: %%
\begin{equation}
h^0 \rightarrow a^0 a^0 \rightarrow (X\bar{X})(X\bar{X}).
\end{equation}
When $m_{a^0} > 2 m_b$, the pseudoscalars each decay into a pair of
b quarks.  This search is strongly constrained by LEP, with $m_{h^0}
\geq 110$ GeV  \cite{Schael:2006cr,Abdallah:2004wy,Abbiendi:2004ww}.
Recent analyses have found that this $4b$ signal can be discovered
over the QCD background with 30 fb$^{-1}$ of data
\cite{Carena:2007jk,Cheung:2007un}.  However, this is highly
dependent on the b-tagging efficiencies that can be achieved at the
LHC and if the efficiency is less than 50\% for $p_T \sim 15 $ GeV,
higher luminosities will be needed.  The $2b 2\tau$ decay mode was
also explored, but found to be less promising.

Below the b-quark threshold, the LEP bounds on the Higgs mass
weaken.  When $m_{a^0} < 2 m_{\mu}$, a $4\mu$ search is appropriate.
When $2 m_{\mu} < m_{a^0} < 2 m_{\tau} $, each pseudoscalar decays
primarily to a pair of taus. %%
\begin{figure*}[tb] %  figure placement: here, top, bottom, or page
   \includegraphics[angle=270,width=2.95in]{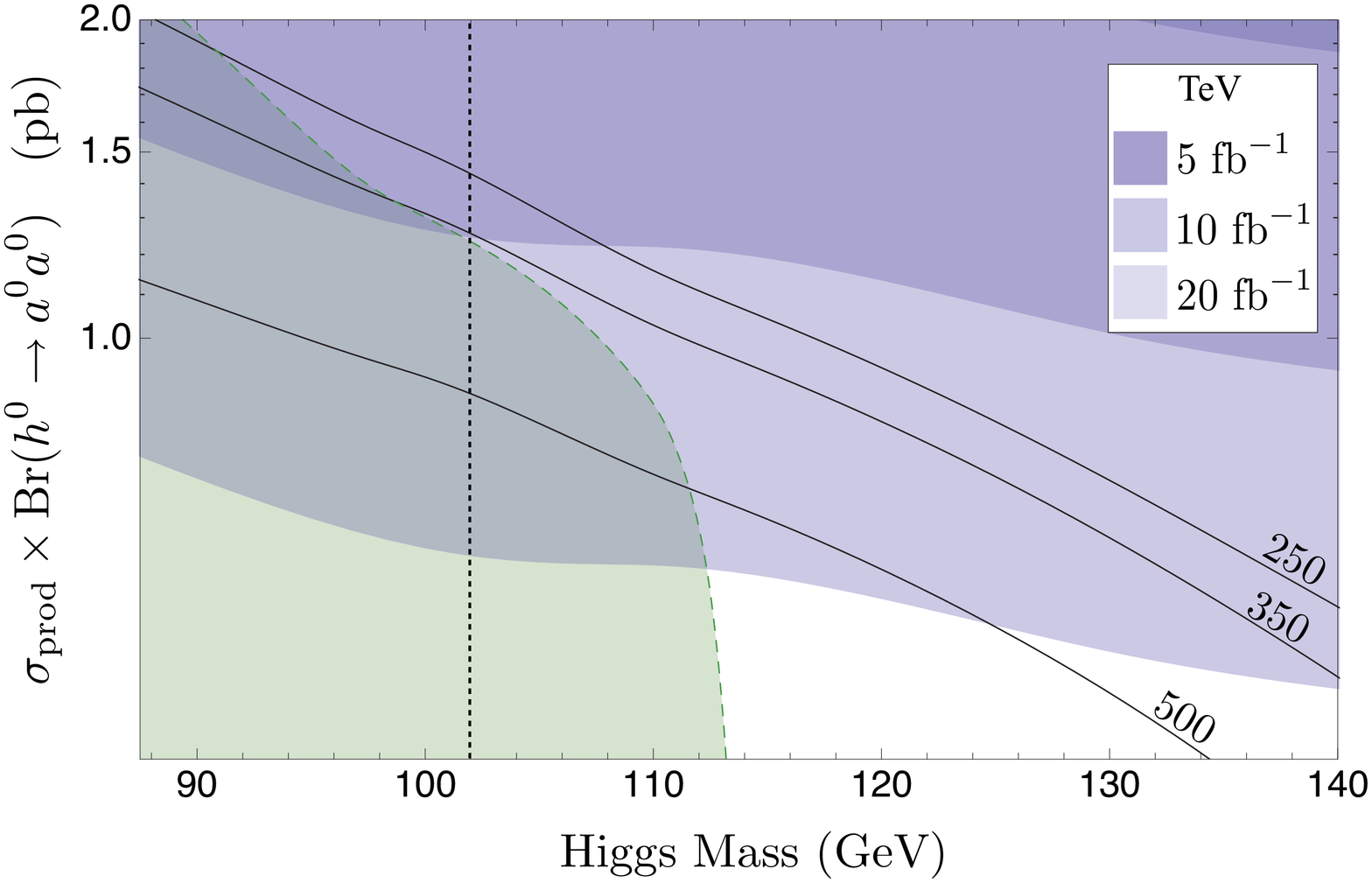}
   \hspace{0.1in}
   \includegraphics[angle=270,width=3.05in]{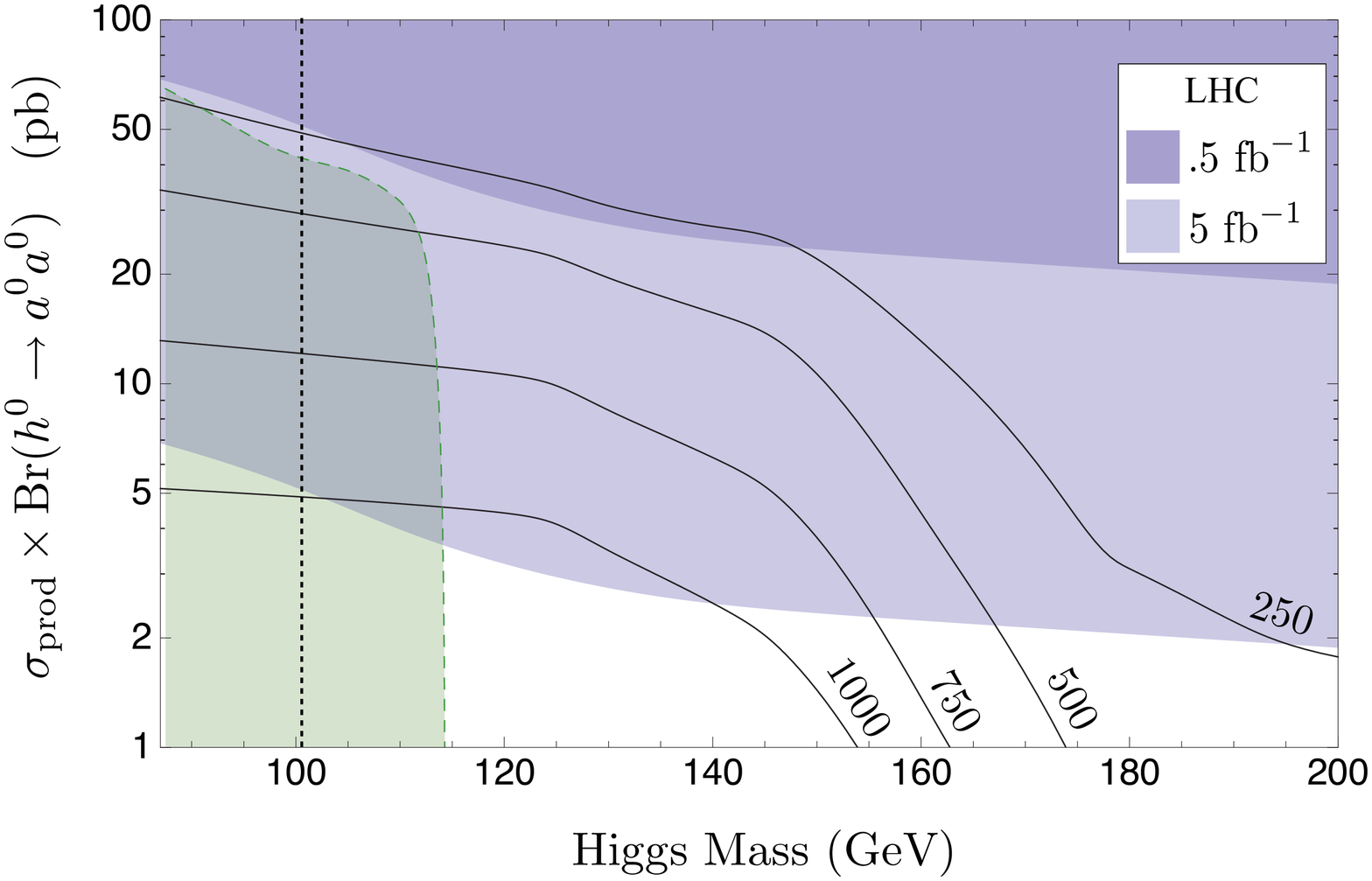}
   \caption{Expected sensitivity to the Higgs production cross section at the Tevatron %for the current 4 fb${}^{-1}$ per experiment and for the eventual combined search with expected, full integrated luminosity
   (left) and LHC (right) for $m_{a^0} = 7$ GeV.  The contour lines indicate the cross sections for several values of $\langle S \rangle /\sin2\beta$ (in GeV), which alters the higgs branching fraction to pseudoscalars.  The light green region is excluded by LEP.  The vertical dashed line indicates the expected limit of a LEP reanalysis of the $h^0 \rightarrow 4\tau$ channel \cite{GGI}.}
   \label{fig: excplot}
\end{figure*}
In comparison to the $4\mu$ search, the $4\tau$ signature is
particularly challenging because  the taus decay to leptons only a
third of the time and the leptons are typically very soft
\cite{Graham:2006tr}.  Currently, there are several proposed
searches for the $4\tau$ signal at both ATLAS and CMS
\cite{Forshaw:2007ra,:2008uu}.  The ATLAS collaboration is exploring
the $4\mu8\nu$ channel and CMS is looking at ($\mu^{\pm}
\tau_h^{\mp}$)($\mu^{\pm} \tau_h^{\mp}$).

We have proposed an alternative search to $h^0 \rightarrow 4\tau$
that takes advantage of the pseudoscalar's subdominant decay to
muons \cite{Lisanti:2009uy}.  In particular, we consider the case
where one pseudoscalar decays to a pair of taus and the other decays
to a pair of muons.  This decay channel has not been previously
explored because the decay into muons is suppressed by
$\OO(m_{\mu}^2/m_{\tau}^2)$.  For example, when $m_{a^0} = 7$ GeV
and $\tan \beta \gtrsim 4$, the branching fraction to taus is
$98\%$, while that into muons is $0.4\%$.  However, the Higgs
production cross section can be large enough to compensate for this
small branching fraction; the gluon-gluon fusion can be as high as 2
pb at the Tevatron or 50 pb at the LHC.  It is therefore possible to
get $\sim 300$ events at the Tevatron with 20 fb$^{-1}$ and $\sim
250 $ events at the LHC (at $\sqrt{s} = 14$ TeV) with 0.5 fb$^{-1}$.

The signal topology for the $h^0 \rightarrow 2 \mu 2\tau$ search is
as follows.  The pseudoscalars are highly boosted and lead to
nearly-collinear acoplanar lepton pairs.  Each tau has a 66\%
hadronic branching fraction, so there is a 44\% chance that both
taus will decay to pions and neutrinos, which the detector will see
as jets and missing energy.  If only one tau decays hadronically,
there will still be a jet and missing energy.  About 3\% of the
time, both taus will decay to muons.  The signal of interest is
therefore %%
\begin{equation}
pp \rightarrow \mu^+ \mu^- + \text{di}\tau + \MET,
\end{equation}
where $\text{di}\tau$ refers to the ditau object.  Because the taus
are nearly collinear, they will often be picked out as a single jet.
The missing energy is in the same direction as the jet.

To reduce the background contributions, all events are required to
have a pair of oppositely-signed muons within $|\eta| < 2$, where
each muon has a $p_T$ of at least 10 GeV.  A jet veto of 15 and 50
GeV for Tevatron and LHC, respectively, is placed on all jets except
the two hardest.  Also, it is required that the hardest muon is
separated from the $\MET$ by $\Delta \phi \geq 140^{\circ}$.  The
three higher level cuts are: $p_T^{\mu \mu} \gtrsim 0.4 m_{h^0}$,
where $p_T^{\mu \mu}$ is the sum of the transverse momentum of the
two muons, $\MET \gtrsim (0.2 -0.5) \times m_{h^0}$, and $\Delta
R(\mu,\mu) \gtrsim 4 m_{a^0}/m_{h^0}$.  It is important to emphasize
that standard lepton isolation must be altered when doing such
searches.  In particular, it is necessary to remove the adjacent
muon's track and energy before estimating the nearby hadronic
activity.

The main backgrounds to this signal are: Drell-Yan muons recoiling
against jets, electroweak processes, and leptons from hadronic
resonances.  The Drell-Yan background is the most important; in
these events, the missing energy arises from jet energy
mismeasurement or neutrinos from heavy semileptonic decays in jets.
The Drell-Yan background dominates over electroweak contributions
from WW and $t\bar{t}$.  Contributions to the hadronic backgrounds
arise from several different sources.  One example is the
possibility of double semi-leptonic decays in jets ($b\rightarrow c
\rightarrow s/d$).  This turns out to be minimal after cuts because
high $p_T$ muons are rare and there is a lot of hadronic activity
surrounding the muons.  Another possibility comes from upsilon
decays into taus, which then decay to muons.  In this case, few
events survive the cuts because the missing energy is in the
direction of the muon pair and the $p_T$ spectrum of the upsilons
falls off rapidly.  The final possibility arises from leptonic
decays of light mesons, such as the $J/\psi$.  This turns out to be
minimal because high-$p_T$ muons only occur out on the Lorentzian
tail of the decay width or on the Gaussian mismeasurement tail.  In
all, the hadronic contribution is $\ll 10\%$ of the Drell-Yan
background.  For a complete discussion of the backgrounds, see
\cite{Lisanti:2009uy}.

Figure~\ref{fig: excplot} shows the 95\% exclusion plot for the
Tevatron and LHC for a pseudoscalar with $m_{a^0} = 7$ GeV.  For
lighter pseudoscalars, the sensitivity can increase by a factor of
two.  The contour lines show the cross sections for values of
$\langle S \rangle/ \sin 2\beta$.  The $h^0 \rightarrow 2\mu 2\tau$
search was recently done at $\DO$ using 4.2 fb$^{-1}$ of data
\cite{Abazov:2009yi}.  From the figure, it is clear that the
Tevatron will start probing $\langle S \rangle/ \sin 2\beta = 250$
GeV with 10 fb$^{-1}$ of data, and can probe up to 500 GeV when the
projected 20 fb$^{-1}$ luminosity is reached.   The LHC will be able
to recover the LEP limit with 1 fb$^{-1}$ of data and has the
potential for higgs discovery with a sub-fb$^{-1}$ data set.

%%%%%%%%%%%%%%%%%%%%%HIGGSCOUPLINGS%%%%%%%%%%%%%%%%%%%%%%%%%%%%%%%%%%%%%%%%%%%

\section{Determination of Higgs-Boson Couplings}
\label{sec:HiggsCouplings}

{\it M.~D\"uhrssen, M.~Rauch, R.~Lafaye, T.~Plehn  and D.~Zerwas}\medskip

%\section{Introduction}
After establishing the presence of a light Higgs boson at the LHC, the
next step will be to measure its properties, in particular its couplings
to other particles and to itself~\cite{couplings,duehrssen,Lafaye:2009vr}.
In the Standard Model these are
completely determined by the measured masses of the particles together
with the gauge coupling $g$ and the electroweak mixing angle. Deviations
from these relations can occur through an extended Higgs sector
such as the MSSM~\cite{mhiggsAEC,susy,susyhiggs}.
Another possibility for modifications are additional particles which can
shift the couplings via loop contributions or lead to different
branching ratios by providing additional decay channels.

We do not
consider channels necessitating more than $30\ \ifb$
which should not feed back
into the leading parameter set.
Our underlying model for the analysis~\cite{duehrssen,Lafaye:2009vr} is
the Standard Model where we let the Higgs couplings float freely around
its Standard Model value.  The main channels for the coupling
measurements can be found in
Refs.~\cite{duehrssennote,Aad:2009wy,Ball:2007zza}.

\subsection{Determination}
%\subsection{Higgs mass dependence}
The study in~\cite{duehrssen} analyses the measurement of Higgs
couplings in the mass range of $110$ to $190$~GeV. The errors are
extracted as deviations on the coupling square. New invisible and/or
undetectable Higgs boson decay modes are allowed.
The $t\bar{t}H (H\to b\bar{b})$ channel is based on older analyses
with higher sensitivities to this channel.

Figure~\ref{coupl:MassScan} shows the precision on the coupling
determination and the total width as function of the Higgs boson mass
with and without systematic uncertainties for the combination of two LHC
experiments at $30\ \ifb$.
\begin{figure}[ht]
 \begin{center}
 \includegraphics[width=0.38\textwidth]{%
     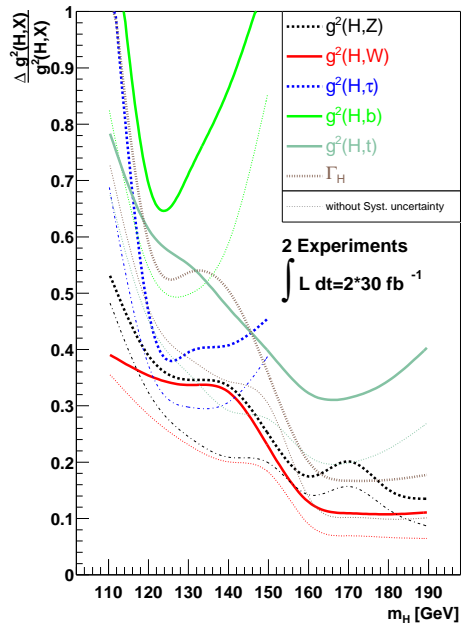}
 \end{center}
 \vspace{-1.1cm}
\caption[]{Relative error on the coupling determination for two LHC
experiments at $30\ \ifb$ with and without systematic experimental and
theory errors as function of the Higgs boson mass.}
 \vspace{-0.5cm}
\label{coupl:MassScan}
\end{figure}
The crucial role of the bottom coupling for low Higgs masses is visible
here, as the errors for all couplings are dominated by the large
uncertainty on the $b\bar{b}$ measurement below the $WW$ threshold mass.
For these masses, errors range from $\approx35$-$100\%$, above $160$~GeV
a precision of $15$-$40\%$ on $\Delta g^2(H,j)$ is reached.

%%% SFITTER
For the remainder of this short overview we will put the focus on a
Higgs boson mass of $120$~GeV. This is the preferred region for a
Standard Model Higgs boson from electroweak precision data and
compatible with the lower limit from direct LEP
searches~\cite{Collaboration:2008ub}.
The by far leading decay channel in this mass range is into a pair of
bottom quarks. Via the total width it enters into the branching ratios
of all other particles. Therefore a precise knowledge is essential,
particularly in the light of the severely reduced $ttH$-production-channel
sensitivity which we now account for, and we include the recent subjet
analysis~\cite{Butterworth:2008iy}. It can greatly improve the accuracy
on this coupling, up to similar levels than the older $ttH$ results,
and is currently under study by both experimental
groups~\cite{atlsubjet}. Also now we do not allow for invisible or
undetectable Higgs decay modes, so the total width is fixed to the sum
of the observable Higgs decay widths.
To perform the analysis we use the
SFitter~\cite{Lafaye:2007vs} framework to map these highly correlated
measurements onto the parameter space. A detailed overview of the
individual channels, its associated experimental and theory errors and
the correlations between the errors is in Ref.~\cite{Lafaye:2009vr}.

We parametrize the Higgs couplings $g_{jjH}$ as deviations from its
Standard Model value $g_{jjH}^{\text{SM}}$ via
\vspace*{-2ex}
\begin{equation}
 g_{jjH} \longrightarrow g_{jjH}^{\text{SM}} \; \left( 1 + \Delta_{jjH}
                                   \right)  \; ,
\label{coupl:SM_Change}
\end{equation}
where the $\Delta_{jjH}$ are independent of each other.
Furthermore we allow for additional contributions to the two important
loop-induced couplings $g_{ggH}$ and $g_{\gamma\gamma H}$
\vspace*{-2ex}
\begin{equation}
 g_{jjH} \longrightarrow
  g_{jjH}^{\text{SM}} \; \left(
   1 + \Delta_{jjH}^{\text{SM}} + \Delta_{jjH} \right) \; ,
\label{coupl:SMeff_Change}
\end{equation}
where $g_{jjH}^{\text{SM}}$ is the loop-induced coupling in the
Standard Model, $\Delta_{jjH}^{\text{SM}}$ the contribution
from modified tree-level couplings to Standard-Model particles, and
$\Delta_{jjH}$ an additional dimension-five contribution, for
example from new heavy states.
The relation to the
definition in Fig.~(\ref{coupl:MassScan}) is
\vspace*{-1ex}
\begin{equation}
\Delta g^2(H,j) \approx 2 \cdot \Delta_{jjH} \; .
\end{equation}

%\section{Determination}
\subsubsection{Profile likelihood}

\begin{figure}[t]
\begin{center}
\includegraphics[scale=0.98]{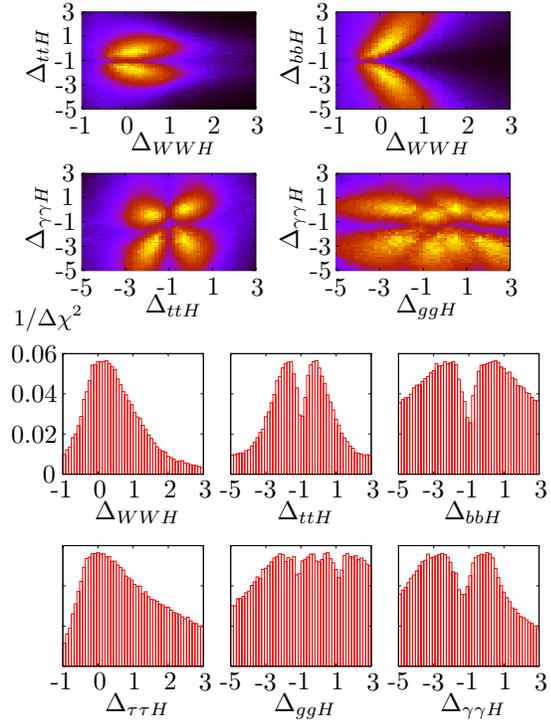}
\end{center}
\vspace*{-10mm} \caption[]{Profile likelihoods for smeared
measurements assuming
 $30~\ifb$.  We include both experimental and theory errors in our
 analysis.}
\vspace{-0.7cm} \label{coupl:profile}
\end{figure}

\begin{table*}
\begin{small}
\begin{tabular}{l|l|l|ll|l|l|ll|l|l|ll}
 &         \multicolumn{4}{c|}{no effective couplings} &
           \multicolumn{4}{c|}{with effective couplings} &
           \multicolumn{3}{c}{ratio $\Delta_{jjH/WWH}$} \\
 & RMS & $\sigma_\mathrm{symm}$ & $\sigma_\mathrm{neg}$ & $\sigma_\mathrm{pos}$
 & RMS & $\sigma_\mathrm{symm}$ & $\sigma_\mathrm{neg}$ & $\sigma_\mathrm{pos}$
 & $\sigma_\mathrm{symm}$ & $\sigma_\mathrm{neg}$ & $\sigma_\mathrm{pos}$
\\\hline
$\Delta_{WWH}$
 & $\pm\,0.31$ & $\pm\,0.23$ & $-\,0.21$ & $+\,0.26$
 & $\pm\,0.29$ & $\pm\,0.24$ & $-\,0.21$ & $+\,0.27$
 &\phantom{$\pm$} --- &\phantom{$-$} ---
&\phantom{$+$} --- \\
$\Delta_{ZZH}$
 & $\pm\,0.49$ & $\pm\,0.36$ & $-\,0.40$ & $+\,0.35$
 & $\pm\,0.46$ & $\pm\,0.31$ & $-\,0.35$ & $+\,0.29$
 & $\pm\,0.41$ & $-\,0.40$ & $+\,0.41$ \\
$\Delta_{ttH}$
 & $\pm\,0.58$ & $\pm\,0.41$ & $-\,0.37$ & $+\,0.45$
 & $\pm\,0.59$ & $\pm\,0.53$ & $-\,0.65$ & $+\,0.43$
 & $\pm\,0.51$ & $-\,0.54$ & $+\,0.48$ \\
$\Delta_{bbH}$
 & $\pm\,0.53$ & $\pm\,0.45$ & $-\,0.33$ & $+\,0.56$
 & $\pm\,0.64$ & $\pm\,0.44$ & $-\,0.30$ & $+\,0.59$
 & $\pm\,0.31$ & $-\,0.24$ & $+\,0.38$ \\
$\Delta_{\tau\tau{}H}$
 & $\pm\,0.47$ & $\pm\,0.33$ & $-\,0.21$ & $+\,0.46$
 & $\pm\,0.57$ & $\pm\,0.31$ & $-\,0.19$ & $+\,0.46$
 & $\pm\,0.28$ & $-\,0.16$ & $+\,0.40$ \\
$\Delta_{\gamma\gamma{}H}$ &
 \phantom{$\pm$} --- &\phantom{$\pm$} --- &\phantom{$-$} ---
&\phantom{$+$} ---
 & $\pm\,0.55$ & $\pm\,0.31$ & $-\,0.30$ & $+\,0.33$
 & $\pm\,0.30$ & $-\,0.27$ & $+\,0.33$  \\
$\Delta_{ggH}$             &
 \phantom{$\pm$} --- &\phantom{$\pm$} --- &\phantom{$-$} ---
&\phantom{$+$} ---
 & $\pm\,0.80$ & $\pm\,0.61$ & $-\,0.59$ & $+\,0.62$
 & $\pm\,0.61$ & $-\,0.71$ & $+\,0.46$
\end{tabular}
\end{small}
\caption[]{Errors on the measurements from 10000 toy experiments. We
  quote errors for Standard Model couplings only and including
  effective $ggH$ and $\gamma\gamma H$ couplings using $30\ \ifb$ of
  integrated luminosity, as well as the error on the ratio of the
  coupling to the $WWH$ coupling. The different $\sigma$ measures we
  define in the text.}
  \label{coupl:errors}
  \vspace{-0.2cm}
\end{table*}

In Fig.~\ref{coupl:profile} we show profile likelihoods for various
parameters, where we smear the set of data input arbitrarily within
their respective errors. We see that in all cases a value of $\Delta=0$,
\ie the Standard Model solution, is compatible with the data.
Furthermore for the tree-level couplings there are solutions at
$\Delta=-2$, corresponding to a flipped sign of the coupling. For the
loop-induced couplings four solutions exist, originating from both
unflipped and flipped sign for the $t\bar{t}H$ coupling and the
additional contribution to the effective coupling. In the $ggH$ case two
solutions coincide at $\Delta ggH = 0$ for exact data; due to the
smearing they get shifted apart and we indeed see all possibilities
distinctly. For $\gamma\gamma H$ the top-quark contribution is
subleading, so all maxima are unique, but a pair of two, corresponding
to flipped sign of the top-quark coupling, is close to each other and
they get smeared into a single one.

\subsubsection{Errors}
In Tab.~\ref{coupl:errors} we show the errors on the extraction of Higgs
coupling parameters. We obtain these errors by running 10000 toy
experiments and smearing the data around the true point including all
experimental and theory errors. The best fits for each parameter we
histogram and extract $\sigma_\mathrm{symm}$ using a Gaussian fit of the
central peak. As we do not expect the errors to be symmetric, we also
fit a combination of two Gaussians with the same maximum and the same
height, but different widths, labeling these ($\sigma_\mathrm{neg}$) and
($\sigma_\mathrm{pos}$).  We also show a root-mean-square (RMS) error.
These are systematically higher as in this case outliers have larger
impact.

In the left column we quote the errors for both additional contributions
to the effective couplings set to zero and in the middle one when we also
allow these couplings to deviate from zero. In the column on the
right-hand side we show the errors on the ratio of the coupling
to the $WWH$ coupling. We define them as the deviation from~$1$ of the
ratios of the coupling constants
\vspace*{-1ex}
\begin{equation}
\frac{g_{jjH}}{g_{WWH}} \rightarrow
\left( \frac{g_{jjH}}{g_{WWH}} \right)^\mathrm{SM}
\left( 1 + \Delta_{jjH/WWH} \right) \; .
\end{equation}

We see immediately that the $ttH$ coupling obtains increased errors once
we allow effective couplings due to its dominant contribution to
$g_{ggH}$.
A similar effect we would expect for $g_{WWH}$ where the photon decay
channel gets effectively removed but here the accuracy of the remaining
measurements is sufficiently well.
Both $\tau\tau H$ and $bbH$ couplings are strongly linked to $g_{WWH}$,
so here we do not see a change either.

In particular the $bbH$ coupling benefits from forming the ratio. This
coupling appears in all rate predictions via the total width which leads
to strong correlations.
For all other couplings we observe minor improvements from the channels
where the production-side $g_{WWH}$ enters the determination of the
decay-side couplings.

\subsection{Conclusions}
In summary we can determine the Higgs couplings with an
accuracy up to $10\%$, and to about $20$-$40\%$ in the
phenomenologically favored
region of $120$~GeV using an integrated luminosity of
$30\ \ifb$. Forming ratios of couplings can slightly improve these
numbers. For a light Higgs boson the determination of its coupling to
bottom quarks is crucial. Due to its large branching ratio it is linked
to all channels via the total width. Therefore a reliable measurement
for example using subjet analyses~\cite{Butterworth:2008iy} is vital.
Given these errors a distinction between the Standard Model and the MSSM
or other decoupling models seems not likely, but dramatic modifications
(like a gluophobic Higgs in the MSSM) will be clearly visible.
%%%%%%%%%%%%%%%%%%%%%HIGGSSTRINGS%%%%%%%%%%%%%%%%%%%%%%%%%%%%%%%%%%%%%%%%%%

\section{On the Possible Observation of Light Higgses $A, H, H^{\pm}$ at the LHC}

{\it D.~Feldman, Z.~Liu and P.~Nath}\medskip

The Higgs patterns in SUGRA models and the connection to the String Landscape
are discussed.

\subsection{Light Higgses in the SUGRA and String Landscape}
An exhaustive survey of sparticle mass hierarchies and their signatures
reveals the presence of light SUSY Higgses ($A, H, H^{\pm}$) in high scale models   \cite{Feldman:2007zn}.
Such light Higgses have been uncovered in mSUGRA and non-universal SUGRA models,
 as well as in a class of D-brane models. The light Higgses are found to exist in specific mass hierarchies
or Higgs Patterns (HPs)   \cite{Feldman:2007zn},
and consist of SUGRA patterns labeled as mSP14, mSP15 and mSP16, which are defined by the following :
\vspace{-.1cm}
\begin{equation}
\begin{array}{rl}
 {\rm mSP14 :}   &\naPN   <  A\sim H < \hcPN,    \nonumber\\
 {\rm mSP15 :}   &\naPN   <  A\sim H < \chaPN,  \nonumber\\
 {\rm mSP16 :}   &\naPN   <  A\sim H  < \staPN,  \nonumber
\end{array}
\end{equation}
where a relative switch
between the ordering of $(A,H)$ and the LSP ($\naPN$) is possible as well.
It is shown in   \cite{Feldman:2007zn} that
these patterns are very stable with respect to nonuniversalities in the  Higgs sector  \cite{Nath:1997qm}.
There are many other patterns that emerge in the survey  of the SUGRA landscape. These
may be classified minimally according to the NLSP, and there are found to be NLSPs which
are the chargino, stau, stop, higgs, gluino and the sneutrino; comprising a total of around 40 4-particle mass patterns
discounting the lightest CP even Higgs boson whose  mass is fixed over a narrow  corridor of about 25 GeV.

\begin{figure}[t]
\begin{center}
\includegraphics[width=6.5cm,height=5.5cm]{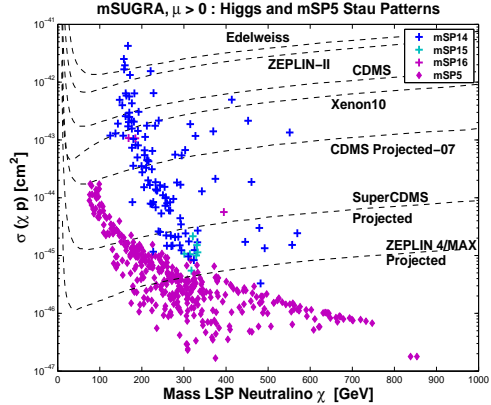}
\caption{
  Constraints on the Higgs Pattern mSP14 and mSP16
from  experiments for the direct  detection of dark matter.
Also shown for comparison are models in  mSP5  where stau is the NLSP.
From   \cite{Feldman:2007zn}. }
 \label{fig:HP1}
  \end{center}
\end{figure}

\begin{figure*}[t]
\begin{center}
\includegraphics[width=6.25cm,height=5.25cm]{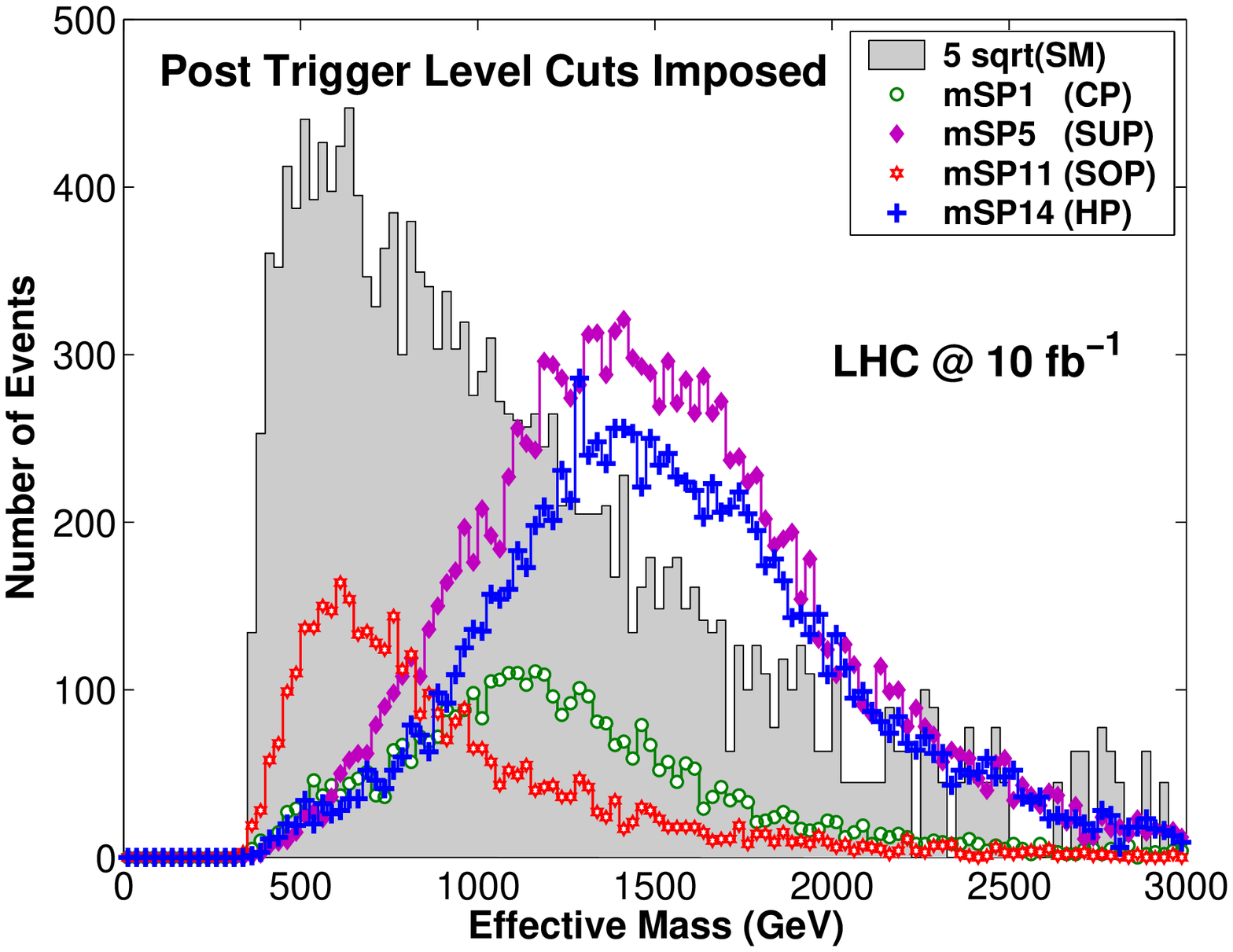}
\includegraphics[width=6.25cm,height=5.25cm]{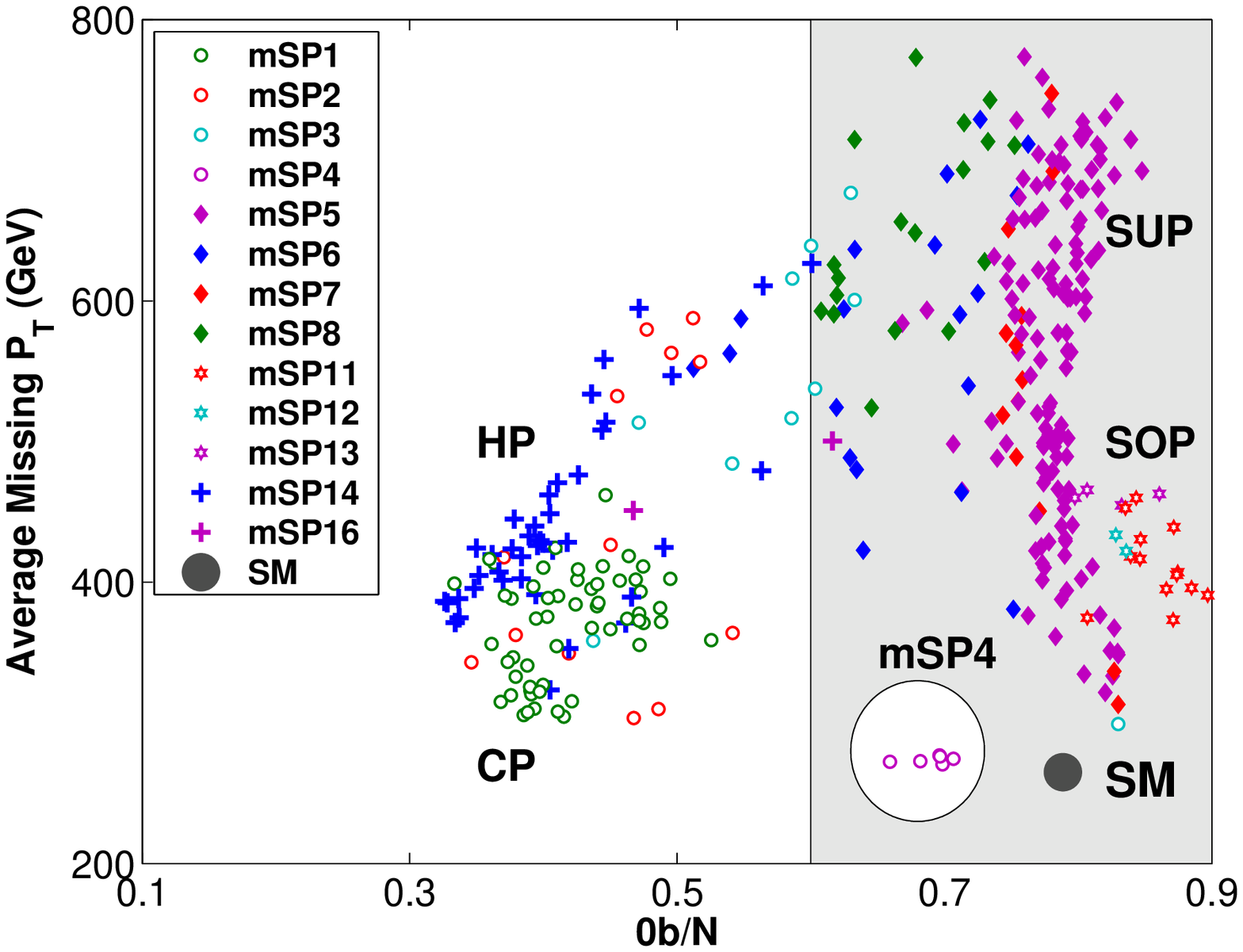}
\caption{Left: Effective mass distribution ($M_{\rm eff}$) of the
light Higgs pattern mSP14 with
 $(m_{A,H},M_{H^\pm})  \sim (156,180)~ \rm GeV$, $m_{\naPN}$ $\sim 158  ~{\rm GeV} > m_{A,H}$.
Also shown for comparison are  $M_{\rm eff}$  for sample benchmarks in  mSP5 (stau NLSP),  mSP11 (stop NLSP),
and mSP1 (chargino NLSP). Right:
Higgs patterns (HP) vs Chargino (CP), Stau (SUP) and Stop Patterns (SOP)
in a correlation of the average missing $P_T$ vs. the number of events with no $b$ tagged jets.
Also shown is the SM contribution.
 The combined analysis, which is based on $\sim$ 900 model points for mSUGRA at the LHC
 with 10fb$^{-1}$ of data, shows  a clear separation of the Higgs Patterns from the other
 patterns when examining both correlations and distributions. From Ref.  \cite{Feldman:2007zn} }
 \label{fig:HP2}
  \end{center}
\end{figure*}
 In fact the Higgs Patterns are found to be one of the more dominant patterns that can arise. These patterns
are likely to be one of the first to be tested at the LHC, as we will discuss further.
 \subsubsection{Dark matter constraints on Higgs Patterns mSP(14-16)}
The Higgs Patterns are found to be constrained by  dark matter experiments
from limits on the spin independent cross section, $\sigma (\chi p)$ ,
as seen in Fig.(\ref{fig:HP1}). These  constraints on the HPs arise specifically for
low values of the neutralino mass. The $\sigma (\chi p)$ for HPs is much larger than, for example,
 the stau pattern (SUP)  mSP5. This is in part  because of  a larger Higgsino component of the
LSP for the HPs relative to the Stau Pattern where the LSP  is more bino like and also because the HPs have lighter
neutral SUSY Higgs being exchanged for the same neutralino mass.
\subsubsection{Discovering light Higgses  $A, H, H^{\pm}$  at the LHC}

The lightness of $A$ (and also of $H$ and $H^{\pm}$)
 in the HPs implies that the Higgs  production cross sections at colliders can be enhanced for these
 patterns. This is especially true at  large $\tan \beta$ in processes   via gluon fusion  and bottom quark annihilations.
Such an enhancement
opens fruitful discovery prospects in the 2 $\tau$ mode   as well as possibilities for the charged Higgs decaying
into $\bar b  t$.  For the case of large $\tan \beta$ one is  led to
corroborating constraints from the recent Tevatron  collider data including constraints on $B_s\to \mu^+\mu^-$
and, as already noted,
  from direct detection dark matter experiments  \cite{Feldman:2007zn}.
In Fig.(\ref{fig:HP2}) we give a relative comparison of the  effective mass distributions
($ M_{\rm eff} \equiv \sum_{\rm jet} P^{\rm jet}_T +{{\not\!\!{P_T}}}$) of a single Higgs
pattern (HP) and of a Chargino Pattern (CP) mSP1,  a Stau Pattern (SUP)  mSP5,
and of a Stop Pattern (SOP) mSP11.
The analysis of  Fig.(\ref{fig:HP2}) shows that the HP and  the
 SUP have a relatively larger $M_{\rm eff}$  while
 the $M_{\rm eff}$ of the CP and of SOP are much smaller.
 This is found more generally to be the case.
The values of the effective mass are intimately tied to the length of the decay chains for the specific patterns and
the overall scale of the colored sector.  The fat distributions from the HPs make them easier to discover at the LHC,
 but can easily be confused with a SUP. Thus additional signatures are needed to discriminate
 among them.
 One such discriminating correlation of signatures is  given in Fig.(\ref{fig:HP2})
 which shows signals of the HPs vs other Patterns (CPs, SOPs, SUPs) in a plane of
 average missing $P_T$ vs. the fraction of events $N_{0b}/N_{\rm SUSY}$.
  Here one finds that the  HPs and the SUPs are, on the whole, significantly well separated.
  In summary the analysis above illustrates that a combination of signatures at the LHC should give rise to the
  resolution of the  Higgs Patterns from other SUSY patterns and allow one to discover the light
  Higgses of the HPs if such particles are indeed light.

%%%%%%%%%%%%%%%%%%%%%%%%%%%%%%%%%%%%%%%%%%%%%%%%%%%%%%%%%%%%%%%%%%%%%%%%%%%%%%
%%%%%%%%%%%%%%%%%%%%%%%%%%%%%%%%%%%%%%%%%%%%%%%%%%%%%%%%%%%%%%%%%%%%%%%%%%%%%%

%\input{biblio}
%\clearpage

%%%%%%%%%%%%%%%%%%%%%%%%%%%%%%%%%%%%%%%%%%%%%%%%%%%%%%%%%%%%%%%%%%%%%%%%%%%%%%
%%%%%%%%%%%%%%%%%%%%%%%%%%%%%%%%%%%%%%%%%%%%%%%%%%%%%%%%%%%%%%%%%%%%%%%%%%%%%%

%\end{document}

%%%%%%%%%%%%%%%%%%%%%%%%%%%%%%%%%%%%%%%%%%%%%%%%%%%%%%%%%%%%%%%%%%%%%%%%%%%%%%%%%%%%%%%%%%%%%%
%%%%%%%%%%%%%%%%%%%%%%%%%%%%%%%%%%%%%%%%%%%%%%%%%%%%%%%%%%%%%%%%%%%%%%%%%%%%%%%%%%%%%%%%%%%%%%
\chapter{CP Violation at the LHC}
\epigraphhead[20]{\epigraph{\large {\em Tarek Ibrahim, Jae Sik Lee,
Pran Nath, Apostolos Pilaftsis}}{\large Apostolos Pilaftsis
(Convener)}}
%
%%%%%%%%%% espcrc2.tex %%%%%%%%%%
%
% $Id: espcrc2.tex 1.2 2000/07/24 09:12:51 spepping Exp spepping $
%
%\documentclass[fleqn,twoside]{article}
%\usepackage{espcrc2}

% change this to the following line for use with LaTeX2.09
% \documentstyle[twoside,fleqn,espcrc2]{article}

% if you want to include PostScript figures
%\usepackage{graphicx}
% if you have landscape tables
%\usepackage[figuresright]{rotating}

% put your own definitions here:
\def\lsim{\:\raisebox{-0.5ex}{$\stackrel{\textstyle<}{\sim}$}\:}
\def\gsim{\:\raisebox{-0.5ex}{$\stackrel{\textstyle>}{\sim}$}\:}

\newcommand{\imag}{\Im {\rm m}}
\newcommand{\real}{\Re {\rm e}}
\renewcommand{\beqn}{\begin{eqnarray}}
\renewcommand{\eeqn}{\end{eqnarray}}
\def\beq{\begin{equation}}
\def\be{\begin{equation}}
\def\eeq{\end{equation}}
%%%%%%%%%%%%%%%%%%%%%%%%%%%%%%%%%%%%%%%%%%%%
%% FRONTMATTER
%%%%%%%%%%%%%%%%%%%%%%%%%%%%%%%%%%%%%%%%%%%%

%\title{\bf CP Violation at the LHC}

%\title{\bf Higgs-Sector CP Violation at the LHC}

%\begin{document}

%\begin{abstract}
We give a brief overview of the hunt for CP violation in
supersymmetric signatures at the LHC. Such signatures could arise
from large CP violating  phases which are still  consistent with the
electric dipole moment (EDM)  experiments for  the  EDMs of the
electron, for the neutron and for the Mercury and Thallium atoms.
Specifically reviewed are the potential signatures that arise from a
supersymmetric CP-violating Higgs sector at the LHC. As particular
examples, we discuss phenomena of Higgs-sector CP violation at the
LHC based on the MSSM scenarios {\it CPX} and {\it Trimixing}. Also
discussed are the possible tests of the Cancelation Mechanism from
various processes such as from studies  of  CP violating effects on
$B_s^0\to \mu^+\mu^-$, and CP effects on sparticle decays as well
effects on dark matter which is strongly interconnected with LHC
physics in supersymmetric unified models of particle interactions.
%\end{abstract}

%\begin{abstract}
%We give a brief overview of the potential signatures that arise from a
%supersymmetric CP-violating Higgs sector at the LHC. As particular
%examples, we discuss phenomena of Higgs-sector CP violation at the LHC
%based on the MSSM scenarios {\it CPX} and {\it Trimixing}.
%\end{abstract}

%\maketitle

%%%%%%%%%%%%%%%%%%%%%%%%%%%%%%%%%%%%%%%%%%%%
%% MAINMATTER
%%%%%%%%%%%%%%%%%%%%%%%%%%%%%%%%%%%%%%%%%%%%

\section{CP violation in Supersymmmetric Theories}
The Standard Model of particle interactions has two sources of CP violation, one that enters in the
Cabibbo-Kobayashi-Maskawa (CKM)  matrix ($V_{ij}$) and the other  that enters in the strong
interaction dynamics via the term  $ \theta \frac{\alpha_s}{8\pi}G\tilde G$, where $G_{\mu\nu}$ is
the gluonic field strength.  These phases are  constrained by the neutron electric dipole moment (EDM)
so that  $\bar\theta =(\theta+  ArgDet(M_uM_d)) <O(10^{-10})$. The phase that enters in the CKM
matrix  is also separately constrained, for example, by the condition

\beq
V_{ud}V^*_{ub} + V_{cd}V^*_{cb} + V_{td} V^*_{tb}=0
\label{ckm}
\eeq
This constraint can be exhibited by a unitarity triangle with angles
$\alpha, \beta, \gamma$ where
$\alpha =arg(-{V_{td}V^*_{tb}}/{V_{ud}V^*_{ub}})$,
$\beta =arg(-{V_{cd}V^*_{cb}}/{V_{td}V^*_{tb}})$,
and $\gamma =arg(-{V_{ud}V^*_{ub}}/{V_{cd}V^*_{cb}})$.
Our current direct knowledge of CP violation hinges on experiments on
the K and B systems. This evidence arises in the neutral Kaon
system in the form of $\epsilon$ and $\epsilon '/\epsilon$ where
experimentally
%measured to be
\beqn
 \epsilon = (2.28\pm 0.02)\times 10^{-3},\nonumber\\
 \epsilon'/\epsilon =(1.72\pm 0.18)\times 10^{-3}.
 \label{cp1}
 \eeqn
 In the B system $B^0_d$ ($\overline{B^0_d}$)$\rightarrow J/\Psi K_s$ decay
 gives a direct measurement of $\sin (2\beta)$ so that
 \beqn
    \sin(2\beta) = 0.75 \pm 0.10 ~~~BaBar;\nonumber\\
    0.99 \pm 0.15 ~~~Belle
    \label{cp2}
\eeqn
Finally another piece of evidence for CP violation comes from the
 baryon asymmetry in the universe so that
 \beq
 n_B/n_{\gamma}= (1.5- 6.3)\times 10^{-10}
 \label{cp3}
 \eeq
The experimental results of  Eqs.~(\ref{cp1}) and Eq.~(\ref{cp2})
are consistent with  CP violation given by the Standard Model.
 However,  an explanation
 of the baryon asymmetry in the universe implies the need of a
CP violation beyond the Standard Model.
Further, it is possible that in the future
  more  accurate measurements of the angles
$\alpha, \beta,\gamma$ may indicate a breakdown of the unitarity
triangle which would point to the existence of
a source of CP violation
  beyond the  standard model. Another possible indication
  of a new source of CP violation would be an experimental observation of
  an EDM of elementary fermions.
    Specifically in the Standard Model the EDM
    of a lepton arises at the multiloop level~\cite{Hoogeveen:1990cb}
    and is too small to be  observed (see Table~\ref{table:leptonedm}).
     However, much larger EDMs can arise
    in   supersymmetric theories.

\begin{center} \begin{tabular}{|c|c|c|}
\multicolumn{3}{c}{ Table 1:  Lepton EDMs } \\
\hline
  & SM (ecm) & Experiment (ecm)\\
 \hline
 $e$ &  $\leq 10^{-38}$ & $< 4.3 \times 10^{-27}$ \\
 \hline
 \hline
  $\mu$ & $\leq 10^{-35}$   & $ < 1.1\times 10^{-18}$ \\
 \hline
 $\tau$ &  $\leq 10^{-34}$  &  $ < 3.1\times 10^{-16}$\\
\hline
 \hline
\end{tabular}\\
\label{table:leptonedm}
\end{center}
%%%%
Thus softly broken supersymmtric  theories contain additional sources of  CP violation.
 For example, in mSUGRA with the inclusion of CP phases $m_{\frac{1}{2}}$, $A$ and $\mu$ become
 complex. However, one of these  phases  can be rotated away and one may choose the remaining
 complex parameters to be  $\mu$ and $A$ so that
$\mu =|\mu| e^{i\theta_{\mu}}$ and  $A =|A| e^{i\alpha_A}$. So in this case  the complex mSUGRA
model (cmSUGRA)  has the following parameter space
\beqn
m_0, m_{\frac{1}{2}}, \tan\beta, |A|, \alpha_A, \theta_{\mu} ~~~({\rm cmSUGRA})
\eeqn
 Such new sources of CP violation may play a role
 in the generation of a baryon asymmetry in the universe. However, the new CP phases are
 constrained by the experimental limits on the  EDMs of the electron, of the neutron, and
 of atoms such as of  Hg, and Thallium.  In non-universal SUGRA models one can get a
 much larger  set  of phases. Thus, e.g., allowing for arbitrary masses ($m_{\alpha}$)
  for the gauginos and for the
 trilinear couplings ($A_{\alpha}$) one has
 \beqn
 m_{\alpha}= |m_{\alpha}| e^{i\xi_{\alpha}}, ~~\alpha=1,2,3\nonumber\\
 A_{a} =|A_a| e^{i\alpha_a}, ~~a=1,2,3.
 \eeqn
However, as mentioned  earlier not all the phases are independent and further  in physical
computations only certain combinations  appear  as exhibited in Table (2).
%Table (\ref{table:phases}).

\begin{figure}
  \includegraphics[height=.25\textheight,width=0.25\textwidth]{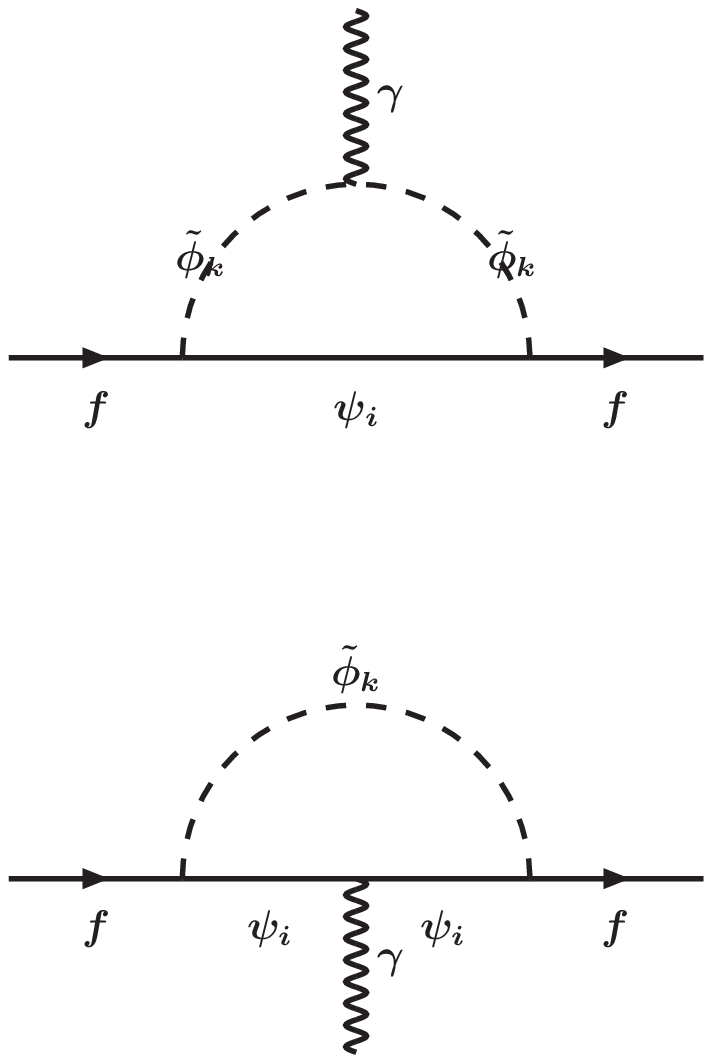}
\hspace{-1cm}
\includegraphics[height=.25\textheight,width=0.25\textwidth]{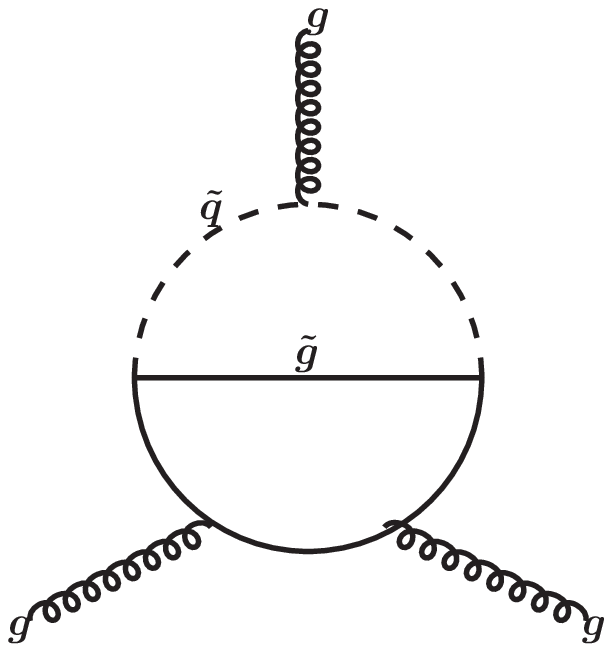}
     \caption{{Left:  One loop correction to EDM of an elementary fermion. Right: Purely gluonic dimension
     six operator~\cite{Weinberg:1989dx}     which contributes to the EDM of the quarks. }}
\label{fig:edm1}
\end{figure}
As mentioned above the current data on the EDM of the electron, of the neutron and of the
 atoms  impose rather severe contraints.   The electron and the quarks receive contributions
 from the electric  dipole operator  (see Eq.~(\ref{edm3})) and  Fig.~(\ref{fig:edm1}))
 while the quarks receive contributions to their  EDMs from the electric dipole operator,
 the chromoelectric dipole operator (see Eq.~(\ref{edm4}),
 and  from the purely gluonic dimension six operator of Weinberg (see Eq.~(\ref{edm5}))  and the
 right panel of Fig.~(\ref{fig:edm1})).
 \beq
{\cal L}_E=-\frac{i}{2}\tilde d_f \bar{f} \sigma_{\mu\nu} \gamma_5  f
 F^{\mu\nu },
 \label{edm3}
\eeq
 \beq
{\cal L}_C=-\frac{i}{2}\tilde d^c \bar{q} \sigma_{\mu\nu} \gamma_5 T^{a} q
 G^{\mu\nu a},
 \label{edm4}
\eeq
  \beq
{\cal L}_G=-\frac{1}{6}d^G f_{\alpha\beta\gamma}
G_{\alpha\mu\rho}G_{\beta\nu}^{\rho}G_{\gamma\lambda\sigma}
\epsilon^{\mu\nu\lambda\sigma},
\label{edm5}
\eeq
where $G_{\alpha\mu\nu}$ is the
 gluon field strength tensor, $f_{\alpha\beta\gamma}$
 are the Gell-Mann coefficients, and $\epsilon^{\mu\nu\lambda\sigma}$
is the totally antisymmetric tensor with $\epsilon^{0123}=+1$.
 In addition, there are important two loop contributions to the EDMs as
 shown in Fig.~(\ref{fig:edm3}).
\begin{figure}
\hspace{-0.75cm}
  \includegraphics[height=.25\textheight,width=0.5\textwidth]{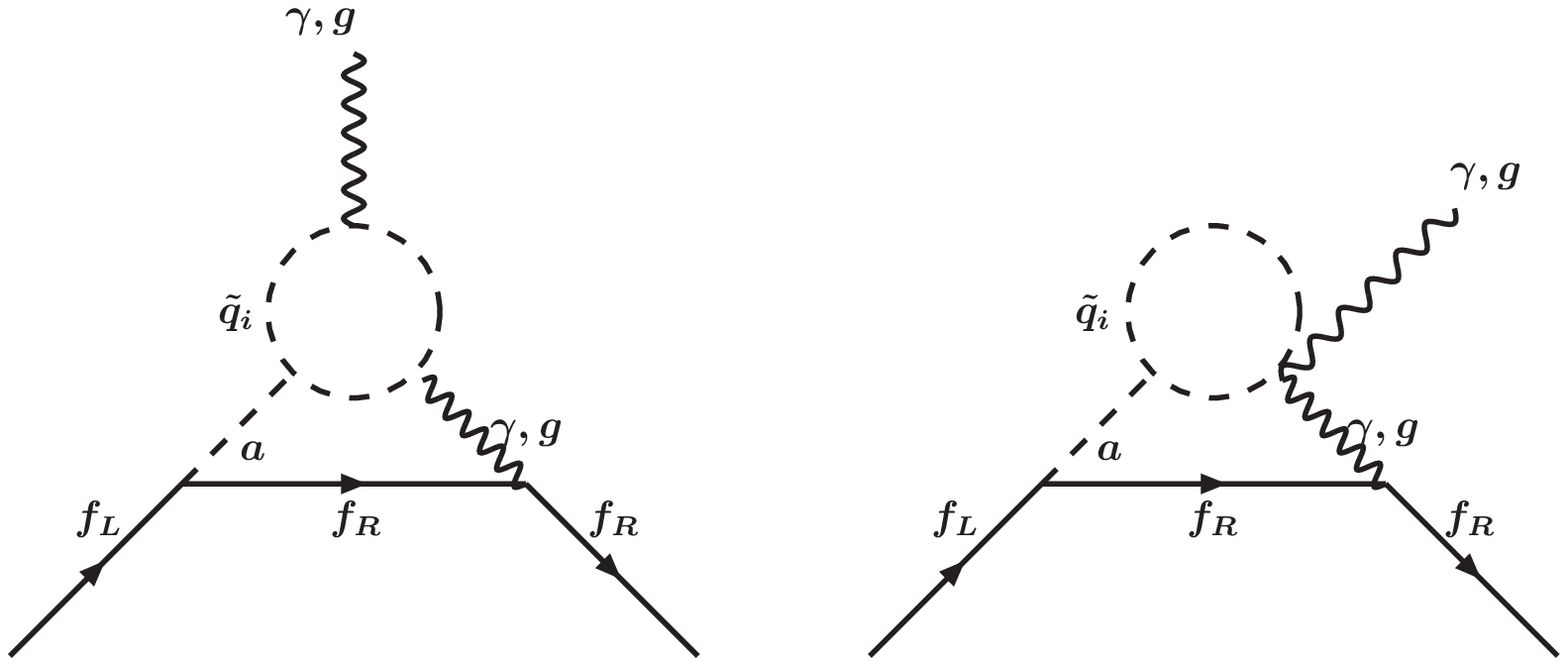}
  \vspace{0cm}
  \caption{{\small Two loop correction to the EDM of an elementary fermion that contribute
  in supersymmetric models~\cite{Chang:1998uc}. }}
\label{fig:edm3}
\end{figure}
    \vspace{0cm}
    \begin{figure}
  \includegraphics[height=.4\textheight,width=0.4\textwidth,angle=270]{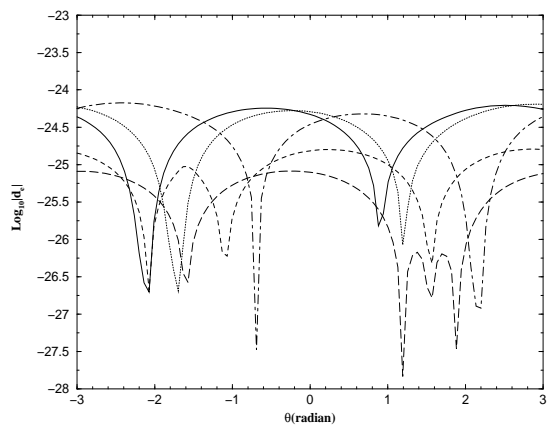}
%    \vspace{-1cm}
   \caption{ An exhibition of the cancellation mechanism for the electron edm with a plot of
    $Log_{10}|de|$ vs $\theta_{\mu}$ where the five curves
   are for the sets of  parameters $\tan\beta$ , $m_0$ , $m_{1/2}$ , $\xi_1$,
   $\xi_2, \xi_3, \alpha_{A0}$ and $A_0$  given by (1): $2, 71,148,
-1.15,-1.4,1.27, -.4,4$ (dotted), (2): $2,71,148, -.87,-1.0,1.78,
-.4,4$ (solid), (3): $4, 550,88, .5,-1.55,1.5, .6, .8$ (dashed),
(4):  $4,750,88, 1.5,1.6,1.7, .6,  .8$ (long dashed), and (5): $2,
71,148, .55, 1 ,1.35, -.4,4$ (dot-dashed). All masses are in GeV and
all phases are in radians. From~\cite{Ibrahim:2000tx}.  }
\label{fig:edm4}
\end{figure}
 As mentioned earlier not all the phases are independent and thus
  the phases enter the EDMs of the leptons and of the quarks only in certain combinations
 which are shown in Table (2).\\
 %Table (\ref{table:phases}).\\
 CP phases also arise in string models.  Recently there has been considerable progress,
 for example, in constructing models based on M theory compactified on   CY $\times S^1/Z_2$
 and models in the framework of Type IIB orientifolds.  Thus  in compactifications on a Type IIB
 theory on a six-torus $T^6=(T^2)^3$, CP violation will arise in F-type breaking which may
 be parametrized by
 \beqn
 F^S=\sqrt{3} m_{\frac{1}{2}} (S+S^*)\sin\theta e^{-i\gamma_s}\nonumber\\
 F^i= \sqrt{3} m_{\frac{1}{2}} (T+T^*)\cos\theta \Theta_i e^{-i\gamma_i},
 \eeqn
 where S is the dilaton field, and $T_i$ the moduli fields, $\theta$ and $\Theta_i$ parametrize
  the Goldstino direction in the $S, T_i$ field space and $\sum_i \Theta_i^2=1$. Here one
  has  four CP violating phases ($\gamma_s, \gamma_i, i=1,2,3$).
  However,
  one problem encountered  both in MSSM and in string based models
    is the so called supersymmetric CP problem, i.e., that the
 phases arising from the soft parameters give too large an EDM for the electron and for
 the quarks. Several  ways have been suggested to control this problem. One solution
 is that CP phases  are small~\cite{ellis}. This requires  a rather artificial fine tuning of phases  since
 in most supersymmetric models and in string models  the CP phases are
 typically not suppressed. Another proposal is that the phases could be $O(1)$ but their
 contributions  are  suppressed due to heavy sparticle masses~\cite{na}. However, from the point
 of view  of discovery of the sparticles very heavy masses are  not preferred.  One possibility
 which allows  for  large phases consistent with a light sparticle  spectrum is the so called
 cancellation mechanism proposed  in\cite{incancel} (For further work see~\cite{incancel2,incancel3}).
 Here one finds  that contributions to the EDMs
  from different sources tend to cancel each other with an appropriate choice of phases.
  Thus, e.g., cancellations can occur in the EDM of the electron between contributions from
  the chargino and from the neutralinos. For the EDMs of the quarks there can be  additionally
  cancellations between the contributions arising from the electric  dipole, chromoelectric  dipole
  and the purely gluonic dimension six  operator.  As a consequence  of these  cancellations
  it is possible to have large CP phases and compatibilty with the experimental EDM data.
  Finally, it is possible that by some symmetry principle phases in the first two generations
  vanish or are highly suppressed and CP phases arise only in the third generation. In this
  case the EDMs will also be suppressed but the CP phases in the third generation could be
  large. For another possible solution to the SUSY CP problem see~\cite{bdm2}.
  \\

  \begin{figure}
%\hspace{1cm}
  \includegraphics[scale=0.3,angle=90]{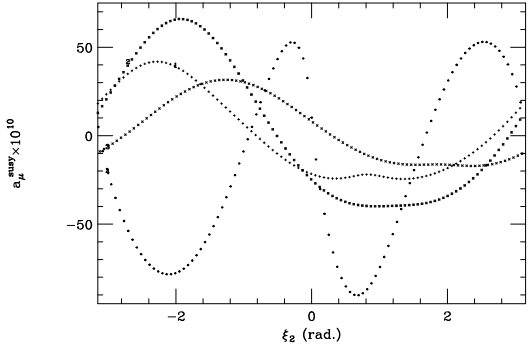}
   \caption{{CP phase dependence of the muon anomalous moment on the phase
  $\xi_2$ with different inputs for the other soft parameters. From~\cite{Ibrahim:1999aj} }}
\label{fig:gmuphase}
\end{figure}
The EDMs of the  quarks and of the leptons are  of course very
sensitive to the phases as they arise directly from the phases. It
is interesting, however, that the anomlaous magnetic moments of
quarks and of  leptons are also sensitive to the phases. In
Fig.~(\ref{fig:gmuphase}) we give a display of the sensitivity of
the supersymmetric contribution to the anomalous magnetic moment of
the muon $a_{\mu}=(g_{\mu}-2)/2$  on the phases. One finds that  the
CP phases can change both the sign and the magnitude of the
supersymmetric contribution. It is then reasonable to ask if the
current experimental data on the muon anomalous moment puts
constraints on the allowed parameter space of the CP violating
phases.  An analysis of this issue shows that this indeed is the
case. As an illustration the constraints arising from the Brookhaven
experiment on  $\theta_{\mu}$ and $\xi_2$  are shown in
Fig.~(\ref{fig:edm5}). One finds that indeed the Brookhaven
experiment on $a_{\mu}$  puts severe constraints on the allowed
region of the CP phases
 for a given set of other soft parameters eliminating large parts of the parameter
 space as exhibited in Fig.~(\ref{fig:edm5}).
   \begin{figure}
\hspace{0.5cm}
  \includegraphics[height=.15\textheight,width=0.15\textwidth,angle=90]{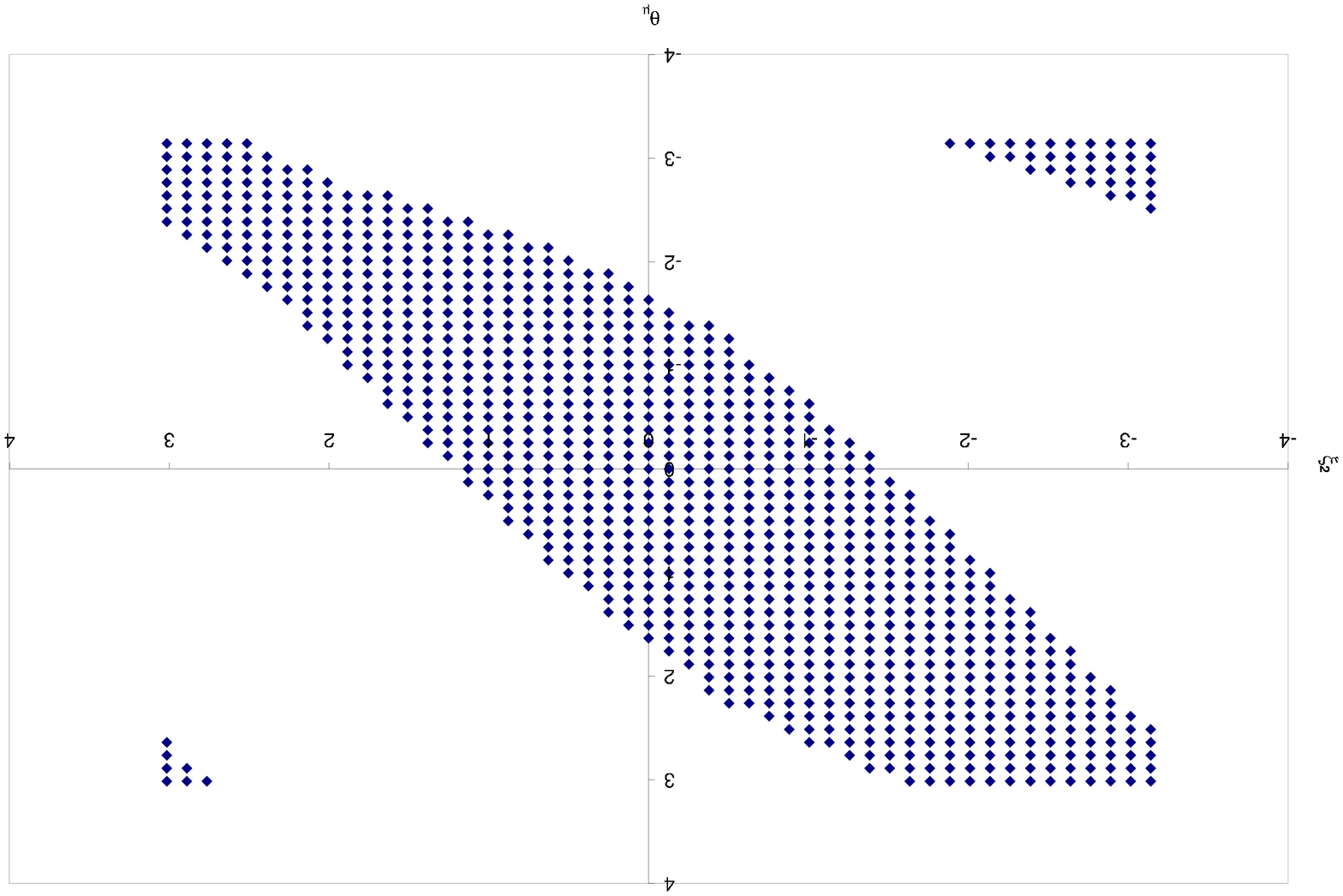}
    \includegraphics[height=.15\textheight,width=0.15\textwidth,angle=90]{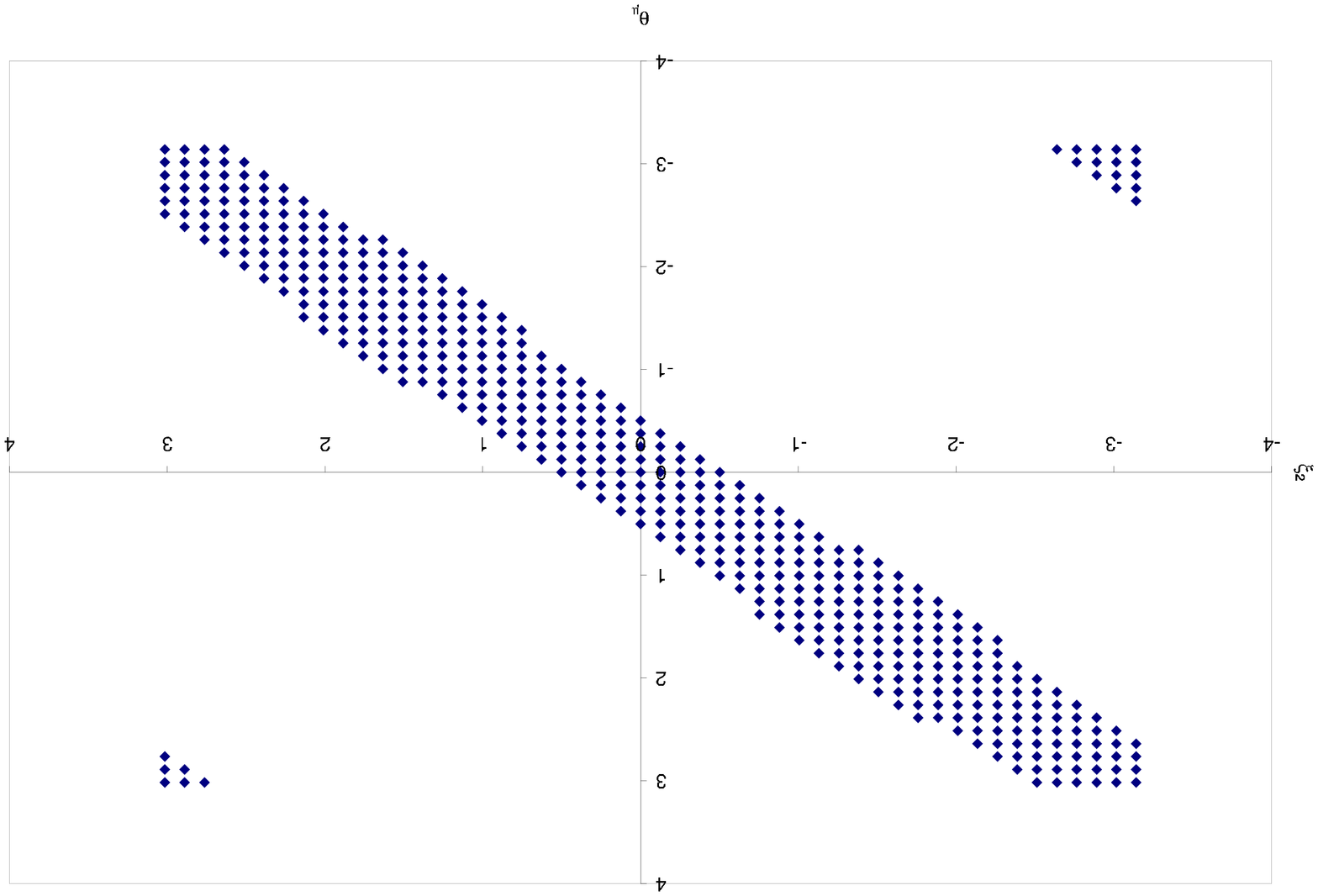}
         \vspace{0cm}
   \caption{{Left panel:
  An exhibition of the allowed  region  of the parameter space in the $\xi_2-\theta_{\mu}$ plane
when $m_0 = 100$, $m_{\frac{1}{2}}=  246$, $\tan\beta =20$, $A_0 =1$, $\xi_1 = .3$, and $\alpha_{A_0} =.5$
where all masses are in GeV using the Brookhaven experimental result on $a_{\mu}$. Right panel: Same as the left panel except that $m_0=400$. From\cite{Ibrahim:2001ym}.} }
\label{fig:edm5}
\end{figure}
 If the CP phases are large then
 many physical processes, some of them directly measurable at the
  LHC,  will be affected and consequently a discovery of such phases can come about at the LHC.
 Thus the  phases appear in a number of processes such as in the decay
  $B_s\to \mu^{+}\mu^{-}$~\cite{inhg199}, in  the production and decay of the sparticles~\cite{kane,zerwas,Ibrahim:2003jm},
  and in neutralino relic density analyses and in the direct detection of
  neutralino dark matter~\cite{Chattopadhyay:1998wb,Falk:1998xj}.
 Further,   large CP phases from
  the soft parameters produce through one loop effective potential induced CP violating phases
  in the Higgs sector allowing for a mixing between the CP even and CP odd Higgs
  neutral Higgs fields.  Such mixing effects can be discerned in the LHC data.
  This topic is discussed in further detail below.

 {\scriptsize
\begin{center} \begin{tabular}{|c|c|}
\multicolumn{2}{c}{Table 2: Examples of CP phases in  SUSY phenomena } \\
\hline
SUSY Quantity  & Combinations of CP violating phases \\
\hline
$m_{\tilde W}$ ($m_{\chi_i}$)  &  $\xi_2+\theta_{\mu}$
($\xi_2+\theta_{\mu}$, $\xi_1+\theta_{\mu}$) \\
\hline
$b\rightarrow s+\gamma$ & $\alpha_{A_t}+\theta_{\mu}$,
$\xi_2+\theta_{\mu}$, $\xi_3+\theta_{\mu}$,
  $\xi_1+\theta_{\mu}$\\
\hline
$\tilde W\rightarrow q_1\bar q_2 + \chi_1$,..& $\xi_2+\theta_{\mu}$,
$\alpha_{A_{q1}}+\theta_{\mu}$,$\alpha_{A_{q2}}+\theta_{\mu}$,
$\xi_1+\theta_{\mu}$,.\\
\hline
$\tilde g\rightarrow q\bar q + \chi_1$,..& $\xi_2+\theta_{\mu}$,
$\alpha_{A_{q}}+\theta_{\mu}$, $\xi_2+\theta_{\mu}$, $\xi_1+\theta_{\mu}$,.\\
\hline
$g_{\mu}-2$ & $\xi_2+\theta_{\mu}$, $\xi_1+\theta_{\mu}$,
$\alpha_{A_{\mu}}+\theta_{\mu}$\\
\hline
$m_{H_i}$(small $\tan\beta$) & $\alpha_{A_t}+\theta_{\mu}$  \\
\hline
 $m_{H_i}$(large $\tan\beta$)
  & $\alpha_{A_t}+\theta_{\mu}$, $\alpha_{A_b}+\theta_{\mu}$,
   $\xi_2+\theta_{\mu}$,  $\xi_1+\theta_{\mu}$\\
   \hline
  $Z^*\rightarrow Z+H_i$ & $\alpha_{A_t}+\theta_{\mu}$, $\alpha_{A_b}+\theta_{\mu}$,
   $\xi_2+\theta_{\mu}$,  $\xi_1+\theta_{\mu}$\\
 \hline
 $d_e$ ($d_{\mu}$)  &   $\xi_2+\theta_{\mu}$, $\xi_1+\theta_{\mu}$,
 $\alpha_{A_e}+\theta_{\mu}$($\alpha_{A_e}+\theta_{\mu}$ )\\
 \hline
 $d_n$ & $\xi_3+\theta_{\mu}$,
 $\xi_2+\theta_{\mu}$, $\xi_1+\theta_{\mu}$,\\
   & $\alpha_{A_{ui}}+\theta_{\mu}$,$\alpha_{A_{di}}+\theta_{\mu}$ \\
 \hline
\end{tabular}
 \label{table:phases}
 \end{center}
}

\section{CP violation in the Higgs sector}
As discussed already
the  Standard  Model  (SM)  has  two  sources  of  CP  violation:  the
Kobayashi--Maskawa  (KM) phase  in  the quark  mixing  matrix and  the
so-called  strong  CP  phase  through  the  QCD  anomaly.   The  Higgs
potential of the Standard Model is CP-invariant to all orders, whereas
a possible mixing  of the $Z$ boson with the SM  Higgs boson can first
occur  at the  3-loop level~\cite{Pilaftsis:1998pe}  and  is therefore
very suppressed.   Significant new sources  of CP violation  emerge in
minimal  Higgs-sector   extensions  of  the   SM,  such  as   the  two
Higgs-doublet model~\cite{TDLee}.

A very predictive model with  an extended Higgs sector and new sources
of CP  violation is the Minimal Supersymmetric  Standard Model (MSSM),
with  supersymmetry (SUSY)  softly broken  at the  TeV scale.   In the
MSSM,   assuming   flavour  conservation,   there   are  12   physical
combinations of CP phases~\cite{Dugan:1984qf,Dimopoulos:1995kn}
\footnote{The relevant soft-SUSY breaking terms are as in
$-{\cal L}_{\rm soft}\supset
\frac{1}{2}
( {M_3} \, \widetilde{g}\widetilde{g}
+ {M_2} \, \widetilde{W}\widetilde{W}
+ {M_1} \, \widetilde{B}\widetilde{B}+{\rm h.c.})
+( \widetilde{u}_R^* \,{A_u}\, \widetilde{Q} H_2
- \widetilde{d}_R^* \,{A_d}\, \widetilde{Q} H_1
- \widetilde{e}_R^* \,{A_e}\, \widetilde{L} H_1
+ {\rm h.c.})\,.$ }
\begin{equation}
{\rm Arg}({M_i\, \mu \, (m_{12}^2)^*})\,, \ \ \
{\rm Arg}({{A_{f}}\, \mu \, (m_{12}^2)^*})\,,
\end{equation}
with $i=1-3$ and $f=e,\mu,\tau;u,c,t,d,s,b$.  In the convention of
real and positive $\mu$ and $m_{12}^2$, the most relevant CP phases
pertinent to the Higgs sector are
\begin{equation}
\Phi_i\equiv {\rm Arg}({M_i})\,; \ \ \
\Phi_{A_{f_3}}\equiv {\rm Arg}({A_{f_3}})\,,
\end{equation}
with $f_3=\tau,t,b$.

\smallskip

The Higgs sector of the MSSM consists of two doublets:
\begin{eqnarray}
H_1=\left(\begin{array}{c}
            {H_1^0} \\
            H_1^-
          \end{array}
\right)\,;~~~~~~
H_2=\left(\begin{array}{c}
            H_2^+ \\
            {H_2^0}
          \end{array}
\right)\,.
\end{eqnarray}
The neutral components can be rewritten in terms of 4 real field as
\begin{equation}
{H_1^0}=\frac{1}{\sqrt{2}}({\phi_1}
-i{a_1})\,, \ \ \
{H_2^0}=\frac{1}{\sqrt{2}}({\phi_2}
+i{a_2})\,,
\end{equation}
where ${\phi_{1,2}}$  and ${a_{1,2}}$  are CP-even and  CP-odd fields,
respectively.     After    the    electroweak    symmetry    breaking,
$\langle{\phi_1}\rangle = v \cos\beta$ and $\langle{\phi_2}\rangle = v
\sin\beta$, we are left with 5  Higgs states: 2 charged and 3 neutral.
The  3 neutral  states  consists  of one  CP-odd  state, ${A}  =-{a_1}
\sin\beta + {a_2}  \cos\beta$, and two CP-even ones,  ${h}$ and ${H}$.
At the tree level, the Higgs  potential is CP invariant and the mixing
between the two CP-even states  is described by the $2\times 2$ matrix
with the mixing angle $\alpha$ as
\begin{equation}
\left(\begin{array}{c}
      {h} \\ {H}
      \end{array} \right) =
\left(\begin{array}{cc}
      \cos\alpha & -\sin\alpha \\ \sin\alpha & \cos\alpha
      \end{array} \right)
\left(\begin{array}{c}
      {\phi_2} \\ {\phi_1}
      \end{array} \right)\,.
\end{equation}

However, the  presence of  the soft CP  phases may  introduce sizeable
CP-violating couplings  in the MSSM Higgs  potential through radiative
corrections~\cite{Pilaftsis:1998pe,Pilaftsis:1998dd}.   In particular,
the  non-vanishing  CP phases  of  third  generation  $A$ terms  could
radiatively induce  significant mixing between the  CP-even and CP-odd
states                                                     proportional
to~\cite{Pilaftsis:1998pe,Pilaftsis:1998dd,Pilaftsis:1999qt,Demir:1999hj,Choi:2000wz,Carena:2000yi,Carena:2001fw,Ibrahim:2000qj,Ibrahim:2007fb}
\begin{equation}
\frac{3m_f^2}{16\pi^2}\,\frac{\imag(A_f\,\mu)}
{(m_{\tilde{f}_2}^2-m_{\tilde{f}_1}^2)}\,.
\end{equation}
The CP phase of the gluino mass parameter also contribute to the CP-violating
Higgs mixing through the so-called threshold corrections to the Yukawa
couplings
\begin{equation}
{h_b}=
\frac{\sqrt{2}\,m_b}{v\,\cos\beta}\,
{\frac{1}{1+{\kappa_b}\,{\tan\beta}}}\,,
\label{eq:threshold}
\end{equation}
where
\begin{eqnarray}
{\kappa_b}&=&\frac{2\alpha_s}{3\pi} \, {M_{3}^* \mu^*}
    I(m_{\tilde{b}_1}^2,m_{\tilde{b}_2}^2,|M_{3}|^2)
\nonumber \\ &+&
\frac{|h_t|^2}{16\pi^2} \, {A_t^* \mu^*}
    I(m_{\tilde{t}_1}^2,m_{\tilde{t}_2}^2,|\mu|^2)\,,
\end{eqnarray}
with
%\begin{equation}
 \beqn
  \label{Ixyz}
\hspace{-0.5cm}
I(x,y,z) \!=~~~~~~~~~~~~~~~~~~~~~~~~~~~~~~~~~~~~~~\nonumber\\
\! \frac{xy\,\ln (x/y)\: +\: yz\,\ln (y/z)\: +\:
             xz\, \ln (z/x)}{(x-y)\,(y-z)\,(x-z)}\,.
  \eeqn
            %\end{equation}
This  is  formally a  two-loop  effect  but  could be  important  when
$\tan\beta$ is large.

\smallskip

The consequences  of the CP-violating  mixing among the  three neutral
Higgs bosons are: $(i)$ the neutral  Higgs bosons do not have to carry
any  definite CP parities,  $(ii)$ the  neutral Higgs-boson  mixing is
described  by the  $3\times 3$  mixing  matrix ${O_{\alpha  i}}$ as  $
(\phi_1,\phi_2,a)^T = {O_{\alpha i}} (H_1,H_2,H_3)^T $ with $H_{1(3)}$
the  lightest (heaviest)  Higgs state,  $(iii)$ the  couplings  of the
Higgs bosons to the SM  and SUSY particles are significantly modified.
In    our    numerical    analysis,    we   use    the    code    {\tt
CPsuperH}~\cite{Lee:2003nt,Lee:2007gn}   which   is   based   on   the
renormalization-group-(RG-)improved effective potential approach.

\section{CPX scenario}

The CPX scenario is defined as~\cite{Carena:2000ks}:
\begin{eqnarray}
&& \hspace{-1.3cm}
M_{\tilde{Q}_3} = M_{\tilde{U}_3} = M_{\tilde{D}_3}
= M_{\tilde{L}_3} = M_{\tilde{E}_3} = M_{\rm SUSY}\,,
\\
&& \hspace{-1.3cm}
|\mu|=4\,M_{\rm SUSY}\,, \ \
|A_{t,b,\tau}|=2\,M_{\rm SUSY} \,, \ \
|M_3|=1 ~~{\rm TeV}. \nonumber
\label{eq:CPX}
\end{eqnarray}
The parameter $\tan\beta$, the charged Higgs-boson pole mass
$M_{H^\pm}$, and the common SUSY scale $M_{\rm SUSY}$ can be varied.
For CP phases, taking a common
phase $\Phi_A=\Phi_{A_t}=\Phi_{A_b}=\Phi_{A_\tau}$ for $A$ terms,
we have two
physical phases to vary: $\Phi_A$ and $\Phi_3={\rm Arg}(M_3)$.

\smallskip

\begin{figure}
  \includegraphics[height=.3\textheight,width=0.48\textwidth]{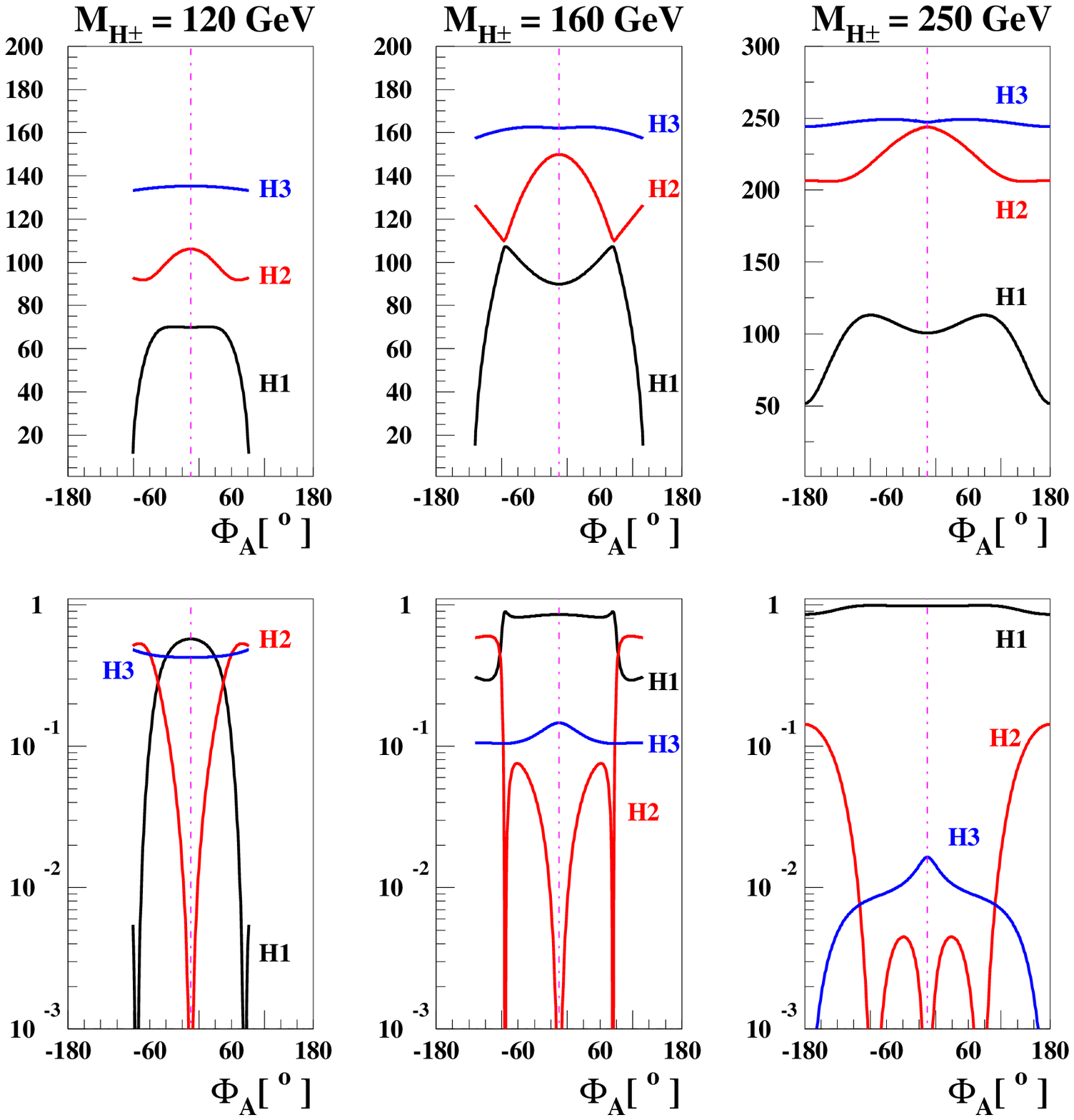}
  \caption{{\small The Higgs-boson masses $M_{H_i}$ (upper frames) in GeV
and $g_{H_iVV}^2$ (lower frames) as functions of $\Phi_A$ for the CPX scenario
for three values of the charged Higgs-boson pole mass
when $\tan\beta=4$, $\Phi_3=0^\circ$, and $M_{\rm SUSY}=0.5$ TeV; from
Ref.~\cite{Accomando:2006ga}.
}}
\label{fig:mhghvv}
\end{figure}
In Fig.~\ref{fig:mhghvv}, we show the Higgs-boson pole masses and their couplings to two
vector bosons normalized to the SM value as functions of $\Phi_A$ for three values of the
charged Higgs-boson pole mass: 120 GeV
(left frames), 160 GeV (middle frames), and 250 GeV (right frames).
We observe, when $M_{H^\pm}=120$ GeV,
$M_{H_1}$ can be as light as a few GeV around $\Phi_A=\pm 90^\circ$
where $H_1$ is almost CP odd with nearly vanishing coupling to two vector bosons.
In the decoupling limit, $M_{H^\pm}=250$ GeV, the lightest Higgs boson
is decoupled from the
mixing  but there could still be a significant CP-violating
mixing between the two heavier states.
\begin{figure}
  \includegraphics[height=.35\textheight]{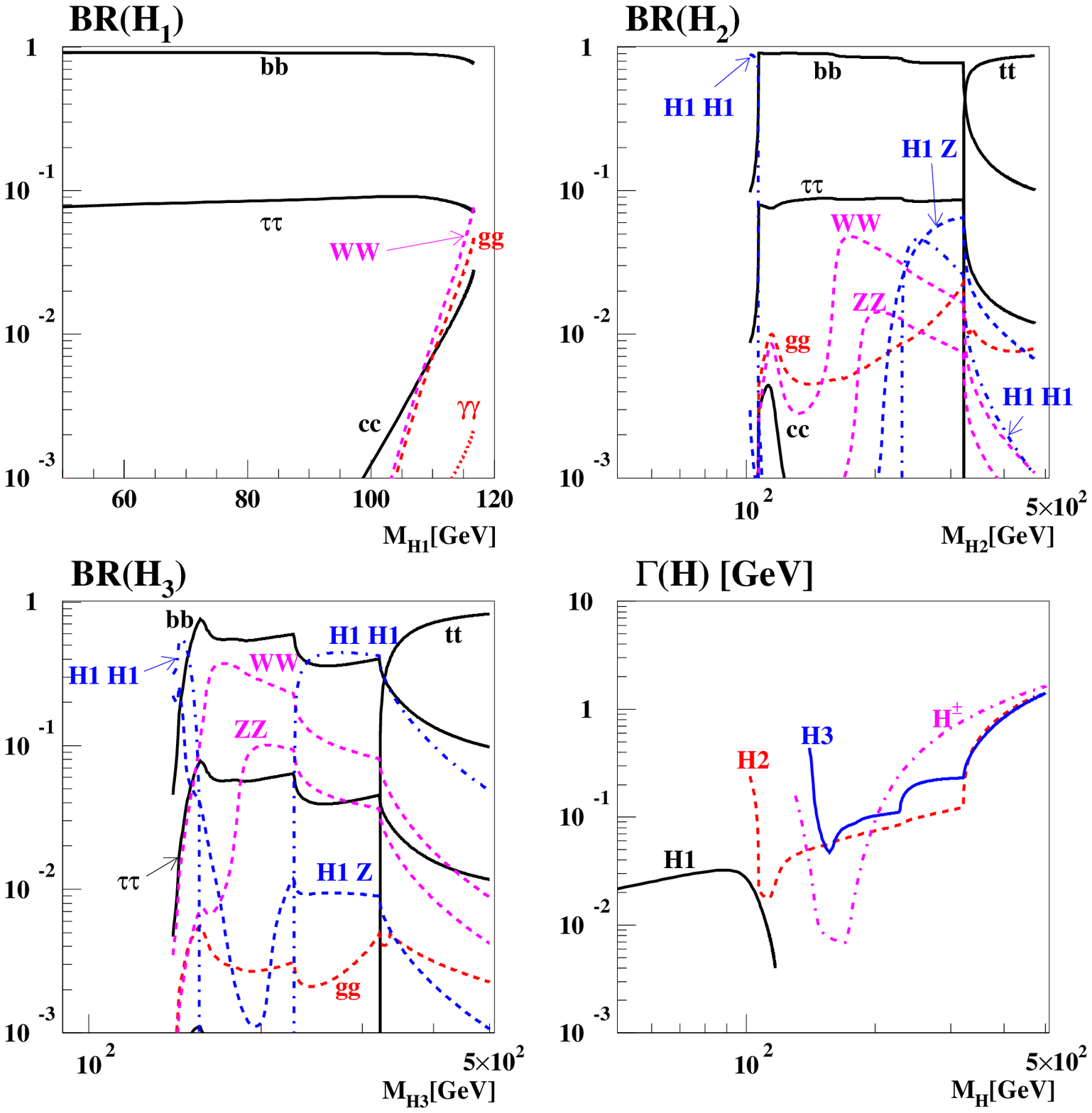}
  \caption{\small The branching fractions and decay widths of the MSSM
Higgs bosons for the CPX scenario with $\tan\beta=4$ and $M_{\rm SUSY}=0.5$ TeV
as functions of their masses when $\Phi_A=\Phi_3=90^\circ$; from Ref.~\cite{Lee:2003nt}.
}
\label{fig:cpx1}
\end{figure}
In  Fig.~\ref{fig:cpx1}, we  show  the branching  fractions and  decay
widths of  the Higgs bosons when  $\Phi_A=\Phi_3=90^\circ$.  The decay
patterns of the heavier Higgs states become more complex than those in
the   CP-conserving   case~\cite{Choi:1999uk,Choi:2001pg,Choi:2002zp}.
If kinematically allowed, the heavier Higgs states decay predominantly
into the two lightest
Higgs  bosons  which increase  their  decay  widths  considerably (see  the
lower-right frame
\footnote{In the case of the charged Higgs boson, it decays dominantly
  into $W^\pm$ and $H_1$.}).
These features combined make the Higgs searches at LEP difficult, resulting in
two uncovered holes
on the $\tan\beta$-$M_{H_1}$ plane when $M_{H_1}\lsim 10$ GeV and
$M_{H_1} \sim 30 - 50$ GeV for intermediate values of $\tan\beta$,
as shown in Fig.~\ref{fig:LEPlimit}.
\begin{figure}
  \includegraphics[height=.35\textheight]{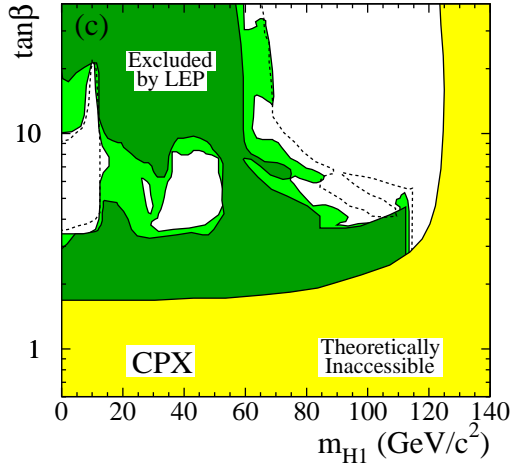}
  \caption{\small The LEP exclusion plot on the $\tan\beta$-$M_{H_1}$ plane
for the CPX scenario when $\Phi_A=\Phi_3=90^\circ$; from
Ref.~\cite{Schael:2006cr}.
}
\label{fig:LEPlimit}
\end{figure}
It seems difficult to cover the holes completely at the LHC~\cite{Carena:2002bb}
without relying on the decay mode $H^\pm \to W^\pm\, H_1$.

\smallskip

In  the scenario  with  large $|\mu|$  and  $|M_3|$ like  CPX,  the
threshold  corrections significantly modify  the relation  between the
down-type  quark  mass  and  the corresponding  Yukawa  coupling  when
$\tan\beta$ is large,  see Eq.~(\ref{eq:threshold}).  The modification
leads to strong CP-phase dependence of the $b$-quark fusion production
of the neutral Higgs bosons.
\begin{figure}
  \includegraphics[height=.15\textheight]{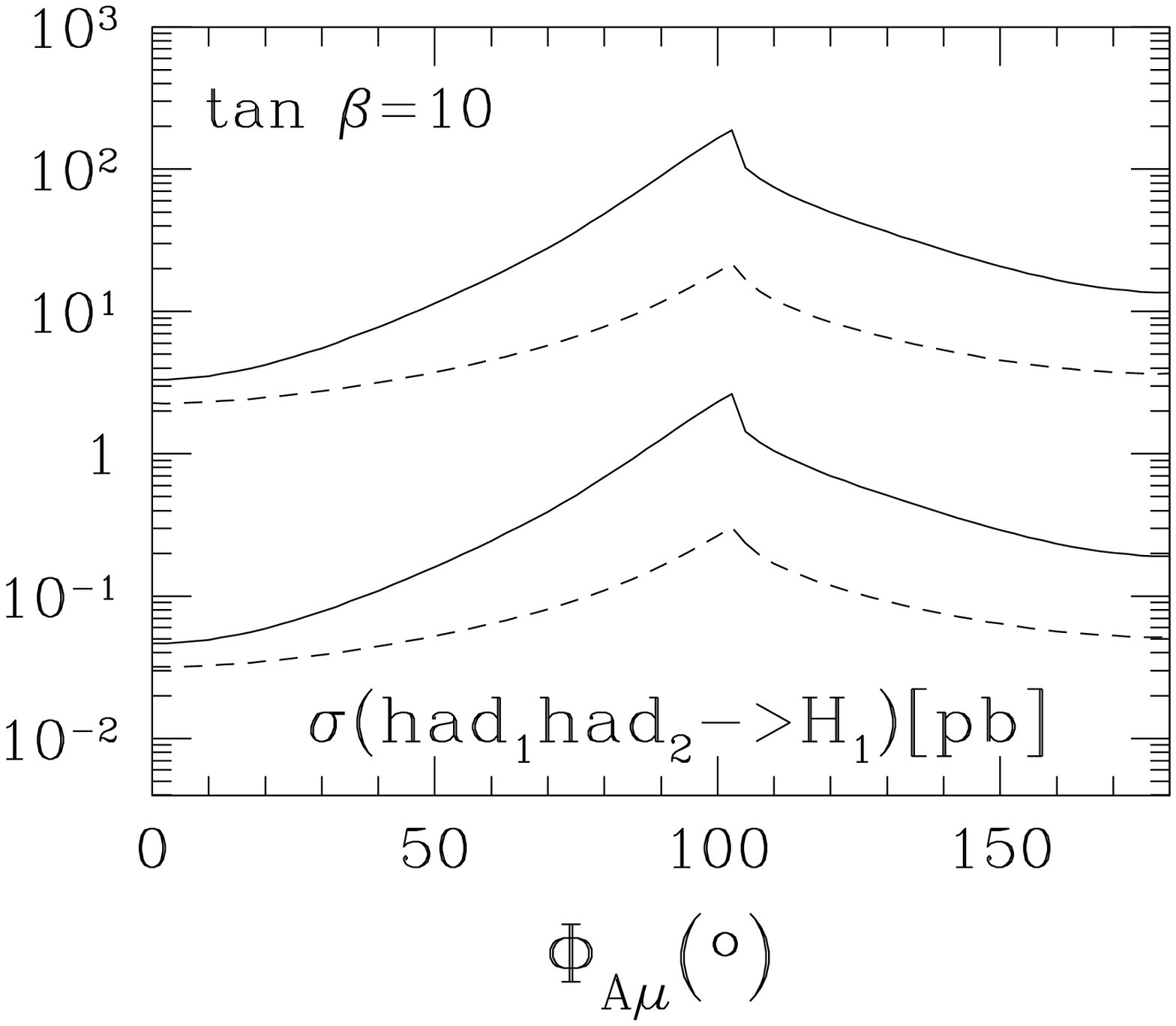}
  \includegraphics[height=.15\textheight]{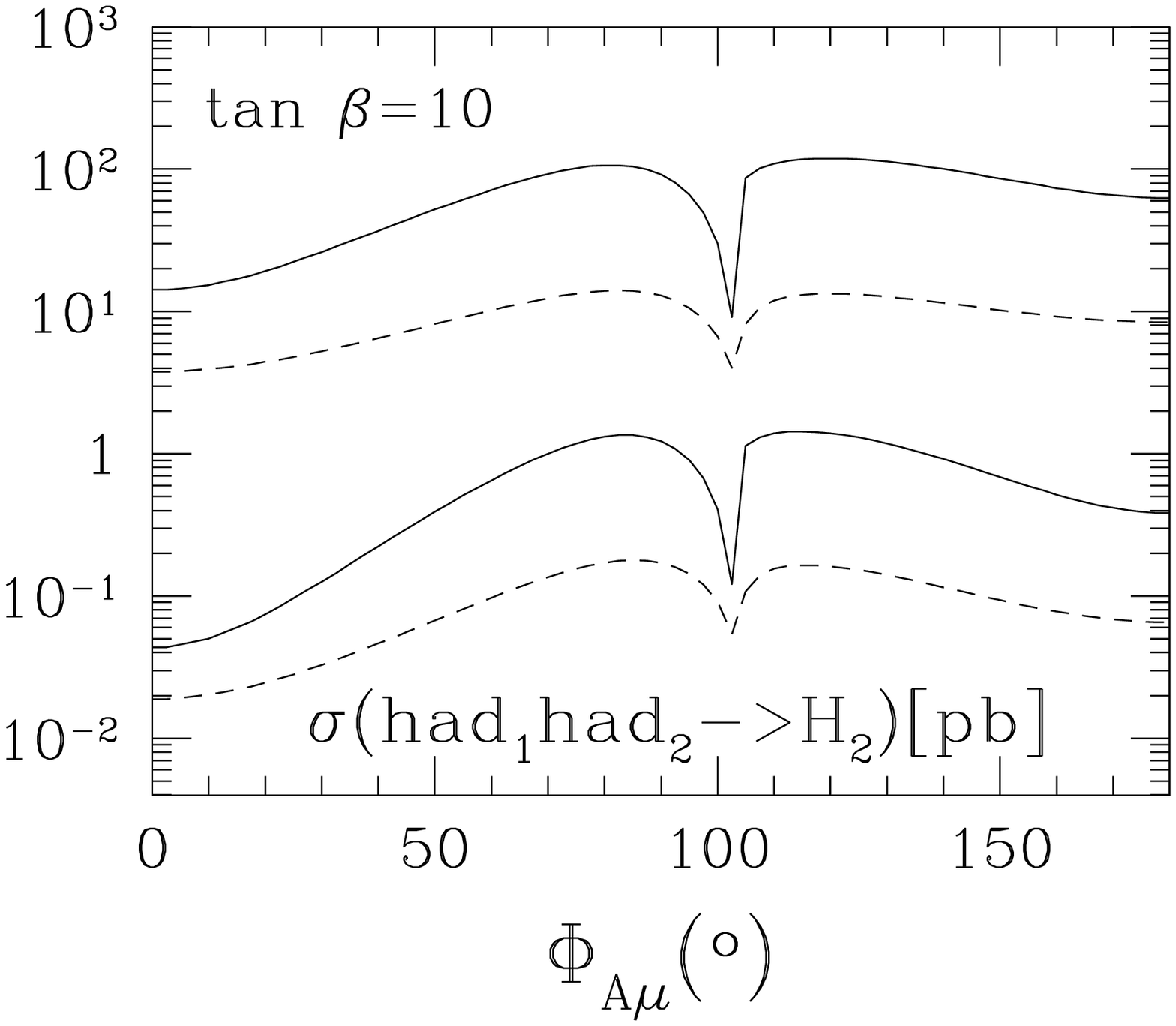}
  \caption{\small The inclusive production cross sections
of $H_1$ and $H_2$ via $b$-quark fusion for the CPX scenario
as functions of $\Phi_A$ when $\tan\beta=10$ at the LHC (upper lines)
and Tevatron (lower lines); from Ref.~\cite{Borzumati:2004rd}.
}
\label{fig:bbh_CPX}
\end{figure}
In Fig.~\ref{fig:bbh_CPX}, we show the inclusive production cross sections
of $H_1$ and $H_2$ via $b$-quark fusion as functions of $\Phi_A$. We see about
a factor 100 enhancement in the $H_1$ production and the
corresponding suppression in the $H_2$ production
around $\Phi_A= 100^\circ$, where the mass difference between $H_1$ and $H_2$ is
only $3 - 5$ GeV. Taking account of the
good $\gamma\gamma$ and $\mu^+\mu^-$ resolutions of
$1-3$ GeV at the LHC~\cite{Aad:2009wy,Ball:2007zza},
the combined analysis of Higgs decays to photons and muons may help to
resolve the two CP-violating adjacent peaks as illustrated in Fig.~\ref{fig:bbh_diff}.
\begin{figure}
  \includegraphics[height=.15\textheight]{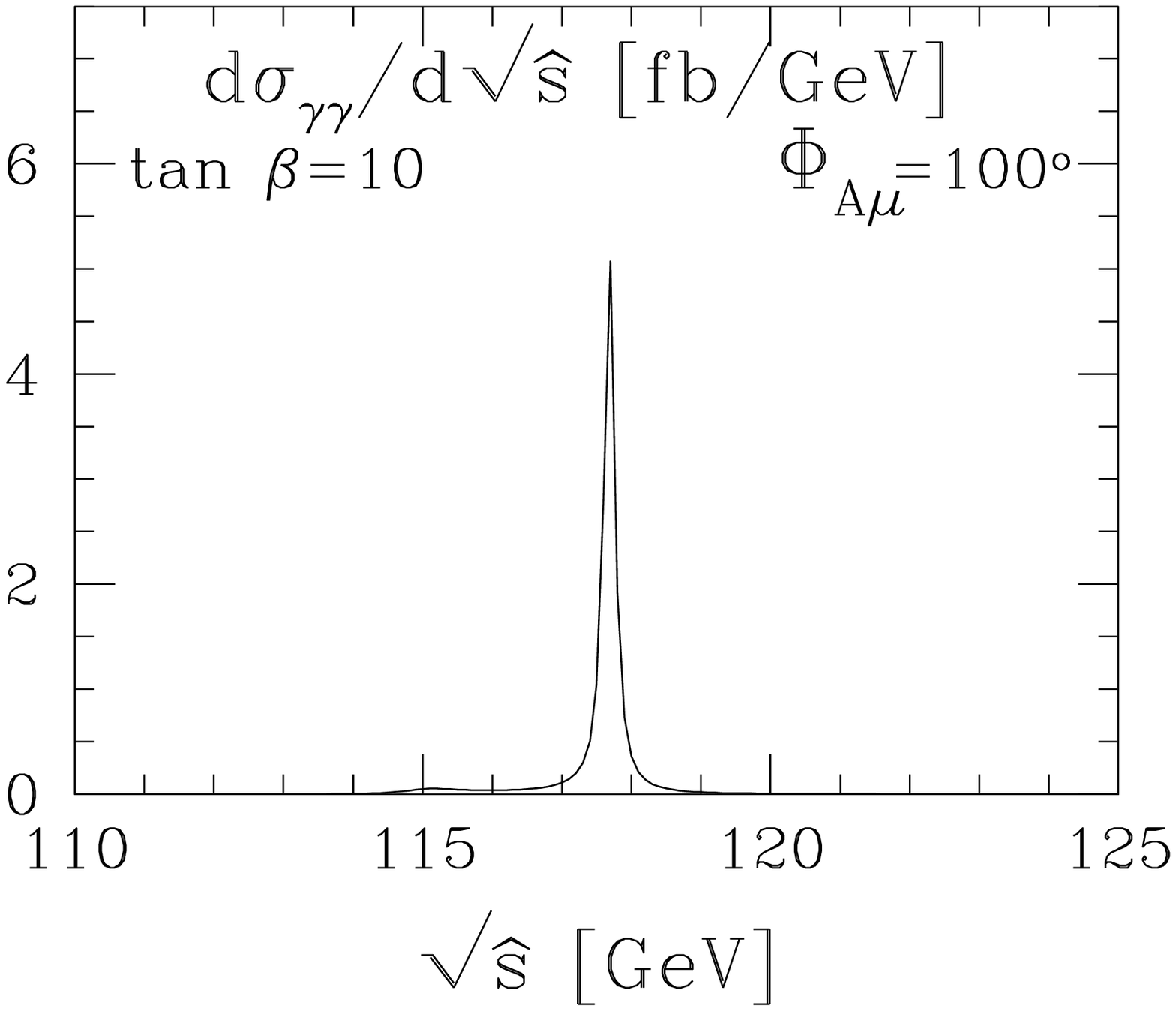}
  \includegraphics[height=.15\textheight]{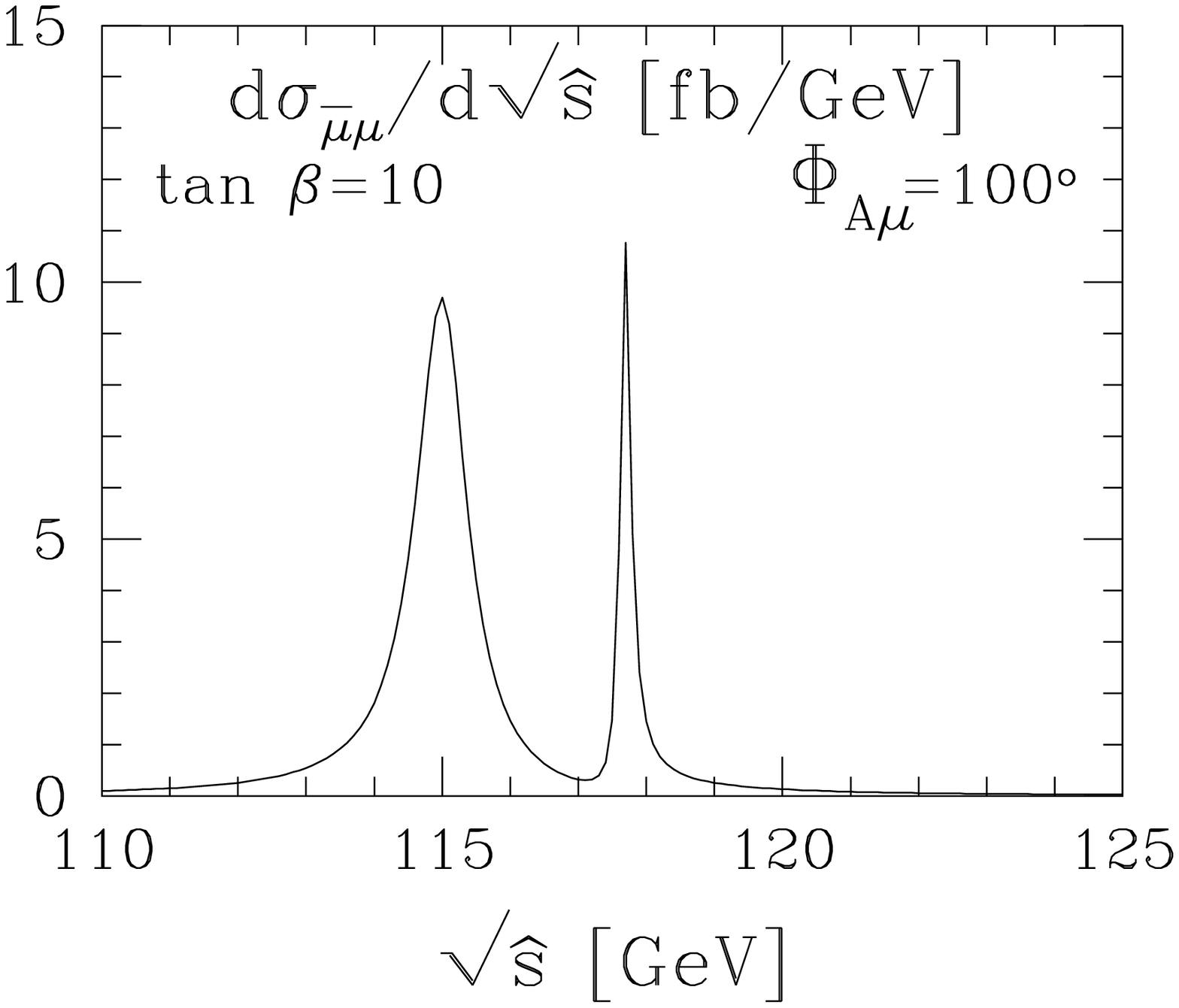}
  \caption{\small The LHC differential production cross sections
of $H_1$ and $H_2$,
produced via $b$-quark fusion and decaying into photons (left) and muons (right),
for the same scenario as in Fig.~\ref{fig:bbh_CPX} but with $\Phi_A=100^\circ$
as functions of the invariant mass of two photons and
two muons. We see only one peak in the photon decay mode (left)
since $H_1$ with 115 GeV mass
is almost CP odd; from Ref.~\cite{Borzumati:2006zx}.
}
\label{fig:bbh_diff}
\end{figure}

\section{Trimixing scenario}

The trimixing scenario is characterized by large $\tan\beta$ and a
light charged Higgs boson, resulting in a strongly mixed system of the
three neutral Higgs bosons with mass differences comparable to their
decay widths~\cite{Ellis:2004fs}.
In this scenario, the neutral Higgs bosons cannot be treated separately and
it needs to consider
the transitions between the Higgs-boson mass
eigenstates induced by the off-diagonal absorptive parts,
$\left.\imag\hat\Pi\right|_{i\neq j}(\hat{s})$.
In Fig.~\ref{fig:dh3}, we show the absolute value of each component of the
dimensionless $4\times 4$ neutral Higgs-boson propagator matrix
\begin{equation}
D^{H_0}_{ij}(\hat{s})\equiv \hat{s}\,\,[(\hat{s}-M_{H}^2)\,{\bf{1}}_{4\times 4}
+ i\,\imag\hat{\Pi}(\hat{s})]^{-1}_{ij}\,,
\end{equation}
with $i,j=1-4$ corresponding to $H_1$, $H_2$, $H_3$, and $G^0$.
Compared to the case without including the off-diagonal elements
(dashed lines in the upper frames), we observe that the peaking
patterns are different (solid lines in the upper frames).  We also
note the off-diagonal transition cannot be neglected (middle frames).

\smallskip

\begin{figure}
  \includegraphics[height=.35\textheight]{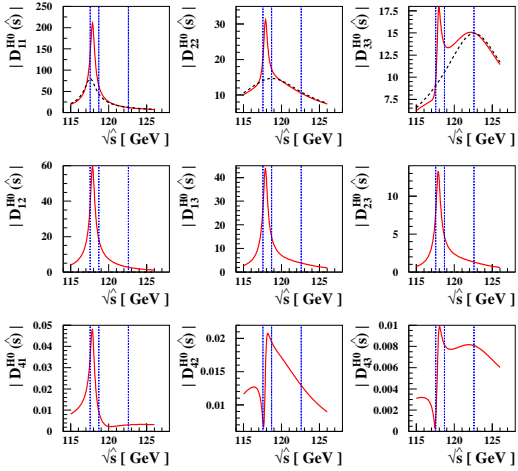}
  \caption{\small The absolute value of each component of the neutral Higgs-boson
propagator matrix
$D^{H^0}({\hat{s}})$ with (red solid lines) and without (black dashed lines)
including off-diagonal absorptive parts in the trimixing scenario with
$\Phi_A=-\Phi_3=90^\circ$. We note that
$|D^{H^0}_{4\,4}({\hat{s}})|=1$.
The three Higgs-boson pole masses are indicated by thin vertical
lines; from Ref.~\cite{Lee:2007gn}.
}
\label{fig:dh3}
\end{figure}

\begin{figure}
  \includegraphics[height=.25\textwidth]{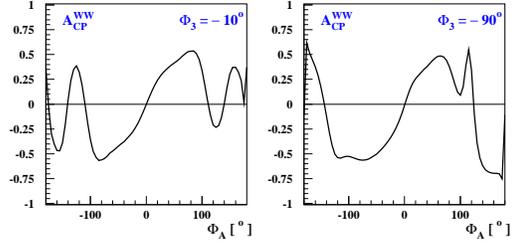}
  \caption{\small The CP  asymmetry
${\cal A}^{WW}_{\rm CP}$ as functions of $\Phi_A =
\Phi_{A_t} =  \Phi_{A_b} =  \Phi_{A_\tau}$ in the trimixing scenario
with $\Phi_3  = -10^\circ$ (left)
and $-90^\circ$ (right); from Ref.~\cite{Ellis:2004fs}.}
\label{fig:acpww}
\end{figure}
At the LHC, there may be a way to probe
CP violation in the trimixing scenario, though it seems challenging.
In the $WW$ fusion production of the Higgs bosons decaying into tau leptons,
the difference between the cross sections into the right-handed and
left-handed tau leptons signals CP violation.
The corresponding CP asymmetry turns out to be
large over the whole range of $\Phi_A$ independently of $\Phi_3$ in
the trimixing
scenario, as shown in Fig.~\ref{fig:acpww}.

 \section{Testing the Cancellation Mechanism at the LHC}
 The experimental data at the LHC will allow one to test the cancellation mechanism
 in a direct  fashion. This could be  done in several ways. One manifestation is of course
 through phenomena related to  the neutral CP even and CP odd mixing discussed above.
 However, there  are other processes where this can be done.
 One such process where the phases play a very discernible role is in the decay
$B_{s}^0\to \mu^+\mu^-$ on which the Tevatron already sets upper
limits and which will also be measurable at the
LHC~\cite{Ibrahim:2002fx}. Here the counterterm diagram shown on the
left panel  of Fig.~(\ref{fig:edm4b}) gives an amplitude which
behaves like $\tan^3\beta$   and thus the branching ratio
$B_{d,s}^0\to \mu^+\mu^-$  can get very large for large $\tan\beta$.
As discussed above in the presence of  CP violation one has  mixing
between the CP even and CP odd Higgs states and the  mass
eigenstates $H_1, H_2, H_3$ exchanged in the left hand panel of
Fig.~(\ref{fig:edm4b}) are linear combination of CP even and CP odd
Higgs fields. Additionally the vertices are also affected by the CP
violating phases.   As a consequence of these two overlapping
effects the $B_{s}^0\to \mu^+\mu^-$  shows a  very strong dependence
on the phases. The right panel of Fig.~(\ref{fig:edm4b}) gives an
analysis of the dependence of the $BR(B_{s}^0\to \mu^+\mu^-)$ on the
phase $\alpha_A$ in the scenario with minimal flavor violation where
the squark mass matrices are assumed flavor-diagonal.  The analysis
shows that $BR(B_{s}^0\to \mu^+\mu^-)$  can vary by as much as  two
orders of magnitude with phases and thus this process is one of the
prime processes at the LHC to look for CP violating phases.
     \begin{figure}
\hspace{0cm}
  \includegraphics[height=.12\textheight,width=0.2\textwidth]{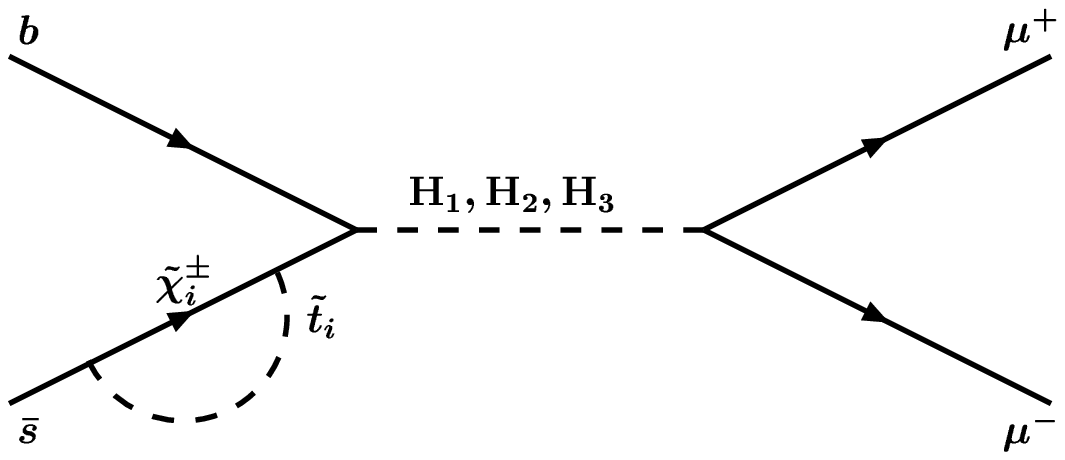}
 \includegraphics[height=.12\textheight,width=0.2\textwidth]{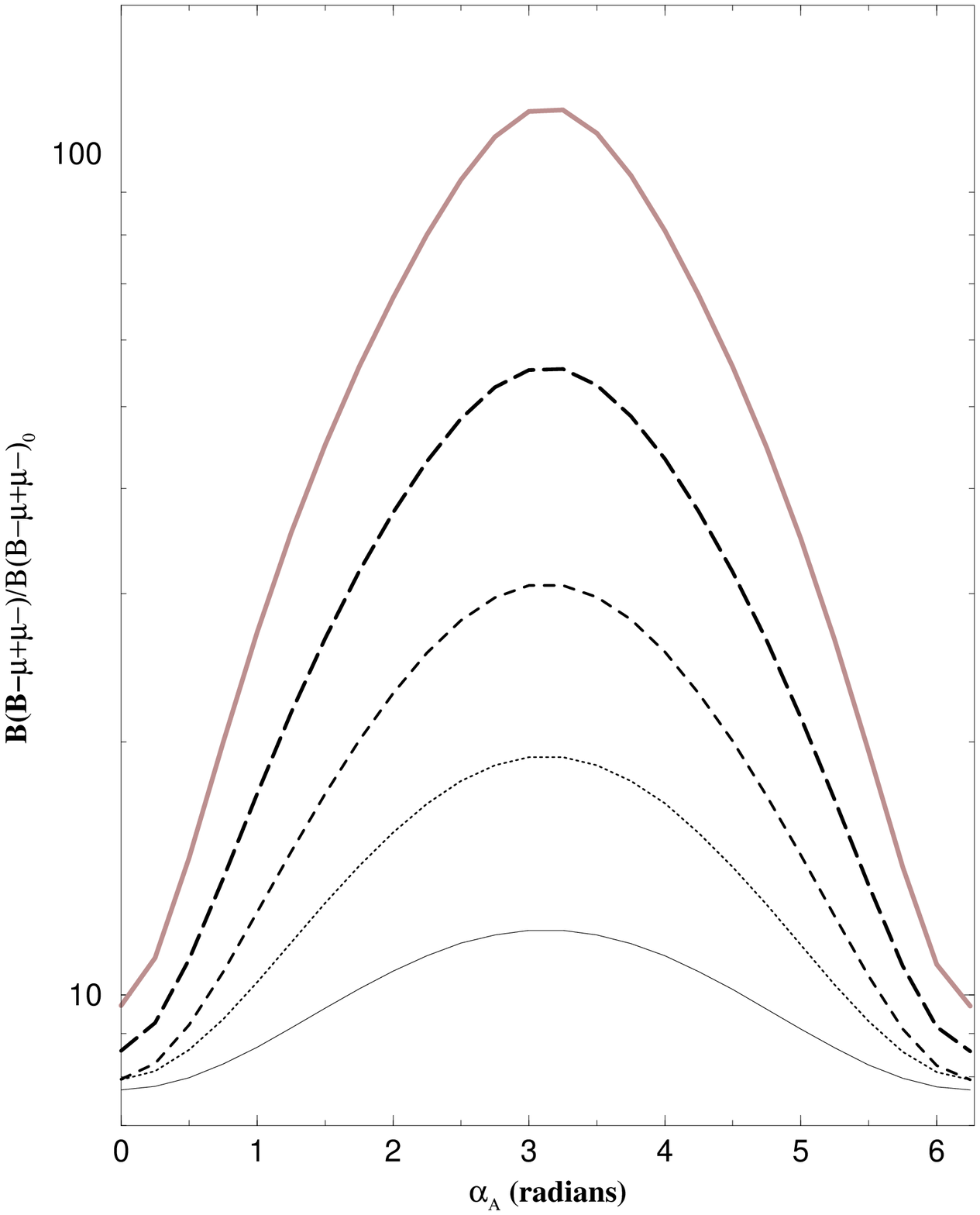}
   \vspace{-1cm}
   \caption{{Left panel:  The diagram that contributes to the $B_s\rightarrow\mu^+\mu^-$ decay.
   Right panel: The Large CP phase dependence of the branching ratio $BR(B_s\to \mu^+\mu^-)$
   on the phase $\alpha_A$.  From~\cite{Ibrahim:2002fx} }}
\label{fig:edm4b}
\end{figure}

In addition to the above it is important to look for CP odd  or T
odd operators~\cite{Christova:1992ju} (assuming CPT invariance)
which are measurable at the LHC.  For example with processes
involving n  particles  with $n>4$ one may form  T odd operator such
as~\cite{Langacker:2007ur}
$\epsilon_{\alpha\beta\gamma\delta}p^{\alpha}_ip^{\beta}_jp^{\gamma}_k
p^{\delta}_l$. One such example is the decay of a squark so that
$\tilde t\to t+l^+l^-+\chi_1^0$, a process which is detectable at
the LHC~\cite{Langacker:2007ur}.  The effect of CP phases in
sparticle decays have been discussed
by~\cite{Alan:2007rp,Bartl:2006hh,Bartl:2006yv,Ibrahim:2004gb} and
in the Higgs sector
by~\cite{Ghosh:2004cc,Choi:2002zp,Ibrahim:2004cf}. Aside from the
LHC, the linear collider is an excellent machine  for the detection
of CP phases and this topic has  been discussed in many
works~\cite{Akeroyd:2001kt,Choi:2004rf,MoortgatPick:2005cw}.

% \begin{figure}[th]
%\hspace{0cm}
% \includegraphics[scale=0.3]{CP/ohtan_EDM.ps}
%     \caption{{ A display of the CP phase dependence of  the relic density.
% The neutralino relic density is
%displayed as a function of $\tan\beta$ for three cases given by: (i)
%$m_0=m_{1/2}$=$|A_0|=300$ GeV, $\alpha_{A_0}=1.0$, $\xi_1=0.5$,
%$\xi_2=0.66$, $\xi_3=0.62$, $\theta_\mu=2.5$; (ii)
%$m_0=m_{1/2}=|A_0|=555$ GeV, $\alpha_{A_0}=2.0$, $\xi_1=0.6$,
%$\xi_2=0.65$, $\xi_3=0.65$, $\theta_\mu=2.5$; \indent (iii)
%$m_0=m_{1/2}=|A_0|=480$ GeV, $\alpha_{A_0}=0.8$, $\xi_1=0.4$,
%$\xi_2=0.66$, $\xi_3=0.63$, $\theta_\mu=2.5$. In all cases the EDM
%constraints are satisfied for $\tan\beta=40$.
%From\cite{Gomez:2004eka}.}} \label{fig:dark}
%\end{figure}

\begin{figure}[th]
\hspace{0cm}
  \includegraphics[scale=0.4]{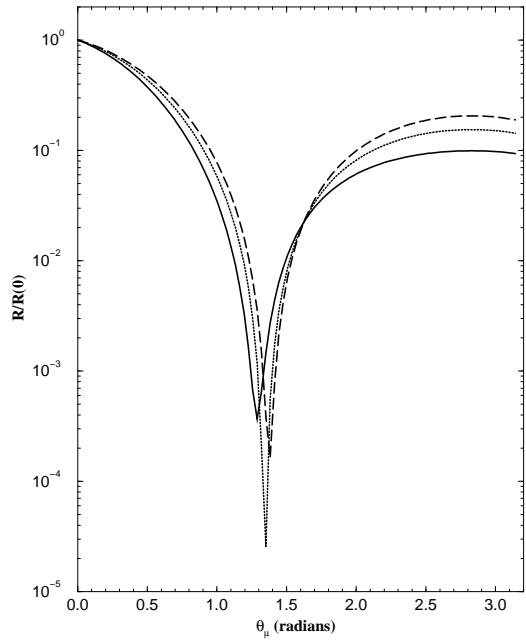}
     \caption{{A display of the CP phase dependence of event rates as a function of the CP phase
     $\theta_{\mu}$ without the imposition of the EDM constraints. From~\cite{Chattopadhyay:1998wb}.}}
\label{fig:dark}
\end{figure}

There are also strong interconnections of  LHC physics  with the
direct and indirect detection of dark matter.  This topic is
discussed in depth in a later section on dark matter. It should be
noticed, however, that CP phases also affect dark matter and thus
there  is a further correlation between LHC physics and dark matter
in this regard. For dark matter, the relic density is much less
sensitive to the CP phases than the neutralino-proton cross
sections which enter directly in the event rates for dark matter
detectors.  In Fig.~(\ref{fig:dark}), a plot is given of the ratio
$R(\theta_{\mu})/R(0)$ where $R$ is the event rate as  a function of
the phase $\theta_{\mu}$. One finds that the ratio changes rather
rapidly with $\theta_{\mu}$.
 The analysis of this plot is without the imposition of the EDM constraints, and the inclusion of
  those constraints will limit the allowed range of $\theta_{\mu}$.

\section{Summary}
While the observed phenomena in the Kaon and B physics appear
consistent with the CP violation arising from the Standard Model, the existence
of a large baryon asymmetry in the universe points to the existence of an additional CP
violation beyond the Standard Model.
Supersymmetric extensions of  the SM  contain several new sources  of CP violation.
 If  the CP violating phases are large,  as often is the case in softly broken supersymmetric theories
 and in string based models, then
 they would affect many supersymmetric phenomena such as production
and decay of sparticles. Further,
the CP-violating phases
could radiatively induce significant mixing between the CP-even and CP-odd Higgs
states.
It turns out that the CP-violating mixing could make
the Higgs boson lighter than 50 GeV elusive at LEP and even at the
LHC, specifically in
the CPX scenario.
In the scenario,
when $\tan\beta$ is intermediate or large, the production cross sections of the
neutral Higgs bosons via $b$-quark fusion strongly
depend on the CP phases due to the threshold corrections and the
CP-violating Higgs mixing.  At the LHC,
it might be possible to disentangle two adjacent CP-violating
Higgs peaks by exploiting its
decays into photons and muons unless the mass difference is smaller
than 1 or 2 GeV.
The constraints on the CPX scenario from the non-observation of the
Thallium, neutron,
Mercury EDMs can be evaded~\cite{Ellis:2008zy} by appealing to
the cancellation mechanism~\cite{incancel,incancel2,incancel3}.
%

%%%
We presented the general formalism for
a coupled system of CP-violating
neutral Higgs bosons at high-energy colliders.
It is suggested to measure the polarizations of the tau leptons in the process
$W^+W^-\rightarrow H_{i\oplus j}\rightarrow \tau^+_{R,L}\tau^-_{R,L}$
to probe the Higgs-sector CP violation at the LHC.
The study of the final state spin-spin correlations of
tau leptons, neutralinos, charginos, top quarks, vector bosons, stops, etc are
crucial for proving SUSY itself as well as for
the CP studies of the Higgs bosons at the LHC.\\
In addition to the CP even and CP odd Higgs mixing phenomena for neutral Higgs discussed above, large CP phases
may be detectable in $B_s^0\to \mu^+\mu^-$ and in the
decays of squarks and of sleptons and well as in the charged Higgs decays.
Study of CP odd  or T odd operators and of  forward-backward  asymmetries  could
also reveal the existence of such phases.

In summary the LHC has the ability to discover the presence of large CP phases. Further, the LHC
data will allow  one to check on the CP even and CP Higgs mixing phenomenon and also allow one
to test the cancellation mechanism.

%%%%%%%%%%%%%%%%%%%%%%%%%%%%%%%%%%%%%%%%%%%%%%%%%%%%%%%%%%%%%%%%%%%%%%%%%%%%%%%%%%%%%%%%%%%%%%
%%%%%%%%%%%%%%%%%%%%%%%%%%%%%%%%%%%%%%%%%%%%%%%%%%%%%%%%%%%%%%%%%%%%%%%%%%%%%%%%%%%%%%%%%%%%%%
\chapter{Connecting Dark Matter to the LHC}
\epigraphhead[20]{\epigraph{\large {\em B.~Altunkaynak, B.~Dutta,
D.~Feldman, M.~Holmes, Z.~Liu, Pran Nath, B.D.~Nelson}}{\large
Bhaskar Dutta (Convener)}}
%
%%%%%%%%%%%%%%%%%%%%%%% file template.tex %%%%%%%%%%%%%%%%%%%%%%%%%
%
% This is a template file for the SUSY07 conference based on the
% EPJ styfile
%
% Copy it to a new file with a new name and use it as the basis
% for your article
%
%%%%%%%%%%%%%%%%%%%%%%%% SUSY07  %%%%%%%%%%%%%%%%%%%%%%%%%%
%

%\documentclass[epj]{svjour}
\newcommand\T{\rule{0pt}{0ex}}
% Remove option referee for final version
%
% Remove any % below to load the required packages
%\usepackage{latexsym}
%\usepackage{graphicx}
%\usepackage{fancyhdr}
\def\roots{{\sqrt s}}
\newcommand{\mett}{\mbox{$E\!\!\!\!/_{T}$}}
\def \Et {\mbox{$E_T$}}
\def \Etg {E_T^{\gamma}}
\def\Z{{ Z^0}}
\def\selectron{\mbox{$\widetilde{e}$}}
\def\goes{\mbox{$\rightarrow$}}
\def\Gravitino{\widetilde{G}}
\newcommand{\eeggmett}{ee\gamma\gamma\mett}
\def\pbarp{p{\bar p}}
\newcommand{\etalnoem}{et al.}
\def\degrees{^\circ}
\def\deg{^\circ}
\newcommand{\squarkr}{\mbox{$\widetilde{q}_{R}$}}
\newcommand{\tops}{\mbox{$\widetilde{t}_{1}$}}
\newcommand{\ett}{\Et}
\newcommand{\ptt}{\mbox{$p_T$}}
\newcommand{\PT}{\ptt}
\newcommand{\NONE}{\mbox{$\widetilde{\chi}_1^0$}}
\newcommand{\NTWO}{\mbox{$\widetilde{\chi}_2^0$}}
\newcommand{\CONE}{\mbox{$\widetilde{\chi}_1^{\pm}$}}
\newcommand{\CONEP}{\mbox{$\widetilde{\chi}_1^{+}$}}
\newcommand{\CONEM}{\mbox{$\widetilde{\chi}_1^{-}$}}
\newcommand{\none}{\NONE}
\newcommand{\ntwo}{\NTWO}
\newcommand{\cone}{\CONE}
\newcommand{\coneb}{\CONEB}
\newcommand{\NTGNO}{\mbox{$\NTWO \rightarrow \gamma \NONE$}}
\newcommand{\NTGTGR}{\mbox{$\NONE \rightarrow \gamma \Gravitino$}}
\newcommand{\tanbeta}{\mbox{$\tan\beta$}}
\newcommand{\nosls}{\mbox{$N_{OS-LS}$}}
\newcommand{\mtt}{\mbox{$M_{\tau\tau}^{\mathrm{peak}}$}}
\def\Journal#1#2#3#4{{#1} {\bf #2}, #3 (#4)}
\def\PrePrint#1{\mbox{hep-ph/#1}}
\def\PRL{\rm Phys. Rev. Lett.}
\def\PRD{{\rm Phys. Rev.} D}
\def\NIM#1#2#3{\rm Nucl. Instr. and Meth A{#1}~(#2)~#3}
\newcommand{\scinotn}[2]{\mbox{${#1}\times 10^{#2}$}}
\def\rphi{r-$\phi$}
\def\rz{r-z}
\newcommand{\gt}{\ifm{>}}
\newcommand{\lt}{\ifm{<}}
\def \gtsim    {\relax\ifmmode{\mathrel{\mathpalette\oversim >}}
                  \else{$\mathrel{\mathpalette\oversim >}$}\fi}
\def \ltsim    {\relax\ifmmode{\mathrel{\mathpalette\oversim <}}
                  \else{$\mathrel{\mathpalette\oversim <}$}\fi}
\def\oversim#1#2{\lower4pt\vbox{\baselineskip0pt \lineskip1.5pt
            \ialign{$\mathsurround=0pt#1\hfil##\hfil$\crcr#2\crcr\sim\crcr}}}
%===============================
%
% Units: Energy, Momentum, Mass
%
\newcommand{\mev}  {\mbox{${\rm MeV}$}}
\newcommand{\mevc} {\mbox{${\rm MeV}/c^2$}}
\newcommand{\mevcc}{\mbox{${\rm MeV}/c^2$}}
\newcommand{\gev}  {\mbox{${\rm GeV}$}}
\newcommand{\gevc} {\mbox{${\rm GeV}/c$}}
\newcommand{\pgev} {\mbox{${\rm GeV}/c$}}
\newcommand{\gevcc}{\mbox{${\rm GeV}/c^2$}}
\newcommand{\mmev} {\mbox{${\rm MeV}/c^2$}}
\newcommand{\mgev} {\mbox{${\rm GeV}/c^2$}}
\newcommand{\tev}  {\mbox{${\rm TeV}$}}
\newcommand{\mtev} {\mbox{${\rm TeV}/c^2$}}
\newcommand{\tevcc}{\mbox{${\rm TeV}/c^2$}}
\newcommand{\cmtwo}{\mbox{${\rm cm}^2$}}
%
% Units: Luminosity
%
\newcommand{\invpb}{\mbox{${\rm pb}^{-1}$}}
\newcommand{\invfb}{\mbox{${\rm fb}^{-1}$}}
\newcommand{\lumin}{\mbox{${\rm cm}^{-2}{\rm s}^{-1}$}}
\newcommand{\lum}  {\mbox{${\cal L}$}}
\newcommand{\intlum}{\mbox{${ \int {\cal L} \; dt}$}}
%===============================
%
% Symbols
%
\newcommand{\degr}{\mbox{$^{\circ}$}}
\newcommand{\Pol} {\mbox{${\cal P}$}}   % Polarization
\newcommand{\Br}  {\mbox{${\cal B}$}}   % Branching ratio
\newcommand{\faketau} {\mbox{$f_{j\rightarrow\tau}$}}
%
% Kinematical variables
%
\renewcommand{\pt}  {\mbox{$p_{\rm T}$}}
\newcommand{\ptvis}  {\mbox{$p_{\rm T}^{\rm vis}$}}
\newcommand{\et}  {\mbox{$E_{\rm T}$}}
\newcommand{\dphi}{\mbox{$\Delta\varphi$}}
\newcommand{\met} {\mbox{${E\!\!\!\!/_{\rm T}}$}}
\newcommand{\menergy} {\mbox{${E\!\!\!\!/}$}}
\newcommand{\mpt} {\mbox{${p\!\!\!/_{\rm T}}$}}
\newcommand{\mtautau}{\mbox{$M_{\tau\tau}$}}
\newcommand{\mtautaupeak}{\mbox{$M_{\tau\tau}^{\rm peak}$}}
\newcommand{\mjtt}{\mbox{$M_{j\tau\tau}$}}
\newcommand{\mjttpeak}{\mbox{$M_{j\tau\tau}^{\rm peak}$}}
\newcommand{\mjt}{\mbox{$M_{j\tau}$}}
\newcommand{\mjtpeak}{\mbox{$M_{j\tau}^{\rm peak}$}}
\newcommand{\meff}{\mbox{$M_{{\rm eff}}$}}
\newcommand{\meffpeak}{\mbox{$M_{{\rm eff}}^{\rm peak}$}}
\newcommand{\meffb}{\mbox{$M_{{\rm eff}}^{(b)}$}}
\newcommand{\meffbpeak}{\mbox{$M_{{\rm eff}}^{(b)\; \rm peak}$}}
\newcommand{\dM}{\mbox{$\Delta M$}}
\newcommand{\mtautaumax}{\mbox{$M_{\tau\tau}^{\rm max}$}}
\newcommand{\mtautauvis}{\mbox{$M_{\tau\tau}^{\rm vis}$}}
%\newcommand{\mtautaupeak}{\mbox{$M_{\tau\tau}^{\rm peak}$}}

%========================================
%
%-----------------------------------------------------------------------------
%
% SM Particles
%
\newcommand{\qbar} {\mbox{$\overline{q}$}}
\newcommand{\cbar} {\mbox{$\overline{c}$}}
\newcommand{\bbar} {\mbox{$\overline{b}$}}
\newcommand{\tbar} {\mbox{$\overline{t}$}}
\newcommand{\ppbar}{\mbox{$p\overline{p}$}}
\newcommand{\qqbar}{\mbox{$q\overline{q}$}}
\newcommand{\ccbar}{\mbox{$c\overline{c}$}}
\newcommand{\bbbar}{\mbox{$b\overline{b}$}}
\newcommand{\ttbar}{\mbox{$t\overline{t}$}}
\newcommand{\Wpm}  {\mbox{$W^{\pm}$}}
\newcommand{\Wp}   {\mbox{$W^{+}$}}
\newcommand{\Wm}   {\mbox{$W^{-}$}}
\newcommand{\Zz}   {\mbox{$Z^0$}}
\newcommand{\wb}   {\ifm{W}}
\newcommand{\zb}   {\ifm{Z}}
\newcommand{\lnu}  {\mbox{$\ell \nu$}}
\newcommand{\emu}  {\mbox{$e\mu$}}
\renewcommand{\ee}   {\mbox{$ee$}}
\newcommand{\mumu} {\mbox{$\mu\mu$}}
\newcommand{\lplm} {\mbox{$\ell^{+} \ell^{-}$}}
\newcommand{\epem} {\mbox{$e^+e^-$}}
\newcommand{\mpmm} {\mbox{$\mu^+\mu^-$}}
\newcommand{\tauh} {\mbox{$\tau_{\rm h}$}}
%
%-----------------------------------------------------------------------------
%
% mSUGRA parameters
%
%\newcommand{\azero}{\ifm{A_{0}}}
%\newcommand{\tanb}{\ifm{\tan\beta}}
%\newcommand{\mzero}{\ifm{m_{0}}}
%\newcommand{\mhalf}{\ifm{m_{1/2}}}

\newcommand{\azero}{\mbox{$A_{0}$}}
\newcommand{\tanb} {\mbox{$\tan\beta$}}
\newcommand{\mzero}{\mbox{$m_{0}$}}
\newcommand{\mhalf}{\mbox{$m_{1/2}$}}

%
% SUSY: Squarks, Gluinos, Gravitinos
%
\newcommand{ \photino}  {\mbox{$\tilde{\gamma}$}}
\newcommand{ \gravitino}{\mbox{$\tilde{G}$}}
\newcommand{ \gluino}   {\mbox{$\tilde{g}$}}
\newcommand{ \squark}   {\mbox{$\tilde{q}$}}
\newcommand{ \squarkL}   {\mbox{$\tilde{q}_{L}$}}
\newcommand{ \squarkR}   {\mbox{$\tilde{q}_{R}$}}
\newcommand{ \squarkb}  {\mbox{$\bar{\tilde{q}}$}}
\newcommand{ \usquark}  {\mbox{$\tilde{u}$}}
\newcommand{ \usquarkL}  {\mbox{$\tilde{u}_{L}$}}
\newcommand{ \usquarkR}  {\mbox{$\tilde{u}_{R}$}}
\newcommand{ \dsquark}  {\mbox{$\tilde{d}$}}
\newcommand{ \csquark}  {\mbox{$\tilde{c}$}}
\newcommand{ \ssquark}  {\mbox{$\tilde{s}$}}
\newcommand{ \scharm}   {\mbox{$\tilde{c}$}}
\newcommand{ \csquarkL} {\mbox{$\tilde{c}_L$}}
\newcommand{ \csquarkLb}{\mbox{$\bar{\tilde{c}}_L$}}
\newcommand{ \bcsquarkL}{\mbox{$\bar{\tilde{c}}_L$}}
\newcommand{ \bsquark}  {\mbox{$\tilde{b}$}}
\newcommand{ \tsquark}  {\mbox{$\tilde{t}$}}
\newcommand{ \sbottom}  {\mbox{$\tilde{b}$}}
\newcommand{ \sbottomb} {\mbox{$\bar{\tilde{b}}$}}
\newcommand{ \sbottomone}{\mbox{$\tilde{b}_{1}$}}
\newcommand{ \sbottomoneb}{\mbox{$\bar{\tilde{b}}_{1}$}}
\newcommand{ \sstop}    {\mbox{$\tilde{t}$}}
\newcommand{ \stopo}    {\mbox{$\tilde{t}_1$}}
\newcommand{ \stopone}  {\mbox{$\tilde{t}_{1}$}}
\newcommand{ \stoponeb} {\mbox{$\bar{\tilde{t}}_{1}$}}
\newcommand{ \stoptwo}  {\mbox{$\tilde{t}_{2}$}}
\newcommand{ \stoptwob} {\mbox{$\bar{\tilde{t}}_{2}$}}
\newcommand{ \sfermion} {\mbox{$\widetilde f$}}

\newcommand{ \sbottomtwo}{\mbox{$\tilde{b}_{2}$}}

\newcommand{ \cscsb}    {\mbox{$\csquarkl\bcsquarkl$}}
\newcommand{ \sqsq}     {\mbox{$\squark\squark$}}
\newcommand{ \sqsqb}    {\mbox{$\squark\squarkb$}}
\newcommand{ \ssb}      {\mbox{$\squark\overline{\squark}$}}
\newcommand{ \ststbone} {\mbox{$\stopone\stoponeb$}}
\newcommand{ \ttbone}   {\mbox{$\stopone\stoponeb$}}
\newcommand{ \ttbtwo}   {\mbox{$\stoptwo\stoptwob$}}
%
% SUSY: Sleptons
%
\newcommand{ \slepton}  {\mbox{$\tilde{\ell}$}}
\newcommand{ \sleptonR}  {\mbox{$\tilde{\ell}_{R}$}}
\newcommand{ \seleR}    {\mbox{$\tilde{e}_{R}$}}
\newcommand{ \seleRp}    {\mbox{$\tilde{e}_{R}^{+}$}}
\newcommand{ \seleRm}    {\mbox{$\tilde{e}_{R}^{-}$}}
\newcommand{ \seleRpm}  {\mbox{$\tilde{e}_{R}^{\pm}$}}
\newcommand{ \seleL}    {\mbox{$\tilde{e}_{L}$}}
\newcommand{ \seleLpm}  {\mbox{$\tilde{e}_{L}^{\pm}$}}
\newcommand{ \sele}     {\mbox{$\tilde{e}$}}
\newcommand{ \selepm}   {\mbox{$\tilde{e}^{\pm}$}}
\newcommand{ \selep}    {\mbox{$\tilde{e}^{+}$}}
\newcommand{ \selem}    {\mbox{$\tilde{e}^{-}$}}
\newcommand{ \smu}      {\mbox{$\tilde{\mu}$}}
\newcommand{ \smuR}     {\mbox{$\tilde{\mu}_R$}}
\newcommand{ \smuRpm}   {\mbox{$\tilde{\mu}_R^{\pm}$}}
\newcommand{ \smupm}    {\mbox{$\tilde{\mu}^{\pm}$}}
\newcommand{ \smup}     {\mbox{$\tilde{\mu}^{+}$}}
\newcommand{ \smum}     {\mbox{$\tilde{\mu}^{-}$}}
\newcommand{ \stauone}  {\mbox{$\tilde{\tau}_{1}$}}
\newcommand{ \stauonep} {\mbox{$\tilde{\tau}_{1}^{+}$}}
\newcommand{ \stauonem} {\mbox{$\tilde{\tau}_{1}^{-}$}}
\newcommand{ \stauonepm}{\mbox{$\tilde{\tau}_{1}^{\pm}$}}
\newcommand{ \stautwo}  {\mbox{$\tilde{\tau}_{2}$}}
\newcommand{ \stau}     {\mbox{$\tilde{\tau}$}}
\newcommand{ \stauL}    {\mbox{$\tilde{\tau}_{L}$}}
\newcommand{ \stauR}    {\mbox{$\tilde{\tau}_{R}$}}
\newcommand{ \staupm}   {\mbox{$\tilde{\tau}^{\pm}$}}
\newcommand{ \staup}    {\mbox{$\tilde{\tau}^{+}$}}
\newcommand{ \staum}    {\mbox{$\tilde{\tau}^{-}$}}
\newcommand{ \snu}      {\mbox{$\tilde{\nu}$}}
\newcommand{ \sneutrino}{\mbox{$\tilde{\nu}$}}
%
% Charginos and Neutralinos
%
\newcommand{\schi}{\mbox{$\tilde{\chi}$}}
\newcommand{\lsp}      {\mbox{$\tilde{\chi}_{1}^{0}$}}
\newcommand{\schionezero}{\mbox{$\tilde{\chi}_{1}^{0}$}}
\newcommand{\schitwozero}{\mbox{$\tilde{\chi}_{2}^{0}$}}
\newcommand{\schithreezero}{\mbox{$\tilde{\chi}_{3}^{0}$}}
\newcommand{\schifourzero}{\mbox{$\tilde{\chi}_{4}^{0}$}}
\newcommand{\schionepm}{\mbox{$\tilde{\chi}_{1}^{\pm}$}}
\newcommand{\schionemp}{\mbox{$\tilde{\chi}_{1}^{\mp}$}}
\newcommand{\schionep} {\mbox{$\tilde{\chi}_{1}^{+}$}}
\newcommand{\schionem} {\mbox{$\tilde{\chi}_{1}^{-}$}}
\newcommand{\schitwopm}{\mbox{$\tilde{\chi}_{2}^{\pm}$}}
\newcommand{\schitwop} {\mbox{$\tilde{\chi}_{2}^{+}$}}
\newcommand{\schitwom} {\mbox{$\tilde{\chi}_{2}^{-}$}}
%\def\makeheadbox{{%  remove EPJC header
% \hbox to0pt{\vbox{\baselineskip=10dd\hrule\hbox
% to\hsize{\vrule\kern3pt\vbox{\kern3pt
% \hbox{\bfseries\@journalname\ manuscript No.}
% \hbox{(will be inserted by the editor)}
% \kern3pt}\hfil\kern3pt\vrule}\hrule}%
% \hss}
% }}
%\setlength{\topmargin}{-0.6cm} \setlength{\headheight}{0.5cm}
%\setlength{\oddsidemargin}{-0.2cm} \setlength{\evensidemargin}{-0.8cm}
%\setlength{\textwidth}{16.9cm} \setlength{\textheight}{24.4cm}
% etc

%%%%%%%DO NOT CHANGE THE FOLLOWING FOUR LINES. MAKE YOUR SELECTIONS BELOW%%%%
%\def\mytitle{My title}
%\def\myauthors{My name}
%\def\mytype{My type of session}
%\def\mysession{My session}
%%%%%%%%%%%%%%%%%%%%%%%%%%%%%%%%%%%%%%%%%%%%%%%%%%%%%%%%%%%%%%%%%%
\def\mg{\mbox{$M_{\tilde{g}}$}}
\def\dm{\mbox{$\Delta{M}$}}

%%%%%%%%%%%%%SELECTIONS FOR PLENARY SPEAKERS%%%%%%%%%%%%%%%%%%%%
%   (uncomment the selections below by removing the %)
%%%%%%%%%%%%%%%%%%%%%%%%%%%%%%%%%%%%%%%%%%%%%%%%%%%%%%%%%%%%%%%%%
%\def\mytitle{Short title of talk} %Put your title here!
%\def\myauthors{Name of Author}    %Put your name here!
%\def\mytype{Review}
%\def\mysession{\myauthors}
%%%%%%%%%%%%%%%%%%%%%%%%%%%%%%%%%%%%%%%%%%%%%%%%%%%%%%%%%%%%%%%%

%%%%%%%%%%%%%SELECTIONS FOR PARALLEL SPEAKERS%%%%%%%%%%%%%%%%%%%%
%   (uncomment the selections below by removing the %)
%%%%%%%%%%%%%%%%%%%%%%%%%%%%%%%%%%%%%%%%%%%%%%%%%%%%%%%%%%%%%%%%%
%\def\mytitle{Short title of talk} %Put your title here!
%\def\myauthors{Name of Author}    %Put your name here!
%\def\mytype{Contributed Talk}
%\def\mysession{Cosmology and Astrophysics}
%\def\mysession{Colliders - Higgs Phenomenology}
%\def\mysession{Colliders - SUSY Phenomenology}
%\def\mysession{Alternatives}
%\def\mysession{Flavor Physics}
%\def\mysession{Theoretical Models}

%%%%%%%%%%%%%%%%%%%%%%%%%%%%%%%%%%%%%%%%%%%%%%%%%%%%%%%%%%%%%%%%

%\pagestyle{fancyplain}

%\renewcommand{\sectionmark}[1]{\markboth{#1}{}}
%\rhead[\fancyplain{}{{\it\mytype}}]           {\fancyplain{}{\it\mytitle}}
%\chead[\fancyplain{}{}]                   {\fancyplain{}{}}
%\lhead[\fancyplain{}{\it\mysession}]         {\fancyplain{}{{\it\myauthors}}}
%\lfoot[\fancyplain{}{}]           {\fancyplain{}{}}
%\cfoot[\fancyplain{}{}]                   {\fancyplain{}{}}
%\rfoot[\fancyplain{}{}]         {\fancyplain{}{}}
%
%\begin{document}
%
\section{Dark Matter at the LHC}

{\it B.~Dutta}\medskip

\subsection{Introduction} We are about to enter  an era of major
discovery. The trouble-ridden Standard Model (SM) of particle
physics  needs a major rescue act. The supersymmetric extension of
SM (MSSM) seems to have all the important virtues. The Higgs
divergence problem is resolved, grand unification of the gauge
couplings can be achieved, the electroweak symmetry can be broken
radiatively. A dark matter candidate can be obtained in
supersymmetric SM. This dark matter candidate  can explain the
precisely measured 23\% of the universe in the WMAP
data~\cite{wmap}.

We need to have a direct proof of the existence of supersymmetry
(SUSY). SUSY particles can be directly observed at the large hadron
collider which is about to start. A large range of SUSY parameter
space can be investigated. The dark matter allowed regions of SUSY
parameter space can be probed and therefore, the connection between
cosmology and  particle physics can be established on a firm
footing. When LHC will be operating, there will be many other
experiments e.g. Fermi, PLANCK,  CDMS, XENON100, LUX, PAMELA, AMS,
ATIC etc, probing indirectly the SUSY models. It will be very
important to have  these different experiments to establish the
complete picture. The next few years could be the most crucial years
to establish the correct theory of nature beyond the SM.

At the LHC, the main production mechanisms for SUSY  are $\tilde
q\tilde g$, $\tilde q \tilde q$, $\tilde g\tilde g$ etc. Typically,
the squarks and gluinos then decay  into  quarks  neutralinos and
charginos. The heavier neutralino and charginos then decay into
lightest neutralino ($\tilde\chi^0_1$ and Higgs, $Z$, leptons etc.
The final state typically has multiple leptons plus multiple jets
plus missing transverse energy.  $\tilde\chi^0_1$ is the dark matter
candidate -since it does not decay into anything. The signal
typically has $\sim 10^5$ events per $fb^{-1}$ of luminosity. There
will be about $\sim 10^{8-9}$ SM events for the same amount of
luminosity which will form the background to our search for SUSY. In
order to see the signal beyond the background, the typical event
selection is made with large amount of  missing energy,  high $p_T$
jets, large numbers  of jets and leptons.

The SUSY models have new masses and therefore many new parameters.
The minimal supersymmetric SM or MSSM has more than hundred
parameters. The attempt will be to measure all these parameters at
the LHC from the decay chains  which is not an easy task. The models
based on new symmetries (e.g., grand unification), however, contains
less number of parameters and can be probed via the characteristic
features of the models. Since these model parameters are also much
less than MSSM,  one may be able to determine them after measuring a
few observables. After we confirm a model  from the real data, the
next step would be to extract the prediction of the model for
cosmology. The parameters of these models will be used to calculate
relic density and then we need to compare them with the WMAP
results~\cite{Allanach:2004xn}. This is very important since from
this exercise, we will  also be able to know if there is any need
for another dark matter candidate or whether we found the right
model for dark matter. When the LHC will be operating, these models
also will be simultaneously searched at many different experiments,
e.g., direct and indirect detection experiments of dark matter,
quark and lepton flavor violating decay modes etc.

In this review, we will concentrate on the specific  LHC signals of
SUSY models starting from the most simplest one, minimal SUGRA
model~\cite{sugra01}.

\subsection{mSUGRA} The mSUGRA model is a simple model which contains
only five parameters:\begin{equation} m_0,\ m_{1/2},\ A_0,\
\tan\beta \ {\rm and}\ sign(\mu ).
\end{equation}
$m_0$ is the universal scalar soft breaking parameter at $M_{\rm
GUT}$; $m_{1/2}$ is the universal gaugino mass at $M_{\rm GUT}$;
$A_0$ is the universal cubic soft breaking mass at $M_{\rm GUT}$;
and $\tan\beta = \langle \hat{H}_1 \rangle / \langle \hat{H}_2
\rangle$ at the electroweak scale, where $\hat{H}_{1}$
($\hat{H}_{2}$) gives rise to up-type (down-type) quark masses. The
model parameters are already significantly constrained by different
experimental results. Most important for limiting the parameter
space are: (i)~the light Higgs mass bound of $M_{h^0} > 114$~GeV
from LEP~\cite{higgs1}, (ii)~the $b\rightarrow s \gamma$ branching
ratio bound of $1.8\times10^{-4} < {\cal B}(B \rightarrow X_s
\gamma) < 4.5\times10^{-4}$ (we assume here a relatively broad
range, since there are theoretical errors in extracting the
branching ratio from the data)~\cite{bsgamma}, (iii)~the 2$\sigma$
bound on the dark matter relic density: $0.095 < \Omega_{\rm CDM}
h^2 <0.129$~\cite{wmap}, (iv)~the bound on the lightest chargino
mass of $M_{\schionepm} >$ 104~GeV from LEP \cite{aleph} and (v) the
muon magnetic moment anomaly $a_\mu$,  where one gets a 3.3$\sigma$
deviation from the SM from the experimental
result~\cite{BNL,dav,hag}. Assuming the future data confirms the
$a_{\mu}$ anomaly, the combined effects of $g_\mu -2$ and
$M_{\schionepm} >$ 104~GeV then only allows $\mu >0$. The  allowed
mSUGRA parameter space, at present, has four distinct
regions~\cite{darkrv}: (i)~the stau neutralino
($\stauone$-$\schionezero$) coannihilation region where
$\schionezero$ is the lightest SUSY particle (LSP)(In
Fig.~\ref{fig:WMAP_allowed_region}, this dark matter allowed region
is the narrow corridor along $m_{1/2}$ for smaller values of $m_0$),
(ii)~the $\schionezero$ having a larger Higgsino component
(hyperbolic branch/focus point) (In
Fig.~\ref{fig:WMAP_allowed_region}, this dark matter allowed region
appears for larger values of $m_0$), (iii)~the scalar Higgs ($A^0$,
$H^0$) annihilation funnel (2$M_{\schionezero}\simeq M_{A^0,H^0}$)
(For the parameter space of the fig.1, this region appears for
larger values of $m_{1/2}$ which is not shown in the figure), (iv) a
bulk region where none of these above properties is observed, but
this region is now very small due to the existence of other
experimental bounds (In Fig.~\ref{fig:WMAP_allowed_region} this
region is eclipsed by the bound from $b\rightarrow s\gamma$).  These
four regions have been selected out by the CDM constraint.   The
allowed parameter space for $\tan\beta$ =40 is shown in
Fig.~\ref{fig:WMAP_allowed_region}.

\begin{figure}
\begin{center}
\includegraphics[width=0.45\textwidth,height=0.45\textwidth,angle=0]{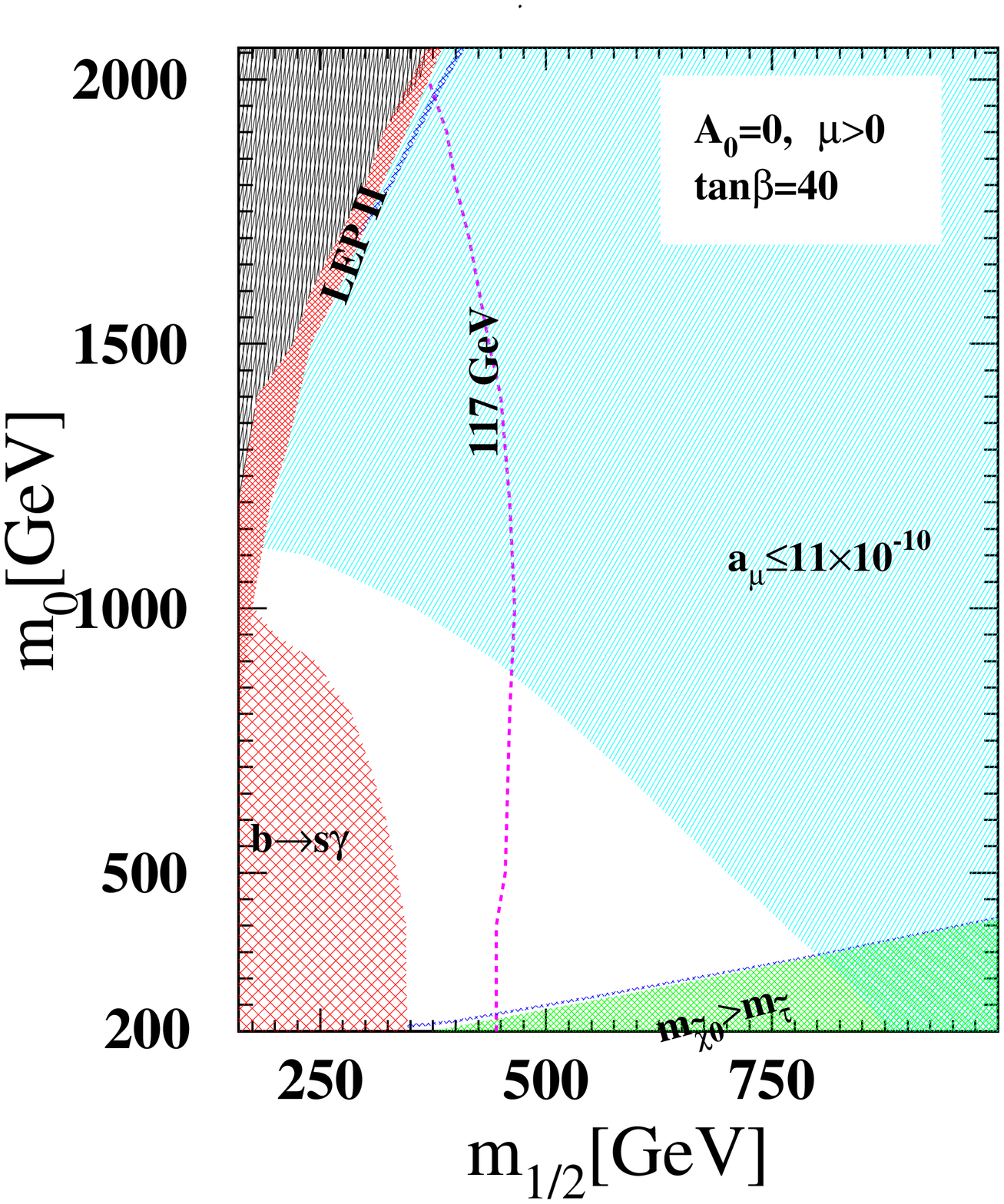}
\caption{The narrow $\dM$ coannihilation band is plotted as a
function of $m_{1/2}$ for $\tanb$ = 40 with  $A_0  = 0$ and
$\mu >0$. The left end of the band is due to the $b\rightarrow s \gamma$
branching ratio bound  and the right end by $a_\mu<11\times 10^{-10}$.
}
\label{fig:WMAP_allowed_region}\end{center}
\end{figure}

\subsection{mSUGRA at the LHC and the Determination of Dark Matter
Content} One of the first analysis for mSUGRA at the LHC will
involve the measurement of $M_{eff}$ which is the scalar sum of the
transverse momenta of the four leading jets and the missing
transverse energy:
\begin{equation}
M_{eff} = p_{T,1} + p_{T,2} + p_{T,3} + p_{T,4} + \met\,.
\end{equation} The requirement for this measurement are the following:
(1)  At least four jets with $p_{T,1}>100\,\rm{GeV}$ and
$p_{T,2,3,4}>50\,\rm{GeV}$, where the jets are numbered in order of
decreasing $p_T$.(2) $M_{eff}>400\,\rm{GeV}$, where
(3) $\met >{\rm{max}}(100\,\rm{GeV},0.2 M_{eff})$.
%\end{itemize}
In Fig.~\ref{pointx}, the distribution of $M_{eff}$ and the background are shown~\cite{hinch-paige}.
\begin{figure}[h]
\includegraphics[width=0.47\textwidth,height=0.45\textwidth,angle=0]{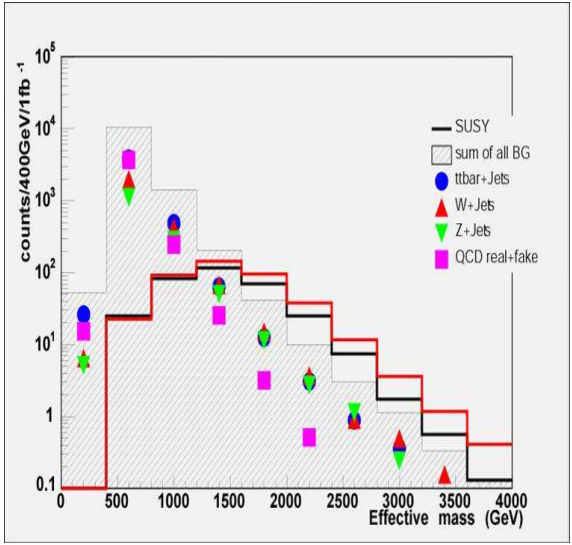}
\caption{LHC Point~1 signal and Standard Model backgrounds. Open
circles: SUSY signal.  Solid circles: $t\bar t$. Triangles: $W\rightarrow\ell\nu$,
$\tau\nu$.  Downward triangles:  $Z\rightarrow\nu\bar\nu$, $\tau\tau$.
Squares: QCD jets.  Histogram: sum of all
backgrounds~\cite{hinch-paige}}\label{pointx}
\end{figure}
The peak of the distribution  varies linearly with the Min[$m_{\tilde q}m_{\tilde g}$]~\cite{hinch-paige,tovey} for the mSUGRA model and
therefore the scale of SUSY can be surmised from this peak measurement

After we establish the existence of SUSY and an overall scale for the SUSY production, we need to measure the masses.
The existence of missing energy in the signal will tell us the possibility of dark matter candidate,
but the calculation of the relic density is based on the parameters of the models which depends on
the measurement of  masses and the mixing matrices.

\begin{figure}[t]
\includegraphics[width=0.45\textwidth,height=0.45\textwidth,angle=0]{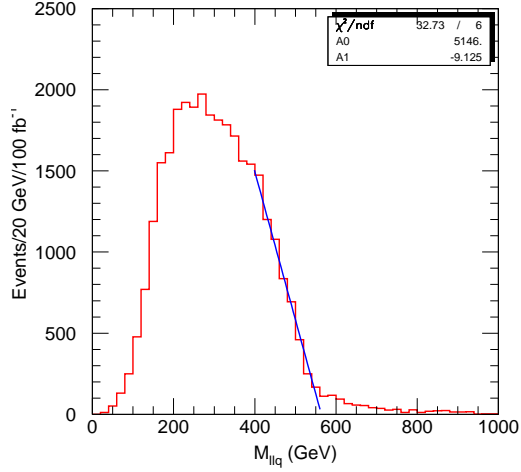}
\caption{Mass distribution for the smaller of the two $\ell^+\ell^-q$
  masses showing a linear fit near the four-body end point~\cite{hinch-paige}.}
\label{mllq}
\end{figure}

%\begin{figure}[t]
%\includegraphics[width=0.45\textwidth,height=0.45\textwidth,angle=0]{c5_100.4_mlq.eps}
%\caption{Distribution of  the larger of the two $\ell^\pm q$ masses
%for $\ell^+\ell^-q$ events~\cite{hinch-paige}}\label{mlq}
%\end{figure}
Now we discuss the mass measurements. Suppose $\tilde q_L$ is pair produced and then $\tilde q_L$  decays into $\tilde\chi_2^0 q$. The $\tilde\chi^0_2$ then decays into a pair of opposite sign leptons (via slepton) and $\tilde\chi^0_1$. It is expected that the two high energy jets will be arising directly
from $\tilde q_L \to \tilde\chi_2^0 q$ as a dominant production process is that
which leads to $\tilde q_L \tilde g$ and hence to pairs of $\tilde q_L$.
 Therefore, the smaller of the two masses
formed by combining the leptons with one of the two highest $p_T$ jets
should be less than the four-body kinematic end point for squark decay,
e.g.,
\begin{equation}
M_{\ell\ell q}^{\rm max} = \left[{ \left(M_{\tilde q_L}^2-M_{\tilde \chi_2^0}^2\right)
\left(M_{\tilde \chi_2^0}^2-M_{\tilde\chi^0_1}^2\right) \over M_{\tilde \chi_2^0}^2}
\right]^{1/2}.
\end{equation}
The distribution of the smaller $\ell^+\ell^-q$ mass is shown  in
Fig.~\ref{mllq} subtracting the opposite flavor  combination from the  same flavor lepton pairs.
The $e^+e^- + \mu^+\mu^- -
e^\pm\mu^\mp$ combination cancels all contributions from two independent
decays and  reduces the
combinatorial background.  % Fig.~\ref{mlq}  shows a linear
%fit near the end point for the distribution of the larger of the two $\ell^\pm q$ masses
%for $\ell^+\ell^-q$ events.
The end points of $\ell^\pm q$, $\ell^+\ell^-$, $higgs +q$, $Z+q$  distributions are also used to determine model
parameters. These types of measurements  can be used to determine
the masses of the SUSY particles without any  choice of model  by solving the algebraic equations.
These measurement methods to determine the parameters of different mSUGRA allowed parameter space.

\subsection{Stau-Neutralinno Coannihilation} In this region the stau
and the neutralino masses are close. The relic density is satisfied
by having both stau and neutralino mass to be close  and thereby
increasing the neutralino annihilation cross section. This
phenomenon occurs for a large region of mSUGRA parameter space for
smaller values of $m_0$.

The crucial aspect of the signal is the low energy tau and in the analysis.
Fig.~\ref{fig:WMAP_allowed_region1} shows the range of allowed $\dM$
values in the coannihilation region as a function of $m_{1/2}$ for
$\tanb$ = 40. We see that $\dM$ is narrowly constrained and varies
from 5-15 GeV. Because of the small $\dM$ value, $\tau$'s from
$\tilde\tau_1\rightarrow \tau \tilde\chi^0_1$ decays are  expected to have
low energy providing the characteristic feature of  the coannihilation
region.

\begin{figure}
\begin{center}
\includegraphics[width=0.45\textwidth,height=0.45\textwidth,angle=0]{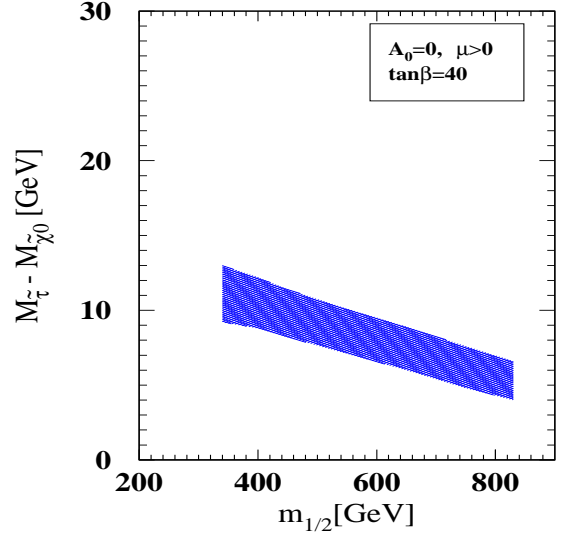}
\caption{The narrow $\Delta M$ coannihilation band is plotted as a
function of $m_{1/2}$ for $\tanb$ = 40 with  $A_0  = 0$ and
$\mu >0$. The left end of the band is due to the $b\rightarrow s \gamma$
branching ratio bound  and the right end by $a_\mu<11\times 10^{-10}$~\cite{stauneutralino1}.
}
\label{fig:WMAP_allowed_region1}\end{center}
\end{figure}
 We are mostly interested in events
from $\schionezero \schitwozero$, $\schionepm \schitwozero$, or
$\schitwozero \schitwozero$ pairs, where the $\schionezero$ in the
first case is directly from the $\squarkR$ decay. The branching
ratio of $\schitwozero  \rightarrow \tau\ \stauone$ is about 97\%
for our parameter space and is dominant even for large $\mhalf$ in
the entire coannihilation region; the same is true for the
$\schionepm \rightarrow \nu\ \stauone$ decay mode. (It should be
noted that both \seleR\ and \smuR\ are lighter than $\schitwozero$
by about 10~GeV . However, the branching ratio for $\schitwozero \to
e \seleR$ or $\mu \smuR$ is much less than 1\%.) Since the stau
decays via $\stauone \rightarrow \tau \schionezero$,  we expect
inclusive $\schitwozero$ events to include at least two $\tau$
leptons plus large \et\ jet(s) and large \met\ (from the
$\schionezero$).\\

{\bf Measurement of relic density at the
LHC~\cite{stauneutralino3}}\\

In order to predict the relic density from the collider measurements
in mSUGRA model, we need to determine the model parameters from the
mass measurements. The main trouble is the determination of $A_0$
and $\tan\beta$. These two parameters should in principle be
measured from the third generation squark masses. However the main
problem in this tactic is the ability to distinguish stop from
sbottom and vice versa. The presence of bottom quarks in the final
states from both these quarks make the individual measurement of
these masses so difficult. Instead of measuring these masses, we
measure observables which depend on both these masses and try to
measure the parameters from them.

We now show the observables (beyond what we already discussed) which
can be used to measure the masses and therefore the model
parameters. We analyze three samples in the final state of large
transverse missing energy (\met) along with jets ($j$'s), $\tau$'s,
and $b$'s: (i) 2$\tau$ + 2$j$ + \met, (ii) 4$j$ + \met, and (iii)
1$b$ + 3$j$ + \met.

%%%%%%%
%%%%%%% Table 1
%%%%%%%

\begin{table*}[t]
\caption{SUSY masses (in GeV) for our reference point
$m_{1/2}$ = 350~GeV, $m_0$ = 210 GeV, $\tan\beta = 40$,
$\azero = 0$, and $\mu > 0$. }
 \label{tab:SUSYmass}
\begin{center}
\begin{tabular}{c c c c c c c c}
\hline \hline
$\gluino$ &
$\begin{array}{c} \usquarkL \\ \usquarkR \end{array}$ &
$\begin{array}{c} \stoptwo \\ \stopone \end{array}$ &
$\begin{array}{c} \sbottomtwo \\ \sbottomone \end{array}$ &
$\begin{array}{c} \seleL \\ \seleR \end{array}$ &
$\begin{array}{c} \stautwo \\ \stauone \end{array}$ &
$\begin{array}{c} \schitwozero \\ \schionezero \end{array}$ &
$\dM$
\\ \hline
831&
$\begin{array}{c} 748 \\ 725 \end{array}$ &
$\begin{array}{c} 728 \\ 561 \end{array}$ &
$\begin{array}{c} 705 \\ 645 \end{array}$ &
$\begin{array}{c} 319 \\ 251 \end{array}$ &
$\begin{array}{c} 329 \\ 151.3 \end{array}$ &
$\begin{array}{c} 260.3 \\ 140.7 \end{array}$ &
10.6
\\ \hline \hline
\end{tabular}
\end{center}
\end{table*}

The primary SM backgrounds for the 2$\tau$ + 2$j$ + \met\, final
state is from $t\bar{t}$, $W$+jets and $Z$+jets production. The
2$\tau$ + 2$j$ + \met\ sample is selected using the following
cuts~\cite{stauneutralino1,LHCtwotau}: (a) $N_{\tau} \geq 2$
($|\eta| < 2.5$, $\ptvis >20\ \gev$; but $>40\ \gev$ for the leading
$\tau$); (b) $N_{j} \geq 2$  ($|\eta| < 2.5$, $\et > 100\ \gev$);
(c) $\met > 180\ \gev$ and $\et^{j\rm{1}} + \et^{j\rm{2}}$ + \met\
$>$ 600 GeV; (d) Veto the event if any of the two leading jets are
identified as $b$. In order to identify $\schitwozero \rightarrow
\tau\stauone \rightarrow \tau\tau\schionezero$ decays, we categorize
all pairs of $\tau$'s into opposite sign (OS) and like sign (LS)
combinations,
 and then use the OS minus LS (OS$-$LS) distributions
to effectively reduce the SM events as well as the
combinatoric SUSY backgrounds.
We reconstruct  the  decay chains of
$\squarkL \rightarrow q \schitwozero
\rightarrow q \tau \stauone \rightarrow q \tau\tau \schionezero$
using the following five kinematic variables:
(1) $\alpha$, the slope of the \ptvis\ distribution for the lower energy $\tau$
in the OS$-$LS di-$\tau$ pairs,
(2) $M_{\tau\tau}^{\rm peak}$, the peak position
of the visible di-$\tau$ invariant mass distribution,
(3) \mjttpeak, the peak position  of the invariant $j$-$\tau$-$\tau$ mass distribution,
and
(4~,5) \mjtpeak, the peak position of the invariant $j$-$\tau$ mass distribution
where each $\tau$ from the OS$-$LS di-$\tau$ pair is examined separately.
Note that we have used the peak positions
instead of the end-points because of the $\tau$'s in the final state.

We follow the recommendation of Ref.~\cite{hinch-paige} for the 4$j$
+ \met\ sample. The variable $\meff \equiv \met\ + \sum_{{\rm
4~jets}} \et^{j}$, which is a function of only the $\gluino$ and
$\squark$ masses, is reconstructed for each event that passes the
following selection cuts: (a) $N_{j} \geq 4$  ($|\eta| < 2.5$, $\et
> 100\ \gev$ for the leading jet; $> 50\ \gev$ for other jets); (b)
$\met > 100\ \gev$; (c) Transverse sphericity $>$ 0.2; (d) Veto on
all events containing an isolated electron or muon with $\pt > 15\
\gev$ and $|\eta| < 2.5$.; (e) $\met > 0.2 \meff$. Again we require
that none of these jets identified as  a $b$ jet. We use
ISAJET~\cite{isajet} and PGS4~\cite{pgs} for our work.

Similar cuts are used to make
the 1$b$ + 3$j$  + \met\ sample.
However here we introduce a new variable,
\meffbpeak, similar to \meffpeak, but requiring
that the leading jet be from a $b$ quark.

%%%%%%%%
%%%%%%%% Figure 1
%%%%%%%%

\begin{figure}
\centering
\includegraphics[width=.55\textwidth,height=0.35\textheight,angle=-90]{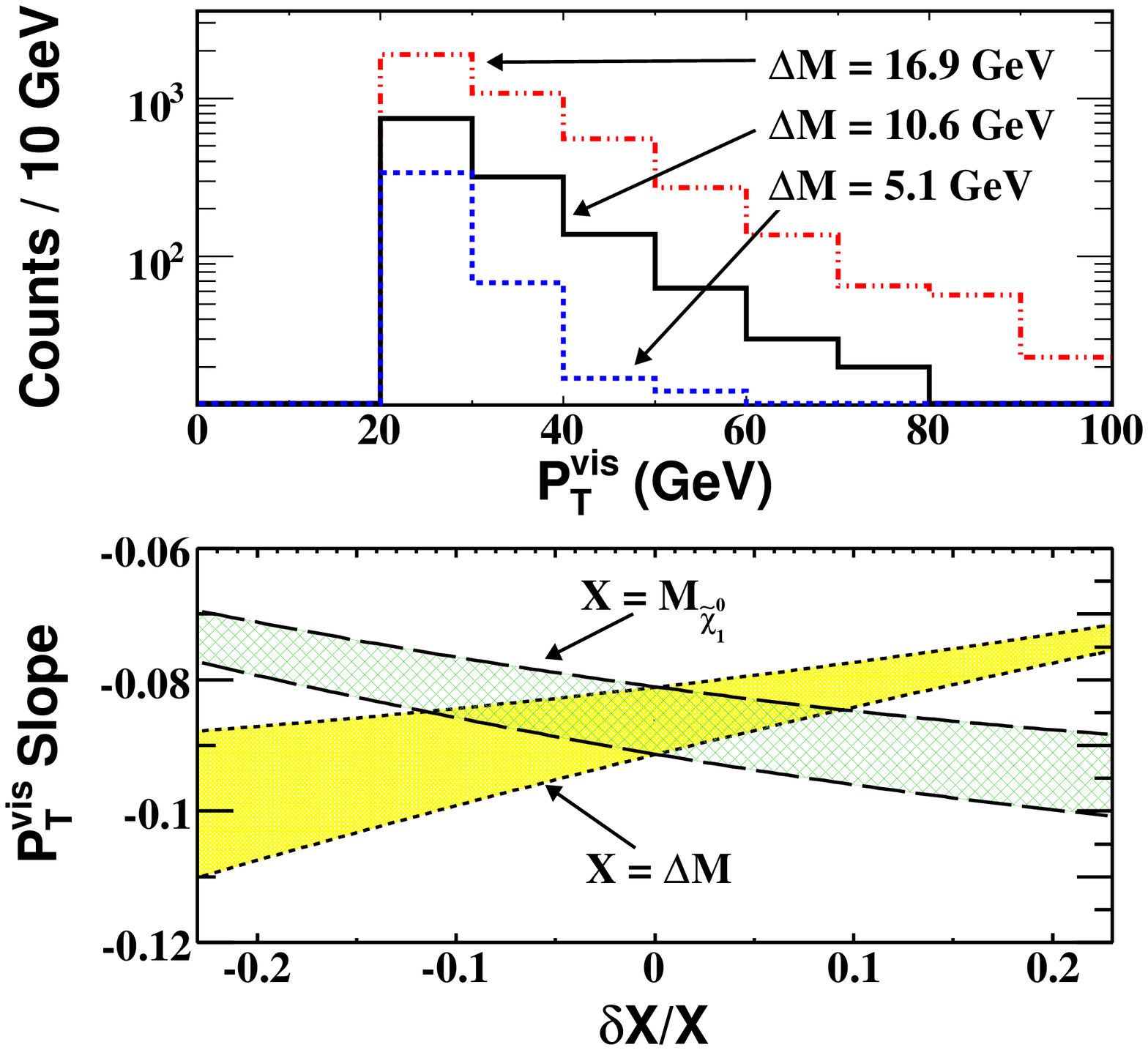}
\caption{[top] The $\ptvis$ distribution  of the lower-energy $\tau$'s
using the OS$-$LS technique in the three samples (arbitrary luminosity) of SUSY events
with \dM\ = 5.1, 10.6 and 16.9 GeV,
where only \stauone\ masses are changed at our reference point.
[bottom] The $\ptvis$ slope (defined as $\alpha$ in the text)
 as a function of the relative change of
 $\dM$ or $M_{\schionezero}$
from its reference value where all other SUSY masses are fixed.
 The bands correspond to estimated uncertainties with 10 \invfb\
 of data.
}
\label{fig:visTauPt}
\end{figure}

%%%%%%%%

The measurement of a small value of $\alpha$  from the 2$\tau$ + 2$j$ + \met\ sample indicates low energy $\tau$'s in the final state
(thus \dM\ is small) and
 provides  a smoking-gun signal for the CA region.
In Fig.~\ref{fig:visTauPt}, we show the \ptvis\ distribution obtained
by the OS$-$LS technique and how it varies as a function of  \dM\  in the CA region.
Note that $\alpha$ only depends on
the \stauone\ and \schionezero\ masses.
The \stauone\ and \schionezero\ dependences are
shown in Fig.~\ref{fig:visTauPt}.

To get a set of measurements of the SUSY particle masses,
we use the remaining variables from
the 2$\tau$ + 2$j$ + \met\ and 4$j$ + \met\ samples.
The variables \mjttpeak\ and \mjtpeak\ probe the
$\squarkL \rightarrow q \schitwozero
\rightarrow q \tau\stauone
\rightarrow q \tau\tau\schionezero$ decay chains.
To help identify these  chains we additionally require
 OS$-$LS di-$\tau$ pairs with
$M_{\tau\tau} < M_{\tau\tau}^{\mbox{\rm end-point}}$
and construct $M_{j\tau\tau}$ for every jet with
$\et > 100\ \gev$ in the event.
With three jets, there are
three masses: $M_{j\tau\tau}^{(1)}$,
$M_{j\tau\tau}^{(2)}$, and
$M_{j\tau\tau}^{(3)}$, in a decreasing order.
We choose
 $M_{j\tau\tau}^{(2)}$ for this analysis~\cite{hinch-paige}.
Figures~\ref{fig:Mjtautau}  shows the $M_{j\tau\tau}^{(2)}$
distributions for two different \squarkL\ masses, and
$M_{j\tau\tau}^{(2)~\rm peak}$ as a function of $M_{\squarkL}$ and
$M_{\schionezero}$,
 keeping \dM\ constant.
Similarly, one can show that
the $M_{j\tau}^{(2)~\rm peak}$ value depends on the \squarkL,
\schitwozero, \stauone\  and  \schionezero\  masses.
The value of \meffpeak,
has been shown to be  a function of
only the \squarkL\ and \gluino\ masses.%~\cite{hinch1}.

%%%%%%%%
%%%%%%%% Figure 2
%%%%%%%%

\begin{figure}
\centering
\includegraphics[width=.45\textwidth,height=0.35\textheight,angle=-90]{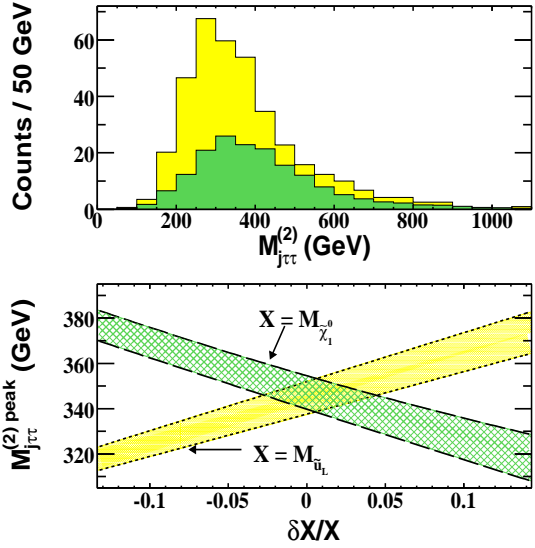}
\caption{[top] $M_{j\tau\tau}^{(2)}$ distributions using the OS$-$LS technique
for SUSY events at our reference point, but with
$M_{\squarkL}$ = 660 GeV (yellow or light gray histogram)
and 840 GeV (green or dark gray histogram), where 748 GeV is
our reference point;
[bottom] The peak position of the mass distribution
 as a function of $M_{\schionezero}$ or  $M_{\squarkL}$.
 The bands correspond to estimated uncertainties with 10 \invfb\
 of data.
}
\label{fig:Mjtautau}
\end{figure}

%%%%%%%%

The determination of the SUSY particle masses
is done by inverting the six functional relationships between the variables
and the SUSY particle masses
to simultaneously solve
for the \gluino, $\tilde\chi^0_{1,2}$, \stauone, and
average \squarkL\ masses and their uncertainties.
The six parametrized functions are:
(1) \mtautaupeak\ = $f_1$($M_{\schitwozero}$, $M_{\schionezero}$, $\dM$),
(2) $\alpha$ = $f_2$($M_{\schionezero}$, $\dM$),
(3) $M_{j\tau\tau}^{(2)\; \rm peak}$ =
 $f_3$($M_{\squarkL}$, $M_{\schitwozero}$,
$M_{\schionezero}$),
(4~,5) $M_{j\tau(1,2)}^{(2)~\rm peak}$ = $f_{4,5}$($M_{\squarkL}$,
 $M_{\schitwozero}$, $M_{\schionezero}$, $\dM$),
 and
(6) \meffpeak\ = $f_6$($M_{\squarkL}$, $M_{\gluino}$).
With 10 \invfb\ of data,
we obtain (in GeV)
$M_{\gluino} = 831 \pm 28$,
$M_{\schitwozero} = 260 \pm 15$,
$M_{\schionezero} = 141 \pm 19$,
$\dM = 10.6 \pm 2.0$, and
$M_{\squarkL} = 748 \pm 25$.
The accurate determination of $\Delta M$ would also confirm  that we are in the CA region.
We also test the universality of the
gaugino masses at the GUT scale which
implies
$M_{\gluino}/M_{\schionezero}=5.91$ and
$M_{\gluino}/M_{\schitwozero}=3.19$
at the electroweak scale.
With the above gaugino masses,
we obtain $M_{\gluino}/M_{\schionezero}=5.9 \pm 0.8$ and $M_{\gluino}/M_{\schitwozero}=3.1 \pm 0.2$, which would
validate the universality relations to $14 \%$ and $6 \%$, respectively.

In order to achieve the  primary goal which is to determine
$\Omega_{\schionezero} h^{2}$ in the mSUGRA model,
we determine $m_0$, $m_{1/2}$, \azero\ and $\tan\beta$.
\meff\ and \mjtt\ depend only on
the \squarkL\ (first two generations),
\gluino, \schitwozero\ and \schionezero\ masses.
This provides a direct handle on \mzero\ and \mhalf\ and is shown in Fig.~\ref{fig:Mjtautau_mSUGRA}.
We note that these values are  insensitive to \azero\ and $\tan\beta$ and therefore require no knowledge of their values.
On the other hand, \mtautaupeak\ and  \meffbpeak\  provide
a direct handle of \azero\ and $\tan\beta$.
\mtautaupeak\ depends on the \stauone\ mass;
\meffbpeak\ depends on the \stopone\ and \sbottomone\ masses,
since both the \stopone\ and \sbottomone\ decays always
produce at least one $b$ jet in the final state.
Figure~\ref{fig:Meffb} shows
the values of  \mtautaupeak\ and \meffbpeak\
as functions of \azero\ and $\tan\beta$.
Combining these four measurements and inverting,
as done to determine the SUSY particle masses,
we find
$m_0 = 210 \pm 4\ \gev$, $m_{1/2} = 350 \pm 4\ \gev$,
$A_0 = 0 \pm 16\ \gev$, and $\tan\beta = 40 \pm 1$
with 10 \invfb\ of data.
Note that all uncertainties are statistical only.

%%%%%%%%
%%%%%%%% Figure 3
%%%%%%%%

\begin{figure}
\centering
\includegraphics[width=.55\textwidth,height=0.35\textheight,angle=-90]{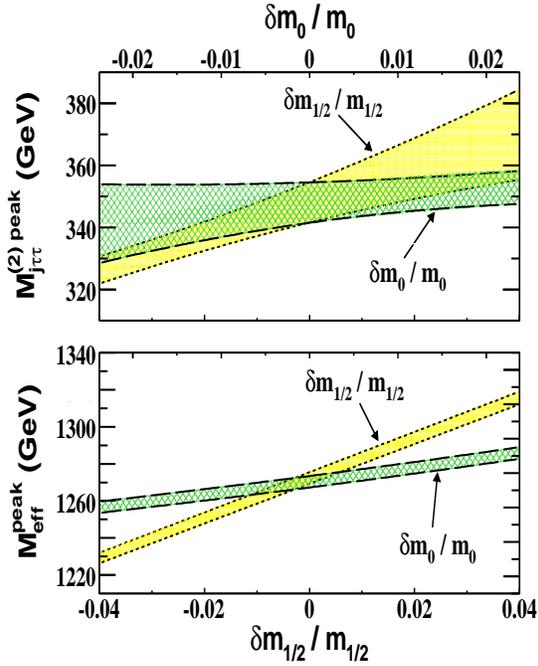}
\caption{The dependence of $M_{j\tau\tau}^{(2)\; \rm peak}$ (top)
and \meffpeak\ (bottom) as a function of $m_{1/2}$ and $m_{0}$.
 The bands correspond to estimated uncertainties with 10 \invfb\
 of data.
}
\label{fig:Mjtautau_mSUGRA}
\end{figure}

%%%%%%%%
%%%%%%%% Figure 4
%%%%%%%%

\begin{figure}
\centering
\includegraphics[width=.55\textwidth,height=0.35\textheight,angle=-90]{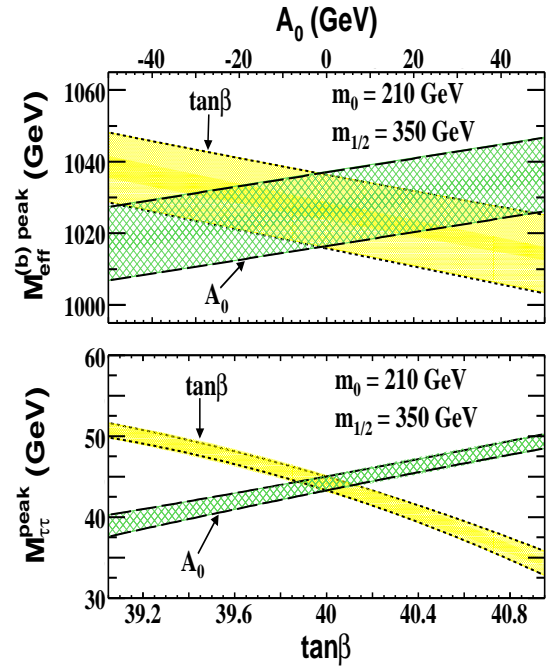}
\caption{The dependence of  \meffbpeak\ (top)  and
\mtautaupeak\ (bottom) as a function of $\tan\beta$ and $A_{0}$.
 The bands correspond to estimated uncertainties with 10 \invfb\
 of data.
}
\label{fig:Meffb}
\end{figure}
%%%%%%%%

After we measure the mSUGRA variables, we  can calculate
$\Omega_{\schionezero} h^2$ using DarkSUSY~\cite{Gondolo:2004sc}. In
the coannihilation region, $\Omega_{\schionezero} h^2$ depends
crucially on $\dM$ due to the Boltzmann suppression factor $e^{-
\Delta M/k_{B}T}$ in the relic density formula~\cite{Griest:1990kh}.
In figure~\ref{fig:Omegah2} we show  contour plots of the 1$\sigma$
uncertainty in  the $\Omega_{\schionezero} h^2$-$\dM$ plane since
the two measurements are highly correlated. The uncertainty on
$\Omega_{\schionezero} h^2$ is 11~(4.8)\% at 10~(50) \invfb. Note
that it is 6.2\% at 30 \invfb, comparable to that of the recent WMAP
measurement~\cite{wmap}.

In summary, we have described a technique to make a precision  measurement of $\Omega_{\schionezero} h^2$
at the LHC in the \stauone-\schionezero\ CA region
of the mSUGRA model.
This is achieved by using  only the model parameters,
determined by the  kinematical analyses
of 3 samples of  \met\ + $j$'s (+ $\tau$'s) events with and without $b$ jets.
The accuracy of the $\Omega_{\schionezero} h^2$ calculation at 30 \invfb\
of data
is expected to be
comparable to that of $\Omega_{\rm CDM} h^2$ by WMAP. This technique of measuring the mSUGRA parameters is general and can
be applied to any SUGRA models.
With these types of measurements at the LHC,
it is possible to confirm
%[that the co-annihilation process was important and]
that the DM we observe today were $\tilde\chi^0_1$'s created in the early universe.

\begin{figure}
\centering
\includegraphics[width=.4\textwidth,height=0.3\textheight,angle=-90]{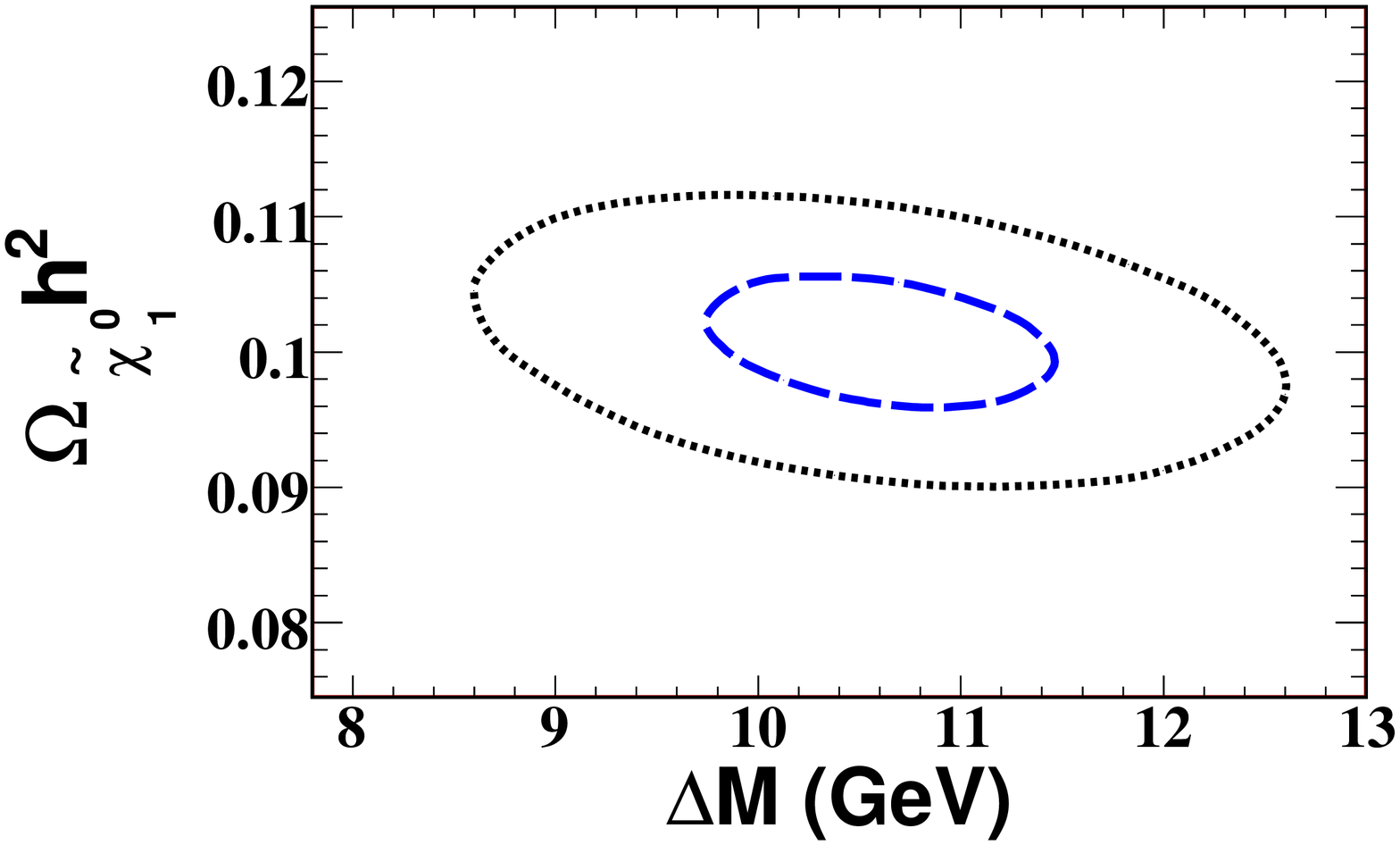}
\caption{Contour plot of the 1$\sigma$ uncertainty
in the $\Omega_{\schionezero} h^2$-$\dM$ plane
with 10 \invfb\ (outer ellipse) and 50 \invfb\ (inner ellipse) of data. }
\label{fig:Omegah2}
\end{figure}

\subsection{Hyperbolic branch/Focus point}In this region, $m_0$ is
very large, but $m_{1/2}$ can be small which means the gaugino
masses can be small. For a fixed value of the parameter $m_{1/2}$ in
the mSUGRA model, if $m_0$ is taken to be of order the weak scale,
then $m_{H_u}^2$ is driven to negative values at the weak scale due
to the large top quark Yukawa coupling in the RGEs, whereas if $m_0$
is taken too large, then the GUT scale value of $m_{H_u}^2$ is so
high that it does not become negative values when the weak scale is
reached in RG running. Intermediate to these two extreme cases there
exists a region where $\mu^2$ is found to be zero, which forms the
large $m_0$ edge of parameter space. If $\mu^2$ is positive, but
tiny, then  light higgsino-like charginos will be generated and one
needs to be worried about the  LEP limit on chargino masses which
require $m_{\tilde\chi^\pm_1}>103.5$ GeV. If $\mu^2$ is large enough
to evade LEP2 limits, then large higgsino-bino mixing occurs in the
chargino and neutralino sectors, the lightest neutralino becomes a
mixed higgsino-bino dark matter particle. A lightest neutralino of
mixed higgsino-bino form has a large annihilation rate, and hence
satisfy the  WMAP measurements. In the WMAP-allowed hyperbolic
branch/focus point region, since squarks have masses in the TeV
range, only three-body decay modes of the gluino are allowed. The
third generation quark-squark-neutralino/chargino couplings are
enhanced by top quark Yukawa coupling terms since the neutralino and
chargino can have large higgsino component.

One search strategy of this region is to study the shape of dilepton
final state. The dileptons are produced from $\tilde\chi^0_3$ and
$\tilde\chi^0_2$ decays. Using the parameter space, $m_0$=3550 GeV;
$m_{1/2}$=300 GeV; $A_0$=0; $\tan\beta$=10 ; $\mu>$0, Tovey etal has
shown that the gaguino mass differences can be measured with an
accuracy of 1 GeV. This error can be improved up to 0.5
GeV~\cite{moortgat}.

In the reference~\cite{barger}, it is shown that  by requiring high
jet and $b$-jet multiplicity, and a high effective mass cut, a
rather pure signal arises over the dominantly $t\bar{t}$ SM
background. Since the signal came almost entirely from gluino pair
production, and the decay branching fractions were fixed by assuming
the neutralino relic density saturated the WMAP
$\Omega_{\tilde\chi^0_1}h^2$ measurement, the total signal rate has
been used to extract an estimate of the gluino mass. It is found in
the reference~\cite{barger} that, $m_{\tilde g}$ could be measured
to a precision of about 8\% for 100 fb$^{-1}$ of integrated
luminosity. In order to make this measurement, the signal contains
$n\geq 7$ jets, $n\geq 2$ b-jets and  $A_T=E_T(miss)+\Sigma
E_T(jet)+\Sigma E_T(lepton)>1300$ GeV with 100$fb^{-1}$ luminosity.
The $A_T$ distribution in events with $\ge 7$ jets and $\ge 2$
$b$-tags, for the model point $m_0=3050$ GeV, $m_{1/2}=400$ GeV,
$A_0=0$,$\tan\beta =30$, $\mu >0$ is shown in
figure~\ref{fig:meff7}.

\begin{figure}[htbp]
\begin{center}
\includegraphics[angle=0,width=0.35\textwidth,height=0.35\textwidth,]{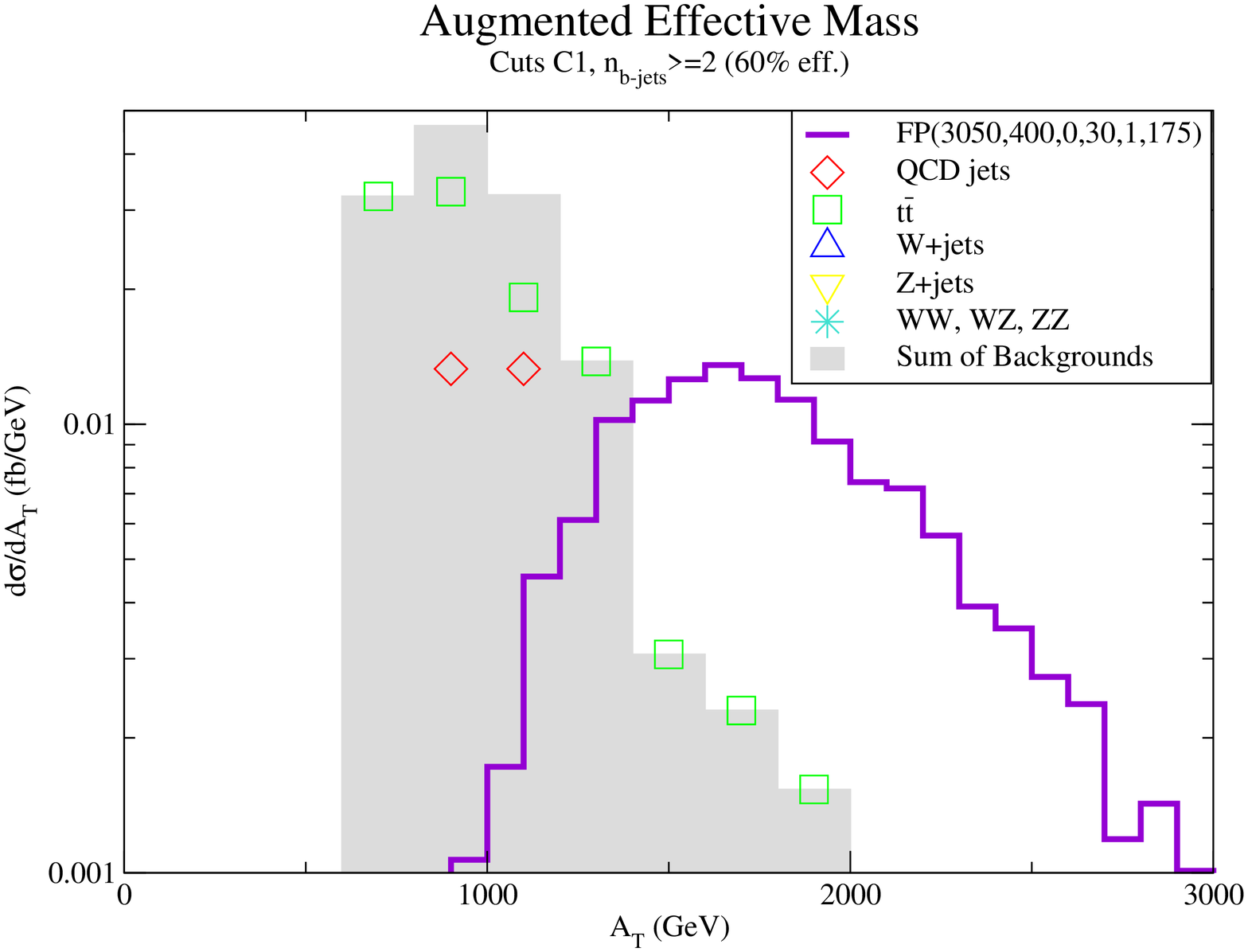}
\caption{Distribution of $A_T$ in events with $\ge 7$ jets and $\ge 2$ $b$-tags, for the model point
$m_0=3050$ GeV, $m_{1/2}=400$ GeV, $A_0=0$,$\tan\beta =30$, $\mu >0$ and $m_t=175$ GeV,
versus various SM backgrounds~\cite{barger}.}
\label{fig:meff7}
\end{center}
\end{figure}

In addition, the signal from this region can be separated as to its
isolated lepton content. The OS/SF dilepton mass distribution
embedded in the hard signal component should exhibit mass edges at
$m_{\tilde\chi^0_2}-m_{\tilde\chi^0_1}$ and also at
$m_{\tilde\chi^0_3}-m_{\tilde\chi^0_1}$, which are distinctive of
this scenario in which the LSP is a mixed bino-higgsino particle.
The mass-difference edges, along with the absolute gluino mass, may
provide enough information to constrain the absolute chargino and
neutralino masses.

Since it is possible to measure the gluino, neutralino  masses and we can solve for parameters like, $\mu$, $m_{1/2}$ and $\tan\beta$ which primarily enter into the calculation of relic density in this region via the chargino and neutralino matrices. Since the sfermions are heavy in this region, the charginos-neutralinos primarily contribute to the dark matter content calculation. We are finding that the DM content can be determined within 30\% accuracy (for 300 fb$^{-1}$ luminosity)~\cite{talk}.

Large $m_0$ region also explains the EGRET excess of diffuse galactic gamma rays by supersymmetric dark matter annihilation. The SUSY parameter space for this region:  $m_0$=1400 GeV, $\tan\beta$=50
$m_{1/2}$=180 GeV, $A_0$=0.5 $m_0$~\cite{wim}.

\subsection{Bulk Region} In this region, the relic density constraint
is satisfied by t channel selectron, stau and sneutrino exchange.
Nojiri et al~\cite{Nojiri:2005ph} has analyzed the bulk region by
measuring the masses from the end point measurements. The parameter
pace point is $m_0$=70 GeV; $m_{1/2}$=250 GeV; $A_0$=-300;
$\tan\beta$=10; $\mu>$0 for the  analysis of the bulk region. The
end points have been determined for the $lq$, $llq$, $ll$ etc
distributions as described before and are given in
Table~\ref{tabsusymasses}.
%The meaning of the different
%observables, and their expression in terms of sparticle masses is
%given in Ref.~\cite{Allanach:2000kt}.

\begin{table}[htb]
\begin{center}
\caption{\label{tabsusymasses} { Table of the SUSY
measurements which can be performed at the LHC with the ATLAS
detector ~\cite{Nojiri:2005ph}.
The central values are calculated with ISASUSY 7.71,
using the tree-level values for the sparticle masses.
The statistical errors are given for the integrated
luminosity of 300~$fb^{-1}$. The uncertainty in the energy scale is
taken to result in an error of 0.5\% for measurements including jets,
and of 0.1\% for purely leptonic mesurements~\cite{Nojiri:2005ph}.}}
\vskip 0.2cm
\begin{tabular}{|l|c|c|}
\hline
Variable & Value (GeV) & Error \\
\hline
\hline
$m_{\ell\ell}^{max}$             &    81.2 &     0.09 \\
$m_{\ell\ell q}^{max}$           &   425.3 &        2.5 \\
$m_{\ell q}^{low}$               &   266.9 &       1.6 \\
$m_{\ell q}^{high}$              &   365.9 &        2.1 \\
$m_{\ell\ell q}^{min}$           &   207.0 &        1.9 \\
$m(\ell_L)-m(\tilde\chi^0_1)$              &    92.3 &        1.6 \\
$m_{\ell\ell}^{max}(\tilde\chi^0_4)$  &   315.8 &        2.3 \\
$m_{\tau\tau}^{max}$             &    62.2 &        5.0 \\
\hline
\end{tabular}
\end{center}
\end{table}
  The  measurement of the
sparticle masses are done from the measured edges. The error is of
$\sim 9$ GeV for the masses of the sparticles. The distribution of
the measured $\tilde\chi^0_1$ masses for a set of Monte Carlo
experiments is shown in Fig.~\ref{fig:mchi01a}.
\begin{figure}[htb]
\begin{center}
\includegraphics[width=0.40\textwidth,height=0.40\textwidth,angle=0]{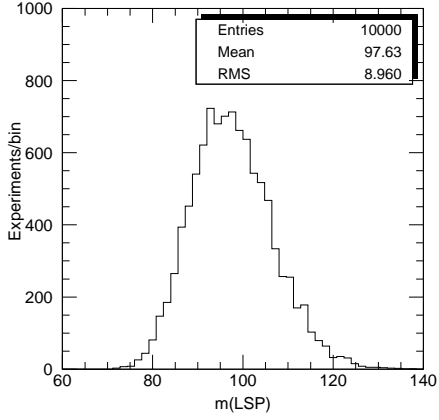}
\caption{\label{fig:mchi01a} {\it
Distribution of the measured value of $m(\tilde\chi^0_1)$
for a set of Monte Carlo
experiments, each corresponding to an integrated statistics of
300~fb$^{-1}$. The $m(\tilde\chi^0_1)$ mass in the model is 97.2 GeV~\cite{Nojiri:2005ph}.}}
\end{center}
\end{figure}
Since the masses are determined from a set of algebraic equations the   errors
are strongly correlated. The mass difference is strongly constrained (e.g., $m(\tilde l_R)-m(\tilde \chi^0_1)$
is $\sim200$~MeV due to the very good precision of the edge measurements, but the absolute error has loose constrained
$sim$ $\sim9$~GeV.
The calculated precision on
$m(\tilde\tau_1)-m(\tilde \chi^0_1)$ is $\sim2.5$~GeV. In this case the stau neutralino mass difference is larger than the neutralino-stau
coannihilation region. The $\tau$s are more energetic in this case.
After putting all the measurements together,  the relic density is calculated in this scenario with an accuracy $0.108  0.01$(stat + sys) with a luminosity of 200 $fb^{-1}$~\cite{talk}.

\subsection{Over-Dense Dark Matter Region in the mSUGRA model} We
investigated  a   region of the mSUGRA parameter space where the DM
content is over-dense, but due to a modification of the Boltzmann
equation  we showed that this region can be  allowed which permitted
us to investigate a larger region of the mSUGRA model parameter
space~\cite{we1} at the LHC.
 We showed that the final states  mostly contains Z and/or Higgs and we developed techniques to extract the
 model parameters by developing observables using  the end points of  $M_{jZ/H}$ distribution~\ref{fig:hj}.
  Using these measurements, the DM content was determined with an accuracy 20\% for 500 fb$^{-1}$ of data.
\begin{figure}[htb]
\begin{center}
\includegraphics[height=7cm,width=7cm]{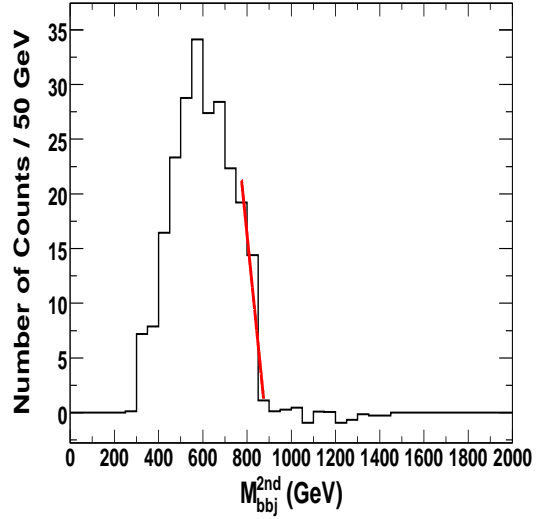}
\end{center}
\vskip -0.4cm \caption{The Higgs (tagged b jet pair) plus jet invariant mass distribution reconstructed
through PGS in a 500 fb$^{-1}$ mSUGRA sample at $m_0$ = 651 GeV, $m_{1/2}$ = 440 GeV, $tan\beta$ = 40,
$A_0$ = 0, and $\mu > 0$.}
\label{fig:hj}
\end{figure}

\subsection{Other Models}

We first  discuss a very important extension of the mSUGRA model:

{\bf Higgs nonuniversality} In these types of models, the Higgs masses are nonuniversal at the GUT scale, $m_{H_1}^2=m_0^2 (1+\delta_1)$ and  $m_{H_2}^2=m_0^2 (1+\delta_2)$, where the $\delta_i$s are nonuniversal  parameters. The constraints on the parameter space of these scenarios are discussed in the references~\cite{Higgsnonuni,Higgsnonuni1}. There can be two different types of Higgs non universality: case (1) $m_{H_u}^2=m_{H_d}^2\ne m_0^2$ at $M_{GUT}$.  In this case,
the parameter space of this one parameter extension of the mSUGRA model
is given by,
\begin{eqnarray}
NUHM1 &:& m_0,\ m_\phi ,\ m_{1/2},\ A_0,\nonumber \\ & &  \tan\beta
\ {\rm and}\ sign(\mu ).
\end{eqnarray}
The second case
is inspired by GUT models where ${H}_u$
and ${H}_d$ belong to different multiplets and $m_{H_u}^2\neq m_{H_d}^2$ at $M_{GUT}$.
The parameter space for this second case is then given by
\begin{eqnarray}
NUHM2 &:& m_0,\ m_{H_u}^2,\ m_{H_d}^2,\ m_{1/2},\ A_0,\nonumber \\
 & & \tan\beta \ {\rm and}\ sign(\mu ) .
\end{eqnarray}
The first case can have two regions of dark matter allowed: Higgsino region and A funnel. In the Higgsino region of the NUHM1 model,
charginos and neutralinos are light,
and more easily accessible to collider searches. In addition,
lengthy gluino and squark cascade decays to the various charginos
and neutralinos occur, leading to the possibility of
spectacular events at the LHC. In the $A$-funnel region, the
$A,\ H$ and $H^\pm$ Higgs bosons are lighter and appear in the final stages of cascades at the CERN LHC.
In the second case, since
$\mu$ and $m_A$ can now be used as input parameters, it is always
possible to choose values such that one lies either in the higgsino
annihilation region or in the $A$-funnel region, for
any value of $\tan\beta$, $m_0$ or $m_{1/2}$ that gives rise to a
calculable SUSY mass spectrum. In the low $\mu$ region, charginos and
neutralinos are again likely to be light, and accessible to
to the LHC searches.
If instead one is in the $A$-annihilation funnel,
then the heavier Higgs scalars may be light enough
to be produced at observable rates.
In addition, new regions are found where consistency with WMAP data is obtained
because either $\tilde u_R,\ \tilde c_R$ squarks or left- sleptons become very light.
The $\tilde u_R$ and $\tilde c_R$ co-annihilation region leads to large rates for
direct and indirect detection of neutralino dark matter. In both models the A annihilation funnel can occur for ant $\tan\beta$.
In Figs.~\ref{higgsnonunix} and~\ref{higgsnonuni2x}, the ranges of $\Omega_{\tilde\chi^0_1}h^2$ together with contours  of
$BF(b\to s\gamma )$ and $\Delta a_\mu$ in the $\mu\ vs.\ m_A$
plane for $m_0=m_{1/2}=300$ GeV is shown for the NUHM2 model.
\begin{figure}
% For example, with the option graphicx use
\includegraphics[width=0.45\textwidth,height=0.50\textwidth,angle=0]{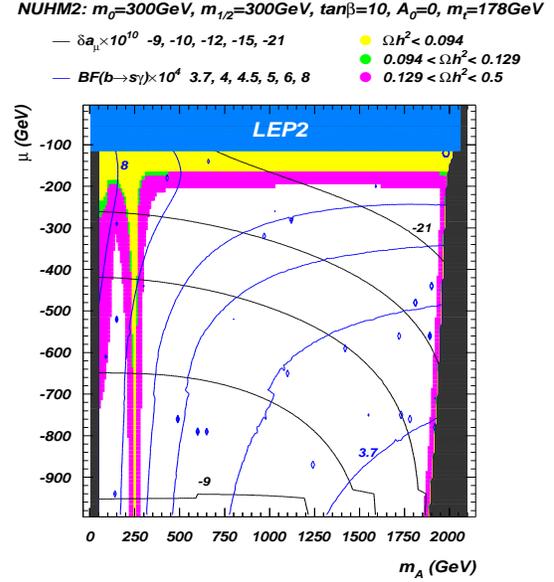}
%\epsfig{file=nuhm2_300_pmu.eps,width=7cm}
%\epsfig{file=nuhm2_300_nmu.eps,width=7cm}
%\vspace*{-0.5cm}
\caption{Ranges of $\Omega_{\tilde\chi^0_1}h^2$ together with contours  of
$BF(b\to s\gamma )$ and $\Delta a_\mu$ in the $\mu\ vs.\ m_A$
plane for $m_0=m_{1/2}=300$ GeV,
$A_0=0$, $\tan\beta =10$ and $m_t=178$ GeV  for  $\mu >0$. For very large values of
$m_A$, the stau co-annihilation region arises~\cite{Higgsnonuni1}.
}\label{higgsnonunix}
\end{figure}

\begin{figure}
% For example, with the option graphicx use
\includegraphics[width=0.45\textwidth,height=0.50\textwidth,angle=0]{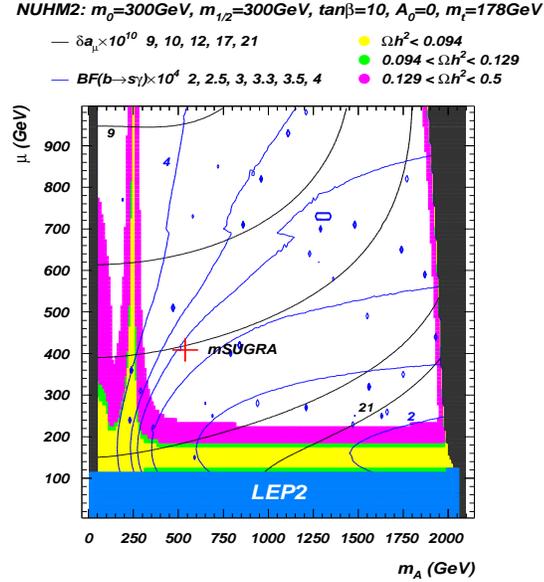}
\caption{ Same as in previous Fig.~\ref{higgsnonunix} for $\mu <0$ case~\cite{Higgsnonuni1}.
}
\label{higgsnonuni2x}
\end{figure}

It is possible to measure dark matter content accurately in these
models. The final states in these NUHM involve more W bosons in this
case (compared to the coannihillation case) which we use to
construct observables (after reconstructing the W boson). Since we
need to extract six parameters (due to two additional parameters in
the Higgs sector) we need to use  multiple end-points (and/or peak
positions) of different mass distributions. For example, the
invariant W-jet mass distribution ($M_{Wj}$) has multiple endpoints
due to decays arising from  $\tilde q\rightarrow
q\tilde\chi^{\pm}_1\rightarrow q W\tilde\chi^0_1$ or $\tilde
q\rightarrow q\tilde\chi^{\pm}_2\rightarrow q W\chi^0_2\rightarrow q
W \tau\tau\tilde\chi^0_1$. Similarly, $M_{W\tau\tau}$,
$M_{j\tau\tau}$ distributions also show multiple end-points. We are
showing various possible end-points of $M_{jW}$ in fig.\ref{Mjw}. We
are reconstructing some of the most visible endpoints in order to
determine the model parameters~\cite{talk}.
\begin{figure}
% For example, with the option graphicx use
\vspace{-2cm}
\includegraphics[width=0.65\textwidth,height=0.3\textwidth,angle=270]{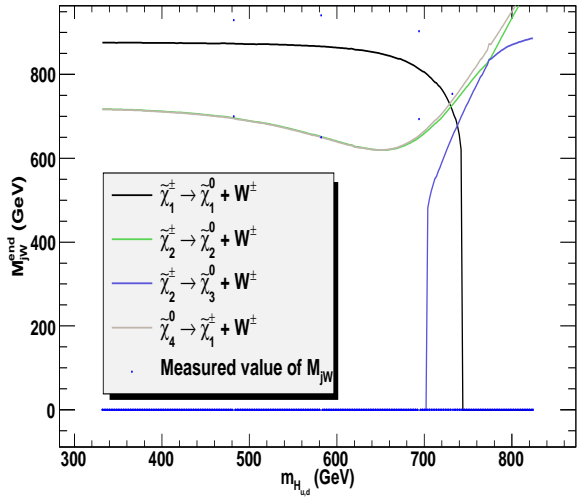}
\vspace*{-2cm}
\caption{Possible end-points of $M_{jW}$ ar shown in a NUHM model.
}
\label{Mjw}
\end{figure}

There exist many more very interesting dark matter allowed SUSY
models. We just mention a few of them below. In KKLT type moduli
mediation~\cite{kklt}, the soft masses have been calculated. The
ratio of anomaly mediation and  modular mediation is given by a
phenomenological parameter $\alpha$. The mass spectrum is different
from the mSUGRA models since the unification of the scalar masses
happen at a scale smaller than the GUT scale. Similar situation also
arise in GUT less model~\cite{gutless}. In these models, the scale
of SUSY breaking soft masses has been assumed to be smaller than the
GUT scale.

The nonminimal models (with an additional singlet) also possess interesting signatures and phenomenologies. These models can have smaller lightest Higgs mass and this Higgs can decay into a pair of psedo-scalar Higgs~\cite{nmssm}.

In Compressed MSSM, the gluino mass is small in order to have smaller $\mu$~\cite{martin}.
These models have many top quarks  in the final states at the LHC.

The flat directions $LLe$ and $udd$ within the minimal supersymmetric
Standard Model provide all the necessary ingredients for a successful
inflation with the right amplitude of the scalar density
perturbations, negligible gravity waves and the spectral tilt within
$2\sigma$ observed range $0.92 \leq n_s \leq 1.0$~\cite{flatdirection}.  Remarkably for the inflaton, which is a
combination of squarks and sleptons, there is a stau-neutralino
coannihilation region below the inflaton mass $500$~GeV for the
observed density perturbations and the tilt of the spectrum.

There also exists models where right sneutrino is a successful dark matter candidate~\cite{matchev}. Inflation can be
explained in such models in terms of flat directions which involves the interaction terms involving
neutrinos~\cite{flatdirection1}. These models have spin zero dark matter. The signal of this model
is similar to what we observe in the regular SUGRA models with neutralino being the dark matter
candidate, only difference is however in the fact that this model has a spin zero dark matter.
The probing of the spin therefore will lead to the discovery of this model.

\subsection{Conclusion} The cosmological connection of the particle
physics models can be established at the the LHC. In order to
achieve this the SUSY model parameters  need to be measured with a
great accuracy. In this review, we discussed the minimal SUGRA model
which is a well motivated minimal model of SUSY. The features of
this model which are associated with the dark matter explanation are
general, i.e., can show up in other models. In the minimal SUGRA
model, the stau neutralino coannihilation region appears
 for smaller values of sparticle massees. In this region however, there exists low energy taus. It is
 possible to measure observables with these taus with good accuracy and therefore, the relic density
 can be measured with good accuracy in this model parameter space. The gaugino masses can be measured
 with less than 10\% accuracy in the focus point region. The bulk region (which is less favored)
 can also be investigated quite precisely with a very accurate determination of the relic density.
 There are many other SUSY models with different characteristic signals. The special features of
 these models will be investigated at the LHC. The relic density can be calculated with good accuracy in the
 non-universal Higgs models based on LHC measurements. One interesting scenario is the right handed sneutrino
 being the dark matter candidate. In this case, the signal could be the same but the spin of the
 dark matter particle is different. The measurement of the spin of the missing particle will establish
 one scenario over the other.

%This work was supported in part by the DOE grant DE-FG02-95ER40917.
%I would like to thank my collaborators Richard Arnowitt, Adam Arusano, Rouzbeh Allahverdi,
%Alfredo Gurrola, Teruki Kamon, Nikolay Kolev, Abram Krislock, Anupam Mazumdar, Yukihiro Mimura  and
%Dave Toback for the works related to this review.

%%%%%%%%%%%%%% BEGIN NATH SECTION %%%%%%%%%%%%%%%%%%%%%%%%
\renewcommand{\GeV}      {~\mathrm{GeV}}
\renewcommand{\pb}      {~\mathrm{pb}}
\def\nj{$n_{jet}$}
\def\njs{$n_{jet}^*$}
\def\co{coannihilation~}
\def\pts{$P_T^{\rm * miss}$}
\def\pt{\not\!\!{P_T}}
\def\no{\nonumber\\}
\def \chan{\widetilde{\chi}}
\def \cha{\widetilde{\chi}^{\pm}_1}
\def \chb{\widetilde{\chi}^{\pm}_2}
\def \na{\widetilde{\chi}^{0}}
\def \nb{\widetilde{\chi}^{0}_2}
\def \nc{\widetilde{\chi}^{0}_3}
\def \nd{\widetilde{\chi}^{0}_4}
\def \g{\tilde{g}}
\def \ql{\widetilde{q}_L}
\def \qr{\widetilde{q}_R}
\def \dl{\widetilde{d}_L}
\def \dr{\widetilde{d}_R}
\def \ul{\widetilde{u}_L}
\def \ur{\widetilde{u}_R}
\def \ccl{\widetilde{c}_L}
\def \ccr{\widetilde{c}_R}
\def \ssl{\widetilde{s}_L}
\def \ssr{\widetilde{s}_R}
\def \ta{\widetilde{t}_1}
\def \tb{\widetilde{t}_2}
\def \ba{\widetilde{b}_1}
\def \bb{\widetilde{b}_2}
\def \sta{\widetilde{\tau}_1}
\def \stb{\widetilde{\tau}_2}
\def \smr{\widetilde{\mu}_R}
\def \ser{\widetilde{e}_R}
\def \sml{\widetilde{\mu}_L}
\def \sel{\widetilde{e}_L}
\def \slr{\widetilde{l}_R}
\def \sll{\widetilde{l}_L}
\def \snl{\widetilde{\nu}_{\tau}}
\def \snm{\widetilde{\nu}_{\mu}}
\def \sne{\widetilde{\nu}_{e}}
\def \hc{H^{\pm}}
\def \lra{\longrightarrow}
\def \ETmiss{${\not\!\!{E_T}}~$}
\def \missET{${\not\!\!{E_T}}$}
\def \mh{m_{1/2}}
\def\.4{\vspace{-.5cm}}
%%%%%%%%%%%%%%%%%%%%%%%%%%%%%%%%%%%%%%%%%%%%%%%%%%%%%%%%%%%%%%%%%%%%%%%
\def\beq{\begin{equation}}
\def\be{\begin{equation}}
\def\beqn{\begin{eqnarray}}
\def\ee{\end{equation}}
\def\eeq{\end{equation}}
\def\eeqn{\end{eqnarray}}
\newtheorem{tab}[table]{Table}
%%%%%%%%%%%%%%%%%%%%%%%%%%%%%%%%%%%%%%%%%%%%%%%%%%%%%%%%%%%%%%%%%%%%%%%
\def \non{nonuniversalities~}
\def\a{GNLSP$_{\rm A}~$}
\def\b{GNLSP$_{\rm B}~$}
\def\c{GNLSP$_{\rm C}~$}

\section{Decoding the Origin of Dark Matter with LHC Data}
% and Dark Matter Direct Detection}

{\it D.~Feldman, Z. ~Liu, P.~ Nath}\medskip

Within the SUGRA framework, it is found that LHC data will allow for
the discrimination of the two dominant branches which lead to dark
matter in the early universe arising  from stau coannihilation and
annihilation on the Hyperbolic branch. Gluino coannihilation is also
discussed. It is seen that a gluino NLSP (GNSLP) can lead to an
early discovery of SUSY at the LHC.

\subsection{Decoding Dark Matter with the LHC}
 It is now known that production of dark matter in the early universe can occur in many
 diverse ways. These include   annihilation due to
 pole enhancement~\cite{Griest:1990kh,Gondolo:1990dk,Arnowitt:1993mg}
    and  coannihilation~\cite{Griest:1990kh,Ellis:1999mm,CNADS}.
   Specific regions which lead to relic densities consistent with WMAP include the so called
 stau  coannihilation region~\cite{Ellis:1999mm,CNADS}
   (Stau-Co)  and the hyperbolic branch/focus point
 (HB/FP) region~\cite{Chan:1997bi,Feng:1999mn}.
 These regions have been studied in depth (see, e.g.,
\cite{Arnowitt:2008bz,Baer:2007ya}  for recent works) and  a
combined analysis on both regions along with their LHC signatures
has been studied in~\cite{Feldman:2007zn,Feldman:2008jy} (for a
recent review see~\cite{Feldman:2009jg}).
\begin{figure*}[t]
\includegraphics[width=9.5 cm,height=6 cm]{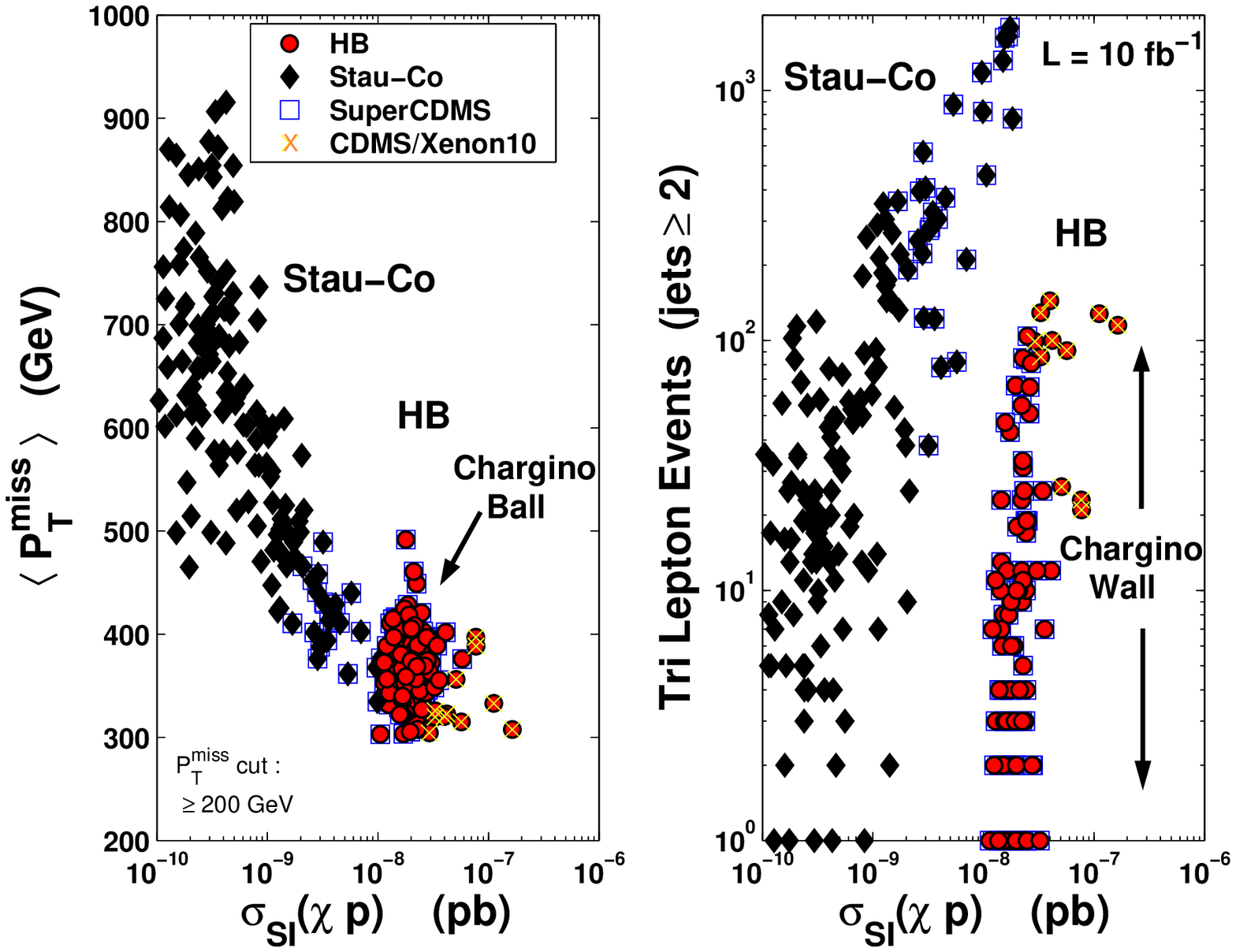}
\hspace{-.8cm}
\includegraphics[width=7.0 cm,height=5.0 cm]{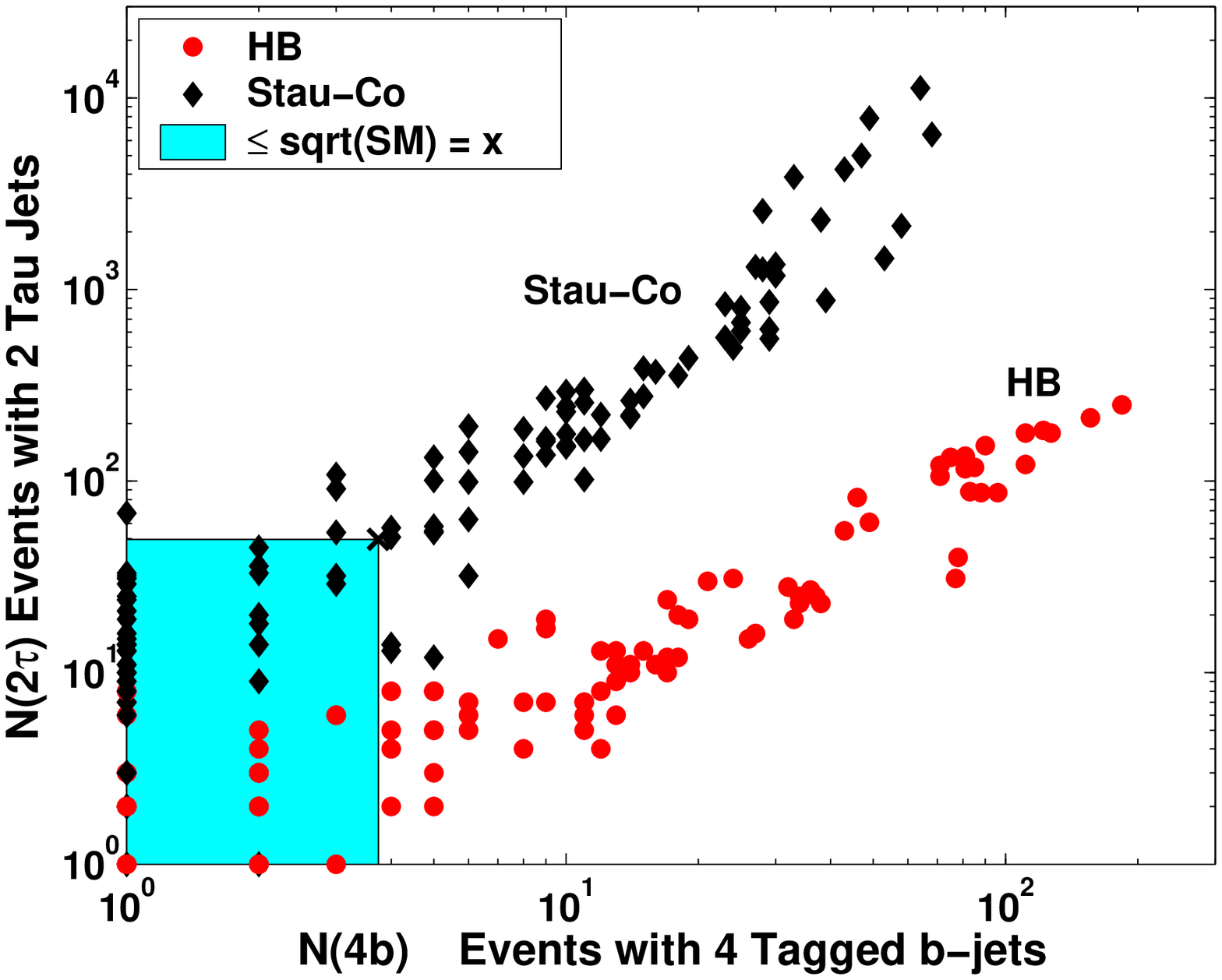}
\caption{ \small Left/Middle: Discriminating the two branches for
the production of dark matter in the early universes with LHC data
and Dark matter direct detection . The Chargino Ball and Chargino
Wall describe the clustering of model for which the chargino is the
NLSP arising on the HB/FP. The CDMS/Xe10
constraints\cite{Ahmed:2008eu} and  constraints expected from
SuperCDMS\cite{Ahmed:2008eu} are also shown. Right : The discovery
limits for model points on the Stau-Co and HB branches in the
signature space spanned
 by multi taus and multi tagged b-jets events.
The predictions  for  the models are constrained by WMAP
measurements \cite{Komatsu:2008hk} , by FCNC limits and by sparticle
mass limits \cite{Feldman:2008jy}. } \label{FLN_decode}
\end{figure*}
 It is interesting to ask if the LHC data will allow us to decipher the possible origin of dark
  matter, i.e., allow one to pin down the precise branch on which the neutralino annihilate.
On  the HB/FP the  presence of a flat region in the $\sigma_{SI}$
plane (where  $\sigma_{SI}$  is the spin-independent
neutralino-nucleon cross section) was first observed in \cite{dmdd}.
This region has sinced been dubbed the Chargino Wall
\cite{Feldman:2007fq} as the analysis of \cite{Feldman:2007fq}
uncovered the fact that this region  is entirely composed of a
chargino NLSP. Here the LSP has a sizeable higgsino component and
the cross section can be approximated in terms of the eigencontent
of the LSP as \cite{Feldman:2008jy} $ \sigma_{SI}({\rm WALL}) \sim
F_p C(h) [(g_Y \tilde B -g_2 \tilde W)({\tilde H}_2+\alpha {\tilde
H}_1)]^2 $ where $C(h)$ is a fixed by the SM up to the light CP even
higgs mass,  $\alpha$ is the CP even Higgs mixing parameter and $
F_p $ depends only on nuclear form factors (see
\cite{Chattopadhyay:1998wb} for the complete expression).

On the Wall one then gets
 $\sigma_{SI}({\rm WALL}) \sim  \mathcal{O}(10^{-8}) $[pb].
Significant  information regarding the Stau-Co and  HB/FP regions
can be obtained by correlating  LHC and dark matter direct detection
signatures. The analysis  of Fig.(\ref{FLN_decode}) illustrates the
resolving power of such a correlation showing  that the \co and the
HB/FP
 regions are well separated in the space spanned
by the trileptonic signature  3L (L=$e$,$\mu$) and $\sigma_{SI}$
where the Wall referred to above can be seen. Further one observes
that in Fig.(\ref{FLN_decode})  models arising on the HB/FP region
in the $ \langle P^{miss}_T \rangle - \sigma_{SI}$ plane are
clustered together in a ball shaped region and well separated from
points in the Stau-Co region which lie on  a slope again providing a
strong discrimination between the Stau-Co and the HB/FP regions.

The typical disparity  between the  $ P_T^{\rm miss}$ on the Stau-Co
and the HB/FP regions can be understood by analyzing the decay
chains of sparticles. Often the sparticle decays for models arising
from the Stau-Co region  involve two body decays, however,  for the
case of the HB/FP sparticles produced in $pp$ collisions have
typically a longer decay chain  which depletes the $ P_T^{\rm miss}$
in this case. Thus on the HB/FP typically the dominant production
modes are $pp \to (\g \g /  \nb \cha / {\widetilde{\chi}^{\pm}_1}
{\widetilde{\chi}^{\mp}_1}$) while squark production is highly
suppressed since the gluino mass is  suppressed relative to the
heavy squarks. The dominantly produced $\tilde g$ decay via the 3
body modes
 ${\rm Br}[\g \to {\widetilde\chi}^{0}_{i} + q + \bar q]$ and ${\rm Br}[\g \to {\widetilde\chi}^{\pm}_{j} + q + \bar q']$.
followed by
 ${\rm Br}[\nb \to {\widetilde\chi}^{0}_{1} + f + \bar f]$
and
 ${\rm Br}[{\widetilde\chi_1}^{\pm} \to {\na} +  f + \bar f']$.
 Thus the decay chain for sparticles produced on the HB/FP tend to be long and moreover
successive three body decay result in reduced $P_T^{miss}$. On the
Stau-Co  mixed squark gluino production and squark sqaurk production
$(\g {\widetilde{q}},{\widetilde{q}}{\widetilde{q}})$ typically
dominate while  $\g\g$ production is relatively suppressed. The
decay chains here can be short, for example,  ${\rm Br}[\qr \to \na
+ q]\sim 100\%$ and large branchings into ${\rm Br}[\ql \to
(\nb,\cha) + (q,q')]$ with subsequent 2 body decays giving
$P_T^{miss}$ from the LSP and neutrinos. Further, the on-shell decay
of the gluino into the squark + quark is open which doubles up the
above results. Since the  decay chain for sparticles  on the Stau-Co
can be short,  proceeding via 2 body decays with large branching
fractions into the gauginos, the resulting $P_T^{miss}$ is less
depleted and can get quite large. The right most panel of
Fig.(\ref{FLN_decode}) shows the discovery prospects of the HB/FP
and of the Stau-Co at the LHC in the b-jet - tau-jet plane.  Here
one observes a clean separation of the signatures of the HB/FP.
 The richness
of b-jets on the HB/FP is governed by the fact that the 3 body
decays are dominated by $b\bar b \na$ while a good amount of b-jets
are also possible on the Stau-Co since the SUSY scale here can be
rather light and the total number of events passing the triggers is
typically larger. In this analysis triggers were designed based on
CMS trigger tables \cite{Ball:2007zza}. ATLAS triggers have recently
been updated and are given in \cite{Aad:2009wy}.
\begin{figure*}[t]
  \begin{center}
\includegraphics[width=5.85cm,height=4.5cm]{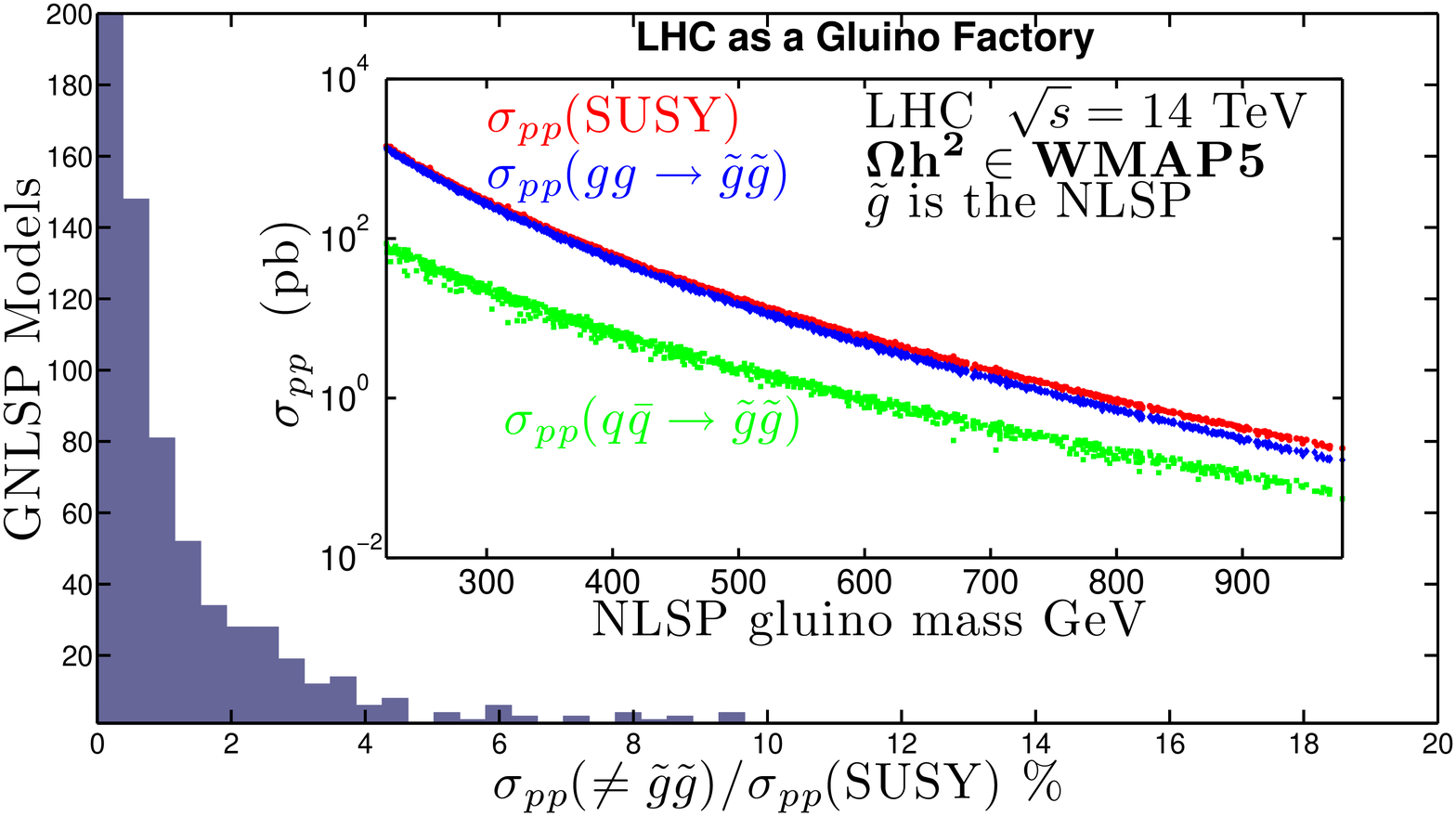}
\includegraphics[width=5.85cm,height=4.5cm]{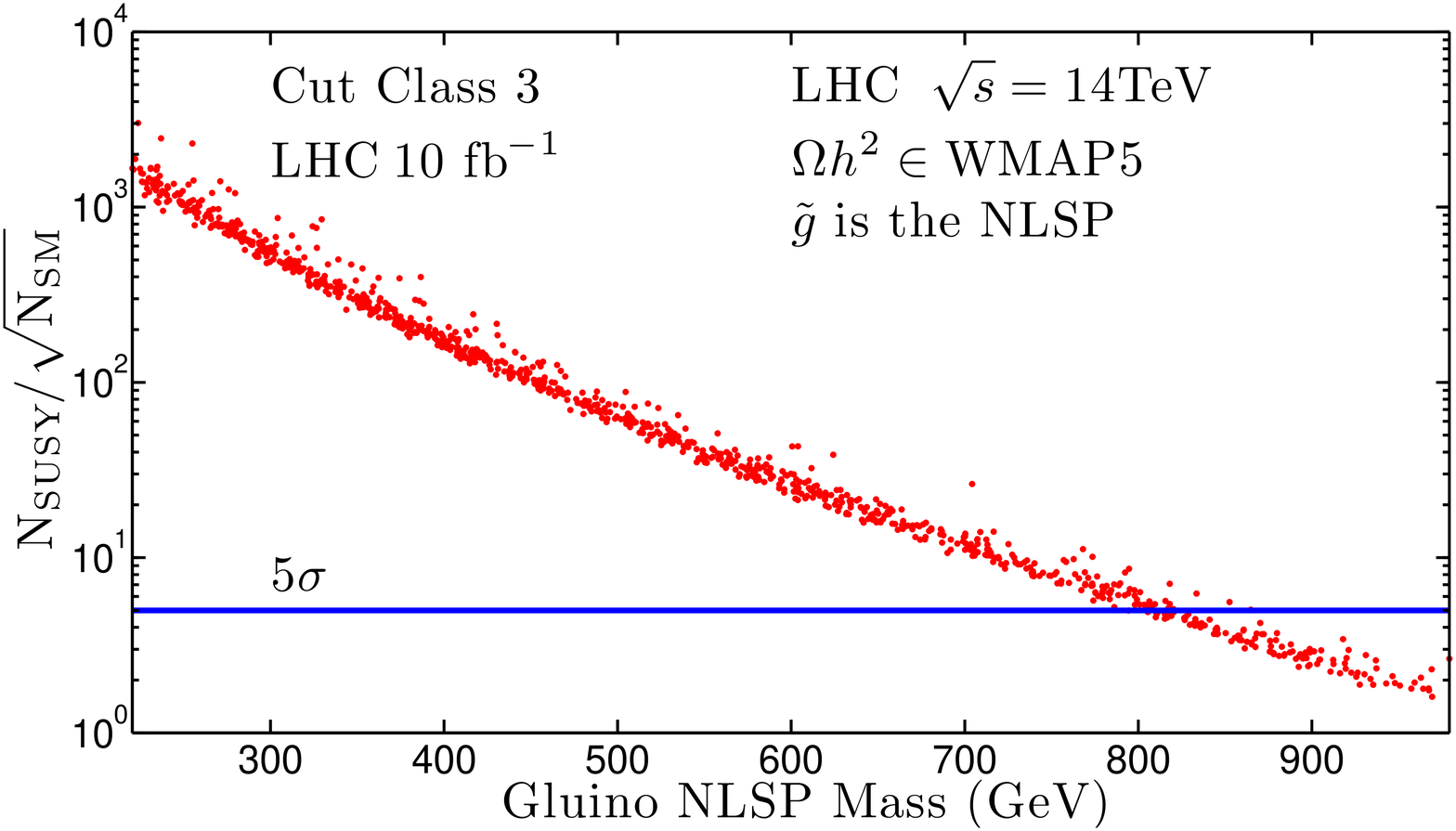}
\includegraphics[width=5.85cm,height=4.5cm]{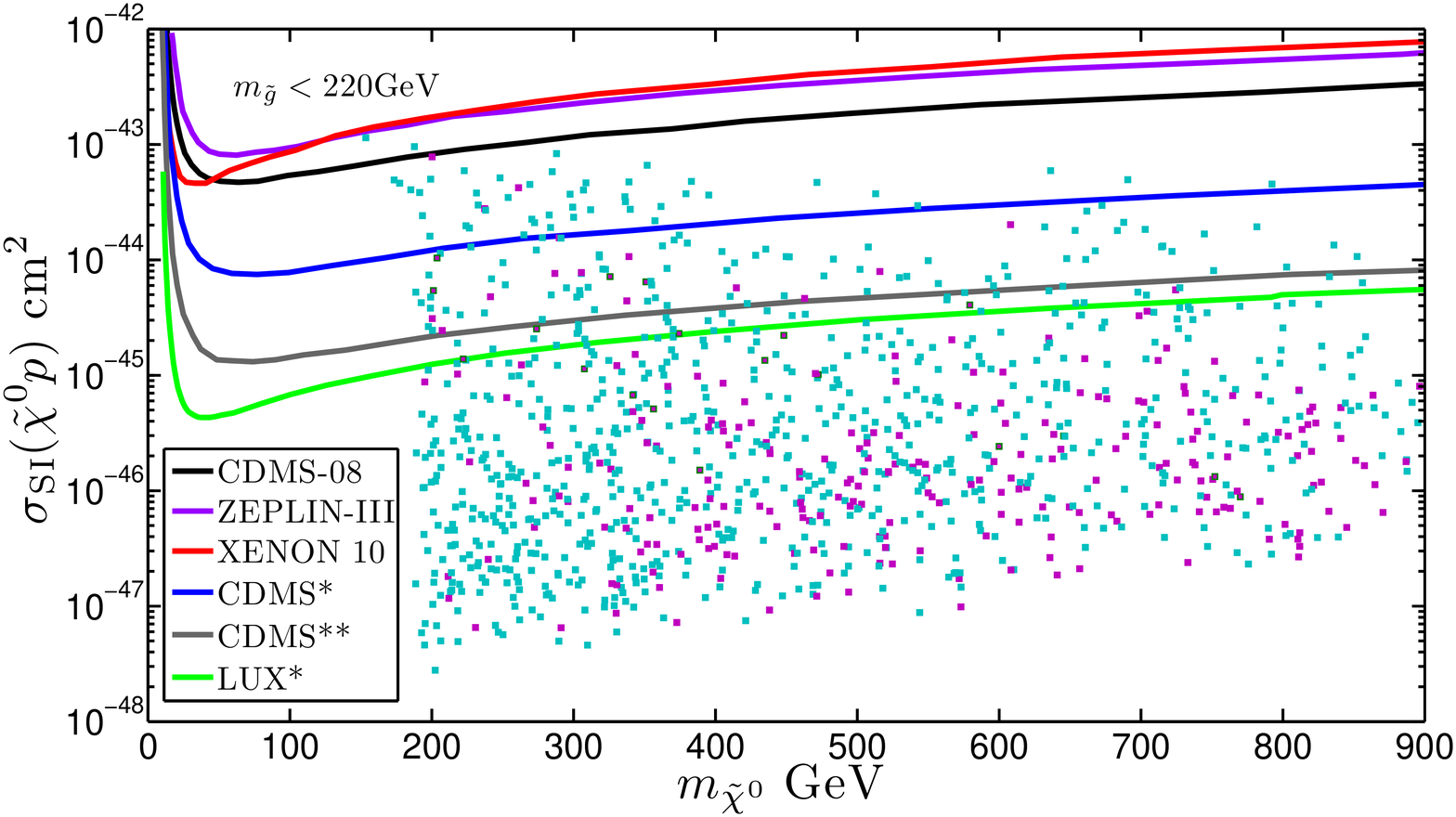}
\includegraphics[width=7.4cm,height=5.1cm]{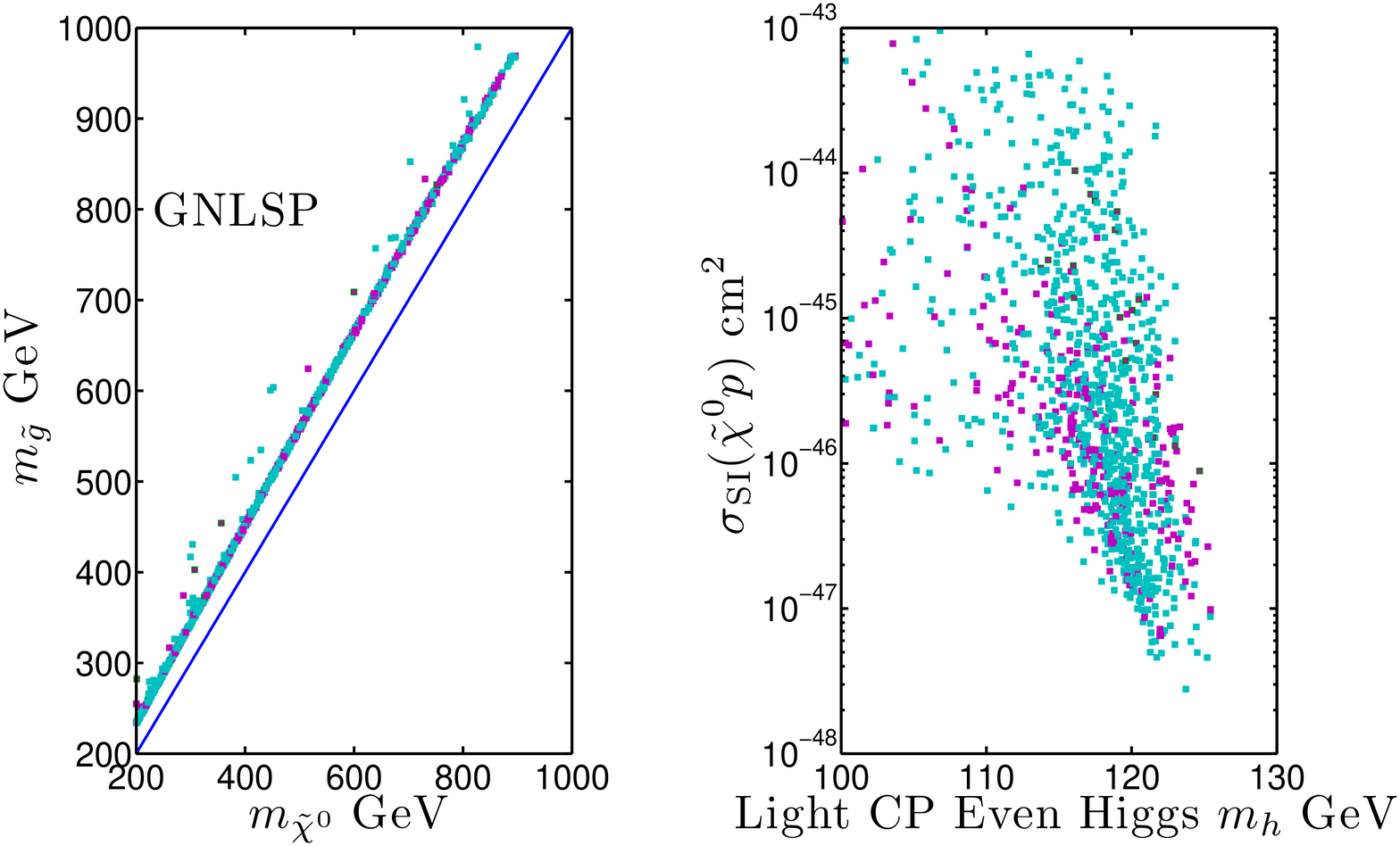}
\caption{\small Top Left:  In the GNLSP model the LHC will turn into
a gluino factory. Top Right: Discovery limit at the LHC in the total
number of SUSY events; the GNLSP can be tested in early runs. Bottom
Left:  $\sigma_{SI}$  vs  $\na$ ; GNLSP models with an LSP with a
significant Higgsino component are detectable  in the direct
detection experiments:  NUSP13 (light blue), NUSP14  dark (magenta).
Bottom Middle: Exhibition of the explicit  scaling relation between
$m_{\na}$ and $m_{\g}$ for the GNLSP models. Bottom Right:
 $\sigma_{SI}$  vs
 the light Higgs boson mass; much of NUSP13 has the light Higgs boson mass
 near 120 GeV. Taken from   \cite{Feldman:2009zc}.
 }
\label{GF}
  \end{center}
\end{figure*}

\subsection{Light Gluinos in SUGRA GUTS and discovery at the LHC} In
SUGRA GUT models, there is in fact a class of models where the
gluino is the NLSP (GNLSP), which arise from a specific set of
hierarchical mass patterns \cite{Feldman:2007zn} ($\equiv$ NUSP =
non universal SUGRA patterns \cite{Feldman:2007fq,Feldman:2009zc})
\beqn {\rm NUSP13}  &:&   \na   < \g    < \cha \lesssim \nb,
\nonumber \\\nonumber {\rm NUSP14}  &:&   \na   < \g    < \ta <
\cha,       \nonumber  \\\nonumber {\rm NUSP15}  &:&  \na   < \g <
A\sim H       ~~(\rm rare~pattern).   \nonumber   \\\nonumber \eeqn
Such models can arise when there is F type breaking of the GUT
symmetry in $SU(5), SO(10)$, and $E_6$ models where the breaking
proceeds with two irreducible representations, namely with a singlet
and a non-singlet F term.
  In this case an  interesting phenomenon arises in that  models with the same
$r\equiv (M_2-M_1)/(M_3-M_1)$ (where $M_i$ are the gaugino masses at
the GUT scale)
  can be made isomorphic under redefinitions and scalings in the gaugino sector.
Therefore, in essence,  models
 with the same value of
 $r$ would in fact be equivalent, or phenomenologically indistinguishable, when taken  in a linear combination of breakings including singlets.
Examples of these isomorphisms are given in  \cite{Feldman:2009zc}
along with generalized sum rules on the gaugino masses and specific
benchmark models.

\begin{figure}[htbp]
  \begin{center}
  \includegraphics[width=5.5cm,height=4.3cm]{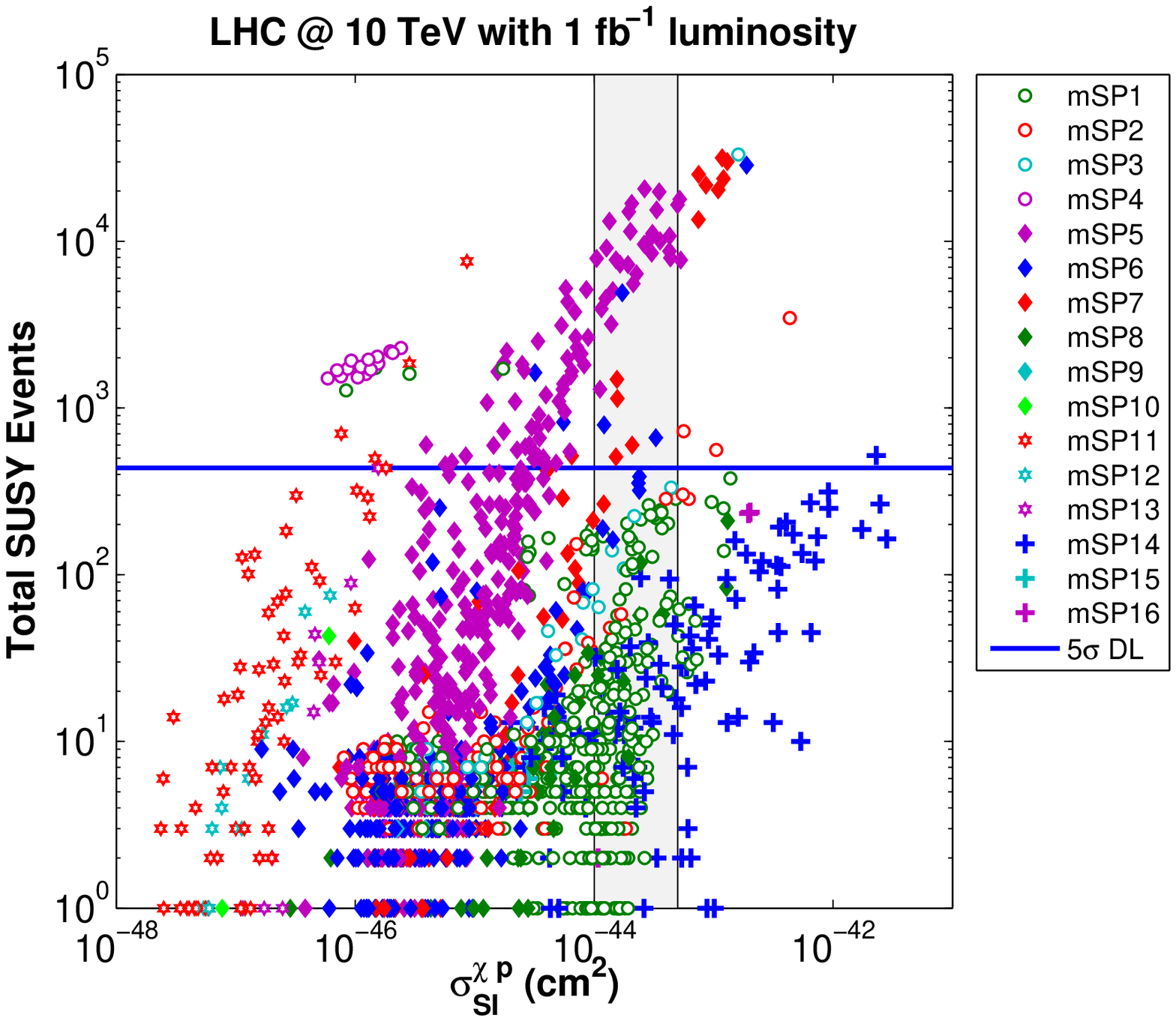}
 \includegraphics[width=5.5cm,height=4.3cm]{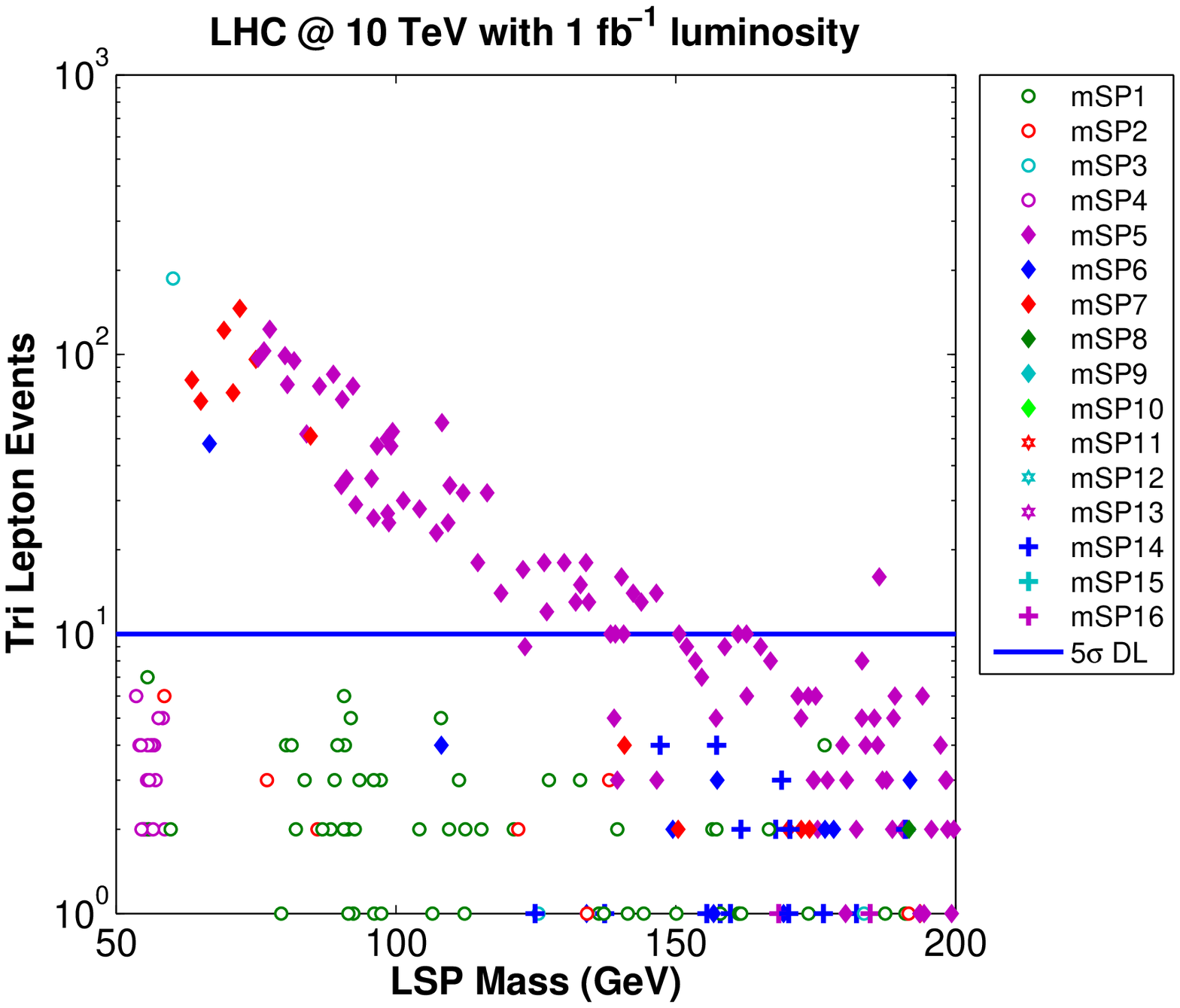}
\caption{(Color Online)
The top panel gives he total number of SUSY events at
10 TeV with 1 fb$^{-1}$ luminosity at the LHC vs the
spin independent neutralino-proton cross section
$\sigma^{\tilde \chi p}_{\rm SI}$. The
 region $\sigma^{\tilde \chi p}_{\rm SI}=(1-5)\times 10^{-44}$ cm$^2$
 is the shaded area.
%Middle panel: The total number of SUSY events vs the LSP mass.
The bottom panel gives the number of trileptonic events \cite{trilep} vs the LSP mass
with the analysis done also at 10 TeV with 1 fb$^{-1}$ luminosity at the LHC.
The horizontal lines in each case are  the $5\sigma$ discovery reach.
To suppress the background  events that have
$\not\!\!{P_T}>200$ GeV and contain at least 2 jets with $P_T>60$ GeV are selected.
 Taken from \cite{Feldman:2009pn}. }
\label{figcdms2a}
  \end{center}
\end{figure}

 %%%%%%%%%%%%%%%%%%%%%%%%%%%%%%%%%%%%%%%%%%%%%%%%%%%%%%
A feature of interest in the GNLSP class of models is that the relic
density of the LSP is controlled by gluino coannihilations, and one
has that effective cross section which enters into $\langle
\sigma_{\rm eff} v \rangle$ can be approximated by
\cite{Feldman:2009zc} $ \sigma_{\rm eff}\simeq \sigma_{\tilde g
\tilde g}\gamma^2_{\na} \left(\gamma^2 + 2\gamma
\frac{\sigma_{\na\tilde g}}{\sigma_{\tilde g \tilde g}} +
\frac{\sigma_{\na\na}}{\sigma_{\tilde g \tilde g} }\right)~,
\label{relic2} $ where $\gamma =\gamma_{\tilde g}/\gamma_{\na}$ and
where $\gamma_i$ are the Boltzmann suppression factors. While many
of the models arise from a bino like LSP \cite{Profumo} one also
finds  models with a sizeable Higgsino component
\cite{Feldman:2009zc} which has important implications for the
direct detection in dark matter experiments of the GNLSP models. One
also finds, that  in the  GNLSP models, there is a  2nd generation
slepton-squark degeneracy and in some cases an inversion where these
sleptons  are heavier than the squarks \cite{Feldman:2009zc}.
%%%%%%%%%%
\begin{figure}[htbp]
  \begin{center}
  \includegraphics[width=5.5cm,height=4.3cm]{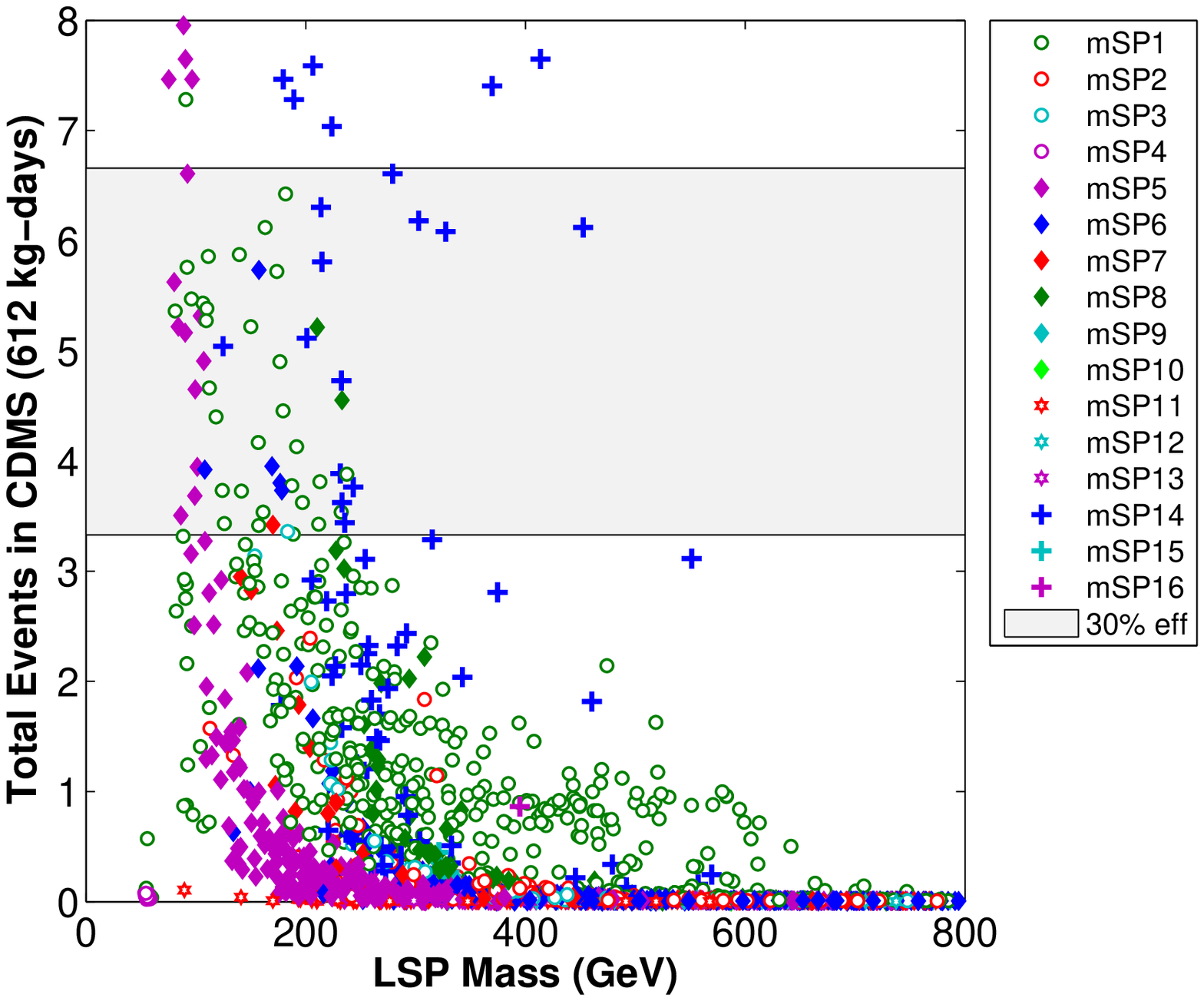}
    \includegraphics[width=5.5cm,height=4.3cm]{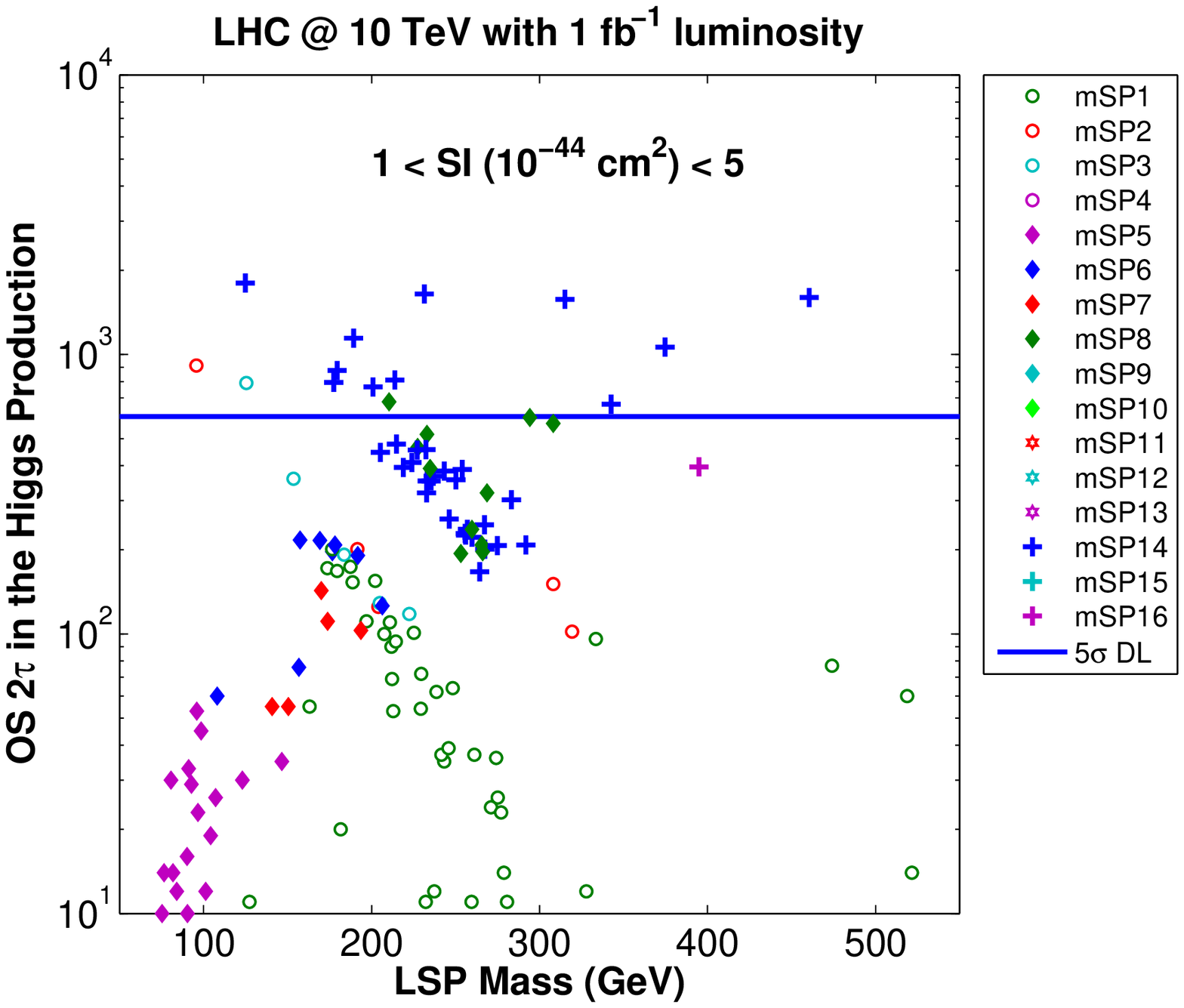}
      \caption{(Color Online) Top panel exhibits  the number of events predicted in the CDMS detector
 with 612 kg-d of data assuming 100\% efficiency.
  The assumption that one (both) events in the CDMS detector are signal events
 gives the lower (upper) horizontal lines where  in drawing the
 lines a 30\% detector efficiency is assumed.
 The Bottom panel gives the Opposite sign (OS) two tau signal. This signal arises from Higgs production
 and in the analysis only trigger level cuts are assumed.
 The $5\sigma$ discovery limit is indicated by the
       horizontal line. Taken from \cite{Feldman:2009pn}. }
\label{figcdms2b}
  \end{center}
\end{figure}
%%%%%%%%%%
The LHC signals of the GNLSP models will dominantly come from large
amounts of multi-jets and thus the SM backgrounds can generally be
reduced by cutting on the azimuthaseparation between the hardest
jets $\Delta\phi ({\rm jet}_1, {\rm jet}_2)$ to suppress the QCD
background due to light quark flavours and $b \bar b$ as well as $t
\bar t$. A rejection of isolated $e/\mu$ from the background $W/Z$
leptonic decays significantly enhances the GNLSP signals over the SM
background. Specific cuts are given in \cite{Feldman:2009zc}.

If the GNLSP class of models are realized, the LHC will turn into a
gluino factory. This can be seen in Fig.(\ref{GF}). The production
cross sections
 are overwhelmingly governed by the gluino production, much as the relic density
is dominated by gluino annihilations. The SUSY model becomes
exceptionally predictive in that the colored sector determines the
LHC signal of multi-jets with the mass splitting of the LSP and the
GNSLP controlling the 3 body gluino decays as well as
 the opening of the radiative decay of the gluino, while the relic density is determined by the inverse
process. Note that in Fig.(\ref{GF}) every model point satisfies the
two sided WMAP relic density constraints. The GNLSP covers the
entire range from $ \sim $ 220 GeV to almost a TeV over the
parameter space investigated and with just 10/fb a GNLSP can be
discovered up to about 800 GeV at $\sqrt s = 14 ~ \rm TeV$
\cite{Feldman:2009zc}.

An analysis of the spin independent  cross sections in dark matter
experiments is given in Fig.(\ref{GF}). It is seen that there are a
class of GNLSP models with sizeable Higgsino components which are
beginning to be constrained by the CDMS and Xenon limits and these
models also have large LHC  $\tilde g\tilde g$ production cross
sections and will be easily visible at the LHC. Thus if a light
gluino is indeed indicated early on  at the LHC, it may also provide
a  hint of the size of the dark matter  signal in the direct
detection of dark matter.
%%%%%%%%%%%
%\begin{figure}[htbp]
%  \begin{center}
%  \includegraphics[width=5.5cm,height=4.3cm]{dark/events.eps}
%    \includegraphics[width=5.5cm,height=4.3cm]{dark/WIMP_higgs_10TeV.eps}
%      \caption{(Color Online) Top panel exhibits  the number of events predicted in the CDMS detector
% with 612 kg-d of data assuming 100\% efficiency.
%  The assumption that one (both) events in the CDMS detector are signal events
% gives the lower (upper) horizontal lines where  in drawing the
% lines a 30\% detector efficiency is assumed.
% The Bottom panel gives the Opposite sign (OS) two tau signal. This signal arises from Higgs production
% and in the analysis only trigger level cuts are assumed.
% The $5\sigma$ discovery limit is indicated by the
%       horizontal line. Taken from \cite{Feldman:2009pn}. }
%\label{figcdms2b}
%  \end{center}
%\end{figure}
%%%%%%%%%%%
%%%%%%%%%%%%%%%%%%%% BEGIN ADDITION   %%%%%%%%%%%%%%%%%%%%%%%%%
%%%%%%%%%%%%%%%%%%%%%%%%%%%%%%%%%%%%%%%%%%%%%%%%%%%%%%%
\subsection{CDMS II and LHC}
   Very recently  the CDMS II  \cite{cdmsnew} has announced a result on the spin indpendent neutralino-proton
   cross section $\sigma_{SI}^{\tilde \chi p}$ with a new upper limit of $3.8\times 10^{-44}$ cm$^2$.
   There is also the tentalizing possibility that the CDMS II may have seen actually one or two
   events in their detector. We investigate the implications of these results for the possible
   observation of sparticles at the LHC.  As already discussed there exists a strong connection between
   experiments for the direct detection of dark matter and new physics at the LHC
   (For additional references see
   \cite{Drees:2000he,Feng:2005gj,Arnowitt:2007nt,Kane:2007pp,Altunkaynak:2008ry,Baer:2008uu,Bottino:2008xc,Feldman:2009wv}).  Several papers have analyzed the implications of the new
  CDMS II data for supersymmetry\cite{Feldman:2009pn}\cite{Kadastik:2009gx}\cite{Bottino:2009km}
  \cite{Ibe:2009pq}\cite{Allahverdi:2009sb}\cite{Holmes:2009uu}.
   Here we discuss one such analysis \cite{Feldman:2009pn}.
   The top panel of Fig.(\ref{figcdms2a})
   gives an analysis of the total number of SUSY events vs the spin independent cross section
   $\sigma_{SI}^{\tilde \chi p}$.  Plotted are  the mSUGRA model points which pass the
   REWSB constraint,  the relic density constraint, and other experimental constraints including those
   from LEP and from the Tevatron. The various points are indicated by the corresponding
   minimal supergravity patterns labeled by mSP1-mSP16.
    One finds that the allowed patterns with cross sections around
   $3.8\times 10^{-44}$ cm$^2$ are the Chargino Patterns (mSP1-mSP4),  the Stau Patterns (mSP5-mSP10),
   and the Higgs Patterns (mSP14-mSP16). Specifically the stop patterns do not produce spin independent
   cross sections of this size.
    The bottom panel of Fig.(\ref{figcdms2a})
   gives an analysis of the total number of trileptonic\cite{trilep}  events vs the LSP mass for the same
   sample of points as in the top panel. To reduce the background  detector cuts so that
   events that have
$\not\!\!{P_T}>200$ GeV and contain at least 2 jets with $P_T>60$ GeV are imposed.

     We turn now to the possibility that one or two events in the CDMS II detector may be
      dark matter events. The CDMS II experiment  after quality cuts had an exposure of
     612 kg-days. The top panel  Fig.(\ref{figcdms2b}) gives an analysis of model points that
     produce one or two CDMS events
     where we have assumed 30\% efficiency. It is seen that the points that survive
     are either Stau Patterns, Chargino Patterns, or Higgs Patterns.
   This means that the NLSP that should be seen at the LHC would either be a
  Stau, a Chargino, or a CP odd/CP even (A/H) Higgs.
     The dark matter  experiments are continuing to improve their sensitivity. Thus by the summer of
     2010 CDMS will have three times more Germanium in their detector. Also the Xenon experiment
     is running and accumulating data. Thus in the near future we can expect to see a sensitivity of
     $1\times 10^{-44}$ cm$^2$ for the spin independent cross section. It is then reasonable to
     explore the parameter points that give a spin independent cross  section in the range
     $(1-5)\times 10^{-44}$ cm$^2$ and investigate their signatures at the LHC.
     An analysis of this type is given in  the top panel of Fig.(\ref{figcdms2a}) (see the shaded region)
     and in the
     bottom panel of Fig.(\ref{figcdms2b}) where
     the number of  opposite  sign taus in the Higgs production vs the LSP mass is
     exhibited at $\sqrt s=10$ TeV with 1 fb$^{-1}$ of integrated luminosity. Thus one finds
     that a significant number of parameter points can be explored even with as little as
      1 fb$^{-1}$ of integrated luminosity in this region of the parameter space.

\section{Lifting LHC Degeneracies Using Dark Matter Observations}
{\it B.~Altunkaynak, M. ~Holmes, B. D. ~Nelson}\medskip
\subsection{Introduction}
Once LHC data taking is underway and evidence of BSM physics is
established the arduous task of reconstructing an underlying
theoretical framework will begin.  Under the assumption that the BSM
physics is SUSY, in \cite{ArkaniHamed:2005px} the authors point out
that even within a reduced 15 dimensional parameter space of the
MSSM many possible candidate models may give rise to
indistinguishable signatures at the LHC.  Sets of parameters may
have many pairs of ``degenerate twins'' which give similar fits to
the data and how to differentiate these degenerate models is the LHC
inverse problem.
\par
\par   Using degenerate pairs of model points from \cite{ArkaniHamed:2005px}
the ability of ILC data \cite{Berger:2007yu} and dark matter
observations \cite{Altunkaynak:2008ry} to lift the degeneracies were
shown to be beneficial in lifting the degeneracies.  % and are  complimentary to the LHC data.
Furthermore it has been shown that the inverse problem can be highly
reduced by combining LHC data with other observations from
astrophysical, collider and low energy measurements
\cite{Balazs:2009it}.  Combining measurements from other arenas
provides further model constraints which are complimentary to LHC
data and thus may resolve model degeneracies.   Here we outline the
utility of dark matter observables to lift LHC degeneracies
following \cite{Altunkaynak:2008ry}. That dark matter observables
can help lift LHC degeneracies is of no surprise as the signals are
sensitive to the make-up of the
LSP~\cite{BirkedalHansen:2001is,BirkedalHansen:2002am}, of which LHC
signals are much less so.
\subsection{Degenerate Pairs}
\par The authors of \cite{ArkaniHamed:2005px} considered MSSM models at the electroweak scale which were defined by the set of 15 SUSY parameters
\beqn
\left\{\begin{array}{c} \tan\beta,\,\,\mu,\,\,M_1,\,\,M_2,\,\,M_3 \\
m_{Q_{1,2}},\,m_{U_{1,2}},\,m_{D_{1,2}},\,m_{L_{1,2}},\,m_{E_{1,2}} \\
m_{Q_3},\,m_{U_3},\,m_{D_3},\,m_{L_3},\,m_{E_3} \end{array} \right\}
\, , \label{paramset} \eeqn
while holding fixed $m_A=850$ GeV, $A=800$ GeV for third generation
squarks and $A=0$ GeV for all others.  Over 43,000 parameter sets as
in Eqn. (\ref{paramset}) were chosen at random and for each set 10
fb$^{-1}$ of LHC data was generated using \texttt{PYTHIA}
\cite{Sjostrand:2006za} and \texttt{PGS} \cite{PGS}.  Common initial
cuts were applied to the data set to reduce Standard Model
backgrounds and then two classes of signatures were investigated for
the models.  These were basic counting type signatures for many
combinations of final state topologies as well as key kinematic
distributions of final state decay products.  The shapes of the
kinematic distributions were parametrized in such a way so to
include both classes of signatures in a $\chi^2$-like variable.  In
total 1808 signatures were considered which defined a SUSY model in
signature space at the LHC.  Using a metric in the signature space
of the models a method for deeming two distinct models as being
degenerate or not was constructed.  It was determined that of the
over 43,000 models considered 283 pairs of models failed to be
distinguished using the 1808 signatures.  Some of the models were
degenerate with more than one other set of parameters and a total of
384 models make up the 283 degenerate pairs.  These degenerate pairs
are used in \cite{Altunkaynak:2008ry} to determine how dark matter
observations can help lift degeneracies.
\par Using \texttt{DarkSUSY} \cite{Gondolo:2004sc} the degenerate pairs are initially classified according to the thermal relic abundance.
The prediction for the thermal relic abundance $\Omega_{\chi}{\rm
h}^2$, as computed by {\tt DarkSUSY} \cite{Gondolo:2004sc}, is
displayed as a function of the LSP mass $m_{\chi}$ and is shown in
Fig. (\ref{fig:Omegavsmass}). The $2\sigma$ band in this quantity
favored by the WMAP three-year data set, $0.0855 < \Omega_{\chi}{\rm
h}^2 < 0.1189$, is indicated by the solid horizontal lines. Although
few models agree with WMAP data, there is no doubt that models could
be found to give similar signatures at the LHC and give reasonable
values of $\Omega_{\chi}{\rm h}^2$.  When calculating observable
quantities that depend on the relic neutralino number density
$n_{\chi}$ present in our galaxy (or the energy density $\rho_{\chi}
= m_{\chi} n_{\chi}$) the assumed local density of $(\rho_{\chi})_0
= 0.3$~GeV/cm$^3$ will be rescaled by the multiplicative factor
$r_{\chi} = {\rm Min}(1,\, \Omega_{\chi}{\rm h}^2/0.025)$. The
shaded region bounded by the dashed line in Fig.
(\ref{fig:Omegavsmass}) represents the set of models for which the
local number density of neutralinos should be rescaled by the
multiplicative factor $r_{\chi}$.

%=============== Relic density versus mass ===================
\begin{figure}[t]
\begin{center}
\includegraphics[scale=0.5,angle=0]{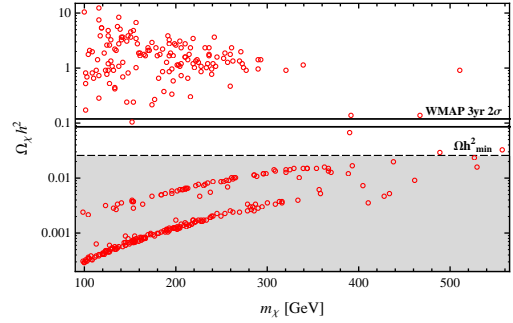}
\caption{\footnotesize Thermal relic abundance of neutralino~LSP for
378~models from~\cite{ArkaniHamed:2005px}. } \label{fig:Omegavsmass}
\end{center}
\end{figure}
%===================================================================
%\par  Models were also further classified by constraints on quantities such as $m_h$, ${\rm Br}(B \to X_s \gamma)$ and $\frac{g_{\mu} -2}{2}$.  A few models were also removed as they give a $\widetilde{\tau}$ LSP when the spectrum is calculated using {\tt DarkSUSY}.  Various subsets of the 378 models considered are also studied and more information can be found in \cite{Altunkaynak:2008ry}.
%
\subsection{Direct Detection Experiments}
\par It is assumed that LHC data has resulted in more than one set of parameters of the form of (\ref{paramset}) which describe the data equally well.  These parameter sets then allow one to predict the resulting dark matter signatures.  For this the focus will be on direct detection of relic neutralinos via their scattering from target nuclei.  These scattering events are signaled by detection of nuclear recoil of elastic scatters or by detecting ionization of target nucleus for inelastic scattering.
\par Two types of detector targets are considered, xenon and germanium, for the simplicity and reliability of background estimations.  Table (\ref{experiments}) lists some relevant experiments along with the target type.  The first three experiments listed (CDMS~II, XENON10 and ZEPLIN~II) have reported limits on neutralino-nucleus interaction rates and the reported fiducial mass is used.  The other experiments listed are planned for the future with the nominal masses given in the table and SCDMS stands for SuperCDMS. Using measured background rates in current experiments one may extrapolate to large scale experiments.  The reach and resolving power of multiple experiments are presented by using exposure time in germanium and xenon targets.
%---------------------- Summary Table -------------------------
\begin{table}[t]
\begin{footnotesize}
\begin{center}
\begin{tabular}{|c|l||c|c|l|} \hline
Ref. & Experiment & Target & Mass (kg) \\% & Detected Object(s) \\
\hline
\cite{Ahmed:2008eu} & CDMS~II & Ge & 3.75 \\% & athermal phonons, ionization charge\\
\cite{Angle:2007uj} & XENON10 & Xe & 5.4 \\% & scintillation photons, ionization charge\\
\cite{Alner:2007ja} & ZEPLIN~II  & Xe & 7.2 \\% & scintillation photons, ionization charge \\ \hline \hline
\hline
\cite{Akerib:2006rr} & SCDMS (Soudan)  & Ge & 7.5 \\% & see CDMS~II\\
\cite{Akerib:2006rr} & SCDMS (SNOlab) & Ge & 27 \\% & see CDMS~II\\
\cite{Akerib:2006rr} & SCDMS (DUSEL) & Ge & 1140 \\% & see CDMS~II\\
\cite{Aprile:2004ey} & XENON100  & Xe & 170 \\% & see XENON10 \\
\cite{Aprile:2004ey} & XENON1T  & Xe & 1000 \\% & see XENON10 \\
\cite{LUX} & LUX  & Xe & 350 \\% & scintillation photons, ionization charge \\
\hline
\end{tabular}
\end{center}
{\caption{\label{experiments}\footnotesize  Direct detection
experiments considered. From \cite{Altunkaynak:2008ry} where more
experiments are listed.}}
\end{footnotesize}
\end{table}
%------------------------- END OF THE TABLE ---------------------
\par  To compute the interaction rate of relic neutralinos with the nuclei of the target material one
considers both spin-dependent (SD) and spin-independent (SI)
interactions. For target nuclei with large atomic numbers the SI
interaction, which is coherent across all of the nucleons in the
nucleus, tends to dominate.  This is true of xenon and, to a
slightly lesser extent, germanium as well.  The SI
cross section $\sigma^{\mathrm{SI}}$ is computed on an arbitrary nuclear target via%~\cite{Gondolo:2004sc}
\begin{equation}\label{ds30}
    \sigma^{\mathrm{SI}}_{\chi i}=\frac{\mu^2_{i\chi}}{\pi}\big|
    ZG^p_s+(A-Z)G^n_s \big|^2\, ,
\end{equation}
where $i$ labels the nuclear species in the detector with nuclear
mass $M_i$, $\mu_{i\chi}$ is the reduced mass of the
nucleus/neutralino system $\mu_{i\chi}=m_{\chi} M_i/(m_\chi + M_i)$,
and $A$ and $Z$ are the target nucleus mass number and atomic
number, respectively. The quantities $G^p_s$ and $G^n_s$ represent
scalar four-fermion couplings of the neutralino to point-like
protons and neutrons. They can be described schematically as
$G^N_s=\sum_{q}\langle N|\bar q q | N\rangle \times A$, where $A$ is
calculable given a SUSY model.  The initial nuclear matrix elements
are as of yet not calculable from first principles and the values
are inferred from pion-nucleon scattering data.  There are
potentially large uncertainties in these parameters (specifically
the $\pi\,N$ $\Sigma$-term) which can result in large uncertainties
in the resulting dark matter cross sections \cite{Ellis:2008hf}.  In
the following default values are used and in the future one hopes
these parameters will be better understood.
\begin{figure}[t]
\begin{center}
\includegraphics[scale=0.5]{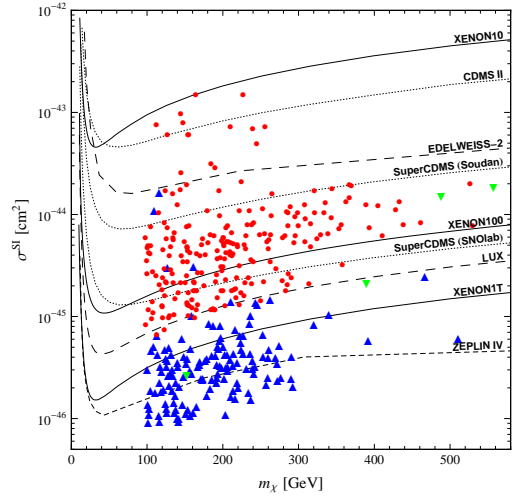}
\caption{\footnotesize Spin independent neutralino-proton
interaction cross-section as a function of $m_{\chi}$ for the
378~models.  From \cite{Altunkaynak:2008ry}. }
 \label{cross_sec}
  \end{center}
\end{figure}
\par In Fig. (\ref{cross_sec}) the spin independent neutralino-proton cross sections are shown.  The 378~models (6 of original 384 removed due to $\widetilde{\tau}$ LSP) are divided into three groupings: those
with $\Omega_{\chi}{\rm h}^2 > 0.1189$ (darker filled triangles), $
0.025 < \Omega_{\chi}{\rm h}^2 < 0.1189$ (lighter inverted
triangles) and $\Omega_{\chi}{\rm h}^2 < 0.025$ (filled circles).
Sensitivity curves for several of the experiments in
Table~\ref{experiments} are also shown.  Based on the figure it
seems that some of the models ought to have given a measurable
signal.  However, experiments don't measure directly the cross
sections, rather they measure counting rates.  The two are related,
although one must make assumptions about the local halo density as
well as the velocity distribution of the relic neutralinos.
Furthermore if one rescales the number density by $r_\chi$, then
none of the models ought to have given signals at experiments.  This
demonstrates why it is important to work with count rates for which
it is also necessary to understand background rates at experiments.
\par To calculate the rates at experiments one starts by considering the differential rate per nuclear recoil energy
 \beqn\label{ds29}
\frac{dR}{dE}=\sum_i c_i \frac{\rho_\chi \sigma_{\chi
i}|F_i(q_i)|^2} {2 m_\chi \mu^2_{i
\chi}}\int_{v_{min}}^\infty\frac{f(\vec v,t)}{v}d^3v \, .
 \eeqn
The sum is over all nuclear species present, with $c_i$ being the
mass fraction of species $i$ in the detector. The quantity $f(\vec
v,t)\,d^3v$ is the neutralino velocity distribution (presumed to be
Maxwellian) with $v=|\vec v|$ the neutralino velocity relative to
the detector. Finally $|F_i(q_i)|^2$ is the nuclear form factor for
species $i$, with $q_i=\sqrt{2M_i E}$ being the momentum transfer
for a nuclear recoil with energy $E$.  The differential rates are
calculated using {\tt DarkSUSY} via~(\ref{ds29}), over a range of
recoil energies relevant to the desired experiment. For a given
experiment there is typically a minimum resolvable recoil energy
$E_{\rm min}$ as well as a maximum recoil energy $E_{\rm max}$ that
is considered. These energies are ${\mathcal{O}}(10-100)$~keV and
represent the nuclear recoil energy of~(\ref{ds29}) inferred from
the observed energy of the detected physics objects. The range of
integration is generally different for each experiment and is
determined by the physics of the detector as well as the desire to
maximize signal significance over background. A numerical
integration of (\ref{ds29}) is performed by constructing an
interpolating function for the differential rate sampled in 0.5 keV
intervals. The integration ranges in this analysis are performed
over ranges similar to those used in the first two experiments
listed in Table (\ref{experiments}).
\beqn \label{rates}
 R_1 &:& 5\,{\rm keV} \leq E_{\rm recoil} \leq 25\,{\rm keV}
\nonumber \\ R_2 &:& 10\,{\rm keV} \leq E_{\rm recoil} \leq
100\,{\rm keV} \, . \eeqn
%
%\par To proceed it helps to understand how well a given experiment can distinguish nuclear recoils due to neutralino scattering from fakes and background events.  This is also relevant to the question of whether two possible signals can reliably be distinguished at any given experiment.  Backgrounds may be roughly classified into two categories.  The first are neutron recoils which are measured but do not originate from neutralino interactions, rather these may come from alpha radiation making up or surrounding the detector or from cosmic ray muons interacting with the neutrons.  The second type is due to electric charge induced from something other than neturon recoil possibly due to residual radioactivity in detector elements like the photomultiplier tubes common to many experiments.  The first type of backgrounds can be reduced to near zero with proper shielding or by a sufficiently subterranean experiment site.  The second type are harder to eliminate and imporovements in background rejection is necessary for scaling up experiments.
\par In what follows a single overall background figure will be used for each type of target.  This is done so that one may use the entire collection of future experiments as an ensemble in order to try to resolve degeneracies.  Projections for large scale germanium-based detectors are for background event rates of no more than a few events per year of exposure. The liquid xenon detectors project a slightly higher rate, but still on the order of 10-20 events per year of exposure (mostly of the electron recoil variety). To be conservative, the following requirements are used on two potentials signals to proclaim them distinguishable:
\begin{enumerate}
\item The count rates for the two experiments ($N_A$ and $N_B$),
obtained from integrating~(\ref{ds29}) over the appropriate range
in~(\ref{rates}), must {\em both} exceed $N$~events when integrated
over the exposure time considered. We will usually consider $N=100$,
but also show results for the weaker condition $N=10$.
\item The two quantities $N_A$ and $N_B$ must differ by at least
$n\,\sigma^{AB}$, where we will generally take $n=5$.
\end{enumerate}
The quantity $\sigma^{AB}$ is computed by assuming that statistical
errors associated with the measurements are purely $\sqrt{N}$
\beqn \sigma^{AB} = \sqrt{(1+f)(N_A + N_B)} \, , \label{oursigma}
\eeqn
and the overall multiplicative factor $(1 + f)$ allows us to be even
more conservative by taking into account a nominal background rate
or allow for uncertainties in the local halo density. The case $f =
0$ would therefore represent the case of no background events and in
all that follows this is the case considered.
 \par Based on this criteria none of the 378 models would have been distinguished already in CDMS~II, XENON10 or ZEPLIN~II.                                         We do find nine models which would have given at least ten events in 316.4 kg-days of exposure time in the Xenon10 experiment, and five that would have given at least ten events in 397.8 kg-days of exposure time in the CDMS II experiment. These are models that could have been discovered at CDMS II (where no signal-like events were observed) or nearly discovered at Xenon10 (where ten signal-like events were reported).  However these models all have $\Omega_{\chi}{\rm h}^2 < 0.025$ and upon rescaling $\rho_\chi$ by $r_\chi$ then none of the models should have been seen at any experiments.
\begin{figure}[t]
\begin{center}
\includegraphics[scale=0.475]{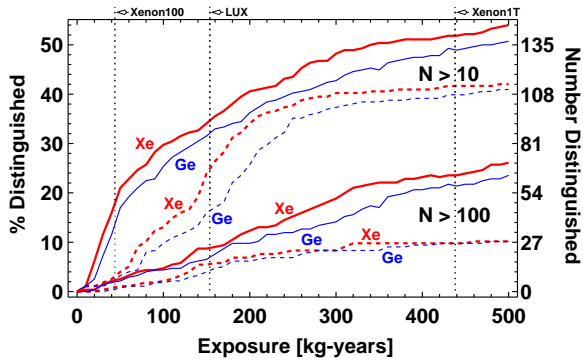}
\caption{\footnotesize The number of degenerate pairs/percentage of
the total that can be distinguished
          as a function of integrated exposure time.
  From \cite{Altunkaynak:2008ry}. }
 \label{dd_rates}
  \end{center}
\end{figure}
\par In Fig.(\ref{dd_rates}) we show the percentage of 276 degenerate pairs which can be distinguished as exposure time is accumulated in xenon and germanium targets.  In the figure we use a seperability criterion of 5$\sigma$ and assume no theoretical uncertainty.  Heavy (red) lines are labeled for xenon, thinner (blue) lines are labeled for germanium. Solid lines have not been rescaled by the relic density ratio $r_{\chi}$, dashed lines have. The upper four lines are obtained by requiring only $N \geq 10$ recoil events for both models. The lower four lines are obtained by requiring $N\geq 100$ recoil events for both models. The predicted exposure after one year for three projected liquid xenon experiments is indicated by the vertical lines as labeled. Note the assumption of 200 days of data-taking per calendar year with 80\% of the mass from Table~\ref{experiments} used as a fiducial target mass.
\par    Generally speaking when two models are visible they are easily distinguished, at least under the idealized assumption of perfect theoretical control over the input nuclear matrix elements.  To investigate the effects of theoretical uncertainties associated with the nuclear matrix elements on the analysis consider errors in the cross sections of the form
\begin{equation} \left(\delta \sigma^{\mathrm{SI}}_{\chi} \right)_{\rm theor}
= \epsilon \times \sigma^{\mathrm{SI}}_{\chi} \label{epsilon}
\end{equation}
which is added in quadrature to the statistical errors in Eqn.
(\ref{oursigma}).  In Table (\ref{theoryunc}) we give the number of
pairs distinguishable after a given accumulated exposure in ton
$\times$ years in xenon when an additional theoretical uncertainty
of the form of~(\ref{epsilon}) is included.  The experimental error
is taken to be purely statistical $f=0$, require 10 signal events
and performed rescaling via $r_\chi$ where necessary.
%---------------- Theoretical Error Table --------------------
\begin{table}[h]
\begin{footnotesize}
\centering
\begin{tabular}{llcccc}
\hline \hline
 & \multicolumn{5}{c}{Require 10 Events, Xenon} \\
\hline
 & & $\epsilon=0$ & $\epsilon=0.1$ & $\epsilon=0.25$ & $\epsilon=0.5$ \\
 & $3\sigma$ & 164 & 118 & 13 & 0 \\
\raisebox{1.5ex}{1 ton-yr} & $5\sigma$ & 112 & 46 & 0 & 0 \\
\hline
 & $3\sigma$ & 217 & 149 & 25 & 0 \\
\raisebox{1.5ex}{5 ton-yr} & $5\sigma$ & 187 & 77 & 0 & 0 \\
\hline%
\\
\end{tabular}
{\caption{\label{theoryunc}\footnotesize  Effect of theoretical
uncertainties associated with nuclear matrix elements. From
\cite{Altunkaynak:2008ry}.}}
\end{footnotesize}
\end{table}
%------------------------- END OF THE TABLE ---------------------
Clearly if theoretical uncertainties stay at their present level
(with roughly 50\% uncertainty in the cross-section predictions)
then it will be impossible to distinguish models with direct
detection experiments  even after five ton-years of exposure and
requiring only 3$\sigma$ separation. If the uncertainty in the
$\pi\,N$ $\Sigma$-term can be reduced so as to generate only a 10\%
theoretical uncertainty in $\sigma^\mathrm{SI}_{\chi p}$ the ability
to distinguish models will still be significantly reduced, but some
hope for separating models will remain. For this reason it is
important for further experimental work aimed towards reducing these
uncertainties.
\subsection{Conclusion}
The utility of direct detection observations for distinguishing
between SUSY models has been investigated.  Using 276 degenerate
model points at the LHC it has been shown that in principle dark
matter observations may be quite useful to separate degenerate
pairs.  However the ability to distinguish models is very dependent
on future theoretical determination on such things as nuclear matrix
elements.  If one assumes perfect knowledge of these theoretical
inputs as well as low background interference on signals the
prospects for untangling degerate models using dark matter
observables is quite promising. One may also consider further dark
matter observables such as gamma ray signals as is also done in
\cite{Altunkaynak:2008ry}.

%\end{document}

% end of file template.tex

%%%%%%%%%%%%%%%%%%%%%%%%%%%%%%%%%%%%%%%%%%%%%%%%%%%%%%%%%%%%%%%%%%%%%%%%%%%%%%%%%%%%%%%%%%%%%%
%%%%%%%%%%%%%%%%%%%%%%%%%%%%%%%%%%%%%%%%%%%%%%%%%%%%%%%%%%%%%%%%%%%%%%%%%%%%%%%%%%%%%%%%%%%%%%
\chapter{Top-Quark Physics at the LHC}
\epigraphhead[20]{\epigraph{\large {\em Tao Han, Seung J. Lee, Fabio
Maltoni, Gilad Perez, Zack Sullivan, Tim M.P.~Tait, Lian-Tao
Wang}}{\large Tao Han (Convener)}}
%
%%%%%%%%%% espcrc2.tex %%%%%%%%%%
%
% $Id: espcrc2.tex 1.2 2000/07/24 09:12:51 spepping Exp spepping $
%
%\documentclass[fleqn,twoside]{article}
%\usepackage{espcrc2}

% change this to the following line for use with LaTeX2.09
% \documentstyle[twoside,fleqn,espcrc2]{article}

% if you want to include PostScript figures
%\usepackage{graphicx}
% if you have landscape tables
%\usepackage[figuresright]{rotating}

% put your own definitions here:
%   \newcommand{\cZ}{\cal{Z}}
%   \newtheorem{def}{Definition}[section]
%   ...
% Zack's commands
\newcommand{\dzero}{D$\slash\!\!\!0$}
\newcommand{\ttbs}{\char'134}

\section{Introduction}

% Intro rewritten by Zack:

As the $SU(2)_L$ partner of the bottom quark, all of the top quark
properties except for its mass, are fully predicted in the Standard
Model (SM): The spin, the QCD and electroweak (EW) charges, and the Lorentz structure of
the couplings are uniquely assigned. On the other hand, physics associated
with the top quark holds
great promise in revealing the secret of new physics beyond the SM.

%is of the order of the
%electroweak scale, the top decays too fast for non-perturbative
%strong interactions to kick in and affect its dynamics.  The large
%mass of the top quark leads to a Yukawa coupling of $1.00$.  For this
%reason, the top is expected to play an important role in the yet
%unexplored electroweak symmetry breaking sector \cite{Hill:2002ap}.
%This behaviour is quite general, and in many
%of beyond the standard model scenarios, the top-quark ends up being
%the most sensitive SM probe of new physics: new interactions or
%particles couple strongly with the top and either modify its SM
%properties via loops, or are produced in its association or as its
%decay products.
%
%The theoretical considerations include the following:
\begin{itemize}
\item With the largest Yukawa coupling $y_t\sim 1$ among the SM fermions,
and a mass at the electroweak scale $m_t\sim v/\sqrt 2$ (the vacuum
expectation value of the Higgs field),  the top quark is naturally related to the yet unexplored
electroweak symmetry breaking (EWSB), and may reveal new dynamics \cite{Hill:2002ap}.
\item The largest contribution to the quadratic divergence of the SM Higgs mass comes
from the top-quark loop, which implies the immediate need for new physics at the Terascale
for a natural EW theory,
% \cite{Giudice:2008bi},
with SUSY and Little Higgs as prominent examples.
\item Its heavy mass opens up a larger phase space for its decay to heavy
states $Wb,\ Zq,\ H^{0,\pm}q$, {\it etc.}
\item Its prompt decay much shorter than the  QCD scale offers the opportunity to explore
the properties of  a ``bare quark", such as its mass, spin and correlations.
\end{itemize}

In anticipation of the LHC era, we review the physics potential associated with the top quark.
For recent reviews on the related topics, see {\it e.g.}, Ref.~\cite{Quadt:2007jk}.

\section{Standard Model Top-Quark Physics}

{\it Z.~Sullivan and F.~Maltoni}\medskip
% beginning of Zack Sullivan and Fabio Maltoni: ($4 \pm 1$ pages)

SM measurements of the top properties have
played a key role at the Tevatron and will continue to play an
important role at the LHC.  Apart from the mass measurement, for which
a dedicated and possibly long experimental effort to control
systematics will be needed to improve on the current impressive
Tevatron extractions  $m_t = 173.1\pm 1.3$~GeV \cite{:2009ec},
many other measurements will be accessible at
the LHC that were statistically limited before.  These include rare
decays, and properties of the $t\bar t$ and single-top production
mechanisms, from total cross to differential distributions, and from
polarization to spin-correlation measurements.

In the search for possible deviations of the SM prediction theoretical
predictions and/or simulation tools will have to be accurate enough at
least to match the expected experimental precision. In the following,
we first give a few selected examples of the most promising and challenging
measurements to be performed at the LHC.

% motivation Vtb, W', background

\subsection{Top Quark Decay}

In the Standard Model, the top quark decays at tree level through its
charged-current weak interaction into a down-type quark and an
(on-shell) $W$ boson.  Assuming purely SM physics, the constraints on
the CKM matrix from its unitarity require $V_{tb} \geq 0.9990$ (at the
$95\%$ confidence limit) \cite{Amsler:2008zzb}, which implies that top
decays into $W$ and a bottom quark with a branching ratio very close
to unity.  The partial width at leading order is,
\begin{eqnarray}
\Gamma \left( t \rightarrow W^+ b \right) & \simeq &
\frac{G_F m_t^3}{\sqrt{2} 8 \pi} |V_{tb}|^2
\end{eqnarray}
where $G_F$ is the Fermi constant.  This formula is valid at tree
level in the limit $m_t \gg m_W, m_b$.  The dominant corrections are
the $m_W^2/m_t^2$ ones, of the order of $10\%$.  The top width is
parametrically larger than that of any other known quark, because the
large top mass allows the decay into an on-shell $W$ boson, and thus
is a $2$-body decay rather than a $3$-body one.  Nonetheless, the top
width is small enough that it is challenging to measure it directly
from top decay products.

The top width $\Gamma_t \sim 1.5$~GeV is significantly larger than the
scale of nonperturbative QCD interactions, $\Lambda_{QCD}$.  As a
result, top decays before it hadronizes, and its spin is passed to its
decay products.  In the Standard Model, the $W$-$t$-$b$ interaction is
exclusively left-handed, implying that the $W$ bosons from top decay
are left- or longitudinally polarized, with the fraction of
longitudinal $W$'s given by \cite{Kane:1991bg},
\begin{eqnarray}
f_ 0 & \simeq & \frac{m_t^2}{2 M_W^2 + m_t^2} \sim 70\%~.
\end{eqnarray}
The $W$ polarization is reflected in the kinematics of the charged
lepton from its decay, allowing one to relatively easily reconstruct
the distribution of $W$ polarizations and providing a test of the
left-handed nature of the $W$-$t$-$b$ vertex \cite{Tait:2000sh}.
Existing measurements from the Tevatron lead to the results,
\begin{eqnarray}
f_0 &=& 0.66 \pm 0.16 \nonumber \\
f_+ & = & -0.03 \pm 0.07
\end{eqnarray}
from CDF  \cite{Aaltonen:2008ei} and
\begin{eqnarray}
f_0 &=& 0.490 \pm 0.106 \pm 0.085 \nonumber \\
f_+ & = & 0.110 \pm 0.059 \pm 0.052
\end{eqnarray}
from D0 \cite{D0:f0fp}.  Measurements at the LHC are expected to reach
the $5\%$ level and are expected to be dominated by systematics
\cite{Cabrera:2009zza}.

Given the SM expectation that the branching ratio into $W b$ is
extremely close to one, top decays are labeled by the $W$ boson decay
mode.  The decays into each of the three charged leptons plus its
neutrino ($\ell \bar\nu_\ell$) have branching ratio close to $1/9$ for each flavor.
The remaining $2/3$ of
the decays go into light quarks ($\bar q q'$), resulting in typically unflavored
jets.

\subsection{$t \bar t$ Production}

The best predictions in QCD for fully inclusive $t \bar t$ cross
sections at the LHC are at NLO accuracy plus next-to-leading-log
corrections~\cite{Cacciari:2008zb} or approximate
NNLO~\cite{Moch:2008qy} both giving consistent results of about $960$
pb with an error of several percent due to unknown higher-order
corrections and a few percent from PDF uncertainties.  For less
inclusive observables, and more experimental-friendly predictions,
Monte Carlo tools such as matrix elements predictions interfaced with
the shower are available at LO~\cite{Alwall:2007fs} and
NLO~\cite{Frixione:2003ei,Frixione:2007nw}.

One of the first aims of the LHC will to rediscover the top and to
confirm the SM expectations for the production rates.  It will take
considerable experimental effort, however, to further improve the
precision on the cross section as this will be dominated by systematic
effects related to the understanding of both the collider (for the
luminosity) and the detector (such as reconstruction efficiencies).  A
more appropriate (and much more promising) goal than accurate mass and
cross section measurements in the earlier data, will be to {\it use}
the Tevatron mass extraction and the SM cross section as calibration
tools for other studies, in $t \bar t$ itself, in single-top or in
other SM and BSM processes.

 \begin{figure}[t]
\includegraphics*[width=\columnwidth]{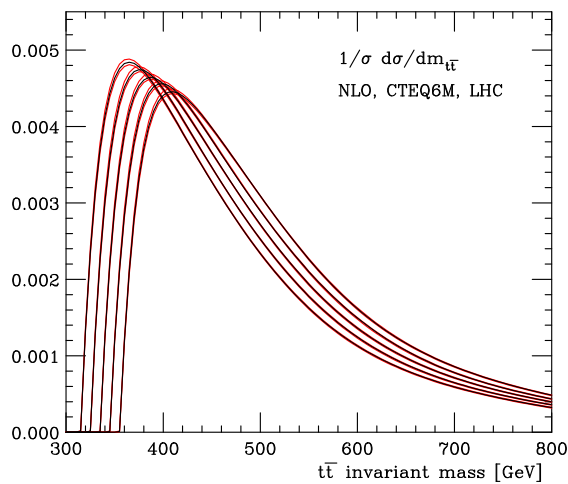}
\caption{
The NLO normalized $t\bar{t}$ production cross section as a function of the
$t\bar{t}$ invariant mass, $m_{t\bar{t}}$, for  the LHC. Solid lines from left to right are
for a top quark mass of $m_t=160,\ldots,180$ GeV in steps of 5 GeV, respectively.
The bands spanned by the red lines show the scale uncertainties. From Ref.~\cite{Frederix:2007gi}.
\label{fig:ttbar-inv-mass}
}
\end{figure}

An interesting example is the study of the differential distributions,
such as the $p_T$ of the tops, or the invariant mass of the $t \bar t$
pair.  These distributions are extremely well predicted already at the
NLO, as the theoretical uncertainties mostly affect the overall
normalization of the cross sections but not the shapes.

 \begin{figure}[t]
\includegraphics*[width=\columnwidth]{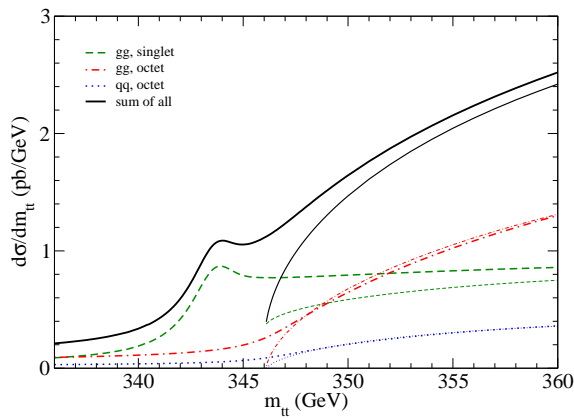}
\caption{Invariant $m_{t\bar t}$ distribution at threshold. The pseudo bound-state enhancement is visible in the form of
a peak. From Ref.~\cite{Hagiwara:2008df}.
\label{fig:ttbar-inv-resonance}
}
\end{figure}

As an example, the $m_{t\bar t}$ distribution is shown in
Fig.~\ref{fig:ttbar-inv-mass} for different top masses, where the
scale uncertainties are displayed as an (almost invisible) red
envelope. The reconstruction strategies for such a quantity vary
depending on the decay mode of the top, but several promising
approaches have been suggested~\cite{Barger:2006hm}. Such a
distribution therefore offers a great observable to both test the SM
and find new physics, as in the search for resonances,
Sec.~\ref{sec:toptquarkres}.
A very challenging and exciting example
of SM physics that has never been observed before and could be visible
in $m_{t\bar t}$ is an enhancement of the cross section at threshold
due to long range Coulomb
interactions~\cite{Hagiwara:2008df,Kiyo:2008bv} see
Fig.~\ref{fig:ttbar-inv-resonance}: The measurement of the position of
the peak could give a top mass determination free from
non-perturbative ambiguities, while the height and width of the peak
would provide a direct measurement of the top width. On the BSM side,
many examples of physics affecting this distribution not directly
related to resonances decaying into $t \bar t$ have been given, for
instance in Refs.~\cite{Han:2008gy,Kumar:2009vs}.

As the statistics accumulated by the LHC increase, with the
detectors better understood and the experimental systematic
uncertainties under control, more studies will be possible. Spin
correlations, for instance, will be eventually measured and the $t
\bar t$ production dynamics clearly identified.

\subsection{Single-top Production}

Once a $t\bar t$ signal is established at the LHC, attention will turn
to measuring the single-top-quark cross section.  Single top quarks
can be produced via three processes at the LHC: $t$-channel,
$s$-channel, and $Wt$-associated production, shown in Fig.\
\ref{fig:singtopmodes}.  Assuming $|V_{tb}| \gg |V_{td}|, |V_{ts}|$,
the independent measurement of each of the cross sections leads to a
direct determination of the CKM matrix element $|V_{tb}|$, in contrast
to top branching measurements which are very weakly dependent on
$|V_{tb}|$.  A combined measurement of the single-top cross sections
could also provide information on the three $|V_{tq}|$ CKM matrix
elements~\cite{Alwall:2006bx}.  The $V-A$ structure of the
Standard-Model charged current vertex leads to highly polarized top
quarks, which in turn produce strongly correlated decay products
\cite{Sullivan:2005ar,Motylinski:2009kt}.  The $t$-channel cross
section is also a significant background to several Higgs production
modes, e.g., $H\to WW$ and $WH\to Wb\bar b$
\cite{Sullivan:2006hb,Bernreuther:2008ju}, charged vector currents
($W^\prime$) \cite{Sullivan:2002jt}, and any process with jets,
leptons, and/or missing transverse energy at the LHC.

\begin{figure}[tb]
\includegraphics*[width=\columnwidth]{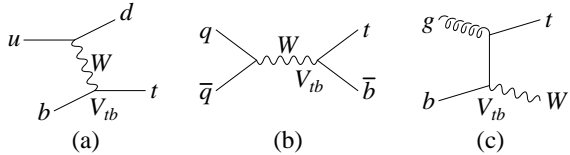}
\caption{Representative Feynman diagrams for (a) $t$-channel, (b)
  $s$-channel, and (c) $Wt$-associated production of a single top quark.
The CKM matrix element $V_{tb}$ appears in the production diagrams.
\label{fig:singtopmodes}}
\end{figure}

% CDF/D0 discovery

The CDF \cite{Aaltonen:2009jj} and \dzero\ \cite{Abazov:2009ii}
Collaborations have published first observations of single-top-quark
production.  The Tevatron Electroweak Working Group (TEWWG) reports a
joint Tevatron analysis of the measured cross section of
$2.76^{+0.58}_{-0.47}$ pb using $\sim$3~fb$^{-1}$ of data, and a
direct measurement of $|V_{tb}|=0.88\pm 0.7$ \cite{Group:2009qk}.
Despite having roughly $1/2$ the cross section of $t\bar t$, a clean
extraction of the signal requires good understanding of the $t\bar t$
and $W+$jets backgrounds.

% LHC cross section and predictions

The theoretical status of single-top-quark production is quite strong.
Several calculations of the NLO cross sections have been performed
\cite{cfwerner}.  The NLO cross sections, updated with newer CTEQ 6.6
parton distribution functions, are shown in Table~\ref{tab:singcs}
\cite{Campbell:2009gj,Nadolsky:2008zw,Kidonakis:2007ej}.  At 10~TeV,
the $t$-channel cross section is roughly $1/2$ what it is at $14$~TeV,
82(45)~pb for $\sigma_t$($\sigma_{\bar t}$).  NLO Monte Carlos with
\cite{Frixione:2005vw}, and without
\cite{Campbell:2004ch,Campbell:2005bb,Campbell:2009ss} showering,
exist for these processes.  Experimental systematic errors are
expected to dominate both in the extraction of the single top cross
sections, and in the estimates of them as backgrounds to new physics
at the LHC.  These systematic errors are especially sensitive to jet
matching schemes and angular correlations
\cite{Sullivan:2004ie,Sullivan:2005ar}.

\begin{table}[tb]
\caption{Single-top-quark cross sections at LHC (14~TeV)
for $m_t=171$~GeV.}
\label{tab:singcs}
\begin{tabular}{llll}
\hline
&$t$-channel & $s$-channel & $Wt$-assoc. \\
\hline
$\sigma_t$ & $152\pm 6$ pb & $7.6\pm 0.7$ pb & $45\pm 5$ pb\\
$\sigma_{\bar t}$ & $90\pm 4$ pb & $4.2\pm 0.3$ pb & $45\pm 5$ pb\\
\hline
\end{tabular}
\end{table}

% ATLAS

Establishing the $t$-channel cross section is the most straightforward
of the single-top processes.  For example, the ATLAS Collaboration
expects to identify $t$-channel production with 1~fb$^{-1}$ of data
via a set of kinematic cuts, and a multivariate Boosted Decision Tree
analysis that utilizes the distinctions in the correlations between
the signal and backgrounds, cf., Fig.\ \ref{fig:singtopatlas}
\cite{Aad:2009wy}.  To achieve this striking signature will require a
solid understanding of physical backgrounds, e.g., $W+$jets and QCD,
as well as detector effects, e.g., $b$-tagging performance, jet energy
scales, and missing energy.  Other single-top channels should be
measurable with 10--30~fb$^{-1}$ of data in both ATLAS
\cite{Aad:2009wy} and CMS \cite{Ball:2007zza}.

\begin{figure}[tb]
\includegraphics*[width=\columnwidth]{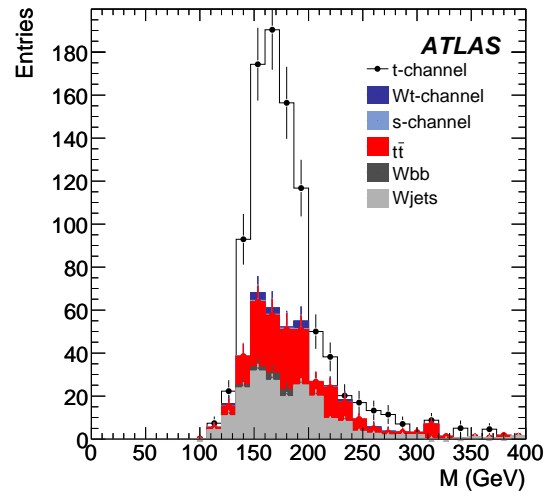}
\caption{ATLAS estimated top quark mass distribution from a Boosted Decision
Tree analysis of $t$-channel events with 1~fb$^{-1}$ of data.
\label{fig:singtopatlas}}
\end{figure}

% beginning of Tim's: (4+-1)

\section{New Physics in Top-Quark Decay}

{\it T.~Tait}\medskip

The LHC is a top factory, whose large statistics allow searches for ultra-rare decays.
Nonstandard top decays can be broadly divided into decays of the top into ordinary Standard Model particles at unexpected rates and decays of top into particles not found in the Standard
Model itself.  Even when new physics does not single top out in particular, the large top mass
permits the second option to take place for exotic particles with weak scale masses.

\subsection{Rare Decays into Standard Model Particles}
\vspace*{0.2cm}

In the Standard Model, top decays into
standard particles other than $W b$ are either suppressed by small CKM elements
(in the cases of $W s$ and $W d$) or occur at loop level (as in the FCNC
decays $V c$ and $V u$ where $V=Z$, $g$, or $\gamma$).  Three- (and higher)
body decays such as $W b \gamma$ occur in the SM
at higher orders in perturbation theory.  The fact that SM rates for these
processes are extremely low (in many cases low enough that the expectation is that the
LHC will see less than one event) makes them potentially very sensitive to physics
beyond the SM.

It is traditional to parameterize the possibility of contributions to nonstandard decays in terms
of ``effective operators" which are added to the Standard Model lagrangian.  Each term respects
the $SU(3) \times SU(2) \times U(1)$ gauge symmetry of the Standard Model, as well as
Lorentz invariance.  Since by definition these operators are non-renormalizable, their effects
are typically more pronounced at high energies, again implying that the top quark is a natural
laboratory to test for their presence.  Each term has a
coefficient which parameterizes its strength.
As a concrete example, consider adding a
term to the SM Lagrangian density such as \cite{Malkawi:1995dm},
\begin{eqnarray}
& &
\frac{\kappa^g_{tc}}{\Lambda^2}
\left( H \overline{Q_2} \right) \sigma^{\mu \nu} G_{\mu \nu} t_R
\end{eqnarray}
where $Q_2$ is the second family quark doublet, $H$ is the Higgs doublet,
$G_{\mu \nu}$ is the gluon field strength, and $t_R$ is
the right-handed top quark.
One could just as easily have chosen $Q_1$ instead of $Q_2$, which would
result in an anomalous coupling of top to the up-quark instead of charm.
The combination
$\kappa^g_{tc} / \Lambda^2$ (with dimension [mass]$^{-2}$)
is the coupling
constant for this new interaction.
This term is consistent with all of the gauge
symmetries of the SM, but is not properly part of the SM because it is
non-renormalizable.

This effective operator
can be understood as the low energy remnant effect of some kind of
high scale physics, produced by a particle whose mass is of order
$\Lambda$.  The size of the dimensionless coupling $\kappa^g_{tc}$ would
depend on the details of this hypothetical new particle.  For example, if
it produces this coupling in loops, one would expect
$\kappa^g_{tc} \sim \alpha_X / 4 \pi$ where $\alpha_X$ is the strength of the
interactions in the loops.  If it generates this operator at tree level or
through strong dynamics ($\alpha_X \sim 4 \pi$), then one would expect
$\kappa^g_{tc} \sim 1$.  At energies above $\Lambda$,
this description will need to be supplemented by a detailed picture of the new
particles and interactions, but at low energies it captures
all of the relevant physics.

Replacing the Higgs doublet by its vacuum expectation value, we arrive
at a new interaction,
\begin{eqnarray}
& &
\frac{v \kappa^g_{tc}}{\Lambda^2}
\overline{c}_L \sigma^{\mu \nu} G_{\mu \nu} t_R \, .
\end{eqnarray}
In this way we see that EWSB converts the dimension six operator into a
dimension five operator, whose vertex now is a flavor-changing neutral
current (FCNC) $g$-$t$-$c$ interaction with coupling constant
$v \kappa^g_{tc} / \Lambda^2$ with a combined dimension of [mass]$^{-1}$.

This FCNC results in an anomalously high branching fraction for $t \rightarrow g c$
\cite{Han:1996ce,Tait:1996dv,Zhang:2008yn,Ferreira:2008cj}.
One can also write down very similar
terms inducing $t \rightarrow \gamma c$ and $t \rightarrow Z c$
\cite{Han:1996ep,Han:1995pk,Fox:2007in} decays, as well as influencing $b$ physics
observables through loops.
Simply replacing the charm quark by the up quark allows for FCNC decays
into up as well.   This family of FCNC operators can be induced by many popular
theories for physics beyond the Standard Model, including the
MSSM \cite{Li:1993mg,Couture:1994rr,Lopez:1997xv,deDivitiis:1997sh,Yang:1997dk,Guasch:1999ve,Cao:2002si,Delepine:2004hr,Cao:2007dk}, models with two Higgs doublets
\cite{Kagan:2009bn},
Technicolor variants \cite{Wang:1994qd,Yue:2001qr,Lu:2003yr,Larios:2006pb},
Little Higgs theories \cite{HongSheng:2007ve},
extra-dimensional models of flavor \cite{Chang:2008zx,Agashe:2006wa},
and models with additional generations of quarks \cite{Herrera:2008yf}.

\begin{table*}[tb]
\caption{Some possible rare decays of the top quark into Standard Model particles, the Standard Model branching ratio predictions \cite{Eilam:1990zc},
existing experimental constraints, and
prospects for experimental measurements at the LHC.}
\label{table:topdecaysummary}
\newcommand{\m}{\hphantom{$-$}}
\newcommand{\cc}[1]{\multicolumn{1}{c}{#1}}
\renewcommand{\tabcolsep}{2pc} % enlarge column spacing
\renewcommand{\arraystretch}{1.2} % enlarge line spacing
{\small
\begin{tabular}{@{}lccc}
\hline
Decay Mode           & SM BR & $95\%$ CL Tevatron &
LHC Prospects~10 fb$^{-1}$\\
\hline
$t \rightarrow b W$ & $\sim 1$  & $> 0.79$~\cite{Abazov:2008yn}$^{*}$ &
$0.998$  \cite{Beneke:2000hk}$^{\dagger *}$ \\
$t \rightarrow s W$ & $1.6 \times 10^{-3}$ & (see above) &
(see above) \\
$t \rightarrow d W$ & $10^{-4}$ & (see above) &
(see above) \\
$t \rightarrow q Z$ $(q=u,c)$ & $1.3 \times 10^{-13}$ & $< 0.037$~\cite{:2008aaa} &
$6.5 \times 10^{-4}$ ~\cite{:1999fr} \\
$t \rightarrow q \gamma$ $(q=u,c)$ & $5 \times 10^{-13}$ & $<0.18$~\cite{cdf:fcnc} &
$1.9 \times 10^{-4}$ ~\cite{:1999fr} \\
$t \rightarrow q g$ $(q=u,c)$ & $5 \times 10^{-11}$ & $<0.12$~\cite{cdf:fcnc} &
$10^{-2}$ (1 fb$^{-1}$) \cite{Cabrera:2009zza} \\
$t \rightarrow q h^0$ $(q=u,c)$ & $8 \times 10^{-14}$ & -- &
$1.4 \times 10^{-4}$ \cite{AguilarSaavedra:2000aj} \\
\hline
\end{tabular} }
\\[2pt]
$^*$Assuming no appreciable FCNC or exotic particle decays for top.  The lower limits
for $t \rightarrow W b$ thus translate into limits on the sum of $t \rightarrow W s$ and
$t \rightarrow W d$.  See the text for more details.\\
$^\dagger$Current estimates include only statistical uncertainties; the actual sensitivity
is likely to be systematics-dominated.
\end{table*}

These rare FCNC decays are most efficiently searched for
in $t \overline{t}$ production, given its large rate and the ability to
tag one of the tops through a standard decay.  One thus tags the event by
looking for a standard (usually semi-leptonic) top decay, and examines the
other side of the event to see how the second top quark decayed.  In the case
of $t \rightarrow Z q$, one can look for leptonic $Z$ decays.  $t \rightarrow \gamma q$
will have a hard photon and jet whose invariant mass reconstructs the top mass.
The decay $t \rightarrow g q$ results in two jets which reconstruct the top mass, and suffers
from much larger backgrounds than the first two modes.
The same operators which produce anomalously large FCNC top decays also  lead to
new channels mediating single top production, allowing cross-checks between
observed anomalies, and further information which can help disentangle which
operator is responsible for a given observation.  Existing bounds from the Tevatron
are already at the few per cent level, considerably higher than the Standard Model predictions, but
beginning to provide information about models of physics beyond the Standard Model.
In Table~\ref{table:topdecaysummary}, we show several possible decay modes of the
top quark,  the Standard model predictions \cite{Eilam:1990zc}, current Tevatron bounds,
and expected LHC sensitivities.

Charged current decays of the top into standard particles include the principle
decay mode $Wb$, as well as the CKM-suppressed modes $Ws$ and $Wd$.
The charged current couplings are generally modified away from the Standard Model
expectations when the top mixes with additional quarks, such as e.g. a chiral
fourth generation \cite{Kribs:2007nz,Chanowitz:2009mz,Bobrowski:2009ng}
(in which case $3 \times 3$ unitarity no longer constrains $V_{tb}$, relaxing the bound
to the measured value from single top production of $V_{tb} \geq 0.78$ at the
$95\%$ CL~\cite{Abazov:2009ii,Aaltonen:2009jj}).

With a sufficiently precise
understanding of the probability to $b$-tag jets coming from top decays, one can use the ratio
of the number of $t \bar{t}$ events with two $b$-tags to the number of events with one $b$-tag to
estimate the ratio,
\begin{eqnarray}
R & = & \frac{BR(t \rightarrow W b)}{BR( t \rightarrow W q)}
\end{eqnarray}
where $q=d,s,b$.  The Standard Model expectation for this quantity is $0.999$, with
Tevatron measurements \cite{Abazov:2008yn}
consistent with this number but with large error bars.  In order to interpret this measurement
as a branching ratio for $t \rightarrow W b$, one must assume that all relevant top decays
are included in $t \rightarrow W q$.

If the Higgs is light enough, the decay $t \rightarrow h^0 c$ may be allowed.  Depending on the
Higgs mass, decays of  $h^0 \rightarrow b \bar{b}$ and $h^0 \rightarrow W^+ W^-$ are possible.
The rate is predicted to be unobservably small in the Standard Model
\cite{Eilam:1990zc,Mele:1998ag}, but may be enhanced in models with multiple
Higgs doublets \cite{Baum:2008qm}, in the minimal supersymmetric standard
model \cite{Guasch:1999jp,Eilam:2001dh,DiazCruz:2001gf}, and in Little Higgs
theories \cite{Tabbakh:2005kf}.
The 10~fb$^{-1}$ LHC sensitivity for $m_{h^0} = 120$~GeV has been estimated to be
$1.4 \times 10^{-4}$ \cite{AguilarSaavedra:2000aj}.

\subsection{Exotic Decays into Nonstandard Particles}
\vspace*{0.2cm}

The second class of rare decay is the top decaying into a non-SM particle.  There are
a plethora of possibilities, so this discussion will be limited to
charged Higgs decay $t \rightarrow H^+ b$.   Additional Higgs
$SU(2)$ doublets are perhaps the most innocuous additions to the Standard Model
Higgs sector from the point of view of precision electroweak constraints, and arise
naturally in the context of supersymmetric and composite Higgs theories.
They inevitably result in physical charged scalars in the spectrum, which inherit a large
coupling to the top.
Provided the $H^+$-$t$-$b$ coupling is large enough, and the mass of $H^+$
is sufficiently smaller than $m_t$ (less than about 150~GeV),
top decays can provide an excellent way to produce charged Higgs bosons.

In a type-II two Higgs doublet model (such as the MSSM), one Higgs doublet gives mass to the
up-type quarks, and one to the down-type quarks.   An important parameter for phenomenology
is the ratio of the vacuum expectation values of the two doublets, $\tan \beta = v_1 / v_2$.
At tree level, the $H^+$-$t$-$b$ vertex is enhanced for either very large or very small
values of $\tan \beta$.  In the first limit, the charged Higgs will dominantly decay into
$\tau^+ \nu$ and in the second into jets, $c \bar{s}$.  The first appears as an anomalously
large branching ratio of top into tau leptons, and the second as a set of top decays for which the
untagged jets have an invariant mass inconsistent with a $W$ boson decay.

Current limits from the Tevatron
vary somewhat with the Higgs mass, but require ($95\%$ CL) the branching ratios for
$t \rightarrow H^+ b$ to be less than $15\%$ when $H^+ \rightarrow \tau \nu$
\cite{Abazov:2009ae} or less than $30\% - 10\%$ (as $m_{H^+}$ ranges from
90~GeV to 150~GeV) when $H^+ \rightarrow c \bar{s}$ \cite{Aaltonen:2009ke}.
At the LHC the expectation is that with $100~{\rm fb}^{-1}$, any mass less than 155 GeV
(for all $\tan \beta$) can be discovered \cite{Beneke:2000hk}.

\section{Top Quarks in New Resonant Production
\label{sec:toptquarkres} }

{\it S.-J.~Lee and G.~Perez}\medskip

%%% Beginning of Gilad Perez:  ($4 \pm 1$ pages)

 %%%%%%%%%%%%%%%%%%%%%%%%%%%%%%%%%
%\subsection{Brief Introduction}
%%%%%%%%%%%%%%%%%%%%%%%%%%%%%%%%%
%As discussed above the top plays a unique role within the standard model (SM).
%
There are good reasons to suspect that the top quark is only a tip of the iceberg
and there is a whole top sector and top dynamics which describes our microscopic
universe, just waiting to be discovered in near future experiments.
One generic possibility is that the top quark field couples to new particles more dominantly
than the other SM fields. Once these new degrees of freedom are produced they will, therefore, predominantly decay
into SM top quarks. If the new particles are bosons, with appropriate gauge quantum numbers, then the simplest decay process would
probably be into two top quarks.
Thus, a natural way to look for the top dynamics beyond the SM is in a form of resonant structure in processes that involve top pairs.
However, due to the very high mass of the top, it is not until very recently that one could directly test whether
$t\bar t$ resonances exist in nature. Such a probe clearly requires production of on shell top pairs, away form threshold.
Direct searches for signal at the Tevatron~\cite{Tevttbar} are now, for the first time, mature enough, and collected enough luminosity, in order
to study precisely the $t\bar t $ differential cross section, at sizable invariant masses,  $m_{t\bar t}$.
The differential $t \bar t$ distribution,
$d\sigma_{t\bar t}/d m_{t\bar t}$ shows no access, up to $m_{t\bar t}$ of about a TeV~\cite{Tevttbar},
as long as it is narrow enough.
To demonstrate the power of this limit concretely, the Tevatron experiments have, this summer, published an independent lower limit on the mass of a leptophobic $Z'$~\cite{Tevttbar},
$M_{Z'}\geq {\cal O}\big(800\,{\rm GeV}\big)\,.$

%%%%%%%%%%%%%%%%%%%%%%%%%%%%%%%%%%%%%%%%%%%%%%%%%%%%%%%%%%%%
\subsection{Emergence of Top Jets}
%%%%%%%%%%%%%%%%%%%%%%%%%%%%%%%%%%%%%%%%%%%%%%%%%%%%%%%%%%%%
The absence of a new physics signal in $d\sigma_{t\bar t}/d m_{t\bar t}$ may not be shocking due
to constraints from indirect electroweak precision tests, which exclude new low-mass states.
Therefore, it is important to consider the possibility of a few TeV resonances decaying dominantly into tops, which, however, pose an experimental and theoretical challenge:
Roughly, the distribution of the outgoing $W-b$ opening angle in the transverse plane will be peaked around
$2 m_t/p_T$. Thus, we see that for a large boost the top decay products are highly collimated. In Fig.~\ref{coll_rate_fig},
we plot the rate of collimation as a function of the top $p_T$
(for related discussions and analyses see~\cite{Baur:2007ck,LHCnotes}), where the collimation rate is defined as the fraction of top quarks which reconstruct to a jet having $140 {\rm\, GeV} \le m_J \le 210  {\rm\, GeV}$~\cite{tjets}, given a fixed $p_T$.
We see for $p_T > {\cal O}(800\,\rm GeV)$ the majority of events would be fully collimated even
if one is to use the smallest, commonly used, cone size $R=0.4$.
In this case the conventional top tagging methods are doomed to be failed since the tops are going to be manifested as a single jet. Hence, these high $p_T$ jets were denoted as top jets~\cite{KKG1} in the context of LHC study of a bulk, Randall-Sundrum (RS), Kaluza-Klein gluon~\cite{KKG1,KKG2}, where the problem of top collimation was first pointed out in this context (see also~\cite{Lillie:2007ve,Frederix:2007gi} for a more general discussion within and beyond the RS framework).
This marks a serious problem since highly boosted top pair events look very similar to QCD di-jet ones and
the corresponding signal to background ratio is worse than 1:100!

Naively, the problem with collimation appears to be merely an artifact which can be resolved by changing the cone size smaller. However, this is problematic since the smallest hadronic calorimeter cell at both ATLAS and CMS is of $0.1\times 0.1$ size. Thus, moving into a smaller cone would not necessarily help and issues related to finite resolution
are expected to arise. Furthermore, there is a fundamental reason for the architecture of the hadronic calorimeter cells: studies for the LHC experiments, of hadron shower size, show that at least one hadronic cell is required to contain 95\% of the energy of a 100 GeV pion
(see {\it e.g.}~\cite{Amaral:1999ki}). In Fig.~\ref{pionnumber_fig} we show that for a TeV top jet we expect five or more energetic pions, thus decreasing the cone size
much below 0.4 would not help, since the momenta of the top daughter products is expected to be smeared by the hadronic showers in the calorimeter.
In addition to the above challenges, one also expect that for boosted tops the efficiency and rejection power for $b$-tagging would also be degraded (for more details see {\it e.g.}~\cite{btag}).
Below we mostly focus on hadronic decaying tops. However, even the semi-leptonic decay modes pose a challenge,
since a conventional isolation cut between the lepton and the $b$-jet would fail~\cite{KKG1}. Studies of semi-leptonic boosted tops  can be found in~\cite{Barger:2006hm,Baur:2007ck,LHCnotes}.

\begin{figure}[tb]
%\vspace{9pt}
\includegraphics*[width=\columnwidth]{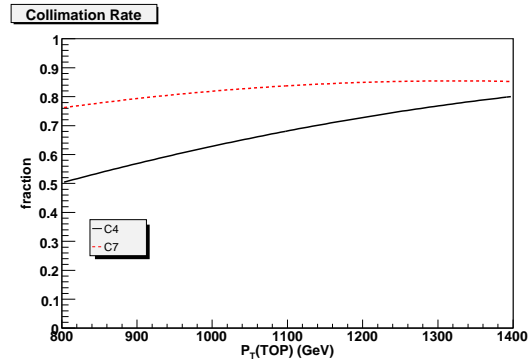}
\caption{The collimation rate for top quarks for 0.4 (black solid curve) and 0.7 (red dashed curve) cone jets.
~\cite{tjets}.}
 \label{coll_rate_fig}
\end{figure}

\begin{figure}[tb]
%\vspace{9pt}
% \includegraphics[width=2.75in]{pions_ave_multp.ps}
\includegraphics*[width=\columnwidth]{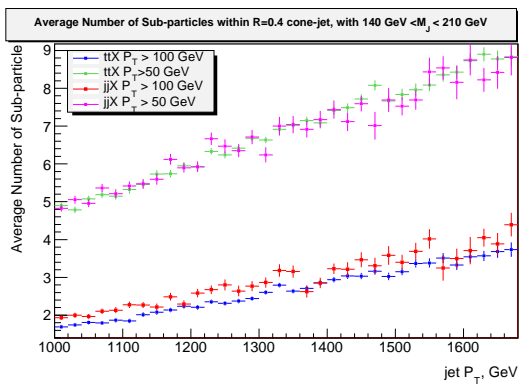}
\caption{Average subparticle multiplicity for a top jet.}
 \label{pionnumber_fig}
\end{figure}

\subsubsection{Jet mass}
%%%%%%%%%%%

Without the ability to conventionally tag top events one is required to look for alternative methods.
Maybe the most direct approach towards distinguishing between the QCD background and the top signal is via jet mass tagging~\cite{Skiba:2007fw,Holdom:2007ap}.
However, the jet mass distributions of both tops and QCD jets are not trivial as we discuss next.
In Fig.~\ref{ttbarmass_fig},
% we see the effects of the detector smearing, using the transfer function, on the $\ttbar$ signal,
we present the top jet mass distributions for a 0.4 cone, with and without detector smearing, for $ p_T^{lead} \ge 1000 \rm\,GeV$.
Due to the finite cone size and gluon radiation, even the top jet mass distribution is far from the naive Breit-Wigner shape.
In cases where the outgoing $b$ quark is outside the cone, we expect the top jet mass to be peaked around the $W$ mass.
In cases where one of the quarks from the $W$ decay is outside the cone we expect
a smooth distribution with masses well below the top one. On the other hand, gluon radiation would make the mass harder (this broadening is not crucial for identifying the top tagging, but very important in order to improve the top mass determination see~\cite{tscet}).
These effects are present even at the particle level, without detector effects.

\begin{figure}[tb]
\includegraphics*[width=\columnwidth]{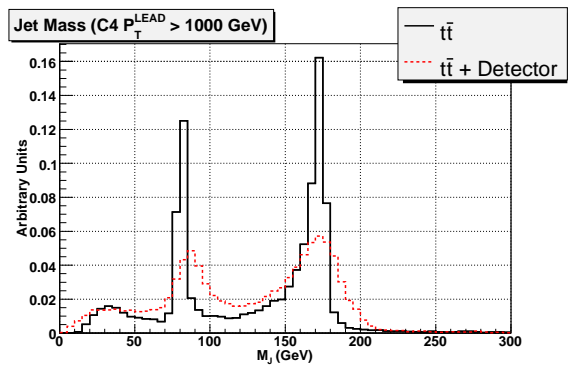}
\caption{Top jet mass distribution $\left( p_T^{lead} \ge 1000 {\rm\,GeV} \right)$
with (the red dotted curve) and without (the black solid curve) leading detector effects~\cite{tjets}.}
 \label{ttbarmass_fig}
\end{figure}

In order to make the mass tagging method viable, characterization of  the dominant QCD jet background is necessary~\cite{Ellis:2007ib}.
The difficulty is that no experimental data of high $p_T$ jets with high masses has been analyzed.
Thus, one needs to be careful when studying these objects only via Monte Carlo (MC) simulation.
In Ref.~\cite{tjets}, a semi-analytic calculation of QCD jet mass distribution was derived based on QCD factorization~\cite{Collins:1989gx,Berger:2003iw},
%based on the expectation that the jet mass is dominated by a single gluon emission.
where the mass is dominantly due to a single gluon emission.
The jet function can be defined via the total differential rate
\begin{equation}
\frac{d \sigma(R)}{d p_T d m_{J} }  =\sum_{q,G} J^{q,G} (m_{J},p_T,R) \, \frac{d\hat{ \sigma}^{q,G}(R)}{d p_T},
\end{equation}
where $\hat{ \sigma}^{q,G}$ is the factorized Born cross section.
Corrections of ${\cal O}\big(R^2\big)$ are neglected and the analysis is applied to the high mass tail, $m_J^{\rm peak}\ll m_J\ll p_T R$ ($m_J^{\rm peak}$ corresponds to the peak of the jet mass distribution). A simple approximation for the full result
is \cite{tjets}
\begin{equation}
J (m_J,p_T,R) \simeq \alpha_s (p_T)  \frac{  4\,  C_{q,G}}{\pi m_J} \log\left(\frac{R\,p_T}{m_J} \right),
\end{equation}
where $\alpha_s(p_T) $ is the strong coupling constant at the appropriate scale and $C_{q,G}=4/3,3$ for quark and gluon jet respectively.
In the absence of real data, the above expression can be only compared with that from the different MC generators.
In Fig.~\ref{jet_mass_theory_vs_data_fig}, $J (m_J,p_T,R)$ is compared with various MC results. The theoretical mass distribution is plotted for 100\% gluon, quark cases which are harder and softer respectively. Hence, it is expected that the MC data,
which consists of admixture of the two would interpolate between the two theoretical curves.
Although it is bothersome that the different generators seem to show non-negligible differences,
we see that roughly all the curves agree with the theoretical predictions.
It is clear from these results that a sizable fraction of high $p_T$ QCD jets is of rather high mass.
Applying  a double mass window cut still yields a problematic signal to background ratio of ${\cal O}(20\%)$.
The situation, however, can be improved by implementing a side band analysis driven by the theoretical jet mass expressions~\cite{tjets}.
A rather detailed study (using transfer function to capture the leading realistic detector smearing~\cite{JoeVirziAtlas},
shown as the red curve of Fig.~\ref{ttbarmass_fig}) presented in~\cite{tjets} showed that detector effects would rather significantly degrade the rejection power.
Inclusion of side band analysis (for the leading jet while the other jet is naively tagged) was shown to improve the situation and yield a rejection power of
${\cal O}(10\%)$.
\begin{figure}
\hspace*{-.5cm}\includegraphics[width=3.in]{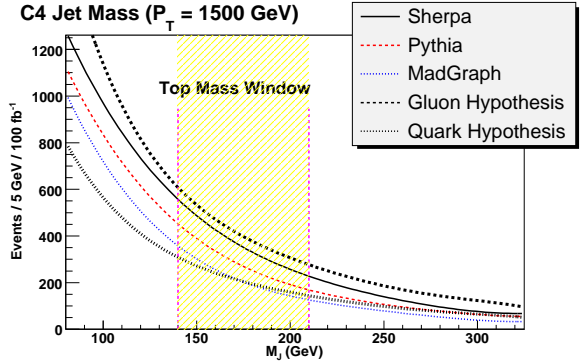}
\caption{The jet mass distributions for Sherpa, Pythia and MG/ME and
the theoretical expression are plotted for QCD jets with $1450 {\rm
\,GeV} \le p_T \le 1550  {\rm \,GeV} $ and $R=0.4$~\cite{tjets}. }
\label{jet_mass_theory_vs_data_fig}
\end{figure}

\subsubsection{Jet shape and substructure}
%%%%%%%%%%%%%%%%%%%%%%%%%%%%%%%%%%%%%%%%%%%%%%%%%%%%%%%%%%%%%%%%%%%%
Top jets and QCD jets fundamentally differ from each other and one should be able to find observables which exploit this
essential difference.
As we have seen, jet mass has a limited, but certainly non-negligible rejection power. Once the jet mass is fixed at a high scale, it is important to note that
a large class of other jet shapes becomes perturbatively calculable.
%perturbative and can be calculated theoretically to leading order.
An interesting way to proceed beyond the jet mass is to look at energy distribution and substructure within the jet itself.
The effort in the literature~(see~\cite{Salam:2009jx} for a recent review)
can be characterized according to two wide classes.
Techniques geared towards two~\cite{Butterworth:2008iy,Butterworth:2002tt,jetshapes}
and three~\cite{Butterworthsparticles,Butterworth:2009qa,tjets,jetshapes,Thaler:2008ju,Kaplan,Krohn:2009wm,LHCnotes} pronged kinematics
%, at leading order in perturbation theory
({\it e.g.} $h\rightarrow b\bar b$ for two-body and $t\rightarrow bq\bar q$ for three-body kinematics).
These two wide classes can be further broken into techniques which are defined via
$1\to2$ splitting and ones defined via energy flow and moments within the jet.

Due to space limitation, we focus on three effective variables to distinguish signal from background~(see~\cite{Salam:2009jx} for a more comprehensive discussion):\\
(i)  $Y$-splitter - analysis based on $k_T$ distance~\cite{KtHH}, $d_{ij} = \min(p_{T_i}^2, p_{T_j}^2) \Delta R_{ij}^2\,.$
The basic observation~\cite{Butterworth:2002tt} is that for jets originated from a decay of a boosted heavy particle,
$d_{ij}$ is of the order of its mass square [$d_{ij}/m_{ij}^2 \sim \min(p_{T_i}, p_{T_j})/ \max(p_{T_i}, p_{T_j})\sim {\cal O}(1)$~\cite{Salam:2009jx}].
This is simply due to the fact that, for a low spin mother particle, the angular
distribution in the rest frame is uniform, so that the daughter particles would roughly have the same momenta in the boosted frame.
On the other hand, due to soft collinear singularities, the QCD background tends to yield an asymmetric momenta distribution
between the mother parton and the showered one. \\

(ii) Angularity - a class of
 jet shapes~\cite{Berger:2003iw,jetshapes},
\begin{eqnarray}
\nonumber
\tau_a(R,p_T) = \frac{2}{m_J} \sum_{i \in jet} \omega_i\, \sin^a \frac{\pi \theta_i}{2R}  \left(\sin\frac{\pi \theta_i}{4R}\right)^{2(1-a)},
\end{eqnarray}
where $\omega_i$ is the energy of a calorimeter cell inside the jet and $a\leq2$ ensures IR safety.
To leading order, the angularity distribution, $d\sigma/d\tau_a$ is similar over a large class of jet definitions (for instance the $k_T$  and anti-$k_T$ variety~\cite{Cacciari:2008gp}) and do no require one to break the jet into subjets~\cite{jetshapes}.
Since angularity basically measures the energy distribution inside the jet, it is particularly sensitive to how symmetric the energy deposition is and
can distinguish jets originated from QCD and boosted heavy particle decay, just as $Y$-splitter can.
As shown in~\cite{jetshapes}, angularity become a rather simple perturbative quantity to evaluate at high masses and, in fact, for two pronged
decay the $Y$-splitter and angularity distributions are in one to one correspondence.
One important point is that inside a fixed high mass window the angularity distribution of the signal and background are similar in shape both
peaked around the symmetric $p_T$ distribution. The difference is only quantitative, the QCD distribution has a broader tail towards larger angularity value (similar conclusion should holds for the $Y$-splitter case). \\
(iii) Planar flow: As mentioned, the above two kinematical variables were used in the two pronged decay and also for the three pronged decay cases.
For signal events, characterized by high mass scale and a three-pronged decay, one can define
another IR-safe jet shape, denoted as planar flow ($Pf$), which can be used to distinguish planar from linear configurations~\cite{jetshapes,tjets}.
The utility of a closely-related observable was emphasized in Ref.~\cite{Thaler:2008ju} (see also~\cite{Krohn:2009wm}).
In Fig.~\ref{PlanarFlow} we show that, given a high mass cut, $Pf$ can help
distinguish QCD jets from top-jets.
QCD jets peak around small values of $Pf$, while the top jet events are more dispersed.
As shown in~Fig.~\ref{PfvsTau_fig}, a $Pf$ cut around 0.4 with a mild angularity cut yield a rejection power of 1:4~\cite{Pfvstau}.
The plot shows that a high mass and $Pf$ cut would yield a similar angularity distribution of signal and background.
The $Y$-splitter distribution was used in~\cite{LHCnotes} in this context, where the expected peaks in both the $W$ and top mass windows were exploited
to reject the QCD signal.

%Denoting a jet shape by $e$, we then have,
%\begin{eqnarray}
%\frac{d\sigma}{dm_J\,de}
%=
%\sum_c \int_{p_{T}^{min}}^\infty dp_{T} \frac{d\hat\sigma_c(p_{T})}{d p_{T}}\ \frac{dJ_c(e,m_J,p_{T},R)}{de}\nonumber,
%\label{diffJ}
%\end{eqnarray}
%where $d\hat\sigma/dp_{T}$ includes the hard scattering and the
%parton distributions of the incoming hadrons, and where
%the jet function for partons $c$
%in the final state is shown in the previous section.

\begin{figure}[tb]
%\vspace{9pt}
\includegraphics*[width=\columnwidth]{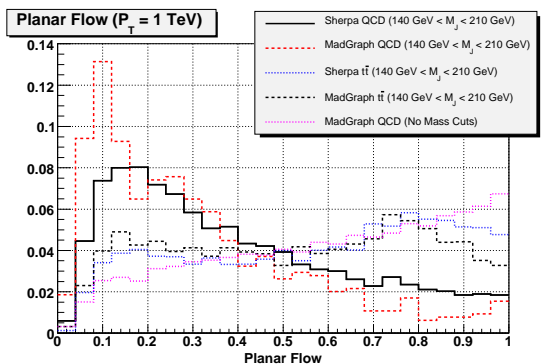}
\caption{The planar flow distribution for QCD and top jets obtained from MadGraph and Sherpa.
Distributions are normalized to the same area \cite{jetshapes}. }
 \label{PlanarFlow}
\end{figure}

\begin{figure}[tb]
%\vspace{9pt}
\includegraphics*[width=\columnwidth]{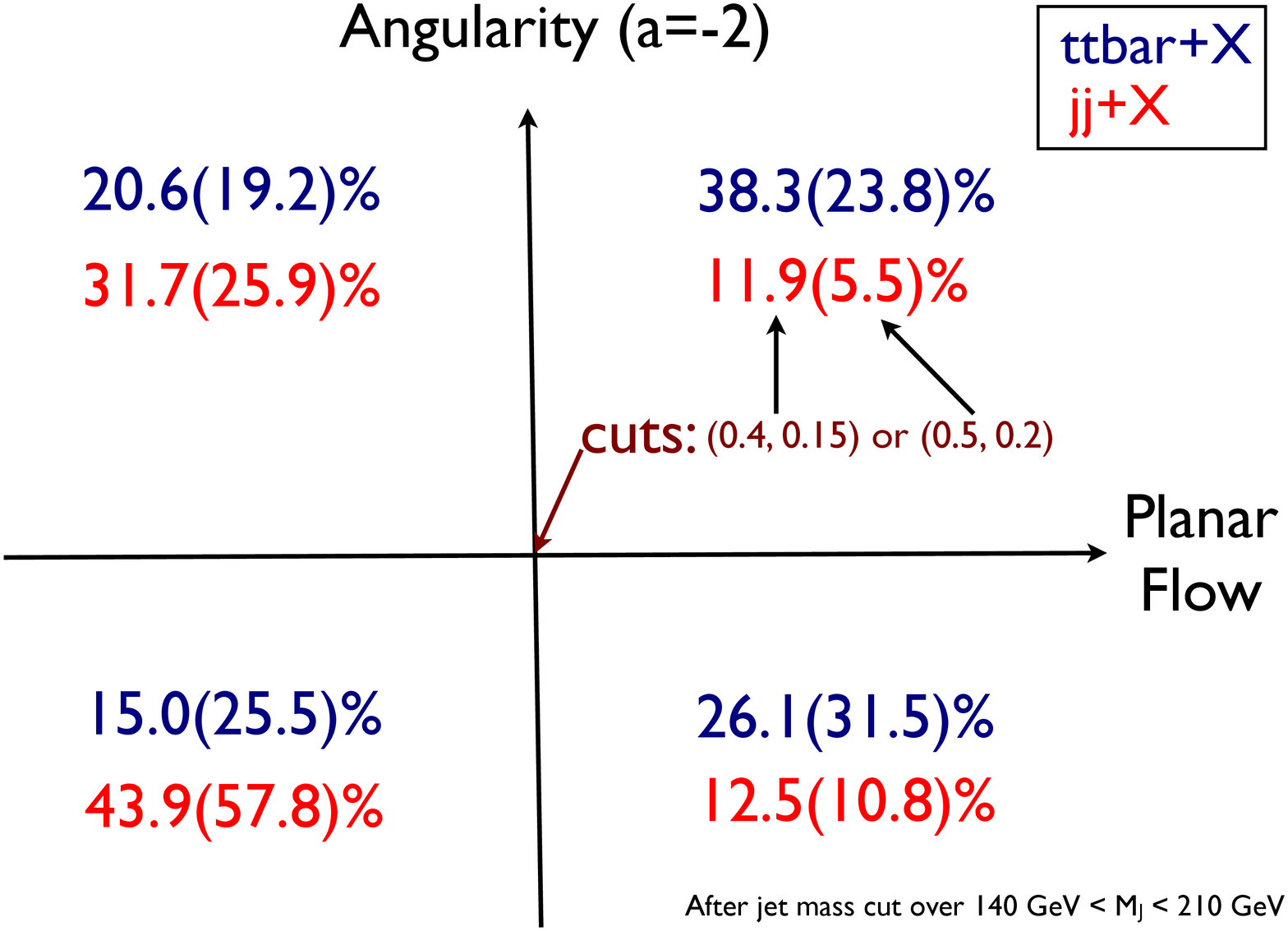}
\caption{The signal and background distributions as a function of two kind of cuts on Planar flow $Pf$  and angularity, after top mass window cut were applied on R=0.4 cone jet \cite{Pfvstau}.}
 \label{PfvsTau_fig}
\end{figure}

% \subsection{Compositeness and boosted tops}
 \subsection{Chiral Coupling to New Particles}
%
%The TeV composite resonances typically decay into very energetic tops $p p \to X \to t \bar{t}$ with
%$p^T_{\rm top} \sim m_X/2$. The decay products of such highly boosted top quarks are collimated, and the top quarks %are difficult to reconstruct using methods designed for Standard Model $t \bar{t}$ production near the threshold. %
%
Variety of techniques, such as using jet substructures,
have been the focus of a number of recent studies \cite{Barger:2006hm,Agashe:2006hk,Fitzpatrick:2007qr,Lillie:2007yh,Skiba:2007fw,Baur:2007ck,Frederix:2007gi,Baur:2008uv,Brooijmans,Thaler:2008ju,Kaplan:2008ie,Almeida:2008yp,Bai:2008sk,Almeida:2008tp,Ellis:2009su,cms-ca}.
% This subject is covered in detail in Section .....  Here, we will mention
%
A recent  study of variables shows the sensitivity to the chirality of the top's coupling to new heavy physics resonances \cite{tjets,Agashe:2006hk,Shelton:2008nq,Krohn:2009wm}. In particular, new variables based on sub-jets are proposed \cite{Krohn:2009wm} which require neither b-tagging nor the reconstruction of the top rest frame,  which is a considerable advantage in the case of boosted top quarks. An earlier study of measuring top polarization by direct reconstruction in stop decay chain can be found in Ref.~\cite{Perelstein:2008zt}.

\section{Top-Rich Events for New Physics}

{\it L.-T.~Wang}\medskip

%%% beginning of L.-T. Wang:   $4 \pm 1$ pages
%\subsection{Motivations}
%There is strong motivation to consider TeV new physics in connection with the top quark because the top quark couples
% strongly to the electroweak symmetry breaking dynamics of the Standard Model. This simple consideration leads to at %least two classes of new physics scenarios.
 %
% First, the heaviness of the top quark leads to the postulation that it is a composite state from new TeV scale strong %dynamics. In this case, the other composite resonances of such a new strong interaction will couple strongly to the top %quark. Therefore, production and decay of such resonances will give rise to very energetic $t \bar{t}$ final states. %Particular realizations of this scenario can be found in, for example, Ref.~\cite{Agashe:2003zs}.
%
%Second, v
%
Virtual effect involving the top quark gives the largest radiative contribution to the Higgs potential. Naturalness arguments, or the insensitivity to ultra-violate physics, demands the existence of ``top" partners with similar gauge quantum numbers as the top quarks.\footnote{Note, however, an important exception to this case in the model of \cite{Chacko:2005pe}.} Well-known examples of top partners include the scalar top in supersymmetry, and fermionic top partner in Little Higgs models \cite{ArkaniHamed:2001nc}. Each top partner will decay into a top quark and additional states, leading to multiple top final-states at the LHC.

The top quark rich new physics signals are exciting, but they can be challenging to identify at the LHC. In the rest of this section, we will outline the general feature of such signals, and summarize some recent progresses on discovering and studying them at the LHC.

 \subsection{Signal of New Top Partners}
% ($4 \pm 1$ pages)

\begin{figure}[tb]
\vspace{9pt}
%\framebox[55mm]{\rule[-21mm]{0mm}{43mm}}
%\includegraphics[scale=0.4]
%\includegraphics*[width=\columnwidth]{t-prod-littlest.eps}
\includegraphics*[width=\columnwidth]{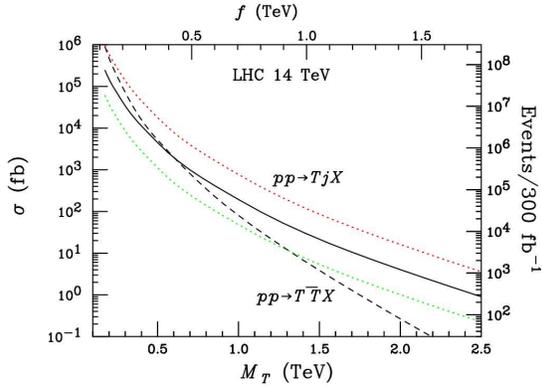}
\caption{The production rate of Little Higgs top partners at the LHC \cite{Han:2003wu}. Both single (solid) and pair productions are included (dash). A vector-like $SU(2)$ singlet $T'$ has been considered in this plot. }
\label{fig:prod-tprime}
\end{figure}

 \begin{figure}[tb]
\vspace{9pt}
%\framebox[55mm]{\rule[-21mm]{0mm}{43mm}}
%\includegraphics[scale=0.3]
\includegraphics*[width=\columnwidth]{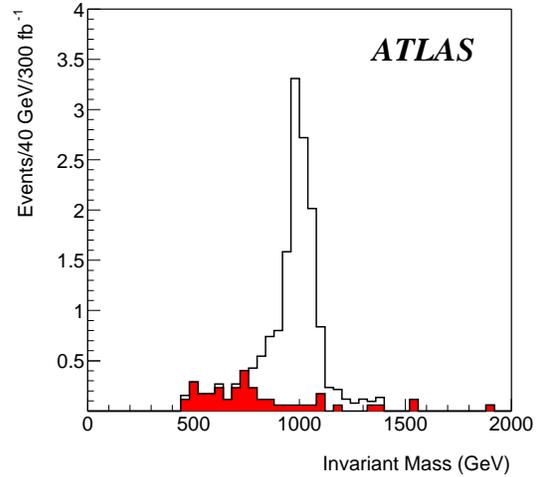}
\caption{The ATLAS study \cite{Azuelos:2004dm} of the reconstruction of the $T'$ in the $tZ$ channel.}
\label{fig:tZ-recon}
\end{figure}

The discovery and study of top partners can be challenging at the LHC. Since top partners typically carry gauge quantum numbers similar to those of the top quark, their production  at the LHC is dominated by analogous QCD processes. The top partner, $T'$, typically decays via $T' \to t + Y$, where $Y$ denotes the additional states. Therefore, the top final states typically have different kinematics in comparison with the Standard Model top quark productions. In the following, we will review several such channels that have been studied recently.

\subsubsection{Single top partner production}

We begin with the case of the top partner in the little Higgs models \cite{ArkaniHamed:2001nc,ArkaniHamed:2002qy}. In this class of models,  the $T'$ can be singly produced through $bW \to T' $ which dominates for large $T'$ mass $m_{T'} \geq 700$ GeV, as shown in Fig.~\ref{fig:prod-tprime} \cite{Han:2003wu}.
Recent NLO cross sections for single $T'$ production can be found Ref.~\cite{Campbell:2009gj}.
Measuring the production in this channel provides a direct probe of the coupling of the top partner to the Higgs field, and it is crucial to understanding its role in the electroweak symmetry breaking \cite{Perelstein:2003wd}. Otherwise, the model independent QCD pair production mode gives the dominant contribution.

If the single $T'$ production channel is allowed, then the Goldstone equivalence theorem dictates the existence of three possible decay channels with fixed branching ratios, BR($T' \to t H$):BR($T' \to tZ$):BR($T'\to bW$) = 1:1:2. The $T'$ reconstruction in these decay channels have been studied \cite{Azuelos:2004dm}. Observation of the $T'$ resonance is possible, although high statistics are necessary $\mathcal{O} (100)$s fb$^{-1}$.  An example of such a reconstruction is shown in Fig.~\ref{fig:tZ-recon}. Reconstruction in the $bW$ and $tH$ channel were also carried out in the same study, but found to be less efficient.

\subsubsection{Pair production}

In various constructions, typically motivated by better consistency with electroweak precision measurements,  it is usually desirable to have the top partner odd under an additional discrete $Z_2$ symmetry, frequently called a new parity. The new physics states in such scenarios also typically include a neutral and stable particle. Well known examples of such new symmetries include T-parity in the Little Higgs models \cite{Cheng:2003ju,Cheng:2004yc,Cheng:2005as}, and KK-parity in the UED model \cite{Appelquist:2000nn}, with the new stable particle denoted as LTP and LKP, respectively. Another well-known top partner is of course the scalar top $\tilde{t}$ in low energy supersymmetry. In this case, a somewhat different motivation (proton decay) leads to the imposition of the R-parity, which predicts the lightest superpartner (LSP) to be stable.

 \begin{figure}[tb]
%\vspace{9pt}
%\framebox[55mm]{\rule[-21mm]{0mm}{43mm}}
%\includegraphics[scale=0.7]
\includegraphics*[width=\columnwidth]{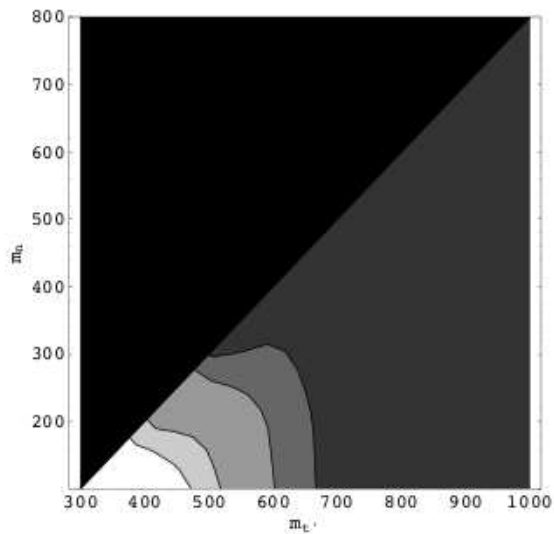}
\caption{The reach of fermionic top partners in the fully hadronic channel \cite{Meade:2006dw} in the parameter region of top partner mass $m_{t'}$ and missing particle mass $m_n$, for an integrated luminosity of 10 fb$^{-1}$.  The contours (from left to right) represent signiÞcance of $> 15\sigma$, $> 10\sigma$, $> 5\sigma$,
$> 3\sigma$, and $< 3\sigma$ }
\label{fig:tprime-reach-had}
\end{figure}

The existence of such a new parity dramatically changes the top partner phenomenology.
First of all, the top partners can only be pair produced. Therefore, the dominant production channel is the QCD process shown in Fig.~\ref{fig:prod-tprime}. Second, the typical decay mode is $T' \to t + $ neutral stable particle. Therefore, the collider signature of the top partner is $t \bar{t} + \not{\!\!E}_T$.

Discovery of top partners in this channel can be challenging.  The existence of additional missing particle implies that there is not enough constraints to  fully reconstruct the top partner kinematics is impossible.  The reconstruction of the top quarks in the final states is of obvious importance since it can help reducing the background and identifying the underlying event topology. However, unlike the Standard Model $t \bar{t}$ production, we can only fully reconstruct top quarks if both top quarks decay hadronically. Discovery in this fully hadronic channel has been studied \cite{Meade:2006dw,Matsumoto:2006ws,Nojiri:2008ir}.  An example of the reach is shown in Fig.~\ref{fig:tprime-reach-had}.

\begin{figure}[tb]
%\vspace{9pt}
%\framebox[55mm]{\rule[-21mm]{0mm}{43mm}}
\includegraphics[scale=0.4]{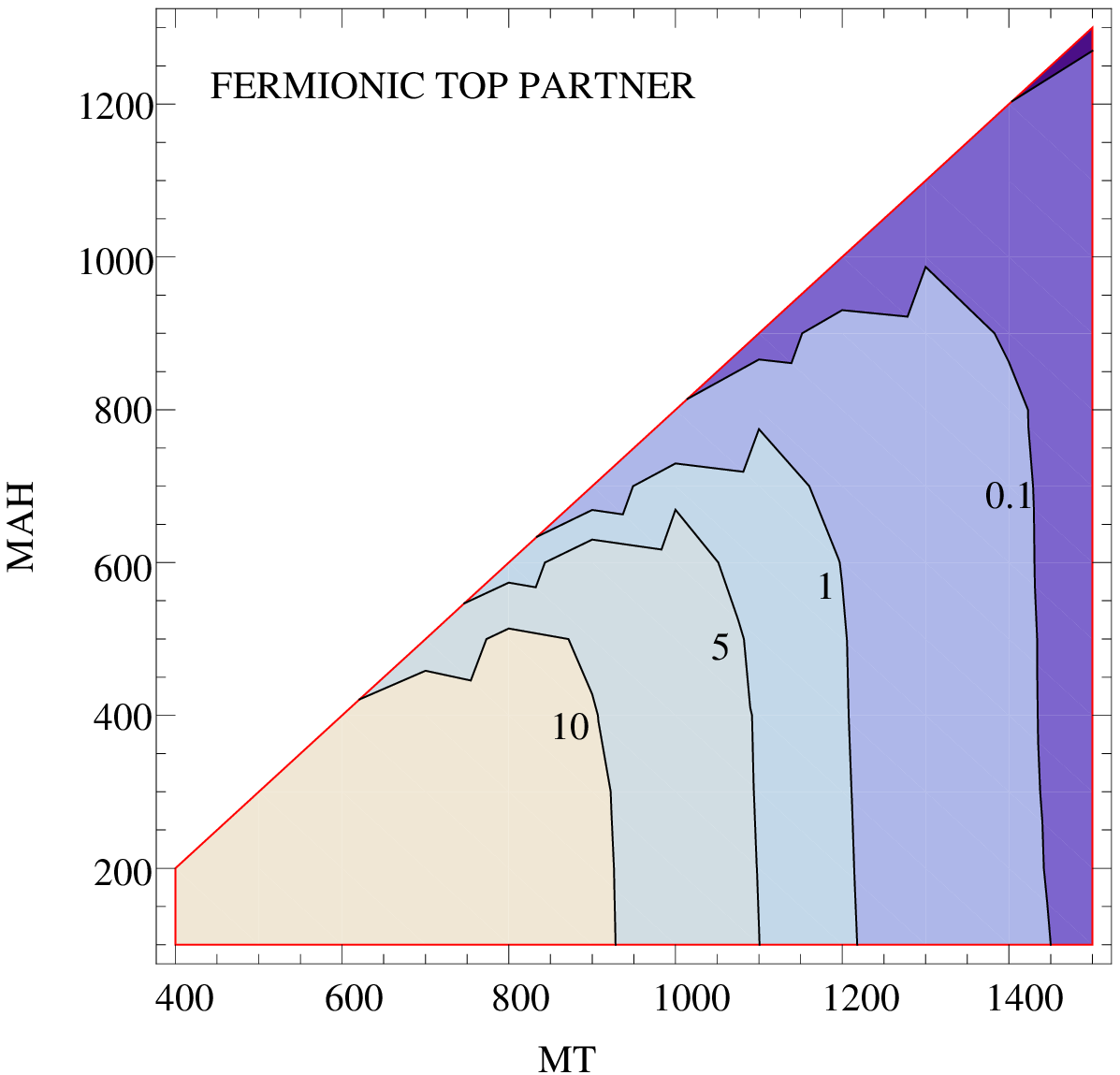}
\includegraphics[scale=0.4]{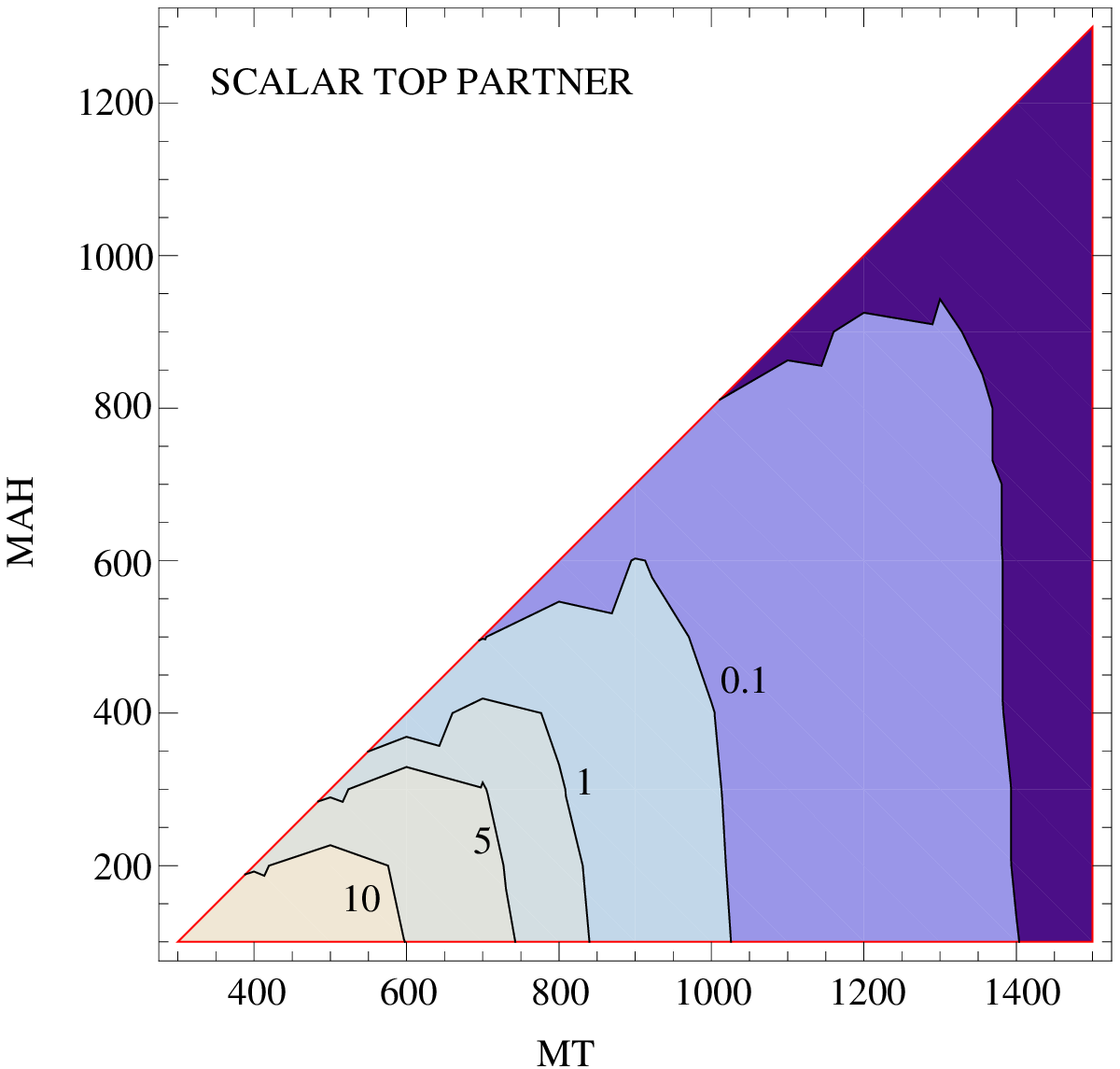}
\caption{The reach of top partners , $T'$ (upper panel), $\tilde{t}_R$ (lower panel), for an integrated luminosity of 100 fb$^{-1}$ \cite{Han:2008gy}. $M_{\rm AH}$ denotes the mass of the missing particle. The contours of several statistical significances (as labelled in the plot) are shown. }
\label{fig:tprime-reach}
\end{figure}

It is also desirable to discover the top partner in the (cleaner) semi-leptonic mode. As commented above, due to the existence of additional missing particles, it no longer possible to reconstruct the top quark directly. As shown in Ref.~\cite{Han:2008gy}, the simple cut on $\not{\!\!E}_T$ is unlikely to be enough as the dominant background comes from the Standard Model $t \bar{t}$ with semi-leptonic decays. However, Ref.~\cite{Han:2008gy} points out that the lack of reconstruction can be used to help us separate signal from background. Both the signal and the semi-leptonic $t \bar{t}$ background give the same final state $b \bar{b} j j \ell+ \not{\!\!E}_T$. We can proceed with reconstruction {\it assuming} they are all from semi-leptonic $t \bar{t}$.  In the background events, we have made the correct assumption and we will reconstruct top up to detector resolutions. On the other hand, reconstruction will {\it fail} for the signal events. Ref.~\cite{Han:2008gy} demonstrated that this can be a powerful discriminant against the Standard Model background.

The discovery reach of $T'$ in the semi-leptonic channel is presented in Fig.~\ref{fig:tprime-reach}, for both fermionic $T'$ and superpanter scalar top $\tilde{t}_R$. The reach for the scalar top is worse due to its smaller production rate, $\sigma_{T' T'} \simeq 8 \sigma_{\tilde{t} \tilde{t}^*}$.  An important parameter in determining the reach is the mass difference between the top partner and the missing particle $\Delta M = M_{T'} - M_{\rm AH}$. As $\Delta M \sim m_{\rm top}$, the kinematics of the top quarks become very similar to the SM QCD $t \bar{t}$ production, and the reach decreases significantly.

Pair production can also be the dominant production mechanism in models with partner quarks
with exotic charges such as $5/3$, for which the single production rate is suppressed by the
negligible top PDF in the proton at LHC energies.  Such
objects decay into $Wt$, leading to like-sign dileptons with low Standard Model backgrounds,
and discovery prospects for masses up to about one TeV with approximately 20~fb$^{-1}$
of integrated luminosity \cite{Dennis:2007tv,Contino:2008hi}.

\subsection
{Multiple Top Production}

\begin{figure}[tb]
\vspace{9pt}
%\framebox[55mm]{\rule[-21mm]{0mm}{43mm}}
\hspace{-1cm}
\includegraphics[scale=0.3]{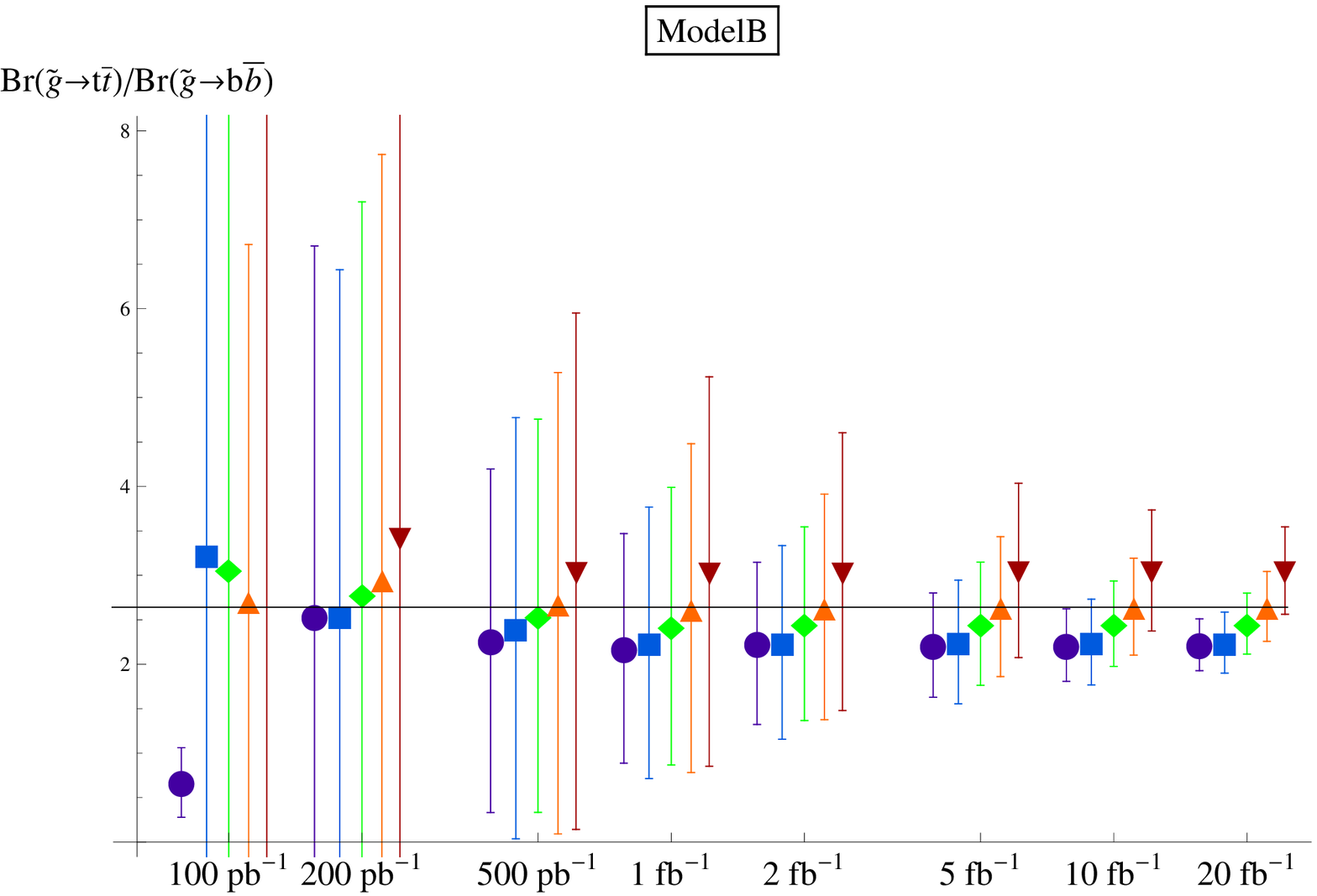}
\includegraphics[scale=0.2]{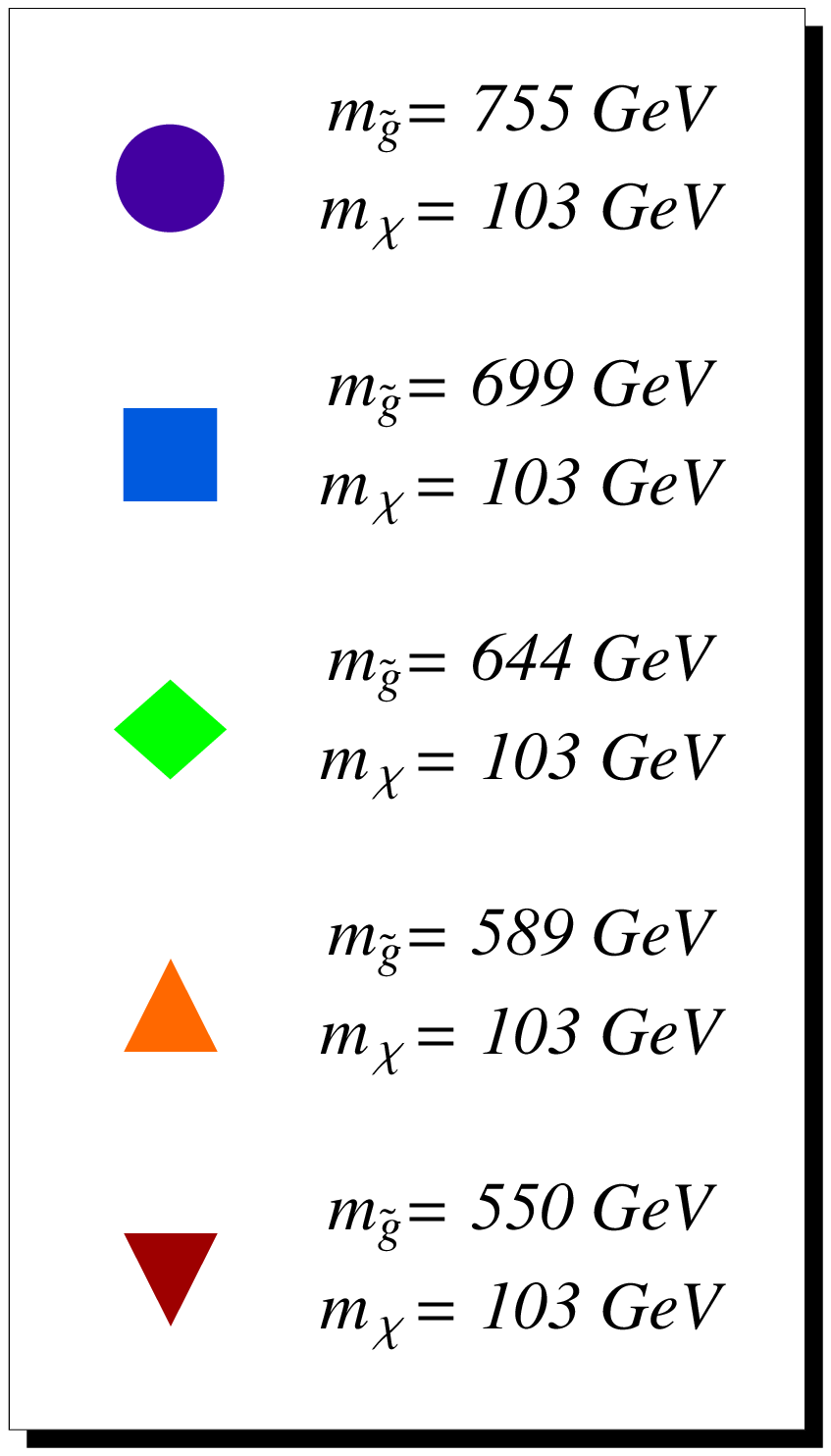}
\caption{The fit of ratio BR($\tilde{g} \to t  \bar{t}$)$/$BR($\tilde{g} \to b \bar{b}$) in a benchmark model, with model parameters $m_{\tilde{g}} = 650$ GeV and bino LSP $m_{\rm LSP}=100$ GeV. The value of this ratio from the underlying model is represented by the solid horizontal line. Different mass hypothesis have been used to demonstrate the robustness of this method. }
\label{fig:gluino-fit}
\end{figure}

New physics final states with more top quarks are possible. A particularly interesting case is gluino cascade decay. In a large class of models, third generation squarks are lighter than the first two due to the RGE evolution effects stemming from their Yukawa couplings. In this case, gluino decay  can be dominated by third generation channels. Depending on the identity of the electroweak-ino in the next stage of the decay chain, the gluino can decay into  $t \bar{t}$, $tb$, and $b \bar{b}$.  Pair production of light gluinos, with large production cross section, can have up to four top quarks in the final state \cite{Baer:1990sc,Hisano:2002xq,Hisano:2003qu,Mercadante:2007zz,Baer:2007ya,Gambino:2005eh,Toharia:2005gm,Acharya:2009gb}. Such bottom rich and lepton rich final states can lead to exciting early discovery at the LHC \cite{Acharya:2009gb}, for example, in the same sign dilepton channel. On the other hand, reconstructing all top quarks in such a busy environment is almost impossible. Therefore, the challenge here is to understanding precise event topology, and distinguish various decay channels. Recently, Ref.~\cite{Acharya:2009gb} demonstrated that such a goal can be achieved through a simple fitting method. First, a set of events templates from various possible decay channels are generated. Then the weight of each channel is obtained by fitting to a set of simple counts of the signal events in various (typically bottom rich and lepton rich) discovery channels. An particular example of such a fit is shown
in Fig.~\ref{fig:gluino-fit}.

Multi-top events are also a generic consequence of models in which the top is composite
\cite{Lillie:2007hd}.  Studies of four top final states in the same-sign dilepton channel indicate
that multi-TeV compositeness scales can be detected with a data sample of order 100~fb$^{-1}$
at the LHC \cite{Kumar:2009vs,Lillie:2007hd,Pomarol:2008bh}.

\section{Summary}

The LHC will be a true top-quark factory. With 80 million top-quark pairs plus
34 million single tops produced
annually at the designed high luminosity, the properties of this particle
will be studied to a great accuracy, such as its large mass, the couplings, and
its polarizations and spin correlations. Theoretical arguments
indicate that it is highly likely that new physics associated with the top quark
at the Terascale will show up at the LHC. This article only
touches upon the surface of the rich top quark physics, and is focused
on possible new physics beyond the SM in the top-quark sector. The layout of
this article has been largely motivated by experimental signatures for the LHC.
Interesting signatures covered here include
\begin{itemize}
\item Rare decays of the top quark to new light states, or to SM particles
via the charged and neutral currents through virtual effects of new physics.
\item Top quark pair production via the decay of a new heavy resonance, resulting
in fully reconstructable kinematics for detailed studies.
\item Top quark pair production via the decay of pairly produced top partners,
usually associated with two other missing particles, making the signal
identification and the property studies challenging.
\item Multiple top quarks, $b$ quarks, and $W^\pm $'s coming from theories
of electroweak symmetry breaking or an extended top-quark sector.
\end{itemize}

The physics associated with top quarks is rich, far-reaching, and exciting. It opens up
golden opportunities for new physics searches, while brings in new challenges as well.
%It should be of high priority in the LHC program for both theorists and experimentalists.

%\end{document}

%\bibitem{Amsler:2008zzb}
%  C.~Amsler {\it et al.}  [Particle Data Group],
%  %``Review of particle physics,''
%  Phys.\ Lett.\  B {\bf 667} (2008) 1.
%  %%CITATION = PHLTA,B667,1;%%

%%\cite{Agashe:2003zs}
%\bibitem{Agashe:2003zs}
%  K.~Agashe, A.~Delgado, M.~J.~May and R.~Sundrum,
%  %``RS1, custodial isospin and precision tests,''
%  JHEP {\bf 0308}, 050 (2003)
%  [arXiv:hep-ph/0308036].
%  %%CITATION = JHEPA,0308,050;%%

%%%%%%%%%%%%%%%%%%%%%%%%%%%%%%%%%%%%%%%%%%%%%%%%%%%%%%%%%%%%%%%%%%%%%%%%%%%%%%%%%%%%%%%%%%%%%%
%%%%%%%%%%%%%%%%%%%%%%%%%%%%%%%%%%%%%%%%%%%%%%%%%%%%%%%%%%%%%%%%%%%%%%%%%%%%%%%%%%%%%%%%%%%%%%
\chapter{\ensuremath{Z'}\ Physics at the LHC}
\setlength{\epigraphrule}{0pt}
\epigraphhead[20]{\epigraph{~~}{\large Paul Langacker (Convener)}}
%
%%%%%%%%%% espcrc2.tex %%%%%%%%%%
%
% $Id: espcrc2.tex,v 1.2 2004/02/24 11:22:11 spepping Exp $
%
%\documentclass[fleqn,twoside]{article}
%\usepackage{latexsym}
%\usepackage{amssymb,amsmath}

% \usepackage{espcrc2}
% Use the option 'headings' if you want running headings
%\usepackage[headings]{espcrc2}

%%%%%%%%%%%%%%%%%%%%%%%%%%%%%%%
%\usepackage{natbib}
%\usepackage{color}

\newcommand{\auth}[1]{(#1)}

\newcommand{\includefigurevh}[4]{\newpage {\centering
   \vspace*{#3}    \hspace*{#4}  \includegraphics*[scale=#1]{#2}}}

\newcommand{\commt}[1]{{\bf [#1]}}

\newcommand{\refl}[1]{(\ref{#1})}
\newcommand{\eeql}[1]{\label{#1}\eeq}
\newcommand{\vp}{\ensuremath{\phi}}
\newcommand{\lag}{\ensuremath{\mathcal{L}}}
\newcommand{\ord}[1]{\ensuremath{\mathcal{O}(#1)}}
%: Standard macros
\newcommand{\gZ}{\ensuremath{g^{\phantom 2}_Z}}

\renewcommand{\real}{\ensuremath{\Re e}}
\renewcommand{\imag}{\ensuremath{\Im m}}
\newcommand{\bbz}{\ensuremath{\beta \beta_{0\nu}}}

\newcommand{\dnn}{\ensuremath{\Delta N_\nu}}
\newcommand{\dms}{\ensuremath{\Delta m^2}}

\newcommand{\ra}{\ensuremath{\rightarrow}}
\newcommand{\Ra}{\ensuremath{\Rightarrow}}

\newcommand{\zpr}{\ensuremath{Z'}}
\newcommand{\mzp}{\ensuremath{M_{Z'}}}

%\newcommand{\upr}{\ensuremath{U(1)'}}
%%\upr conflicts with macro in Feynman diagram package
\newcommand{\uprm}{\ensuremath{U(1)'}}

\newcommand{\delhad}{\ensuremath{\Delta \alpha_{\rm had}^{(5)}(M_Z)}}

\newcommand{\msb}{\ensuremath{\overline{MS}}}
\newcommand{\mt}{\ensuremath{m_t}}
\renewcommand{\mh}{\ensuremath{M_H}}
\newcommand{\mz}{\ensuremath{M_Z}}
\newcommand{\mw}{\ensuremath{M_W}}
\newcommand{\alsz}{\ensuremath{\alpha_s(M^2_Z)}}
\newcommand{\als}{\ensuremath{\alpha_s}}
\newcommand{\suf}{\ensuremath{SU(5)}}
\newcommand{\arrowsub}[1]{\begin{array}[t]{c}
\rightarrow   \vspace{-2mm} \\{#1}\end{array}}
\newcommand{\stacksub}[2]{\ensuremath{\ _{\stackrel{\textstyle
#1}{\scriptstyle #2}}\ }}
\newcommand{\skipblk}[1]{}
\def\bqa{\begin{eqnarray}}
\def\eqa{\end{eqnarray}}
\renewcommand{\ee}{\ensuremath{e^+ e^-}}
\newcommand{\mm}{\ensuremath{\mu^+ \mu^-}}
\newcommand{\ttt}{\ensuremath{\tau^+ \tau^-}}
\newcommand{\douba}[2]{\ensuremath{
\left( \begin{array}{c} #1    \\ #2
        \end{array}\right)}}
\newcommand{\pri}{\ensuremath{^{\prime}}}
\newcommand{\vect}[1]{\ensuremath{
\left( \begin{array}{c} #1    \end{array} \right) }}

\newcommand{\doub}[3]{\ensuremath{
\left( \begin{array}{c} #1    \\ #2  \end{array} \right)_#3}}
%%%%%%%%%%%%%%%%%%%%%%%%%
\newcommand{\ac}{\bf}
\newcommand{\etal}{{\em et al., }}
\newcommand{\cl}{{\em c.l.}}
\newcommand{\cf}{{\em cf.\ }}
\newcommand{\etc}{{\em etc.}}
\renewcommand{\eg}{{\em e.g., }}
\renewcommand{\ie}{{\em i.e., }}
\newcommand{\sthto}{\ensuremath{SU(3) \x SU(2) \x U(1)}}
\newcommand{\stto}{\ensuremath{SU(2)_{L} \x SU(2)_{R} \x U(1)}}
\newcommand{\sto}{\ensuremath{SU(2) \x U(1)}}
\newcommand{\st}{\ensuremath{SU(2)}}
\newcommand{\sth}{\ensuremath{SU(3)}}
\newcommand{\stf}{\ensuremath{SU(5)}}
\newcommand{\x}{\ensuremath{\times}}
\newcommand{\rep}[1]{\ensuremath{\underline{#1}}}
\newcommand{\lfms}{\ensuremath{\Lambda^{(4)}_{\overline{MS}}\ }}
\newcommand{\lms}{\ensuremath{\Lambda_{\overline{MS}}\ }}
\newcommand{\sinn}{\ensuremath{\sin^2\theta_W\,}}
\newcommand{\sinna}{\ensuremath{\sin^2\theta_W}}
\newcommand{\pr}{\ensuremath{^{\prime}}}
\renewcommand{\snu}{\ensuremath{\stackrel{(-)}{\nu}}}
\renewcommand{\beq}{\begin{equation}}
\renewcommand{\eeq}{\end{equation}}
\newcommand{\rt}{\ensuremath{\sqrt{2}}}
\newcommand{\oh}{\ensuremath{\frac{1}{2}}}
\newcommand{\snh}{\ensuremath{\sin^2\hat{\theta}_W}}
\newcommand{\sih}{\ensuremath{\sin^2\hat{\theta}_W(M^2_W)\ }}
\newcommand{\siz}{\ensuremath{\sin^2\hat{\theta}_W(M^2_Z)\ }}
\newcommand{\RA}{\ensuremath{\rightarrow}}
\newcommand{\p}[1]{\ensuremath{10^{#1}}}
\newcommand{\pp}[1]{\ensuremath{10^{#1}}}
\newcommand{\degree}{\ensuremath{^{\circ}}}

%%%%%%%%%%%%%%%%%%%%%%%%%%%%%%%%%%%%%%%%%

%\usepackage{graphicx}
%\title{\zpr\ Physics at the LHC}

%\author{Paul Langacker\address{Institute for Advanced Study
%,Princeton, NJ 08540}\thanks{This work was supported by  NSF PHY-0503584 and by the IBM Einstein Fellowship.}}

%\usepackage{ifthen}
%\newboolean{hyperlinks}
%\setboolean{hyperlinks}{false}
%\setboolean{hyperlinks}{true}
%\ifthenelse{\boolean{hyperlinks}}{\usepackage{hyperref}\bibliographystyle{h2-elsevier}
%%\ifthenelse{\boolean{hyperlinks}}{\usepackage{hyperref}\bibliographystyle{href-elsevier}
%}{\newcommand{\href}[2]{#2}
%\newcommand{\url}[1]{}\bibliographystyle{h2-elsevier} }
%\begin{document}

%\begin{abstract}
%The existing limits on \zpr\ gauge bosons and prospects for discovery and diagnostic studies at the LHC are briefly reviewed.\vspace{1pc}
%\end{abstract}

%\maketitle

\section{Introduction}\label{ints}
Additional \zpr\ gauge bosons  occur frequently in extensions of the
standard model (SM) or its minimal supersymmetric extension (MSSM), usually emerging as
an unbroken ``remnant''  of a larger gauge symmetry. Examples include superstring constructions,
grand unified theories, extended electroweak groups, or alternatives to the minimal Higgs mechanism
for electroweak breaking. Kaluza-Klein excitations  of the SM gauge bosons also occur in models involving large and/or warped extra dimensions provided the gauge bosons are free to propagate in the bulk, with
$M\sim R^{-1}\sim 2 {\ \rm TeV }\x (10^{-17} {\rm  cm }/R)$ in the large dimension case.
The new \zpr s may occur at any mass scale, but here we concentrate on  the
TeV scale relevant to the LHC, which is especially motivated by
supersymmetric \uprm\ models (in which both the electroweak and \uprm\ breaking scales are usually
set by the soft supersymmetry breaking parameters) and by alternative models of electroweak symmetry breaking. We first briefly review the formalism and the existing constraints from precision electroweak
(weak neutral current, $Z$-pole, LEP 2, and FCNC) measurements and from direct searches at the Tevatron, and then comment on the prospects for a \zpr\ discovery, diagnostics of its
couplings, and related issues such as the associated extended Higgs and neutralino sectors at the LHC. Much more extensive discussions of specific models and other implications, along with a more
complete set of references, are given in several reviews~\cite{Langacker:2008yv,Rizzo:2006nw,Leike:1998wr}. Other recent developments, especially the possibility of a \zpr\ as a ``portal''
to a quasi-hidden sector, such as may be associated with dark matter or supersymmetry breaking,
were reviewed in~\cite{Langacker:2009im,Goodsell:2009xc}.

\section{Formalism}\label{formalism}
The interactions of the photon ($A$), $Z$ (i.e., $Z^0_1$) and other flavor-diagonal neutral gauge bosons
with fermions  is
\beq -\lag_{NC}= \underbrace{eJ^\mu_{em} A_\mu+g_1 J^\mu_1 Z^0_{1 \mu}}_{SM}+ \sum_{\alpha=2}^{n+1} g_\alpha J^\mu_\alpha Z^0_{\alpha \mu}, \eeql{f1}
where $g_\alpha$ are the gauge couplings  (with $g_1=g/\cos\theta_W$), and the currents are
\beq J^\mu_\alpha= \sum_i \bar f_i \gamma^\mu[\epsilon_L^{{\alpha}}(i)P_L+\epsilon_R^{{\alpha}}(i)P_R] f_i. \eeql{f2}
 $\epsilon_{L,R}^{{\alpha}}(i)$ are the $U(1)_\alpha$  charges of the left- and right-handed components of  fermion $f_i$, and the theory is chiral for $\epsilon_{L}^{{\alpha}}(i)\ne \epsilon_{R}^{{\alpha}}(i)$.
We also define the vector and axial couplings
 \beq g_{V,A}^{{\alpha}}(i) =\epsilon_L^{{\alpha}}(i)\pm\epsilon_R^{{\alpha}}(i). \eeql{f1a}
 It is often convenient to instead
 specify the charges $Q_\alpha$ for  the left-chiral
fermion $f_L$ {and}  and left-chiral antifermion  $f^c_L$,
\beq Q_{\alpha f}=\epsilon_L^{{\alpha}}(f) \quad  \qquad Q_{\alpha f^c}=-\epsilon_R^{{\alpha}}(f). \eeql{f3}
For example, the SM charges for the $u_L$ and $u^c_L$ are
$Q_{1u}=\oh - \frac{2}{3} \sinn$ and
$Q_{1u^c}=+ \frac{2}{3} \sinn$.
One can similarly define the $U(1)_\alpha$  charge of the scalar field $\vp$  as $Q_{\alpha\vp}$.

For a single extra \zpr, the $Z-\zpr$ mass matrix after symmetry breaking is
\beq M^2_{Z-Z'} =   \left(
\begin{array}{cc}
 M_{Z^0}^2 & \Delta^2\vspace*{2pt} \\  \Delta^2 & M_{Z'}^2
\end{array} \right). \eeql{f4}
If, for example, the symmetry breaking is due to an \st-singlet $S$ and two doublets
$\phi_u= \douba{\phi^0_u}{\phi^-_u}, \  \phi_d= \douba{\phi^+_d}{\phi^0_d}$,
then
\beq
\begin{split}
M_{Z^0}^2 =&  \frac{1}{4} g_1^2 \bigl(|\nu_u|^2+|\nu_d|^2\bigr)  \\
 \Delta^2 = & \oh g_1 g_2 \bigl(Q_u |\nu_u|^2-Q_d |\nu_d|^2 \bigr) \\
M_{Z'}^2 =& g_2^2 \bigl( Q_u^2 |\nu_u|^2 + Q_d^2|\nu_d|^2+Q_S^2 |s|^2 \bigr),
\end{split}
\eeql{f5}
where
\beq \begin{split}
 \nu_{u,d} &\equiv \sqrt{2} \langle  \phi^0_{u,d} \rangle, \qquad s=\sqrt{2}  \langle S \rangle\\
\nu^2&=|\nu_u|^2+|\nu_d|^2\sim (246 {\rm\ GeV})^2 .
\end{split}
\eeql{f6}
The physical mass eigenvalues are $M_{1,2}^2$, the physical gauge particles are $Z_{1,2}$, and the  mixing angle
 $\theta_{Z\zpr}$ is given by
 $ \tan^2\theta_{Z\zpr} ={(M_{Z^0}^2-M_1^2)}/{(M_2^2-M_{Z^0}^2)}$.
In the important special case  $M_{Z'} \gg (M_{Z^0}, |\Delta|)$ one finds
\beq \begin{split}
M_1^2 & \sim M_{Z^0}^2 - \frac{\Delta^4}{M_{Z'}^2}\ll M_2^2, \qquad M_2^2 \sim M_{Z'}^2 \\
\theta_{Z\zpr} & \sim -\frac{\Delta^2}{M_{Z'}^2}\sim C \frac{g_2}{g_1} \frac{M_1^2}{M_2^2}
\\ C& \equiv
2\left[  \frac{Q_u |\nu_u|^2-Q_d |\nu_d|^2}{|\nu_u|^2+|\nu_d|^2} \right].
\end{split} \eeql{f8}

A \uprm\ can yield a natural solution to the supersymmetric $\mu$ problem~\cite{Kim:1983dt}
(unless the charges are obtained from $B-L$ and $Y$), by forbidding
an elementary $\mu$ term but allowing the superpotential term $W \sim \lambda_S S H_u H_d$, where $S$ is a
SM singlet charged under the \uprm. This induces an effective $\mu$ parameter $ \mu_{eff} = \lambda_S \langle S \rangle$,
which is usually of the same scale as the soft supersymmetry breaking parameters~\cite{Suematsu:1994qm,Cvetic:1995rj,Cvetic:1997ky}. This mechanism is similar to the NMSSM (e.g.,~\cite{Accomando:2006ga,Maniatis:2009re}), but is automatically free of induced tadpole and domain wall problems.

We have so far implicitly assumed canonical kinetic energy terms for the $U(1)$ gauge bosons. However,  $U(1)$ gauge invariance allows a more general kinetic mixing~\cite{Holdom:1985ag},
\beq\begin{split}
 \lag_{kin} \ra & -\frac{1}{4} F^{0\mu\nu}_1 F^0_{1\mu\nu}  -\frac{1}{4} F^{0\mu\nu}_2 F^0_{2\mu\nu}
\\  &\qquad -\frac {\sin \chi}{2} F^{0\mu\nu}_1 F^0_{2\mu\nu}
\end{split}\eeql{f9}
for $U(1)_1\x U(1)_2$.
Such terms are usually absent initially, but a (usually small) $\chi$ may be induced by loops, e.g., from nondegenerate heavy particles, in running couplings
if heavy particles decouple, or at the string level. The kinetic terms may be put in canonical form by the non-unitary transformation
\beq  \douba{Z^0_{1\mu}}{Z^0_{2 \mu}}=
\left(
\begin{array}{cc} 1 &   -\tan \chi   \\ 0 &   1/\cos \chi\end{array}\right)
\douba{\hat Z^0_{1\mu}}{\hat Z^0_{2\mu}},
\eeql{f10}
where  the $\hat Z^0_{\alpha}$  may still undergo ordinary mass mixing, as in \refl{f4}.
The kinetic mixing has a negligible effect on masses for  $|M_{Z_1}^2| \ll |M_{Z_2}^2|$
and $|\chi|\ll 1$, but the current coupling to the heavier boson is shifted,
\beq  -\lag \ra  g_1 J^\mu_1\hat Z_{1 \mu}+  (g_2 J^\mu_2 -g_1 \chi J^\mu_1)\hat Z_{2 \mu} .
\eeql{f11}

The \zpr\ mass and mixing may also be generated by the St\"uckelberg
mechanism~\cite{Stueckelberg:1900zz,Kors:2004dx,Feldman:2006ce,Nath:2008ch}.

\section{Existing Limits}\label{existing}
\zpr s with electroweak coupling are mainly  constrained by precision electroweak data,
direct searches at the Tevatron, and searches for flavor changing neutral currents (FCNC).
 Low energy weak neutral current
 processes, which are still very important,  would be affected by  $Z_2$ exchange and by $Z-\zpr$ mixing~\cite{Erler:2009jh,Durkin:1985ev,Amaldi:1987fu,Costa:1987qp,Langacker:1991pg,Amsler:2008zzb}.
The effective four-Fermi WNC interaction becomes
\beq
 -\lag_{eff}=\frac{4G_F}{\sqrt{2}} (\rho_{eff} J^2_1+2wJ_1J_2+yJ^2_2), \eeql{a1}
where
\beq \begin{split} \rho_{eff}=&\rho_1 \cos^2 \theta_{Z\zpr} + \rho_2 \sin^2 \theta_{Z\zpr} \\
w=&\frac{g_2}{g_1}\cos\theta_{Z\zpr}\sin\theta_{Z\zpr} (\rho_1-\rho_2) \\
y =&\left(  \frac{g_2}{g_1} \right)^2 (\rho_1 \sin^2 \theta_{Z\zpr} + \rho_2 \cos^2 \theta_{Z\zpr}),
\end{split}
\eeql{a2}
with
\beq
\rho_\alpha\equiv M_W^2/(M^2_\alpha \cos^2 \theta_W).
\eeql{a2a}
 The $Z$-pole experiments at LEP and  SLC~\cite{lep:2005ema} are extremely sensitive to $Z-\zpr$ mixing, which
  shifts $M_1$ downward from the SM expectation and also affects the $Z_1$ vector and axial
 vertices, which become
\beq \begin{split}
V_i =&\cos \theta_{Z\zpr}  g^1_V(i) + \frac{g_2}{g_1} \sin \theta_{Z\zpr}   g^2_V(i) \\
A_i =&\cos \theta_{Z\zpr}  g^1_A(i) + \frac{g_2}{g_1} \sin \theta_{Z\zpr}   g^2_A(i) .
\end{split}
\eeql{a3}
However, the $Z$-pole experiments have little sensitivity
 to $Z_2$ exchange. At LEP2~\cite{Alcaraz:2006mx} virtual $Z_2$ exchange leads to a four-fermi operator, analogous to the
 $\rho_2$ part of $\lag_{eff}$ in \refl{a1}, which interferes with the $\gamma$ and $Z$.

The CDF~\cite{Aaltonen:2008ah,Aaltonen:2008vx} and D\O~\cite{do:2009} collaborations at the Tevatron have
searched for Drell-Yan resonances, especially  $ \bar p p \ra e^+ e^-, \mu^+ \mu^-$~\cite{Robinett:1981yz},
as illustrated in Figure \ref{CDFmuons}.
 In the narrow width approximation,
the tree-level rapidity distribution for $AB\ra Z_\alpha$ is
\beq \begin{split}
\frac{d\sigma_{\zpr}}{dy}=& \frac{4\pi^2 x_1x_2}{3M_\alpha^3} \sum_{i} \bigl(f_{q_i}^A(x_1)f_{\bar q_i}^B(x_2)\bigr. \\
&+\bigl. f_{\bar q_i}^A(x_1)f_{q_i}^B(x_2)\bigr)
\Gamma (Z_\alpha \ra q_i \bar q_i),
\end{split}
\eeql{a4}
where the $f$'s are the parton distribution functions, the partial widths are
\beq \Gamma (Z_\alpha \ra f_i \bar f_i)= \frac{g_\alpha^2 C_{f_i} M_\alpha}{24\pi}
\bigl( \epsilon_L^{{\alpha}}(i)^2+\epsilon_R^{{\alpha}}(i)^2 \bigr),
\eeql{a5}
$x_{1,2}=(M_\alpha/\sqrt{s}) e^{\pm y}$, and $ C_{f_i}$ is the color factor.
More detailed estimates for the Tevatron and LHC are given in
\cite{Leike:1998wr,Amsler:2008zzb,Weiglein:2004hn,Dittmar:2003ir,Carena:2004xs,Kang:2004bz,Fuks:2007gk,Petriello:2008zr,Baur:2001ze,Papaefstathiou:2009sr,Diener:2009vq},
including discussions of parton distribution functions, higher order QCD and electroweak effects, fermion mass corrections, decays into bosons or Majorana fermions,  width effects,
resolutions, and backgrounds.
%: figure  cdf limits
\begin{figure}[htb]
\includegraphics*[angle=270,width=3.25in]{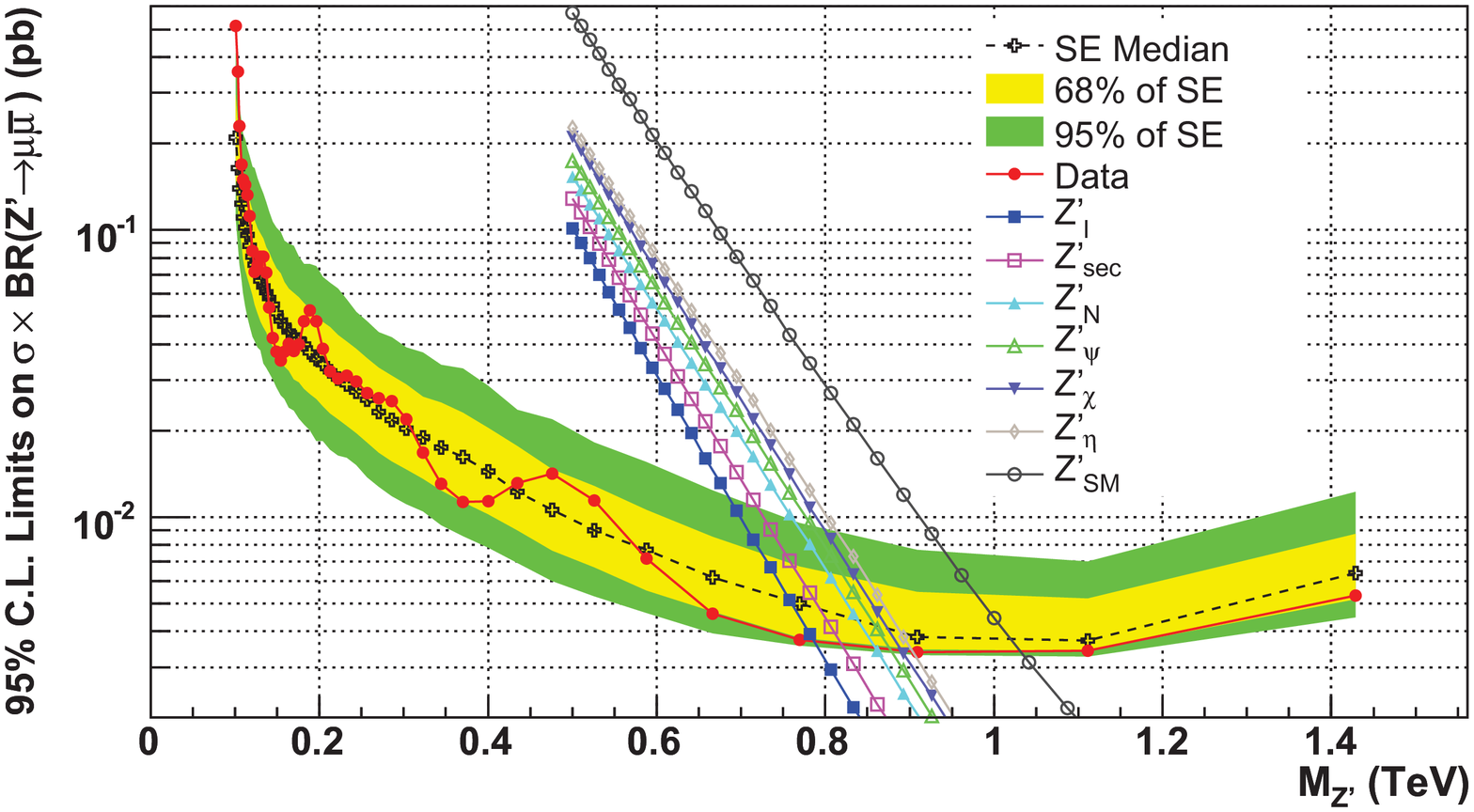}
\caption{CDF limits on various \zpr\ models from the dimuon channel, from \cite{Aaltonen:2008ah}.}
\label{CDFmuons}
\end{figure}

Other search  channels relevant to hadron colliders include $\zpr \ra e^\pm\mu^\mp$~\cite{Abulencia:2006xm}; $\tau^+\tau^-$~\cite{Acosta:2005ij}; $jj$, where $j=$ jet~\cite{Weiglein:2004hn,Aaltonen:2008dn}; $b\bar b$; and  $t\bar t$~\cite{Han:2004zh,cdf:2007dz,Baur:2008uv}.
Another important probe is
the forward-backward asymmetry for  $p p (\bar p p) \ra\ell^+\ell^-$
(as a function of rapidity, $y$, for $pp$) due to $\gamma,Z,\zpr$ interference below the \zpr\
peak~\cite{Langacker:1984dc,Rosner:1995ft,Dittmar:1996my,Abulencia:2006iv}.

All of these existing limits are listed for a variety of models in
Table \ref{table:1zp} and the allowed regions in mass and mixing are
displayed in Figure \ref{chipsi} for two examples
 in the often studied \zpr\ models ~\cite{Langacker:1984dc,Hewett:1988xc,Langacker:2008yv} based on the
 $E_6$ decomposition  $E_6\ra SO(10)\x U(1)_\psi \ra SU(5)\x U(1)_\chi \x U(1)_\psi$.
%: figure  chi psi limits
\begin{figure*}[htb!]
\begin{minipage}[t]{5cm}
\includegraphics*[angle=0,scale=0.36]{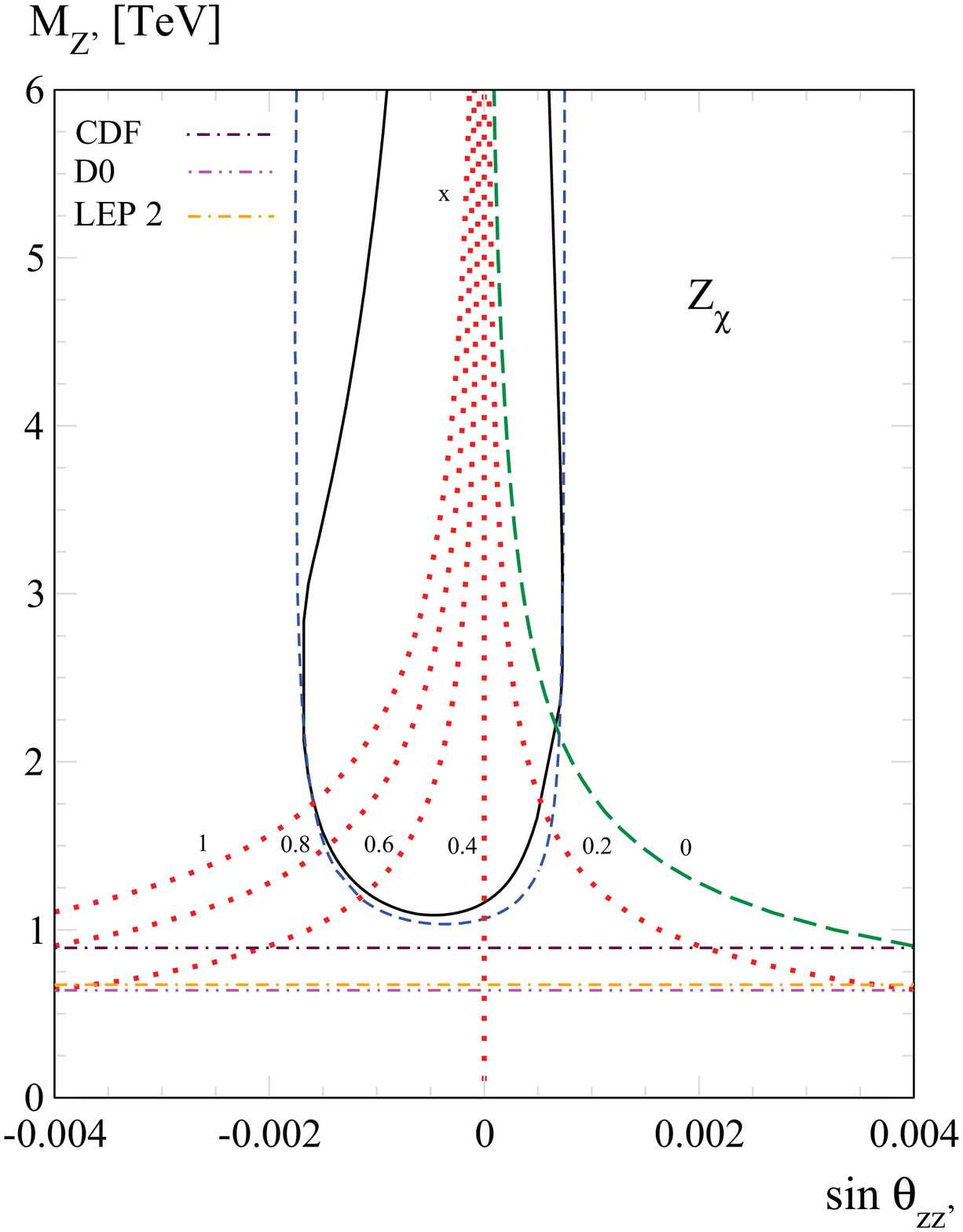}
\end{minipage}
\hspace*{3.0cm}
\begin{minipage}[t]{5cm}
\includegraphics*[angle=0,scale=0.36]{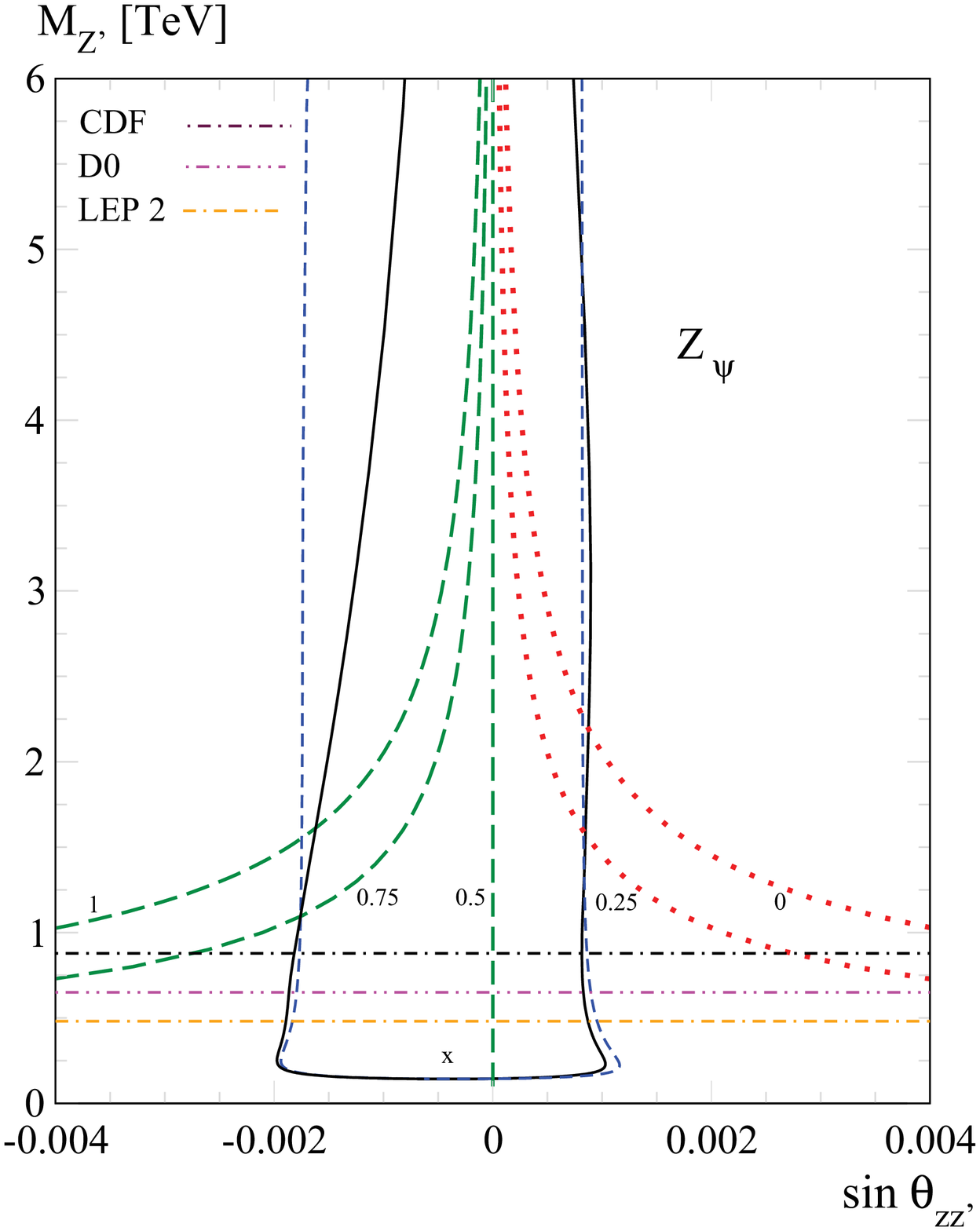}
\end{minipage}
\caption{Experimental constraints on the mass and mixing angle for the $Z_\chi$
 and  $Z_\psi$,
 from~\cite{Erler:2009jh}.  The solid lines show the regions allowed by precision electroweak data at 95\% C.L. assuming Higgs doublets and singlets, while the dashed regions allow arbitrary Higgs. The labeled curves assume specific
ratios of Higgs doublet VEVs.}
\label{chipsi}
\end{figure*}

There are also significant constraints on \zpr s with family nonuniversal couplings, which lead to
FCNC when fermion mixing is turned on. Such nonuniversal couplings often occur in string constructions,
or for Kaluza-Klein excitations in extra-dimensional models. Constraints from $K$ decays and mixing, and
from $\mu$ decays and interactions, are usually sufficiently stringent to exclude such effects for the first
two families for a TeV \zpr\ with electroweak couplings~\cite{Langacker:2000ju}. However, the third family could be nonuniversal,
and \zpr-mediated effects could account for possible anomalies in the $B$ system~\cite{Barger:2009qs,Barger:2009eq,He:2006bk,Cheung:2006tm,Baek:2008vr,Mohanta:2008ce,Chang:2009wt,susyfcnc:2009}.

There has recently been considerable discussion of a possible
light \zpr\ in the MeV-GeV range (referred to as a $U$-boson~\cite{Fayet:2006sp,Fayet:2007ua})
which only couples to ordinary matter through  kinetic mixing with the photon.
Such a particle, which is motivated by dark matter considerations~\cite{ArkaniHamed:2008qn}, could have implications for or is constrained by, e.g.,
$g_\mu-2$, $e^+e^-\ra U\gamma\ra e^+e^-\gamma$,  the HyperCP events,
and a variety of laboratory and collider experiments~\cite{Borodatchenkova:2005ct,Pospelov:2008zw,Gopalakrishna:2009yz,Baumgart:2009tn,Reece:2009un,Morrissey:2009ur,Essig:2009nc,Batell:2009yf,Bjorken:2009mm,Batell:2009di,Schuster:2009au,ArkaniHamed:2008qp,Cheung:2009su}.

%: table
\newlength{\tspace}
\setlength{\tspace}{3mm}
\begin{table*}[tb!]
\caption{95\% C.L. limits on \mzp\ and central values and 95\% C.L.
upper and lower limits on $\sin \theta_{Z\zpr}$ for a variety of
models. The results are updated from~\cite{Erler:2009jh}, where the
models are defined.} \label{table:1zp}
%\centering
\begin{tabular}{|c||r|r|r|r||r|r|r||c|}
\hline
 $Z'$ & \multicolumn{4}{c||}{$M_{Z'}$ [GeV]} & \multicolumn{3}{c||}{$\sin\theta_{Z\zpr}$}  & $\chi^2_{\rm min}$ \\ \hline
& electroweak& CDF & D\O\ & LEP~2 & $\sin\theta_{Z\zpr}$ & $\sin\theta_{Z\zpr}^{\rm min}$ & $\sin\theta_{Z\zpr}^{\rm max}$ & \\
\hline\hline
$Z_\chi$         & \hspace{\tspace}1,141 &    892 & \hspace{\tspace} 800 &    673 & \hspace{\tspace} $-0.0004$ & \hspace{\tspace} $-0.0016$ & \hspace{\tspace} 0.0006 & 47.3 \\ \hline
$Z_\psi$         &    147 &    878 & 763 &    481 & $-0.0005$ & $-0.0018$ & 0.0009 & 46.5 \\ \hline
$Z_\eta$         &    427 &    982 & 810 &    434 & $-0.0015$ & $-0.0047$ & 0.0021 & 47.7 \\ \hline
$Z_I$              & 1,204 &    789 & 692 &            & $ 0.0003$ & $-0.0005$ & 0.0012 & 47.4 \\ \hline
$Z_S$            & 1,257 &    821 & 719 &            & $-0.0003$ & $-0.0013 $& 0.0005 & 47.3 \\ \hline
$Z_N$            &    623 &    861 & 744 &            & $-0.0004$ & $-0.0015$ & 0.0007 & 47.4 \\ \hline
$Z_R$            &    442 &            &         &            & $-0.0003$ & $-0.0015$ & 0.0009 & 46.1 \\ \hline
$Z_{LR}$       &    998 &    630 &         &    804 & $-0.0004$ & $-0.0013$ & 0.0006 & 47.3 \\ \hline
$Z_{\not{L}}$ &   (803) & (740) \hspace{-8pt} & & & $-0.0015$ & $-0.0094$ & 0.0081 & 47.7 \\ \hline
$Z_{SM}$      & 1,403 & \hspace{\tspace} 1,030 & 950 & \hspace{\tspace} 1,787 & $-0.0008$ & $-0.0026$ & 0.0006 & 47.2 \\ \hline
$Z_{string}$  & 1,362 &            &         &            & $ 0.0002$ & $-0.0005$ & 0.0009 & 47.7 \\ \hline\hline
SM                  & \multicolumn{4}{c||}{$\infty$}                               & \multicolumn{3}{c||}{0}                    & 48.5 \\ \hline
\end{tabular}
\end{table*}

\section{The LHC}\label{lhcpotential}
\subsection{Discovery}
The LHC should ultimately have a discovery reach for \zpr s with electroweak-strength couplings to
$u,d,e,$ and $\mu$  up to $\mzp\sim 4-5$ TeV~\cite{Weiglein:2004hn,Dittmar:2003ir,Kang:2004bz,Diener:2009vq}. This is based on decays into $\ell^+\ell^-$ where $\ell=e$ or $\mu$, and  assumes $\sqrt{s}= 14$ TeV and $ \mathcal{L}_I= \int \mathcal{L} dt =100$ fb$^{-1}$.
The reach for a number of models is shown for  various energies and integrated luminosities
in Figure \ref{discreach}. A recent detailed study emphasized the \zpr\ discovery potential in early LHC running at
 lower energy and luminosity for couplings
 to $B-L$ and $Y$~\cite{Salvioni:2009mt}.

%: figure  lhc reach
\begin{figure}[htb!]
\includegraphics*[angle=0,scale=0.4]{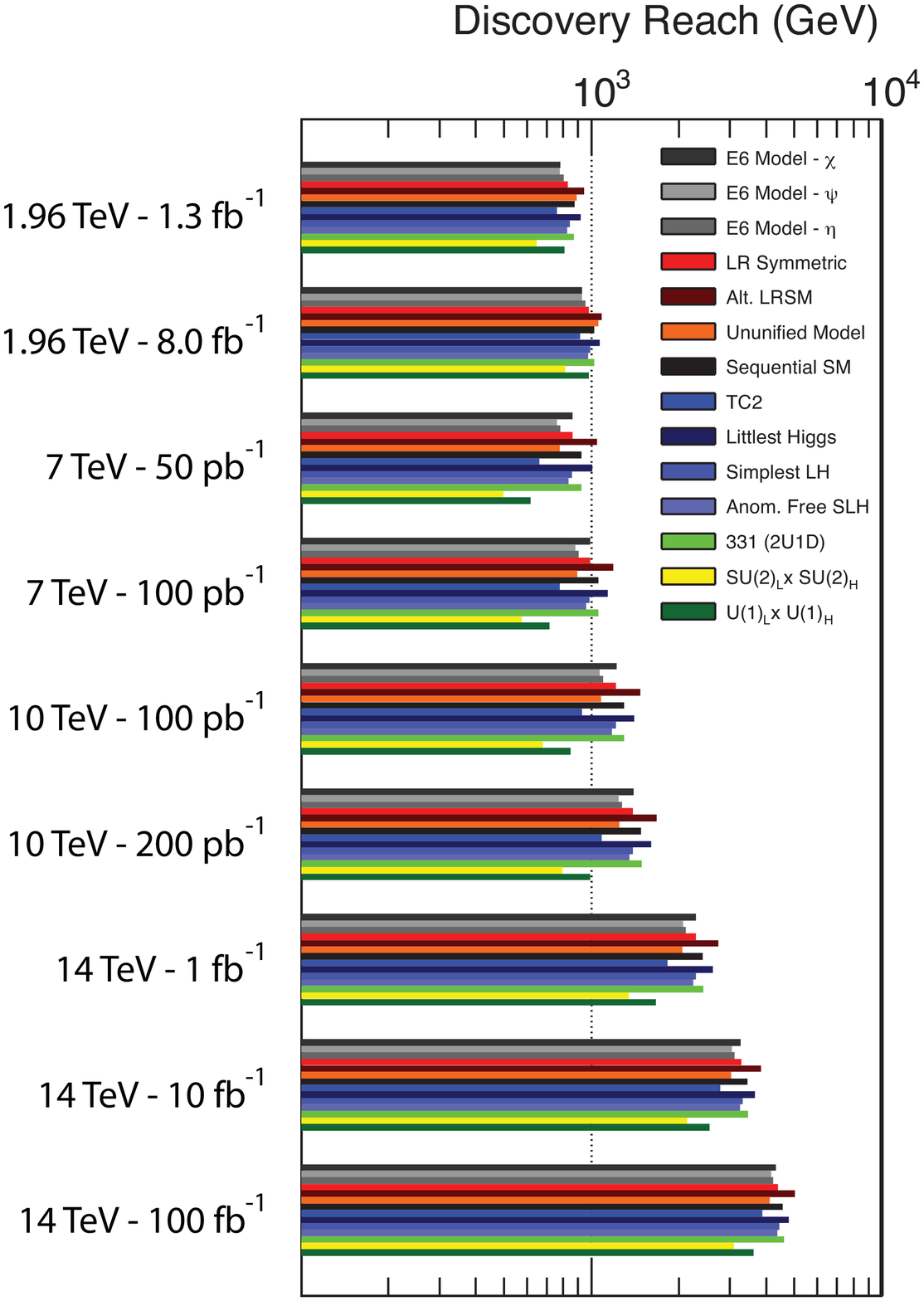}
\caption{LHC discovery reach, based on 5 dilepton events, for
typical \zpr\ models as a function of energy and integrated
luminosity, from \cite{Diener:2009vq}.}\label{discreach}
\end{figure}

The cross section for $pp\ra f\bar f$ (or $ \bar p p \ra  f\bar f$) for a specific final fermion $f$
is just
\beq
\sigma_{\zpr}^f \equiv \sigma_{\zpr} B_f =N_f/ \mathcal{L}_I ,
\eeql{crosf}
where $B_f=\Gamma_f/\Gamma_{Z'}$ is the branching ratio into $f\bar{f}$,
$\sigma_{\zpr}=\int  \frac{d\sigma_{\zpr}}{dy} dy$,
 and $N_f$ is the number of produced $f\bar f$ pairs for integrated luminosity $ \mathcal{L}_I $.
 For given couplings to the SM particles, $\sigma_{\zpr}^f$ and therefore the discovery reach
 depend on the total width $\Gamma_{Z'}$. For example,
 in the $E_6$ models
 $\Gamma_{Z'}/\mzp$ can vary from $\sim 0.01-0.05$ depending on whether the important open channels
 include light (compared to \mzp) superpartners and exotics in addition to the SM fermions~\cite{Kang:2004bz}.
 The consequences for the discovery reaches at the Tevatron and LHC are illustrated in
 Figure   \ref{widtheffect}, where it is seen, e.g., that the LHC reach can be reduced by $\sim 1$  TeV if there are many open channels.

There are a number of other potential two-fermion discovery channels,
such as $\tau^+\tau^-$ and $t \bar t $, as mentioned in Section \ref{existing}, while multibody channels will be touched on  in Section \ref{diagnostics}. In principle, the LHC reach in the Drell-Yan dilepton channels can be extended by
using virtual \zpr\ interference effects (cf., the observation of $Z$-propagator effects below the $Z$-pole at TRISTAN~\cite{Mori:1989py}),
though this is difficult in practice~\cite{Rizzo:2009pu}.

%: figure  width effect
\begin{figure*}[htb!]
\begin{minipage}[t]{5.0cm}
\includegraphics*[angle=270,width=2.9in]{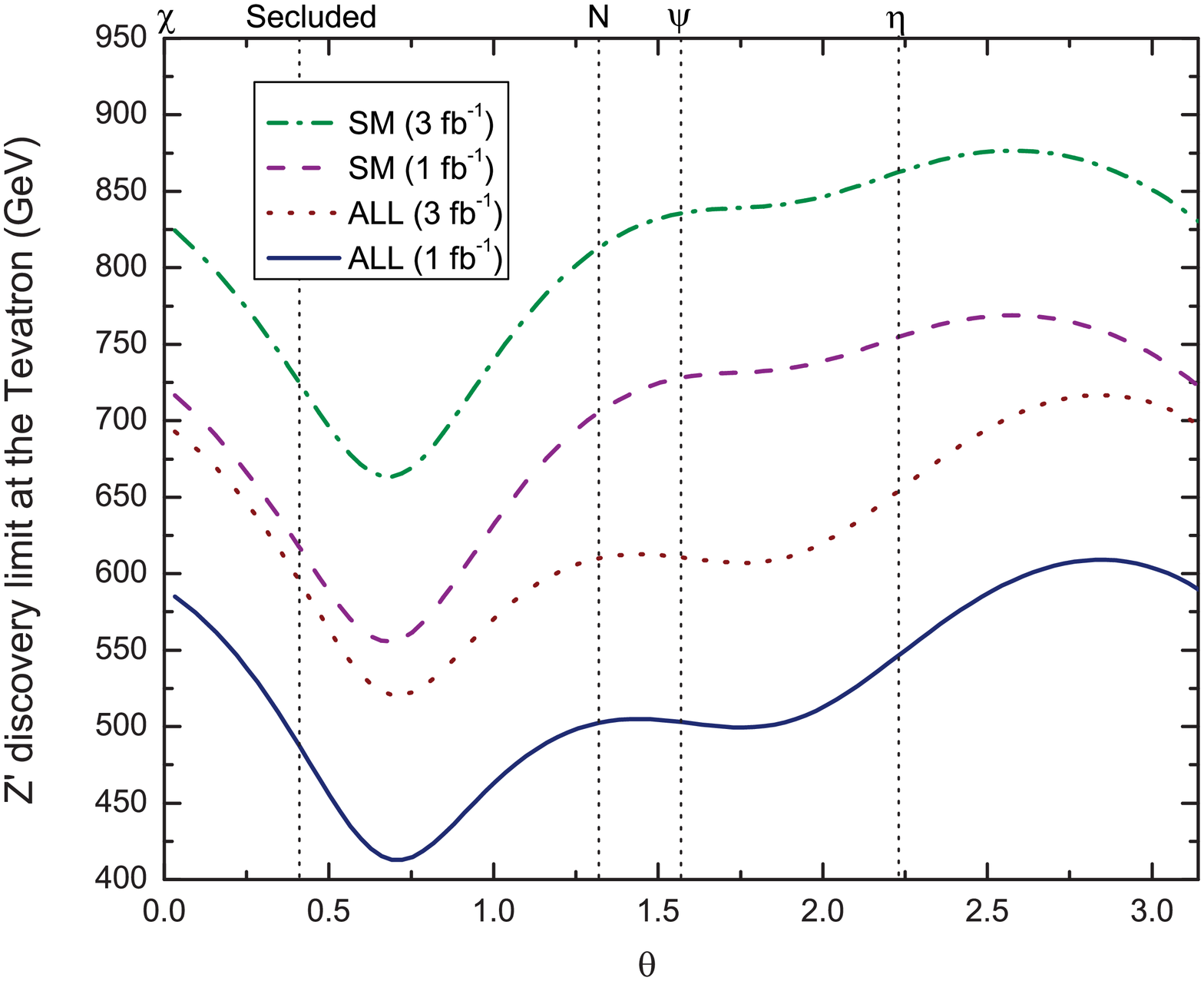}
\end{minipage}
\hspace*{2.5cm}
\begin{minipage}[t]{5.0cm}
\includegraphics*[angle=270,width=2.95in]{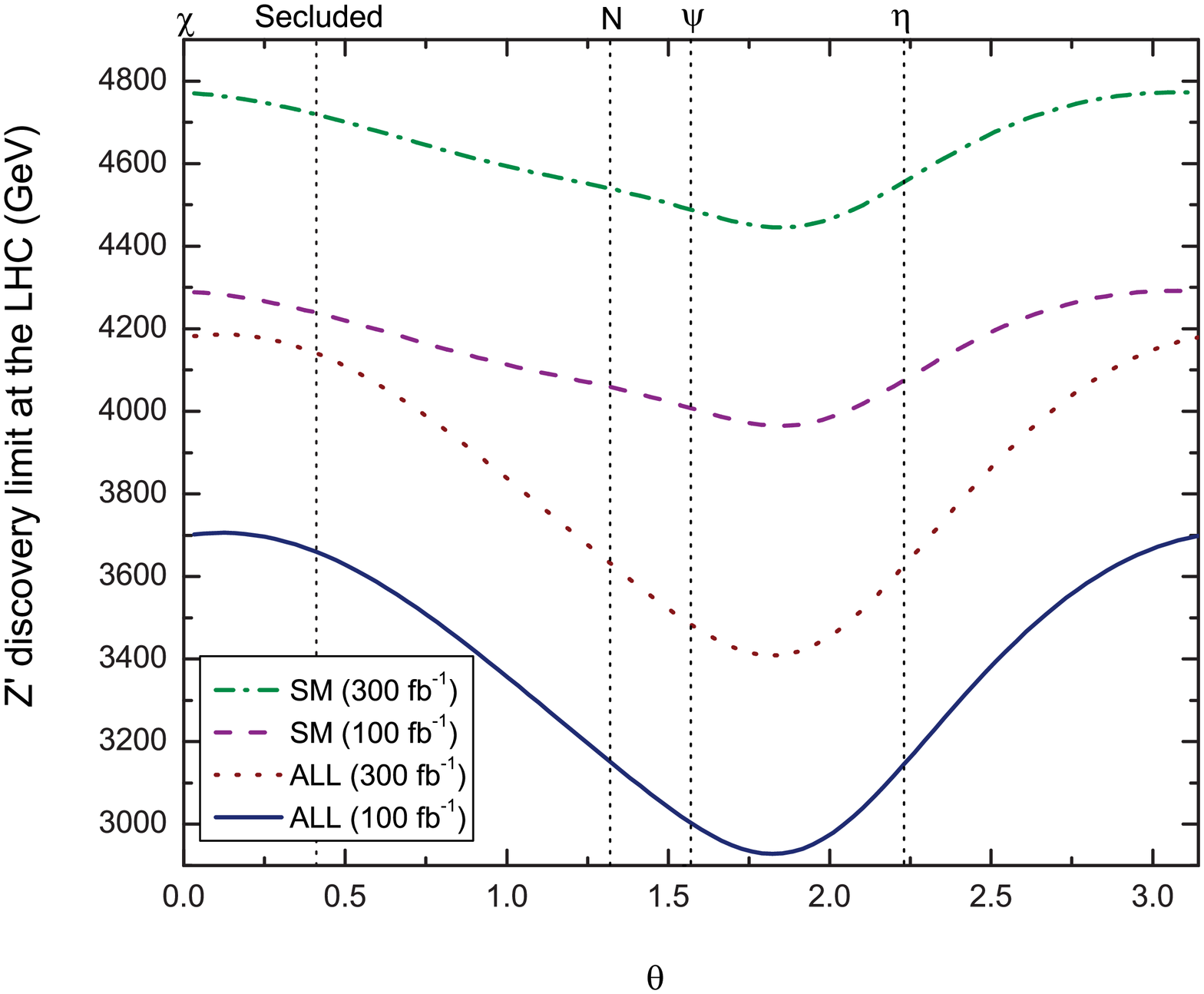}
\end{minipage}
\caption{Discovery reach of the Tevatron and LHC (at 14 TeV) for $E_6$ models,
 assuming decays (a) into SM particles only (SM) and (b) allowing
unsuppressed decays into exotics and sparticles (ALL), based on 10 dilepton events.
The charges are $Q= Q_\chi \cos \theta + Q_\psi \sin \theta$, where $Q_\chi$ and $Q_\psi$ are
associated with $SO(10)$ and $E_6$, respectively. From \cite{Kang:2004bz}.}
\label{widtheffect}
\end{figure*}

\subsection{Diagnostics}\label{diagnostics}
The spin of a resonance in the $\ell^+\ell^-$ channel would distinguish a  \zpr\ or other vector from,
e.g., a spin-0 Higgs resonance or a spin-2
Kaluza-Klein graviton excitation. The spin can be determined by the angular distribution in the
resonance rest frame, which for the spin-1 interactions in \refl{f1} is
\beq
\frac{d \sigma_{\zpr}^f }{d \cos \theta^\ast}  \propto \frac{3}{8} (1+\cos^2\theta^\ast)+A_{FB}^f \cos \theta^\ast,
\eeql{angdist}
where $\theta^\ast$ is the angle between the incident quark  and the $\ell$.
(Magnetic or electric dipole interactions lead to a different distribution~\cite{Chizhov:2009fc})
One does not know which hadron is the source of the $q$
and which the $\bar q$ on an event by event basis, but
the ambiguity washes out in the determination
of the $1+\cos^2\theta^\ast$ distribution~\cite{Langacker:1984dc,Dittmar:1996my}.
See~\cite{Osland:2009tn} for a recent detailed study. The \zpr\  spin can also be probed in $t\bar t$ decays~\cite{Barger:2006hm}.

Useful diagnostic probes of the chiral couplings to the quarks, leptons, and other particles, which would help discriminate between \zpr\  models, should be possible for masses up to $\sim$ $2-2.5$ TeV at the LHC, assuming  typical couplings.
(The gauge coupling $g_2$ can be fixed to the value
$ g_2=\sqrt{\frac{5}{3}} g \tan \theta_W\sim 0.46$ suggested by some grand unified theories, or alternatively can  be taken as a free parameter if the charges are normalized by some other convention.)

 For  $pp \to Z'\to \ell^+
\ell^-$ ($\ell=e,\mu$), one would be able to measure
the mass $M_{Z'}$,  the leptonic cross section
 $\sigma_{\zpr}^{ \ell}= \sigma_{\zpr} B_\ell$, and possibly the width $\Gamma_{\zpr}$
 (if it is not too small compared to the detector resolutions). The expected dilepton lineshape
 is illustrated in Figure \ref{dilepton}.
 By itself, $\sigma_{\zpr}^{ \ell}$  is { not}   a useful diagnostic
for the $Z'$ couplings to quarks and leptons:
while $\sigma_{\zpr}$   can be
calculated to within  a few percent for
given $Z'$ couplings,
$B_\ell$
depends strongly on the contribution of exotics
 and sparticles to $\Gamma_{\zpr}$~\cite{Kang:2004bz}.
However, $\sigma_{\zpr}^{ \ell}$
would be a useful indirect probe for the existence of the exotics
 or superpartners.
 The absolute magnitude of the quark and lepton couplings
 is probed by the product
$\sigma_{\zpr}^{ \ell}\Gamma_{\zpr} =
\sigma_{\zpr} \Gamma_\ell$.
%: figure  dilepton lineshape
\begin{figure}[htb]
\begin{minipage}[t]{5.0cm}
\includegraphics*[angle=270,width=3.15in]{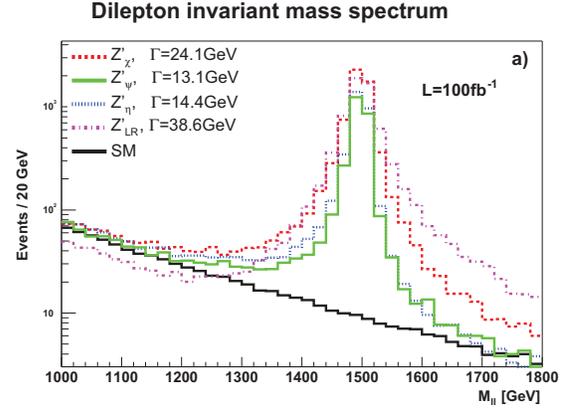}
\end{minipage}
\caption{Dilepton mass spectrum at the LHC
for typical models with $\mzp= 1.5$ TeV, $\sqrt{s}=14$ TeV,  and an integrated luminosity of 100 fb$^{-1}$, from \cite{Dittmar:2003ir}.}
\label{dilepton}
\end{figure}

The most useful diagnostics   involve the {relative
strengths} of $Z'$
couplings to ordinary quarks and leptons. The
forward-backward asymmetry as a function of the $Z'$
rapidity, $A_{FB}^f(y)$~\cite{Langacker:1984dc,Rosner:1995ft,Dittmar:1996my},
avoids the $q\bar q $ ambiguity in {Eq.~\ref{angdist}}.
For $AB\ra\zpr\ra f\bar f $, define $\theta_{CM}$ as the angle of fermion $f$
with respect to the direction of hadron $A$
in the \zpr\ rest frame, and let $F$ ($B$) be the cross section for fixed rapidity $y$
with $\cos \theta_{CM} >0$ ($<0$). Then, $A_{FB}^f(y)\equiv(F-B)/(F+B)$, with
\begin{align}
&F\pm B  \sim    \left[\begin{array}{c} 4/3 \\ 1\end{array} \right]     \notag   \\
 &\x  \sum_{i} \Bigl( f_{q_i}^A(x_1)f_{\bar q_i}^B(x_2)\pm f_{\bar q_i}^A(x_1)f_{q_i}^B(x_2) \Bigr)\label{abfy}   \\
& \x  \Bigl( \epsilon_L(q_i)^2\pm\epsilon_R(q_i)^2 \Bigr)
 \Bigl( \epsilon_L(f)^2\pm\epsilon_R(f)^2 \Bigr).\notag
\end{align}
 Clearly, $A_{FB}^f(y)$ vanishes for $pp$ at $y=0$,
but can be nonzero at large $y$ where there is more likely a valence $q$ from the first proton
and sea $\bar q$ from the other. The leptonic forward-backward asymmetry is
sensitive to a combination of quark and lepton chiral couplings and is
a powerful discriminant between models, as can be seen in Figure \ref{diagnosticdists}.
An variant definition of the asymmetry based on the pseudorapidities of the leptons is another
possibility~\cite{Diener:2009ee}.

The
ratio of cross sections  for $\zpr \ra \ell^+\ell^-$ in different rapidity bins~\cite{delAguila:1993ym} gives
information on the relative $u$ and $d$ couplings (Figure \ref{diagnosticdists}). Possible observables in other two-fermion final state channels
include the polarization of produced $\tau$'s~\cite{Anderson:1992jz};
the $pp\to Z'\to j  j$ cross section~\cite{Rizzo:1993ie,Weiglein:2004hn}; and branching ratios,
forward-backward asymmetries, and spin correlations for $b\bar b$ and $t \bar t$~\cite{Barger:2006hm,Godfrey:2008vf,Arai:2008qa,Jung:2009jz}.
There are no current plans for polarization at the LHC, but polarization asymmetries
at a future or upgraded hadron collider would provide another useful diagnostic~\cite{Fiandrino:1992fa}.
Family nonuniversal but flavor conserving effects are discussed in~\cite{Chen:2008za,Salvioni:2009jp}.

 In four-fermion final state  channels the
rare decays $\zpr\ra V f_1 \bar f_2$, where $V=W$ or $Z$ is radiated from the \zpr\ decay products, have a double logarithmic enhancement.
In particular, $Z'\to W\ell\nu_{\ell}$ (with $W\ra$ hadrons and an  $\ell\nu_{\ell}$ transverse
mass  $> 90$ GeV to separate from SM background) may be observable
and projects out the left-chiral lepton couplings~\cite{Rizzo:1987rw,Cvetic:1991gk,Hewett:1992nf}.
Similarly, the associated productions $pp\rightarrow Z' V$
with $V=(Z,W)$~\cite{Cvetic:1992qv}  and $V=\gamma$~\cite{Rizzo:1992sh}
could yield information on the quark chiral couplings. The processes $pp\ra \zpr Z$ or $ \zpr\gamma$
with the \zpr\ decaying invisibly  into neutrinos or hidden sector particles
may also be observable and could serve as a discovery mode
if the \zpr\ does not couple to charged leptons~\cite{Petriello:2008pu,Gershtein:2008bf}.
The importance of the width for invisible \zpr\ decays for constraining certain extra-dimensional models has been
emphasized in~\cite{Gninenko:2008yq}.

Decays into two bosons, such as $\zpr\ra W^+ W^-, Zh,$ or $W^\pm H^\mp$,
can usually occur only by $Z-\zpr$ mixing or with amplitudes related to the mixing.
However, this suppression may be compensated for the longitudinal
modes of the $W$ or $Z$ by the large polarization vectors, with components scaling as
$\mzp/M_W$~\cite{Robinett:1981yz,Rizzo:1985kn,Nandi:1986rg,delAguila:1986ad,Barger:1987xw,Baer:1987eb,Gunion:1987jd,Deshpande:1988py}.
For example, $\Gamma(\zpr \ra W^+ W^-)\sim \theta_{Z\zpr} ^2,$ which appears to be hopelessly
small to observe. However, the enhancement factor is $\sim (\mzp/M_W)^4$. Thus, from
{Eq.~\ref{f8}}, these factors compensate, leaving a possibly observable rate that
in principle could give information on the Higgs charges. In the limit of $\mzp \gg M_Z$ one has
\beq
\begin{split}
\Gamma(\zpr \ra W^+ W^-) =& \frac{g_1^2 \theta_{Z\zpr} ^2\mzp}{192\pi} \left(\frac{\mzp}{M_Z}\right)^4\\
=& \frac{g_2^2 C^2 \mzp}{192\pi}.
\end{split}\eeql{zpww}
%\beq
%\Gamma_{W^+ W^-} = \frac{g_1^2 \theta_{Z\zpr} ^2\mzp}{192\pi} \left(\frac{\mzp}{M_Z}\right)^4
%= \frac{g_2^2 C^2 \mzp}{192\pi}.
%\eeql{zpww}
The decay $\zpr\ra Z Z$ has recently been considered~\cite{Keung:2008ve}. The Landau-Yang theorem~\cite{Yang:1950rg} can
be evaded by anomaly-induced or $CP$-violating operators involving a longitudinal $Z$.
The LHC reach of spin-1 resonances associated  with electroweak symmetry breaking and the associated
$\zpr\ra W^+W^-$ or $W' \ra ZW$ decays have been studied in~\cite{Alves:2009aa}, and
more complicated decays such as $\zpr \ra ggg$ or $gg\gamma$  in~\cite{FloresTlalpa:2009jh}.

An alternative source of triple gauge vertices involves anomalous \uprm\ symmetries, which often occur in string constructions.
The anomalies must
be cancelled by a generalized Green-Schwarz mechanism.
The \zpr\ associated with the \uprm\ acquires a  string-scale mass
by what is essentially the St\"uckelberg mechanism, and effective trilinear vertices may be generated between the \zpr\ and the SM
gauge bosons~\cite{Coriano:2005js,Anastasopoulos:2006cz}. If there are  large extra dimensions
the string scale and therefore the \zpr\ mass may be very low, e.g., at the TeV scale,
with  anomalous decays into $ZZ$, $WW$, and $Z\gamma$, e.g.,~\cite{Armillis:2008vp,Kumar:2007zza,Anastasopoulos:2008jt}.

Some \zpr\ models lead to distinctive multi-lepton decay modes at a possibly observable rate that are almost entirely free of SM backgrounds.
For example, a \zpr\ could decay into $\ell \bar \ell \ell \bar \ell $ via intermediate sneutrinos
in an $R$-parity violating supersymmetric model~\cite{Lee:2008cn}, or $\zpr\ra 3\ell 3 \bar \ell$
by an intermediate $ZH\ra 3Z$ in some models with extended Higgs structures~\cite{Barger:2009xg}.
The latter could occur even in leptophobic models (i.e., with no direct coupling to leptons).
A light (GeV scale) \zpr, suggested by some recent dark matter models, would be highly boosted
at the LHC, leading to  narrow ``lepton jets''
from $Z' \ra \ell^+ \ell^-$ and possible displaced vertices, e.g.,~\cite{ArkaniHamed:2008qp,Cheung:2009su,Langacker:2009im}.

Global studies of the possible LHC diagnostic possibilities for determining ratios
of chiral charges in a model independent way and discriminating models are given in~\cite{Dittmar:2003ir,Petriello:2008zr,Diener:2009vq,Dittmar:1996my,delAguila:1993ym,Cvetic:1995zs,Li:2009xh}. The complementarity of LHC and ILC observations is
especially emphasized in~\cite{Cvetic:1995zs,delAguila:1993rw,delAguila:1995fa,Weiglein:2004hn}.

%:figure fb and rapidity
\begin{figure*}[htb]
\begin{minipage}[t]{5.0cm}
\includegraphics*[angle=270,width=3.1in]{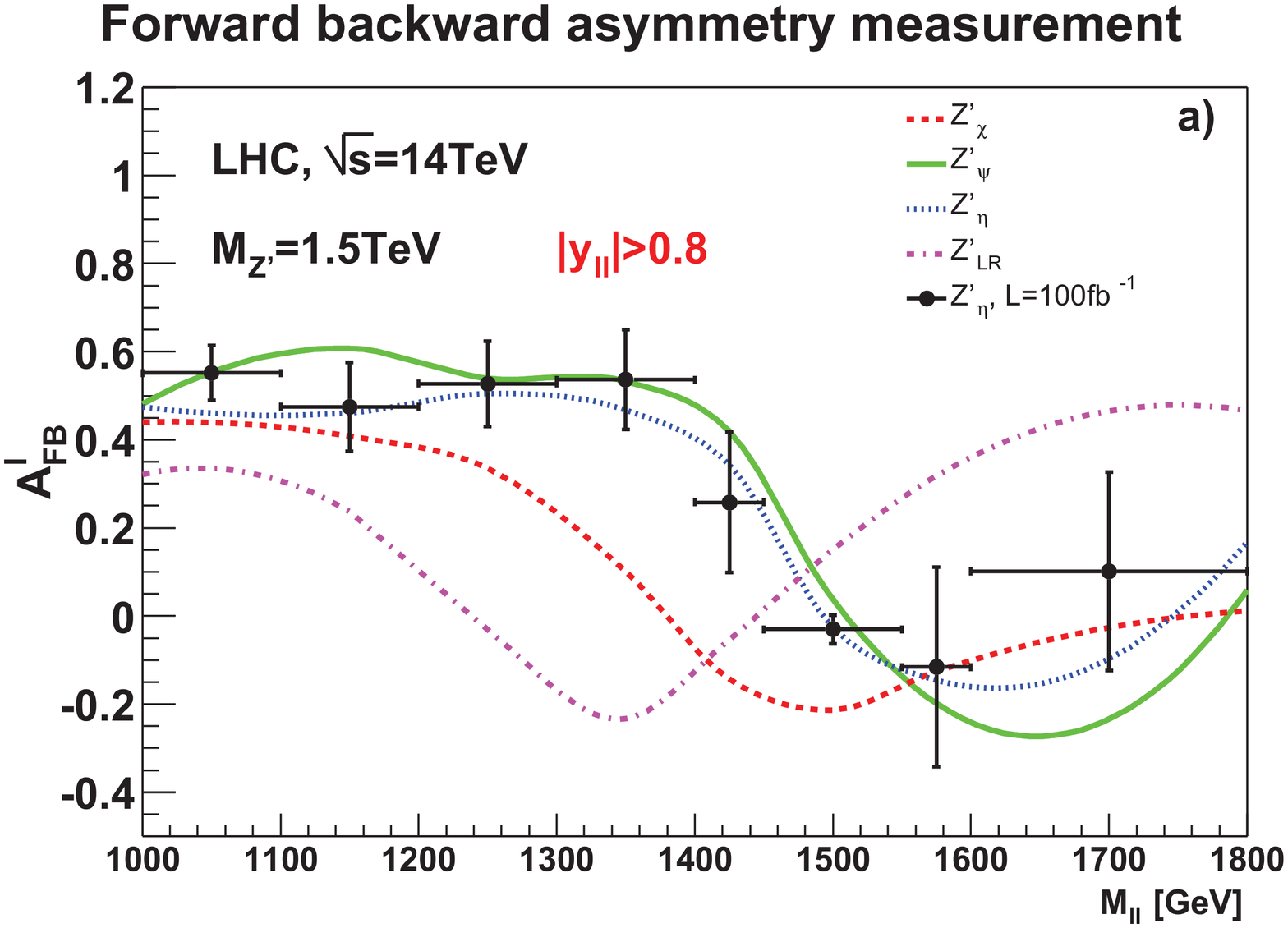}
\end{minipage}
\hspace*{3cm}
\begin{minipage}[t]{5.0cm}
\includegraphics*[angle=270,width=3.2in]{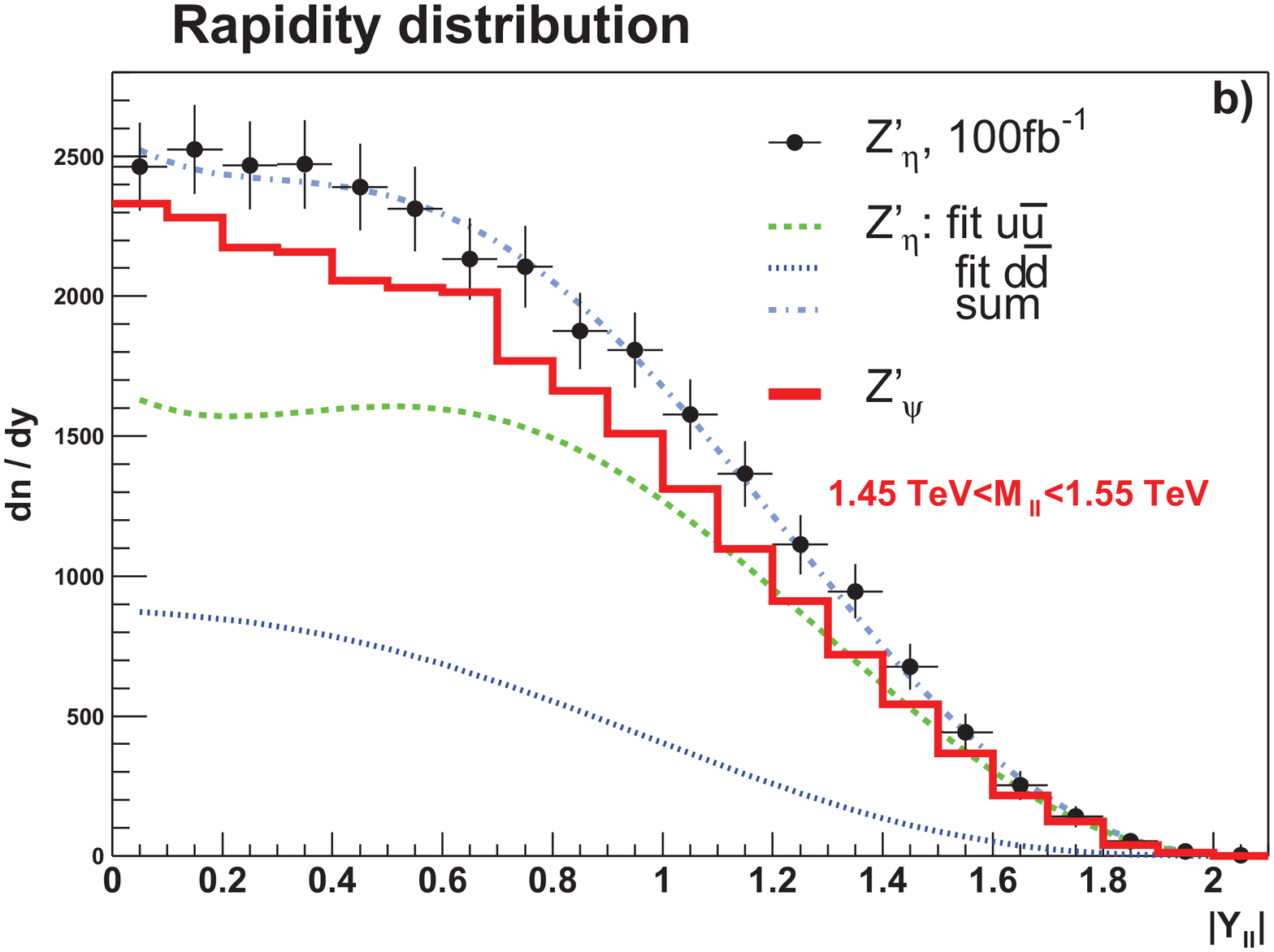}
\end{minipage}
\caption{Forward backward asymmetry and rapidity distributions
for typical models with $\mzp= 1.5$ TeV, $\sqrt{s}=14$ TeV, and an integrated luminosity of 100 fb$^{-1}$, from \cite{Weiglein:2004hn,Dittmar:2003ir}.}
\label{diagnosticdists}
\end{figure*}

\section{Other LHC Implications}\label{other}
There are several other implications of a \zpr\ for the LHC. For example,
TeV scale \uprm\ models generally involve an extended Higgs sector, requiring at least a SM singlet $S$
to break the \uprm\ symmetry.  New $F$ and $D$-term contributions can relax the theoretical upper
limit of $\sim 130$ GeV on the lightest Higgs scalar in the MSSM up to $\sim 150$ GeV, and
smaller values of $\tan \beta$, e.g. $\sim 1$, become possible. Conversely, doublet-singlet
mixing can allow a Higgs lighter than the direct SM and MSSM limits. Such mixing as well as
the extended neutralino sector can lead to non-standard collider signatures, e.g.,~\cite{Accomando:2006ga,Barger:2006dh,Ham:2009bu}.

\uprm\ models also have extended neutralino sectors~\cite{Barger:2007nv,Choi:2006fz}, involving at least the
$\tilde Z'$ gaugino and the $\tilde S$ singlino, allowing  non-standard couplings (e.g., light singlino-dominated),
extended cascades, and modified possibilities for cold dark matter, $g_\mu-2$, etc.

Most \uprm\ models (with the exception of those involving $B-L$ and $Y$) require new exotic fermions to cancel
anomalies. These are usually non-chiral with respect to the SM (to avoid precision electroweak constraints) but chiral under the \uprm. A typical example is a pair of \st-singlet colored quarks $D_{L,R}$  with charge $-1/3$.
Such exotics may decay by mixing, although that is often forbidden by $R$-parity. They may also decay by diquark or leptoquark couplings, or they be quasi-stable, decaying by higher-dimensional operators~\cite{Kang:2007ib,King:2005jy,Athron:2009bs}.

A heavy  \zpr\ may decay efficiently into sparticles, exotics, etc., constituting a ``SUSY factory''~\cite{Kang:2004bz,Lee:2008cn,Baumgart:2006pa,Cohen:2008ni,Ali:2009md}.

 For other theoretical, experimental, and cosmological/astrophysical \zpr\  implications  see~\cite{Langacker:2008yv,Langacker:2009im}.

%\cite{*}
%\bibliographystyle{h-elsevier}
%\bibliographystyle{href-elsevier}
%\bibliographystyle{h2-elsevier}
%:Bibliography

%\end{document}

%%%%%%%%%%%%%%%%%%%%%%%%%%%%%%%%%%%%%%%%%%%%%%%%%%%%%%%%%%%%%%%%%%%%%%%%%%%%%%%%%%%%%%%%%%%%%%
%%%%%%%%%%%%%%%%%%%%%%%%%%%%%%%%%%%%%%%%%%%%%%%%%%%%%%%%%%%%%%%%%%%%%%%%%%%%%%%%%%%%%%%%%%%%%%
\chapter{Visible Signatures from Hidden Sectors}
\setlength{\epigraphrule}{1pt}
\epigraphhead[20]{\epigraph{\large {\em Daniel Feldman, Zuowei Liu,
Lian-Tao Wang and Kathryn~Zurek}}{\large Daniel Feldman \& Zuowei
Liu (Conveners)}}
%
%%%%%%%%%% espcrc2.tex %%%%%%%%%%
%
% $Id: espcrc2.tex 1.2 2000/07/24 09:12:51 spepping Exp spepping $
%
%\documentclass[fleqn,twoside]{article}
%\usepackage{espcrc2}

% change this to the following line for use with LaTeX2.09
% \documentstyle[twoside,fleqn,espcrc2]{article}

% if you want to include PostScript figures
%\hyphenation{author another created financial paper re-commend-ed Post-Script}
%\usepackage{amssymb,graphics,graphicx, epsfig}
%\usepackage{amsfonts}
%\usepackage{hyperref}
%\usepackage{multicol}
%\usepackage{bm}
%\usepackage[figuresright]{rotating}

% put your own definitions here:
%   \newcommand{\cZ}{\cal{Z}}
%   \newtheorem{def}{Definition}[section]
%   ...

\def\Lag{\mathcal{L}}
\def\be{\begin{equation}}
\def\beqn{\begin{eqnarray}}
\def\ee{\end{equation}}
\def\eeqn{\end{eqnarray}}
\def \ETmiss{${\not\!\!{E_T}}~$}
\def \missET{${\not\!\!{E_T}}$}
\def\nj{$n_{jet}$}
\def\njs{$n_{jet}^*$}
\def\co{co-annihilation~}
\def\pts{$P_T^{\rm * miss}$}
\def\pt{$P_T^{\rm miss}$}
\def\st{Stueckelberg~}
\def\sm{Standard Model~}
\def\stm{StMSSM~}
\def\s1{$s_{\alpha}$}
\def\s2{$s_{\gamma}$}
\def\s3{$s_{\delta}$}
\def\c1{$c_{\alpha}$}
\def\c2{$c_{\gamma}$}
\def\c3{$c_{\delta}$}
\def\st{Stueckelberg\ }
\def\s{Stueckelberg~}
\def\za{\Gamma_Z^2 M^2_Z}
\def\zb{\Gamma^2_{Z'} M^2_{Z'}}
\def\y{Y_{\phi}}
\def\s1{$s_{\alpha}$}
\def\s2{$s_{\gamma}$}
\def\s3{$s_{\delta}$}
\def\c1{$c_{\alpha}$}
\def\c2{$c_{\gamma}$}
\def\c3{$c_{\delta}$}
\newcommand{\sellpm}{\widetilde{\ell}^{\pm}}
\newcommand{\sell}{\widetilde{\ell}}
\newcommand{\el}{\widetilde{e}_L}
\newcommand{\ml}{\widetilde{\mu}_L}
\newcommand{\nuni}{non-universalities~}
\newcommand{\alpd}{\dot{\alpha}}
\newcommand{\alp}{\alpha}
\renewcommand{\qbar}{\bar{\theta} }
\newcommand{\q}{{\theta} }
\newcommand{\om}{\overline{m}}
\newcommand{\vol}{{\cal V}}
\newcommand{\tom}{{\tilde \omega}}
\newcommand{\G}{{\mathcal G}}
\renewcommand{\non}{\nonumber}
\renewcommand{\te}{\theta}
\newcommand{\thb}{\bar\theta}
\renewcommand{\vp}{\varphi}
\newcommand{\mathsym}[1]{{}}
\newcommand{\unicode}{{}}
\def\dmu{\partial_{\mu}}
\def\gmu{\gamma_{\mu}}
\def\gnu{\gamma_{\nu}}
\def\g5{\gamma_5}
\def\ab{\bar{\alpha}}
\def\eslt{\not\!\!{E_T}}
\def\Dsl{\not\!\!{D}}
\def\psl{\not\!\!{\partial}}
\def\to{\rightarrow}
\def\Phat{\hat{\Phi}}
%%% particles
\def \cha{\widetilde{\chi}^{\pm}_1}
\def \chb{\widetilde{\chi}^{\pm}_2}
\def \na{\widetilde{\chi}^{0}_1}
\def \nb{\widetilde{\chi}^{0}_2}
\def \nc{\widetilde{\chi}^{0}_3}
\def \nd{\widetilde{\chi}^{0}_4}
\def \g{\widetilde{g}}
\def \ql{\widetilde{q}_L}
\def \qr{\widetilde{q}_R}
\def \dl{\widetilde{d}_L}
\def \dr{\widetilde{d}_R}
\def \ul{\widetilde{u}_L}
\def \ur{\widetilde{u}_R}
\def \ccl{\widetilde{c}_L}
\def \ccr{\widetilde{c}_R}
\def \ssl{\widetilde{s}_L}
\def \ssr{\widetilde{s}_R}
\def \ta{\widetilde{t}_1}
\def \tb{\widetilde{t}_2}
\def \ba{\widetilde{b}_1}
\def \bb{\widetilde{b}_2}
\def \sta{\widetilde{\tau}_1}
\def \stb{\widetilde{\tau}_2}
\def \smr{\widetilde{\mu}_R}
\def \ser{\widetilde{e}_R}
\def \sml{\widetilde{\mu}_L}
\def \sel{\widetilde{e}_L}
\def \slr{\widetilde{l}_R}
\def \sll{\widetilde{l}_L}
\def \snl{\widetilde{\nu}_{\tau}}
\def \snm{\widetilde{\nu}_{\mu}}
\def \sne{\widetilde{\nu}_{e}}
\def \hc{H^{\pm}}
\def \lra{\longrightarrow}
\def \W{{\mathcal{W}}}
\def\y{Y_{\phi}}

%\title{
%%Hidden Symmetries
%Visible Signatures from Hidden Sectors}

%\begin{document}

%\begin{abstract}
%The concept of a hidden sector (HS) is generic to a broad class of models which predict  physics
%beyond the standard model. Reviewed here are recent developments in the hidden sector models
%which provide a portal to the visible sector through kinetic mixings, mass mixings, and higher dimensional operators.
%\vspace{1pc}
%\end{abstract}

%\maketitle

\section{Introduction} \label{introHS}

%The idea of the hidden sector (HS) has its modern roots in local supersymmetry , or supergravity.
%In this paradigm, the HS matter is uncharged under the standard model gauge group [the visible sector (VS)].
%The HS matter is responsible for the breaking of supersymmetry where communication, or connection, to the VS
%occurs due to the presence of gravity (See Sec.(1)).
%In the braneworld picture with extra dimensions, fields propagating in the bulk act as the connectors
%linking  different branes (See Sec.(2)).

In this section  we discuss a broad class of models with visible signatures due to the presence of
hidden gauge symmetries. %; sometimes referred to as ``dark force'' models.
%or ``dark forces.''
 The specific classes of  models  we review each  have a hidden sector, a visible
sector and a communication between the hidden and the visible sectors.
While there are many hidden sector models which have been discussed, we will focus here on communication via Stueckelberg mass mixing \cite{KN1,KN2,FLN1,FLN2},  higher dimension operators mediated by heavy states in Hidden Valleys \cite{HV1,HV2,SUSYHV,Han:2007ae},
models with mediation via kinetic mixing
\cite{Holdom:1985ag,MR,Wells,FLN2,MeV,PRV} and specifically kinetic mixing in the class  of
 dark force models discussed in \cite{Hamed,LTW,ew_prod,Cheung}. We also discuss
generalized portals
occurring due to hidden-visible sector couplings arising from both
kinetic and mass mixings  \cite{FLN2,Desy,Burgess}.

% Popular mechanisms for communication include kinetic mixing  \cite{Holdom:1985ag,MR,FLN2,MeV,PRV,Hamed,LTW,ew_prod,Cheung}, Stueckelberg mass mixing \cite{KN1,KN2,FLN1}, as well as higher dimension operators mediated by heavy states \cite{HV1,HV2,SUSYHV,Han:2007ae,UP1,UP2}.

The concept of the hidden sector has a long history and its modern roots lie in supersymmetry where hidden sectors are responsible for the breaking of supersymmetry.  However, typically the fields in the hidden sectors are very massive. Thus while the consequences of the hidden sectors have direct bearing on the building of phenomenologically viable models whose experimental signatures will be probed at the LHC and in dark matter experiments, the actual internal dynamics of the hidden sector are unreachable directly with colliders or cosmology.  However, more recently it has been shown that hidden sectors can give rise to unique signatures at colliders when the mass scale in the hidden sector is well below a TeV, as in Hidden Valleys, Stueckelberg extensions  and Unparticle models.  In particular, confining dynamics in the hidden sector \cite{HV1,HV2,UP1,UP2} give rise to exotic signatures such as high jet multiplicity events \cite{Han:2007ae} and lepton jets, and such events multiplicities
 are also a feature of the models of Refs. \cite{Hamed,LTW,ew_prod,Cheung}.
Thus in models with extended hidden sectors, the cascades and dynamics can become rich and complex.
Rich event topologies arise in models of  \st mass generation and kinetic mixings, where multi-lepton jet signals and missing energy  are a consequence of
of gauged hidden sector vector multiplets. Here one has complex susy cascades and heavy flavor jet signatures from new scalars \cite{KN2},
multilepton production and jet production \cite{FLN1,FLN2,FLNN}  as well as the possibility of mono-jet and mono-photon signatures \cite{KCTC}; where the latter signatures also arise in the models of \cite{Petriello,Dudas:2009uq,ew_prod}.

There are indeed many recent developments in hidden sector models,
and by no means will we be able to cover all models, which include Higgs mediators, light gauged mediators and axion mediators, see e.g.,
\cite{Wells,FKN,KCTC,FLN2,Kumar,Kim,Desy,Burgess,Lu:2007jj,Petriello,PRV,MeV,Hamed,LTW,Ibarra:2008kn,LGM,models_axion,BreitWigner},
as well as investigations of their phenomenological implications \cite{nima_lhc,Posp,Baek:2008nz,Chun,ew_prod,Cheung,Katz,Ross,models_comp,Reece,GeVSpectra,Dudas:2009uq,Feng,toro,Mambrini:2009ad,Goodsell,NA,Dedes,FLNN}.  We aim instead to outline some of the possibilities, and refer the reader to these references for further details.

These classes of models
also lead to astrophysical predictions offering several explanations
to the recent positron anomaly seen in the PAMELA satellite data. Such proposals include
 multi-component dark matter \cite{LGM}, a
boost in positrons from a sommerfeld enhancement \cite{Hamed}
and a Breit-Wigner enhancement of dark matter annihilations \cite{BreitWigner}
(see also \cite{FLN2}). Further, the presence of hidden sector states degenerate with the
dark matter particle can lead
instead  to a boost in the relic density via coannihilation effects
  \cite{FKN,FLNN}.
We discuss now some of the models in further detail.

\section{ \st Extensions} \label{StueckelbergExt}
\subsection{Massive \st vector bosons } The Stueckelberg mechanism
allows for mass generation for a $U(1)$ vector field without the
benefit of a Higgs mechanism. The $U(1)_X$ Stueckelberg extensions
of the  Standard Model (SM)\cite{KN1}, i.e., $SU(3)_C \times SU(2)_L
\times U(1)_Y \times U(1)_X$, involve a non-trivial mixing of the
$U(1)_Y$ hypercharge gauge field $B^{\mu}$ and the $U(1)_X$
Stueckelberg   field $C^{\mu}$. The Stueckelberg
 field $C^{\mu}$ has no couplings with the visible sector
fields, while it  may couple with a hidden sector, and thus the
physical $Z'$ gauge boson connects with the visible sector only via
mixing with the  SM gauge bosons. These  mixings,
however, must be small because of the LEP electroweak constraints\cite{FLN1}.

 The $U(1)_X$ Stueckelberg extension of the Standard Model (SM)  can be generalized further
 to include a gauge kinetic mixing
(StkSM)\cite{FLN2}. In the gauge vector boson sector, the
effective Lagrangian is then given by  $\mathcal{L}_{\rm StkSM} = \mathcal{L}_{\rm SM} +\Delta  \mathcal{ L}$ \cite{FLN2}
\begin{eqnarray}
\Delta\mathcal{L} & \owns &
-\frac{1}{2}(\partial_{\mu}\sigma+ M_1 C_{\mu}+M_2 B_{\mu})^2
\\
 & &  -\frac{1}{4} C_{\mu \nu}C^{\mu \nu}
- \frac{\delta}{2}C_{\mu\nu}B^{\mu \nu} + g_{X}J^{\mu}_{X}C_{\mu}\nonumber \nonumber
\label{stksm}
\end{eqnarray}
\begin{figure}[t]
\centering
\includegraphics[width=6cm,height=5cm]{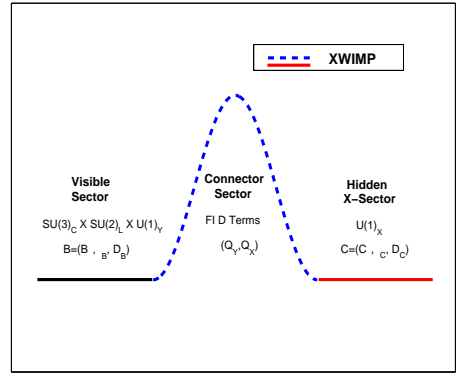}
\vspace{-.7cm}
\caption{\small An XWIMP contains a combination of fields both from the VS and the HS which communicate
due to the presence of a connector sector (CS). Suppressed interactions in the HS  leads
to a boost in the relic density relative to what would be obtained  without the presence of the
HS states\cite{FKN,FLNN}.  (Figure from \cite{FKN}).}
 \label{connector}
\end{figure}
\begin{figure*}[t]
\center
\includegraphics[width=6.5cm,height=4.5cm]{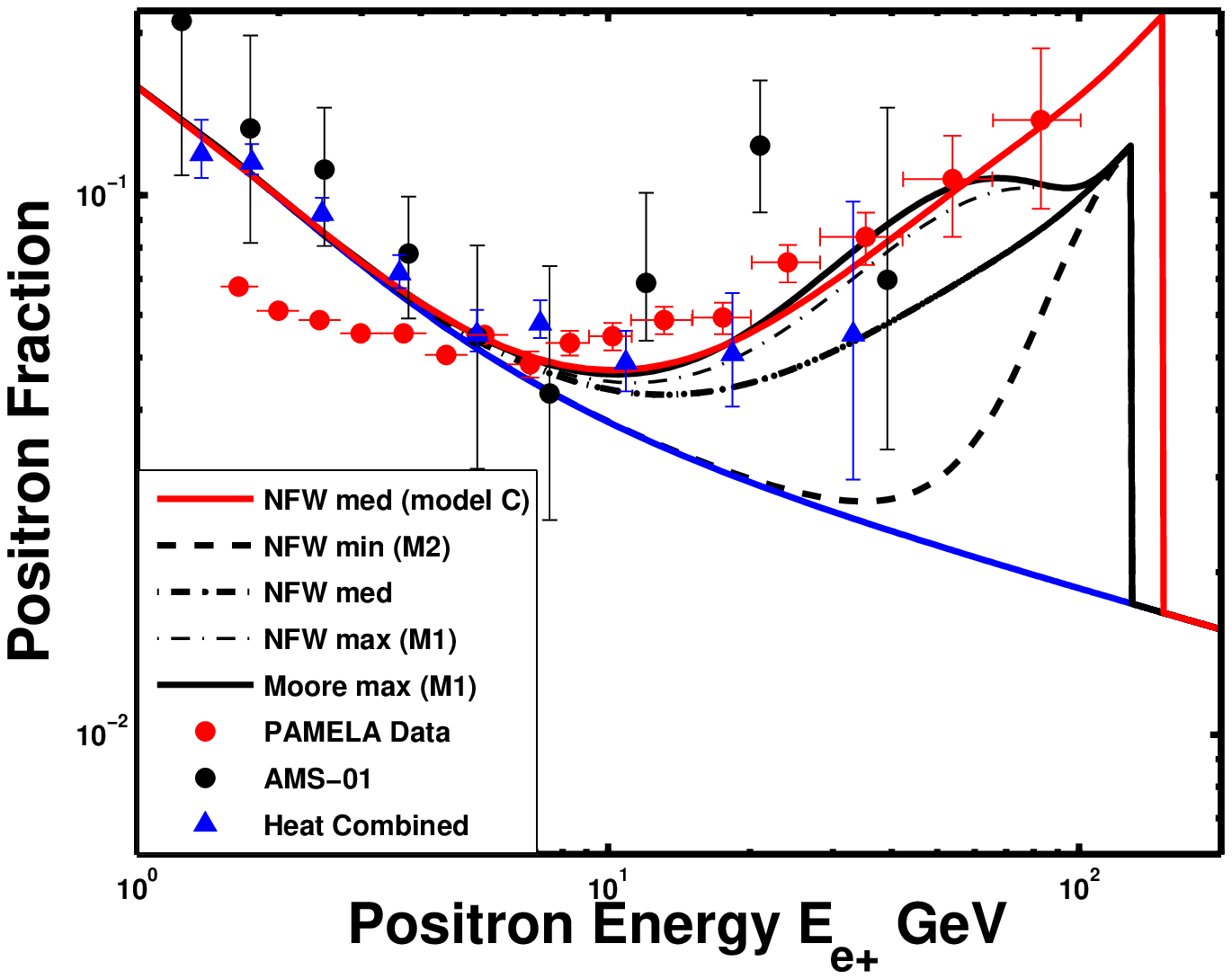}
\includegraphics[width=6.0cm,height=4.0cm]{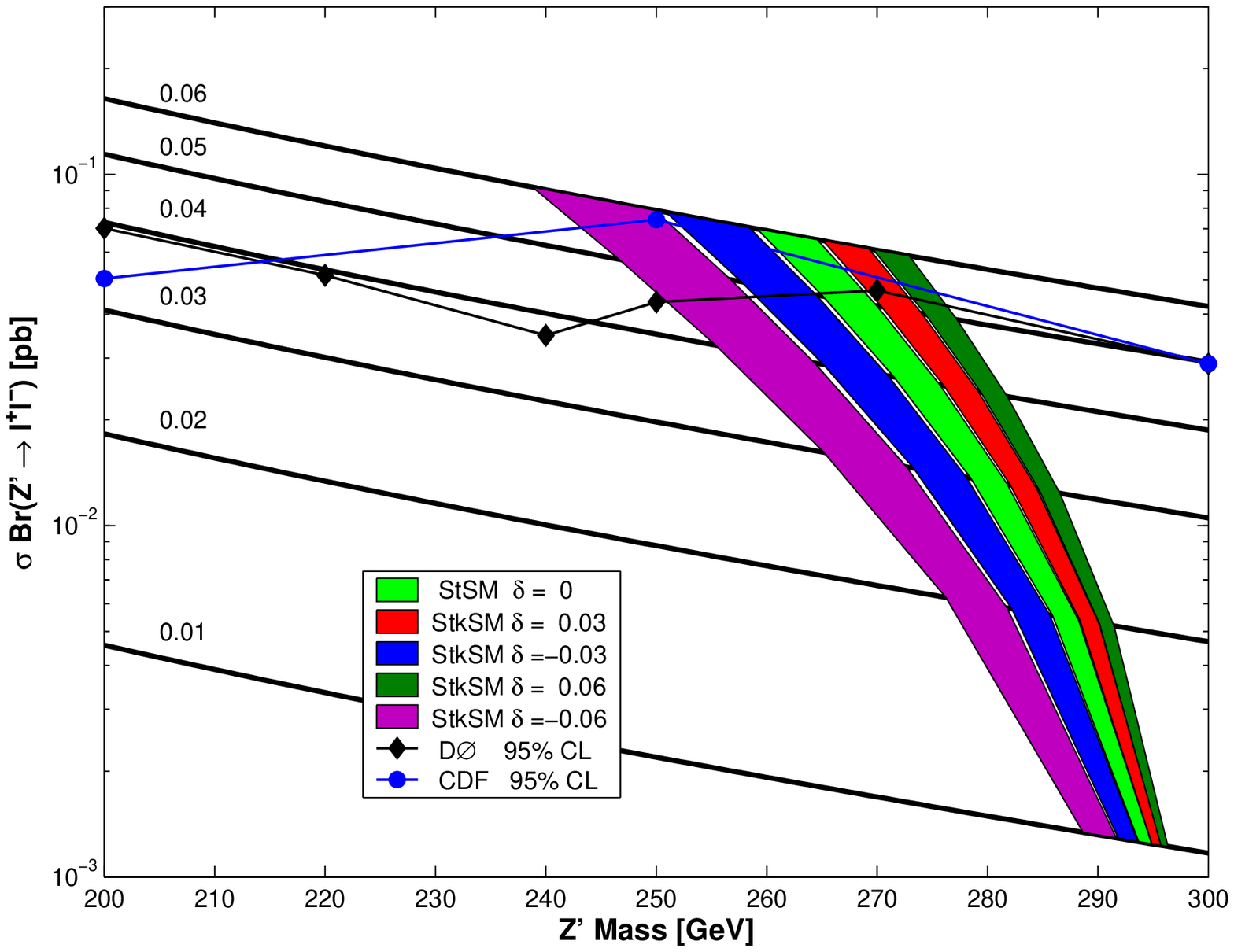}
\vspace{-.7cm} \caption{\small Left: A dark Dirac fermion $(m_{D})$
which couples to the \st $Z'$ produces fits to the PAMELA positron
data \cite{Adriani:2008zr} due to the presence of a Breit-Wigner
Pole \cite{FLN2}. Right: The  \st $Z'$ produces a detectable signal
in the dilepton channel consistent with electroweak constraints
(black curves) and  simultaneously produced the correct relic
abundance of dark matter in the vicinity of the Breit-Wigner Pole
\cite{FLN1} (shaded/colored bands). The \st $Z'$ can therefore be
tested at low mass ranges where   $Z'$ from GUT models are  already
eliminated \cite{FLN1,findST}.}
 \label{fig:FLN}
\end{figure*}
\noindent where $\delta$ is the gauge kinetic mixing parameter and
 $M_1$, $M_2$ are the \st mass parameters \cite{FLN2,Desy,Burgess}.
Here $\sigma$ is a psuedoscalar axion which transforms under $U(1)_X$ as
 well as under $U(1)_Y$ so that $\Delta\mathcal{L}$ is gauge invariant.

  Upon coupling to the SM, the HS and VS mix
through the neutral vector boson sector  and one finds a massless
photon $A$, the $Z$ boson, and a $Z'$ boson, the latter of which  is
dominantly composed of $C$ . In the {\it absence} of kinetic mixing,
this arises from diagonalizing the mass$^2$ matrix in the neutral
vector sector (in the basis $(C,B,A^3)_{\mu}$)~\cite{KN1,KN2,FLN1}
\be \left[
\begin{array}{ccc}
    M_1^2             & M_1^2    \epsilon                            & 0 \\
    M_1^2 \epsilon        & M_1^2 \epsilon^2 + \frac{1}{4}v^2g_Y^2  & -\frac{1}{4}v^2g_2g_Y \\
      0                   & -\frac{1}{4}v^2g_2g_Y            & \frac{1}{4}v^2g_2^2
\end{array}
\right]\ , \label{vmatrix} \ee where the effective parameter is the
ratio $\epsilon \equiv M_2/M_1$, which is constrained by the
electroweak data such that $ |\epsilon| \lesssim 0.061  \sqrt{
1-(M_Z/M_1)^2}$. This constraint was derived in \cite{FLN1} and a
very similar constraint  appears in the Refs(1,2) of \cite{Wells}.
Consequently the couplings of the $Z'$ boson to the visible matter
fields are extra weak, leading to a very narrow $Z'$ resonance when
decays to hidden sector matter are forbidden\cite{KN1,FLN1}. The
physical width of such a boson could be as wide as ${\cal O} (100~
\rm MeV)$  or as narrow as a  few MeV or even narrower and lie in
the sub-MeV range \cite{FLN1}, provided that the $Z'$ does not decay
into hidden sector matter \cite{KCTC}. These widths are much smaller
than those that  arise for  the $Z$ primes in GUT models (see
Ref.~\cite{Langacker:2008yv}) for  a recent review of $Z'$ models,
as well as an overview of other models with  Stueckelberg mass
mixing \cite{Kiritsis,Anastasopoulos}). In the presence of kinetic
mixing along with Stueckelberg mass mixing but with no matter fields
in the hidden sector, it is shown in \cite{FLN2} that the analysis
of the electroweak sector depends not on $\epsilon$ and $\delta$
separately but on the rescaled parameter $\bar \epsilon = (\epsilon
-\delta)/(1-\delta^2)^{1/2}$  and it is therefore
 $\bar \epsilon$  that is constrained rather than $\epsilon$ by the electroweak data.
However, in the presence of matter in the hidden sector the analysis
in the electroweak sector will depend both on $\epsilon$ and on
$\delta$. Further, it is easily seen that all matter in the hidden
sector acquires a milli
charge~\cite{KN1,FLN1,KCTC,FLN2,Kim,Baek:2008nz,Chun,Feng}.

%%%%
\subsection{Explaining PAMELA Positron Data} The Dirac fermion in
the hidden sector discussed above is a natural candidate for  dark
matter and explicit analyses show this to be the case
\cite{KCTC,FLN2}. Further, the recent PAMELA positron excess anomaly
can be naturally explained by a  Breit-Wigner enhancement of the
annihilation cross sections of these Dirac fermions in the galaxy
when they annihilate in the vicinity of the $Z'$ pole. Such
enhancement can only be achieved when $M_{Z'} \lesssim 2 M_{D}$
 \cite{BreitWigner} where $M_D$ is the mass of the Dirac fermion.
This phenomenon is shown in Fig.(\ref{fig:FLN}).
Thus, the interaction of the hidden sector matter with the \st field given in \cite{KN2} produces upon diagonalization
$g_X Q_X J^{\mu}_{X} C_{\mu} \to   \bar D \gamma^{\mu}  [c_A A_{\mu}+c_Z  Z_{\mu} +c_{Z'} Z'_{\mu}]  D $, while
$c_{Z'}/c_Z \approx  c_{Z'}/c_A  \sim 30  $ for $\epsilon =.06$; i.e. for $(Q_X, g_X)= (1, g_Y)  \rightarrow  c_{A, Z}  \sim 1/100$, while $c_{Z'} \sim g_Y $.
One may then obtain the integrated cross section for $\sigma ( D \bar D \to f \bar f)$ \cite{KCTC,FLN2,BreitWigner},
\beqn
\sigma_{f \bar f} \simeq \frac{N_f s}{32\pi} \frac{\beta_f}{\beta_D} [( |\xi_{L}|^2 +|\xi_{R}|^2)\cdot F_{1} + Re(\xi_L^*\xi_R)\cdot F_{2}],\nonumber
\eeqn
where
$F_{1}=    1+  \beta_D^2 \beta_f^2/3
+{4 M_D^2}{s}^{-1} \left(1-{2m_f^2}/{s}\right )$,
$F_{2}  =   8  {m^2_f}{s}^{-1} \left(1+{2M_D^2}/{s}\right )$,
 $\beta_{f,D}=(1-4m^2_{f,D}/s)^{1/2}$,
 $s=4m^2_{D}/(1-v^2/4)$ and $\xi_{L,R}$  include the
 $(\gamma, Z, Z')$ poles.  The dominant effect in the mass range
 of interest arise from the Breit-Wigner  $Z'$ pole
\be
         \xi^{Z'}_{L,R} =\frac{C^{Z'}_{D} C_{f_{L,R}}^{Z'} }{s - M_{Z'}^2 + i \Gamma_{Z'} M_{Z'}}\nonumber
\label{7} \ee where the explicit expressions for the couplings are
given in \cite{FLN2,BreitWigner}. The Breit-Wigner enhancement
allows for the satisfaction of the relic density consistent with the
WMAP data as shown in Fig.(\ref{fig:FLN}).

\subsection{Stueckelberg Extension of MSSM  } The   \st extension of MSSM (StMSSM) is
constructed from a
 Stueckelberg chiral multiplet mixing vector superfield multiplets for
the $U(1)_Y$ denoted by $B=(B_\mu,\lambda_B,D_B)$
and for the $U(1)_X$ denoted by  $C=(C_\mu,\lambda_C,D_C)$
 and a chiral supermultiplet
$S=(\rho+i\sigma,\chi,F_S)$\cite{KN1,FLN1}
\begin{equation} {\cal L}_{\rm St} = \int d^2\te d^2\thb\ (M_1C+M_2B+  S +\bar
S )^2 . \label{stmssm1} \end{equation}
The Lagrangian of Eq.(\ref{stmssm1}) is invariant under the
supersymmetrized gauge
transformations: $\delta_Y (C, B, S) =(0,
\Lambda_Y + \bar\Lambda_Y,- M_2 \Lambda_Y)$ and $\delta_X
(C,B,S)=(\Lambda_X + \bar\Lambda_X,0,- M_1 \Lambda_X)$.
In the above, the superfield $S$  contains a scalar
$\rho$ and an axionic pseudo-scalar $\sigma$.
\begin{figure*}[t]
\includegraphics[width=5.3cm,height=4.5cm]{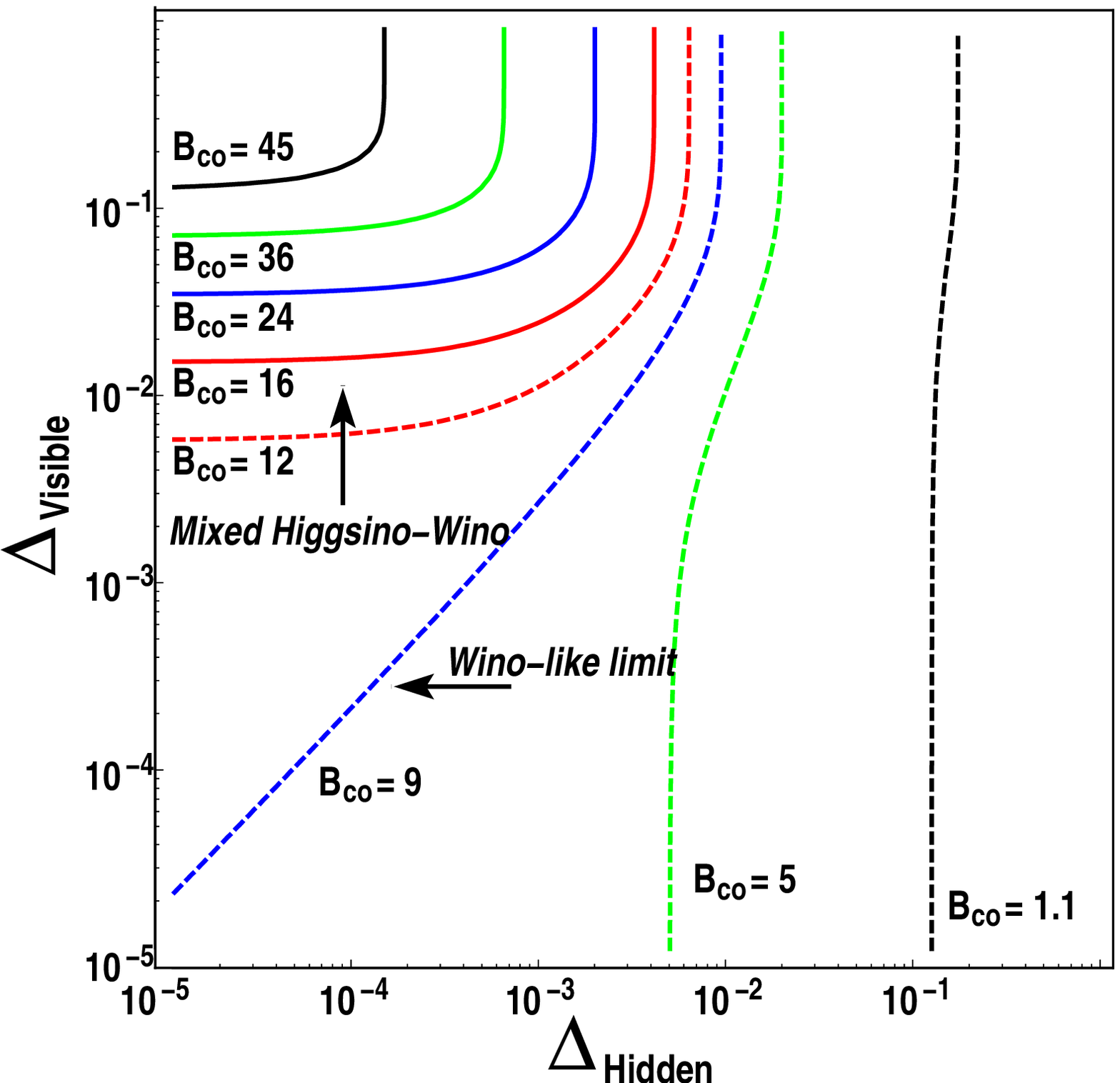}
\includegraphics[width=5.3cm,height=4.5cm]{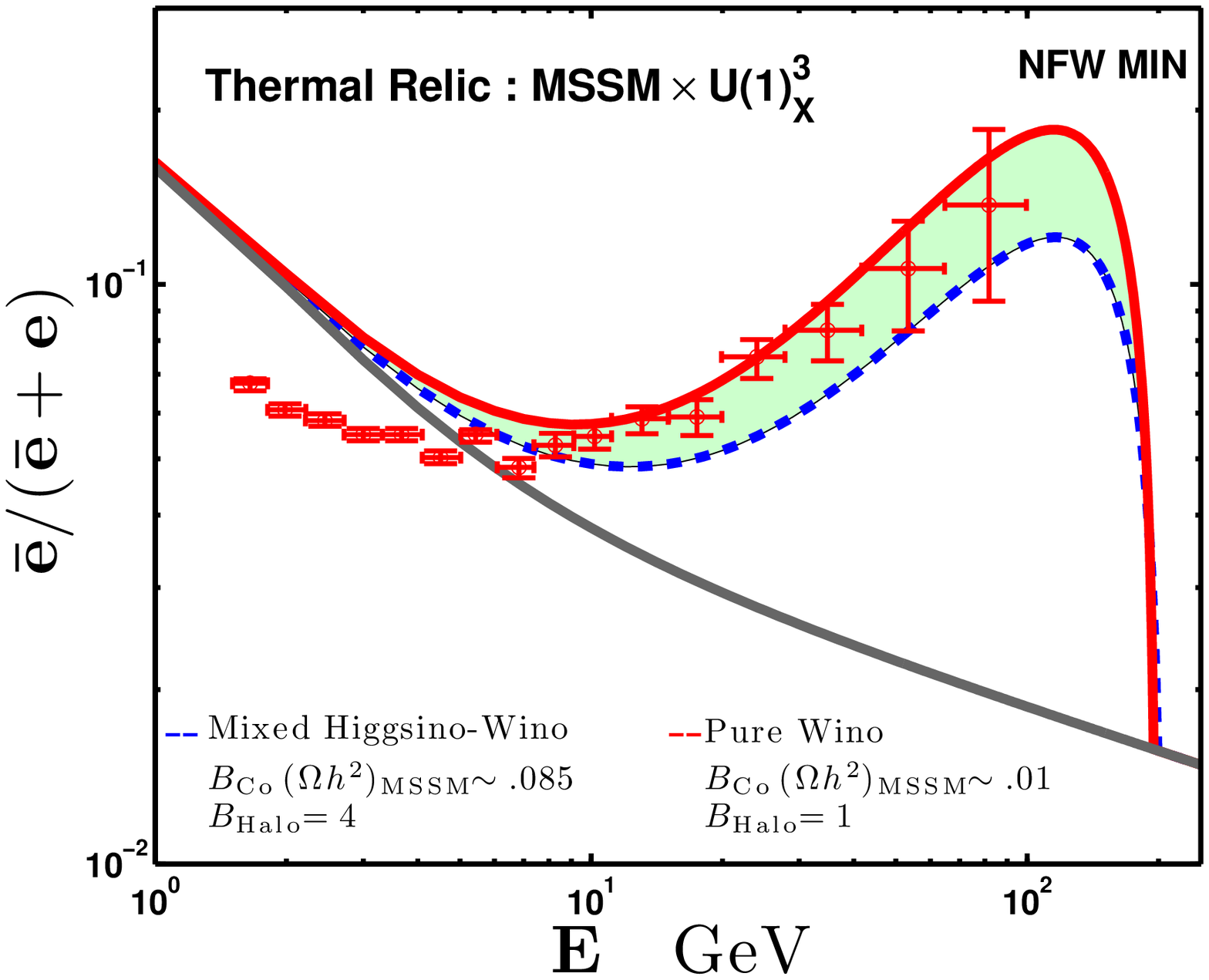}
\includegraphics[width=5.3cm,height=4.5cm]{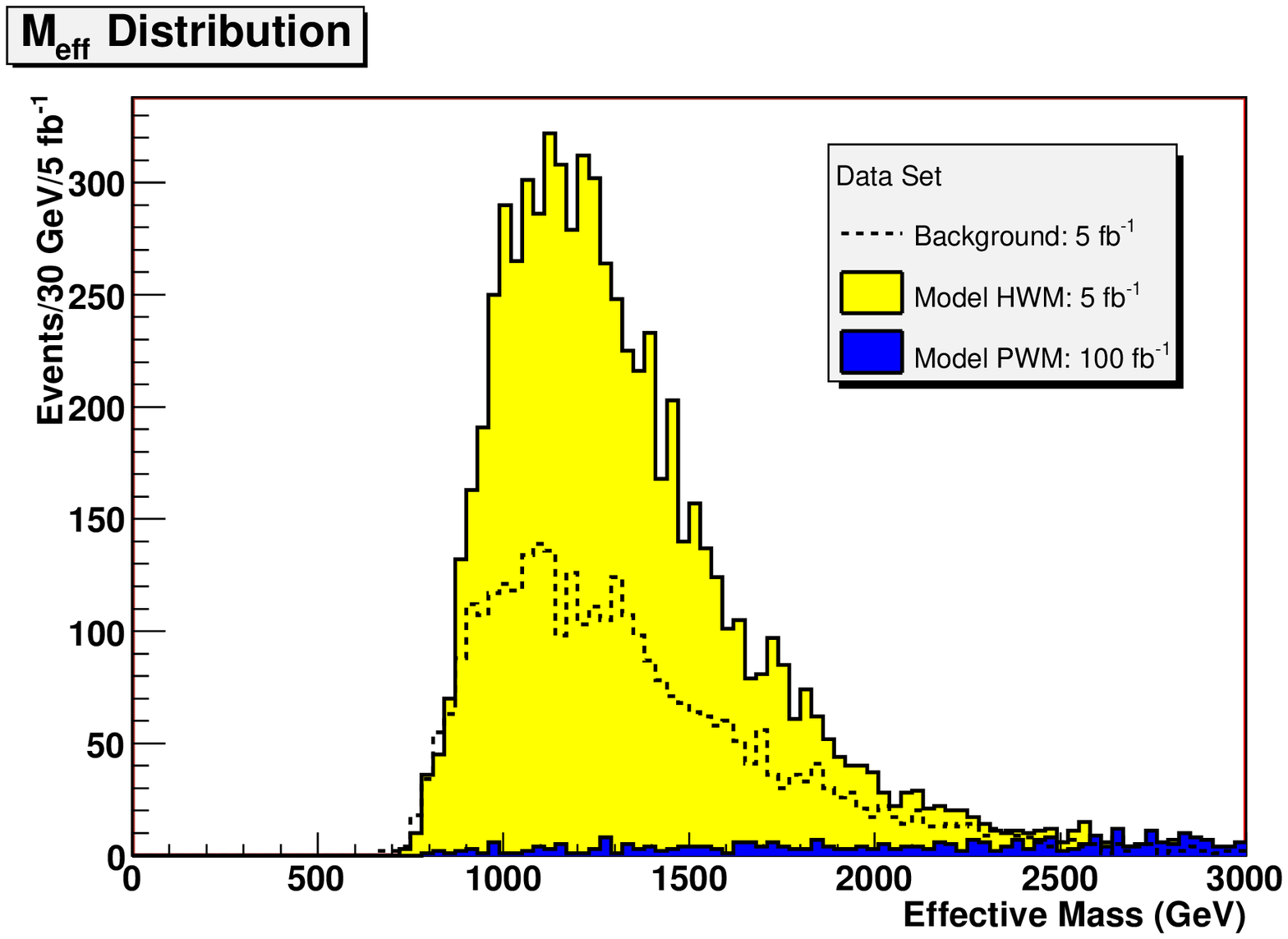}
\vspace{-.7cm} \caption{\small Left: Enhancement of the relic
density via the presence of spectator states in the HS . Right:
Neutralino dark matter producing the PAMELA positron excess for a
pure wino and mixed Higgsino-Wino model(HWM) . With three residual
$U(1)_X$ gauge symmetries the Higgsino-wino model can lead to the
WMAP relic density. Far Right: A strong LHC signal manifests for the
HWM, while the pure wino model has a suppressed LHC signal. From
Ref. (\cite{FLNN}) [similar fits as in the middle panel in both the
shape and normalizations can be seen in \cite{Grajek:2008pg}].}
\label{fig3panel}
\end{figure*}
The StMSSM model class also provides an example
of a model where the astrophysical implications for a wino LSP (a wino LSP in the MSSM  has been re-emphasized in \cite{Grajek:2008pg,FLNN}) as
well as a mixed Higgsino
wino LSP \cite{FLNN} have important effects on observables.
A new feature of this extension (for techincal details see \cite{KN2,FKN,FLNN}) is that it expands the neutralino sector
of the MSSM. The neutralino sector  consists of the Majorana spinors $(\chi_1^0,\chi_2^0, \chi_3^0,\chi_4^0)$
and minimally new Majorana fields labeled $(\xi_1^0, \xi_2^0)$ formed out of the $U(1)_X$ gaugino
and the chiral fermion from the chiral fields $S$ and $\bar S$.

\subsection{Enhancement of Relic Density via Coannihilation with
Hidden Matter} We discuss now an interesting phenomenon in that
matter in the hidden sector can coannihilate with the LSP which has
the effect of enhancing the relic density for the LSP by as much an
order of magnitude or more. This enhancement can occur through the
presence of
 $n$ $U(1)_X$ gauge symmetries in the hidden sector and  $n$ sets of new scalars with
 Stueckelberg masses generated for the $n$ $U(1)_X$ gauge bosons\cite{FLNN}.
 This model then leads to $2n+4$ Majorana states: $(\chi_1^0,(\xi_1^0, \xi_2^0 \ldots \xi_{2n}^0 ),\chi_2^0,\chi_3^0,\chi_4^0)$
where $\chi_i^0$ $(i=1,2,3,4)$ are essentially the four neutralino
states of the MSSM and  $\xi_{\alpha}^0$, $(\alpha =1,...,2n)$ are the
additional states \cite{FLNNdub}. Assuming that the Majorana fields of the hidden sector interact extra weakly,
one finds that there is an enhancement of the relic density by a factor $B_{Co}$ through
coannihilation effects. This enhancement is given by
\beqn
 B_{\rm Co} \hspace{-.3cm} &=&
 \simeq  \frac{\sum_{a,b}\int_{x_f}^{\infty} \langle \sigma_{ab}
 v \rangle \gamma_a \gamma_b \frac{dx}{x^2}}{\sum_{A,B}\int_{x_f}^{\infty} \langle
 \sigma_{AB}  v \rangle  \Gamma_A \Gamma_B \frac{dx}{x^2}},\nonumber \\
 \gamma_a &=&\frac{g_a(1+\Delta_a)^{3/2} e^{-\Delta_a x}}
 {\sum_{b} g_b(1+\Delta_b)^{3/2} e^{-\Delta_b x}}, ~ \rm  MSSM \nonumber \\
 \Gamma_A  \hspace{-.15cm} &=&\frac{g_A(1+\Delta_A)^{3/2} e^{-\Delta_A x}}
 {\sum_{A} g_A(1+\Delta_A)^{3/2} e^{-\Delta_A x}}, \rm\ MSSM \otimes Hid \nonumber .
 \eeqn
  Here $a$ runs over the channels which coannihilate in the MSSM sector,
  while $A$ runs over channels both in the MSSM sector
and in the hidden sector ({\em i.e.}, $A$ =1,..,$n_v+n_h$).
  In the limit when the Majoranas in the hidden sector
  are essentially degenerate with the LSP in the visible sector
  one has for the case of $n$  hidden sector $U(1)$s the result
   $B_{\rm Co} =(1+ {d_h}/{d_v})^2$, where $d_{s} =\sum_{s} g_{s}$, for $s=(v,h)$, i.e.
%%%%%
%\beqn
% {(\Omega h^2)}_{\chi^0}  \simeq {(1+ \frac{d_h}{d_v})^2}(\Omega h^2)_{\rm MSSM},\nonumber\\
% B_{\rm Co} =(1+2n)^2.   \nonumber
% \label{3.6}\eeqn
\begin{equation}
 {(\Omega h^2)}_{\chi^0}  \simeq {(1+ \frac{d_h}{d_v})^2}(\Omega h^2)_{\rm MSSM}.
\end{equation}
When coannihilation effects are negligible in the MSSM sector, one finds that
\begin{equation}
 B_{\rm Co} =(1+2n)^2.
 \label{3.6}
\end{equation}
 Thus  a large enhancement of the relic density can occur even for a modest value of $n$, i.e.,
 $n=3$ leads  to $B_{Co}=49$ in the degenerate limit\cite{FLNN}.  The above phenomenon gives rise to
 viable models which would otherwise be disallowed due to WMAP constraints.
The left panel of Fig.(\ref{fig3panel}) shows this effect which is more pronounced
when the LSP has a non-neglible Higgsino components. The middle panel of
Fig.(\ref{fig3panel}) shows the fit the PAMELA data for two model classes
with hidden sector LSP components (a pure wino and mixed Higgsino-wino LSP)
and the right panel Fig.(\ref{fig3panel}) shows the effective mass distributions
for these models at the LHC with low luminosity sitting high above the background
for the specific Higgsino-wino mixed model.

\vspace{-.1cm} \subsection{Narrow Resonances at the LHC}
\label{stlhc} As discussed above the \st extension of SM and of MSSM
lead to a narrow $Z'$ resonance. Indeed the LHC has the capability
to detect resonances of such small widths as the di-lepton
production can produce a significant number of events above the SM
backgrounds\cite{FLN1}. The analysis of
Fig.(\ref{fig:narrowzprimelhc}) shows that even with 5 fb$^{-1}$ of
integrated luminosity one will be able to discriminate a narrow
Stueckleberg $Z'$ resonance from the standard model background. The
leading order cross section  (before trigger level cuts) for the
model given in Fig.(\ref{fig:narrowzprimelhc}) reported by Pythia is
$\sigma(pp \to {Z'} \to e^{+}e^{-})  =  (0.45 {\rm pb})(0.13)$ for
$(M_1/{\rm GeV},\epsilon) = (500,0.06)$. NLO enhancements are
expected to introduce an enhancement by a factor of $1.3-1.5$. The
result is in excellent accord with predictions given in
\cite{FLN1,FLN2} where previous analyses of the di-lepton cross
sections over large mass ranges are given along with general
expressions and numerical results for the vector and axial vector
couplings of the $Z'$ with SM fermions.
\begin{figure}[h]
\includegraphics[width=6cm,height=5cm]{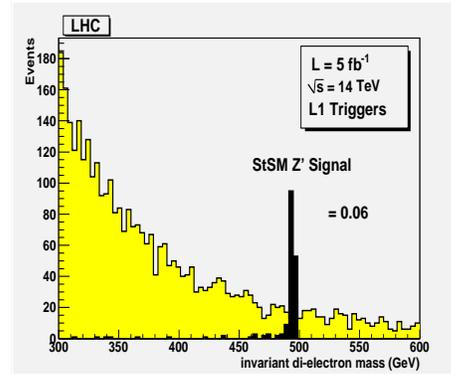}
\vspace{-.7cm}
\caption{\small Narrow  \st $Z'$ at the LHC standing well above the SM backgrounds;
the analysis uses PGS4 with L1 triggers only.
The Drell-Yan Cross section from Pythia agree with the studies of \cite{FLN1}.}
\label{fig:narrowzprimelhc}
\end{figure}
\vspace{-.4cm} \subsection{Summary: Stueckelberg Extensions} The
Stueckelberg extensions of the SM and of the MSSM give rise to
testable signatures of new physics. The minimal model produces a
narrow vector resonance that is detectable in the di-lepton channel
at the Tevatron and at the LHC  \cite{FLN1,FLN2}. At a linear
collider the forward-backward asymmetry near the $Z'$ pole can also
provide a detectable signal \cite{KN2}.
 Further,
if the $Z'$ decays dominantly into the hidden sector,
the mono-jet signatures can also provide a discovery mode \cite{KCTC}.
 The supersymmetric extension  also predicts the presence of
a sharp scalar resonance in the Higgs sector (see \cite{KN2}).

The predictions in the fermionic sector are also rich with implications
for dark matter and for the LHC.
The extensions gives rise to three classes of dark matter (a) milli-weak
(b) milli-charged (c) neutralino-like with extra hidden sector degrees of
freedom.
Thus, the models provide a dirac dark matter candidate \cite{KCTC,FLN2}
that can fit the WMAP data when
integrating over the Breit-Wigner Poles \cite{FLN2} and can also fit the PAMELA
data due the Breit-Wigner enhancement \cite{BreitWigner} from the $Z'$ pole.
The extensions also lead to a fit on the WMAP and PAMELA data for an LSP with a significant
wino component with supressed hidden sector components\cite{FLNN}. Quite generally
the presence of extra weakly interacting hidden sector states provide a boost to the relic density of dark matter due to the presence
of extra degrees of freedom in the hidden sector  \cite{FKN,FLNN}. These models can also yield
large LHC signatures of supersymmetric event rates for a mixed Higgsino-wino LSP in a significant part of
the parameter space.
For further related reviews of the \st extensions we refer the reader to
\cite{Kors:2004iz,Feldman:2007nf,Langacker:2008yv,Nath:2008ch,Langacker:2009im,Liu:2009pt}.

\section{ Hidden Valleys}\label{HVM} We review a few hidden sector
dark matter models, from those that arise in Hidden Valley models,
to solutions to the baryon dark matter coincidence.

\subsection{Overview and basic framework}

Over the past several decades a dominant paradigm for dark matter has emerged at the weak scale.  In theories that stabilize the Higgs mass at the weak scale, there are often new symmetries that give rise to stable particles.  Computing the thermal relic abundance of the weak scale mass particles gives rise, in many of these models, to a dark matter density in accord with what is observed.  This remarkable coincidence has been termed the ``WIMP miracle,'' and is perhaps the most compelling reason to focus theoretically and experimentally on dark matter at the weak scale.

It has been realized in recent years, however, that extensions to the Standard Model can be weakly interacting with the Standard Model while the masses of such states are much lighter than the weak scale, and that in these models the phenomenology can be quite distinct and difficult to uncover at the LHC.  This was the focus of the Hidden Valley models \cite{HV1,HV2}, where a light gauged hidden sector communicates the the Standard Model through weak scale states, as illustrated in Fig.~(1).  These models also bear similarities and connection to ``quirk'' models \cite{quirk} and unparticles \cite{UP1}.

In these models, states at the TeV scale are often unstable to decay to lighter particles in the hidden sector.  This includes, for example, weak scale supersymmetric states that were previously dark matter candidates.  Often the lightest $R$-odd state will reside in the hidden sector, and the MSSM dark matter candidate will decay to such a light state, modifying the dark matter dynamics and the freeze-out calculation \cite{SUSYHV}.

Is the WIMP miracle thus destroyed in the context of these low mass
hidden sectors?  In many cases no.  This can be for one of two
reasons.  First, the same annihilation rate for thermal freeze-out
can be naturally maintained in these hidden sectors.  The
annihilation cross-section needed to obtain the observed relic
abundance is $\langle \sigma_{weak} v \rangle \simeq 3 \times
10^{-26} \mbox{ cm}^3/\mbox{s}$, logarithmically sensitive to the
dark matter mass.   This relation is particularly naturally obtained
for weak scale dark matter, since $g^4/m_{X}^2 \simeq 3 \times
10^{-26} \mbox{ cm}^3/\mbox{s}$ for an ${\cal O}(1)$ gauge coupling
$g$ and weak scale dark matter mass $m_X$.  However, if $g \ll 1$
and $m_{X} \sim g^2 m_{weak}$, the relation still holds for much
lighter dark matter masses. This is particularly well motivated in
the context of gauge mediation, where the dark hidden sector mass
scale, $m_{DHS}$, is set via two loop graphs, $m_{DHS}^2 \simeq g^4
F^2/(M^216 \pi^2)^2 \log(m_{weak}/m_{DHS})$.  Since $m_{DHS}$ scales
with $g^2$, the WIMP miracle still holds for dark matter masses well
below the TeV scale, a ``WIMPless miracle''\cite{wimpless}. For
$10^{-2} \lesssim g \lesssim 0.1$, dark matter in the 0.1 GeV-1 TeV
range is naturally obtained.  On the other hand, if kinetic mixing
is involved , even lower mass scales, such as an MeV, may naturally
be induced \cite{MeV} (though there are strong experimental
constraints on such MeV gauged hidden sectors \cite{toro}).  (For
hidden sectors communicating to the Standard Model through kinetic
mixing where supersymmetry breaking does not set the mass scale in
the hidden sector, see \cite{FLN2,PRV}.  These models are discussed
in the previous section.)  Depending on whether supersymmetry is
predominantly communicated to the hidden sector through a D-term or
gauge mediated two loop graphs, the mass scale in the hidden sector
is $\sqrt{\epsilon g g_Y} m_{weak}$ or $\epsilon g g_Y m_{weak}$,
where $g_Y$ is the hypercharge gauge coupling.  For $\epsilon \simeq
10^{-2}-10^{-4}$ GeV dark forces are obtained as studied recently in
\cite{Hamed,LTW,GeVSpectra}.  For smaller $\epsilon$, lower mass
dark forces may be obtained.  We describe some of these models in
more detail in the next section.

The second case where the observed relic abundance is naturally obtained with dark matter mass well below the weak scale is via solutions to the baryon-dark matter coincidence problem.  In these cases a light hidden sector is, in many cases, {\em required} to reproduce the observed relic abundance.  The baryon-dark matter coincidence is the fact that observationally $\Omega_{DM}/\Omega_b \simeq 5$, while for the standard thermal freeze-out and baryogenesis models, these two quantities are set by unrelated parameters in the model (as in the MSSM, for example, where the dark matter and baryon asymmetries are set largely by dark matter mass and CP asymmetries, respectively).

\vspace{-.1cm}
\begin{figure}[htb]
\includegraphics[scale=0.35]{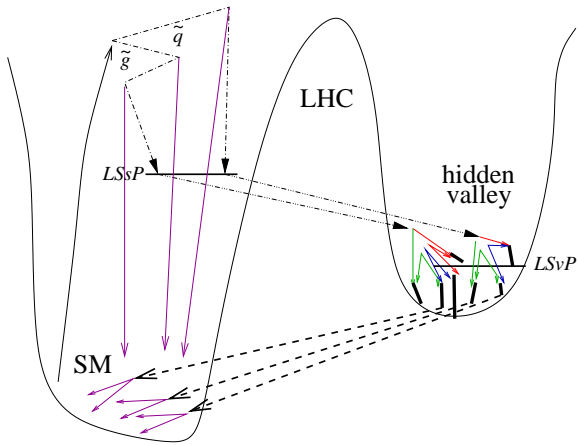}
\caption{A schematic of the Hidden Valley type dark sectors under
consideration.  From Ref.~\cite{SUSYHV} } \label{fig:scheme}
\vspace{-.1cm}
\end{figure}

Solutions to this problem often relate the asymmetric number densities of the dark matter, $n_X - n_{\bar{X}}$, to the baryons (or leptons), $n_X - n_{\bar{X}} \sim n_b - n_{\bar{b}}$, where the exact relations are ${\cal O}(1)$ and depend on the particular operator transferring the asymmetries.  This relation in turn implies a connection between the baryon (proton) mass and the dark matter mass: $m_X \sim 5 m_p$, where again the precise factor will depend on the particular operator transferring the asymmetry.  In this case the dark matter is low mass and weakly coupled to the Standard Model, residing in a Hidden Valley.

In the remainder of this section, we describe an illustrative model of each type, the kinetic mixing type and the baryon-dark matter coincidence type.  We also describe the effects of strong dynamics in particular on the latter type, and lastly turn to discussing collider implications.  This discussion is not meant to be in any sense a complete description of these models, but rather a broad overview of the types of hidden sectors that have been constructed.  We refer the reader to the appropriate references for details on their construction.

\section{Models of hidden dark matter}

\subsection{Low mass dark sectors mediated by kinetic mixing}

As we indicated above, low mass dark forces may be particularly well motivated in the context of gauge mediation with kinetic mixing of a new $U(1)_x$ with hypercharge, as considered in \cite{MeV,Hamed,LTW,GeVSpectra}.  What happens to the dark force in the hidden sector?  As we show here, SUSY breaking effects will induce a vev for the dark Higgses, breaking the dark force and giving it a mass set by the size of the SUSY breaking mass scale in the hidden sector, typically much lower than the TeV scale.

Hypercharge $D$-terms will induce a vev for a dark Higgs, $\phi_i$ in the hidden sector through the potential
\be
V_D = \frac{g_x^2}{2} \left( \sum_i x_i |\phi_i|^2 - \frac{\epsilon}{g_x} \xi_Y \right)^2,
\ee
where $x_i$ is the $U(1)_x$ charge of the Higgs, $g_x$ the gauge coupling and $\xi_Y = -\frac{g_Y}{2} c_{2\beta} v^2$ is the hypercharge $D$-term, with $v = 246 \mbox{ GeV}$ and $\beta$ the mixing between up and down-type Higgses.  This potential induces a vev for the dark Higgs
\be
\langle \phi_i\rangle \simeq \left( \frac{\epsilon \xi_Y}{g_x x_i}\right)^{1/2}.
\ee
For $\epsilon \sim 10^{-3} - 10^{-4}$ the dark $U(1)_x$ gauge boson acquires a GeV scale mass.  For smaller kinetic mixings, smaller gauge boson masses are obtained, even into the MeV range.

There is a subdominant effect, termed Little Gauge Mediation \cite{LGM,GeVSpectra}, which communicates a soft mass to the hidden Higgs of size $m_{soft}^{hid} \sim \epsilon m_{soft}^{vis}$ through the usual two loop gauge mediation diagrams. More precisely this gives rise to a dark Higgs mass
\be
m_{\phi_i}^2 = \epsilon^2 x_I^2 \left( \frac{g_x}{g_Y}\right)^2 m_{E^c}^2,
\ee
where $m_{E^c}$ is the SUSY breaking mass of the right-handed selectron.  These terms are almost always important for determining the precise spectrum of the hidden sectors, particularly when the hypercharge $D$-term is zero.

The spectrum in the hidden sector will depend on the precise matter content, however taking a simple anomaly free dark sector
\be
W_d = \lambda S \phi \bar{\phi},
\ee
results in one stable, $R$-odd fermion, whose mass is either $ \lambda \langle{\phi}\rangle$ or $\sqrt{2} x_H g_x \langle \phi \rangle$.

In these models the dark matter mass is set by thermal freeze-out, and for some ranges of parameters and mass spectra a ``WIMPless miracle'' for dark matter in the MeV to tens of GeV mass range naturally results \cite{GeVSpectra}.  While in some classes of these low mass hidden sector models, thermal freeze-out naturally results in the right relic abundance, we now turn to a class of models where GeV mass states will automatically give the correct relic abundance: solutions to the baryon-dark matter coincidence.

\subsection{Low mass dark sectors as solutions to the baryon-dark
matter coincidence}

There are a number of solutions to to the baryon-dark matter coincidence in the literature \cite{baryonDMcoincidence}, especially in the context of technicolor \cite{technicolor}.  We focus here on a particularly simple class which fits the paradigm of the low mass hidden sector, or Hidden Valley.  This particular class of models is termed Asymmetric Dark Matter \cite{ADM}, and in these cases the dark matter candidate is not derived from models designed to stabilize the weak scale.

The idea behind these models is to write an effective field theory
which describes the interactions between the hidden sector and
visible sector (integrating out the fields residing at the ``pass''
in Fig.~(1), which transfers a Standard Model baryon or lepton
asymmetry to the dark sector.  The dark matter in these models must
be sterile, so this limits the number of operators which can be
constructed to accomplish this purpose.  In particular, in the
context of supersymmetry, the lowest dimension operators carrying
lepton or baryon number which are sterile are $udd$ and $LH$. If
these operators are connected to the hidden sector containing the
dark field $\bar{X}$ to transfer an asymmetry, we have
\begin{eqnarray}
\label{ops}
W = \frac{\bar{X}^2 udd}{M^2} \\ \nonumber
W = \frac{\bar{X}^2 LH}{M}.
\end{eqnarray}
The second operator, for example, enforces $2(n_X - n_{\bar{X}}) = n_{\bar{\ell}} - n_{\ell}$, and a detailed calculation relating the lepton asymmetry to the baryon asymmetry (through sphalerons) consequently shows that this model predicts $m_X \simeq 8 \mbox{ GeV}$.  Note that we added $\bar{X}^2$ and not $X$, since the additional $Z_2$ symmetry ensures DM stability.  In some other cases \cite{displaced}, $R$-parity may be utilized instead to stabilize the dark matter

Now once the Standard Model baryon or lepton asymmetry has been transferred to the dark sector, the symmetric part of the dark matter (which is much larger than the asymmetric part, $n_X + n_{\bar{X}} \gg n_X - n_{\bar{X}}$) must annihilate, leaving only the asymmetric part.  There are a variety of mechanisms to do this, but the difficulty here is having a mechanism which is efficient enough annihilate away the whole of the symmetric part through $X \bar{X} \rightarrow SM$.  Such a process, through a dimension six operator has a cross-section
\be
\sigma v = \frac{1}{16 \pi} \frac{m_X^2}{M'^4}.
\ee
This cross-section must be bigger than approximately 1 pb in order to reduce the dark matter density to its asymmetric component, implying $M' \lesssim 100 \mbox{ GeV}$, a rather severe constraint for any new electroweak state coupling to Standard Model states.

Here confinement in the hidden sector can be a useful tool.  If the dark matter consists of symmetric and asymmetric bound states of elementary dark sector fermions, the symmetric states may decay through the same dimension six operators, while the asymmetric states would remain stable.  For example, suppose in the operator Eq.~(\ref{ops}), we replaced the operator $\bar{X}^2$ with $\bar{v}_1 v_2$, and supposing these $v_1$ and $v_2$ constituents are charged under a hidden sector confining gauge group, such that bound states $\bar{v}_1 v_2$, $\bar{v}_2 v_1$ and $\bar{v}_1 v_1 + \bar{v}_2 v_2$ are the relevant degrees of freedom at low energies.  When Eq.~(\ref{ops}) freezes out, the asymmetric $\bar{v}_1 v_2$ states remain stable, while the symmetric $\bar{v}_1 v_1 + \bar{v}_2 v_2$ states decay rapidly through less suppressed operators (that is, we take $M' \ll M$).  In the next section we describe a related class of confinement models where the constituents of the dark matter bound states carry electroweak charges.  In these models sphalerons rather than higher dimension operators such as Eq.~(\ref{ops}) to transfer the asymmetry.

\subsection{Dark sectors with confinement}

We now illustrate a dark sector model with confinement recently considered in \cite{QDM}.  We note that these models bear some similarity to models constructed earlier in the context of technicolor \cite{technicolor}.  The new defining characteristic of this hidden sector model is the presence of a new non-abelian gauge group which confines {\em at a low scale}.  The dark matter candidate is a {\em charge neutral} composite of electroweak charged, weak scale mass, ``quirks.''  These quirks, $U$ and $D$ are analogous to quarks except they carry a new global charge that keeps one combination, $UD$, stable ($U$ and $D$ carry opposite electric charge).  That is, analogous to the proton, the dark matter is a composite dark baryon. In the language of Fig.~(1), the low mass dark glueballs resides in the hidden sector, while the dark matter constituents are themselves heavy weak scale fields and act as the connectors between the Standard Model and dark glue sector.

Since the constituents are electroweak charged, they can be processed by sphalerons.  In particular, the sphalerons will violate some linear combination of $B$, $L$ and dark baryon number, $DB$.  Thus an asymmetry in $B$ and $L$ (produced from some leptogenesis or baryogenesis mechanism) will be converted to an asymmetry in $DB$.  The $DB$ asymmetry then sets the dark matter relic density.  Since the dark matter mass is around the mass of the weak scale quirk constituents, there must be a Boltzmann suppression in $DB$ to achieve the observed relation $\Omega_{DM} \simeq 5 \Omega_b$.  This can be naturally achieved when the sphalerons decouple just below the dark matter mass:
\be
\Omega_{DM} \sim \frac{m_{DM}}{m_p} e^{-m_{DM}/T_{sph}} \Omega_b,
\ee
where $T_{sph}$ is the sphaleron decoupling temperature, and the exact proportions are worked out in \cite{QDM}.

These dark sectors with confinement have also effectively been used to achieve the mass splittings necessary to realize the inelastic \cite{inelastic,Chang:2008xa} and exciting \cite{exciting} dark matter scenarios \cite{models_comp}.  In these models the dark matter is again a weak scale composite with the confinement scale of the gauge group binding the constituents at the 100 keV-MeV.  The result is mass splittings between the dark matter ground state and excited states set by the confinement scale, and these mass splittings are phenomenologically of the size to fit DAMA \cite{DAMA} and INTEGRAL \cite{INTEGRAL} observations through the excitation of the dark matter ground state to one of the higher states, which then decays back to the ground state, producing $e^+ e^-$ or resulting in an inelastic scattering of dark matter off nuclei.

\subsection{Collider signatures}

The collider signatures for these models can be as diverse as the dark sectors themselves.  These include displaced vertices from hidden sector decays, dark hadronization jets and lepton jets.  We draw attention here to some of the collider signatures which are not discussed elsewhere in this paper.

First, as pointed out in \cite{SUSYHV}, the presence of Hidden Valleys with supersymmetry causes the MSSM lightest supersymmetric partner to decay to hidden sector states.  For example, through the operators Eq.~(\ref{ops}), the neutralino can have exotic decays to light dark matter states, such as $\chi^0 \rightarrow \nu \bar{X}\bar{X},~h^-\ell^+ \bar{X}\bar{X}$ or $\chi^0 \rightarrow XX qqq$.  The phenomenology of these models remains to be studied further.

In other models discussed here, once the dark states are produced, cascade decays in the hidden sector are completely invisible, and as a result the only signature is missing energy.  In this case, how does one ascertain the nature of the dark sector and extract key information?  Information about the hidden sector must be obtained in this case utilizing initial state radiation (ISR), as a jet or photon radiated off an initial state quark or lepton \cite{FLN2,KCTC,Petriello}.   An example is the invisible $Z'$: in these models a $Z'$ couples to both the hidden and visible sectors, so that the $Z'$ has a significant hidden decay branching fraction.  One way to search for these  models is to simply do a counting experiment \cite{Petriello}: look for an excess of mono-photon or mono-jet plus missing energy.  As shown in Fig.~(2), for a $Z'$ mass below $\sim 1 \mbox{ TeV}$, even such a simple counting experiment can uncover new physics.  The signal required to discover an invisibly decaying $Z'$ with initial state photon for 10, 30 and 100 $\mbox{ fb}^{-1}$ is compared against typical invisible decay signals for sequential and $U(1)_\chi$ $Z'$'s.  Further signal separation could be achieved using more sophisticated event shape variables.

\vspace{-.5cm}
\begin{figure}[htb]
\includegraphics[scale=0.33,angle=90]{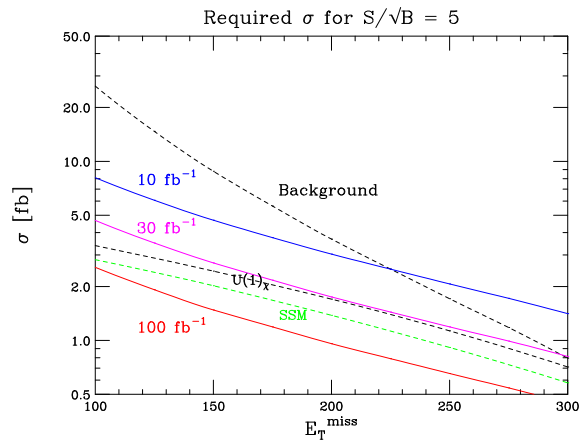}
\caption{Required cross-section to discover at 5$\sigma$ an
invisibly decaying $Z'$ with mono-photon plus missing energy for 10,
30 and 100 fb$^{-1}$, along with the expected signals from two
invisibly decaying $Z'$, from the sequential standard model and
$U(1)_\chi$.  From Ref.~\cite{Petriello}.} \label{fig:scheme}
\vspace{-.5cm}
\end{figure}

\subsection{Summary of Low Mass Dark Sectors}

As we have discussed, there are broad classes of models of low mass hidden dark matter that retain many of the phenomenological successes of weak scale, weakly interacting particles.  We have outlined three such classes, the first where the light dark sector communicates to the Standard Model through light states which have, however, small interactions with Standard Model states through kinetic mixing (or simply small gauge couplings).  Though the dark matter is much lighter than the weak scale, in some of these models the WIMP miracle is still obtained, and the observed relic density of dark matter is produced.  Second, we looked at Asymmetric Dark Matter models, where a dark matter mass near the proton mass is necessary to give rise to the observed relic abundance.  Third, we examined cases where these dark sectors feature new confining gauge groups with a low confinement scale as in a Hidden Valley, quirk or unparticle model.

In summary, we are beginning to learn that the dark sector could be complex -- it may not be simply be a single, stable, weakly interacting particle.  There may be multiple resonances in the hidden sector with an array of new forces that govern their interactions, from confining gauge groups to a dark $U(1)$.  And this new dynamics need not reside at the weak scale, opening new avenues for exploration.

\section{Probing the GeV dark sector at the LHC} \label{pheno}
Dark matter can carry GeV$^{-1}$ scale self-interactions. The GeV
force carrier and associated states constitute a so-called dark
sector.   We outline the LHC signals of such a dark sector.
\subsection{Overview}
Motivated by astrophysical observations, it has been proposed \cite{Hamed} (see also \cite{PRV} ) that electroweak scale dark matter ($m_{\rm DM} \sim $ TeV) have  GeV$^{-1}$ range self-interactions. The force carrier and associated states are collectively referred to as ``the dark sector".  In order to account for the excesses in the cosmic ray observations, the dark sector generically also couples the Standard Model states. To satisfy the experimental constraints, such couplings  (the``portal"),  are expected to be tiny.  More specific model buildings for the dark sector have been carried out in \cite{LGM,NA,LTW,Chun,Cheung,Katz,GeVSpectra,Feng,Goodsell,models_comp,models_axion}. We also note that this class of models can be regarded as a distinct possibility of the hidden valley scenario \cite{HV1,HV2}.\\

\subsection{Basic framework}
%\vspace{cm}
\begin{figure}[htb]
\includegraphics[scale=0.5]{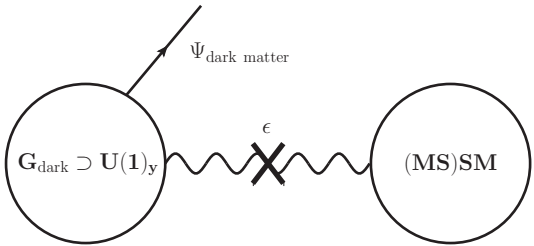}
\caption{ A schematic of the setup under consideration. Dark matter
carries GeV$^{-1}$ range self-interaction $G_{\rm dark}$. The GeV
dark sector couples to the SM via some small coupling $\epsilon$.}

\label{fig:scheme}
\vspace{-1cm}
\end{figure}
A schematic setup for the dark sector model is shown in Fig.~\ref{fig:scheme}. Different choices of $G_{\rm d}$ and the portal to the Standard Model have been considered \cite{models_axion}. In the following, we will focus on the case in which  $G_{\rm d}$ is a gauge interaction, and the portal is generated by  kinetic mixing between an $U(1)_y$ factor of $G_{\rm d}$ and the hypercharge $U(1)_Y$. In the following, we will discuss the most  relevant part of the  Lagrangian from which the most generic signals can be derived. The kinetic mixing can be parameterized as \cite{ew_prod}
\beqn
\tiny
\label{eqn:kinmix}
{\cal L}_\textrm{gauge mix} &=& -\frac{1}{2}\epsilon_1 b_{\mu\nu} A^{\mu\nu} - \frac{1}{2} \epsilon_2  b_{\mu\nu} Z^{\mu\nu} \nonumber \\
&=&-\frac{1}{2}\epsilon_1^\prime b_{\mu\nu} B^{\mu\nu} - \frac{1}{2} \epsilon_2^\prime  b_{\mu\nu} W_3^{\mu\nu} \nonumber \\
\eeqn
where $b_{\mu\nu}$ denotes the field strength for the dark gauge boson and $\epsilon_{1,2}$ and $\epsilon_{1,2}^\prime$ are related by the Weinberg angle. In particular, when only $\epsilon_1^\prime$ is present, we have $\epsilon_1 = \epsilon_1^\prime\cos\theta_W$ and $\epsilon_2 = \epsilon_1^\prime\sin\theta_W$\footnote{ \tiny $\epsilon'_2$ can arise from higher dimensional operators such as $b_{\mu\nu} tr(H^\dag W^{\mu\nu} H)/\Lambda^2$. We will not focus on this situation here, as it will not qualitatively change the phenomenology.}. In supersymmetric scenarios, there is also an identical mixing between the gauginos
\beqn
\tiny
{\cal L}_{\textrm{gaugino mix}} = -2i\epsilon_1' {\tilde b}^\dagger \bar\sigma^\mu \partial_\mu {\tilde B}   -2i\epsilon_2' {\tilde b}^\dagger \bar\sigma^\mu \partial_\mu {\tilde W_3}%+{\rm h.c.}
\nonumber \\
\eeqn
The kinetic mixings can be removed from by appropriate field redefinitions, which lead to the portal couplings
\beqn
\tiny
\label{eq:portal}
{\cal L}_{\rm portal} &=& \epsilon_1 b_\mu J_{\rm EM}^\mu + \epsilon_2 Z_\mu J_b^\mu \nonumber \\ &+& \epsilon_1' {\tilde B} \tilde J_{\tilde b}+\epsilon_2' {\tilde W}_3 \tilde J_{\tilde b} ,
\eeqn
\beqn
 J_b^\mu &=& g_d \sum_i q_i\left(i (h_i^\dagger \partial^\mu h_i -h_i \partial^\mu h_i^\dagger ) + \tilde h_i^\dagger \bar{\sigma}^\mu \tilde h_i\right)
\nonumber \\
\tilde{J}_{\tilde b} &=& -i\sqrt{2} g_d \sum_i q_i \tilde{h}_i^\dagger h_i
\eeqn
where $J_{\rm EM}$ is the SM electromagnetic current. $J_b$ contains dark scalar and dark fermion bilinears, and $\tilde J_{\tilde b}$ contains mixed dark scalar-fermion bilinears. We will consider couplings in the range  $\epsilon_i \sim 10^{-3} - 10^{-4}$, which satisfies all the constraints (For recent studies, see \cite{Posp,Reece} and references therein.) and can arise naturally in models.

We will focus on the simplest case $G_{\rm d} = U(1)_y$  (and denote $b_\mu$ as dark photon) for the rest of note, which encapsulates the main features of dark sector phenomenology \cite{LTW,ew_prod,Cheung}.  We will highlight the new features from a more complicated dark sector.

\subsection{Production at the LHC}

We will discuss in this section relevant production channels for the LHC search for the GeV dark sector. The relevant rates are shown in Fig.~\ref{fig:rates}. Such GeV dark sector states will decay back to Standard Model light states, such as leptons, and produce distinct signals which we will discuss in detail in the next section.

{\bf \em Prompt ``dark photon".} We see from the first term in  Eq.~\ref{eq:portal} that the dark $U(1)_y$ couples just like the Standard Model photon, except with a coupling suppressed by $e_{\rm eff}/e \equiv \epsilon_1$. Therefore, the dark photon, $\gamma' \equiv b_\mu $, should be produced just like the Standard Model photon (with a much smaller rate), for example, through the prompt photon process $p p \to \gamma' +$X.

{\bf \em Rare Z decay.} The second term in Eq.~\ref{eq:portal} implies that the Standard Model $Z^0$ has a rare decay mode into the dark sector, with a branching ratio proportional to $\epsilon_2^2$.\\
\begin{figure}[h!]
\vspace{-0.5cm}

\includegraphics[scale=0.24,angle=270]{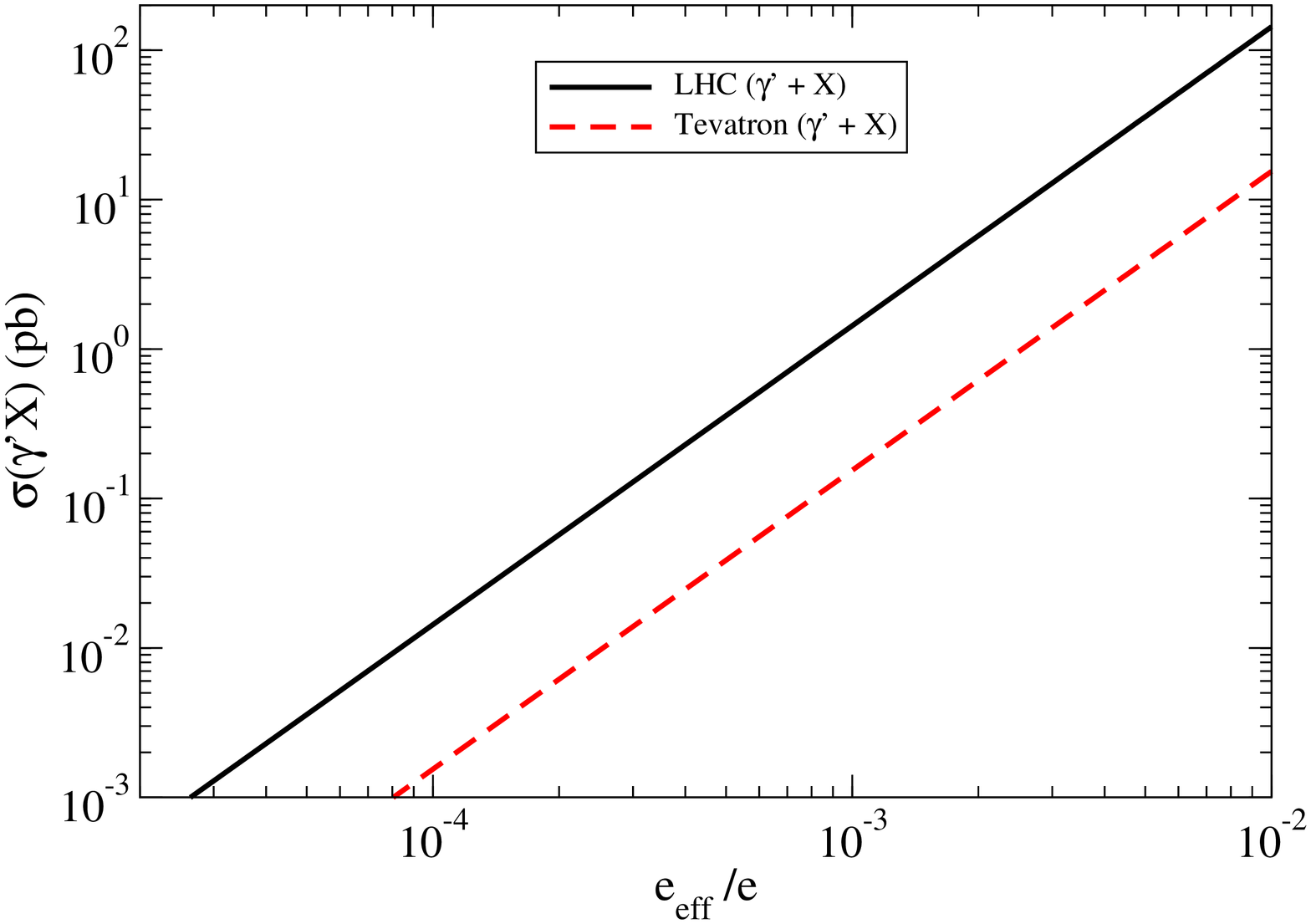}\\

\includegraphics[scale=0.36]{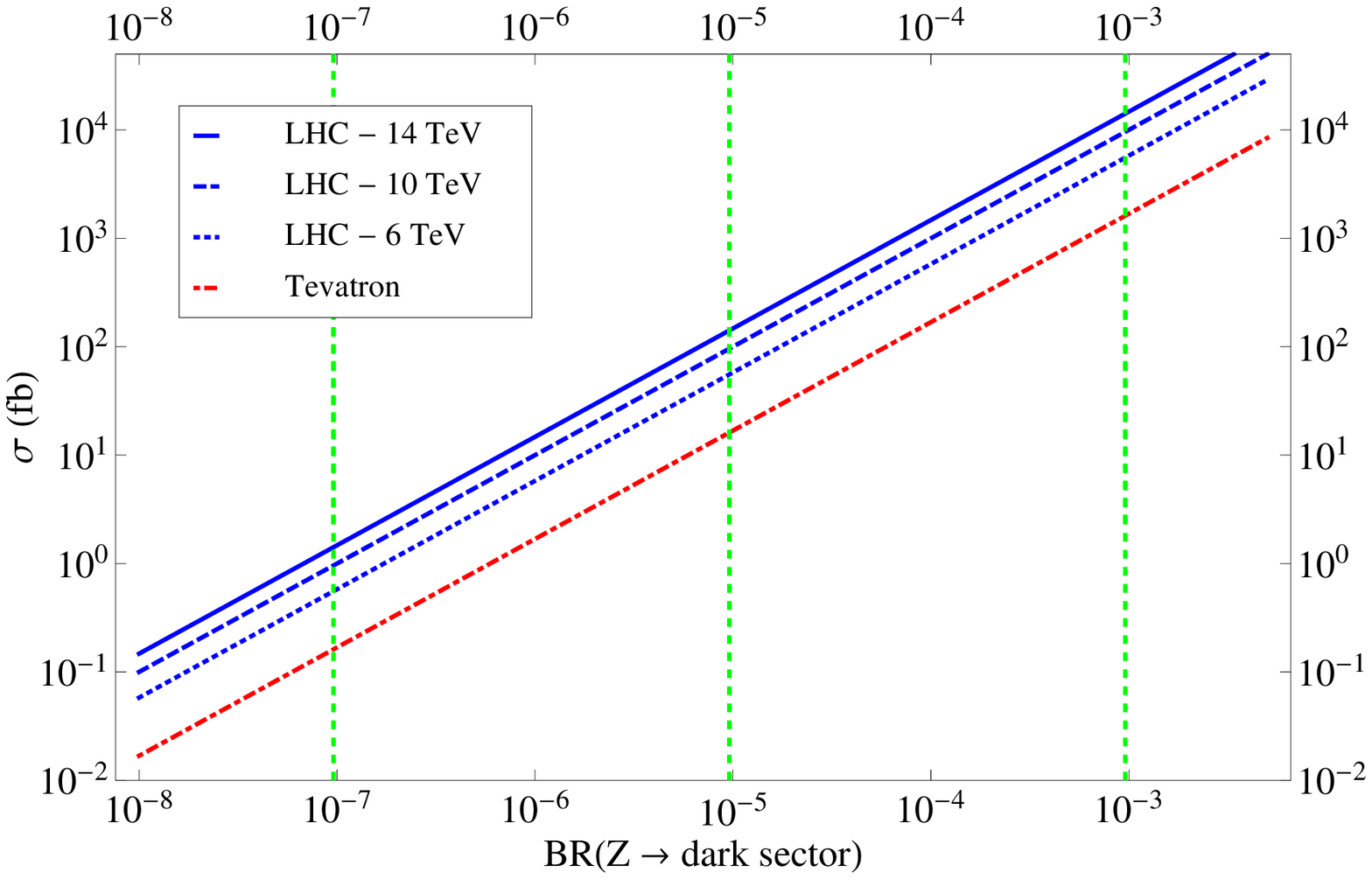} \\

\vspace{0.2cm}
\includegraphics[scale=0.24]{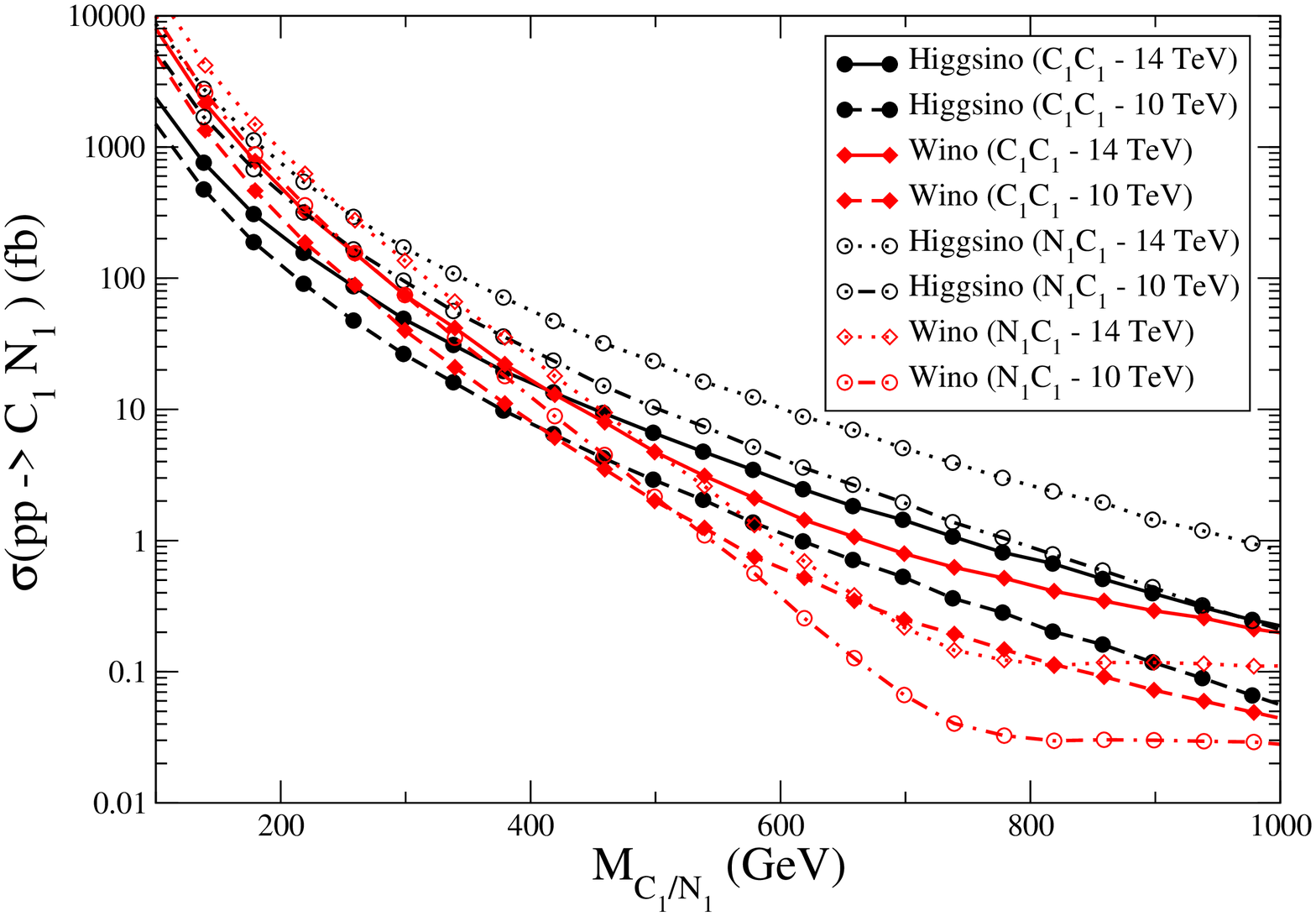}

\vspace{-1cm}
\caption{Rates of dark sector production processes. Top: prompt dark photon at the LHC ($E_{\rm cm}=14$ TeV); middle: rare Z decay at the LHC, $\alpha_d=1/127$; bottom: some important SUSY electroweak-ino production processes. See text for detailed explanation.  }
\label{fig:rates}
\vspace{-0.7cm}
\end{figure}
{\bf \em SUSY electroweak-ino production.}  Supersymmetry provides natural setups of the GeV dark sector, in which both the GeV scale and small portal coupling are generated in very simple models \cite{LTW,ew_prod,Cheung}. The presence GeV dark sector dramatically changes the SUSY phenomenology \cite{nima_lhc,LTW,ew_prod}. In particular, LSP will decay into the dark sector through the last two couplings in Eq.~\ref{eq:portal}, the subsequent  decay of the dark sector states will result in collimated Standard Model charged leptons. As the LSP is alway present at the end of any SUSY decay chain, the production rates for dark sector states are just the production rates of the electroweak-inos.
Of course, the dark sector states can also be produced in longer SUSY decay chains starting with colored superpartners, with hard jets. Although not as clean as the direct electroweak-ino production, it can certainly be a very useful channel given the larger production rate of the colored superpartners.

{\bf \em Dark sector cascade and parton shower.} The dark sector typically has at least several states. Heavier dark sector states, after being produced through one of channels mentioned above, will cascade down to lighter states. In addition, if the dark sector gauge coupling is not so small, dark sector state can have ``dark radiations" similar to the QCD and QED radiations.

{Signals: lepton jets and beyond}:

We begin by describing the decay of the dark sector states back to the Standard Model.

{\bf \em Dark photon and lepton jet}

The first term in Eq.~\ref{eq:portal} implies that the dark photon will decay into charged particles of the Standard Model. Since $m_{b_{\mu}} \sim $ GeV, typically the dominant channels  are $e^+ e^-$, $\mu^+  \mu^-$, and $\pi^+ \pi^-$, with significant branching ratios into the leptonic channels (for recent studies see \cite{Posp,Reece},\cite{PDG-R}).  Since the dark photon are produced at the LHC typically with large boost, for example $\gamma=m_{\rm Z} /2 m_{b_{\mu}} \sim 50 $ from Z decay, the resulting decay products are highly collimated. This leads to a class of unique objects, {\em lepton jets} \cite{nima_lhc,LTW,ew_prod}, which are high collimated energetic leptons. The typical multiplicity of the leptons in a lepton jet is model dependent. A dark photon decays into a pair of leptons. Cascade, and parton showering, in the dark sector can lead to higher multiplicities (possibly 4 or more). For the range of $\epsilon$s under consideration, the decay of dark photon is almost always prompt.

{\bf \em Dark Higgs}

The dark gauge interaction must be spontaneously broken at around a GeV, which can be achieved by introducing a dark Higgs sector. The dark Higgs particles can be produced at the LHC through Z and LSP decay, and possibly through a dark sector cascade. Heavier dark Higgses will cascade down to the lighter ones and possibly lighter dark gauge bosons. The LHC signal of  the dark higgs sector depends on the mass of the lightest dark higgs  in comparison with $m_{b_\mu}$.  If $m_{h_{d}} > 2 m_{b_\mu}$, we have $h_d \to b_\mu b_\mu$, followed by $b_\mu$ decay, giving rise to  multiple ($>$ 4) lepton  final states which reconstruct 2 dark photon resonances and the dark Higgs.  If $m_{b_\mu}< m_{h_d} < 2 m_{b_\mu}$, we have $h_d \to b_\mu^{*} b_\mu$. The final state is similar to the previous case with less reconstructed resonances. There is also a possibility of having displaced vertices in this case.  If $m_{h_d} < m_{b_\mu}$, dark Higgs will decay either to a 4 body final state through 2 off-shell $b_\mu^*$,  or to 2 body final states through a loop process. In either case, the decay lifetime is much longer than the detector time scale, and the dark higgs will leave its trace as missing energy.

{\bf \em More details of the Signal}

A more detailed study of lepton jets from electroweak processes, including Z and LSP decay, has been carried out in Ref.~\cite{ew_prod}. In this more realistic study, an isolation criterion is adopted.  We require that: Two or more leptons each with $p_T>10 $ GeV inside a cone of $\Delta R < 0.1$ with hadronic/leptonic isolation cut of $\sum p_T < 3 $ GeV in an annulus of $0.1< \Delta R < 0.4$ around the lepton jet. We have included the effect of dark sector parton showering (in the simple case of $G_{\rm d} =U(1)$).  The decay branching ratios of dark photon into leptonic and pion final states have been properly taken into account.  We find that

\begin{figure}[htb]
\includegraphics[scale=0.26]{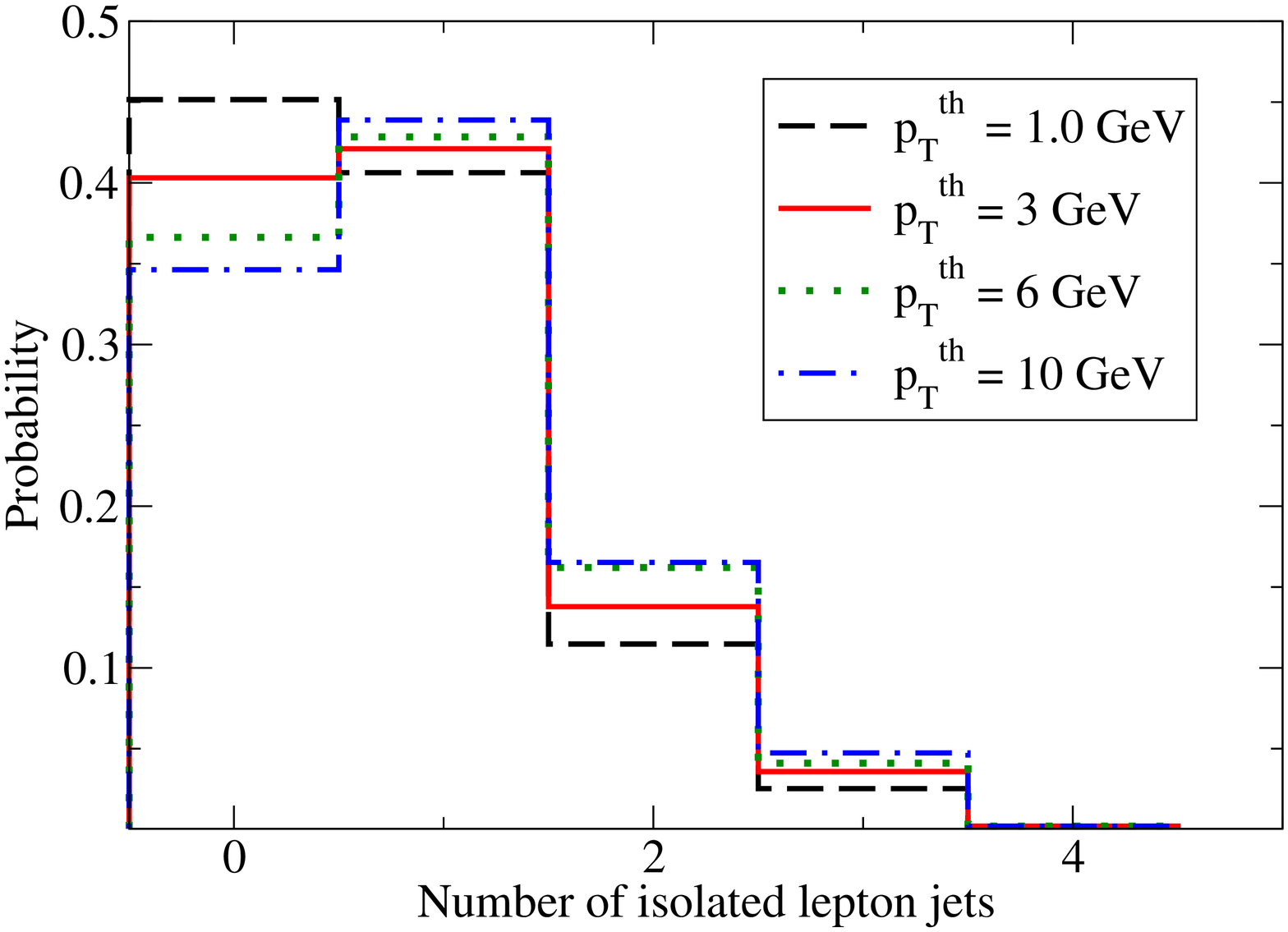}\\

\vspace{0.25cm}
\includegraphics[scale=0.26]{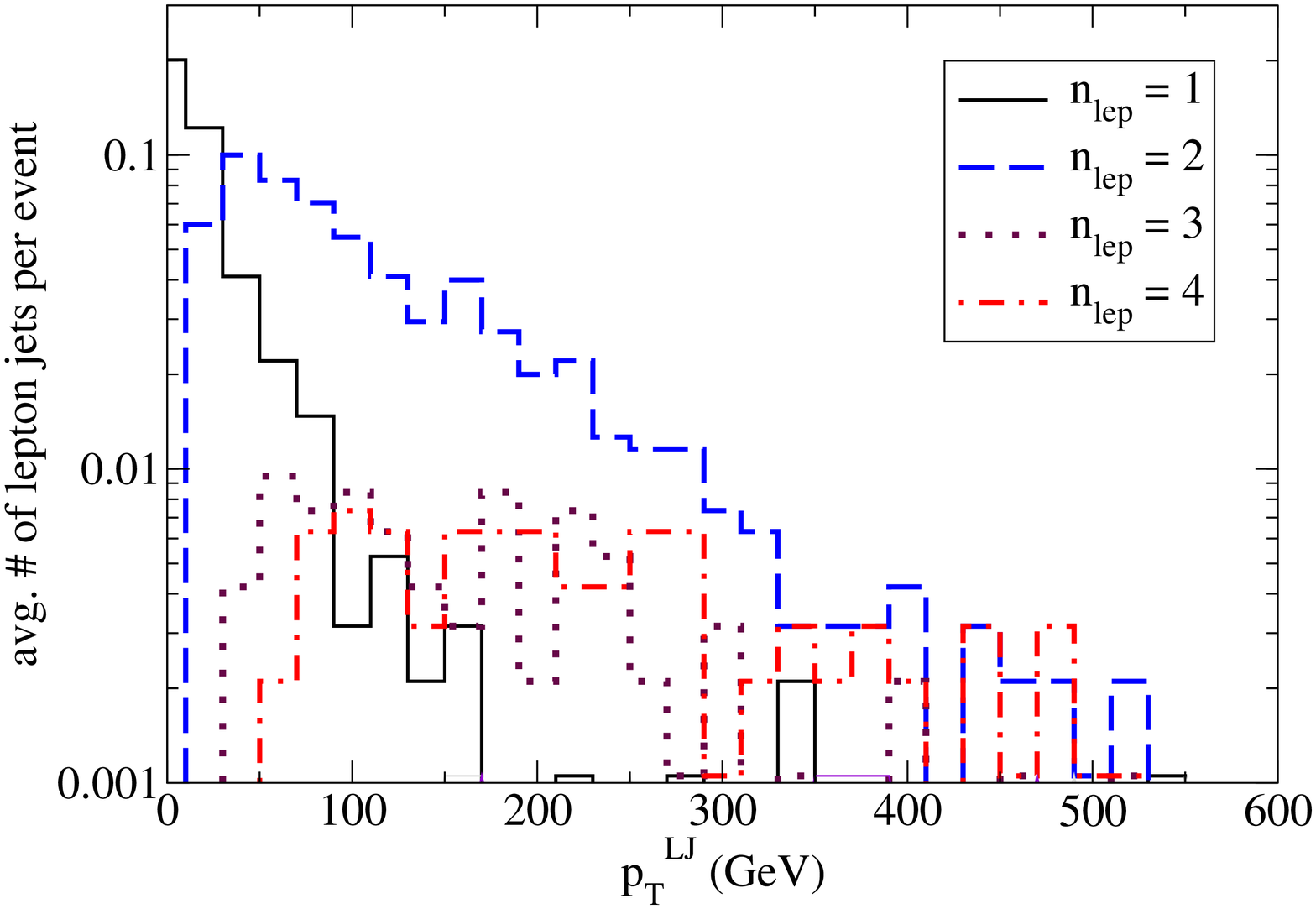} \\

\vspace{-1cm}
\caption{Top: lepton jet efficiency; bottom: lepton multiplicity and lepton jet $p_T$. $m_{\rm LSP} = 300$ GeV, $\alpha_d=0.1$. }
\label{fig:LJ}
\vspace{-0.7cm}
\end{figure}

\noindent{\it 1.} The efficiency of having well isolated lepton jet(s) is significant. We see from Fig.~\ref{fig:LJ} that for electroweak-ino production, more than half of the event will have at least one well isolated lepton jet.

\noindent{\it 2.} The hardest leptons are dominantly from the decay of the dark photon coming directly from the decay of the LSP (or Z), while the radiated dark photons (in the weakly coupled models) typically contribute a number of soft (several to 10s of GeV) leptons. Lepton jets with 2 leptons receive contributions  from both direct decay and radiation. Lepton jets with 3 or more leptons are dominated by the direct decay, as a result, the leptons are more energetic, see Fig.~\ref{fig:LJ}.

\noindent{\it 3.} There are indeed a large number of isolated leptons. Typically coming from the decay of soft dark photons, they are less energetic.  A significant fraction of them could still be hard enough, $\geq 10 $ GeV,  to be useful.

\noindent{\it 4.} The results shown here is for a particular choice of dark gauge coupling and leptonic decay branching ratio. See Ref.~\cite{ew_prod} for more detailed studies with different choices of parameters. Generically, the effect of radiation decreases (increase) linearly with smaller (larger) dark gauge couplings, from almost no radiation (with small coupling) to the case where there is no clear distinction between direct decay and radiation.

\subsection{Summary of GeV Dark Sector Signatures} \noindent{\bf
$\bullet$} Lepton jet recoiling against a QCD-jet would be an
inclusive search for a prompt dark photon production.

\noindent{\bf $\bullet$} Two lepton jets recoiling against each other and reconstructing the $Z^0$ would be an interesting signal of rare $Z^0$ decays into the dark sector and can be looked for at LEP, Tevatron, and LHC.

\noindent{\bf $\bullet$} Two (or more) lepton jets together with missing energy and possibly other isolated final states (e.g. a muon, an electron, and etc.) can be the result of electroweak-ino production and their eventual cascade into the dark sector.

\noindent{\bf $\bullet$} Lepton jets in association with QCD-jets could be the result of strong production of colored particles which eventually cascade into the dark sector.

\section{Conclusions} \label{concl}
%The hidden sector introduced in supergravity unified models to achieve a phenomenologically
%consistent breaking of supersymmetry turns out to be a generic feature of string  models.
%It is thus interesting to investigate what the implications of such hidden sectors may be for
%LHC physics.

 The analyses presented here show that in a variety of settings the presence of a hidden sector gives rise to unique signatures in both collider physics and in the hunt for dark matter. The mechanisms for communication between the hidden and visible sectors, aside from by gravity, could be via $U(1)$ gauge fields in the hidden sector which mix with the
gauge fields in the visible sector via kinetic mixings or via mass mixing by the Stueckelberg
mixing mechanism, or via higher dimensional operators.

 Specifically,
in Sec.(\ref{StueckelbergExt}) hidden sector extensions with \st mass and kinetic mixing were discussed which
lead to several new models of dark matter and a host of new physics signatures both in dark matter experiments and at the  LHC;
 the most striking of which at hadron colliders would be a very a narrow $Z$ prime
resonance in the di-lepton channel accompanied by an excess of positrons from the galactic halo due to a Breit-Wigner pole enhancement. These phenomena would help
 pinpont the mass of the dark matter particle.
In Sec.(\ref{HVM}) classes of hidden sector models with low mass dark matter were reviewed which can arise via kinetic mixings, as well as via asymmetric dark matter models, and dark sectors with a new confining gauge groups which are natural in a Hidden Valley, a quirk or unparticle model. Collider
implications of a invisibly decaying $Z$ prime was also re-emphasized.
In Sec.(\ref{pheno}) photon, lepton and jet signatures of dark sectors with a GeV mass $Z'$ particle were reviewed in both supersymmetric and non-supersymmetric models
with kinetic mixings. Discovery prospects at the LHC in several channels were discussed in detail.

In summary, the models
discussed here provide visible signatures of hidden symmetries. With the turn on of the LHC and forthcoming data
from several dark matter experiments, the hidden sector models of the type discussed above can be put to the
test on both fronts.

%%%%%%%%%%%%%%%%%%%%%%%%%%%%%%%%%%%%%%%%%%%%%%%%%%%%%%%%%%%%%%%%%%%%%%%%%%%%%%%%%%%%%%%%%%%%%%
%%%%%%%%%%%%%%%%%%%%%%%%%%%%%%%%%%%%%%%%%%%%%%%%%%%%%%%%%%%%%%%%%%%%%%%%%%%%%%%%%%%%%%%%%%%%%%
\chapter{Probing the Origin of Neutrino Mass at the LHC}
\setlength{\epigraphrule}{1pt}
\epigraphhead[20]{\epigraph{\large {\em J.A.~Aguilar-Saavedra, Borut
Bajc, F.~de~Campos, O.J.P.~\'Eboli, Pavel Fileviez P{\'e}rez,
W.~Grimus, Tao Han, M.~Hirsch, L.~Lavoura, M.B.~Magro, M.~Malinsky,
R.N.~Mohapatra, S.~Morisi, B.~Mukhopadhyaya, Werner Porod,
D.~Restrepo, Goran Senjanovi\'c, J.C.~Rom\~ao, J.~Schechter,
J.W.F.~Valle}}{\large R.N.~Mohapatra \& J.W.F.~Valle (Conveners)}}

\newcommand{\nn}{\nonumber}

%\definecolor{brown}{rgb}{0.6,0.3,0}
%\newcommand {\black} {\color{black}}
%\newcommand {\red} {\color{red}}
%\newcommand {\blue} {\color{blue}}
%\newcommand {\green} {\color{green}}
%\newcommand {\brown}[1] {{\color{brown}#1}}
%\definecolor{magenta}{rgb}{0.5,0,0.5}
%\newcommand {\magenta}[1] {{\color{magenta}#1}}

\renewcommand{\baselinestretch}{1.125}

\newcommand{\CL}   {C.L.}
\newcommand{\dof}  {d.o.f.}
\newcommand{\eVq}  {\text{eV}^2}
\newcommand{\Sol}  {\textsc{sol}}
\newcommand{\SlKm} {\textsc{sol+kam}}
\newcommand{\Atm}  {\textsc{atm}}
\newcommand{\Chooz}{\textsc{chooz}}
\newcommand{\Dms}  {\Delta m^2_\Sol}
\newcommand{\Dma}  {\Delta m^2_\Atm}
\newcommand{\Dcq}  {\Delta\chi^2}
\newcommand{\EtAl}  {{\it et al.\/}}
\newcommand{\eps}  {\varepsilon}
\newcommand{\epp}  {\varepsilon'}
\newcommand{\plumin}[2]{^{+#1}_{-#2}}
\renewcommand{\ttbs}{\char'134} %%

\def\rpv{R_p \hspace{-1.0em}/\;\:}
\def\lsim{\raise0.3ex\hbox{$\;<$\kern-0.75em\raise-1.1ex\hbox{$\sim\;$}}}
\def\gsim{\raise0.3ex\hbox{$\;>$\kern-0.75em\raise-1.1ex\hbox{$\sim\;$}}}
\def\e6{$\mathrm{E(6)}$ }
\def\10{$\mathrm{SO(10)}$ }
\def\21{$\mathrm{SU(2) \otimes U(1)}$ }
\def\321{$\mathrm{SU(3) \otimes SU(2)_L \otimes U(1)_Y}$ }
\def\SM{$\mathrm{SU(3) \otimes SU(2)_L \otimes U(1)_Y}$ }
\def\LR{$\mathrm{SU(3) \otimes SU(2)_L \otimes SU(2)_R \otimes
    U(1)_{B-L}}$ }
\def\lr{$\mathrm{SU(3) \otimes SU(2)_L \otimes SU(2)_R \otimes U(1)}$ }
\def\3211{$\mathrm{SU(3) \otimes SU(2)_L \otimes U(1)_R \otimes U(1)_{B-L}}$ }
\def\422{$\mathrm{SU(4) \otimes SU(2)_L \otimes SU(2)_R}$ }
\def\321{SU(3) $\otimes$ SU(2) $\otimes$ U(1)}
\def\vev#1{\left\langle #1\right\rangle}
%% pavel
\def\md{M_\Delta^{}}
\def\dl{\Delta L}
\def\vd{v_\Delta^{}}
%% pavel
\def\bsig{\mbox{\boldmath $\sigma$}}
\def\bsig{\mbox{\boldmath $\Sigma$}}
\def\bgam{\mbox{\boldmath $\gamma$}}
\def\bgam{\mbox{\boldmath $\Gamma$}}
\def\bphi{\mbox{\boldmath $\phi$}}
\def\bphi{\mbox{\boldmath $\Phi$}}
\def\btau{\mbox{\boldmath $\tau$}}
\def\btau{\mbox{\boldmath $\Tau$}}
\def\btau{\mbox{\boldmath $\partial$}}
\def\Delc{{\Delta}_{\circ}}
\def\bp{\mid {\bf p} \mid}
\def\al{\alpha}
\def\bet{\beta}
\def\gam{\gamma}
\def\del{\delta}
\def\Del{\Delta}
\def\te{\theta}
\def\nua{{\nu}_{\alpha}}
\def\nui{{\nu}_i}
\def\nuj{{\nu}_j}
\def\nue{{\nu}_e}
\def\num{{\nu}_{\mu}}
\def\nut{{\nu}_{\tau}}
\def\2te{2{\theta}}
\def\chic#1{{\scriptscriptstyle #1}}
\def\chicl{{\chic L}}
\def\lam{\lambda}
\def\SU{SU(2)_{\chic L} \otimes U(1)_{\chic Y}}
\def\Lam{\Lambda}
\def\sig{\sigma}
\def\'#1{\ifx#1i\accent19\i\else\accent19#1\fi}
\def\O{\Omega}
\def\o{\omega}
\def\s{\sigma}
\def\D{\Delta}
\def\d{\delta}
\def\df{\rm d}
\def\8{\infty}
\def\ld{\lambda}
%
%\tableofcontents

\newcommand{\dis}{\displaystyle} \newcommand{\alfad}{\frac{\dis \bar
    \alpha_s}{\dis \pi}} \newcommand{\bra}{\mbox{$<$}}

\newcommand {\ignore}[1]{}
\def\VEV#1{\left\langle #1\right\rangle}
\let\vev\VEV
\newcommand{\ket}{\mbox{$>$}}
\newcommand{\AHEP}{AHEP Group, Instituto de
  F\'{\i}sica Corpuscular --
  C.S.I.C./Universitat de Val{\`e}ncia, \\
  Campus de Paterna, Aptdo 22085, E--46071 Val{\`e}ncia, Spain}
\newcommand{\UMD}{Maryland Center for Fundamental Physics and
  Department of Physics, University of Maryland, College Park, MD,
  20742}
\newcommand{\UVienna}{University of Vienna, Faculty of
        Physics, Boltzmanngasse 5, A--1090 Vienna, Austria}
\newcommand{\ULis}{Technical University of Lisbon,
        Centre for Theoretical Particle Physics, \\
        1049-001 Lisbon, Portugal}
\newcommand{\IFUSP}{Instituto de F\'{\i}sica,
             Universidade de S\~ao Paulo, S\~ao Paulo -- SP, Brazil.}
\newcommand{\Wisc}{University of Wisconsisn, Madison, WI-53706.}

%\preprint{IFIC/09-vv,UMD-PP-09-xx}

%\title{Probing the Origin of Neutrino Mass at the CERN
%Large Hadron Collider}
%%
%

 %%%%%%%%%%%%%%%%%%%
%\begin{document}
%\vskip 2cm
%\begin{abstract}
%  We discuss how the CERN Large Hadron Collider (LHC) can shed light
%  on the origin of neutrino masses. We focus mainly on various TeV
%  scale seesaw schemes, as well as R-parity broken, intrinsically
%  supersymmetric neutrino masses, and their signatures at the LHC.
%\end{abstract}

%\keywords{13.15.+g,14.60.St,12.20.Fv,vv}
%
%\maketitle

\section{Introduction}
\label{sec:introduction}

Current solar and atmospheric neutrino data in conjunction with data
from reactors and accelerators show that neutrinos change flavor in
their propagation and that the oscillation mechanism is the only one
that provides a consistent picture of the
observations~\cite{Maltoni:2004ei,Schwetz:2008er}~\footnote{The
  oscillation solution has been shown to be mainly robust against
  possible astrophysical and neutrino physics
  uncertainties~\cite{Loreti:1994ry,nunokawa:1996qu,Nunokawa:1997dp,Burgess:2003fj,burgess:2002we,Burgess:2003su,Fogli:2007tx,miranda:2000bi,Miranda:2003yh,Miranda:2004nb}.}.
Although theoretically expected, nonstandard effects can only play a
sub-leading role, their amplitude being effectively constrained by
terrestrial laboratory data.
The observation of neutrino oscillations confirms the existence of
nonzero mass for the neutrinos providing the first evidence for
physics beyond-the-standard-model. The nature of this physics is
however far from clear although there are several plausible
scenarios. Here we discuss possible ways to test for some of these
scenarios using the LHC.  For the LHC to be relevant to this study,
the new physics scale clearly must lie in the TeV region. Luckily,
several scenarios fall into this category and they have already been
discussed in many reviews~\cite{Valle:2006vb,He:2009xd}.
One can broadly classify them as follows: (i) low-scale seesaw
scenarios; (ii) radiative models for neutrino masses and (iii)
supersymmetry with R-parity violation.  We now give brief overview of
the salient features of these different scenarios and focus on their
implications and signals at the LHC.

\section{Seesaw Mechanisms}
\label{sec:seesaw-mechanisms}

The basic idea of the seesaw mechanism is to generate the dimension-5
operator $\lambda L \Phi L \Phi$ (here $L$ denotes a lepton
doublet)~\cite{Weinberg:1980bf} by the tree-level exchange of heavy
states~\cite{Minkowski:1977sc,gell-mann:1980vs,yanagida:1979,mohapatra:1980ia}
in different models. The smallness of its strength is understood by
ascribing it to the violation of lepton number at a high mass scale,
namely the scale at which these states acquire masses.
One may, however, lower the seesaw scale if in the underlying theory
the corresponding Dirac Yukawa couplings $Y_D$ are assumed to be very
small.
In any case the most general description of the seesaw is in terms of
the standard \321 gauge structure, where the most general seesaw
mechanism~\cite{Schechter:1980gr,Schechter:1981cv} is formulated in
terms of $n$ left-handed SU(2) doublet neutrinos $\nu_L$ plus any
number $m$ of right-handed neutrinos $\nu_L^c$. In the basis $\nu_L$,
$\nu_L^c$, the resulting $(n+m)\times (n+m)$ mass matrix is given
as~\cite{Schechter:1980gr}
\begin{equation}
\label{ss1}
M_{\nu}=\left(
\begin{array}{ccc}
M_L&M_D\\
M_D^T & M_{N}
\end{array}
\right).
\end{equation}
Its entries $M_L, M_D, M_N$ transform as SU(2) triplet, doublet and
singlet, respectively~\cite{cheng:1980qt}. For example, the $n \times
n$ mass matrix $M_L$ arises when a scalar SU(2)-triplet takes a vacuum
expectation value (vev), see below. Several cases can be envisaged:

\subsection{ Type-I seesaw}
\label{sec:type-i-seesaw}

The type-I seesaw is the simplest realization of the dimension five
operator, where $M_L=0$.  $M_D$ is a $n\times m$ Dirac mass matrix
and $M_N$ is a Majorana $m\times m$ mass matrix and are given as
\begin{equation}
{\cal L} = Y_{D ij}\overline{l}_{L_i} \tilde{\phi} \nu_{R_j} + M_{N_{ij}}\overline{\nu}_{R_i}\nu_{R_j}^c
\end{equation}
where $\phi=(\phi^+,\phi^0)^T$ is the Standard Model Higgs scalar
doublet and $\vev{\phi^0} \equiv v_2$ then $M_D=Y_{D} v_2$.  Note that
$M_\nu$ is in general symmetric and complex. It is diagonalized by
means of a unitary $(n+m)\times (n+m)$ matrix $ U^T M_\nu
U=\mbox{Diag}(m_i,M_j) $ yielding to $n$ light mass eigenstates with
mass $m_i$ and $m$ heavy with mass $M_j$.  The effective light
$n\times n $ neutrino mass matrix is given by
\begin{equation}\label{ssone}
m_\nu=-M_D M_N^{-1} M_D^T.
\end{equation}
For $M_D\sim 100$~GeV, $M_N\sim 10^{13}$~GeV the resulting neutrino
mass is $m_\nu\sim$~eV. Note that (barring
fortuitous~\cite{Kersten:2007vk} or symmetry-driven~\cite{Gu:2008yj}
cancellations) the smallness of the observed neutrino masses requires
a very large isosinglet mass or a very small Yukawa $Y_D$. As we will
see below, it is the latter case which is relevant for the LHC
discussion.

\subsection{Type-II seesaw}
\label{sec:type-ii-seesaw}

In the presence of a complex SU(2)-triplet~\cite{Konetschny:1977bn} of
Higgs scalar bosons $\Delta~=~(H^{++}, H^{+}, H^{0})$ with
$Y_{\Delta}=2$ one can implement the Type-II seesaw
mechanism~\cite{Schechter:1980gr,Schechter:1981cv,cheng:1980qt}
~\cite{mohapatra:1981yp,Lazarides:1980nt}. Its main feature is the
appearance of a Majorana bilinear term $M_L$ in the neutrino mass
matrix in Eq.~(\ref{ss1}) which emerges from the Yukawa interaction
\begin{equation}
{\cal L} = Y_{L_{ij}} l_i^T \Delta C^{-1} l_{j} \;,
\end{equation}
where C stands for the charge conjugation matrix and the SU(2)
structure has been suppressed.  The neutral triplet component $H^{0}$
can acquire a small\footnote{Note that a non-zero vev of an SU(2)
  scalar triplet affects the SM $\rho$-parameter, hence it is
  constrained as $v_{3}\lesssim 1$GeV.} vev $v_3 \equiv \vev{H^{0}}$
giving rise to a left-handed neutrino mass term
\begin{equation}
M_{L}=    Y_{L} v_3\,.
\end{equation}
This scheme has been considered both in the SM with ungauged lepton
number~\cite{Schechter:1981cv} as well as in the
left-right~\cite{mohapatra:1981yp} or \10
context~\cite{Lazarides:1980nt}.

At the \321 gauge theory level the generic structure (\ref{ss1}) can
be obtained in a scheme featuring an SU(2) triplet $\Delta$, an SU(2) doublet
$\phi$ and an SU(2) singlet $\sigma$ which yields (neglecting for the moment
the flavour structure)
\begin{equation}
M_L\sim v_3, \quad M_D\sim v_2, \quad M_N\sim v_1,
\end{equation}
provided $v_1\equiv\vev{\sigma}$.
From the minimization of a relevant  \321 invariant scalar potential one finds
\begin{equation}
\label{eq:ss2}
v_3 v_1 \sim v_2^2.
\end{equation}
Since $v_2$ is fixed at around the electroweak scale, the induced
triplet vev is inversely proportional to $v_1$ and thus naturally tiny
for large $M_{N}$. The same mechanism works in the left-right
symmetric context where the neutrino mass generation is also linked to
parity violation~\cite{mohapatra:1981yp}.

\subsection{Type-III seesaw}
\label{sec:type-iii-seesaw}

This case is similar to type-I seesaw except that the right-handed
neutrinos $\nu_L^c$ are replaced by the neutral component of an
$SU_L(2)$-triplet $\Sigma$ with zero hypercharge $Y_{\Sigma}=0$
given by~\cite{Foot:1988aq,Ma:1998dn,Perez:2007iw}
\begin{equation}
\Sigma=
\left(
\begin{array}{cc}
\Sigma^0/\sqrt{2}&\Sigma^+\\
\Sigma^-&-\Sigma^0/\sqrt{2}
\end{array}
\right).
\end{equation}
For $m$ different fermion triplets, the minimal type-III seesaw model
is described by the Lagrangian
\begin{equation}\label{lagrt3}
\mathcal{L}=
Y_{D_{ij}} \phi^T  \overline{\Sigma}^c_i L_j
-\frac{1}{2}M_{\Sigma_{ij}}\mbox{Tr}(\overline{\Sigma}_i \Sigma^c_j) +\mbox{h.c.}
\end{equation}
The effective neutrino mass matrix is
a $(n+m)\times (n+m)$ matrix
\begin{equation}
\label{eq:mlep}
M_{\nu}=\left(
\begin{array}{ccc}
0&M_D\\
M_D^T & M_{\Sigma}
\end{array}
\right)
\end{equation}
leading to three light neutrinos
\begin{equation}\label{t3}
  m_\nu = -M_D^T M_\Sigma^{-1} M_D,
\end{equation}
where, as before, $M_D= Y_D v_2$.  The Eq.~(\ref{t3}) is fully
analogous to the type-I relation Eq.~(\ref{ssone}) and the smallness
of the observed neutrino masses requires a very large isotriplet
fermion mass or a very small Yukawa $Y_D$.

\subsection{Double seesaw}
\label{sec:double}

A very different way to understand the small neutrino masses is to add
another set of three SM singlet fermions in addition to the right
handed neutrinos discussed in the case of type-I seesaw. In the
context of grand unified (GUT) or left-right models, this extra
singlet $S$ should be a left-right or SO(10) singlet unlike the RH
neutrino.  Assuming for simplicity there is one such an extra partner
for each of the right-handed neutrinos, the relevant ($9\times 9$)
analogue of the neutrino mass matrix (\ref{ss1}) reads
\begin{equation}\label{minv0}
M_\nu=\left(
\begin{array}{ccc}
0&M_D&0\\
M_D^T&0&M\\
0&M^T&\mu
\end{array}
\right),
\end{equation}
where the zero entries can be justified in the context of string
models~\cite{Mohapatra:1986aw,Valle:1987sq,Mohapatra:1986bd}. %
For $M\gg M_{D}$
the effective light neutrino mass matrix reads
\begin{equation}\label{doubss}
m_\nu=M_DM^{T^{-1}}\mu M^{-1}M_D^T~.
\end{equation}
For $\mu\gg M$ the extra scalar $S$ decouples and the structure $M
\mu^{-1} M^T$ can be viewed as an effective RH neutrino mass matrix
governing a subsequent type-I seesaw in the $\nu_{L}-\nu_{L}^{c}$
sector. Note that in this context $M_{N}\sim M \mu^{-1} M^T$ can be
used as a ``bridge'' over the typical gap between the GUT-scale
$M_{GUT}\sim 10^{16}$GeV and the usual seesaw scale at around
$M_{B-L}\sim 10^{13}$GeV.

\subsection{Inverse seesaw}
\label{sec:inverse-sees}

Note that in Eq.~(\ref{minv0}) when $\mu =0$ the $U(1)_L$ global
lepton number is conserved and neutrinos are massless.  Neutrinos get
masses only when $U(1)_L$ is broken. The latter can be arranged to
take place at a low scale, for example through the $\mu SS$ mass
term.
After $U(1)_L$ breaking the effective light neutrino mass matrix is
given by
\begin{equation}\label{inv}
m_\nu=M_DM^{T^{-1}}\mu M^{-1}M_D^T,
\end{equation}
so that, when $\mu$ is small, $m_\nu$ is also small, even when $M$
lies at the electroweak or TeV scale.  In other words, the smallness
of neutrino masses follows naturally in t'Hooft's sense since as $\mu
\to 0$ the lepton number becomes a good symmetry~\cite{'tHooft:1979bh}
without need for super-heavy physics. The fact that neutrino mass
vanishes as $\mu\to 0$ is just the opposite of the type-I seesaw,
hence the name inverse seesaw.
It may be worth noting that the parameter $\mu$ may ``calculable''
from a very small gauge singlet vev, whose smallness arises
dynamically~\cite{Bazzocchi:2009kc}. A supersymmetric model with this
feature has been presented in Ref.~\cite{Bazzocchi:2009kc}.
Alternatively, $\mu$ may arise spontaneously in a majoron-like scheme
with $\mu\sim \vev{\sigma}$ where $\sigma$ is a \321 singlet
\cite{GonzalezGarcia:1988rw}.

\subsection{Linear seesaw}
\label{sec:linear-seesaw}

An interesting low-scale seesaw variant is the {\it linear seesaw}
that arise from $SO(10)$~\cite{Malinsky:2005bi},
where the $\nu,\,\nu^c, \, S$ mass matrix takes the form
\begin{equation}\label{mlin}
M_\nu=\left(
\begin{array}{ccc}
0&M_D&M_L\\
M_D^T&0&M\\
M_L^T&M^T&0
\end{array}
\right).
\end{equation}
Here the lepton number is broken by the  $M_L\,\nu S$ term,
and the effective light neutrino mass is given by
\begin{equation}\label{lin}
M_\nu=M_D(M_L M^{-1})^T+(M_L M^{-1}){M_D}^T.
\end{equation}
As for the $\mu$ parameter in the inverse seesaw, the smallness of
$M_L$ is natural in t'Hooft's sense since neutrinos become massless as
$M_L\to 0$. In the class of supersymmetric \10 model given in
\cite{Malinsky:2005bi} the neutrino mass can be arbitrarily small
irrespectively of how low is the $B-L$ breaking scale. Apart from
suggesting a plausible leptogenesis scenario~\cite{Hirsch:2006ft} the
model allows for a light Z' that can be produced at the LHC,
say, by the Drell-Yan mechanism.

\subsection{Inverse type-III seesaw}
\label{sec:inverse-type-iii}

One can also combine together inverse seesaw with type-III seesaw
%. We call
(call it {\it inverse type-III} seesaw
\cite{Ma:2009kh,Ibanez:2009du}).  In the basis $\nu_L$, $\Sigma$ and
$S$ one finds from Eq.~(\ref{lagrt3}) that the effective neutrino mass
matrix is
\begin{equation}
\label{eq:invt3}
M_\nu=\left(
\begin{array}{ccc}
0&M_D&0\\
M_D^T&M_\Sigma&M\\
0&M^T&\mu
\end{array}
\right).
\end{equation}
As in the inverse type-I version, for small $\mu$ the neutrino mass is
suppressed.  Note that Dirac Yukawa coupling strength may be of order
one in contrast to the case of normal type-III seesaw.

From Eq.~(\ref{lagrt3}) one also finds that the charged lepton
  mass matrix is a $(n+m)\times (n+m)$ matrix given as
\begin{equation}
\label{eq:mlep2}
M_{lep}=\left(
\begin{array}{ccc}
M_l&M_D\\
0 & M_{\Sigma}
\end{array}
\right)~.
\end{equation}
After diagonalization one finds that the $n$ by $n$ coupling matrix
entering into the charged lepton piece of the NC Lagrangian in the
mass basis is not unitary, similarly to what happens in the case of
neutrinos~\cite{Schechter:1980gr}. This violates the
Glashow-Iliopoulos Mainani mechanism and gives rise to sizeable tree
level flavor-changing neutral currents in the charged lepton
sector~\cite{Ibanez:2009du}.

\subsection{Nesting of seesaw mechanism}
\label{sec:nest-sees-mech}

The general idea for achieving a type-II seesaw for Higgs doublet vevs
was presented in Ref.~\cite{Grimus:2009mm}. Assuming $\left| v_1
\right|$ of the order of the electroweak scale $m_\mathrm{ew}$, then a
suppression factor
\begin{equation}
\left| \frac{v_2}{v_1} \right| \sim
\left( \frac{m_\mathrm{ew}}{m_H} \right)^2
\end{equation}
can be achieved with a heavy mass $m_H$ of $\phi_2$. The original
proposal~\cite{Ma:2000cc} uses two Higgs doublets and a $U(1)$
symmetry $\phi_2 \to e^{i \alpha} \phi_2$ softly broken by the term
$\mu^2 \phi_1^\dagger \phi_2$ in the scalar potential.
Further proposals and applications can be found in
Refs.~\cite{Grimus:2009mm,Davidson:2009ha,Mantry:2007ar,Randall:2007as}.

Here we discuss a simple example of nested seesaw mechanisms for light
Majorana neutrino masses, namely a type-II sessaw mechanism for the
Higgs doublet $\phi_2$ nested within the usual type-I seesaw
mechanism.  One assumes that there are neutrino singlets fields
$\nu_R$ with Majorana mass terms given by the mass matrix $M_R$, and
Dirac mass terms generated by the Yukawa couplings
\begin{equation}
\label{yuk}
\mathcal{L}_\mathrm{Yukawa} =
\bar \nu_R Y \left( \phi_2^0 \nu_L - \phi_2^+ \ell_L \right)
+ \mathrm{H.c.},
\end{equation}
where $Y$ is the matrix of Yukawa coupling constants.  (One needs a
symmetry such that the Yukawa couplings of the $\nu_R$ in
equation~(\ref{yuk}) involve only the Higgs doublet $\phi_2$.)

We want the matrix elements of $\mathcal{M}_\nu$ to be of order
eV. Taking $m_R$ to the TeV scale while keeping $v_2 \sim
m_\mathrm{ew}$ requires (avoiding cancellation mechanisms) the
Yukawa couplings to be of order $10^{-5}$.  But if we let $\left|
v_2 \right| \sim m_\mathrm{ew}^3 / m_H^2$ be suppressed by a
type-II seesaw mechanism for Higgs
doublets~\cite{Ma:2000cc,Adulpravitchai:2009re}, then we obtain
$1 \, \mathrm{eV} \sim m_\mathrm{ew}^6 / \left( m_R m_H^4 \right)$
and this represents a \emph{fivefold} suppression of the neutrino
masses. Assuming for simplicity $m_R = m_H$, one obtains
\begin{equation}
m_H \sim \sqrt[5]{10^{66}}\, \mathrm{eV} \approx 16\, \mathrm{TeV}.
\end{equation}
Other cases of nested seesaw mechanisms are discussed
in~\cite{Grimus:2009mm}.

\subsection{Loop models}
\label{sec:loop-models}

Another interesting class of models are loop
models~\cite{zee:1980ai,babu:1988ki,FileviezPerez:2009ud} where a
clever choice of new particles beyond the standard model instead of
the RH neutrino can lead to small neutrino masses at the radiative
level (one or two loop depending on the model) with particles at the
TeV scale. Typically these particles can be scalar or fermionic and
since in general they have SM quantum numbers, they can be produced at
LHC. For recent discussions see, e.~g.
Refs.~\cite{AristizabalSierra:2006gb,Nebot:2007bc,AristizabalSierra:2007nf}.

%%%%%%%%%%%%%%%%%%%%%%%%%%%%%%%%%%%%%%%%

\section{Phenomenology at LHC}
\label{sec:sees-coll-phen}

The seesaw mechanism responsible for neutrino masses can be realized
at the TeV scale.  In such case the states underlying the different
schemes discussed above can be produced at the LHC if the relevant
cross sections are large enough. In order to distinguish between
various {\em scenaria} one should compare the relevant production
rates in the proton-proton ($pp$) collisions and extract the
expected decay signals from the typically large SM background. As we
shall discuss below, in the simplest type-I seesaw the production of
TeV-scale RH neutrinos at the LHC is neutrino-mass-suppressed.
However, even in such case the new type-I scalars (or gauge bosons
emerging in unified models with low $B-L$ scale) may lead to
detectable signals at the LHC \cite{delAguila:2008iz}.  Furthermore,
the very specific signatures inherent to type-II, type-III and
certain variants of double seesaw make these schemes also testable
and distinguishable from each other as well as from the type-I
seesaw. Note that none of these claims is in conflict with the
smallness of light neutrino masses which can be ascribed to either
an overall suppression of lepton number violation and/or a smallness
of the relevant Yukawa couplings; the latter may give rise to
displaced-vertex events.

\subsection{Type I seesaw}
%\section{RH neutrinos in type-I seesaw}

The RH neutrinos underpinning the TeV-scale type-I seesaw can be
produced in $pp$ collisions via virtual $W,Z$-bosons, but only through
the mixing with the SU(2) doublets.
The general structure of the relevant RH neutrino couplings in the
Standard Model is given in Ref.~\cite{Schechter:1980gr}.  In order to
keep $M_N$ in the TeV region, one typically needs $Y_{D}\sim
10^{-5.5}$ to account for the light neutrino masses, implying that the
mixing between $\nu$ and $N$ is suppressed by $\sqrt{M_D/M_N}\sim
10^{-6}$.

Thus, in a generic type-I seesaw framework the RH neutrino
production cross section is neutrino-mass-suppressed and hence
unobservable at the LHC.
The story may change, however, in specific models. One way is if there
are new gauge bosons at around the TeV scale, such as the $Z'$
associated to the $U(1)_{B-L}$ gauge symmetry and/or $W'$ of
$SU(2)_{R}$, inherent to a wide class of extensions of the SM (like
e.g. left-right models~\cite{Mohapatra:1980qe} and its higher group
embeddings such as, SO(10) or $E_{6}$ models.). Such local symmetries
are any way motivated if one tries to understand why an SM singlet
right-handed neutrino does not have Planck mass.  Since these gauge
bosons naturally couple to quarks they can be produced at the LHC and
their subsequent decay into a
single~\cite{Schechter:1980gr}~\cite{Dittmar:1990yg} or a pair of RH
neutrinos may be observable in the channels
\begin{eqnarray}\label{ec:Nprod}
q\bar q' & \to & W'^{\pm}\to\ell^{\pm}N \nn \\
q\bar q & \to & Z'\to N N~~~\mathrm{or}~~~ \nu N~.
\end{eqnarray}
The fact that right handed neutrinos are Majorana fermions implies
that it can decay with equal probability to both leptons and
anti-leptons. At the collider, this means that a $W'$ production will
be accompanied by no missing energy like-sign
di-leptons~\cite{Keung:1983uu}.  Note that it is quite natural to
expect both the RH neutrinos and the extra neutral gauge boson(s) at
around the same scale because the masses in these sectors are
typically associated to the same [$U(1)_{B-L}$] symmetry breakdown.
The rates of the processes in Eq.~(\ref{ec:Nprod}) above depend
mainly on the gauge boson mass while the decay involves the amount
of admixture of the $SU(2)_{L}$- doublet components within the
heavy neutrinos. For instance, if the neutrino mixing is tiny
(i.e., less than about $10^{-3}$), the single RH neutrino produced
in the first case decays predominantly through an off-shell $W'$
yielding a di-lepton signal with observation ranges stretching up
to about $m_{N}\lesssim 1.8$~TeV, $M_{W'}\lesssim 3.2$~TeV for 30
fb$^{-1}$~\cite{Ferrari:2000sp}\cite{Gninenko:2006br}. On the
other hand, for a larger neutrino mixing the RH neutrino decay is
driven by the SM gauge bosons, leading to di-lepton and tri-lepton
signals. The combined sensitivity across all channels is higher in
this case, reaching up to about $m_{N}\lesssim 2.4$~TeV,
$M_{W'}\lesssim 3.5$~TeV for the same
luminosity~\cite{delAguila:2009bb}. Let us remark that in this
case the single heavy resonance behavior allows for a
reconstruction of the $W'$ mass.

In contrast, the second process in Eqs. (\ref{ec:Nprod}) relies only
on the presence of a light-enough $Z'$ emerging under various
conditions in many popular scenarios (see
e.g.~\cite{Valle:1987sq,delAguila:2007ua,Huitu:2008gf} and references
therein). Moreover, a light $Z'$ does not necessarily require a light
$W'$ counterpart. In fact, it has been shown
e.g. in~\cite{Malinsky:2005bi,Hirsch:2006ft} that even unified gauge
models such as SO(10) GUTs may naturally accommodate a TeV-scale $Z'$
without conflict with gauge coupling unification, neutrino masses or
leptogenesis~\cite{Hirsch:2006ft,Blanchet:2009bu} if $W'$ remains
heavy, killing the first signature in Eqs. (\ref{ec:Nprod}).  The
decay of the $NN$ pair gives rise to di-lepton and tri-lepton final
states, the latter offering the best discovery potential stretching up
to $m_{N} \lesssim 850$ GeV and $M_{Z'} \lesssim 2.1$
TeV~\cite{AguilarSaavedra:2009ik} for the leptophobic $Z'_\lambda$
model in Ref.~\cite{delAguila:2007ua}.  Other models featuring the
beyond-SM Abelian gauge sector yield similar results weighted namely
by the relevant quark and lepton $U(1)'$-charge assignments.  Let us
also remark that the $Z'$-mediated heavy neutrino production can be
distinguished from e.g. the type-III seesaw signals
(c.f. sect. \ref{sec:type-III-pheno}) by the $Z'$ mass reconstruction
and the potential smallness of the four lepton signal from the $Z' \to
NN$ channel.

In order to discuss details of right handed neutrino production and
decay one needs to characterize the structure of their gauge
couplings~\cite{Schechter:1980gr}. For our simplified discussion it is
convenient to use the parametrization given in
Ref.~\cite{Casas:2001sr}. One may first note that the three light
neutrino masses can be expressed in the following way
\begin{equation}
m = V^\dagger \ M_\nu \ V^*,
\end{equation}
where $m=diag (m_1, m_2, m_3)$ and $V$ can be taken as the leptonic
mixing matrix for the three light neutrinos.  Working in the basis
where the heavy neutrino mass matrix is diagonal one can write $m_D$
as
\begin{equation}
m_{D}= V \ m^{1/2} \ \Omega \ M^{1/2},
\label{Dirac}
\end{equation}
where  $M=diag(M_1, M_2, M_3)$ for heavy neutrino masses,
and $\Omega$ is a complex matrix which satisfies the orthogonality
condition $\Omega^T \Omega = 1$.  It can be shown that using the
seesaw formula and the relation between the leptonic mixing one
can find a formal solution for the mixing between the SM
charged leptons ($\ell=e, \mu, \tau$) and heavy neutrinos
($N=1,2,3$):
\begin{eqnarray}
V_{\ell N}= \ V \ m^{1/2} \ \Omega \ M^{-1/2}.
\label{mixing1}
\end{eqnarray}
Therefore, for a given form of $\Omega$, one can establish the
connection between the heavy neutrino decays and the properties of
the light neutrinos~\cite{Perez:2009mu}.
Unfortunately, since the explicit form of this matrix is unknown one
cannot predict the decay pattern of the heavy neutrinos with respect
to the spectrum for light neutrinos.  It is important, however, to
realize that an underlying theory would pick only one specific form of
$\Omega$. This (yet unknown) form would have definite prediction for
the $N$ decay patterns, through which the underlying theory could be
revealed.

%%%%%%%%%%%%%%%%%%%%%%%%%%%%%%%%%%%%%%%%%%%%%%%%%%%%%%%%%%%%%%%%%%%%%%%
\subsubsection{ Heavy Neutrino Decay Modes}
%%%%%%%%%%%%%%%%%%%%%%%%%%%%%%%%%%%%%%%%%%%%%%%%%%%%%%%%%%%%%%%%%%%%%%%

The leading decay channels for the heavy neutrinos include $N_i \to
e^{\pm}_j W^{\mp}$, $N_i \to \nu_j Z$~\cite{Schechter:1980gr} as well
as $N_i \to \nu_j h(H)$. The amplitude for the two first channels are
proportional to the mixing between the leptons and heavy neutrinos
given in Eq.~(\ref{mixing1}), while the last one is proportional to
the Dirac-like Yukawa terms given in Eq.~(\ref{Dirac}).

The partial decay widths of the heavy Majorana neutrinos $N_i$ are given by
\begin{eqnarray}
\Gamma(N_i \to \ell^-W_L^+)= {g^2\over 64\pi
M_W^2}|V_{\ell i}|^2M_i^3(1-\mu_{iW})^2, \nonumber \\
\Gamma(N_i \to \ell^-W_T^+)={g^2\over
32\pi}|V_{\ell i}|^2M_i(1-\mu_{iW})^2, \nonumber \\
\Gamma(N_i\to \nu_\ell Z_L)={g^2\over
64\pi M_W^2}|V_{\ell i}|^2M_i^3(1-\mu_{iZ})^2, \nonumber \\
\Gamma(N_i \to \nu_\ell Z_T)={g^2\over
32\pi c_W^2}|V_{\ell i}|^2M_i(1-\mu_{iZ})^2, \nonumber
\end{eqnarray}
where $\mu_{ij}=M_j^2/M_i^2$. If $N_i$ is heavier than the Higgs
bosons $h$ and $H$, one has the additional channels
\begin{eqnarray}
\Gamma(N_i \to \nu_\ell h)={g^2\over
64\pi M_W^2}|V_{\ell i}|^2M_i^3(1-\mu_{ih})^2 c^2_0,
\nonumber \\
\Gamma(N_i \to \nu_\ell H)={g^2\over
64\pi M_W^2}|V_{\ell i}|^2M_i^3(1-\mu_{iH})^2 s^2_0.
\nonumber
\end{eqnarray}
where $s^2_0 \equiv \sin^2\theta_0$, $\theta_0$ denoting a Higgs
mixing angle.  At a high mass of $M_N$, the branching ratios of the
leading channels go like
\begin{equation}
  \Gamma(\ell^-W_L^+) \approx
  \Gamma(\ell^+W_L^-) \approx
  \Gamma(\nu Z_L) \approx
  \Gamma(\nu h+\nu H).
\nonumber
  \textbf{}\end{equation}
As discussed above, the lepton-flavor content of $N$ decays will be
different in each neutrino spectrum. In order to search for the events with
best reconstruction, we will only consider the $N$ decay to charged
leptons plus a $W^\pm$.

%%%%%%%%%%%%%%%%%%%%%%%%%%%%%%%%%%%%%%%%%%%%%%%%%%%%%%%%
{\bf \textit{ Degenerate Heavy Neutrinos}}
%%%%%%%%%%%%%%%%%%%%%%%%%%%%%%%%%%%%%%%%%%%%%%%%%%%%%%%%

%
\begin{figure}[!ht]
\begin{center}
\includegraphics[width=60mm]{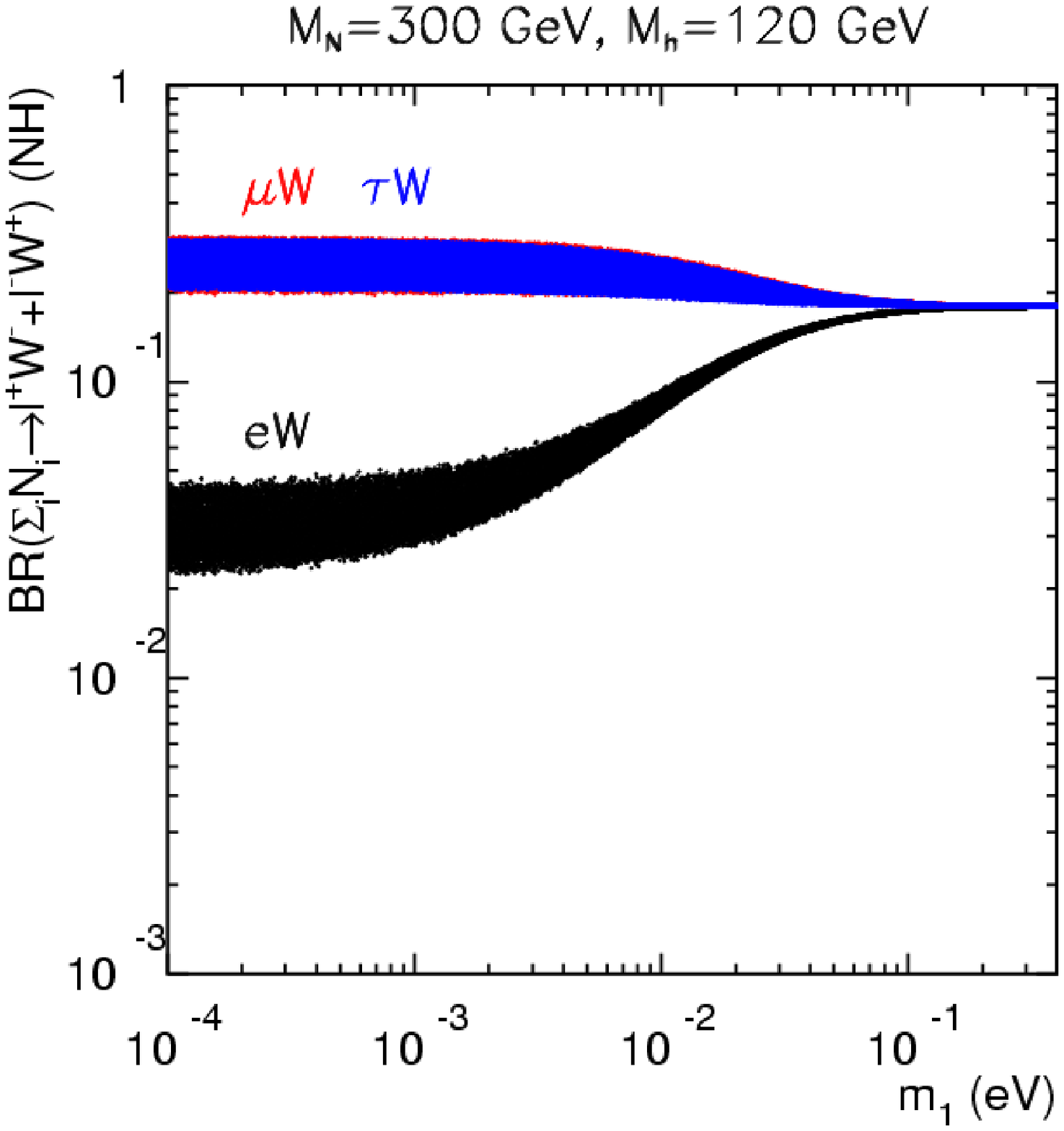}
\includegraphics[width=60mm]{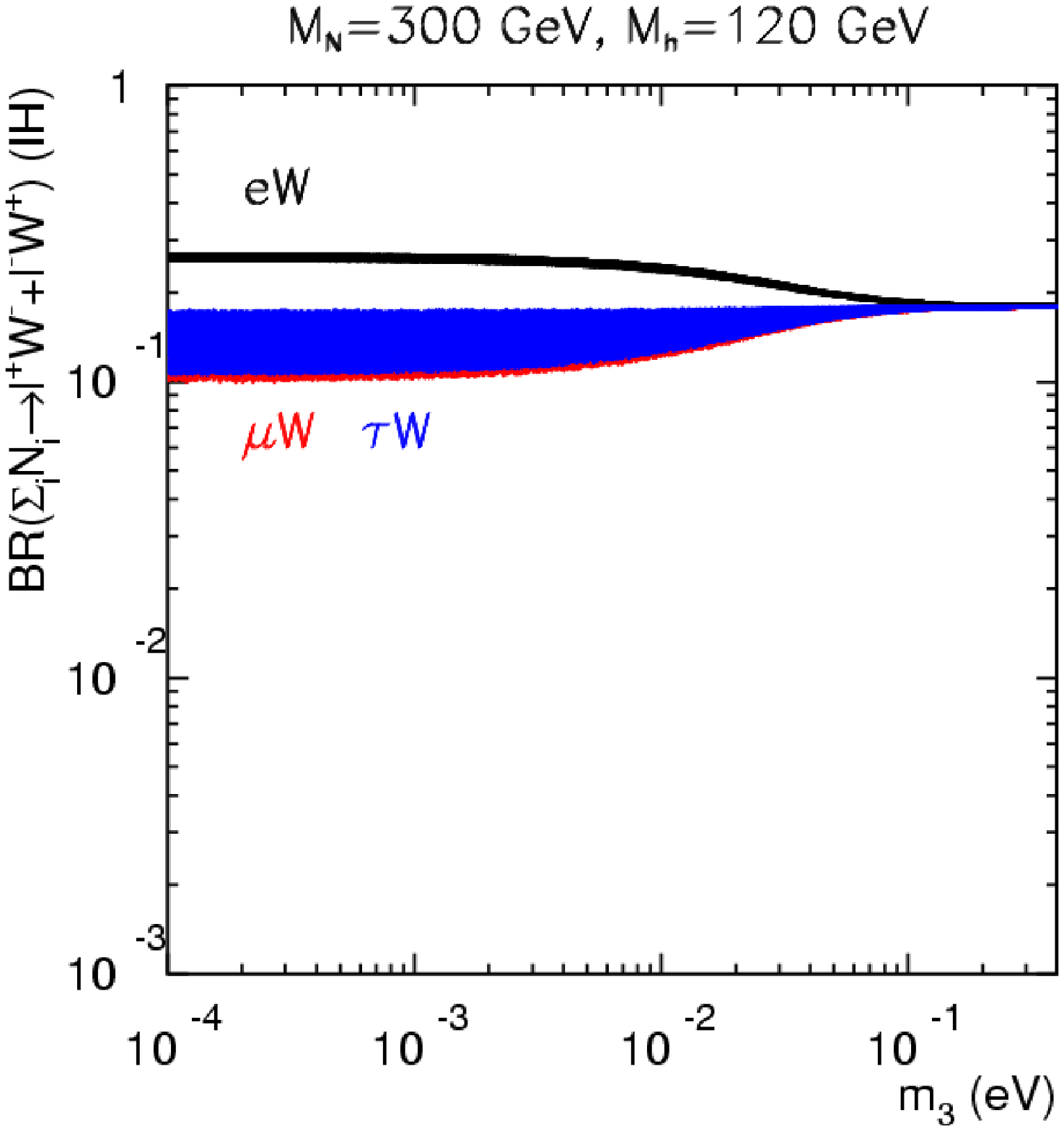}
\end{center}
\caption{Branching fractions of degenerate neutrinos $\sum_iN_i \to
\ell^+ W^-+\ell^-W^+ \ (\ell=e,\mu,\tau)$ for NH and IH versus
lightest neutrino mass with $M_N=300~{\rm GeV}$ and $M_{h}=120~{\rm
GeV}$, assuming vanishing Majorana phases (from Ref.~\cite{Perez:2009mu}).}
\label{nbr}
\end{figure}
In Fig.~\ref{nbr} we show the impact of the neutrino masses and mixing
angles on the branching fractions of the sum of the degenerate
neutrinos $N_i \ (i=1,2,3)$ decaying into $e,\mu,\tau$ lepton plus $W$
boson, respectively, for the Normal Hierarchy (NH) and the Inverted
Hierarchy (IH), assuming vanishing Majorana phases.  Qualitatively, it
follows the pattern
\begin{eqnarray}
&& BR(\mu^\pm W^\mp),BR(\tau^\pm W^\mp)\gg BR(e^\pm W^\mp) \nonumber \\
% \ \ \ {\rm for \ \ NH},\nonumber \\
&& BR(e^\pm W^\mp)>BR(\mu^\pm W^\mp),BR(\tau^\pm W^\mp) \nonumber
%\ \ \ {\rm for \ \ IH}.
\end{eqnarray}
for NH and IH, respectively.  The branching fraction can differ by one
order of magnitude in NH case; and about a factor of few in the IH
spectrum. Note that all these channels are expected to be quite
similar when the neutrino spectrum is quasi-degenerate, $m_1\approx
m_2\approx m_3\geq 0.05$ eV.

Therefore, in this simple case one can hope that if the heavy neutrino
decays are observed in future experiments one should be able to probe
the neutrino spectrum.

%%%%%%%%%%%%%%%%%%%%%%%%%%%%%%%%%%%%%%%%%%%%%%%%%%%%%%%%%%%%%%%%%%%%%%%%
{\bf \textit{Non-Degenerate Heavy Neutrinos}}
%%%%%%%%%%%%%%%%%%%%%%%%%%%%%%%%%%%%%%%%%%%%%%%%%%%%%%%%%%%%%%%%%%%%%%%%

For non-degenerate neutrino spectra we once again study the simple choice:

Case (a) $\Omega=I$.
In this simple case all $|V_{\ell i}|^2 \ (\ell=e,\mu,\tau)$ are
proportional to $m_i$. Therefore the branching ratio of $N_i\to
\ell^\pm W^\mp$ for each lepton flavor is independent of neutrino mass
and thus universal for both NH and IH. Although we cannot distinguish
the neutrino mass hierarchy, we still can tell the difference of the
three heavy Majorana neutrinos according to different SM lepton
flavors in the final states of their dominant decay channels. One has
\begin{eqnarray}
& BR(e^\pm W^\mp) > BR(\mu^\pm W^\mp), BR(\tau^\pm W^\mp) \nonumber \\
% \ \ \ {\rm for}  \ \ N_1,\nonumber \\
& BR(e^\pm W^\mp)\approx BR(\mu^\pm W^\mp)\approx BR(\tau^\pm W^\mp) \nonumber \\
%\ \ \ {\rm for}  \ \ N_2,\nonumber \\
& BR(\mu^\pm W^\mp), BR(\tau^\pm W^\mp)\gg BR(e^\pm W^\mp) \nonumber
%\ \ \ {\rm for}  \ \ N_3.\nonumber
\end{eqnarray}
for $N_i \ (i=1,2,3)$, respectively.
This follows closely to the mixing strengths of the light neutrinos in
the previous section.

Case (b) $\Omega = I_{\rm off}$ is identical to the above if we
identify $N_1 \leftrightarrow N_3$.  A more involved case for $\Omega$
would be some form of superposition of the three decay patterns, that
is to be tested experimentally by the flavor combinations.

\subsubsection{ Heavy Neutrino Decay Lengths}

To complete this section about the heavy Majorana neutrino properties,
we study their total decay widths, which are proportional to $M_\nu
M_N^2/M_W^2$.
\begin{figure}[ht]
\begin{center}
\begin{tabular}{cc}
\includegraphics[width=60mm,angle=90]{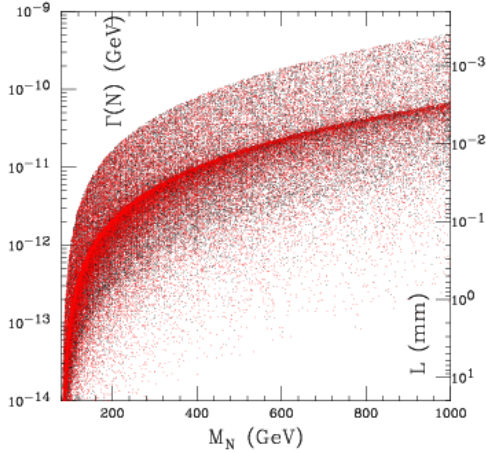}
\end{tabular}
\end{center}
\caption{The total width and decay length of $N$ in the general
non-degenerate case, when the lightest neutrino mass $10^{-4}~{\rm
eV}\leq m_{1(3)}\leq 0.4~{\rm eV}$, $M_{h}=120~{\rm GeV}$ and
$\Omega=R_{12}R_{13}R_{23}$ with random selection of the matrix
elements, from Ref.~\cite{Perez:2009mu}.} \label{totw}
\end{figure}

In Fig.~\ref{totw}, we plot the total width (left axis)
and decay length (right axis) for $N$ versus $M_N$ under the general
non-degenerate case with random selection of the $\Omega$ matrix
elements (similar for NH and IH).
There is a large spread for the possible ranges of the decay lengths,
governed by the mixing parameters.  Although not generally considered
as long-lived for large mass, the $N$ decay lengths may be typically
in the range of $\mu$m$-$cm, so their decays could lead to a visible
displaced vertex in the detector at the LHC.

It is also worth pointing out that when type I seesaw is embedded
in the minimal supersymmetric left-right symmetric model, even
with high scale seesaw one obtains doubly charged scalars at the
collider energies \cite{Aulakh:1997fq}~\cite{Chacko:1997cm}
coupling to right handed electrons. Their collider signatures are
similar to the doubly charged Higgs boson in the type II case
discussed below.

\subsection{ Type II Seesaw at the LHC}

The Higgs sector of the Type II seesaw scenario is composed of the SM
Higgs $H \sim (1,2,1/2)$ and a scalar triplet $\Delta \sim (1,3,1)$.
The crucial terms for the neutrino mass generation in the theory are
\begin{eqnarray}
 - Y_\nu \ l_L^T \ C \ i \sigma_2 \ \Delta \ l_L  +
  \mu \ H^T \ i \sigma_2 \ \Delta^\dagger H \ + \ \text{h.c.}
\label{Yukawa}
\end{eqnarray}
where the Yukawa coupling $Y_\nu$ is a $3\times 3$ complex symmetric
matrix.  The lepton number is explicitly broken by two units due to
the simultaneous presence of the Yukawa coupling $Y_\nu$ and the Higgs
term proportional to the $\mu$ parameter. From the minimization of the
scalar potential one finds a vev $v_3$ for ${\Delta}$ given as
$v_3=\mu v_2^2/\sqrt{2} M_{\Delta}^2$, where $v_2$ is the usual
doublet vev, see also Eq.~(\ref{eq:ss2}).  Therefore, neutrinos acquire
a Majorana mass given by
\begin{eqnarray}
M_{\nu}= \sqrt{2} \ Y_\nu \ v_3 = Y_\nu\  {\mu \ v_2^2}/{  M_{\Delta}^2}.
\label{type2}
\end{eqnarray}
This equation is the key relation of the type-II seesaw scenario. The
neutrino mass is induced by the electroweak symmetry breaking (EWSB)
and its smallness is associated with a large mass scale $\md$. With
appropriate choices of the Yukawa matrix elements, one can easily
accommodate the neutrino masses and mixing consistent with the
experimental observation. For the purpose of illustration, we adopt
the values of the masses and mixing at $2\sigma$ level from a recent
global fit~\cite{Schwetz:2008er}.

%%%%%%%%%%%%%%%%%%%%%%%%%%%%%%%%%%%%%%%%%%%%%%%%%%%%%%%%%
\subsubsection{Properties of the Higgs Sector}
%%%%%%%%%%%%%%%%%%%%%%%%%%%%%%%%%%%%%%%%%%%%%%%%%%%%%%%%

After the EWSB, there are seven massive physical Higgs bosons: two
CP-even neutral Higgs bosons $H_1, H_2$, one CP-odd neutral Higgs $A$,
as well as the singly and doubly charged states $H^{\pm}$,
$H^{\pm\pm}$. Here $H_1$ is SM-like and the rest of the Higgs states
are $\Delta$-like.  Neglecting the Higgs quartic interactions one
finds $M_{H_2}\simeq M_{A} \simeq M_{H^+} \simeq
M_{H^{++}}=M_{\Delta}$.  Since we are interested in a mass scale
accessible at the LHC, we thus focus on $110~{\rm GeV} < \md < 1~{\rm
  TeV}$, where the lower bound is from direct
searches~\cite{Amsler:2008zzb}. Working in the physical basis for the
fermions we find that the Yukawa interactions can be written as
\begin{eqnarray}
& & \nu_L^T \ C \ \Gamma_+ \ H^+ \ e_L, \\
& &  e_L^T \ C \ \Gamma_{++} \ H^{++} \ e_L,
\end{eqnarray}
where
\begin{eqnarray}
&& \Gamma_+  = \frac{ c_{\theta_+} m_\nu^{diag} V^\dagger}{v_{3}},
\\
&& \Gamma_{++} =  \frac{V^* m_{\nu}^{diag} V^{\dagger}}{\sqrt{2} \ v_{3}},
\end{eqnarray}
where $c_{\theta_+}= \cos \theta_+$, $\theta_+$ is the mixing
angle in the charged Higgs sector and $v_{\Delta} \lesssim 1$ GeV
from the $\rho$-parameter constraints.  Here $V$ denotes the
leptonic mixing matrix which may be written as $V_{l}
(\theta_{12}, \theta_{23}, \theta_{13}, \delta) \times K_M$ where
$K_M=\text{diag} (e^{i\Phi_1/2}, 1, e^{i \Phi_2/2})$ accounts for
the Majorana phases~\cite{Schechter:1980gr}. The values of the
physical couplings $\Gamma_+$ and $\Gamma_{++}$ are thus governed
by the spectrum and mixing angles of the neutrinos, and they in
turn characterize the branching fractions of the $\dl=2$ Higgs
decays. For a previous study of the doubly charged Higgs decays
see~\cite{Azuelos:2005uc}\cite{Akeroyd:2005gt}\cite{Chun:2003ej}.

The two leading decay modes for the heavy Higgs bosons are
the $\dl=2$ leptonic mode and the (longitudinal) gauge boson
pair mode. The ratio between them for the $H^{++}$ decay reads as
\begin{equation}
{\Gamma(H^{++}\to \ell^+\ell^+) \over \Gamma(H^{++}\to W^+W^+) }
\approx { | \Gamma_{++} |^2 v_2^4 \over M_\Delta^2 v_3^2}
\approx \left({m_\nu \over M_\Delta^{}}\right)^2
 \left({v_2 \over v_3}\right)^4,\nonumber
\end{equation}
using $m_\nu/\md\sim$ 1 eV/1 TeV, one finds that these two decay modes
are comparable when $v_3^{} \approx 10^{-4}\ {\rm GeV}$.  It is thus
clear that for a smaller value of $v_3$ (a larger Yukawa coupling),
the leptonic modes dominate, while for larger values, the gauge boson
modes take over. In the case of the singly charged Higgs, $H^{\pm}$,
there is one additional mode to a heavy quark pair. The ratio between
the relevant channels is
\begin{equation}
{\Gamma(H^{+}\to t \bar b) \over \Gamma(H^{+}\to W^+Z) }
\approx {3 (v_3 m_t/v_2^2)^2 \md \over M_\Delta^3   v_3^2 /2 v_2^4 } =
6 \left( { m_t \over M_\Delta^{}  } \right)^2.\nonumber
\end{equation}
Therefore, the decays $H^+ \to W^+ Z,\ W^+ H_1$ dominate over $t\bar
b$ for $\md > 400$ GeV~\cite{Perez:2008zc,Perez:2008ha}. In our
discussions so far, we have assumed the mass degeneracy for the Higgs
triplet. Even if there is no tree-level mass difference, the SM gauge
interactions generate the splitting of the masses via radiative
corrections, leading to $\Delta
M=M_{H^{++}}-M_{H^{+}}=540~\text{MeV}$~\cite{Cirelli:2005uq}. The
transitions between two heavy triplet Higgs bosons via the SM gauge
interactions, such as the three-body decays $H^{++} \to H^+ W^{+*},
H^{+} \to H^0 W^{+*}$ may be sizable if kinematically
accessible. However these transitions will not have a significant
branching ratio unless $\Delta M > 1 $
GeV~\cite{Perez:2008zc,Perez:2008ha}. In fact, our analyzes will
remain valid as long as $H^{++}$ and $H^+$ are the lower-lying states
in the triplet and they are nearly degenerate. We will thus ignore the
mass-splitting effect in the current discussion.

%%%%%%%%%%%%%%%%%%%%%%%%%%%%%%%%%%%%%%%%%%%%%%%%%%%%%%%%%%%
\subsubsection{ Higgs Decays}
%%%%%%%%%%%%%%%%%%%%%%%%%%%%%%%%%%%%%%%%%%%%%%%%%%%%%%%%%%%

\begin{figure}[!ht]
\begin{center}
\includegraphics[width=6cm]{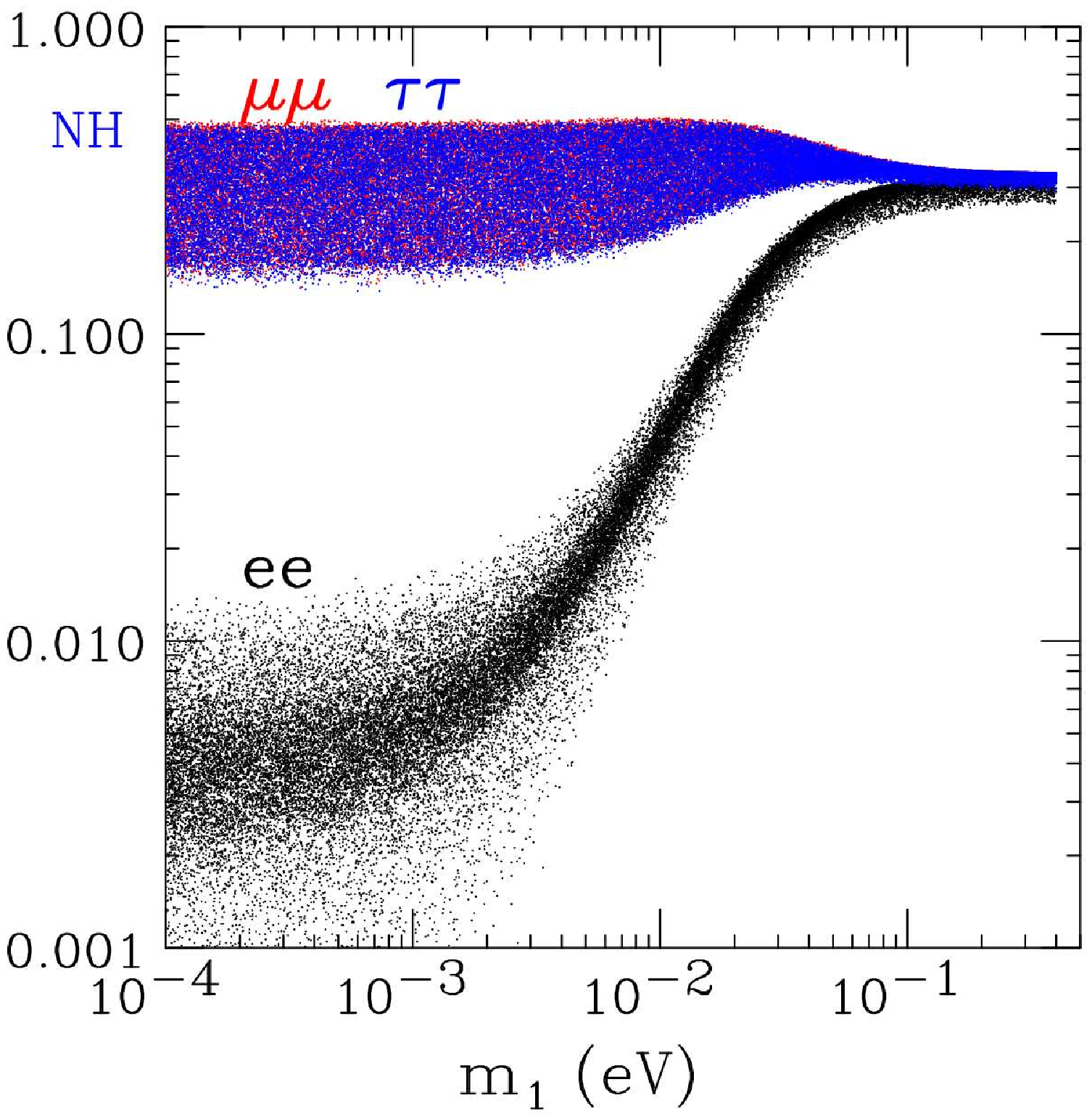}
\includegraphics[width=6cm]{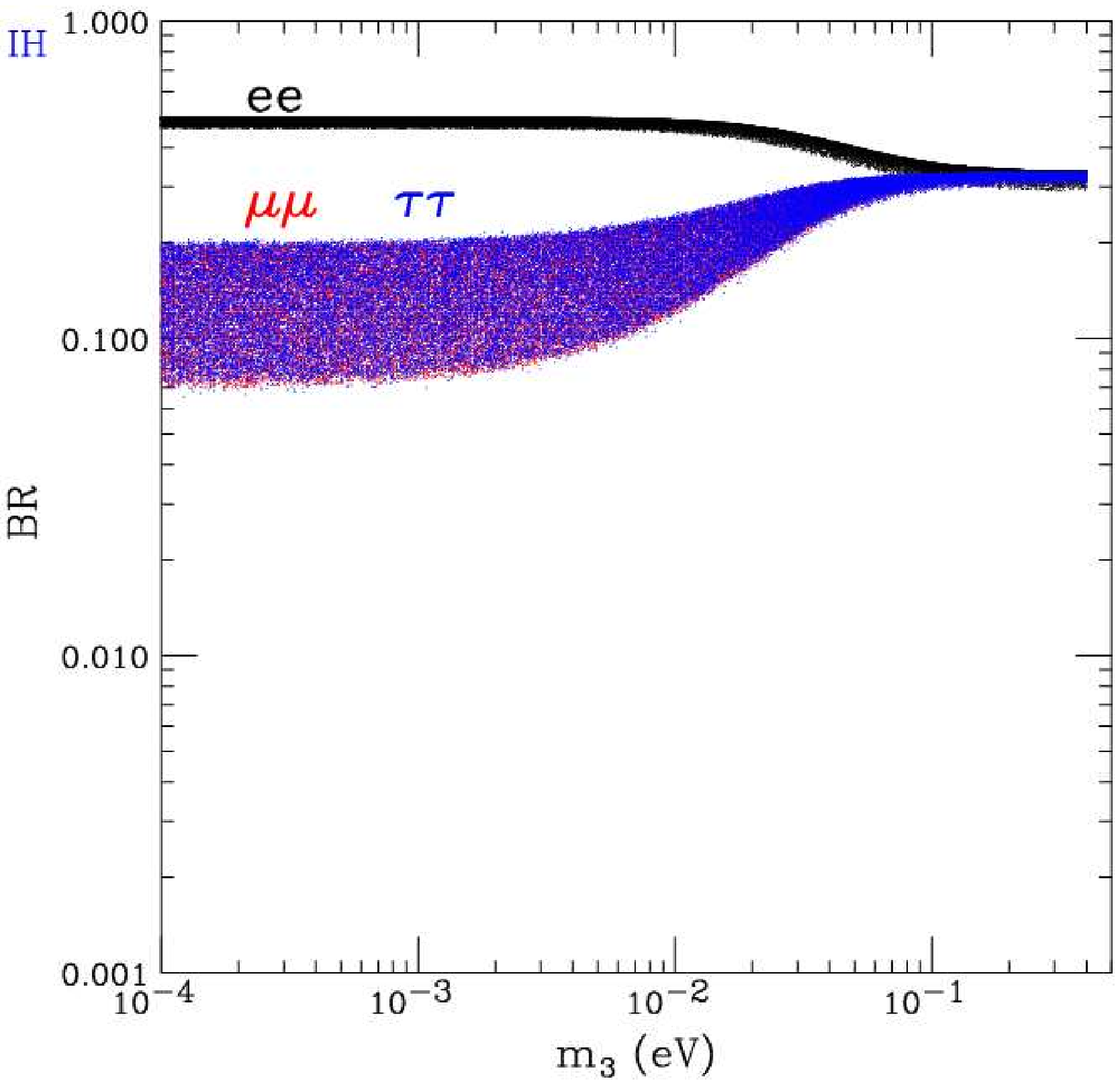}
\end{center}
\caption{Scatter plots for the $H^{++}$ decay branching fractions to
  the flavor-diagonal like-sign di-leptons versus the lowest neutrino
  mass for NH (top) and IH (bottom) with $\Phi_1 = \Phi_2 = 0$, from
  Ref.~\cite{Perez:2008ha}.}
\label{brii}
\end{figure}
For $v_3^{} < 10^{-4}$ GeV, the dominant channels for the heavy
Higgs boson decay are the $\dl=2$ di-leptons.
In Fig.~\ref{brii} we show the predictions for the representative
decay branching fractions (BR) to flavor diagonal di-leptons versus
the lightest neutrino mass.
The spread in BR values is due to the current errors in the neutrino
masses and mixing. Fig.~\ref{brii}(top) is for the $H^{++}$ decay to
same-sign di-leptons in the Normal Hierarchy (NH) ($\Delta m_{31}^2
>0$), and Fig.~\ref{brii}(bottom) for the $H^{++}$ decay in the
Inverted Hierarchy (IH) ($\Delta m_{31}^2 < 0$).
In accordance with the NH spectrum and the large atmosphere mixing
($\theta_{23}$), the leading channels are $H^{++}\to \tau^+ \tau^+,\
\mu^+ \mu^+$, and the channel $e^+ e^+$ is much smaller.  When the
spectrum is inverted, the dominant channel is $H^{++} \to e^+ e^+$
instead.  Also is seen in Fig.~\ref{bri}(top) the $H^{+}\to
\tau^+\bar\nu$ and $H^{+}\to \mu^+\bar\nu$ dominance in the NH and
$H^{+}\to e^+\bar\nu$ in the IH. In both cases the off-diagonal
channel $H^{++} \to \tau^+ \mu^+$ is dominant due to the nearly
maximal atmospheric mixing angle. In the limit of Quasi-Degenerate
(QD) neutrinos one finds that the three diagonal channels are quite
similar, but the off-diagonal channels are suppressed.
\begin{figure}[!h]
\begin{center}
\includegraphics[width=6cm]{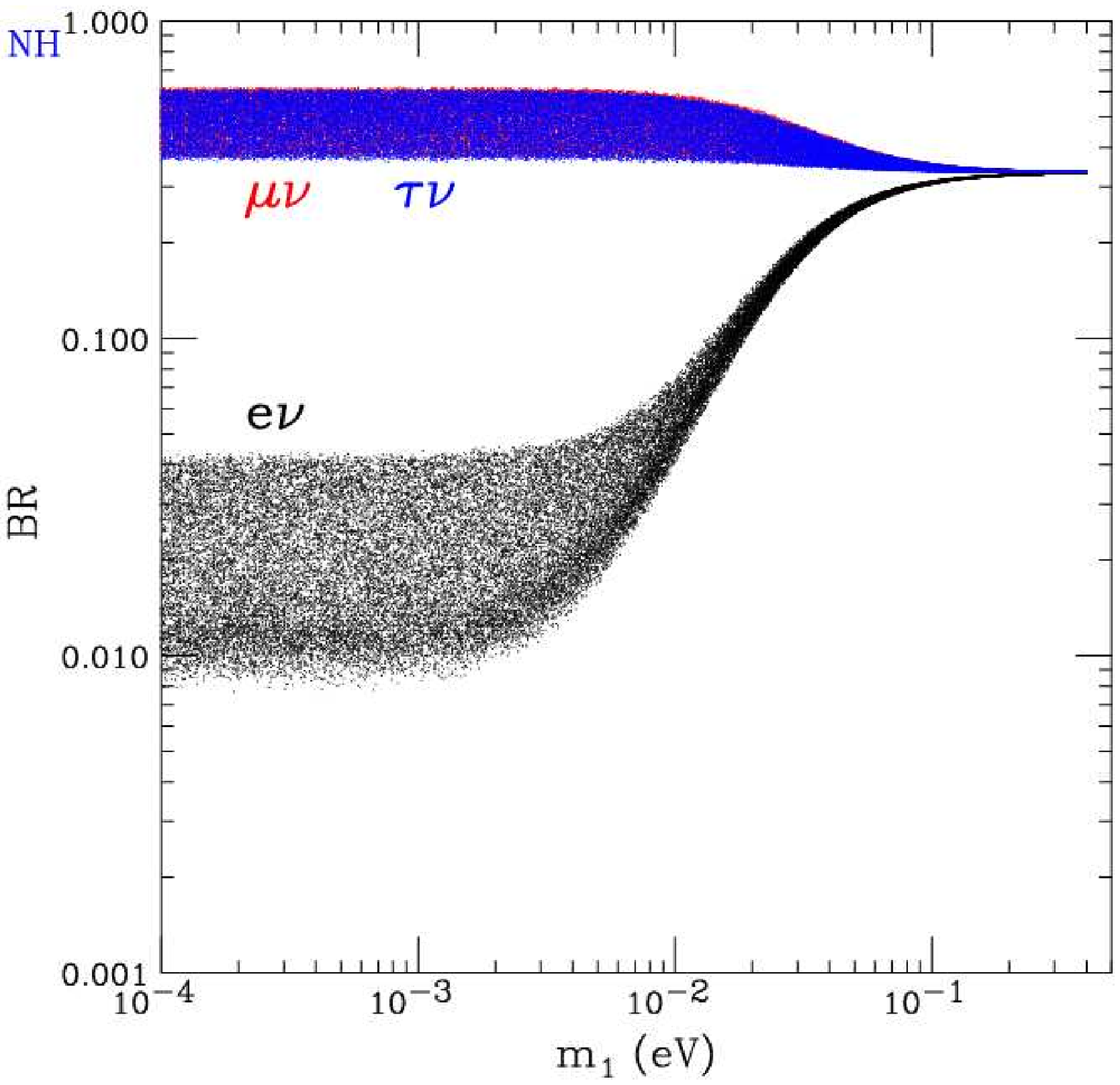}
\includegraphics[width=6cm]{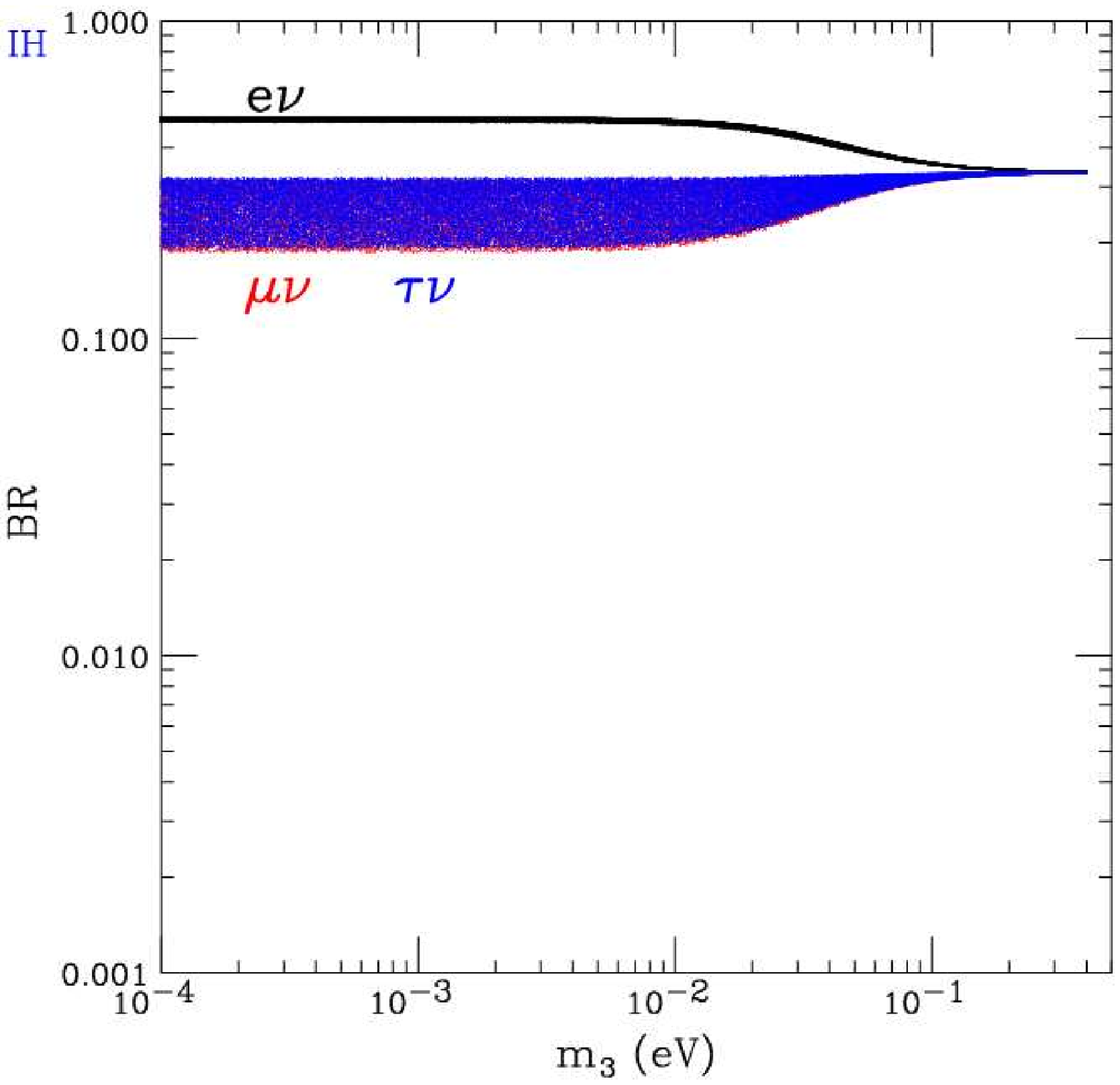}
\end{center}
\caption{Scatter plots for the $H^{+}$ decay branching fractions to
  leptons versus the lowest neutrino mass for NH (top) and IH
  (bottom), from Ref.~\cite{Perez:2008ha}.}
\label{bri}
\end{figure}
The properties of all leptonic decays of the charged Higgs bosons are
summarized in Table.~\ref{Tab1}.
\begin{table}[tb]
\begin{small}
  \caption{\label{contributions} {\small{Relations for the $\dl=2$
        decays of $H^{++},\ H^+$ in three different neutrino mass
        patterns when $\Phi_1= \Phi_2= 0$.}}}
%\begin{ruledtabular}
\begin{tabular}{lcc}
\text{Spectrum} & Relations \\
\hline
NH   & Br$(\tau^+ \tau^+ )$, Br$(\mu^+ \mu^+) \gg$ Br$(e^+ e^+ )$ \\
$\Delta m_{31}^2 > 0$  & Br$(\mu^+ \tau^+) \gg $ Br$(e^+ \tau^+)$, Br$(e^+ \mu^+)$\\
     & Br$(\tau^+ \bar{\nu})$, Br$(\mu^+ \bar{\nu}) \gg $ Br$(e^+ \bar{\nu})$ \\
\hline
IH & Br$(e^+ e^+) > $ Br$(\mu^+ \mu^+)$, Br$(\tau^+ \tau^+)$\\
$\Delta m_{31}^2 < 0$ & Br$(\mu^+ \tau^+) \gg $ Br$(e^+ \tau^+)$,  Br$(e^+ \mu^+)$\\
 & Br$(e^+ \bar{\nu}) > $ Br$(\mu^+ \bar{\nu})$, Br$(\tau^+ \bar{\nu})$ \\
 \hline
QD & Br$(e^+ e^+) \approx$ Br$(\mu^+ \mu^+) \approx$ Br$(\tau^+ \tau^+)$\\
 & Br$(e^+ \bar{\nu}) \approx$ Br$(\mu^+ \bar{\nu}) \approx$ Br$(\tau^+ \bar{\nu})$\\
 & Br$(\mu^+ \tau^+) \approx$ Br$(e^+ \tau^+) \approx$ Br$(e^+ \mu^+)$ \\
\hline
\end{tabular}
%\end{ruledtabular}
\label{Tab1}
\end{small}
\end{table}
Note that the decay in the last row is suppressed.

The effects of the Majorana phases have been neglected so far. They
can only affect lepton number violating
processes~\cite{Schechter:1980gr}~\cite{Schechter:1981gk}, such as
the decays of the doubly charged Higgs boson.
However, these are not very sensitive to the phase $\Phi_2$, with a
maximal reduction of $H^{++}\to \tau^+\tau^+, \mu^+\mu^+$ and
enhancement of $\mu^+\tau^+$ up to a factor of two in the NH case.
However, as shown in Fig.~\ref{Majorana2}, the Majorana phase $\Phi_1$
has a dramatic impact on the $H^{++}$ decay in the IH case.  We see
that for $\Phi_1 \approx \pi$ the dominant channels switch to $e^+
\mu^+,\ e^+ \tau^+$ from $e^+ e^+,\ \mu^+\tau^+$ as in the zero phase
limit.  This provides the best hope to probe the Majorana phase.  The
decays $H^{\pm} \to e^+_i \bar{\nu}$, on the other hand, are
independent of the unknown Majorana phases, leaving the BR predictions
robust. Therefore, using the lepton violating decays of the singly
charged Higgs one can determinate the neutrino spectrum without any
ambiguity.
\begin{figure}[tb]
\begin{center}
\includegraphics[width=6cm]{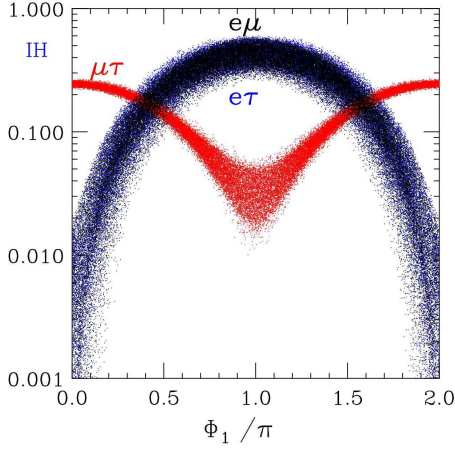}
\end{center}
\caption{\small{Leptonic branching fractions of $H^{++}$ decay versus
    the Majorana phase $\Phi_1$ in the IH for $m_3 \approx 0$, from
    Ref.~\cite{Perez:2008ha}.}}
\label{Majorana2}
\end{figure}

%%%%%%%%%%%%%%%%%%%%%%%%%%%%%%%%%%%%%%%%%%%%%%%%%
\subsubsection{ Testing the Model at the LHC}
%%%%%%%%%%%%%%%%%%%%%%%%%%%%%%%%%%%%%%%%%%%%%%%%%
We consider the following production channels $ q \bar{q} \to
\gamma^*,Z^*\to H^{++} H^{--}$, and $ q \bar{q'} \to W^* \to
H^{\pm\pm} H^{\mp} $.  The total cross sections versus the mass at the
LHC are shown in Fig.~\ref{total}.  The cross sections range in $100 -
0.1$ fb for a mass of 200$-$1000 GeV, leading to a potentially
observable signal with a high luminosity.  The associated production
$H^{\pm\pm} H^{\mp}$ is crucial to test the triplet nature of
$H^{\pm\pm}$ and $H^{\pm}$.

{\bf \textit{Purely Leptonic Modes}}

For $v_3 < 10^{-4}$ GeV, we wish to identify as many channels of
leptonic flavor combination as possible in order to study the neutrino
mass pattern. The $e$'s and $\mu$'s are experimentally easy to
identify, while $\tau$'s can be identified via their simple charged
tracks (1-prong and 3-prongs). We make use of the important feature
that the $\tau$'s from the heavy Higgs decays are highly relativistic
and the missing neutrinos are collimated along the charged tracks, so
that the $\tau$ momentum $p(\tau)$ can be reconstructed effectively.
In fact, we can reconstruct up to three $\tau$'s if we assume the
Higgs pair production with equal
masses~\cite{Perez:2008zc,Perez:2008ha}.  The fully reconstructible
signal events are thus
$$
H^{++} H^{--} \to
\ell^+ \ell^+  \ell^-\ell^-,   \ell^\pm \ell^\pm  \ell^\mp \tau^\mp,
 \ell^\pm \ell^\pm  \tau^\mp  \tau^\mp,$$
$$
 \ell^+\tau^+  \ell^-\tau^-,   \ell^\pm \tau^\pm  \tau^\mp\tau^\mp,
$$
$$
 H^{\pm\pm} H^{\mp}  \to  \ell^\pm\ell^\pm \ell^\mp \nu, \ \
 \ell^\pm\ell^\pm\ \tau^\mp \nu,  \nonumber
$$
where $\ell=e,\mu$. We have performed in
Ref.~\cite{Perez:2008zc,Perez:2008ha} a full kinematical analysis for
those modes, including judicious cuts to separate the backgrounds,
energy-momentum smearing to simulate the detector effects, and the
$p(\tau)$ and $\md$ reconstruction.  We find our kinematical
reconstruction procedure highly efficient, with about $50\%$ for
$\md=200$ GeV and even higher for a heavier mass.  With a 300
fb$^{-1}$ luminosity, there will still be several reconstructed events
in the leading channels up to $\md\sim$ 1 TeV with negligible
backgrounds.

We summarize the leading reconstructible channels and their achievable
branching fractions in Table~\ref{tab:llll}. The $H^\pm$ decays are
robust to determinate the mass pattern since they are independent of
the Majorana phases, more details in~\cite{Perez:2008zc,Perez:2008ha}.
\begin{figure}[tb]
\begin{center}
\includegraphics[width=7cm]{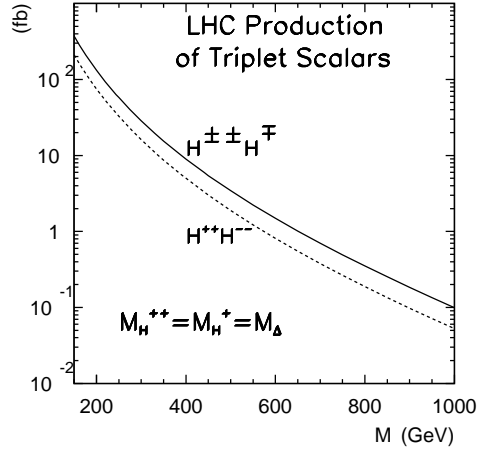}
\end{center}
\caption{\small{Total cross sections in units of fb for $pp\to H^{++}
    H^{--}$ and $H^{\pm\pm}H^{\mp}$ production versus its mass at
    $\sqrt s=14$ TeV, from Ref.~\cite{Perez:2008ha}.}}
\label{total}
\end{figure}
\begin{table}[tb]
\begin{center}
\begin{tabular}{| l| l|}
\hline
Channels & Modes and BR's(NH)  \\
\hline
$H^{++} H^{--}$ & \\
& $\mu^+\mu^+ \mu^- \mu^-\ (40\%)^2$   \\
& $\mu^+\mu^+ \mu^- \tau^-\   40\%\times 35\% $\\
& $\mu^+\mu^+ \tau^- \tau^-\   (40\%)^2$   \\
& $\mu^+\tau^+  \mu^- \tau^-\   (35\%)^2$   \\
$\Phi_1\approx \pi$ & same as  above  \\
$ \Phi_2\approx \pi$ &$\mu\mu,\tau\tau: \times 1/2, \ \mu \tau: \times 2$  \\
\hline
$H^{\pm\pm} H^{\mp}$ &   \\%&  \\
$\Phi_1,\Phi_2=0$& $\mu^+\mu^+ \mu^- \nu\   40\%\times 60\%$   \\
& $\mu^+\mu^+  \tau^- \nu \   40\%\times 60\%$   \\
$\Phi_1\approx \pi $ & same as  above  \\
$\Phi_2\approx \pi$ &$\mu\mu: \times 1/2$  \\% &  same as above \\
\hline
\end{tabular}
\vglue .5cm
\caption{
Leading fully reconstructible leptonic channels and their achievable
branching fractions for NH, from Ref.~\cite{Perez:2008ha}.}
\label{tab:llll}
\end{center}
\end{table}

\begin{table}[tb]
\begin{center}
\begin{tabular}{| l| l|}
\hline
Channels & Modes and BR's (IH)\\
\hline
$H^{++} H^{--}$  & \\
$\Phi_1,\Phi_2=0$   & $e^+e^+ e^-e^-\  (50\%)^2$ \\
&  $e^+e^+ \mu^-\tau^-\  50\%\times 25\%$ \\
 &  $\mu^+\tau^+ \mu^-\tau^-\ (25\%)^2$ \\
$\Phi_1\approx \pi$ & $ee, \mu \tau \to e\mu, e\tau\  (50\%)^2$ \\
$ \Phi_2\approx \pi$ &  same as above  \\
\hline
$H^{\pm\pm} H^{\mp}$   &  \\
$\Phi_1,\Phi_2=0$  & $e^+e^+ e^- \nu\   (50\%)^2$ \\
$\Phi_1\approx \pi $ & $ee \to e\mu, e\tau\  60\%\times 50\%$ \\
$\Phi_2\approx \pi$   &  same as above \\
\hline
\end{tabular}
\vglue .5cm
\caption{
Leading fully reconstructible leptonic channels and their achievable
branching fractions for IH, from Ref.~\cite{Perez:2008ha}.}
\label{tab:llll}
\end{center}
\end{table}

A global analysis of $H^{++} H^{--}$ and $H^{\pm \pm} H^\mp$
production including fast detector simulation and the relevant SM
backgrounds has been performed in Ref.~\cite{delAguila:2008cj} for
multi-leptonic final states with one, two, three and four charged
leptons. In particular, $\tau$ decays giving electrons, muons or
jets are properly included. These decays are specially important
for NH, where the decay of $H^{++}$ and $H^+$ mainly give tau
leptons, which in turn originate secondary electrons and muons
which constitute a combinatorial background for scalar triplet
searches, even larger than the SM one.

The discovery potential strongly depends on the neutrino mixing
parameters determined by oscillation
experiments~\cite{Maltoni:2004ei,Schwetz:2008er} (the dependence on
$s_{13}$ is small), as well as on the neutrino mass hierarchy.  It is
found~\cite{delAguila:2008cj} that for NH (IH) scalar masses up to 600
GeV (800 GeV) can be discovered with 30 fb$^{-1}$. The trilepton
channel is the one where signals are largest both for NH and IH, and
offers the best sensitivity to scalar triplets. It is followed by the
four lepton and like-sign dilepton channels.  Opposite-sign dilepton
signals with a tagged $\tau$ jet are hard (but not impossible) to see,
while the ones with a single charged lepton and three $\tau$ jets seem
hopeless due to the large background from $W$ production plus jets,
misidentified as taus.

With a sufficient luminosity, the trilepton and four lepton channels
can provide evidence of the scalar nature of $H^{\pm \pm}$ with the
analysis of the opening angle distribution. The non-singlet nature
of $H^{\pm \pm}$, $H^\pm$ can also be established in the trilepton
final state with the identification of events with large missing
energy and small hadronic activity.\\

{\bf \textit{Other decay modes}}

For $v_3 > 2\times 10^{-4}$ GeV, the dominant decay modes of the heavy
Higgs bosons are the SM gauge bosons. The decay $H^{\pm\pm}\to W^\pm
W^\pm$ is governed by $v_3$ and $H^\pm \to\ W^\pm H_1,\ t\bar b$ by
the mixing $\mu$, and $H^\pm \to W^\pm Z$ by a combination of
both. Therefore, systematically studying those channels would provide
the evidence of the triplet-doublet mixing and further confirm the
seesaw relation $v_3=\mu v_2^2 / \sqrt{2} M_{\Delta}^2$. We have once
again performed detailed signal and background analysis at the LHC for
those channels. We are able to obtain a $20\%$ signal efficiency and a
signal-to-background ratio $1:1$ or better.  With a 300 fb$^{-1}$
luminosity, we can achieve statistically significant signals up to
$\md\approx 600$ GeV~\cite{Perez:2008zc}.

\subsection{Charged fermions in type-III seesaw}
\label{sec:type-III-pheno}

In the type-III seesaw mechanism in Sec.~(\ref{sec:type-iii-seesaw})
the members of the heavy fermionic $SU(2)_{L}$-triplet\footnote{Recall
  that the hypercharge of this triplet is different from the one of
  the scalar triplet in the type-II seesaw.} (here we denote them
generically as $E^{\pm}\equiv \Sigma^{\pm}$, $N\equiv \Sigma^{0}$)
couple to the SM gauge bosons. There are arguments based on grand
unification which suggest that these new light fermions can have
masses in the TeV range and hence accessible at the LHC
\cite{Bajc:2006ia}. We give a brief description below:

\subsubsection{SU(5) theory motivation }

The minimal SU(5) theory fails for two important reasons: (a) gauge
couplings do not unify $\alpha_2$ and $\alpha_3$ meet at about
$10^{16}$ GeV but $\alpha_1$ meets $\alpha_2$ too early, at $\approx
10^{13}$ GeV; (b) neutrinos remains massless as in the SM.  The $d=5$
Weinberg operator is not enough: neutrino mass comes out too small
($\lesssim 10^{-4} eV$) since the cut-off scale M must be at least as
large as $M_{GUT}$ due to SU(5) symmetry.  In any case, one must first
make sure that the theory is consistent and the gauge couplings unify.
A simple extension cures both problems: add just one extra fermionic
$24_F$ \cite{Bajc:2006ia}. This requires higher dimensional operators
just as in the minimal theory, but can be made renormalizable as usual
by adding extra $45_H$ scalar \cite{Perez:2007rm}.  Under \321 the
adjoint is decomposed as:
$24_F=(1,1)_0+(1,3)_0+(8,1)_0+(3,2)_{-5/6}+(\bar 3,2)_{5/6}$.
Unification works as follows: triplet fermion (like wino in MSSM)
slows down $\alpha_2$ coupling without affecting $\alpha_1$. In order
that they meet above $10^{15}$ GeV to ensure proton stability, the
triplet must be light, with a mass below TeV. Then in turn $\alpha_3$
must be slowed down, which is achieved with an intermediate scale mass
for the color octet in $24_F$ around $10^7$ GeV or so.
This theory behaves effectively as the MSSM with a light wino, heavy
gluino ($10^7 \text{GeV}$), no Higgsino, no sfermions (they are
irrelevant for unification being complete representations). This shows
how splitting supersymmetry~\cite{ArkaniHamed:2004fb} opens a
Pandora's box of possibilities for unification. Unlike the case of
supersymmetry, where the scale was fixed by a desire for the
naturalness of the Higgs mass, and then unification predicted, in this
case the SU(5) structure demands unification which in turn fixes the
masses of the new particles in $24_F$. The price is the fine-tuning of
these masses, but a great virtue is the tightness of the theory: the
low mass of the fermion triplet (and other masses) is a true
phenomenological prediction not tied to a nice but imprecise notion of
naturalness.

With the notation singlet $S=(1,1)_0$, triplet $T=(1,3)_0$, it is
evident that we have mixed Type I and Type III seesaw.

\subsubsection{Type-III seesaw phenomenology }

In type-III seesaw models one expects the following processes
\begin{eqnarray}
q \bar q & \to & Z^*/\gamma^* \to E^+ E^- \,, \nonumber \\
q \bar q' & \to & W^* \to E^\pm N~,
\label{ec:Tprod}
\end{eqnarray}
with typically electroweak cross sections, as seen in
Fig.~\ref{fig:Franceschini}.
\begin{figure}[!ht]
\vspace{9pt}
\begin{center}
\includegraphics[width=60mm,height=45mm]{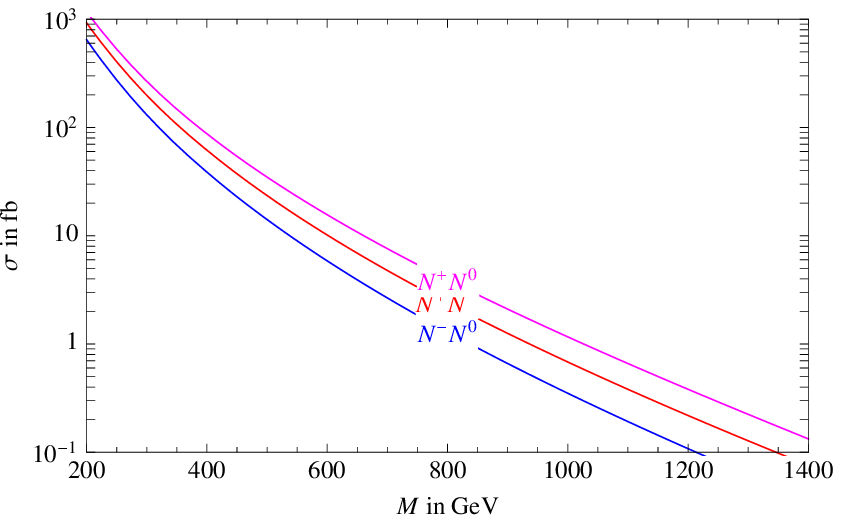}
\includegraphics[width=70mm,height=50mm]{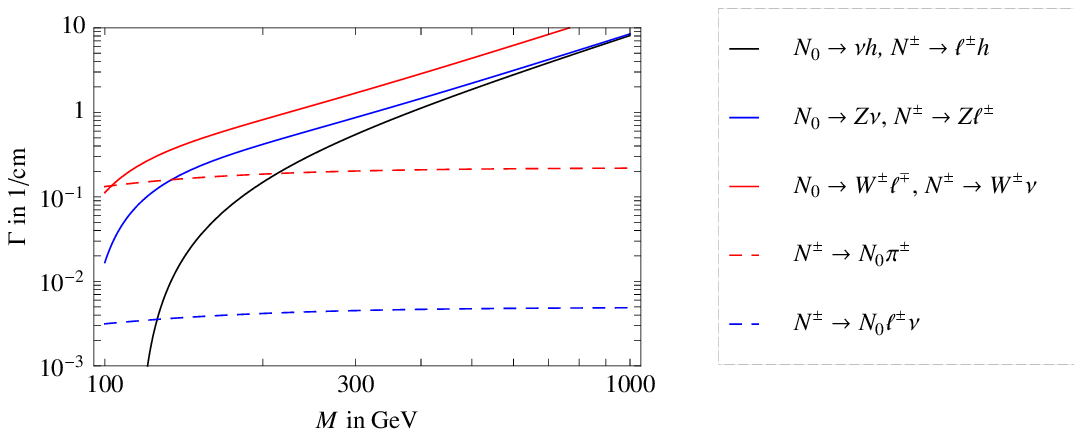}
\end{center}
\caption{Total production cross section of triplet leptons in type-III
  seesaw at LHC (top) and decay widths (bottom), from
  Ref.~\cite{Franceschini:2008pz}, which adopts the convention
  $N^{\pm} \equiv E^{\pm} \equiv \Sigma^{\pm}$.}
\label{fig:Franceschini}
\end{figure}
The heavy leptons subsequently decay into the SM gauge bosons and
leptons, and their pair production in Eqs.~(\ref{ec:Tprod}) yields
several interesting multi-leptonic signals with four, three and two
leptons in the final state. However, the kinematics here is very
different from the analogous type-II signals and the interplay among
various branching ratios admits to distinguish this scenario (either
with Majorana or Dirac heavy neutrinos) also from the other options
like e.g. $Z' \to NN$ production~\cite{AguilarSaavedra:2009ik} in the
left-right symmetric type-I seesaw case discussed above,
c.f. (\ref{ec:Nprod}).

The $\ell^\pm \ell^\pm \ell^\mp$ final state offers the best
discovery potential and allows to reconstruct the heavy neutrino
mass and identify its charge.  The $\ell^+ \ell^+ \ell^- \ell^-$
final state allows to reconstruct the heavy charged lepton mass
also determining its charge. Finally, the presence or absence of
like-sign di-leptons establishes the Majorana or Dirac nature of
the neutrino, distinguishing a minimal type-III seesaw from an
inverted one. Note that these heavy triplet fermions decay
predominantly
\cite{Franceschini:2008pz,Arhrib:2009mz,Bajc:2006ia,Li:2009mw}
into $W,\, Z$, Higgs and a SM lepton.  With a luminosity of 30
fb$^{-1}$, the mass reach for lepton triplets is up to $m_{E,N} =
750$ (700) GeV for the Majorana (Dirac) case, assuming decays to
electrons or muons.
Note that, since the relevant Yukawa couplings should be small in
order to retain light enough neutrino mass $m_\nu\lesssim\,$eV, this
decay can lead to displaced vertices, for mixings $|V| \leq 10^{-8}$,
which may be observable~\cite{Franceschini:2008pz}.

\subsection{Low-scale seesaw schemes}

In the minimal type-I inverse and linear seesaw schemes discussed in
sections (\ref{sec:inverse-sees}) and (\ref{sec:linear-seesaw}) the
Yukawa couplings of a TeV scale RH neutrino need not be suppressed by
the smallness of the neutrino masses, hence it may be directly
produced in collider
experiments~\cite{Dittmar:1990yg,Gonzalez-Garcia:1990fb}.
However, due to the quasi-Dirac nature~\cite{valle:1983yw} of the
heavy states the striking same-sign lepton signals observable e.g. in
the type-II case are generally lost and the opposite-sign signals tend
to be buried in the SM background.  Nevertheless, the scheme may have
other phenomenological implications, inducing for instance lepton
flavor violating (LFV) decays such as $l_i \to l_j \gamma$. These may
proceed either through the exchange of RH
neutrinos~\cite{Bernabeu:1987gr} or as a result of supersymmetric
contributions~\cite{Deppisch:2004fa}. Models leading to tri-bimaximal
mixing~\cite{Harrison:2002er} lead to very specific predictions for
these processes~\cite{Hirsch:2009mx}.

Essentially the same happens also in the inverse type-III seesaw model
in sec~(\ref{sec:inverse-type-iii}) -- the Yukawa couplings of a TeV
scale fermion triplet need not be suppressed by small neutrino masses
so, once produced at the LHC through SM gauge interactions, the
triplet will typically decay with very short decay length. This,
however, contrasts with the ``standard'' type-III seesaw which will be
more likely to lead to displaced vertices.

\section{R-parity violation: Theory}

\subsection{R-parity violating supersymmetry}

We now turn to the exciting possibility that low-energy supersymmetry
itself may provide the origin of neutrino
mass~\cite{Aulakh:1982yn,Ellis:1984gi,Ross:1984yg,hall:1984id,santamaria:1989ic},
for a review see Ref.~\cite{Hirsch:2004he}.  In the simple class of
supersymmetric models widely discussed in the literature, it is
assumed that R-parity, defined as $(-1)^{3B+L+2S}$, is an exact
symmetry under which all superpartners are odd and SM particles even.
However the terms that break R-parity are allowed by supersymmetry as
well as the SM gauge invariance. In the language of superfields, they
have the form $LH_u$, $LLe^c$, $QLd^c$ and $u^cd^cd^c$. If all four
terms are present proton decay becomes very rapid. This problem is
circumvented by simply forbidding the last term, e.g. by using baryon
triality or a similar symmetry \cite{Dreiner:1997uz,Dreiner:2005rd}.
The remaining three terms have the property that they break lepton
number explicitly (LNV). Indeed, a combination of tree and loop
diagrams in these models can lead to realistic neutrino masses and
mixings.  In these models, the lightest superpartner is unstable
unlike the minimal R-conserving MSSM, implying the need for other dark
matter candidates, such as the axion or, in specific scenaria, like
gauge-mediated SUSY breaking, the gravitino~\cite{Hirsch:2005ag}. From
the point of view of collider physics, there is an important
implication of LSP decay, namely, it can lead to observable signatures
and crucial tests that can be performed at the LHC in order to
establish or rule out the supersymmetric origin of neutrino mass. This
is possible since the same couplings governing neutrino physics also
lead to visible decays of the lightest supersymmetric particle (LSP).

Here we will focus on bilinear $R_P$ breaking, for discussion of
tri-linear $\rpv$ see for example
\cite{allanach:1999bf,Barbier:2004ez}.  The absence of tri-linear terms
could be explained, for example, if bilinear R-parity breaking is the
effective low-energy limit of some spontaneous $\rpv$ model, see
below.

\subsection{Explicit bilinear R-parity violation}

The superpotential of the bilinear $\rpv$ model can be written as
\begin{equation} \label{spot}
{\cal W} = \epsilon_i \widehat L_i \widehat H_u + W_{MSSM}.
\end{equation}
In addition, one must include bilinear $\rpv$ soft supersymmetry
breaking terms
\begin{equation} \label{vsoft}
V_{\rm soft} = \epsilon_i B_i {\tilde L}_i H_u + V_{\rm soft}^{MSSM}.
\end{equation}
The above defines the minimal bilinear model~\cite{Diaz:1997xc}. It
contains only 6 new parameters with respect to the MSSM. While all six
parameters could be complex, neutrino physics require a strong
correlation between these, such that effectively only three phases
remain \cite{Hirsch:2002tq}.

The terms in Eq.~(\ref{vsoft}) induce mixing between the MSSM Higgs
bosons and the left scalar neutrinos. Thus, once electro-weak symmetry
is broken, scalar neutrinos acquire vacuum expectation values and one
non-zero neutrino mass is generated at tree-level.  The effective
neutrino mass matrix at tree-level can then be cast into a very simple
form
\begin{equation}
 -{\bf m_{\nu\nu}^{\rm eff}}_{ij} =
\frac{m_{\gamma}}{4 det(M_{\chi^0})} \Lambda_i \Lambda_j
\label{eq:eff}
\end{equation}
The ``photino'' mass parameter is defined as $m_{\gamma} = g^2M_1
+g'^2 M_2$, $det(M_{\chi^0})$ is the determinant of the ($4,4$)
neutralino mass matrix and $\Lambda_{\alpha}$ is the ``alignment
parameter'' given as $\Lambda_{\alpha} \equiv
\epsilon_{\alpha}v_D+v_{\alpha}\mu$, with $v_{\alpha}$ the scalar
neutrino vev of generation $\alpha$.

Due to the projective nature of Eq.~(\ref{eq:eff}) the other two
neutrino masses are generated only at 1-loop order.  Generally the
most important contributions come from loops with scalar bottom quarks
and scalar taus \cite{Diaz:2003as}. However, there exist also
parameter regions in the general $\rpv$ MSSM where the
sneutrino-antis-neutrino loop gives a sizeable contribution
\cite{Grossman:2000ex,Dedes:2006ni,Dedes:2007ef}. For a fully
numerical study of neutrino masses within bilinear $\rpv$, see for
example \cite{Hirsch:2000ef}.  One finds that in order to explain the
observed neutrino mixing angles one requires certain relations among
the $\rpv$ parameters to be satisfied~\cite{Hirsch:2000ef}. For
example, the maximal atmospheric angles requires $\Lambda_{\mu}\simeq
\Lambda_{\tau}$.
\begin{figure}[htb]
\vspace{9pt}
\includegraphics[width=75mm,height=60mm]{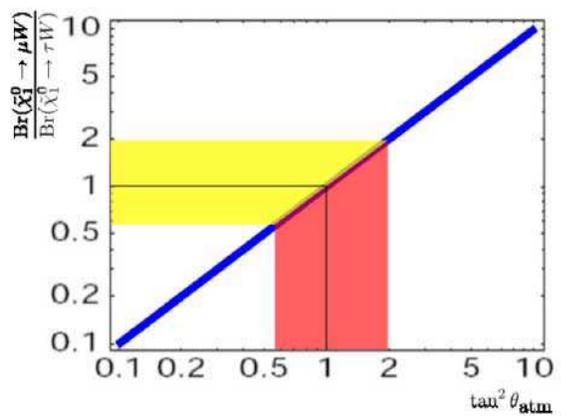}
\caption{Ratio of semi-leptonic branching ratios, Br$(\chi^0_1\to
\mu
  q'{\bar q})$ over Br$(\chi^0_1\to \tau q'{\bar q})$ as a function of
  the atmospheric neutrino angle calculated within bilinear $\rpv$
  SUSY, see Ref.~\cite{Porod:2000hv}.}
\label{fig:ntrl}
\end{figure}

Once R-parity is broken the LSP decays. The decays of a neutralino LSP
have been studied in \cite{mukhopadhyaya:1998xj,Porod:2000hv}. Decay
lengths for the neutralino are approximately fixed once the neutrino
masses are fitted to experimental data. Typical lengths range from
tens of cm for very light neutralinos to sub-millimeter for
neutralinos of several hundred GeV \cite{Porod:2000hv}.  One of the
most exciting aspects of bilinear $\rpv$, however, is the fact that
once neutrino angles are fitted to the values
required~\cite{Maltoni:2004ei,Schwetz:2008er} by the neutrino
oscillation data, the ratios of LSP decay branching ratios are fixed
and correlate with the observed neutrino mixing angles, as illustrated
for example in fig. (\ref{fig:ntrl}).  Measurements at the LHC should
allow to test this prediction, if signals of SUSY are found.

Within $\rpv$ SUSY any supersymmetric particle can be the LSP. The
decays of charged scalar have been studied in \cite{Hirsch:2002ys},
while the case of stop LSP was considered in
Ref.~\cite{Restrepo:2001me}.  For an overview of possible LSP
candidates see Ref.~\cite{Hirsch:2003fe}.  It has been shown that
within bilinear $\rpv$ correlations between the measured neutrino
angles and ratios of LSP decays can be found for all LSP candidates
\cite{Hirsch:2003fe}. Thus, it is possible to {\em exclude} the
minimal bilinear $\rpv$ model experimentally at the LHC.

\subsection{Spontaneous RPV}

In spontaneous R-parity violation (SRPV)
models~\cite{Aulakh:1982yn,Ross:1984yg,santamaria:1989ic} R-parity
violation results from the minimization of the Higgs potential through
nonzero sneutrino vacuum expectation values. If lepton number is
ungauged, as in the \321 model, this implies the existence of a
Nambu-Goldstone boson - the majoron. However, a doublet majoron is
ruled out since by LEP measurements of the Z
width~\cite{Amsler:2008zzb}. Hence, viable spontaneous R-parity
breaking models must be characterized by two types of sneutrino vevs,
those of right and left-handed sneutrinos, singlets and doublets under
\321
respectively~\cite{masiero:1990uj,romao:1992vu,Romao:1996xf}. These
obey the ``vev-seesaw'' relation $v_L v_R \sim h_\nu m_W^2$ where
$h_\nu$ is the small Yukawa coupling that governs the strength of
R--parity violation~\cite{masiero:1990uj,romao:1992vu,Romao:1996xf}.

In this case the majoron is so weakly coupled that bounds from LEP and
astrophysics~\cite{Raffelt:1996wa} are easily satisfied.  For example,
the superpotential of \cite{masiero:1990uj} can be written as
\begin{eqnarray} %
{\cal W} &=& h_U^{ij}\widehat Q_i \widehat U_j\widehat H_u
          +  h_D^{ij}\widehat Q_i\widehat D_j\widehat H_d
          +  h_E^{ij}\widehat L_i\widehat E_j\widehat H_d \nonumber
\\
        & + & h_{\nu}^{i}\widehat L_i\widehat \nu^c\widehat H_u
          - h_0 \widehat H_d \widehat H_u \widehat\Phi
          + h \widehat\Phi \widehat\nu^c\widehat S +
          \frac{\lambda}{3!} \widehat\Phi^3 . \nonumber
\label{eq:Wsuppot}
\end{eqnarray}
The first three terms are the usual MSSM Yukawa terms. The terms
coupling the lepton doublets to $\widehat\nu^c$ fix lepton number.
The coupling of the field $\widehat\Phi$ with the Higgs doublets
generates an effective $\mu$-term a l\'a Next to Minimal
Supersymmetric Standard Model (NMSSM)
\cite{Barbieri:1982eh,Nilles:1982dy,Chamseddine:1982jx}.  The last two
terms, involving only singlet fields, give mass to $\widehat\nu^c$,
$\widehat S$ and $\widehat\Phi$, once $\Phi$ develops a vev.

Note, that $v_R \ne 0$ generates effective bilinear terms $\epsilon_i
= h_{\nu}^i v_R/\sqrt{2}$ and that $v_R$, $v_S$ and $v_{L_i}$ violate
lepton number and R-parity spontaneously. The profile of the majoron
in this model is given approximately as (the imaginary part of)
\begin{equation}
\frac{\sum_i v_{Li}^2}{Vv^2} (v_u H_u - v_d H_d) +
\sum_i \frac{v_{Li}}{V} \tilde{\nu_{i}}
+\frac{v_S}{V} S
-\frac{v_R}{V} \tilde{\nu^c}~,
\label{maj}
\end{equation}
where $V=\sqrt{v_S^2+v_R^2}$. Neutrino oscillation data enforce
$v_{L_i}^2 \ll v_R^2$ and $v_{L_i}^2 \ll v^2$, where $v^2=v_D^2+v_U^2$.
Thus the majoron is mainly a singlet in this model, as required.

The model as specified above produces two non-zero neutrino masses at
tree-level. Whether loop corrections are important or not depends on
the unknown singlet parameters and can not be predicted in general.
However, if the singlets exist around the electro-weak scale the
tree-level contributions are sufficient to explain all oscillation
data.

SRPV models can in principle be distinguished from the explicit $\rpv$
models at colliders, due to the existence of the majoron. It has long
been noted~\cite{romao:1992zx,deCampos:1996bg} that the lightest Higgs
can decay invisibly within SRPV as has been shown in detail in
Refs.~\cite{Hirsch:2004rw,Hirsch:2005wd}. Also the decays of the
lightest neutralino are affected, since the new decay channel
$\chi^0_1\to J \nu$ is invisible at colliders. As shown in
\cite{Hirsch:2006di}, if the scale of $\rpv$ is very low, SRPV might
look very MSSM-like and large statistics might be necessary at the LHC
to establish that R-parity is broken \cite{Hirsch:2008ur}. In this
context it is interesting to mention that it has been pointed out long
ago that majoron emitting charged lepton decays occur in
SRPV~\cite{romao:1991tp}. One can show that these decays are
correlated with the decay $\chi^0_1\to J \nu$, thus allowing to probe
for a complementary part of parameter space \cite{Hirsch:2009ee}.

\begin{figure}[htb]
\vspace{9pt}
\includegraphics[width=75mm,height=60mm]{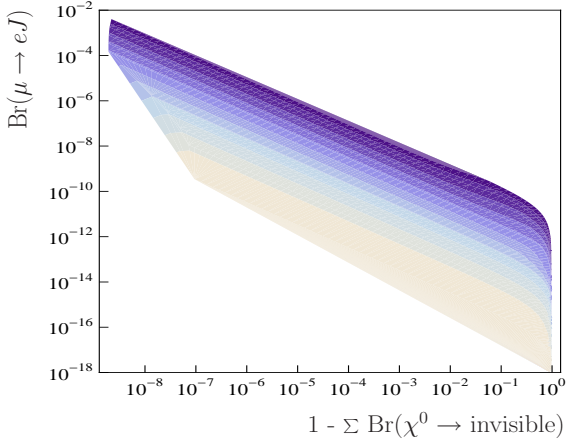}
\caption{Branching ratio Br($\mu\to eJ$) versus visible lightest
neutralino decay. $\mu\to eJ$ and $\chi^0_1\to J \nu$ are correlated,
see \cite{Hirsch:2009ee}.}
\label{fig:VisMuEJ}
\end{figure}

Spontaneous $R$-parity violation can also be obtained by enlarging the
gauge group by an extra U(1), suggested in some superstring models
based on \e6~\cite{Valle:1987sq} \cite{gonzalez-garcia:1991qf}, or by
a full SU(2)$_R$ in left-right symmetric $SU(2)_L \otimes SU(2)_R
\otimes U(1)_{B-L}$ models
\cite{Kuchimanchi:1993jg,Huitu:1994zm,Huitu:1997qu,FileviezPerez:2008sx,Everett:2009vy,Barger:2008wn,Ji:2008cq}. In
fact, in the case of minimal SUSY left-right model with B-L=2 triplets
the only parity violating electric charge conserving minimum breaks
R-parity spontaneously by the non-zero vev of the right handed
sneutrino field\cite{Kuchimanchi:1993jg}. Thus R-parity breaking is
spontaneous and dynamical.  One class of such theories includes
two triplet- (left and right) and two bi-doublet Higgs superfields in
the Higgs sector, with the following $SU(2)_L \otimes SU(2)_R \otimes
U(1)_{B-L}$ quantum numbers:
\begin{eqnarray}
\label{higgses}
&&\widehat\Delta =\left( \begin{array}{cc}
\widehat\Delta^-/\sqrt{2} & \widehat\Delta^{0}\\
\widehat\Delta^{--}&-\widehat\Delta^{-}/\sqrt{2} \end{array} \right)
\sim ({\bf 1,3,}-2), \nonumber
\end{eqnarray}
\begin{eqnarray}
&&\widehat\delta =\left( \begin{array}{cc}
\widehat\delta^{+}/\sqrt{2}& \widehat\delta^{++}\\
 \widehat\delta^{0} &-\widehat\delta^{+}/\sqrt{2} \end{array} \right)
\sim ({\bf 1,3,}2), \nonumber
\end{eqnarray}
\begin{eqnarray}
&& \widehat\phi =\left( \begin{array}{cc}\widehat\phi_1^0&
\widehat\phi_1^+\\\widehat\phi_2^-&\widehat\phi_2^0
\end{array} \right) \sim ({\bf 2,2,}0), \nonumber
\end{eqnarray}
\begin{eqnarray}
&&\widehat\chi =\left( \begin{array}{cc}\widehat\chi_1^0&
\widehat\chi_1^+\\\widehat\chi_2^-&\widehat\chi_2^0
\end{array} \right)  \sim ({\bf 2,2,}0).
\end{eqnarray}
In the fermion sector the `right-handed' matter superfields are
combined to $SU(2)_R$ doublets which requires the existence of
right-handed neutrinos.  The corresponding superfields are denoted by
$\widehat Q^c_R$ and $\widehat L^c_R$ for quark and lepton superfield
respectively. Also in this case all neutral components of the Higgs
fields and all sneutrinos get vevs. However, the majoron now becomes
the longitudinal component of the extra $Z'$ gauge boson.
However, as noted in~\cite{FileviezPerez:2008sx}, the triplet fields
are not mandatory for a realistic theory.

\subsection{The $\mu\nu SSM$}

The superpotential of the MSSM contains a mass term for the Higgs
superfields, $\mu {\widehat H_d}{\widehat H_u}$, phenomenologically
required to lie at the electro-weak scale. However, if there is a
larger scale in the theory, like the grand unification scale, the
natural value of $\mu$ lies at this large scale. This is, in short,
the $\mu$-problem of the MSSM \cite{Kim:1983dt}.  The Next-to-Minimal
SSM (NMSSM) provides a solution to this problem
\cite{Barbieri:1982eh,Nilles:1982dy}, at the cost of introducing a new
singlet field. The vev of the singlet produces the $\mu$ term, once
electro-weak symmetry is broken.

The $\mu\nu$SSM \cite{LopezFogliani:2005yw} proposes to use the same
singlet superfield(s) which generate the $\mu$ term to also generate
Dirac mass terms for the observed left-handed neutrinos:
\begin{eqnarray} %
{\cal W} &=& h_U^{ij}\widehat Q_i \widehat U_j\widehat H_u
          +  h_D^{ij}\widehat Q_i\widehat D_j\widehat H_d
          +  h_E^{ij}\widehat L_i\widehat E_j\widehat H_d \nonumber
\\
        & + & h_{\nu}^{is} \widehat L_i \widehat \nu^c_s \widehat H_u
          - \lambda_s \widehat \nu^c_s \widehat H_d \widehat H_u
 +\frac{1}{3!}\kappa_{stu}\widehat\nu^c_s \widehat\nu^c_t \widehat\nu^c_u~.
\nonumber
\label{eq:WmunuSSM}
\end{eqnarray}
Lepton number in this approach is broken explicitly by the last two
terms.  $R_p$ is broken also and Majorana neutrino masses are
generated once electro-weak symmetry is broken.

Three recent papers have studied the $\mu\nu$SSM in more detail. In
\cite{Escudero:2008jg} the authors analyze the parameter space of the
$\mu\nu$SSM, putting special emphasis on constraints arising from
correct electro-weak symmetry breaking, avoiding tachyonic states and
Landau poles in the parameters. The phenomenology of the $\mu\nu$SSM
has been studied also in \cite{Ghosh:2008yh} including tree-level
neutrino masses and two-body ($W$-lepton) final-state neutralino LSP
decays~\cite{Ghosh:2008yh}. In \cite{Bartl:2009an} detailed study of
the LHC phenomenology of the $\mu\nu{}SSM$ has been carried out.

\begin{figure}[htb]
\vspace{9pt}
\includegraphics[width=75mm,height=60mm]{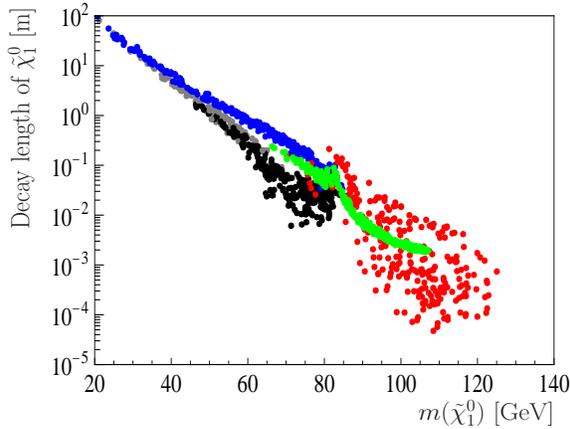}
\caption{Lightest neutralino decay lengths within the $\mu\nu{}SSM$.
From \cite{Bartl:2009an}.} \label{fig:length}
\end{figure}

As pointed out in this work \cite{Bartl:2009an} there are different
variants of the $\mu\nu{}SSM$ which can explain neutrino oscillation
data. The simplest variant has only one generation of singlets and
produces only one non-zero neutrino mass at tree-level. Thus loop
corrections need to be included in this case, just as in the explicit
bilinear RPV model. In case more than one generation of singlets
exist, all neutrino data can be explained at tree-level. The LHC
phenomenology of the $\mu\nu{}SSM$ is similar to bilinear $\rpv$ as
far as decay branching ratios of the LSP are concerned. Correlations
with neutrino angles exist unless (a) there are 3 singlets and (b) the
lightest singlet gives a sub-dominant contribution to the effective
neutrino mass matrix.  Decay lengths of the lightest neutralino, see
fig. (\ref{fig:length}), depend mostly on the mass of the neutralino,
once neutrino masses are fixed.  Since very light singlino-like
neutralinos are possible in this model, however, rather long decay
lengths are not excluded. This might make the search for $\rpv$ from
neutralino decays quite difficult at the LHC.  However, the model
contains also new (singlet) Higgs states, which should be light
whenever the singlinos are light. This offers the possibility to
search for $\rpv$ in the Higgs sector, see \cite{Bartl:2009an}.

\section{R-parity: LHC studies}

In the following we mainly focus on minimal supergravity-type models
denoted by BRpV--mSUGRA where (i) we impose mSUGRA relations to reduce
the number of R-parity conserving parameters, and (ii) we add BRpV
terms at the electroweak scale.
Hence the BRpV--mSUGRA model contains eleven free parameters, namely
\begin{equation}
m_0\,,\, m_{1/2}\,,\, \tan\beta\,,\, {\mathrm{sign}}(\mu)\,,\,
A_0 \,,\,
\epsilon_i \: {\mathrm{, and}}\,\, \Lambda_i\,,
\end{equation}
where $m_{1/2}$ and $m_0$ are the common gaugino mass and scalar soft
SUSY breaking masses at the unification scale, $A_0$ is the common
tri-linear term, and $\tan\beta$ is the ratio between the Higgs field
vevs. We trade the soft parameters $B_i$ by the
``alignment'' parameters $\Lambda_i=\epsilon_iv_d+\mu v_i$ which are
directly related to the neutrino--neutralino
properties~\cite{Hirsch:2000ef}.

In order to fit current neutrino oscillation data, the effective
strength of R--parity violation must be small.
This implies that supersymmetric particle spectra are expected to be
the same as in the conventional R-conserving model, and that processes
involving single production of SUSY states~\cite{nogueira:1990wz} are
negligible at the LHC, thanks to the required smallness of R--parity
violation.
Similarly, processes such as $b \to s\gamma$ and g-2 are essentially
the same in BRpV--mSUGRA as in mSUGRA and hence the resulting
constraints for the latter still hold. The smallness of R--parity
violation also implies that the study of charge breaking minima in the
broken R-parity minimal supersymmetric standard model leads similar
results as the conventional model~\cite{Hirsch:2004hr}.
Last, but not least, SUSY particle pair-production cross sections are
expected to be the same as in the conventional model. Using this one
may perform a robustness check of the supergravity parameter reach
estimates against the presence of ``perturbative'' BRpV terms.
The basic difference is that in the BRpV--mSUGRA scenario the lightest
supersymmetric particle is no longer stable and, thus, decays
typically inside the detector.

\begin{figure}[h]
  \begin{center}
  \includegraphics[width=6cm]{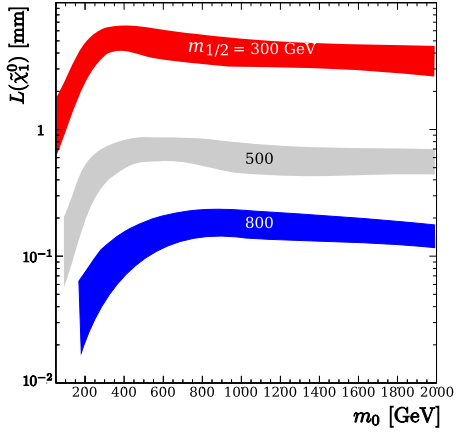}
  \end{center}
  \vspace*{-8mm}
  \caption{ $\tilde\chi_1^0$ decay length versus $m_0$ for $A_0=-100$
    GeV, $\tan\beta=10$, $\mu > 0$, and several values of $m_{1/2}$.
    The widths of the three (colored) bands around
    $m_{1/2}=300,~500,~800$ GeV correspond to the variation of the
    BRpV parameters in such a way that the neutrino masses and mixing
    angles fit the required values within $3\sigma$, taken from
    Ref.~\cite{deCampos:2007bn}. }
\label{fig:ldec}
\end{figure}

\subsection{LSP decays}

As mentioned above the BRpV interactions induce sneutrino
vevs. One--loop radiative corrections are needed to explain
consistently the neutrino data
~\cite{Hirsch:2000ef,Diaz:2003as,Dedes:2006ni} (for a review and more
references see, e.~g. Ref.~\cite{Hirsch:2004he}) and it has been shown
in Refs.~\cite{Hirsch:2000ef,Diaz:2003as} that the neutrino masses and
mixings are approximately given by\\[-1cm]

\begin{equation}
 \begin{array}{lll}
 \Delta m_{12}^2\propto |\vec{\epsilon}|;
&\Delta m_{23}^2\propto |\vec{\Lambda}|&
\\[.2cm]
 \tan^2\theta_{12}\sim \frac{\epsilon_1^2}{\epsilon_2^2};
 &\tan^2\theta_{13}\approx \frac{\Lambda_1^2}{\Lambda_2^2+\Lambda_3^2};
 & \tan^2\theta_{23}\approx  \frac{\Lambda_2^2}{\Lambda_3^2}
 \end{array}
%\label{nuap}
\nonumber
\end{equation}
where we denoted
$|\vec{\Lambda}|=\sqrt{\Lambda_1^2+\Lambda_2^2+\Lambda_3^2}$ and
similarly for $|\vec{\epsilon}|$.

\begin{figure}[!ht]
 \begin{center}
        \includegraphics[width=80mm,height=70mm]{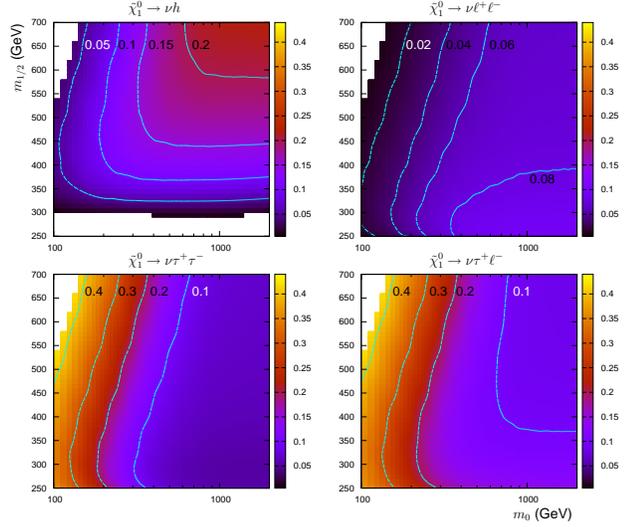}
 \end{center}
 \caption{Lightest neutralino branching ratios as a function of $m_0$
   and $m_{1/2}$ for $A_0=-100$ GeV, $\tan\beta=10$, and $\mu>0$. The
   upper left (right) panel presents the branching ratio into $\nu h$
   ($\nu\ell^+\ell^-$) while the lower left (right) panel is for
   $\nu\tau^+\tau^-$ ($\nu\ell^\pm\tau^\mp$),
   see~\cite{deCampos:2008re,deCampos:2008ic}. }
\label{fig:brlep}
\end{figure}

Apart from generating neutrino masses, neutralino--neutrino mixing
also leads to decay of the LSP into Standard Model particles.
While the BRPV parameters $\epsilon_i$ and $\Lambda_i$ have no effect
in the production cross sections of supersymmetric states, they
determine the LSP decay properties. As an example, the decay length is
illustrated in Fig.~\ref{fig:ldec}.
Concerning the LSP decay modes, the main decay channels of the
lightest neutralino are
$\tilde{\chi}^0_1 %\to \nu Z^*
\to \nu \ell^+ \ell^-$ with $\ell =e$, $\mu$ denoted by $\ell \ell$;
$\tilde{\chi}^0_1 %\to \nu Z^*
\to \nu \tau^+ \tau^-$, called $\tau \tau$;
$\tilde{\chi}^0_1 %\to \tau W^*
\to \tau \nu  \ell$, called $\tau \ell$.
$\tilde{\chi}^0_1 %\to \nu Z^*
\to \nu q \bar{q}$ denoted $jj$;
$\tilde{\chi}^0_1 %\to \tau W^*
\to \tau q^\prime \bar{q}$, called $\tau jj$;
$\tilde{\chi}^0_1 %\to \ell W^*
\to \ell q^\prime \bar{q}$, called $\ell jj$;
$\tilde{\chi}^0_1 %\to \nu Z^*
\to \nu b \bar{b}$, which we denote by $bb$;
$\tilde{\chi}^0_1 %\to \nu h^*
\to \nu b \bar{b}$, which we denote by $bb$;
$\tilde{\chi}^0_1 %\to \nu Z^*
\to \nu \nu \nu$.

We depict in Figure \ref{fig:brlep} the main branching ratios of the
lightest neutralino in the $m_0 \otimes m_{1/2}$ plane. As can be seen
the leptonic decay $\nu \ell^+ \ell^-$ with $\ell^\pm=\mu^\pm$ is of
the order of a few to 10\%, while the decay modes $e^\pm$, $\nu \tau^+
\tau^-$ and $\nu \tau^\pm \ell^\mp$ vary from $\approx 40$\% at small
$m_0$ to a few percent at large $m_0$. At moderate and large $m_0$,
these decays originate from the lightest neutralino decaying into the
two--body modes $\tau^\pm W^\mp$, $\mu^\pm W^\mp$ and $\nu Z$,
followed by the leptonic decay of the weak gauge bosons. In general,
semi-leptonic decays of the LSP are suppressed at small $m_0$
but dominate at large $m_0$ ~\cite{deCampos:2008re}.

\subsection{Three and multi-lepton channels}

In ref.~\cite{deCampos:2007bn} a comparison has been performed between
the reach of LHC for R-parity violating SUSY using the same cuts as
for R-parity conserving models \cite{Baer:2000bs}.
The main topologies are: Inclusive jets and missing transverse
momentum; Zero lepton, jets and missing transverse momentum; One
lepton, jets and missing transverse momentum; Opposite sign lepton
pair, jets and missing transverse momentum; Same sign lepton pair,
jets and missing transverse momentum; Tri-leptons, jets and missing
transverse momentum; Multi-leptons, jets and missing transverse
momentum.  Due to the reduced missing energy the all-inclusive channel
will have a reduced reach in the parameter space. However, the decays
of the neutralino increase the multiplicities of the multi-lepton
channel. As an example we display in Fig.~\ref{fig:di:3lep100} the LHC
reach in the three-- and multi--lepton channels with/without R--parity
conservation for an integrated luminosity of 100 fb$^{-1}$. These
constitute the best standard channels for BRpV discovery.
\begin{figure}[!ht]
  \begin{center}
 \includegraphics[width=6cm]{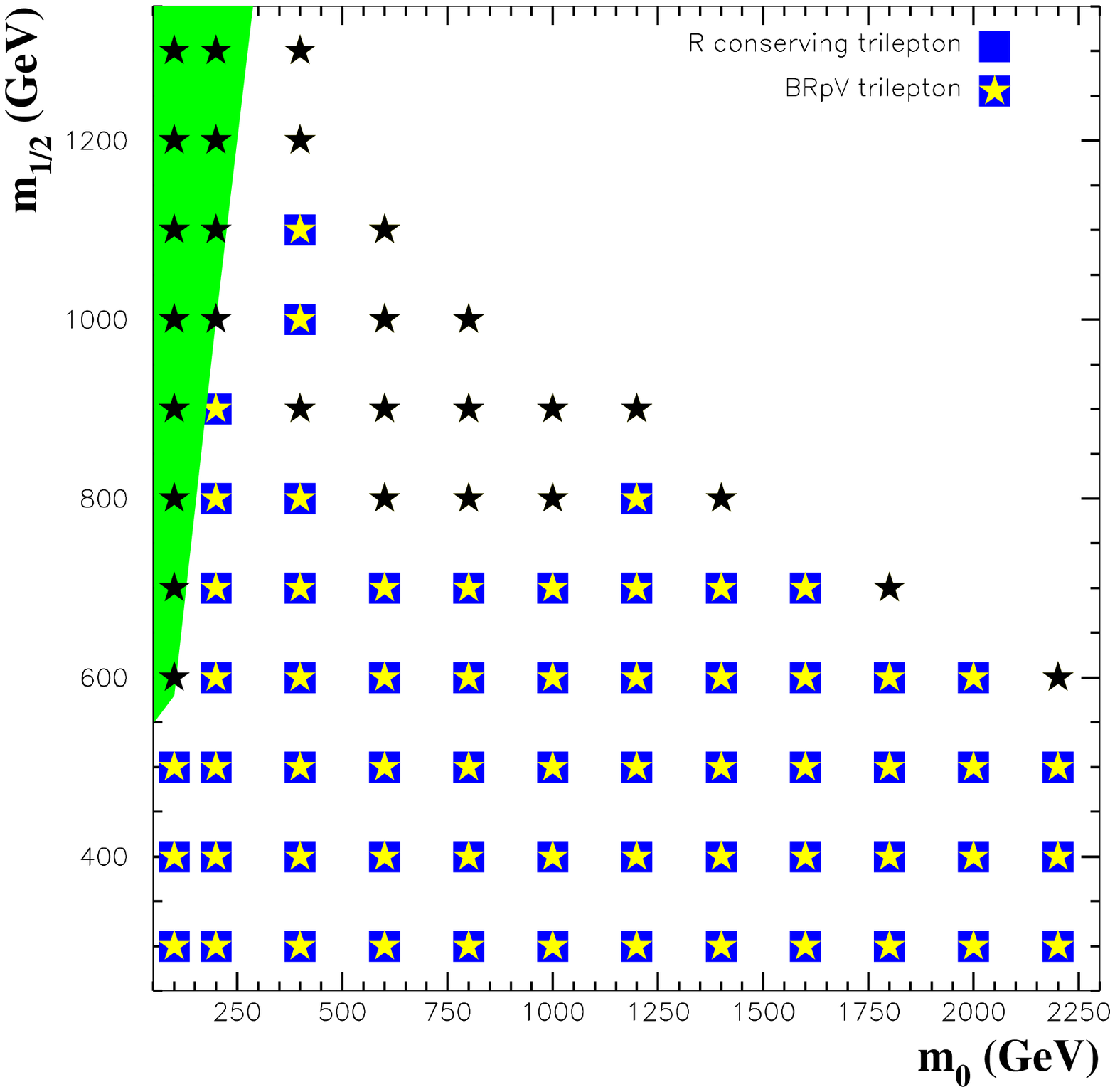}
 \includegraphics[width=6cm]{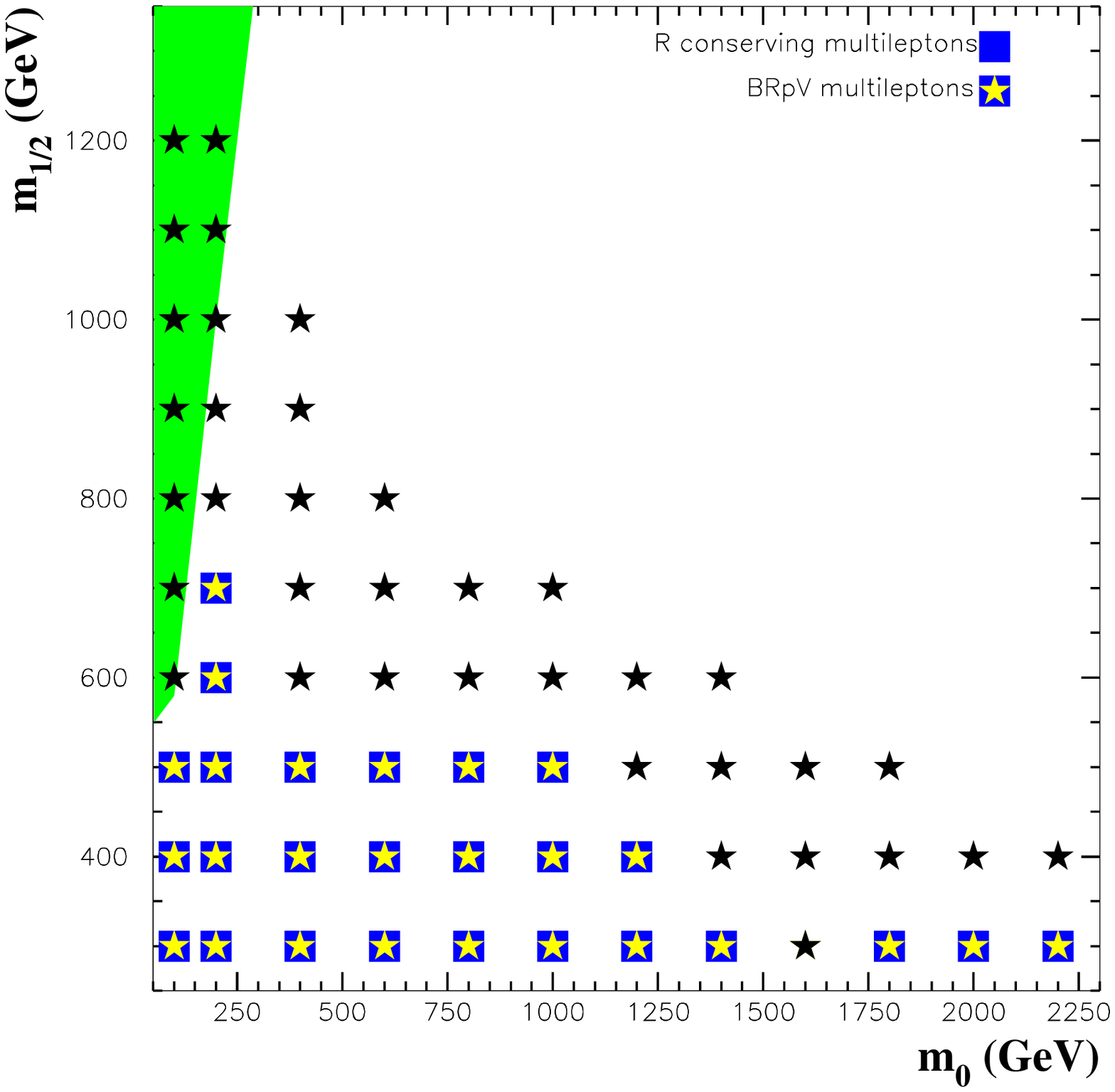}
  \end{center}
  \vspace*{-8mm}
  \caption{LHC discovery potential in the three lepton channel (top
    panel) and the multi-lepton one (bottom panel) for the parameters
    used in Fig.~\ref{fig:brlep} and an integrated luminosity of $100$
    fb$^{-1}$,  taken from
    Ref.~\cite{deCampos:2007bn}. }
\label{fig:di:3lep100}
\end{figure}

\subsection{Displaced LSP decays}

The rather large decay length of the neutralino is quite useful as
this topology has little, if any, background expected at the LHC. This
feature has been exploited in
ref.~\cite{deCampos:2007bn,deCampos:2008re}, where a comparison has
been performed between the reach of LHC for R-parity violating SUSY
including explicitly the displaced vertex topologies.

In Figure~\ref{fig:ms} we present the displaced vertex reach.  As one
can see form this figure, the LHC will be able to look for the
displaced vertex signal up to $m_{1/2} \sim $800 (1000) GeV
($m_{\chi_{1}^{0}}\sim 340$ (430) GeV) for a large range of $m_0$
values and an integrated luminosity of 10 (100) fb$^{-1}$. Notice that
the reach in this channel is rather independent of $m_0$ as expected
from Fig.\ \ref{fig:ldec}.
However, this signal for BRpV--mSUGRA disappears in the region where
the stau is the LSP due to its rather short lifetime.

\begin{figure}[th]
  \begin{center}
  \includegraphics[width=6cm]{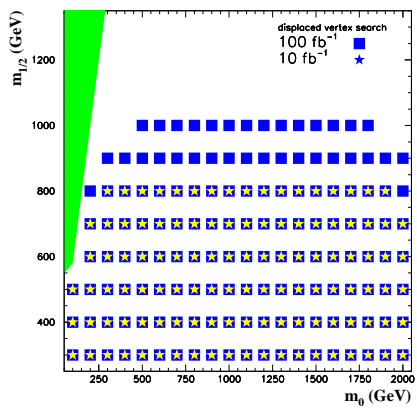}
  \end{center}
  \vspace*{-8mm}
  \caption{Discovery reach for displaced vertices channel in the
    $m_{0}\otimes m_{1/2}$ plane for $\tan\beta = 10$, $\mu > 0$,
    $A_0=-100$ GeV. The stars (squares) stand for points where there
    are more than 5 displaced vertex signal events for an integrated
    luminosity of 10 (100) fb$^{-1}$.  The marked grey (green) area on
    the left upper corner is the region where the stau is the LSP and
    the displaced vertex signal disappears.
    Points already excluded by LEP and Tevatron searches are below the
    $m_{1/2}$ values depicted in this figure,  taken from
    Ref.~\cite{deCampos:2007bn}. }
 \label{fig:ms}
\end{figure}
In ref.~\cite{deCampos:2008re} the possibilities of LHCb have been
investigated and compared to ones of the ATLAS and CMS detectors.

\subsection{Displaced b-jets from Higgs  decay}

Here we discuss a tantalizing possibility, namely a double discovery
at the LHC: (i) find evidence for supersymmetry, and (ii) uncover the
Higgs boson.
Indeed, from the top left panel of Fig~\ref{fig:brlep} one sees that
the LSP $\tilde{\chi}_1^0$ may have a sizeable branching ratio up to
22\% into the channel $\nu h$ where $h$ is the lightest Higgs boson.
This would lead to displaced vertices containing two b--jets as a
characteristic signature for Higgs production at the
LHC~\cite{deCampos:2008ic}.

This possibility has been investigated quantitatively in the simplest
BRpV--mSUGRA model, which accounts for the observed pattern of
neutrino masses and mixings seen in neutrino oscillation experiments.

The displaced vertex signal implies that also LHCb will have good
sensitivity for such scenarios in particular in case of final states
containing muons such as $\tilde{\chi}^0_1 \to \nu \mu^+ \mu^-$.%
Figure~\ref{fig:reach} demonstrates that the ATLAS and CMS experiments
will be able to look for the signal up to $M_{1/2} \sim 700$ $(900)$
GeV for a LHC integrated luminosity of 10 (100) fb$^{-1}$.
The hatched region in Fig.~\ref{fig:reach} indicates the LHCb reach
for 10 fb$^{-1}$. Due to the strong cut on the pseudo--rapidity
required by this experiment the reach for 2 fb$^{-1}$ is severely
depleted and only a small region of the parameter space is covered.

In the analyzes discussed so far the main production channels of the
neutralinos have been via cascade decays of squarks and gluinos.
However, there are regions in parameter space where signals of
neutralinos and charginos from gauge boson fusion are sizable
\cite{Datta:2001cy}.

 \begin{figure}[htb]
  \begin{center}
\includegraphics[width=6cm]{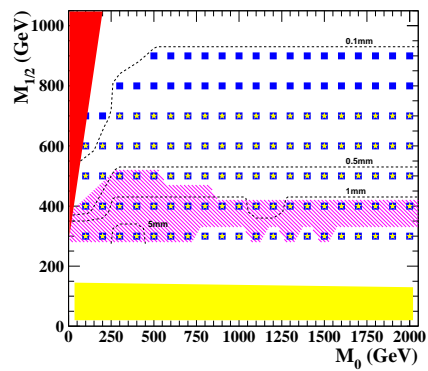}
  \end{center}
  \vspace*{-8mm} \vglue -.5cm
  \caption{LHC reach for Higgs search in displaced vertices for the
    BRpV--mSUGRA model in the plane $M_{1/2}\otimes M_0$ assuming
    $\tan\beta = 10$, $A_0=-100$ GeV, and $\mu > 0$. The yellow stars
    (blue squares) represent the reach for an integrated luminosity of
    10 (100) fb$^{-1}$ while the hatched region corresponds to the
    reach of the LHCb experiment for an integrated luminosity of 10
    fb$^{-1}$.  The (yellow) shaded region in the bottom stands for
    points excluded by direct LEP searches, while the (red)
    upper--left area represents a region where the stau is the
    LSP. Note that the black lines delimit different regimes of LSP
    decay length, taken from Ref.~\cite{deCampos:2008ic}.}
\label{fig:reach} \vglue -.5cm
\end{figure}

\subsection{Discussion}

Bilinear $R$-parity violation is essentially equivalent to tri-linear
$R$-parity breaking with the superpotential
\begin{eqnarray}
W_{tri} = {\textstyle \frac{1}{2}}
 \lambda_{ijk}  \widehat L_i  \widehat L_j\widehat E_k
 + \lambda_{ijk}' \widehat L_i\widehat Q_j\widehat D_k
\end{eqnarray}
where the tri-linear couplings have the following structures
\begin{eqnarray}
\lambda_{ijk} \simeq \frac{\epsilon_i}{\mu} h^{jk}_E \,\,,\,\,
\lambda_{ijk}' \simeq \frac{\epsilon_i}{\mu} h^{jk}_D
\end{eqnarray}
Obviously the phenomenology will be very similar if tri-linear
$R$-parity violation is close to this structure. In the case of
significant deviations from this structure is realized, one
gets new interesting signatures. For example there exists light
stau LSP scenarios where $\tilde \tau_1$ decays dominantly  via
4-body decays such as $ \tilde \tau_1 \to \tau^- \mu^- u \bar{d}$
with long lifetimes leading to displaced vertices \cite{Dreiner:2008rv}.
Another interesting signals are the resonant production of sleptons
as discussed in \cite{Dreiner:2000vf,Moreau:2000bs} or
associated production of single sleptons with $t$-quarks
\cite{Bernhardt:2008mz}.

An interesting question is to which extent one can measure deviations
from the hierarchical structure above, e.g.~the coupling $\lambda_{211}'$
can still be of order $0.1$. It has been shown in \cite{Choudhury:2002av}
that in such a case one LHC will be able to measure such couplings of such
a strength with an accuracy of about 10\%.

\section{Conclusions}
\label{sec:conclusions}

We have considered two broad classes of models where neutrino masses
arise at the TeV scale.  In the simplest seesaw type I scenarios, TeV
scale right handed neutrinos are typically accommodated through very
small Yukawa couplings to account for the lightness of neutrinos. This
has the advantage that it may produce displaced vertex signatures for
the TeV states underlying neutrino mass generation. An attractive
version of such theories has $W_R$ and/or $Z'$ bosons, which can be
produced at LHC and lead to like-sign dilepton, as well as trilepton
signals that can be used to search for them. We have also presented
TeV scale type II models which also have characteristic collider
signals. An advantage of these models in contrast to type I case is
that collider signals can throw direct light on the neutrino masses
and mixings.

Generic collider-accessible seesaw scenarios require small couplings
(say $\sim 10^{-5.5}$), although this is technically quite
natural. Some unified gauge models based on SO(10) may naturally
accommodate a TeV-scale RH neutrinos and $Z'$ without conflict with
gauge coupling unification nor neutrino masses. Another exception is
provided by SUSY left-right seesaw models, where accidental symmetries
may lead to TeV scale doubly charged Higgs and Higgsino fields even
though when seesaw scale lies at $10^{10}$ GeV. These doubly charged
Higgs and Higgsinos couple only to the RH lepton and slepton fields
and can be pair produced at LHC via Z-mediated Drell-Yan diagrams.
In contrast, intrinsically low-scale seesaw models do not require tiny
Yukawa couplings to accomodate small neutrino masses. In this case one
does not expect displaced vertices, a fact which would require more
detailed simulation studies in order to establish the detectability of
the resulting signals.

The second class of models we have considered is based on the idea
that the origin of neutrino masses is intrinsically supersymmetric.
We have considered mainly the lightest neutralino, characteristic of
minimal supergravity, whose decays typically lead to displaced
vertices and branching ratio predictions that correlate with the
atmospheric mixing angle.
However, at a phenomenological level, if R-parity breaks any SUSY
state can be the LSP.
Given our current ignorance of the ultimate mechanism responsible for
supersymmetry breaking, all LSP possibilities should be regarded as
viable, so that staus, stop, chargino or even gravitino may be the LSP
and should be taken up seriously.
Again here it is likely that the decay lengths are short enough that
one typically looses the characteristic displaced vertex signal
arising from neutralino LSP decays predicted in mSUGRA, especially
when only 2-body decay channels exist.  As a result, detailed
detector simulations will be required.

To summarize our main points, the tell-tale signs of neutrino mass
generation at the TeV scale are:

\begin{itemize}

\item New gauge bosons $W_R$, $Z'$, which will lead to like-sign
dilepton with no missing $E_T$ or trileptons with missing $E_T$;

\item Doubly charge Higgs bosons which decay to various like-sign
dilepton channels;

\item No missing energy supersymmetric signals with displaced vertices
  due to LSP decays.

\item LSP decays correlating with the value of neutrino mixing angles
  which would then be redetermined at the LHC.

\end{itemize}

%\section*{Acknowledgments}
%
%Work supported by the US National Science Foundation under grant
%No. PHY-0652363, the European Union RTN UNILHC (PITN-GA-2009-237920),
%Spanish grants FPA2008-00319/FPA and PROMETEO/2009/091, by German
%Ministry of Education and Research (BMBF) under contract 05HT6WWA, and
%Colombian grant UdeA Sostenibilidad 2009-2010.

\renewcommand{\baselinestretch}{1}

% \bibliographystyle{h-physrev4}
% % \bibliography{valle-ref,bibt3,morisi-ref}%%,soko,snova,nu-rev06,parke-ref}
% %\bibliography{valle-ref,grimus}%%,soko,snova,nu-rev06,parke-ref}
% \bibliography{join_new,porod_refs}%%,soko,snova,nu-rev06,parke-ref}
%\end{document}

%%%%%%%%%%%%%%%%%%%%%%%%%%%%%%%%%%%%%%%%%%%%%%%%%%%%%%%%%%%%%%%%%%%%%%%%%%%%%%%%%%%%%%%%%%%%%%
%%%%%%%%%%%%%%%%%%%%%%%%%%%%%%%%%%%%%%%%%%%%%%%%%%%%%%%%%%%%%%%%%%%%%%%%%%%%%%%%%%%%%%%%%%%%%%
\chapter{Extra Dimensions}
\epigraphhead[20]{\epigraph{\large {\em M.~Carena, C.~Cs\'aki,
H.~Davoudiasl, U.~Haisch, K.~Kong, G.~Landsberg, R.~Mahbubani,
P.~Nath, M.~Neubert, E.~Pont\'{o}n, T.G.~Rizzo, J.~Santiago,
M.~Toharia, C.E.M.~Wagner}}{\large Hooman Davoudiasl (Convener)}}
%
%%%%%%%%%% espcrc2.tex %%%%%%%%%%
%
% $Id: espcrc2.tex 1.2 2000/07/24 09:12:51 spepping Exp spepping $
%
%\documentclass[fleqn,twoside]{article}
%\usepackage{espcrc2}
%\usepackage[square,comma,sort&compress]{natbib}
% change this to the following line for use with LaTeX2.09
% \documentstyle[twoside,fleqn,espcrc2]{article}
% if you want to include PostScript figures
%\usepackage{graphicx}
% if you have landscape tables
%\usepackage[figuresright]{rotating}
% put your own definitions here:
%   \newcommand{\cZ}{\cal{Z}}
%   \newtheorem{def}{Definition}[section]
%   ...

%\input{ED/newcommands}

%\title{Extra Dimensions}

%\begin{document}

%\begin{abstract}
%A brief survey of several extra dimensional scenarios and some of their key phenomenological features is presented.
%\vspace{1pc}
%\end{abstract}

%\maketitle

Models with spatial extra dimensions, proposed to address outstanding
questions near the weak scale,
have been the subject of much research for the past decade or so.
These models are expected to be testable, say, at TeV-scale colliders, and
provide a plethora of new and interesting signals.
The following provides a brief survey of the main features of several such
extra dimensional proposals, their current experimental status, and their
discovery prospects.  Section \ref{add} contains a  brief introduction to models with large extra dimensions and
section \ref{bh} focuses on black hole signals at high energy colliders, in models with weak scale quantum
gravity.  Sections \ref{s1z2} and \ref{ued} discuss models with 1/TeV compactification radii.  Collider
and precision aspects of warped 5D models are the subjects of sections \ref{rskk}, \ref{rsew}, \ref{rsf}, and \ref{radion}.
Higgssless models are briefly introduced in section \ref{hless}.

\section{A Short Overview of Large Extra Dimensions}\label{add}

{\it T.G.~Rizzo}\medskip

The scenario of ADD{\cite {ADD1}} was proposed as a solution to the hierarchy
problem, {\it i.e.}, why the Planck scale, $\bar M_{\rm Pl} \simeq 2.4 \cdot 10^{18}$ GeV, is so
much larger than the weak scale $\sim 1$ TeV. ADD propose that we live on a
brane while gravity is allowed to propagate in a (4+n)-dimensional `bulk'
which is, {\it e.g.}, an $n-$torus, $T^n$, with a volume $V_n=(2\pi R)^n$.
This brane is located at the origin in the extra dimensions, {\it i.e.}, {\bf y}=0.
Einstein's Equations tells us that the Planck scale we measure in 4D, $\bar M_{\rm Pl}$, is related
to the true (4+n)-dimensional fundamental scale, as $\bar M_{\rm Pl}^2=V_nM_*^{n+2}$.
$M_*$ can be thought of as the {\it true} Planck scale since it appears in
the higher dimensional Einstein-Hilbert action which
is assumed to describe General Relativity in (4+n)-dimensions. It is possible that $M_*$ could
be $\sim$ a few TeV thus `eliminating' the hierarchy problem. Knowing $\bar M_{\rm Pl}$ and assuming $M_*\sim$ a few TeV we can estimate the value of the radius $R$.
$n=1$ is excluded as then $R\sim 10^{8}$~km; for $n=2$ one obtains $R\sim 100~\mu$m
which is close to the limit of current table top experimental searches for
deviations from Newtonian Gravity. If $n$ is further increased $R$ becomes too
small to probe for directly. Note that if we believe in superstring
theory at high scales then we might expect that $n\leq 6$ or 7.

The Feynman rules for the KK gravitons of the ADD model can be found in
Ref.{\cite {big}}. Note that all of the states in
the graviton KK tower couple to SM matter on the brane with the same strength as does the
ordinary zero-mode graviton: ${\cal L}=-{1/{\bar M_{\rm Pl}}} \sum_n G^{\mu\nu}_n T_{\mu\nu}$
where $G^{\mu\nu}_n$ are the KK graviton fields and $T_{\mu\nu}$ is the stress-energy
tensor of the SM fields.

There are two important signatures for ADD extra dimensions at
colliders. [For a more detailed review see {\cite {JM}}].
The first signature is the emission of graviton KK tower states during the collision of two
SM particles. Consider, {\it e.g.}, the collision of $q \bar q$ to make a gluon; during this process
the SM fields can emit a tower KK graviton states.  These gravitons will then appear as missing
energy since the KK states are coupled too weakly to interact again in the detector. While {\it no one}
KK graviton state yields a large cross section the sum over many KKs does yield a potentially
large rate which only depends on the specific values of $n$ and $M_*$.
Present limits from LEP and the Tevatron require $M_* \geq 1-1.6$ TeV depending upon $n$.
Fig.~\ref {fig1} from Vacavant and Hinchliffe{\cite {vh}} shows the missing $E_T$
spectrum at the LHC assuming an integrated luminosity of 100 fb$^{-1}$ for
the process $pp\to$ jet plus missing energy in the SM and the excess from  ADD graviton emission
assuming different values of $n=\delta$ and $M_*=M_D$. Once the rather large SM
backgrounds are well understood this excess would be clearly visible in these cases.

\begin{figure}[htb]
\includegraphics[width=7.2cm,angle=0]{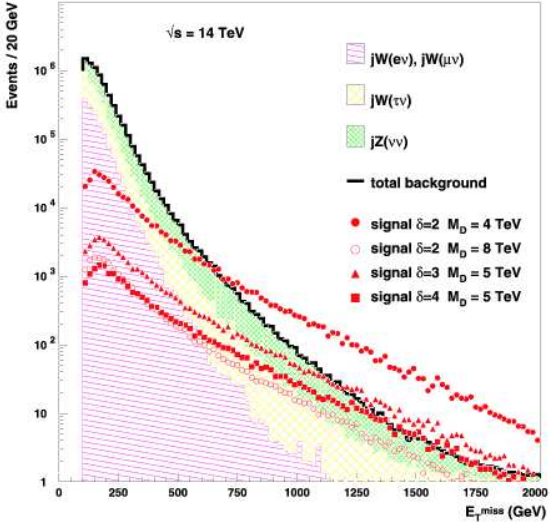}
\vspace*{-0.60cm}
\caption{Missing transverse energy spectrum for the monojet plus missing $E_T$
signature at the 14 TeV LHC assuming an integrated luminosity of 100 fb$^{-1}$ from
Ref{\cite {vh}}. Both the
SM backgrounds and the signal excesses from graviton emission in the ADD model
are shown. Here $M_D=M_*$ and $\delta=n$.}
\label{fig1}
\end{figure}

Another way to see the effect of the KK gravitons is to note that they can be exchanged
between colliding SM particles. This means that
processes such as $q\bar q \to \mu^+\mu^-$ can proceed through
graviton KK tower exchange as well as through the usual SM fields. The
amplitude for one KK intermediate state is quite tiny but we must again sum over all
their exchanges thus obtaining a potentially large result.
One problem with this is that this KK sum is divergent once $n>1$
as is the case here. The conventional approach to this problem
is to cut off the sum near $M_*$ yielding a set of
effective dim-8 operators. (The reasoning here is that we do not know the physics
beyond the scale $M_*$ as this requires an understanding of quantum gravity.)
In the notation of Hewett{\cite {big}}, these interactions are described by
${\cal L}={{4 \lambda }/{\Lambda_H^4}} T_{\mu\nu}^i T^{\mu\nu}_f$
where $\Lambda_H \sim M_*$ is the cutoff scale, $\lambda=\pm 1$ and $T^{\mu\nu}_{i,f}$
are the stress energy tensors for the SM fields in the initial and final state.
This is just a contact interaction albeit of dim-8 and with an
unconventional tensor structure owing to the spin-2 nature of the gravitons being
exchanged. Graviton exchange contributions to SM processes can lead to substantial
deviations from conventional expectations. An example of this at the LHC for the case
discussed above is seen in Fig.~\ref{fig2} from Hewett{\cite {big}}.

\begin{figure}[htb]
\includegraphics[width=5.2cm,angle=-90]{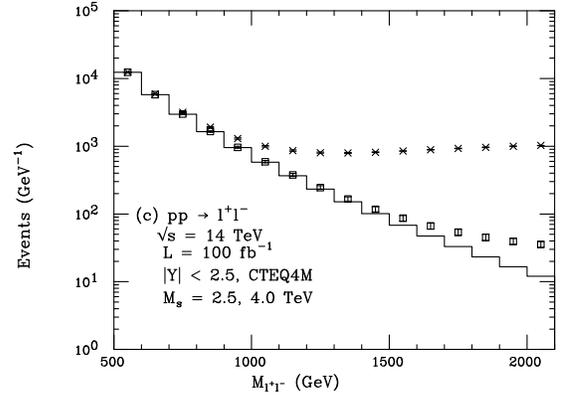}
%\vspace*{-1.50cm}
\caption{ADD contribution to the Drell-Yan process at the LHC.}
\label{fig2}
\end{figure}

It is also possible to constrain the ADD model in other ways, {\it e.g.}, the emission of ADD KK
gravitons can be constrained by astrophysical processes as reviewed in Ref.{\cite {JM}}.
These essentially disfavor values of $M_*$ less than several hundred TeV for $n=2$ but
yield significantly weaker bounds as $n$ increases.

\section{Mini-Black Holes at Modern Colliders}\label{bh}

{\it G.~Landsberg}\medskip

%{\it
%One of the most exciting consequences of models with low-scale gravity is the possibility to produce mini-black holes at the %LHC and future colliders. We discuss phenomenological and experimental aspects of searches for these mini-black holes.
%}

\subsection{Introduction}

Recently a new class of solutions to the infamous hierarchy problem
of the Standard Model (SM) has been proposed, by lowering the scale
of quantum gravity from the Planck energies of $M_{\rm Pl} \sim
10^{16}$ TeV to the electroweak symmetry breaking scale $\sim
1$~TeV. The large extra dimensions model~\cite{ADD1,ADD} achieves
lowering the Planck scale by introducing several ($n$) compact extra
spatial dimensions, in which gravity can propagate. The Planck scale
is lowered in this multidimensional space as the apparent weakness
of gravity is not due to 1/$M_{\rm Pl}^2$ suppression but due to the
enormous volume of extra-dimensional space, thus allowing the
fundamental $(4+n)$-dimensional Planck scale to be of the order of 1
TeV. The Randall-Sundrum model~\cite{RS} embeds the SM in the 5D
anti-deSitter space-time with the ``warped'' metric. The suppression
of the Planck scale to the EWSB energies is the achieved due to
exponential suppression of Planck-scale operators due to the above
warp factor.

It has been suggested~\cite{BHearly} that if the center-of-mass
energy of two colliding particles exceeds the fundamental Planck
scale, a mini-black hole (BH) with the mass of the order of the
collision energy could be produced. More recently~\cite{dlgt}, this
idea has been developed and quantified by showing that the cross
section for BH production is given by the geometrical cross section
of its event horizon. Thus, in both the ADD and RS models, black
holes can be potentially produced with high rates at the LHC and
future multi-TeV colliders.

\subsection{Mini-black hole production and decay}

Consider two partons with the center-of-mass energy $\sqrt{\hat s} =
M_{\rm BH}$ colliding head-on. If the impact parameter of the collision is less than the (higher dimensional) Schwarzschild radius, corresponding to this energy, a black hole (BH) with the mass $M_{\rm BH}$ is formed. Therefore the total cross section of black hole production in particle collisions can be estimated from pure geometrical arguments and is of order $\pi R_S^2$, where $R_S$ is the Schwarzschild radius of a multidimnesional black hole, given by~\cite{mp}. BH production is expected to be a threshold phenomenon and the onset is expected to happen for a minimum black hole mass $\sim M_D$ (ADD) or $\Lambda_\pi$ (RS). (In what follows we will use $M_D$ to denote the scale of TeV gravity.) The total production cross section above this threshold at the LHC can be estimated using standard parton luminosity approach and is given by~\cite{dlgt}:
$$
    \frac{d\sigma(pp \to \mbox{BH} + X)}{dM_{\rm BH}} =
    \frac{dL}{dM_{\rm BH}} \hat{\sigma}(ab \to \mbox{BH})
    \left|_{\hat{s}=M^2_{\rm BH}}\right.,
$$
where the parton-level cross section is given by:
$$
    \hat\sigma \approx \pi R_S^2 = \frac{1}{M_D^2}
    \left[
      \frac{M_{\rm BH}}{M_D}
      \left(
        \frac{8\Gamma\left(\frac{n+3}{2}\right)}{n+2}
      \right)
    \right]^\frac{2}{n+1},
$$
and the parton luminosity $dL/dM_{\rm BH}$ is defined as the sum over
all the types of initial partons:
$$
    \frac{dL}{dM_{\rm BH}} = \frac{2 M_{\rm BH}}{s}
    \sum_{a,b} \int_{M^2_{\rm BH}/s}^1
    \frac{dx_a}{x_a} f_a(x_a) f_b(\frac{M^2_{\rm BH}}{s x_a}),
$$
and $f_i(x_i)$ are the parton distribution functions. The cross section ranges between 15 nb and 1 pb for the Planck scale between 1 TeV and 5 TeV and minimum BH mass equal to the Planck scale, and varies by less than a factor of two for $n$ between 1 (RS) and 7. Note that this cross section is comparable with, e.g., $t\bar t$ production cross section, which result in $\sim 1$~Hz signal event rate at the nominal LHC luminosity, thus potentially qualifying the LHC as a BH factory.

Once produced, mini black holes quickly ($\sim 10^{-26}$~s) evaporate via Hawking
radiation~\cite{Hawking} with a characteristic temperature given by~\cite{dlgt}:
$$
    T_H = M_D
    \left(
      \frac{M_D}{M_{\rm BH}}\frac{n+2}{8\Gamma\left(\frac{n+3}{2}\right)}
    \right)^\frac{1}{n+1}\frac{n+1}{4\sqrt{\pi}} = \frac{n+1}{4\pi R_S}
$$
of $\sim 100$~GeV. The average multiplicity of particles
produced in the process of BH evaporation is given by~\cite{dlgt} and
is of the order of half-a-dozen for typical BH masses accessible
at the LHC. Since gravitational coupling is flavor-blind, a BH emits all the
$\approx 120$ SM particle and antiparticle degrees of freedom with
roughly equal probability. Accounting for color and spin, we expect
$\approx 75\%$ of particles produced in BH decays to be quarks and gluons,
$\approx 10\%$ charged leptons, $\approx 5\%$ neutrinos, and $\approx 5\%$
photons or $W/Z$ bosons, each carrying hundreds of GeV of energy.

A relatively large fraction of prompt and energetic photons, electrons,
and muons expected in the high-multiplicity BH decays would
make it possible to select pure samples of BH events, which are also
easy to trigger on~\cite{dlgt}. The reach of a simple
counting experiment extends up to $M_D \approx 9$~TeV ($n=2$--7),
for which one would expect to see a handful of BH events with negligible
background.

This simple picture is modified in general relativity (GR) if the
black hole has an initial spin or quantum numbers different from
those of vacuum. The process of evaporation generally expected to
consist of three stages. The first stage, known as {\it balding\/},
is when the BH rapidly loses its non-trivial quantum numbers (color,
electric charge, etc.); the second, {\it spindown} stage is when the
black hole loses its angular momentum; at the final, Hawking stage
the BH loses its mass via black-body radiation at Hawking
temperature. This last stage may terminate via formation of a stable
remnant with the mass $\sim M_D$. In GR, the emissivities of the
particles with different spin by a BH are somewhat different,
especially at the spindown stage. The emissivity is parameterized as
``grey body factors,'' which modify the perfect black-body spectrum.
Generally grey-body factors are known for spin 0, 1/2, 1, and 2
particles for non-rotating black holes and for spin 0, 1/2, and 1
particles for rotating ones. For a review of grey-body factors, see,
e.g., Ref.~\cite{BHreviews}. The caveat is that all these GR-based
calculations can be relied upon only for $M_{\rm BH} \gg M_D$, which is
hardly the case for the mini-BH at the LHC. For light black holes
quantum corrections become more and more important, and eventually
would dominate the classical picture. Hence, we do not focus on
these corrections for the purpose of this brief review.

\subsection{Monte Carlo generators}

A number of Monte Carlo generators are available nowadays for
studies of mini-black hole production and decay. The original
generator TRUENOIR~\cite{TRUENOIR} captured most basic aspects of
the black hole phenomenology. A number of advanced modern generators
have appeared since: CHARYBDIS~\cite{CHARYBDIS},
BLACKMAX~\cite{BLACKMAX}, and CATFISH~\cite{CATFISH}. They are
capable of simulating fine properties of black holes, such as spin,
back-reaction, effects of the brane tension, grey-body factors, and
a possibility of a sub-Planckian remnant. While all these effects
are incorporated using classical GR, expected to be modified
drastically for black holes with the mass close to the Planck scale,
they still may be used to study possible modifications of the final
state particle spectra and other aspects of the black hole
production and decay.

\subsection{Experimental studies}

No dedicated searches for black holes have been carried so far. The
Tevatron energy is not sufficient to produce black holes, given
current limits on the fundamental Planck scale coming from the other
way of searching for low-scale gravity (see, e.g., Ref.~\cite{ED}
for a recent review of current constraints). Moreover, generic
searches for high-$p_T$ phenomena at the Tevatron~\cite{generic} did
not reveal any anomalies in the multijet or lepton/photon+jet final
states.

A number of sensitivity studies have been performed for the
LHC~\cite{ATLASBH,CMSBH}. It is anticipated that intense searches
for this phenomenon will start with the first collisions at the LHC
expected later this year.

\section{On the Possible Observation of KK Excitations of SM states at the LHC}
\label{s1z2}

{\it P.~Nath}\medskip

\subsection{Introduction}

The basis model in D=5 (and in extra dimension $d=1$) is
 \begin{eqnarray}
L_5& = & -\frac{1}{4} F_{MN}F^{MN}
-(D_MH)^{\dagger}(D^MH) \nonumber\\
& & -\bar\psi\frac{1}{i}\Gamma^{\mu} D_{\mu}\psi-V(H)+\ldots .
\end{eqnarray}
Here $F^{MN}$ is the field strength of the gauge bosons $A^M$ where  $M,N$ run over $0,1,2,3,4$,
 $D_M$ is the gauge covariant derivative, and $H$
 is the Higgs doublet in 5D.  The theory is compactified on $S^1/Z_2$ with the radius of compactification
 $R=1/M_R$.   It is assumed that the gauge fields and the Higgs fields lie in the bulk while the quark and
 the lepton fields
  lie on the 4D wall. Breaking of the electroweak symmetry in 5D\cite{Nath:1999fs} and the combination of compactification
  and of the breaking of the electroweak symmetry lead to Kaluza-Klein (KK) modes for the photon, for the W boson and for the Z boson.
  Thus the decomposition on  $S^1/Z_2$ indicates  mass terms for the W bosons
 of
 \begin{equation}
 m_W^2+n^2/R^2,  ~n=0,1,2,..,\infty,
 \end{equation}
where the first term arises from spontaneous breaking of the
electro-weak symmetry and the second term arises from the
compactification.
Very similar relations hold for the KK excitations of the
Z boson and for the Higgs boson.
  The 4D gauge fields $A_{\mu}$   and their KK modes $A_{\mu}^n$
  have gauge couplings of the form\cite{Nath:1999fs}
  \begin{equation}
{\it L_{int}}=g_ij_i^{\mu}(A_{\mu i}+\sqrt 2\sum_{n=1}^{\infty} A_{\mu i}^n).
\end{equation}
One may note that the KK modes couple more strongly with the source
than the corresponding zero modes.

  \subsection{Precision constraints}
  The existence of the KK modes have important implications for precision physics.
  Thus, e.g., the exchange of the KK modes of the W boson give
  corrections\cite{Nath:1999fs,Marciano:1999ih}     to the
  Fermi constant which is one of the most accurately measured quantities in physics.
  Thus, for example, for the case of one extra dimension the KK excitations lead to
a correction to the Fermi constant  so that\cite{Nath:1999fs}
\begin{equation}
G_F^{eff}= G_F^{SM}  (1+ \frac{\pi^2}{3} \frac{M_W^2}{M_R^2}).
\end{equation}
  Using the precision data on $G_F^{eff}$ (and identifying  it as the experimental value) one can put
  a limit on $M_R$ so that $M_R>1.6$ TeV $(90\%)$.
  Other  precision electroweak parameters such as $(g_{\mu}-2)$ are also
  affected by the extra dimension\cite{Nath:1999aa}.
   The constraints on $d=2$ and higher dimensions
  depend strongly on the nature of the compactification and they can vary rather significantly.

  Next we discuss the implications of the KK excitations at the
  LHC\cite{Antoniadis:1999bq,Nath:1999mw,Rizzo:1999br}.
    There are a number of interesting
 signatures that arise from the KK excitations such as\cite{Nath:1999mw}
 \begin{equation}
 pp\to (l^+l^-, l^{\pm}\nu_{l}, jj) +X.
 \end{equation}
 The dilepton production via the Drell-Yan process is one of the optimal channels for the
 discovery of
 the KK modes. An  analysis of this phenomenon  is given in Fig.(\ref{smkk1}) where
 a plot of  $d\sigma/{dm_{ll}}$ vs the
dilepton invariant mass  $m_{ll}$ is exhibited for the case $d=1$.
 The analysis shows clear peaks from the KK resonances which may be compared with
 the SM background which is rather smooth. It is to be noticed that the KK resonances
 are not of simple Breit-Wigner type but rather, distorted ones. The distortion is due to
 interference  effects arising from the exchange of  the standard model spin 1 bosons, i.e.,
  $\gamma$  and Z boson, and their KK excitations.

In Fig.(\ref{smkk2}) an analysis is given of
$d\sigma_{ll}/dm_{ll}$ vs $m_{ll}$ for compactifications for both  the d=1 case
and for the  d=2 case. For the d=2 case, the analysis for two different types of orbifolding
is shown. One of these is the $Z_2\times Z_2$ orbifolding while the other is $Z_3$ orbifolding.
The analysis shows that not only can one discriminate between $d=1$ and $d=2$
compactifications but also among different types of compactifications for the $d=2$ case.
\begin{figure}[t]
\begin{center}
\includegraphics[width=5.5cm,height=5.5cm,angle=270]{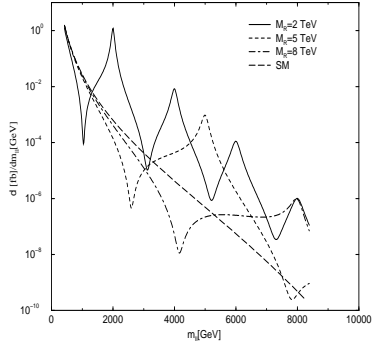}
\caption{\small
 $d\sigma_{ll}/dm_{ll}$ vs $m_{ll}$ when $M_R=2, 5, 8$ TeV. The
 SM case is  shown for comparison.
From  \cite{Nath:1999mw}. }
 \label{smkk1}
  \end{center}
\end{figure}
\begin{figure}[t]
\begin{center}
\includegraphics[width=5.5cm,height=5.5cm,angle=270]{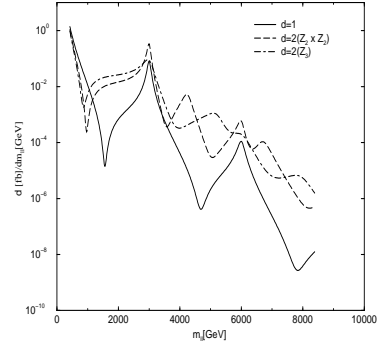}
\caption{\small
 A comparison of compactifications for  $d=1$ and $d=2$ with $Z_2\times Z_2$ and $Z_3$ orbifolding  in the
  $d\sigma_{ll}/dm_{ll}$ vs $m_{ll}$ plot when $M_R=3$ TeV.
  From \cite{Nath:1999mw}. }
 \label{smkk2}
  \end{center}
\end{figure}
     \subsection{Conclusion}
      The KK excitations of the Standard Model are constrained severely by the precision electroweak
      data. It is shown that the precision data on the Fermi constant puts a stringent bound
      on the scale of the extra dimension for  the case d=1.
       For compactifications with  $d=2$ the low energy effects of the KK modes
       depend on the precise nature of the compactification.
          One of the very clear signatures of the extra dimensions is the appearance of
      KK resonances in the Drell-Yan process with two leptons in the final state at the LHC.  Here one
      finds  resonances which are distorted Breit-Wigner.
      Further, the detailed features of  the resonances  contain
    information on the number of extra dimensions as well as on the type of compactification.

\section{Probing Universal Extra Dimensions at Colliders}
\label{ued}

{\it K.~Kong and R.~Mahbubani}\medskip

\subsection{One and Two Universal Extra Dimensions}
\label{sec:ued}

Models of Universal Extra Dimensions (UED)
\cite{Appelquist:2000nn} are characterized by Standard Model (SM)
fields that propagate throughout a flat bulk, i.e. along
all $x^{3+i}$ $(i=1,\ldots,N)$ spatial directions in a
$(4\!+\!i)$-dimensional theory.  In order to be consistent with
observations the extra
dimensions in such models must be compactified on a manifold of size
smaller than the smallest scale that has been resolved by
experiment. Due to space constraints we
shall limit our discussion to theories with one or two UEDs.
Extended reviews of the phenomenology of UED models can be
found in Refs.~\cite{Macesanu:2005jx}.

In order to implement chiral
fermions in $N\!\!=\!1$ UED, where the extra
dimension is compactified on a circle of radius $R$, the opposite
sides of the circle must be identified (the orbifold $S^1/Z_2$), as shown in
Fig.~\ref{fig:UED}(a).
Several possibilities exist for
compactification of $N\!\!=\!2$ UED in a manner that allows for
chiral fermions.  One of these, the
`chiral square' \cite{Dobrescu:2004zi},  shown
in Fig.~\ref{fig:UED}(b), is a square of side $R$ with adjacent
sides identified, also known as the orbifold $T^2/Z_4$.
%
%%%%%%%%%%%%%%%%%%% BEGIN FIGURE %%%%%%%%%%%%%%%%%%%
%\begin{center}
\begin{figure}[tb]
\centerline{\includegraphics[width=7.3cm]{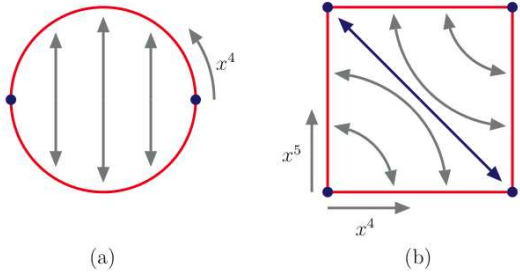}}
\vspace*{-1.2cm} \caption{(a) Compactification of $N\!\!=\!1$ UED on
a circle with opposite points identified. (b) Compactification of
$N\!\!=\!2$ UED on a square with adjacent sides identified (the
`chiral square'). The black arrows indicate the identification, and
the blue dots represent fixed (boundary) points.} \label{fig:UED}
\end{figure}
%\end{center}
%%%%%%%%%%%%% END OF FIGURE ################

The spectrum of UED models consists of an infinite tower of heavy KK
partners for each SM particle.   All KK particles
with a given ${\bf n}=\{n_i\}$, which enumerates the particles'
quantized extra-dimensional momenta, have squared masses ${\bf n}^2/R^2$, for ${\bf n}^2=\sum_{i=1}^N n_i^2$.
An important property of these models is a symmetry known as Kaluza-Klein
(KK) parity, whose conservation can be traced back to the geometrical
symmetries of the full theory and compactification.  It is defined for
a particle by $(-1)^n$ where the level number
  $n=\sum_{i=1}^Nn_i$, is the total number of units of extra-dimensional
momentum carried by that particular particle.  SM particles have $n
\!\!=\!0$ and hence positive KK parity, with alternating levels of
KK partners having even parity.  This symmetry accounts for
the stability of the lightest KK-odd particle (LKP), which
cannot decay into SM particles, and thus determines to a
large extent both the collider and astroparticle phenomenology of
models of UED.
\subsubsection{Mass Spectrum}
\label{sec:flatUEDmass}

Although the masses of the partner particles at each level are exactly
degenerate at leading order, they receive corrections from several
sources which lift this degeneracy. The largest contributions come
from one-loop mass renormalization effects due to SM
interactions in the bulk \cite{Cheng:2002iz,Ponton:2005kx}.  These are
logarithmically enhanced, both with respect to tree-level corrections
arising from electroweak symmetry breaking, and boundary term
contributions at the orbifold fixed points \cite{Cheng:2002iz}
(depicted as blue dots in Fig.~\ref{fig:UED}). The latter are
usually assumed to be negligible in minimal models of UED.
Due to the strength of the
color gauge coupling as well as the multiplicity of colored fermions,
radiative corrections are largest for colored particles (KK quarks and
gluon).

Fig. \ref{fig:signatures} contains a qualitative sketch of the
corrected mass spectrum of $N\!\!=\!1$ minimal UED.
The LKP, denoted by
the symbol $\gamma_1$, is a linear superposition of the KK modes of the hypercharge
gauge boson $B_1$ and
the neutral component of the  $SU(2)$ gauge boson $W^0_1$.

$N\!\!=\!2$ UED introduces to the spectrum a `spinless adjoint', which
is a partner particle with no analogue in the SM or
$N\!\!=\!1$. This spin-0 particle originates in the 6D
gauge boson, which has two
extra-dimensional polarizations, one of which gets `eaten'
by the 4D gauge boson in the effective theory.  It
is the remaining polarization that constitutes the
spinless adjoint, and there is one of these for each SM gauge
boson.  Spinless adjoints get negative
radiative contributions to their masses \cite{Ponton:2005kx}, resulting
in an LKP that is the spinless adjoint partner of the photon, $B_H$
(see Fig \ref{fig:signatures}).

\subsection{Collider signals}
\label{sec:flatUEDcol}

The collider phenomenology of minimal $N\!\!=\!1$ and $N\!\!=\!2$ UED
has been extensively investigated at linear colliders
\cite{Battaglia:2005zf,Freitas:2007rh}, as well as
hadron colliders \cite{Cheng:2002ab,Datta:2005zs,Burdman:2006gy,Dobrescu:2007xf}.
Both models contain an electrically neutral, weakly interacting LKP, which
escapes the detector, giving rise to a missing energy signal.
Since the total parton-level energy in the
collision is a priori unknown at hadron colliders, the presence of
LKPs must be inferred from an imbalance in the total transverse
momentum in the event.
\subsubsection{Pair-production of level-1 modes}

Due to KK parity conservation, level-1 KK modes are always produced in
pairs, subsequently undergoing cascade decays to the
LKP \cite{Cheng:2002ab,Dobrescu:2007xf}.
The main decay modes and products are illustrated in Fig.~\ref{fig:signatures}.
Typical signatures include a number of jets, leptons and photons, plus
missing energy, $\not \!\!\!\! E_T$. Note that cascade decay patterns in UED look very similar to those
arising in R-parity conserving supersymmetry \cite{Cheng:2002ab},
except the former generally have softer decay products, due to the
near-degeneracy of the spectrum at each level.
Possible ways to discriminate between them include invariant mass methods
\cite{Burns:2008cp}.
Due to the addition of the spinless adjoints, which decay mostly
to three-body final states as well as lengthening cascade decays in two extra
dimensions, $N\!\!=\!2$ UED yields events with
a very high lepton multiplicity, a smoking-gun signature for this model.
\begin{figure}[tb]
%\vspace{9pt}
\includegraphics[width=7.4cm]{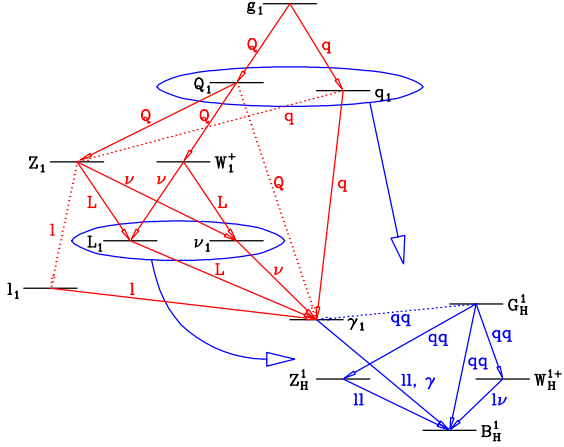}
\vspace*{-0.7cm}
\caption{Qualitative sketch of the level-1 KK spectroscopy depicting
the dominant (solid) and rare (dotted) transitions and
the corresponding decay product for minimal $N\!\!=\!1$ (red) and
additional decay modes for $N\!\!=\!2$ (in blue) UED.  Uppercase
$(Q,L)$ stand for $SU(2)_W$-doublet quarks and leptons respectively,
while lowercase $(q,l)$ represent their singlet partners.  Particles
with a subscript $H$ are `spinless adjoint' partners of the SM gauge
bosons, only present in the $N\!\!=\!2$ case.}
\label{fig:signatures}
\end{figure}

Which mode affords the best prospects for discovery depends on the
interplay between the predicted signal rates and the expected SM
background.  Since SM backgrounds are firmly under control at lepton
colliders, the most promising channels are typically those with the
largest signal rates, associated with production of the lightest
particles in the spectrum (level-1 leptons and electroweak gauge bosons).
In contrast, the dominant production at hadron colliders is through
strong interactions, making the largest cross-sections those of
colored KK particles, which typically decay through jets.
Unfortunately, the SM QCD backgrounds for these modes are significant;
there is therefore a substantial benefit to searching for any
leptons that accompany these jets. One example of an interesting process is pair
production of KK quarks, $Q_1$, which decay through $SU(2)$
KK gauge bosons $W^\pm_1$ and $Z_1$, and may yield up to 4 leptons
(up to 8 leptons, or 4 leptons plus 2 photons for $N=2$) plus missing energy.
The discovery reach for minimal $N\!\!=\!1$ UED in the $4\ell + \not \!\! \!E_T$ channel at the
Tevatron and LHC was discussed
in Ref. \cite{Cheng:2002ab}.  A CDF search in the multilepton channel,
based on 100 pb$^{-1}$ of Tevatron data yielded a lower limit on the
UED scale $R^{-1}$ of 280 GeV at the 95\% C.L. \cite{Lin:2005ix}.

\subsubsection{Production of level-2 modes}
Level-2 modes, which are KK-parity even, also give rise to promising signatures at
hadron colliders. For example, production of the level-2 gluon in $N\!\!=\!1$ UED,
in association with another level-2 colored particle yields $N j + \not \! \! \! E_T$,
a process with good prospects for discovery at the LHC \cite{Cheng:2002ab},
while single production appears as a dijet resonance without $\not \! \! \! E_T$.
Even more interesting is single production of level-2 gauge boson partners in
$N\!\!=\!2$ UED.  Their production cross sections are
larger than those in $N\!\!=\!1$ since their masses are smaller
by a factor of $\sqrt{2}$.  Moreover they have an enhanced branching
fraction to $t\bar{t}$ pairs, a distinctive signature that might even
be visible at the Tevatron \cite{Burdman:2006gy}.
\section{Signals of a Warped New Dimension at Colliders}\label{rskk}

{\it H.~Davoudiasl}\medskip

The Randall-Sundrum (RS) model \cite{RS} was introduced to explain the
hierarchy between the Standard Model (SM) Higgs mass $m_H\sim 100$~GeV and the
reduced Planck mass $\bar M_P\sim 10^{18}$~GeV.  This problem is a
manifestation of the quadratic sensitivity of $m_H$ to quantum
corrections from an arbitrarily high mass scale.

The RS background \cite{RS}  is a slice
of AdS$_5$ (5D spacetime with a negative cosmological constant),
bounded along the fifth dimension $y$ by two 4D Minkowski walls:
the UV brane at $y=0$ and IR brane at $y=\pi r_c$. The RS metric is given by
\begin{equation}
ds^2 = e^{-2 \sigma(y)} \eta_{\mu \nu} \,dx^\mu dx^\nu - dy^2,
\label{RSmetric}
\end{equation}
where $\sigma(y) = k y$ and $k$ is the 5D curvature scale.
One has $\bar M_P^2 \simeq M_5^3/k$, with $M_5$ the 5D fundamental
scale; naturalness implies $k\sim M_5\sim \bar M_P$.
Mass scales get exponentially
redshifted by $e^{-k r_c \pi}$ at the IR brane,
in this background.  Hence, if the Higgs is IR-brane-localized,
the hierarchy problem is resolved, even for a 5D Higgs mass of ${\cal O}(k)$,
as long as $k r_c \pi \approx 36$; $\tilde{k}\equiv e^{-k r_c \pi} k \sim$~TeV .  The size
of the fifth dimension can be stabilized at the required value
without extra fine-tuning \cite{Goldberger:1999uk}.
Warped 5D models discussed in
the following here are generally based on the above setup.

The most distinct signature of the original RS model is a tower
of spin-2 resonances, the Kaluza-Klein (KK) states $G^n$, $n\geq 1$, of the
5D graviton, with masses and couplings set by the
TeV scale.  The production and decay of the KK gravitons give rise to striking
signals at collider experiments \cite{Davoudiasl:1999jd}.
The Tevatron experiments CDF (2.3~fb$^{-1}$) \cite{Aaltonen:2008ah} and 
D0 (1~fb$^{-1}$) \cite{Abazov:2007ra}
have searched for $G^1$ in the original model.
Roughly speaking, the current data disfavors
a $G^1$ lighter than 300 (900)~GeV, for $k/\bar M_P = 0.01 (0.1)$, at 95\% confidence level.
With 100~fb$^{-1}$ and $k/\bar M_P = 0.1$, the ATLAS experiment
\cite{Allanach:2002gn} expects to be able to
discover $G^1$ of the original model, in the $e^+ e^-$
channel, up to a mass of 3.5~TeV.
The CMS reach is somewhat better (about 4~TeV),
in the di-muon channel \cite{CMS_RS}.

The SM gauge fields \cite{Davoudiasl:1999tf,Pomarol:1999ad}
and fermions \cite{Grossman:1999ra} can be moved to
the 5D bulk, leading to realistic 4D flavor patterns if light fermions are UV-localized
and heavy fermions are IR-localized \cite{Gherghetta:2000qt},
along the extra dimension.  In these setups, the KK couplings to light SM fields
({\it e.g.} light quarks, $\mu^\pm$), important for collider discoveries,
are suppressed, while the strongest KK couplings are
to heavy (IR-localized) SM fields
({\it e.g.} top quarks, the Higgs).  We will briefly summarize the
discovery reach for simple
models of this type
(for a more detailed survey of warped collider
phenomenology and additional references, see, for example,
Ref.~\cite{Davoudiasl:2009cd}).  Here, only the SM decay modes
of the KK states are considered, however,
the KK widths can receive important contributions
from non-SM fermions in some extended models
\cite{Carena:2007tn,W'}.

\underline{The KK gluon:} With 100~fb$^{-1}$, the lightest KK gluon $g^1$
up to masses of 3-4~TeV can be discovered at the LHC (from initial $q {\bar q}$ states)
\cite{Agashe:2006hk,Lillie:2007yh}.  The dominant decay channel is into top quarks,
whose polarization can provide a handle on the signal.  Note that models with a bulk custodial
symmetry \cite{Agashe:2003zs}
can accommodate gauge KK masses $m_{KK} \simeq 2.45\, \tilde{k}$ above
$\sim 2-3$~TeV \cite{Carena:2007ua} (see the
discussion in section \ref{rsew}).

\underline{The KK graviton:}
Refs.~\cite{Fitzpatrick:2007qr,Agashe:2007zd} revisited the LHC prospects for the discovery
of $G^1$ (produced from gluon initial states).
Ref.~\cite{Fitzpatrick:2007qr} focused on the top decay channel and concluded that for top
reconstruction efficiencies ranging over 1-100\%, the reach can be 1.5-2~TeV, with 100~fb$^{-1}$.  Ref.~\cite{Agashe:2007zd}
considered the process  $gg \to G^1\to Z_L Z_L \to 4 \ell$, with $\ell=e,\mu$ (clean signal,
but with a small branching fraction), and
found that with 300~fb$^{-1}$, the LHC reach is about 2~TeV.
The $G^1$ is predicted to be $3.83/2.45\simeq 1.56$ times heavier
than the lightest gauge KK state, making its discovery a difficult challenge
at the LHC.

\underline{The electroweak sector KK modes:} The 5D bulk is assumed to have
a custodial $SU(2)_L\times SU(2)_R\times U(1)_X$ gauge symmetry \cite{Agashe:2003zs}.
Thus, at the lowest KK level, there are 3 neutral and 4 charged states,
collectively denoted by $Z'$ and  $W'$, respectively.

Ref.~\cite{Agashe:2007ki} considered the reach for the $Z'$, with main decay channels
$t {\bar t}$, $W_L W_L$, and $Z_L H$.  Due to the near degeneracy of the KK gluon and $Z'$
masses, the top decay channel is dominated by the KK gluon ``background.''  This work concluded that in
the $Z^\prime \to W^+_L W^-_L\to \ell^+ \ell^- E_T\!\!\!\!/$ channel, the reach for the $Z'$
at the LHC is about 2~TeV, with 100~fb$^{-1}$.  Ref.~\cite{W'} found the LHC reach for
the $W'$ to be similar to that for $Z'$.

Many of the above conclusions about the reach of the LHC for new resonances can be
improved by having better control over the reducible backgrounds associated with
the collimated decay products (such as merged dijets) of highly boosted heavy SM states
(for some discussion and references see Ref.~\cite{Davoudiasl:2009cd}).

\underline{Truncated models:}
Some unwanted effects
become suppressed with decreasing $k r_c \pi$ \cite{LRS}.  The truncated volume
can still accommodate natural ``Little Randall-Sundrum'' models of flavor,
with ${\rm TeV}\ll M_5 \ll \bar M_P$, for $k r_c \pi \,\,\rlap{\raise 3pt\hbox{$>$}}{\lower 3pt\hbox{$\sim$}}\,\, 7$
(but much smaller than $\sim 36$) \cite{LRSepsK}.
\begin{figure}[htb]
\includegraphics[width=0.45\textwidth]{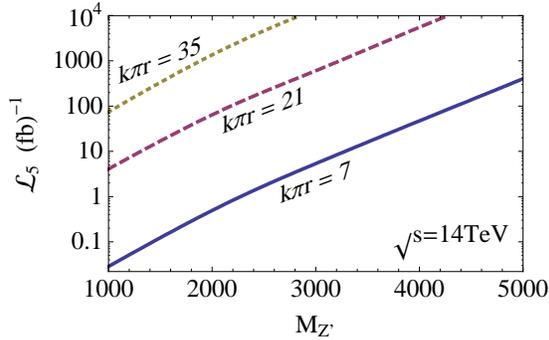}
\caption{The required integrated
luminosity for a 5$\sigma$ signal from
$pp\to Z'\to \ell^+\ell^-$ ($\ell=e$ or $\mu$, not both) with at least 3 events,
as a function of $M_{Z'}$.  The LHC center
of mass energy $\sqrt{s}=14$~TeV is assumed; from
Ref.~\cite{LRSatLHC}.}
\label{10/14TeV}
\end{figure}
Volume-truncation can enhance KK mode discovery prospects.
For example, the $Z'$ discovery reach at the LHC, in the clean dilepton mode  \cite{LRSatLHC},
is displayed in Fig.~\ref{10/14TeV}.  Thus,
certain properties of the TeV-scale KK states can shed light on
the fundamental 5D scale $M_5 \gg$~TeV.

\section{
Precision Measurement Constraints on Warped Extra
Dimensions}\label{rsew}
%%%%%%%%%%%%%%%%%%%%%%%%%%%%%%%%%%%%%%%%%%%%%%%%%

{\it M.~Carena, E.~Pont\'on, J.~Santiago, and C.E.M.~Wagner}\medskip

    Five dimensional (5D) warped extra dimensions provide a very
attractive beyond  the standard model physics scenario, in which the
weak scale-Planck scale hierarchy may be explained in a natural way
~\cite{RS} (see Ref.~\cite{Davoudiasl:2009cd} for a
recent review).
The observed light quark and lepton masses, as well
as the suppression of flavor-violating operators, is naturally
satisfied provided the quark and gauge fields propagate in the bulk and the
first and second generation quark wave functions are localized away from
the infrared brane (IR brane), where the Higgs is
localized and where the natural scale of energies is of the order of the weak
scale ~\cite{ArkaniHamed:1999dc,Gherghetta:2000qt,Huber:2000ie}.

The propagation of gauge and fermion fields in the bulk leads to
Higgs induced mixing
of zero-modes with Kaluza Klein (KK) modes, which result in important
tree-level effects on precision electroweak
observables~\cite{Davoudiasl:1999tf,Chang:1999nh}. This happens
specially for gauge bosons and
third generation quarks~\cite{Csaki:2002gy,Hewett:2002fe,Burdman:2002se,Carena:2002dz,Davoudiasl:2002ua,Carena:2003fx,Carena:2004zn},
%--\cite{Carena:2004zn},
which
tend to be localized close to the IR brane in order to generate the
large top-quark mass. Suppression of these large tree-level effects can
be achieved by
either large KK mode masses, beyond the reach of the LHC, or the presence of
brane kinetic terms, which can diminish the KK particle wave functions at the
infrared brane~\cite{Carena:2002dz,Davoudiasl:2002ua,Carena:2003fx,Carena:2004zn}
%--\cite{Carena:2004zn}
(see Ref.~\cite{Casagrande:2008hr} for an alternative approach to this
question).

The introduction of a custodial $SU(2)_R$ symmetry
together with a discrete left-right symmetry leads to reduced
corrections to the $T$ parameter~\cite{Agashe:2003zs}
and helps protect the bottom-quark coupling to the $Z$ gauge boson
against large tree-level corrections~\cite{Agashe:2006at} (see
also Ref.~\cite{Djouadi:2006rk}).
The above requirements may be satisfied in a natural way by embedding the
Standard Model gauge $SU(2)_L\times U(1)_{Y}$ group and the global
custodial $SU(2)_R$ group into an $SO(5)\times U(1)_{X}$ gauge symmetry
group~\cite{Agashe:2006at}. The $SO(5)\times U(1)_{X}$ symmetry is broken
by boundary conditions at the IR brane down
to $SU(2)_L \times SU(2)_R\times U(1)_{X}$ and
to $SU(2)_L \times U(1)_{Y}$ at the ultraviolet brane
(UV brane), respectively.
The five dimensional components of the gauge bosons associated
with the broken gauge symmetries at the IR brane have the proper
quantum numbers of the Higgs doublet, leading to a natural implementation of
the Gauge-Higgs unification
mechanism~\cite{Agashe:2006at,Djouadi:2006rk,Carena:2006bn,Agashe:2004rs,Medina:2007hz,Contino:2006qr,Agashe:2005dk,gauge:Higgs:unification}.
%--\cite{gauge:Higgs:unification}.

We shall therefore introduce in the quark sector three $SO(5)$
multiplets per generation : Two {\bf 5}'s; the first one, with
localization mass parameter $c_1$, containing in its $SU(2)_L \times SU(2)_R$
bidoublet component the zero modes of the left-handed doublets, and the
second one, with localization mass parameter $c_2$,
containing in its singlet component the right-handed up-quark
zero mode. Finally, the right-handed down quark zero mode is included in a
{\bf 10} of $SO(5)$. Effective up-quark Yukawa couplings are induced by an
IR brane mass terms which couple the left-handed singlet component of the
first {\bf 5} with the right-handed singlet component of the second one.
Down-quark Yukawa couplings are induced in a similar way.

In spite of the suppression of the tree-level contributions,
important corrections to the precision electroweak observables subsist at the
one loop-level, and agreement with data for KK masses at the reach
of the LHC may only be obtained in a certain region of fermion mass bulk
parameters of the third generation
quarks~\cite{Carena:2006bn,Carena:2007ua}. In particular, the
bidoublet containing the left-handed third generation zero modes induces
negative contributions to $\Delta T$
which tend to cancel the positive top quark mass contributions.
%a tendency that is enhanced for non-vanishing value of $M_Q$. We shall
%therefore concentrate in the case $M_Q = 0$.
%For a given localization of the two multiplets,
%the localized mass $M_u$ is then fixed by the requirement of obtaining
%a proper top-quark mass.
A positive value of $\Delta T$ can be obtained if there are
positive contributions induced by the
$SU(2)_L$  singlet KK modes of the top-quark~\cite{Carena:2006bn},
which compete successfuly
against the negative bidoublet contributions.
In Fig.~\ref{Fig1} we plot the $T$ parameter as a function
of $c_{2}$ for several values of $c_{1}$.
We see that
$T$ is negative for
most values of $c_{2}$, and increases rapidly as $c_{2}$ approaches
$-1/2$, for which a  light $SU(2)_L$ singlet KK mode
of the top quark appears, providing the necessary positive $\Delta T$
contributions.
\begin{figure}
[!h]
%\centerline{
\includegraphics[width=0.5\textwidth]{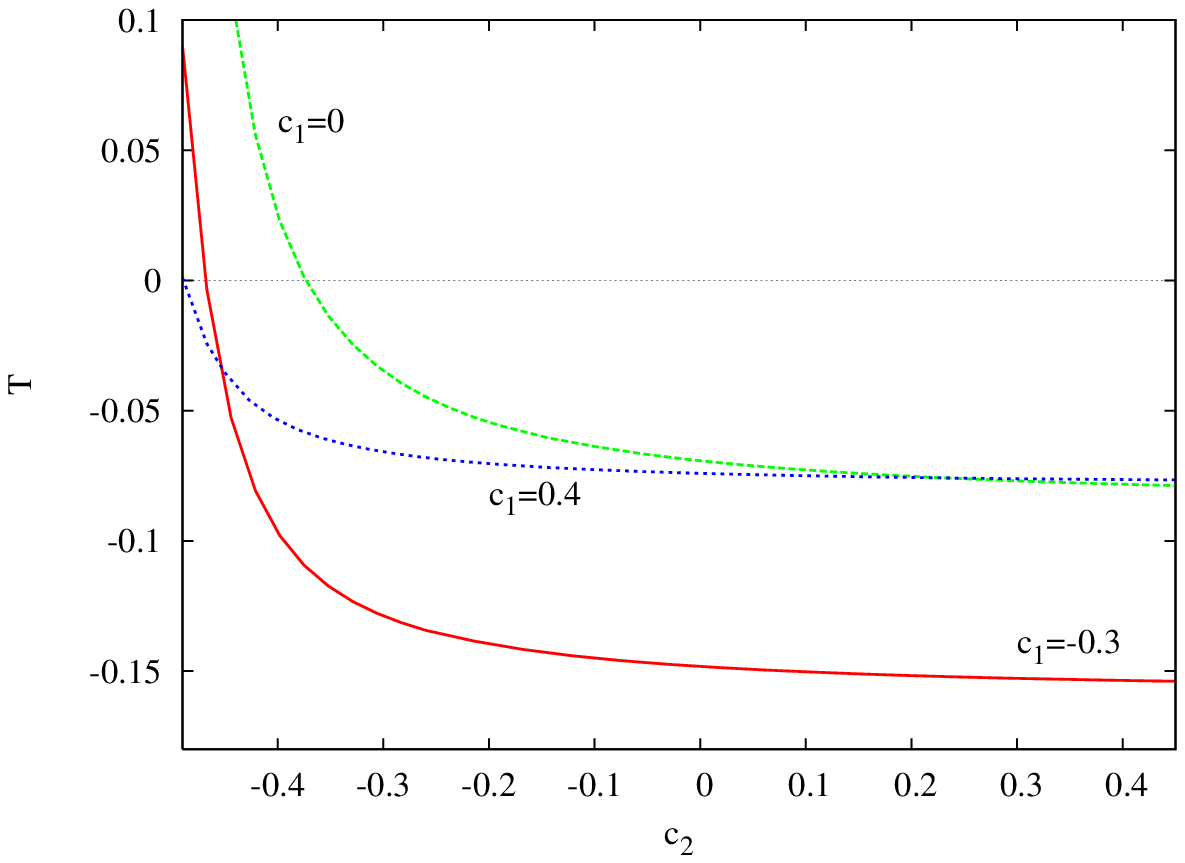}
%}
%\end{figure}
%\begin{figure}[t]
%\centerline{
\includegraphics[width=0.5\textwidth]{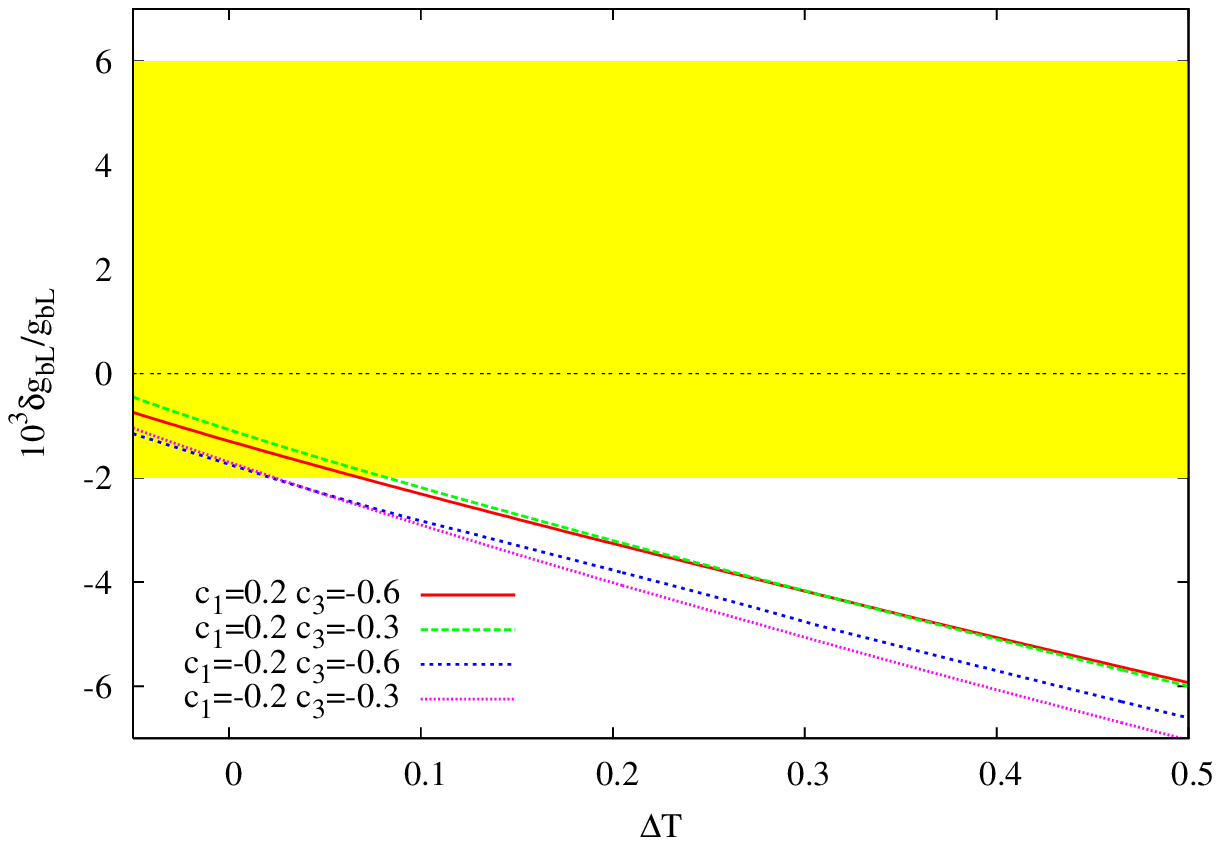}
%}
\caption{(a) Contribution to the $T$ parameter.
We use $\tilde{k} = 1.2~{\rm TeV}$ and $m_{\rm
top} = 167~{\rm GeV}$ (left panel). (b)
Correlation between the one-loop contributions to the $T$
parameter, denoted by $\Delta T$, and the one-loop contributions to
$\delta g_{b_{L}}/g_{b_{L}}$ (right panel).
We show representative curves for a few values of the left-handed top
quark localization parameter, $c_{1}$, and the bottom quark
localization parameter, $c_{3}$, as the right-handed top localization
parameter, $c_{2}$, is varied.
The band corresponds to the 2-$\sigma$ bound on
$\delta g_{bL}/g_{bL}$, assuming no large corrections to the $Z b_R
\bar{b}_R$ coupling.We take the mass of the first KK
excitation of the $SU(2)_L$ gauge bosons $m_1^\mathrm{gauge} =
3.75~{\rm TeV}$.
}
\label{Fig1}
\end{figure}
When the first two families are localized near the UV brane
the prediction for $S$ is
$S \approx 9 \, v^{2}/\tilde{k}^{2}
+ \Delta S_f~$,
where $\tilde{k}$ is the natural scale on the IR brane and $\Delta S_f$
is the relatively small contribution from the fermion loops.
For a light Higgs
with $m_{H} \simeq 115~{\rm GeV}$ (Gauge-Higgs unification
models typically predict a light Higgs),
in order to be consistent
with the $2\sigma$ $S$-$T$ bounds
a positive contribution to $T
\approx 0.3$ is also required~\cite{PDG},
which, as explained above, can only be achieved
for $c_2 \sim -0.5$.
For the above values of the parameters, one also finds
potentially important
loop-level corrections to the coupling of the left-handed bottom quarks
to the $Z$ gauge boson
$\delta g_{b\,L}/g_{b\,L}$, induced by the light KK modes
of the top-quark sector. In
Fig.~\ref{Fig1} we show the correlation between $\Delta T$ and
$\delta g_{b\,L}/g_{b\,L}$. We see that for the region of parameters for
which positive corrections to  $\Delta T$ are found, the corrections to
the bottom quark coupling become significant, pushing $g_{b\,L}$ away from
the experimentally allowed values.  Therefore, the preferred
parameter space can only be defined by a global fit to all EW measurements.
This was done in Ref~\cite{Carena:2007ua}, following the method presented
in Ref.~\cite{Han:2004az}. This work confirmed the preference of values
of $c_2 \simeq -0.5$, and found a lower
bound on the KK scale, $\tilde{k}\gtrsim 1.2$~TeV for first and second
generation fermions close to the conformal point (left-handed quarks
acquiring $c_L \simeq 0.5$), and increasing to
$\tilde{k} \gtrsim 1.4$~TeV when these fermions are localized towards
the UV brane. Interestingly enough, in Ref.~\cite{Medina:2007hz} it was shown
that the region of parameters consistent with
precision electroweak observables is in good agreement with that required to
obtain the breakdown of the electroweak symmetry, with the proper
values of the top-quark, bottom-quark and weak gauge boson masses.

The points leading to a good fit to the precision electroweak constraints
tend to also induce a positive correction to the value of the charged gauge
boson mass $M_W$, something preferred by data. Figure~\ref{MWmt} shows, in
green, the values of $M_W$ predicted in this class of models of
warped extra dimensions.
For comparison, we show, in red, the Standard Model
(SM) predictions for different
values of the Higgs mass. The ellipse shows the experimentally
preferred region~\cite{:2009ec},\cite{Group:2009nu}.

\begin{figure}[!h]
%[tbp]
        \centering
        \includegraphics[scale=0.7, angle=0]{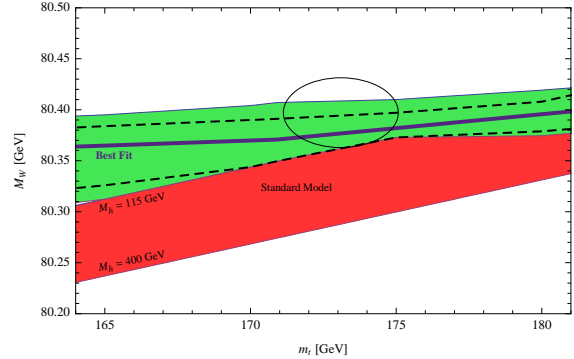}
        \caption{
Predictions for the $W$ mass as a function of the top quark mass
for models of warped extra dimensions with custodial
symmetries (as explained in the text) consistent with precision
measurements (green area) at the 95\% C.L.
SM predictions are shown in red. The ellipse
shows the 68\% C.L.
experimentally preferred region. The black solid and dashed lines
show the best fit to the data, and the area selected at the 68\% C.L.,
respectively.
}
        \label{MWmt}
\end{figure}

The above results have been obtained for a particular implementation
of Gauge-Higgs Unification models
but there are generic properties
that appear in any warped extra dimensional model protected by
custodial symmetries of the kind presented above (see, for instance,
Ref.~\cite{Panico:2008bx} for a model containing a Dark Matter
candidate and Ref.~\cite{Carena:2009yt} for the inclusion of a similar
candidate in the lepton sector).
One of the most important generic properties of these type of models is
the existence of light excited states of the top quark, necessary to
obtain positive values of $\Delta T$.
These quarks are strongly coupled to the gauge bosons KK modes so that the
first KK mode of the gluon, $G^1$, tends to decay into them. This,
in turn,  leads to a reduced decay branching ratio of $G^1$ into
top-quarks. These properties, together with an increase in the
width of $G^1$ make the $G^1$ detection via decay into top quarks
more challenging than in the models which had been previously
analyzed in the
literature~\cite{Agashe:2006hk,Lillie:2007yh}.
In Ref.~\cite{Carena:2007tn} the production of the first
excited state of the top quark  $t^1$, at the LHC was analyzed.
It was shown that the
presence of $G^1$ leads to an important enhancement of the $t^1$
production cross section, allowing an LHC reach up to
masses of about 1.5~TeV, far beyond the ones that can be
tested via direct QCD production.
The analyses of Refs.~\cite{Azuelos:2004dm},\cite{Han:2005ru} in
a slightly different context, show that
single $t^1$ production may provide an interesting alternative for
the detection of these light KK-mode excitations of the top-quark.
Top partners, with charge 5/3, as well as vector-like KK modes of
the first and second generation quarks, which can be present in
models with custodial
protection of the $T$ parameter and the $Zb\bar{b}$ coupling, can be
also discovered at the Tevatron and the LHC in pair and single
production~\cite{Atre:2008iu,Contino:2008hi,AguilarSaavedra:2009es,Mrazek:2009yu}.
%--\cite{Mrazek:2009yu}.

Finally, these models lead also to interesting flavor
signatures~\cite{Csaki:2008zd,Blanke:2008yr,Gori:2009em},
%--\cite{Gori:2009em},
in particular in rare K-decays and $B_s$ CP-violating asymmetries,
that can be tested at current (Tevatron) and future
(LHCb, JPARC, Project X) flavor physics experiments.

\section{Flavor physics in models with warped extra dimensions}
\label{rsf}

{\it U.~Haisch and M.~Neubert}\medskip

Basically all attempts to stabilize the electroweak scale
envision new degrees of freedom at or not far above the TeV
scale. New dynamics at scales required for a natural solution to
the gauge hierarchy problem would however generically lead to
extra flavor- and CP-violating interactions of an amount that is
experimentally ruled out. Insisting on the theoretical prejudice
that new physics has to emerge in the TeV range therefore leads
one to conclude that the new flavor interactions possess a highly
non-generic, close to universal structure, which in turn excludes
the possibility of finding a testable solution to the fermion
mass hierarchy problem within the same framework.

Models with a warped extra dimension, proposed first by Randall and
Sundrum (RS) \cite{RS}, provide a new avenue to flavor physics.
Allowing gauge \cite{Davoudiasl:1999tf,Pomarol:1999ad,Chang:1999nh} and matter fields
\cite{Grossman:1999ra,Gherghetta:2000qt} to spread in the AdS$_5$
bulk not only avoids dangerous higher-dimensional operators
suppressed only by scales of ${\cal O} ( \rm{few \ TeV})$, but also
admits a natural explanation of the hierarchical structures observed
in the masses and mixings of the SM fermions \cite{Huber:2000ie,Huber:2003tu} via geometrical sequestering
\cite{ArkaniHamed:1999dc}. Since the fermion zero modes are
exponentially localized near either the infra-red (IR) or
ultra-violet (UV) brane, the effective Yukawa couplings resulting
from their wave-function overlap with the Higgs boson naturally
exhibit exponential hierarchies. In this way one obtains a
five-dimensional (5D) realization \cite{Casagrande:2008hr,Blanke:2008zb} of the Froggatt-Nielsen mechanism
\cite{Froggatt:1978nt}, in which the flavor structure is accounted
for apart from ${\cal O}(1)$ factors. Addressing the flavor
hierarchies via warping in an extra dimension makes distinctive
predictions for flavor-changing processes as well. Various new
sources of flavor violation arise in RS models as a consequence of
non-trivial overlap factors between fermions and gauge (or Higgs)
bosons, which generically are non-diagonal in the mass basis. While
the new flavor-changing effects generically arise already at tree
level, a dynamical mechanism referred to as RS-GIM mechanism
\cite{Gherghetta:2000qt,Agashe:2004ay,Agashe:2004cp} ensures that
these effects are suppressed, for most observables, to an acceptable
level.

During the past years miscellaneous studies of the flavor structure of
the
quark~\cite{Huber:2003tu,Casagrande:2008hr,Blanke:2008zb,Agashe:2004ay,Agashe:2004cp,Burdman:2002gr,Burdman:2003nt,Agashe:2006wa,Cacciapaglia:2007fw,Fitzpatrick:2007sa,Cheung:2007bu,Chang:2007uz,Csaki:2008zd,Santiago:2008vq,Csaki:2008eh,Agashe:2008uz,Bauer:2008xb,Blanke:2008yr,Azatov:2008vm,Davoudiasl:2009xz,Csaki:2009bb,Buras:2009ka,Gedalia:2009ws,Agashe:2009di,Azatov:2009na,Csaki:2009wc}
and lepton
\cite{Huber:2003tu,Kitano:2000wr,Moreau:2006np,Agashe:2006iy,Chen:2008qg,Perez:2008ee,Csaki:2008qq,Agashe:2008fe,Agashe:2009tu}
sectors in RS models have been performed. An early survey of $\Delta F
= 2$ ({\it i.e.}, neutral meson mixing) and $\Delta F = 1$ ({\it
  i.e.}, rare weak decays) processes in the RS framework was presented
in \cite{Agashe:2004ay,Agashe:2004cp}. The first complete study of
all operators relevant to $K$--$\bar K$ mixing was presented in
\cite{Csaki:2008zd}. Comprehensive analyses of $B_{d,s}$--$\bar
B_{d,s}$ mixing \cite{Blanke:2008zb}, rare $Z$-mediated leptonic $K$-
and $B$-meson decays \cite{Blanke:2008yr} as well as of the
dipole-operator contributions to $B \to X_s \gamma$
\cite{Agashe:2008uz} and $\epsilon_K^\prime/\epsilon_K$
\cite{Gedalia:2009ws} have been performed quite recently. Higgs
\cite{Agashe:2009di,Azatov:2009na} and radion-mediated
\cite{Azatov:2008vm,Davoudiasl:2009xz} flavor-changing neutral
currents (FCNCs) have also been investigated. A first detailed study
of rare, lepton flavor-violating (LFV) decays has been presented in
\cite{Agashe:2006iy}.

One key observation gleaned from the analyses of $\Delta F = 2$
observables \cite{Csaki:2008zd,Santiago:2008vq,Agashe:2008uz,Bauer:2008xb,Blanke:2008yr} is that the four-quark operators
induced by Kaluza-Klein (KK) gluon exchange give the by far dominant (leading)
contributions to the effective weak Hamiltonians describing $K$--$\bar K$ ($B_{d,s}$--$\bar B_{d,s}$ and
$D$--$\bar D$) mixing. This implies that mixing phenomena mainly probe
the extra-dimensional aspects of the strong interactions, but are
to first approximation insensitive to the precise embedding of
the electroweak gauge symmetry in the higher-dimensional
geometry.

The predictions for $\Delta F = 1$ observables, on the other
hand, depend strongly on the exact realization of both the gauge
and fermionic sectors, because they receive the dominant
contribution from tree-level exchange of the $Z$ boson and its KK
excitations \cite{Casagrande:2008hr,Agashe:2004ay,Agashe:2004cp,Blanke:2008yr}. While these effects are
enhanced by the logarithm of the warp factor, $L = \ln (M_{\rm
Planck}/M_{\rm weak}) \approx 37$, in models with $SU(2)_L \times
U(1)_Y$ gauge symmetry \cite{Burdman:2002gr,Casagrande:2008hr,Agashe:2006at}, it is possible to protect the left-handed
$Z$-boson couplings from
$L$-enhanced corrections \cite{Blanke:2008zb,Agashe:2006at,Agashe:2009tu,inprep} by
extending the bulk gauge group to $SU(2)_L
\times SU(2)_R\times U(1)_X \times P_{LR}$ \cite{Agashe:2003zs} and choosing an
appropriate embedding of the down-type quarks \cite{Agashe:2006at} (if the
right-handed up-type quarks transform as $(\bm{1},
\bm{1})_{2/3}$ under the custodial symmetry the $Z u_R^i \bar
u_R^j$ vertices are protected too \cite{Buras:2009ka,Agashe:2006at}). No custodial protection mechanism can however be
tailored for the subleading effects in $L$ that arise from the
different boundary conditions of the $Z_2$-odd and -even
gauge and fermionic fields \cite{inprep}. If the right-handed down-type
quarks are embedded into $(\bm{1}, \bm{3})_{2/3}$, which is
necessary to arrive at an $U(1)_X$ invariant Yukawa coupling,
then the $Z d_R^i \bar d_R^j$ couplings are enhanced by one order
of magnitude relative to the minimal RS model. Despite this
enhancement, right-handed currents in the $b \to d, s$ sector remain small
in the custodial RS model \cite{Blanke:2008yr}, since the
involved right-handed quark wave functions are naturally more UV-localized
than their left-handed counterparts. Larger effects are possible in the $s
\to d$ sector \cite{Blanke:2008yr}, but this would require the bulk mass
parameter of the right-handed top quark to be (at least) of
${\cal O} (1)$ \cite{inprep}. While the pattern of new-physics
effects in processes such as $B_{d, s} \to \mu^+ \mu^-$, $B \to
X_{d,s} \nu \bar \nu$, $K_L \to \mu^+ \mu^-$, $K \to \pi \nu \bar
\nu$, and $K_L \to \pi^0 \ell^+ \ell^-$ is hence model dependent,
order of magnitude enhancements of the branching fractions of
rare $B$- and $K$-meson decays are only possible in the minimal
RS scenario \cite{Blanke:2008yr,inprep}, after satisfying the $Z
\to b \bar b$ constraints by tuning. On the other hand, the
experimental prospects for observing FCNC top-quark decays like
$t \to c Z$ \cite{Agashe:2006wa,Casagrande:2008hr,inprep} seem
more favorable in the custodial RS model.

In spite of the RS-GIM mechanism, a residual ``little CP problem''
is found in the form of excessive contributions to the neutron
electric dipole moment (EDM) \cite{Agashe:2004ay,Agashe:2004cp},
and to the CP-violating parameters $\epsilon_K$ \cite{Csaki:2008zd,Santiago:2008vq,Agashe:2008uz,Bauer:2008xb,Blanke:2008yr,Bona:2007vi,Davidson:2007si} and $\epsilon_K^\prime/\epsilon_K$
\cite{Gedalia:2009ws,inprep} in the neutral kaon system, which for
anarchic choices of parameters turn out to be too large unless the
masses of the lightest KK gauge bosons lie above (10--20)~TeV. This
would prevent the direct discovery of KK excitations at the Large
Hadron Collider (LHC). The little CP problem might be accidentally
solved if a combination of various unrelated CP-violating parameters
just happen to be small, which in the case of $\epsilon_K$ requires
a tuning at the percent level. Since the new CP-odd phases appearing
in the $s \to d$, $b \to s$, and $c \to u$ transitions are highly
uncorrelated, large new CP-violating effects in $B_s$--$\bar B_s$
\cite{Blanke:2008zb,inprep} and $D$--$\bar D$ \cite{inprep,Gedalia:2009kh} mixing are still possible in such a case. An
acceptable amount of indirect CP violation in the kaon sector can be
achieved for masses of the first KK excitation in the ball park of 5
TeV by allowing for larger down-type Yukawa couplings. While this
reduces the chirally enhanced tree-level corrections to $\epsilon_K$
arising from the left-right four-quark operator \cite{Csaki:2008zd},
loop contributions to $B \to X_s \gamma$ \cite{Agashe:2008uz} and
$\epsilon_K^\prime/\epsilon_K$ \cite{Gedalia:2009ws} associated to
dipole operators are enhanced in this limit, making it impossible to
fully decouple flavor-violating effects.

In view of the little CP problem, several modifications of the
quark-flavor sector of warped extra-dimensional models have been
proposed. Most of them try to implement the notion of minimal
flavor violation \cite{Chivukula:1987py,Gabrielli:1994ff,Ali:1999we,Buras:2000dm,D'Ambrosio:2002ex} or
next-to-minimal flavor violation \cite{Agashe:2005hk,Ligeti:2006pm} into the
RS framework by using (gauged) flavor symmetries
\cite{Cacciapaglia:2007fw,Fitzpatrick:2007sa,Santiago:2008vq,Csaki:2008eh,Csaki:2009bb,Csaki:2009wc,Rattazzi:2000hs}. An
important distinction of the suggested solutions is whether
flavor issues are addressed solely by Planck-scale physics on the
UV brane or whether bulk physics participates in the flavor
dynamics as well. In \cite{Cacciapaglia:2007fw,Csaki:2009bb,Rattazzi:2000hs} it was proposed to break the flavor symmetries
only on the UV brane. The downside of these constructions is that
they no longer try to explain the fermion mass hierarchy, but
only accommodate it with the least amount of flavor structure,
making this class of models hard to probe via flavor precision
tests. Other recent proposals \cite{Fitzpatrick:2007sa,Santiago:2008vq,Csaki:2008eh} try to solve the little CP problem
without giving up on addressing the flavor problem and thus may
be probed at the LHC. The basic idea is to align the down-type
quark sector, which includes the bulk masses and the 5D down-type
Yukawa couplings, such that the constraint from $\epsilon_K$ is
satisfied. Potential problems of this idea are loop-induced
misalignment and additional flavor violation from both IR and UV
brane kinetic terms and new gauge bosons. In order to circumvent the latter
problems, the construction in \cite{Csaki:2009wc} makes use of the
mechanism of shining \cite{ArkaniHamed:1998sj,ArkaniHamed:1999yy,ArkaniHamed:1999pv}, {\it i.e.}, the transmission of a symmetry-breaking effect from the UV brane through the bulk
by almost marginal scalar operators
\cite{Rattazzi:2000hs}. One of the most robust predictions of the
proposals featuring an (approximate) alignment in the down-type
quark sector is that the up-quark sector is anarchical, which
suggests a discovery of CP violation in the $D$--$\bar D$ system at
around the current experimental upper bounds. Interesting effects
could also emerge in top-quark FCNCs, but these are more
model dependent. The problem of too large EDMs has been addressed
using the idea of spontaneous CP violation in the context of
warped extra dimensions \cite{Cheung:2007bu}.

In order to accommodate simultaneously the non-hierarchical
neutrino mixing angles and the absence of LFV processes such as $\mu^\pm \to e^+ e^- e^\pm$ and
$\mu \to e \gamma$ \cite{Huber:2003tu,Kitano:2000wr,Moreau:2006np,Agashe:2006iy,Davidson:2007si} for new-physics
scales below 10 TeV also requires additional model-building
\cite{Perez:2008ee,Chen:2008qg,Csaki:2008qq,Agashe:2008fe,Agashe:2009tu}. The simplest constructions \cite{Perez:2008ee,Chen:2008qg,Csaki:2008qq} are 5D realizations of minimal flavor violation  in the
lepton sector.  More recent proposals stick to the anarchic
flavor approach, but utilize a bulk Higgs \cite{Agashe:2008fe} or
new lepton representations under the extended $SU(2)_L \times
SU(2)_R \times U(1)_X$ gauge group \cite{Agashe:2009tu} to
ameliorate the constraints from LFV. Like in the quark sector,
there exists however a tension between loop-induced and
tree-level LFV processes, since they depend in the opposite way
on the 5D Yukawa couplings \cite{Agashe:2006iy}. As a result it
is not possible to decouple all flavor-violating effects, so that upcoming LFV
experiments should see a signal if warped extra dimensions with a
KK mass scale of ${\cal O} (5 \, {\rm TeV})$ are realized in
nature.

\section{Radion Phenomenology in Warped Extra Dimensions}\label{radion}

{\it M.~Toharia}\medskip

We will consider 5D scenarios in which the Standard Model (SM) matter
is allowed to propagate in the bulk. The particular spacetime
structure we are interested in takes the form given in Eq.~(\ref{RSmetric})~\cite{RS}.
%\begin{equation}
%ds^2=e^{-2\sigma} \eta_{\mu\nu} dx^\mu dx^\nu -dy^2 \label{RS}
%\end{equation}
%where $\sigma(y)=ky$, and $k$ is the 5D curvature.
Gravitational perturbations around this metric contain a scalar mode,
the radion $r(x)$ \cite{Charmousis:1999rg}
\begin{equation}
ds^2= e^{-2\sigma}(1+r(x)) dx^2 -\left(1+2 e^{2\sigma}r(x)\right) dy^2
\label{metricpert}.
\end{equation}
It cannot be gauged away due to the presence of the two brane boundaries at
$y=0$ and $y=y_{{}_{IR}}$, whose location remains unfixed. This
makes the radion a massless degree of freedom, but a simple way
to address this potential problem is to add a 5D
scalar field with a nontrivial vev which fixes the inter-brane distance
~\cite{Goldberger:1999uk}.
If this produces a small backreaction on the metric, the radion will be relatively
light with respect to the rest of KK excitations~\cite{Csaki:2000zn}.
\begin{figure}[t]
\vspace{-0pt}
\includegraphics[height=10pc,width=17pc]{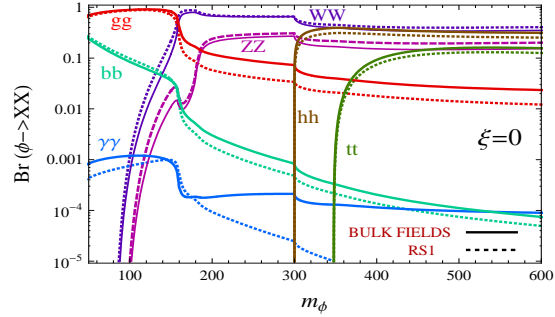}
\vspace{-1cm}
\caption{Branching fractions for the Radion as a function of its mass
  $m_\phi$ (in GeV)
  in the RS1 scenario and the SM Fields in the Bulk scenario.}
\label{fig:branchings}
\vspace{-20pt}
\end{figure}
The interactions of the radion are gravitational in nature and after extracting
its couplings with the lightest modes of the 5D bulk matter, i.e the
SM massive gauge bosons and fermions, one obtains~\cite{Csaki:2007ns}
\vspace{-.1cm}
\begin{equation}
{M^2_V} \left(1-6\ k y_{{}_{IR}}{M^2_V\over\Lambda_\phi^2}
\right)\   {\phi_0\over \Lambda_\phi} V^\alpha V_\alpha, \label{rVV}
\end{equation}
\vspace{-.6cm}
\begin{equation}
{ m_f}(c_L-c_R)\ {\phi_0\over
  \Lambda_\phi}\bar{f}_{UV}f_{UV},\label{rff1}
\end{equation}
\vspace{-.6cm}
\begin{equation}
{ m_f}\ {\phi_0\over \Lambda_\phi}\bar{f}_{IR}f_{IR}\label{rff2},
\end{equation}
where $\phi_0(x)$ is the 4D canonically normalized radion, defined by
$r(x)=-{2\over \Lambda_\phi}\phi_0(x)$ and such that $\Lambda_\phi =
\sqrt{6}\ M_{Pl}\ e^{-k y_{{}_{IR}}}$ is a TeV scale. The
fermions
$f_{UV}$ and $f_{IR}$ are localized near each of the two boundaries
respectively, with $c_L$ and $c_R$ the
corresponding left and right handed 5D fermion mass parameters.
\begin{figure}[t]
\vspace{-0pt}
\includegraphics[height=13pc,width=17pc]{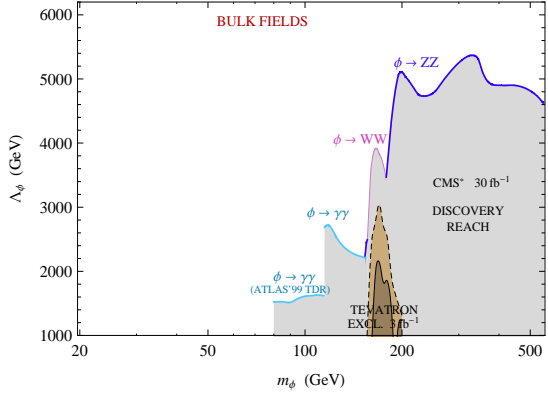}
\vspace{-1cm}
\caption{LHC discovery reach for the radion $\phi$ using ``translated'' Higgs projections
from CMS (and ATLAS in the lower mass region) for $30$ fb$^{-1}$ of luminosity.   }
\label{fig:LHCreach}
\vspace{-25pt}
\end{figure}
In the case of massless gauge bosons, i.e. gluons and photons, the
interactions with the radion appear at the loop-level~\cite{Csaki:2007ns}
\begin{equation}
\left[{1%-4 \pi  \alpha (\tau_{UV}^0+\tau_{IR}^0)
\over 4 k y_{{}_{IR}}}\!
+\! {\alpha\over 8 \pi}\left(b-\sum_i \kappa_i F_i(\tau_i)\right)
\right] {{ \phi_0}\over \Lambda_\phi} F_{\mu\nu}F^{\mu\nu},\label{rgg}
\end{equation}
where $\sum_i \kappa_iF_i$ are the contributions from one-loop diagrams
and $b$ is the beta function coefficient of corresponding gauge group,
appearing in the radion coupling due to the trace anomaly.

If one replaces $\Lambda_\phi$ by the Higgs $vev$, these interactions
become very much Higgs-like. Indeed, Figure
\ref{fig:branchings} shows that the radion decay branching fractions are very
similar to those of the Higgs. A key difference lies in the
larger branching into gluons, due to the enhanced relative coupling of
radion to gluons caused by the large trace anomaly contribution
(the term proportional to $b$ in Eq.~\ref{rgg}). At the LHC
this is a crucial point since it means that radion
production will almost exclusively come from gluon fusion, with
all other production processes comparatively
suppressed \cite{Cheung:2000rw}.
Higgs searches in the gluon fusion channel will
then apply to radion searches in a straightforward way, with some care to
be taken due to the much narrower width of the radion, suppressed by about $(v/\Lambda_\phi)^2$
relative to the Higgs width \cite{Giudice:2000av}. Figure \ref{fig:LHCreach} shows the projected
LHC reach after $30$ fb$^{-1}$ of integrated luminosity.
A radion beyond the $ZZ$ mass threshold will be easily discovered in the four
lepton channel, but the case of a lighter radion, $m_\phi< 150$,
becomes harder with the $\gamma\gamma$ channel and quite uncertain
for a very light radion.

The radion and the Higgs can also mix~\cite{Giudice:2000av}, and
the phenomenological consequences can be important as some of the
dominant channels could become irrelevant and
vice-versa~\cite{Hewett:2002nk}.
On the other hand, a large mixing can cause dangerous contributions to electroweak precision
observables~\cite{Csaki:2000zn} and so one should treat with care that region of parameter
space. Nevertheless, even for small mixing, the radion phenomenology
(but not the Higgs) can still receive important
corrections~\cite{Toharia:2008tm}. Figure \ref{fig:hrmix} shows contours of the
ratio of discovery significances
$R_S^{\gamma\gamma}={S(gg\to\phi\to\gamma\gamma)/S(gg\to
  h_{SM}\to\gamma\gamma)}$,  in the presence of Higgs-radion mixing,
parametrized by $\xi$ for a Higgs mass of $m_h=150$ GeV and for
$\Lambda_\phi=2$ TeV. %% Observability of the $\gamma\gamma$ signal can be
%% very sensitive to the amount of mixing even in the small mixing region.

Finally, it was recently pointed out that one should also
expect to have some amount of
flavor violating couplings of the radion with
fermions \cite{Azatov:2008vm}. In the case of a heavy enough radion,
this might lead to its decaying into top and charm quarks, which might be searched for at
the LHC as an interesting probe of the flavor structure of the scenario.

\begin{figure}[t]
\vspace{-0pt}
\includegraphics[height=9pc,width=15pc]{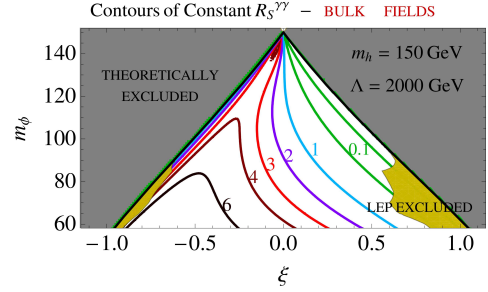}
\vspace{-1cm}
\caption{Contours of the relative discovery significance in the
  $\gamma\gamma$ channel between radion and Higgs, in the presence of
  Higgs-radion mixing parametrized by $\xi$, with varying $m_\phi$ (in GeV).}
\label{fig:hrmix}
\vspace{-20pt}
\end{figure}

\section{A Brief Review of Higgsless Models}\label{hless}

{\it C.~Cs\'aki}\medskip

One of the interesting ways extra dimensions can be used for TeV scale phenomenology is to break the electroweak symmetry via boundary conditions (rather than by a Higgs VEV). In this case the unitarization of the WW and WZ scattering amplitudes would not be due to the exchange of the physical Higgs, but rather due to the exchange of Kaluza-Klein modes of the Z and W bosons. In order for this unitarization to actually happen, the following sum-rules among the couplings and masses of the KK modes have to be satisfied~\cite{CGMPT}:
\begin{equation}
g_{WWWW}=g^2_{WW\gamma}+g^2_{WWZ}+
\sum_i g^2_{WWZ^i} \end{equation}
\begin{equation}
\frac{4}{3} g_{WWWW}M_W^2= g^2_{WWZ}
M_Z^2+\sum_i g^2_{WWZ^i} M_{Z^i}^2
\end{equation}
where $g_{WWWW}$ is the quartic self-coupling of the $W$ bosons, the $g_{WWA}$ are the cubic couplings between two $W$'s and a neutral gauge boson $A=\gamma ,Z, Z',\ldots$, while $M_{W,Z,Z^i}$ are the masses of the respective gauge bosons. The first sum rule will ensure that the terms proportional to $E^4$ in the scattering amplitudes cancel, while the second will eliminate the $E^2$ growth. Similar sum rules can be obtained for the unitarization of the $WZ$ scattering process. One can show that these sum rules are automatically satisfied for a higher dimensional gauge theory, if there is no hard breaking of gauge invariance.

In order for these sum rules to be efficient, the lowest KK modes should show up before the unitarity violation of the SM without a higgs hits, that is below the scale of $4\pi M_W/g \sim 1.5$ TeV. Thus the existence of a $W'$ and $Z'$ with significant cubic couplings to the SM gauge bosons is a robust prediction of higgsless models~\cite{CGMPT}.

For the concrete implementation of the higgsless models one can either use warped extra dimensions~\cite{CGMPT}, or deconstructed versions of that~\cite{deconstruct}. Warped extra dimensions are useful in order to enforce a custodial SU(2) symmetry on the model, which will be implemented as a bulk SU(2)$_L\times$SU(2)$_R\times$U(1)$_{B-L}$ gauge symmetry. The way the proper symmetry breaking is achieved is by breaking the gauge group down to the SM group SU(2)$_L\times$U(1)$_Y$ on the UV brane, thus ensuring that the additional gauge symmetry only manifests itself as a global symmetry in the low energy spectrum. The electroweak symmetry breaking is then achieved via breaking SU(2)$_L\times$ SU(2)$_R\to$SU(2)$_D$ on the TeV brane. All of these breakings are done by imposing the appropriate boundary conditions. The basic parameters of the warped extra dimensional model are the 5D gauge couplings of the 3 gauge groups $g_{5L},g_{5R}$ and $\tilde{g}_{5}$, the AdS curvature $R$ and the IR scale $R'$. In addition one can also introduce brane localized kinetic terms for the gauge fields. For the simplest model the leading order expression for the gauge boson masses will be (for $g_{5L}=g_{5R}$):

\begin{equation}
M_W^2 = \frac{1}{R^{\prime 2} \log \left(\frac{R^\prime}{R}\right)}  ,
M_Z^2  = \frac{g_5^2+2 \tilde g_5^{2}}{g_5^2+ \tilde g_5^{2}}
\frac{1}{R^{\prime 2} \log \left(\frac{R^\prime}{R}\right)}   .
\end{equation}
While the Weinberg angle is given by
\begin{equation}
\sin \theta_W = \frac{ \tilde g_5}{\sqrt{ g_5^2+2 \tilde g_5^2} },
\label{SM3}
\end{equation}
leading to the correct SM masses and couplings to leading order in $\log R'/R$. One can also calculate the first corrections to the electroweak precision observables~\cite{EWP}, to find (assuming that the fermions are localized around the Planck brane)
\begin{equation}
S \approx \frac{6 \pi}{g^2 \log \frac{R^\prime}{R}}, \ T\approx 0
\end{equation}
Thus while T is protected by the built-in custodial symmetry the S-parameter is too large. This conclusion is insensitive to the choices of the parameters of the gauge sector. However, the S-parameter can be canceled by changing the localization properties of the fermions~\cite{deloc}. The relevant quantity that controls the localization of the fermions in warped space is the bulk mass $c$ (measured in units of the AdS curvature). For $c>1/2$ the left handed fermions are localized around the Planck brane and for $c<1/2$ around the TeV brane. The S-parameter will have the following dependence on the mass $c$ of the left-handed SM fermions, assuming that $c$ is close to 1/2:
\begin{equation}
S =  \frac{2 \pi}{g^2\, \log \frac{R'}{R}} \left( 1 + (2 c -1)\, \log
\frac{R'}{R} \right)~.
\end{equation}
Thus for a particular value around $c=1/2$ the S-parameter can be made to vanish. Constructions for eliminating flavor changing neutral currents have been presented in~\cite{higgslessflavor}. A typical mass spectrum and set of couplings is given in Table~\ref{tab:newcustodian}.
 \begin{table} {\small \caption{Typical particle spectrum and couplings for a realistic model with a custodial protection for the $Zb\bar{b}$ vertex from~\cite{newcustodian}. The couplings are in the units of the corresponding SM couplings.\label{tab:newcustodian}}
  \begin{tabular}{c|l}
    $M_{t'}$ & 450 GeV \\
    $M_{b'}$ & 664 GeV \\
    $M_{W'}$ & 695 GeV  \\
    $M_{Z'}$ & 690 GeV  \\
    $M_{Z''}$ & 714 GeV  \\
    $M_{G'}$ & 714 GeV  \\
    $g_{W' u \bar d}$ & $0.07 \; g $  \\
    $g_{Z' q \bar q}$ & $0.14 \; g_{Z q \bar q}$  \\
    $g_{G' q \bar q}$ & $0.22 \; g_c$
   \end{tabular}
  \begin{tabular}{c|l}
    $g_{Z' t_L \bar t_L}$ & $1.83 \; g_{Z t_L \bar t_L} $   \\
    $g_{Z' t_R \bar t_R}$ & $4.02 \; g_{Z t_R \bar t_R}$ \\
   $g_{Z' b_L \bar b_L}$ & $3.77 \; g_{Z b_L \bar b_L}$  \\
    $g_{Z' b_R \bar b_R}$ & $0.26 \; g_{Z b_R \bar b_R}$  \\
    $g_{ZWW}$  & $1.018 \; g\, c_W$\\
    $g_{ZZWW}$  & $1.044 \; g^2 c_W^2$\\
    $g_{WWWW}$ &  $1.032 \; g^2$ \\
    $g_{Z'WW}$  & $0.059 \; g\, c_W$\\
    $g_{ZW'W}$  & $0.051 \; g\, c_W$
  \end{tabular} }
  \end{table}

The experimental observability of these models has been investigated in~\cite{Maxim,Dieter,Veronica}. Refs.~\cite{Maxim,Dieter} studied the vector boson fusion production of the lightest Z' and W' KK modes. A characteristic plot for the transverse mass in WZ fusion from~\cite{Dieter} is shown in Fig.~\ref{fig:WZfusion}. The most recent comprehensive study in~\cite{Veronica} included also the possibility of Drell-Yan production of the KK gauge bosons via the suppressed but non-negligible of the KK gauge fields to the SM fermions. A representative plot of the dilepton mass spectrum is reprinted from~\cite{Veronica} in Fig.~\ref{fig:DYZp}.
Ref.~\cite{Veronica} concluded that about 10 fb$^{-1}$ of luminosity is necessary for the discovery of the resonances in the 700 GeV mass range.

\begin{figure}
\begin{center}
\includegraphics[scale=0.55]{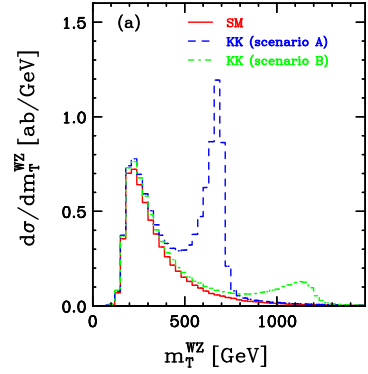}
\end{center}
\vspace*{-1cm}\caption{The transverse mass distribution of the WZ in a higgsless model with a light W' boson from~\cite{Dieter}.\label{fig:WZfusion}}
\end{figure}

 \begin{figure}[!h]
\begin{center}
\includegraphics[scale=0.2]{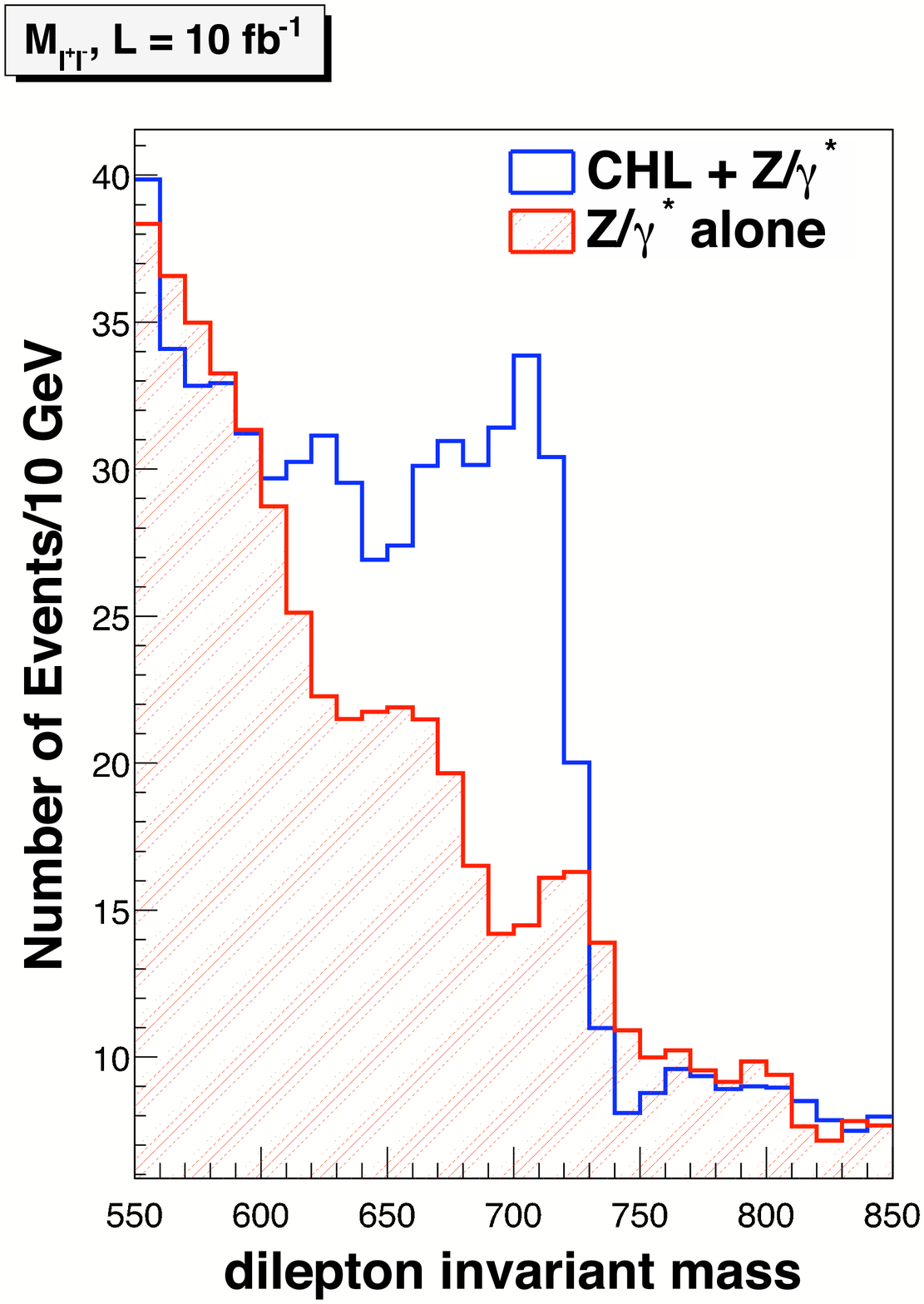}
\end{center}
\vspace*{-1cm}\caption{The dilepton mass for Drell-Yan Z' production from~\cite{Veronica}.\label{fig:DYZp}}
\end{figure}

%\end{document}

%%%%%%%%%%%%%%%%%%%%%%%%%%%%%%%%%%%%%%%%%%%%%%%%%%%%%%%%%%%%%%%%%%%%%%%%%%%%%%%%%%%%%%%%%%%%%%
%%%%%%%%%%%%%%%%%%%%%%%%%%%%%%%%%%%%%%%%%%%%%%%%%%%%%%%%%%%%%%%%%%%%%%%%%%%%%%%%%%%%%%%%%%%%%%
\chapter{String Phenomenology and the LHC}
\epigraphhead[20]{\epigraph{\large {\em Shehu AbdusSalam, Benjamin
Allanach, Luis~A. Anchordoqui, Daniel Feldman, Haim Goldberg, Gordon
Kane, Zuowei Liu, Dieter L\"ust, Pran Nath, B.D.~Nelson, Jing Shao,
Stephan Stieberger, Tomasz~R. Taylor, Fernando Quevedo}}{\large
Brent~D.~Nelson (Convener)}}
%
%%%%%%%%%% espcrc2.tex %%%%%%%%%%
%
% $Id: espcrc2.tex 1.2 2000/07/24 09:12:51 spepping Exp spepping $
%
%\documentclass[fleqn,twoside]{article}
%\usepackage{espcrc2}

% change this to the following line for use with LaTeX2.09
% \documentstyle[twoside,fleqn,espcrc2]{article}

% if you want to include PostScript figures
%\usepackage{graphicx}
% if you have landscape tables
%\usepackage[figuresright]{rotating}

% put your own definitions here:

\def\[{\left [}
\def\]{\right ]}
\def\({\left (}
\def\){\right )}
\renewcommand{\beqn}{\begin{eqnarray}}
\renewcommand{\eeqn}{\end{eqnarray}}
\renewcommand{\beq}{\begin{equation}}
\renewcommand{\eeq}{\end{equation}}
\renewcommand{\bea}{\begin{eqnarray}}
\renewcommand{\eea}{\end{eqnarray}}
\newcommand{\lang}{\left\langle}
\newcommand{\rang}{\right\rangle}
\newcommand{\lbr}{\left\{}
\newcommand{\rbr}{\right\}}
\newcommand{\order}{\mathcal{O}}
\newcommand{\GUT}{\mathsc{gut}}
\renewcommand{\GeV}{~\mathrm{GeV}}
\newcommand{\TeV}{~\mathrm{TeV}}
\newcommand{\LSP}{\mathsc{LSP}}
\newcommand{\UV}{\mathsc{uv}}
\newcommand{\EW}{\mathsc{ew}}
\newcommand{\PL}{\mathsc{pl}}
\newcommand{\DS}{(\Delta S)^2}
\newcommand{\DSaa}{(\Delta S_{AA})^2}
\newcommand{\DSab}{(\Delta S_{AB})^2}
\renewcommand{\met}{\not{\hspace{-.05in}{E_T}}}
\newcommand{\SUSY}{\mathsc{susy}}
\renewcommand{\SM}{\mathsc{sm}}
\newcommand{\slashed}[1]{\not{\hspace{-.05in}#1}}
\newcommand{\wtd}[1]{\widetilde{#1}}
\newcommand{\gappeq}{\mathrel{\rlap {\raise.5ex\hbox{$>$}}
{\lower.5ex\hbox{$\sim$}}}}
\newcommand{\lappeq}{\mathrel{\rlap{\raise.5ex\hbox{$<$}}
{\lower.5ex\hbox{$\sim$}}}}
\def \missET{${\not\!\!{E_T}}$}
\newcommand{\lumint}{5~fb$^{-1}$}
%%%%%%%%%%%%%%

\def\beq{\begin{equation}}
\def\be{\begin{equation}}
\def\beqn{\begin{eqnarray}}
\def\ee{\end{equation}}
\def\eeq{\end{equation}}
\def\eeqn{\end{eqnarray}}

\def \cha{\widetilde{\chi}^{\pm}_1}
\def \chb{\widetilde{\chi}^{\pm}_2}

\def \na{\widetilde{\chi}^{0}_1}
\def \nb{\widetilde{\chi}^{0}_2}
\def \nc{\widetilde{\chi}^{0}_3}
\def \nd{\widetilde{\chi}^{0}_4}

\def \g{\widetilde{g}}
\def \ql{\widetilde{q}_L}
\def \qr{\widetilde{q}_R}

\def \dl{\widetilde{d}_L}
\def \dr{\widetilde{d}_R}
\def \ul{\widetilde{u}_L}
\def \ur{\widetilde{u}_R}

\def \ccl{\widetilde{c}_L}
\def \ccr{\widetilde{c}_R}
\def \ssl{\widetilde{s}_L}
\def \ssr{\widetilde{s}_R}

\def \ta{\widetilde{t}_1}
\def \tb{\widetilde{t}_2}
\def \ba{\widetilde{b}_1}
\def \bb{\widetilde{b}_2}

\def \sta{\widetilde{\tau}_1}
\def \stb{\widetilde{\tau}_2}

\def \smr{\widetilde{\mu}_R}
\def \ser{\widetilde{e}_R}
\def \sml{\widetilde{\mu}_L}
\def \sel{\widetilde{e}_L}

\def \slr{\widetilde{l}_R}
\def \sll{\widetilde{l}_L}

\def \snl{\widetilde{\nu}_{\tau}}
\def \snm{\widetilde{\nu}_{\mu}}
\def \sne{\widetilde{\nu}_{e}}

\def \hc{H^{\pm}}

\def \lra{\longrightarrow}

%%%%%%%
\newcommand{\postscript}[2]{\setlength{\epsfxsize}{#2\hsize}
   \centerline{\epsfbox{#1}}}
\newcommand{\comment}[1]{}
%%%%%%%%%%%% Local definitons %%%%%%%%%%%%%%%%%%%%%%%%%%%%%%%%%%%%
\def\ss#1{{\textstyle #1}}
\def\sss#1{{\scriptstyle #1}}
\def\INT{{\int\limits_0^1\hskip-0.1cm dx\hskip-0.15cm\int\limits_0^1
\hskip-0.1cm dy\hskip-0.15cm\int\limits_0^1 \hskip-0.1cm dz}}
\def\INTT{{\int\limits_0^1\hskip-0.1cm dx\hskip-0.15cm\int\limits_0^1
\hskip-0.1cm dy\hskip-0.15cm\int\limits_0^1 \hskip-0.1cm dz
\hskip-0.15cm\int\limits_0^1 \hskip-0.1cm dw}}
\def\MET{\mbox{${\hbox{$E$\kern-0.6em\lower-.1ex\hbox{/}}}_T$}}
\def\mbb{}
\def\gs{g_{\rm string}}
\def\tb{type $IIB$\ }\def\ta{type $IIA$\ }\def\ti{type $I$\ }
\def\f{\phi}
\def\a{\alpha}
\def\thalf{\tfrac{1}{2}}
\def\cf{{\it cf.\ }}
\def\ie{{\it i.e.\ }}
\def\eg{{\it e.g.\ }}
\def\eqq{{\it Eq.\ }}
\def\eqqs{{\it Eqs.\ }}
\def\th{\theta}
\def\Th{\Theta}
\def\eps{\epsilon}
\def\al{\alpha}
\def\de{\delta}
\def\si{\sigma}
\def\Om{\Omega}
\def\om{\omega}
\def\bet{\beta}\def\be{\beta}
\def\Om{\Omega}
\def\Si{{\Sigma}}
\def\comment#1{{}}
\def\F{{_2F_1}}
\def\FF#1#2{{_#1F_#2}}
\def\tfrac#1#2{{\textstyle\frac{#1}{#2}}}

%%%%%%%%%%%%%%%%%%%%%%%%%%%%%%%%%%%%%%%%%%%%%%%%%%%%%%%%%%%%%%%%
%%%%%   Dirac-Slash
%%%%%%%%%%%%%%%%%%%%%%%%%%%%%%%%%%%%%%%%%%%%%%%%%%%%%%%%%%%%%%%%
\def\slashchar#1{\setbox0=\hbox{$#1$}           % set a box for #1
   \dimen0=\wd0                                 % and get its size
   \setbox1=\hbox{/} \dimen1=\wd1               % get size of /
   \ifdim\dimen0>\dimen1                        % #1 is bigger
      \rlap{\hbox to \dimen0{\hfil/\hfil}}      % so center / in box
      #1                                        % and print #1
   \else                                        % / is bigger
      \rlap{\hbox to \dimen1{\hfil$#1$\hfil}}   % so center #1
      /                                         % and print /
   \fi}
%%%%%%%%%%%%%%%%%%%%%%%%%%%%%%%%%%%%%%%%%%%%%%%%%%%%%%%%%%%%%%%%%
%%%%% Referencing  %%%%%%%%%%%%%%%%%%%%%%%%%%%%%%%%%%%%%%%%%%%%%%
%%%%%%%%%%%%%%%%%%%%%%%%%%%%%%%%%%%%%%%%%%%%%%%%%%%%%%%%%%%%%%%%%
\newif\ifnref
\def\rrr#1#2{\relax\ifnref\nref#1{#2}\else\ref#1{#2}\fi}
\def\ldf#1#2{\begingroup\obeylines
\gdef#1{\rrr{#1}{#2}}\endgroup\unskip}
\def\nrf#1{\nreftrue{#1}\nreffalse}
\def\multref#1#2#3{\nrf{#1#2#3}\refs{#1{--}#3}}
\def\multrefvi#1#2#3#4#5#6{\nrf{#1#2#3#4#5#6}\refs{#1{--}#6}}
\def\multrefv#1#2#3#4#5{\nrf{#1#2#3#4#5}\refs{#1{--}#5}}
\def\multrefiv#1#2#3#4{\nrf{#1#2#3#4}\refs{#1{--}#4}}
\def\multrefvii#1#2#3#4#5#6#7{\nrf{#1#2#3#4#5#6}\refs{#1{--}#7}}
\def\doubref#1#2{\refs{{#1},{#2} }}
\def\threeref#1#2#3{\refs{{#1},{#2},{#3} }}
\def\fourref#1#2#3#4{\refs{{#1},{#2},{#3},{#4} }}
\def\Fourref#1#2#3#4{\nrf{#1#2#3#4}\refs{#1{--}#4}}
\def\fiveref#1#2#3#4#5{\refs{{#1},{#2},{#3},{#4},{#5} }}
\nreffalse

\def\refout{\listrefs}
\def\lref{\ldf}
%%%%%%%%%%%%%%%%%%%%%%%%%%%%%%%%%%%%%%%%%%%%%%%%%%%%%%%%%%%%%%%%%%
%%%%%%%%%%%%%%%%%   Stuff for Figures  %%%%%%%%%%%%%%%%%%%%%%%%%%%
%%%%%%%%%%%%%%%%%%%%%%%%%%%%%%%%%%%%%%%%%%%%%%%%%%%%%%%%%%%%%%%%%%
\font\smallrm=cmr8
\def\figin{\epsfcheck\figin}\def\figins{\epsfcheck\figins}
\def\epsfcheck{\ifx\epsfbox\UnDeFiNeD
\message{(NO epsf.tex, FIGURES WILL BE IGNORED)}
\gdef\figin##1{\vskip2in}\gdef\figins##1{\hskip.5in}% blank space instead
\else\message{(FIGURES WILL BE INCLUDED)}%
\gdef\figin##1{##1}\gdef\figins##1{##1}\fi}
\def\DefWarn#1{}
\def\figinsert{\goodbreak\midinsert}
\def\ifig#1#2#3{\DefWarn#1\xdef#1{fig.~\the\figno}
\writedef{#1\leftbracket fig.\noexpand~\the\figno}%
\figinsert\figin{\centerline{#3}}\medskip\centerline{\vbox{\baselineskip12pt
\advance\hsize by -1truein\noindent\footnotefont{\bf Fig.~\the\figno } #2}}
\bigskip\endinsert\global\advance\figno by1}

%%%%%%%%%%%%%%%%%%%%%%%%%%%%%%%%%%%%%%%%%%%%%%%%%%%%%%%%%%%%%%%%%%%%%
%%%%%%%%%%%%%%%   Standard alltime definitions   %%%%%%%%%%%%%%%%%%%%
%%%%%%%%%%%%%%%%%%%%%%%%%%%%%%%%%%%%%%%%%%%%%%%%%%%%%%%%%%%%%%%%%%%%%
\def\app#1{\goodbreak\vskip2.cm\centerline{{\bf Appendix: #1}}}
\def\appA{A}\def\appAi{A.1.}\def\appAii{A.2.}\def\appAiii{A.3.}\def\appAiv{A.4.}
\def\appB{B}\def\appBi{B.1.}\def\appBii{B.2.}\def\appBiii{B.3.}\def\appBiv{B.4.}
\def\appC{C}\def\appCi{C.1.}\def\appCii{C.2.}\def\appCiii{C.3.}\def\appCiv{C.4.}
\def\appD{D}\def\appDi{D.1.}\def\appDii{D.2.}\def\appDiii{D.3.}\def\appDiv{D.4.}
\def\tilde{\widetilde}

\def\h {{1\over 2}}

\def\ov {\overline}
\def\o {\over}
\def\fc#1#2{{#1 \o #2}}

\def\IZ{ {\bf Z}}
\def\IP{{\bf P}}\def\IC{{\bf C}}\def\IF{{\bf F}}
\def\IR{ {\bf R}}
\def\hat{\widehat}
\def\E {\hat E}      % For Eisenstein E2
\def\Li {{\cal L}i}  % For Polylogarithm
\def\nihil#1{{\sl #1}}
\def\br{\hfill\break}
\def\tr {{\rm tr}}
\def\det {{\rm det}}
\def\mod {{\rm mod}}
\def\lf {\left}
\def\ri {\right}
\def\ra {\rightarrow}
\def\lra {\longrightarrow}
\def\re {{\rm Re}}
\def\im {{\rm Im}}
\def\p {\partial}

\def\Bc {{\cal B}}  \def\Nc{{\cal N}}\def\Wc{{\cal W}}
\def\Zc {{\cal Z}} \def\Qc{{\cal Q}}
\def\Fc {{\cal F}} \def\Gc{{\cal G}}
\def\Cc {{\cal C}} \def\Oc {{\cal O}}
\def\Lc {{\cal L}} \def\Sc {{\cal S}}
\def\Mc {{\cal M}} \def\Ac {{\cal A}}
\def\Pc {{\cal P}} \def\Tc {{\cal T}}
\def\Rc {{\cal R}} \def\Uc {{\cal U}}
\def\Ic {{\cal I}} \def\Jc {{\cal J}}
\def\Kc {{\cal K}} \def\Ec{{\cal E}}
\def\Vc{{\cal V}}  \def\Kc{{\cal K}}
%%%%%%%%%%%%%%%%%%%%%%%%%%%%%%%%%%%%%%%%%%%%%%%%%%%%%%%%%%%%%%%%%%%
%%%%%%%%%%%%%%%%%%%%%%%%%%%%%%%%%%%%%%%%%%%%%%%%%%%%%%%%%%%%%%%%%%%

% declarations for front matter
%\title{String Phenomenology and the LHC}
%\author{
%Shehu AbdusSalam, Benjamin Allanach, Luis A. Anchordoqui, Daniel Feldman, Haim Goldberg, Gordon Kane, Zuowei Liu, Dieter Lust, Pran Nath, Brent D. Nelson,
%Jing Shao, Stephan Stieberger, Tomasz R. Taylor, Fernando Quevedo.
%}
%\begin{document}

%\begin{abstract}
%A broad overview is given of recent progress in string
%phenomenology and its LHC implications including
%heterotic string model-building, D-brane models,
% M-theory models on $G_2$ manifolds,
% F-theory, Large Volume String compactifications,
%and signals of TeV scale Strings.
%\end{abstract}

%\maketitle

Phenomenological model building generally begins by assuming a particular field content, such as the states of the Minimal Supersymmetric Standard Model. These states may  be {\em motivated} by certain considerations such as the desire to solve a particular problem or explain a particular phenomenon -- or perhaps simply for elegance or other subjective considerations. But it is not possible, within such models themselves, to ask {\em whence} these particles came. It merely becomes the task of the experimentalist to find these states
and enumerate their properties.

Within string theory, however, the issue of the particle content is an {\em internal} issue which must be addressed. So too is the low-energy gauge group and the Yukawa interactions which dictate their interactions. To make concrete statements about phenomena relevant at low energies, all string models eventually must be considered in the supergravity limit in which it is possible to use an effective field theory to describe the dynamics of the fundamental fields. Prior to compactification, the field content of string theory is simply that of supergravity in ten or eleven dimensions, and this field content is remarkably unique. The famous variety of low energy outcomes in string model-building is the result of compactifying the theory to four dimensions. The resulting fields can often be determined via powerful and elegant mathematical means and the issue of spectra has been the primary focus of a large fraction of the string phenomenology community.

But there is also the issue of supersymmetry breaking and predicting the masses of the superpartners, as well as the dynamical breaking of additional gauge groups and possible discrete symmetries -- in other words, the very problems that consume the four-dimensional model builder. This second set of issues can be addressed in the low-energy four-dimensional effective field theory and can therefore be formally separated from the spectrum calculation that involves compactification. This is often the path taken by string phenomenologists who choose to focus on one or the other of these issues. Despite the simplicity of this approach, in a string-theoretic consideration to low energy phenomena the two sides are inherently intertwined. Illuminating these relations is the task of the experimental project at the LHC. Here we will survey just a few examples of specific models motivated from a variety of string constructions and the LHC signatures they imply.

\section{New States and New Interactions}

In this section we briefly describe extended supersymmetric models motivated by string theory, particularly of heterotic string theory compactified on orbifolds~\cite{Bailin:1999nk}.

\subsection{Anomalous Vector Boson Couplings}

Explicit string constructions often have one or more anomalous $U(1)$ gauge factors. By this we mean that the charges of the chiral states of the low energy theory do not satisfy the naive anomaly-cancellation conditions. In string models the low energy theory is nevertheless made mathematically consistent by the Green-Schwarz mechanism~\cite{Green:1984sg,Lopes Cardoso:1991zt}. The intricacies of this phenomenon are not relevant for our purposes here. It is sufficient merely to remark that the gauge bosons associated with these anomalous $U(1)$ factors typically acquire a mass via the Green-Schwarz mechanism. While very large masses (at or near the string scale) are common, particularly in explicit orbifold constructions, the masses of these $Z'$-bosons can in principle lie anywhere between the string scale and the scale of supersymmetry breaking.
If these $Z'$ bosons are relatively light (see i.e. \cite{Ghilencea:2002da,Feldman:2006ce}), and the states of the MSSM carry charges under the anomalous gauge factors, then the phenomenology for LHC physics will be similar to that of more conventional $Z'$ scenarios, such as those arising in grand unified theories \cite{Langacker:2008yv}.
Yet even in cases where the MSSM states are uncharged under anomalous $U(1)$ factors, the non-decoupling nature of anomalies implies that observable consequences may still exist.

The authors of~\cite{Antoniadis:2009ze} were motivated to consider cases in which non-vanishing mixed anomalies are present between a single anomalous $U(1)_X$ factor and the electroweak sector $SU(2)_L \times U(1)$ of the Standard Model. Integrating out heavy $U(1)_X$-charged fermions which run inside triangle diagrams results in new effective operators in the low energy Lagrangian. Among the new operators are those which produce triple gauge-boson vertices involving the anomalous $Z'$-boson and
gauge bosons of the Standard Model electroweak sector.

An interesting consequence for the LHC is the case of associated production of the $Z'$ with a vector boson of the Standard Model. The production cross-section depends on the mass of the $Z'$ as well as the type of Standard Model gauge boson with which it is produced. For example, for certain model parameters the associated production of such an anomalous $Z'$ with a photon has a cross section $\sigma \sim \order(1\,{\rm fb})$ for $M_{Z'} \sim 400\GeV$, while the cross-section for associated production with a $Z$ or $W^{\pm}$ drops below 1~fb at $M_{Z'} \sim 700\GeV$. Once produced, the $Z'$-boson decays back into Standard Model gauge boson pairs, producing a distinctive three-boson intermediate state before subsequent decays into leptons and/or jets. In a simple model the production cross-sections and branching fractions to Standard Model gauge bosons are controlled by only two phenomenological parameters.

%The Tevatron should have some sensitivity up to masses of order 750
%GeV with 10 fb$^{-1}$ of integrated luminosity.

One intriguing possibility for the LHC is the channel which involves
the decay $Z' \to \gamma Z$, which produces a prompt photon. This
can then be combined with the associated Standard Model gauge boson
from the production diagram to obtain either $Z\,Z\,\gamma$ or
$Z\,W^{\pm}\,\gamma$ intermediate states. The latter case is
particularly interesting for its unique topology and utility as a
discovery mode for this interaction.
For decays of the $Z$-boson to lepton pairs, the invariant mass of
the combination $\gamma \ell^+ \ell^-$ can be used to infer the mass
of the $Z'$-boson. The additional $W^{\pm}$ can be used for
triggering (by requiring a third lepton) or, if it decays
hadronically, by requiring two jets whose invariant mass reconstruct
the $W$-mass. For the study performed in~\cite{Antoniadis:2009ze}
only the $\gamma \ell^+ \ell^- \ell^{\pm}$ final state was
considered. Photons and leptons were required to have a
pseudorapidity $\eta < 2.5$ and minimum $p_T$ values of 10~GeV for
leptons and 50~GeV for the photons. The invariant mass of the
opposite-sign lepton pair was required to reconstruct the $Z$-mass
to within 5~GeV, and that of the system formed from the opposite
sign leptons and the photon was required to satisfy $m_{\gamma
\ell^+ \ell^-} > 500 \GeV$. A missing energy cut of \missET $\geq
10\GeV$ was also imposed. With these requirements the LHC reach for
such anomalous triple gauge-boson vertices was estimated to be in
the range $2\TeV \leq M_{Z'} \leq 4\TeV$ (depending on the model
parameters) in just 10~fb$^{-1}$ of integrated luminosity.

\subsection{Fractionally-Charged Exotics}

We often expect additional states charged under the Standard Model
gauge group to arise in the low-energy massless spectrum. If such
states come in vector-like pairs, that is if each state is
accompanied by a charge conjugate state in the supersymmetric
spectrum, then a gauge-invariant mass term for these exotics can be
constructed. The mass itself may be the result of the vacuum
expectation value of some Standard Model singlet, in much the same
way that a $\mu$-parameter can be generated from dynamical symmetry
breaking in theories such as the NMSSM~\cite{Balazs:2007pf}. In
principle these states can be of any mass provided they would have
escaped detection through direct production at colliders or through
the indirect effects of these states on rare processes.

In~\cite{Raby:2007hm,Wingerter:2007gg} a number of possible sets of
particles were identified that allow for gauge coupling unification
in the standard sense, but without requiring complete GUT
representations such as a $\mathbf{5}+\bar{\mathbf{5}}$ of $SU(5)$.
Just as with adding complete GUT multiplets, these states only
change the value of the unified gauge coupling at the high scale,
and not the scale of unification itself.

An example of such new states would be an ensemble of Standard Model
analogs $\lbr Q,\,L,\,E,\,E'\rbr$ plus their charge conjugate
superfields. The charge assignments under the Standard Model gauge
groups would be $Q\,(3,1)_{1/3}$, $L\,(1,2)_0$, $E\,(1,1)_{-1}$ and
$E'\,(1,1)_{\pm1}$ where the notation gives the representations
under $SU(3)\times SU(2)$ with the hypercharge given by the
subscript. The normalization here is such that the electric charge
of these states is given by $Q = T_3 + Y/2$, implying that these new
objects will all carry fractional electric charges. Such
fractionally-charged exotics are often consider a ``smoking gun''
for string-theoretic models~\cite{Schellekens:1989qb,Lykken:1996kc}.

While the ensemble of states given above are not complete
representations of $SU(5)$, they {\em do} transform as $(6,1) +
(1,2)$ + cc. under the product group $SU(6)\times SU(2)$. This
higher-rank symmetry group arises explicitly in certain
constructions of heterotic string theory on $Z_{6-II}$ orbifolds,
prior to breaking to the MSSM via the Pati-Salam group via Wilson
lines. The above states arise in one of the twisted sectors
associated with a $Z_2$ orbifolding of the larger internal dimension
of the $T^6$ compactification
manifold~\cite{Kobayashi:2004ud,Kobayashi:2004ya,Buchmuller:2005jr}.

Bound states comprising of these exotic quarks and those of the
Standard Model will also have fractional charges. One therefore
expects such a model to produce exotic baryons and mesons, similar
in nature and phenomenology to the R-hadrons of split
supersymmetry~\cite{Kraan:2005te}. Supersymmetry breaking effects
tend to make the scalars heavier than the fermions in the exotic
supermultiplets, and thus the lower mass fermions can be
approximately stable, allowing such hybrid hadrons to form. If such
states have masses greater than 200~GeV or so they may have evaded
current search limits~\cite{Fairbairn:2006gg}, but they may be
produced copiously at the LHC via Drell-Yan processes.

\subsection{$E_6$-based Exotics}

Many string constructions -- and almost all heterotic string
constructions -- proceed to the Standard Model gauge group through
an intermediate stage in which a residual $E_6$ symmetry is present.
Compactification effects break this $E_6$ structure and destroy
unification of Yukawa couplings, among other effects. But the field
content and superpotential interactions may still reflect an
underlying $E_6$ framework~\cite{Hewett:1988xc}. Such models provide
a natural embedding of the NMSSM framework for generating the
$\mu$-parameter~\cite{Langacker:1998tc,King:2005my,King:2005jy} and
have interesting consequences at the LHC.

Of particular interest are iso-singlet $SU(3)$ triplets $(D,D^c)$
which arise in vector-like pairs under the decomposition of the
fundamental $\mathbf{27}$ representation of $E_6$ under the Standard
Model gauge group. Depending on the discrete symmetries imposed on
the model (necessary to prevent rapid proton decay), these states
can mix with the Standard Model states, can behave as diquarks, or
can behave as leptoquarks. We emphasize that here we have both the
scalar and the fermion in the multiplet, and thus the phenomenology
of such objects at the LHC can be much richer than in traditional
scenarios of leptoquarks and diquarks.

The phenomenological consequences of such exotic states were
considered as part of the Constrained Exceptional MSSM
model~\cite{Athron:2008np,Athron:2009ue,Athron:2009bs,Athron:2009ed}.
In these studies the iso-singlet $SU(3)$ triplets were taken to be
supersymmetric leptoquarks or diquarks which couple only to the
third-generation states of the Standard Model.
Pair production of scalar exotics would give rise to processes such
as $pp \to t\bar{t} b\bar{b}$ for diquark couplings and $pp \to
t\bar{t} \tau\bar{\tau}$ for leptoquark couplings. The Standard
Model particles will undergo there own decays, giving rise to some
\missET in the final state. Such events will prove more challenging
to identify and reconstruct than decays directly to $e$,$\mu$ final
states, as is often assumed in scalar leptoquark searches. Fermionic
exotics in these models are able to decay to the two-body final
states such as $b\tilde{t},\,t\tilde{b}$ for diquarks and $\tau
\tilde{t},\,\tilde{\tau} t,\,\tilde{b} \nu_{\tau},\,b
\tilde{\nu}_{\tau}$ for leptoquarks. The superpartners will then
decay via normal cascade chains, producing final states such as
$b\bar{b}$, $t\bar{t}b\bar{b}$ and $t\bar{t}\tau^+ \tau^-$ but now
with substantial \missET signals. Thus this particular scenario
suggests a very b-jet rich and tau-rich environment at the LHC.
Separating the two exotic components from one another -- and from
the production of non-exotic MSSM states -- may be challenging at
the LHC.

This issue was studied in detail for iso-singlet $SU(3)$ triplets
in~\cite{Kang:2007ib}. In this analysis the exotics were assumed to
couple only to the first two generations of the Standard Model.
Given the much larger production cross-sections for fermionic states
charged under $SU(3)$ than scalars of the same mass, we would expect
fermionic exotics to be produced copiously at the LHC even for
relatively large masses ($M_{D_{1/2}} \lappeq 2\TeV$), while direct
production of scalar exotics (either in pairs or in associated
production with a Standard Model lepton) will generally require much
lower masses ($M_{D_0} \lappeq 800\GeV$).

%------------------ The five benchmark cases -----------------
\begin{table}
{\caption{\label{benchmarks} \footnotesize Five benchmark mass
patterns designed to illustrate the possible collider signatures of
exotic supermultiplets. All values are in GeV at the electroweak
scale.}}
\renewcommand{\tabcolsep}{0.8pc} % enlarge column spacing
\renewcommand{\arraystretch}{1.2} % enlarge line spacing
\begin{tabular}{@{}lccccc} \hline
Mass & A & B & C &
D & E \\
\hline
$M_{D_{1/2}}$ & 300 & 300 & 300 & 600 & 1000 \\
%
%$m_{D_{0}}$ & 400 & 400 & 1000 & 400 & 400 \\
%
%$m_{D_{0}^c}$ & 400 & 400 & 1000 & 400 & 400 \\
%
%$A_5$ & 350 & 150 & 100 & 600 & 1050 \\ \hline
%
%\multicolumn{1}{c}{} & \multicolumn{5}{c}{$U(1)_{\eta}$ Model} \\
%\hline
%
$M_{D_0^1}$ & 367 & 441 & 1024 & 388 & 318 \\
$M_{D_0^2}$ & 587 & 553 & 1053 & 932 & 1482 \\ \hline
\end{tabular}
\end{table}
%-----------------------------------------------------------------

Five benchmark scenarios were studied and the values of the exotic
scalar and fermion masses are given in Table~\ref{benchmarks}. For
cases~A-C the fermion was the lightest exotic particle, while for
cases~D and~E it was the scalar which was the lightest. The
phenomenology at the LHC depends greatly on which of these mass
orderings arises. All models were simulated at the LHC using {\tt
PYTHIA} + {\tt PGS4} for detector response with an integrated
luminosity of 5~fb$^{-1}$. In addition, an appropriately weighted
sample of Standard Model background as well as {\em supersymmetric}
background (in the form of Snowmass point~1A~\cite{Allanach:2002nj})
were included. As expected, the total production cross-section for
the supersymmetric exotics nearly equalled that of the total
production rate for MSSM states when the fermion was the lightest
exotic. Rates dropped by more than a factor of ten for cases~D
and~E. Supersymmetry discovery channels involving jets plus leptons
with \missET will significantly enhanced for all five scenarios. In
addition, events with high lepton multiplicity will favor pair
production of exotics, particularly for cases~A-C.

%========== Invariant mass of lepton1 + jet2 ==============
\begin{figure}[t]
\caption{\footnotesize \textbf{Invariant mass of hardest lepton
paired with softest jet in two jet, OS dilepton events.} Precisely
two jets, neither being B-tagged, were required, as were two
opposite-sign leptons. For the four cases where scalar production
was non-negligible a mass peak can be reconstructed near the
physical mass value for the lightest scalar.} \label{fig:jjll}
\begin{center}
\includegraphics[scale=0.4,angle=0]{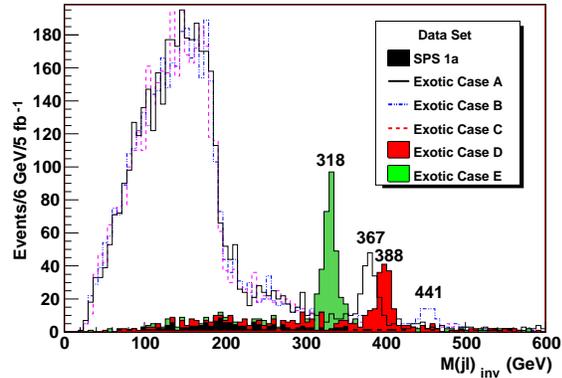}
\end{center}
\end{figure}
%%===================================================================

Figure~\ref{fig:jjll} gives the invariant mass distribution of the
hardest lepton and softest jet in events with precisely two jets and
two (opposite-sign) leptons. Jets were required to have at least 50
GeV of transverse momentum and events were vetoed if either jet was
B-tagged. A cut was made on the $p_T$ of the leading lepton of
50~GeV, and 20~GeV for the trailing lepton. Finally, we require the
events to be somewhat collimated along the event axis, so we require
the transverse sphericity to be no greater than~0.7. This final cut
significantly reduced the contamination from both Standard Model
processes and SPS~1a events (an acceptance rate of approximately
0.04\% for each).

For cases~A-C the invariant mass of the jet/lepton pair shows an
end-point just below 200~GeV. This correctly measures the mass
combination
\begin{equation}
M_{\rm inv}^{\rm edge}(\ell \, j) = \sqrt{\frac{(M^2_{D_{1/2}} -
M_{\tilde{\ell}}^2)(M^2_{\tilde{\ell}} -
M_{\chi_1^0}^2)}{M^2_{\tilde{\ell}}}} \, , \label{edge}
\end{equation}
via the on-shell cascade decay $D_{1/2} \to q \tilde{\ell} \to q
\ell \chi_1^0$. For cases A-C this happens to be very near the mass
difference between the fermionic LEP and the lightest neutralino.
Mass peaks arising from the scalar pair production with $D_0 \to q
\ell$ can be reconstructed for all scenarios in which there is
significant scalar production (case~C had only 38 scalar events in
\lumint~of data). The true mass value for the lighter scalar is
given over the corresponding peak in Figure~\ref{fig:jjll}. We note
that if a cut on missing energy of $\met \geq 50 \GeV$ were applied,
the scalar mass peaks would vanish from the distributions in
Figure~\ref{fig:jjll}, though the end-point in the distribution
associated with fermion pair-production would still be visible.

These peaks can be isolated and sharpened by making stricter cuts on
the data set, such as demanding $\met \leq 25 \GeV$, requiring the
scalar sum of $p_T$ values from the two jets and two leptons sum to
at least 400~GeV, and requiring the invariant mass of the lepton
pair to be at least 100~GeV. An important cross-check is to find the
same peak in the jet/lepton invariant mass distribution in
associated production of scalar leptoquarks through the process
$g\,q \to D_0 q$. We can isolate this process by requiring (a) at
least two jets without B-tags, the hardest jet having at least
200~GeV of transverse momentum and all others having $p_T \geq 50
\GeV$, (b) precisely one isolated lepton with $p_T \geq 50 \GeV$,
and (c) $\met \leq 20 \GeV$. Pairing the second hardest jet with the
single lepton gives a clear peak at the same mass values as those in
Figure~\ref{fig:jjll}.

Thus, in every one of the scenarios of Table~\ref{benchmarks} there
should be at least one exotic state, and occasionally two such
states, which can be identified at the LHC -- even with limited
initial data. With additional statistics it should be possible to
measure the masses of low-lying scalar mass eigenstates in all five
scenarios. Reconstruction of cascade decays with additional
integrated luminosity should also allow a determination of the
exotic fermion mass in all five cases.

\section{Heterotic Orbifold Compactifications}

\subsection{Spectra in Semi-Realistic Orbifold Models}

Recent years have seen a great deal of progress in the calculation
of the initial conditions for the low-energy effective supergravity
theories associated with heterotic orbifold models. These include
the particle spectrum, Yukawa couplings and low-energy gauge groups.
Most, but not all, of these models contain extra matter beyond the
MSSM field content, as alluded to in the previous section. If this
matter comes in vector-like representations then there is no
gauge-invariance argument to forbid a (supersymmetric) mass term for
these exotic states and a reasonable phenomenology can ensue.

%%%%%%%%%%%%%%%%%%%%%%%%%%%%%%%%%
Some of the recent results which are most economical in particle
content and of greatest interest phenomenologically involve
compactification of heterotic string theory on the $Z_2 \times Z_2$
orbifold~\cite{Blaszczyk:2009in}, the $Z_{12}$
orbifold~\cite{Kim:2006hv} and the $Z_6$
orbifold~\cite{Kobayashi:2004ud,Kobayashi:2004ya,Buchmuller:2005jr,Buchmuller:2006ik,Lebedev:2007hv}.
The latter is the most intensely studied and has been shown to have
a number of desirable phenomenological properties: the existence of
realistic three-family models, the ability to give mass to
vector-like exotics along flat directions, the presence of R-parity
in the low energy superpotential and sufficiently long-lived proton,
the possibility of generating Majorana mass terms for right-handed
neutrinos, and the consistency of the construction with such things
as gauge coupling unification and third-generation Yukawa/gauge
unification~\cite{Ratz:2007my}. The $Z_6$-II orbifold models
considered here are unusually efficient at generating realistic
low-energy initial conditions, suggesting that they constitute a
`fertile patch' in the string theory
landscape~\cite{Lebedev:2006kn,Lebedev:2008un}.

%%%%%%%%%%%%%%%%%%%%%%%%%%%%%%%%%%%%%%%%%%%%%%%%%%%%%%%%%
\subsection{Electroweak Symmetry Breaking}

Within orbifold compactification in heterotic string models one has
a so called large radius- small radius symmetry $R\rightarrow
\alpha'/R$. More generally one has an $SL(2,Z)$ symmetry and such a
symmetry is valid even non-perturbatively which makes it rather
compelling that this symmetry survives in the low energy theory.
Thus in order to simulate as much of the symmetry of the underlying
string theory as possible in a low energy effective theory one may
consider low energy effective theories with modular
invariance~\cite{fmtv,gnw}. The above leads one to consider an
effective four dimensional theory arising from string theory assumed
to have a target space modular $SL(2,Z)$ invariance
 $T_i\rightarrow T'_i=\frac{{a_iT_i-ib_i}}{{ic_iT_i+d_i}}$, $\bar
T_i\rightarrow \bar T'_i=\frac{a_i\bar T_i+ib_i}{-ic_i\bar
T_i+d_i}$, $(a_id_i-b_ic_i)=1,~~~ (a_i,b_i,c_i,d_i \in Z)$. While
the superpotential and the K\"ahler potential undergo transform the
 the  scalar potential $V$ defined by
$ V= e^{G}((G^{-1})^i_jG_iG^j+3) +V_D $, where
$G=K+\ln(WW^{\dagger})$  ($K$ is the K\"ahler potential and $W$ is
superpotential) is invariant  under the above  modular
transformations. Thus  one may require that modular invariance be
preserved even when supersymmetry is broken and specifically that
$V_{\rm soft}$ be modular invariant. Under  modular transformations
the chiral fields,  i.e., quark, leptons and Higgs fields will
transform and their transformations are fixed by their modular
weights. The low energy effective K\"ahler potential has the form
$K=D(S,\bar S) -\sum_i \ln(T_i+\bar T_i) + \sum_{i\alpha}(T_i+\bar
T_i)^{n_{\alpha}^i} C_{\alpha}^{\dagger}C_{\alpha}$ where
$C_{\alpha}$ are the chiral fields.

It is often useful to define dilation fraction $\gamma_S$ and moduli
fractions $\gamma_{T_i} $ such that $\gamma_s= (S+\bar S)G_S/\sqrt
3$,  $\gamma_{T_i}= (T_i+\bar T_i)G_{T_i}/\sqrt
3$~\cite{Brignole:1993dj,Brignole:1997dp}. The condition for the
vanishing of the vacuum energy gives one relation between the
dilaton and moduli fractions, i.e., $|\gamma_S|^2+\sum_{i=1}^3
|\gamma_{T_i}|^2 =1$. The soft breaking potential can now be
computed and takes the form
\beqn V_{soft}= m_{3/2}^2
\sum_{\alpha}(1+3\sum_{i=1}^3n^i_{\alpha}|\gamma_{T_i}|^2)
c_{\alpha}^{\dagger}c_{\alpha}+ \nonumber\\
(\sum_{\alpha\beta}B^0_{\alpha\beta}w^{(2)}_{\alpha\beta} +
\sum_{\alpha\beta\gamma}
A_{\alpha\beta\gamma}^0w_{\alpha\beta\gamma}^{(3)}+ H.c.) \nonumber \\
\eeqn
where  $n_{\alpha}^i$ are the modular weights for $C_{\alpha}$,
$w^{(2)}_{\alpha\beta} = \mu_{\alpha\beta} C_{\alpha}C_{\beta}$, and
$w^{(3)}_{\alpha\beta\gamma} = Y_{\alpha\beta\gamma}
C_{\alpha}C_{\beta}C_{\gamma}$. For the case when one assume
$\gamma_{T_i}=\gamma_T$, the vanishing of the vacuum energy
condition determines, $\gamma_T$ given $\gamma_S$, and thus
$\gamma_{T_i}$ are no longer independent variables. In this case,
one has only two independent parameters (aside from phases) which
are $m_{3/2}$ and $\gamma_S$. An interesting result that follows is
that $A^0$ and $B^0$ both have a dilaton front factor
$e^{D/2}$~\cite{Nath:2002nb}, and   this front factor can be
directly related to string gauge coupling constant so that $e^{-D}=
\frac{2}{g_{string}^2}$. Now in electroweak symmetry breaking one
typically eliminates $B^0$ in favor of $\tan\beta$. However, in an
effective field theory arising from strings, $B^0$ is determined in
terms of the moduli, and consequently $\tan\beta$ gets determined.

\begin{figure}
\centering
\includegraphics[height=4cm]{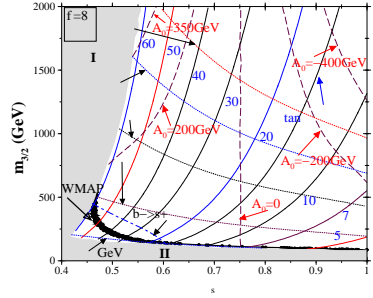}
\caption{ An exhibition of the
contours of constant  $A_0, \mu, \tan\beta$ in the
 $(\gamma_s-m_{3/2})$ plane for the case $\mu>0$.
% Right  panel:   An exhibition of the
%variation of sparticle masses with
%$m_{3/2}$ with $\gamma_s=0.75$ for the case when $\mu>0$.
 Taken from Ref.~\cite{Nath:2002nb}.}
\label{stringtanbeta}
\end{figure}

Figure~(\ref{stringtanbeta}) exhibits the determination of $A_0,
\mu$ an $\tan\beta$  for given values of $m_{3/2}$ and $\gamma_S$
the under constraints of radiative  breaking.
Figure~(\ref{stringspectrum}) gives an illustration of the sparticle
spectrum in this scenario for the case $\gamma_S=0.75$.  It is to be
emphasized that the phenomenon that $\tan\beta$ is determined is not
just specific to the class of models  discussed above but is a more
generic feature of string models.

\begin{figure}
\centering
\includegraphics[height=4cm]{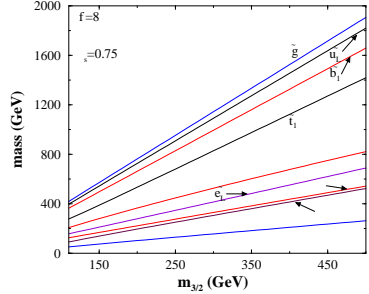}
\caption{
%Left panel: An exhibition of the
%contours of constant  $A_0, \mu, \tan\beta$ in the
% $(\gamma_s-m_{3/2})$ plane for the case $\mu>0$. Right  panel:
  An exhibition of sparticle masses with
 $\gamma_s=0.75$ for the case when $\mu>0$.
 Taken from Ref.~\cite{Nath:2002nb}.}
\label{stringspectrum}
\end{figure}

\subsection{Supersymmetry Breaking}

To complete the process of making contact with low-energy
observations the above ingredients must be brought together with
supersymmetry breaking in order to make meaningful predictions at
the LHC. In string-motivated models this supersymmetry breaking
generally involves non-vanishing auxiliary fields for the various
moduli in the theory, such as the dilaton $S$ and K\"ahler moduli
$T^i$ above. Thus supersymmetry breaking becomes intimately related
to the generating of a scalar potential for these fields and to the
issues of moduli stabilization generally.

The last two years has seen a return of attention to issues of
moduli stabilization and supersymmetry breaking in heterotic string
theory~\cite{Lebedev:2006qq} following on the recent progress in
building the effective Lagrangian for Type~IIA and Type~IIB
$D$-brane models -- both with and without additional flux
contributions to the Lagrangian~\cite{Grana:2005jc}. These studies
involve a number of different mechanisms for achieving moduli
stabilization and supersymmetry breaking while simultaneously
generating a vanishing (or slightly positive) vacuum energy.
Intriguingly, many (but not all) of these constructions share the
property that contributions to supersymmetry breaking from the
auxiliary fields of the moduli are comparable in magnitude to the
contributions arising from the superconformal
anomaly~\cite{Lowen:2008fm}. Such ``mixed-modulus/anomaly
mediation'' arises in other string contexts as
well~\cite{Choi:2004sx}, and was noted in K\"ahler stabilization of
heterotic orbifolds a decade ago~\cite{Gaillard:1999yb}.

The pattern of soft supersymmetry-breaking masses which arise in
these contexts is determined by a single parameter, $\alpha$, which
is related to the relative sizes of the two contributions to
supersymmetry breaking. The pattern has been named the ``mirage
pattern'' and takes the following approximate form at low energies
\begin{equation}
M_1:M_2:M_3 \simeq  (1+0.66\alpha) : (2+0.2\alpha) :(6-1.8\alpha)
\label{mirage_ratios} \end{equation}
where the case $\alpha=0$ is precisely the unified mSUGRA limit. We
note that for $\alpha>0$ one expects a gluino which is much lighter,
relative to the other gauginos, than is expected from the mSUGRA
paradigm. This has significant implications for LHC physics,
implying much higher event rates for events involving multiple jets
and missing transverse energy. The importance of measuring these
Lagrangian parameters for the goal of distinguishing amongst string
scenarios was recently emphasized in Ref.~\cite{Choi:2007ka}. An
initial study of the feasibility of measuring the parameter $\alpha$
at the LHC using targeted combinations of inclusive signatures
appeared in Ref.~\cite{Altunkaynak:2009tg}.

\section{D-Branes}
\label{db}

\begin{figure*}[t]
  \begin{center}
\includegraphics[width=6.0cm,height=5.5cm]{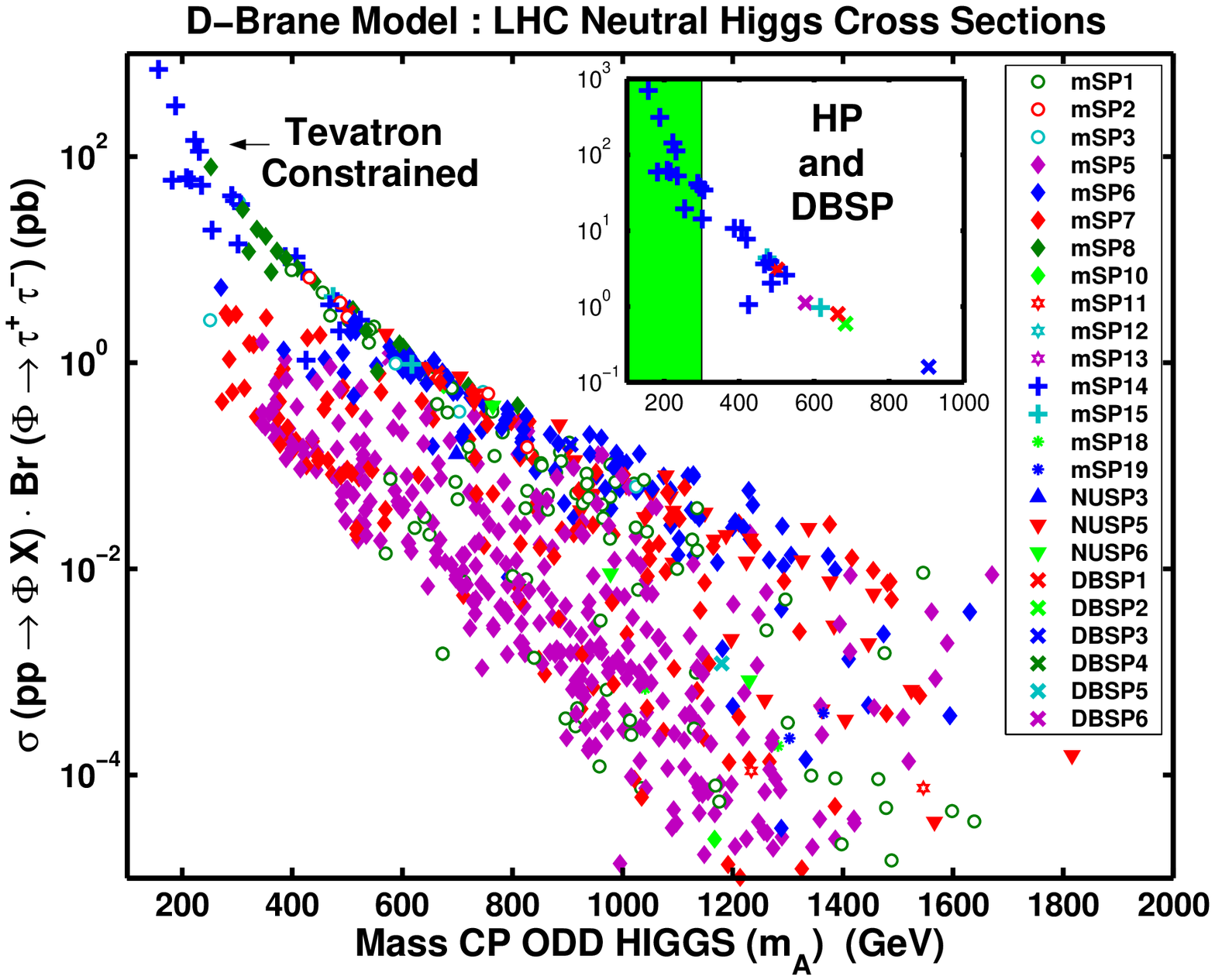}
\includegraphics[width=6.0cm,height=5.5cm]{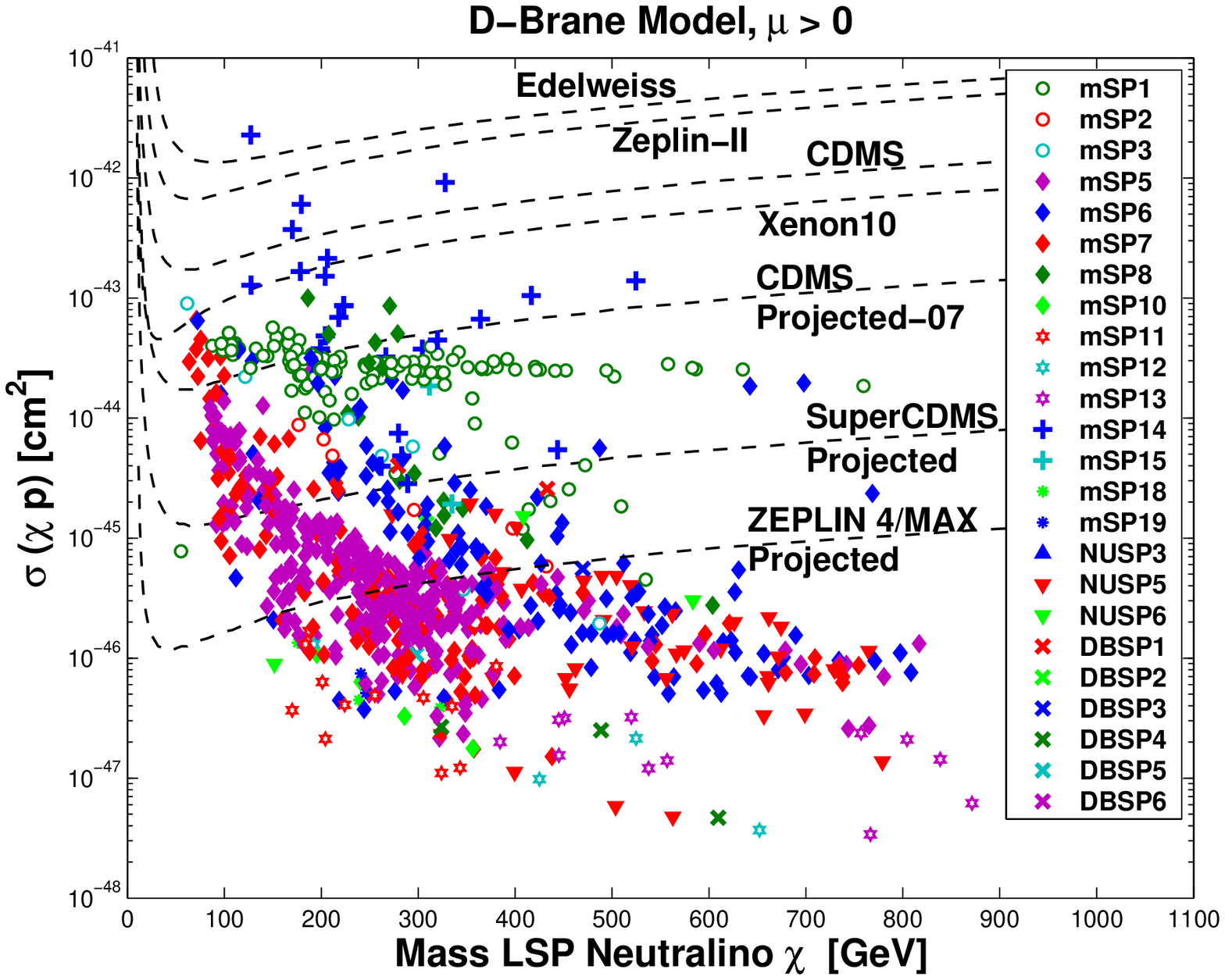}
\caption{\small  Predictions  in D-Brane models for $\mu>0$: The
Higgs  production cross section $\sigma_{\Phi\tau\tau}(pp)$ at the
LHC as a function of the CP odd  Higgs mass $m_A$ and the dark
matter direct detection signature space. (From
\cite{Feldman:2007fq}).} \label{KN}
  \end{center}
\end{figure*}

Recently there has been considerable progress in exploring the
phenomenology of D-brane models
with~\cite{Kors:2003wf,dsoft2,Kane:2004hm,Font:2004cx,Chen:2007zu,Feldman:2007fq}.
Of special relevance to low energy phenomenology is the nature of
soft breaking in D branes [For  recent reviews which include
discussions of soft breaking in D-brane models
see~\cite{Blumenhagen}].
 Here we discuss in the context of D-branes the possibility of Light Higgses, compressed SUSY
 spectra,  and implications for neutralino dark matter. In the first class
 of models we discuss it is found that the nature of soft breaking and constraints on the
 relic abundance of dark matter tend to favor the possibility of a light stau, chargino and CP-Odd/Even Higgs Bosons.
  As mentioned above the phenomenology of
D-branes is governed by the nature of D-branes and here there are
two main sectors  of the theory, the so called BPS 1/2 sector and
the 1/4 sector. We first discuss the analyses based on the BSP 1/2
class~\cite{Kors:2003wf}. This  early example of an
 effective string model where the soft terms have been computed is based on
toroidal  orbifold compactifications based on  ${\cal
T}^6/{Z}_2\times {Z}_2$ where ${\cal T}^6$ is taken
to be  a product of 3  ${\cal T}^2$ tori.   Here the moduli
sector consists of volume moduli $t_m$,  shape moduli $u_m$  $(m
=1,2,3)$  and the axion-dilaton field $s$. The K\"{a}hler metric of the $m^{\rm th}$ component of open strings
are split between common brane stacks $[a,a]$ and twisted open
strings connecting different brane
stacks $[a,b]$. The K\"{a}hler metric is deduced from dimensional reduction
and $T-$duality generalizing the previous known result for the heterotic string \cite{Kors:2003wf}
\beqn
{\mathcal K}^{[a,a]}_{m \bar m}= [(s+ {\bar s})(t_m+ {\bar t}_{\bar m})(u_m+ {\bar u}_{\bar m})]^{-1} \frac{4 \Re(f_a)}{ (1+\Delta^m_a)} \nonumber
\\\nonumber
{\mathcal K}^{[a,b]}_{\alpha \bar \beta} = \delta_{\alpha \bar \beta} (2 \Re(s))^{\theta_{a b}} \prod_{m=1}^3(2 \Re(t_m))^{\theta^m_{ab}}(2 \Re(u_m))^{\theta^m_{ab}} \nonumber
\eeqn
where $\Delta^m_a$ is a known function of the moduli and the background gauge fluxes, and the angles
$\theta^m_{ab} = \nu^m_{ab}-(1+\theta_{ab})$, $\theta_{ab} =\nu_{ab}/2-1$, parametrize the supersymmetry
preserving constraint in the open string sector $\nu_{ab} = \sum^3_{m=1}\nu^m_{ab} \in [0,2]$, with $\nu^m_{ab}$ defining the
relative angle between branes. Here this constraint can align the moduli vevs of $s,t_m$, and such a case
leads to the simplest situation of universal gaugino masses at the GUT scale (though of course this is not generic to the model).
Considered here is a 4-generation model where the brane stacks and associated winding numbers are well known~\cite{Blumenhagen}.
%
%\beqn
%M^2_{Q_L} &=&
% c^2\left( \frac{1}{3} -\sin^2(\theta) -2\alpha \cos^2(\theta)
%+\alpha \cos^2(\theta) [1+F_2]  \right) \ , \\
%M^2_{\bar U, \bar D} &=&
% c^2 \left( \frac{1}{3} - \frac{1}{2} \cos^2(\theta)
% + \cos^2(\theta) [  F_1 (\alpha-\frac{1}{2}) + (1-\alpha) F_2] \right) \ , \\
%M^2_{U,D} &=&
% c^2\left(-\frac{1}{6} + \frac{1}{2}\cos^2(\theta)
% + \cos^2(\theta) [-\alpha F_1  +(\alpha-\frac{1}{2})F_2] \right) \ , \\
%M^2_{H_U,H_D} &=&
% c^2\left(-\frac{1}{6} + (\frac{1}{2}-\alpha)\cos^2(\theta)
% + \cos^2(\theta) [\alpha F_1  +(-\frac{1}{2}+2\alpha)F_2] \right) \ , \\
%
%M^2_{S_U,S_D} &=&
% c^2\left( -\frac{1}{6} + \left( \frac{1}{2}-\alpha' \right)\cos^2(\theta)
% + \cos^2(\theta) [ (-\frac{1}{2} + 2\alpha' )
%  F_1  +  \alpha' F_2] \right) \ , \\\nonumber
%\eeqn
The soft scalars are then simple functions
of the graviton mass, the stack angle, and moduli vevs ($\sum_{i=1}^3 F_i=1,~~F_i = |\Theta_{t_i}|^2+ |\Theta_{u_i}|^2 $)and are given in full in  \cite{Kors:2003wf}.
%\be
%\sum_{i=1}^3 F_i=1,~~F_i = |\Theta_{t_i}|^2+ |\Theta_{u_i}|^2 \ , \quad i=1,2,3.
%\ee

%
%The soft scalars terms are related via
%\be
%M^2_{Q_L}= M^2_{L}\ , \quad
%M^2_{U,D}=M^2_{\nu,E}\ ,\quad
%M^2_{\bar U,\bar D}=M^2_{\bar \nu,\bar E} \ ,
%\ee
%In order to hammer home the general non vanishing of the tri-linear couplings we display
%\be
%A^0= -c\frac{e^{-\rho +\frac{D}{2}}}{\sqrt f} \cos (\theta)
%(\Theta_{t_2}e^{-i\gamma_{t_2}} +\Theta_{u_2} e^{-i\gamma_{u_2}}) \ .
%\ee
%However one may observe that a purely dilaton dominated scenario with $\theta=\pi/2$ would not have any soft
%tri-linear couplings in this model.
%Here, the parametrization $e^{-i\rho} = \langle \hat W \rangle / |\langle  \hat W \rangle|$
%where $\hat W $ is related to the gravitino mass through the standard
%relations of Ibanez et. al in \cite{nuni} and the $\gamma_{t_2,u_2}$
%are phases that may contribute to CP violation.  The moduli enter explicitly through  \cite{Nath:2002nb}
%\beqn
%D &=&-ln(s+\bar s) ~, \\
%f&=& \prod^{3}_{m=1} (t_{m}+{\bar t}_m)\prod^{3}_{m=1} (u_{m}+{\bar u}_m)~.
%\eeqn
In
the analysis we ignore the exotics, set $F_3=0, 0\leq F_1\leq  1$,
 and use
 the naturalness assumptions motivated by SUGRA analyses  with $\mu>0$.
The specific parameter space consists of the of the gravitino mass
$m_{3/2}$, the gaugino mass $m_{1/2}$, the tri-linear coupling $A_0$
(which is in general non-vanishing),  $\tan\beta$, the stack angle
$\alpha$ ($0\leq  \alpha \leq \frac{1}{2}$), the  Goldstino $\theta$
angle , and the the moduli  VEVs $\Theta_{t_i}$, $\Theta_{u_i}$
$(i=1,2,3)$~\cite{Feldman:2007fq}. It is found that the models is
dominated by mSPs (mass supergravity patterns) similar to those seen
in minimal and non-universal SUGRA models~\cite{Feldman:2007zn}.
However six new patterns (at isolated points ) emerge. Specifically
all the Higgs Patterns~\cite{Feldman:2007zn} (HPs where the next to
lightest mass is that of the CP-odd Higgs denoted by mSP14-mSP16)
are seen to emerge in good abundance.  Regarding the new  patterns
we label these patterns D-Brane SUGRA Patterns (DBSPs) since the
patterns arise in the SUGRA field point limit of the D-Branes.
Regarding the new  patterns we label these patterns D-brane Sugra
Patterns (DBSPs) since the patterns arise in the SUGRA field point
limit of the D-branes. Specifically we find six new patterns  $\rm
DBSP(1-6)$ as follows \begin{equation}
\begin{array}{ll}
{\rm DBSP1:}  ~\na < \sta < \snl  <A/H ~;\\
{\rm DBSP2:}  ~\na < \sta < \snl <\slr ~;\\
{\rm DBSP3:}  ~\na < \sta < \snl  <\snm ~;\\
{\rm DBSP4:}  ~\na < {\tilde t}_1  < \sta  <\snl ~;\\
{\rm DBSP5:}  ~\na < \snl  < \sta <\snm ~;\\
{\rm DBSP6:}  ~\na < \snl  < \sta <\cha ~.
\end{array}
\end{equation}

 The analysis of the Higgs  production cross section
$\sigma_{\Phi \tau \tau}(pp)$ in the D-Brane models  at the LHC is
given in the left panel of left panel of Fig.(\ref{KN}).
 The
analysis shows that the HPs dominate the Higgs production cross
sections.  One also finds that the $B_s\to \mu^+\mu^-$ experiment
constraints the HPs in this model~\cite{Feldman:2007fq}. The scalar
dark matter   cross  sections are  given  in the right panel of
Fig.(\ref{KN}).  Here also one finds  that the Higgs Patterns
typically give the largest scalar cross sections followed by the
Chargino Patterns (mSP1-mSP3)
 and then by  the Stau Patterns.
Further, one finds a  Wall  of  Chargino Patterns developing which
enhance the discovery possibilities of the chargino patterns
(see~\cite{Feldman:2007fq} for further details).

\section{Compressed Spectra in Intersecting D-Brane Models}
\label{db2} Another interesting class of intersecting D-Brane models
is motivated by the analyses
of~\cite{dsoft2,Font:2004cx,Bertolini:2005qh}. The specific class of
models considered here is with $u$ moduli breaking. The chiral
particle spectrum arises from intersecting branes with supporting
gauge groups $SU(3)_C \times SU(2)_L$ and $U(1)_a$,$ U(1)_c$,
$U(1)_d$ and $U(1)_Y$, wherein the the anomalous $U(1)= U(1)_a +
U(1)_d$  is assumed cancelled by a Green-Schwarz mechanism  giving a
Stueckelberg mass to the U(1) gauge boson. The K\"{a}hler metric for
the twisted moduli arising from strings stretching between stacks
$P$ and $Q$ for the BPS $1/4$  sector is taken in the form similar
to~\cite{Font:2004cx,Bertolini:2005qh,Kane:2004hm} \beqn \tilde{\cal
K}_{PQ} & = & \tilde{\cal K}_{\phi} \prod_{j=1}^3 \;
\biggr[\frac{{\Gamma}(1-{\theta}^j_{PQ})}{{\Gamma}({\theta}^j_{PQ})}\biggr]^{\sigma/2}
(t^j+\bar{t}^j)^{-{\theta}^j_{PQ}}~, \nonumber \\
 \tilde{\cal K}_{\phi} & = & e^{{\phi}_4}\,e^{{\gamma}_E\,\sum_{j=1}^{3}
{\theta}^j_{PQ}}
\eeqn
and he K\"{a}hler metric for $1/2$ BPS brane configurations is given by
\beqn\tilde{\cal K}^{\tiny \rm Higgs}_{PQ} =
\left((s+\bar{s})(u^1+\bar{u}^1)(t^2+\bar{t}^2)(t^3+\bar{t}^3)\right)^{-1/2}
\nonumber
\eeqn
\noindent where ${\theta}^j_{PQ} = {\theta}^j_{P}-{\theta}^j_{Q}$
is the angle between branes in the $j^{th}$ torus and ${\phi}_4$
is the four dimensional dilaton  and is a logarithmic function of $\Re{(s)}\prod^{3}_{i=1} \Re{(u^i)}$
while $\sigma$ is set to unity in what follows.
The gauge kinetic function is given in terms of products of the brane integers and the $s,u$ moduli
\begin{equation}
f_P=k_P^{-1} (n_P^1n_P^2n_P^3s-n_P^i m_P^j m_P^k u^i )~~i,j,k~{\rm cyclic},
\end{equation}
where the brane integers are given in~\cite{Kane:2004hm}. A useful
parametrization of the soft parameters is in terms of angle $\alpha$
(the free angle between the $P^{th}$ brane and the orientifold plane
of and the $j^{th}$ torus which  is assumed factorized ) and the
real parts of the $u_1,t_2,t_3$ moduli, and $\Theta_2,\Theta_3$ for
the choice $\rho=1$. The soft terms depend logarithmically on  the
moduli and poly-gamma functions of the angle $\alpha$. The
generalized unification constraints on gaugino masses  are  as
follows~\cite{Kane:2004hm,Feldman:2009jg} \beqn
M_{\tilde g}&=&\frac{9 \rho ^4\sqrt{3}m_{3/2}\Theta _1e^{-i \gamma_1}\Re(u^1)}{\Re (s)+9 \rho^4 \Re(u^{1} )}\\
M_{\tilde W}&=&\sqrt{3}m_{3/2}\Theta _2e^{-i \gamma _2}
\eeqn
\beqn
M_{\tilde B} &=\frac{3\sqrt{3}m_{3/2}\rho ^2\left(12 \rho ^2 \Theta _1e^{-i \gamma _1} \Re(u^1)+\Theta _3e^{-i \gamma _3}\Re(u^3)\right)}{4\Re(s)+36 \rho ^4\Re(u^1)+3\rho ^2\Re(u^3)}.
\nonumber
\eeqn
\begin{table}[t]
   \begin{center}
\begin{tabular}{|c||c|c|}
\hline\hline
Sparticle &  D6  & mSUGRA     \\
  type    &   Mass/GeV   &  Mass/GeV
\\\hline\hline
$m_h$      & 113.9      & 113.6    \cr
 $\na$    & 209.0       & 208.8      \cr
 $\cha$    & 229.1     & 388.6   \cr
 $\nb$    &  229.5      & 388.8   \cr
 $\sta$    & 404.2       & 433.3   \cr
\hline
 $\ser,\smr$  & 464.4   & 637.8\cr
 $\sta$      & 547.6       & 929.2\cr
 $\g$      & 760.4        & 1181.4 \cr
$ m_{{\rm max} ={\tilde{s},\tilde{d}}_L}$ &882.2  &  $ m_{{\rm max} ={\tilde{s},\tilde{d}}_L}$ 1210.4 \cr
\hline\hline
 \end{tabular}
\caption{\small  Intersecting D-Brane model (D6) and mSUGRA; a comparison.
The LSP neutralino mass and light Higgs masses are almost identical, yet there is a
a very different pattern of gaugino mass scaling seen in the D6 model relative to that which is expected
in mSUGRA, and further, there is a compressed spectra in the D6 model case.
The hierarchical mass pattern for the first 4 sparticles are the same. (\cite{Feldman:2009jg}.)}
\label{compress}
\end{center}
 \end{table}
\begin{table}[t]
   \begin{center}
\begin{tabular}{|c||c|c|}
\hline\hline
      D6  &  mSUGRA \\
$(\tilde{B},\tilde{W},\tilde{H}_1,\tilde{H}_2)$ &   $(\tilde{B},\tilde{W},\tilde{H}_1,\tilde{H}_2)$     \cr  (0.985,-.133,.104,-.0399)&  (0.994,-.017,.101,-.041)    \cr
 $\sigma^{\rm SI}_{\na p}= 7.4 \times 10^{-9}$ pb    & $\sigma^{\rm SI}_{\na p}= 1.4 \times 10^{-8}$ pb\cr
 $\Omega h^2= 0.099$ co-annh. &  $\Omega h^2= 0.095$ $b \bar b,\bar\tau\tau$\cr
\hline\hline
 \end{tabular}
\label{dist}
\caption{\small Same two models given in Table \ref{compress}; the Intersecting D-Brane model (D6) and a minimal SUGRA model.
Both models produce the  correct relic density, but through very different means,
the D6 model co-annihilated through both gaugino co-annihilations and slepton co-annihilations.
The mSUGRA model annihilated into heavy flavors.  The wino component is substantial in the D6 case. (\cite{Feldman:2009jg}.) }
\end{center}
 \end{table}
In Table(\ref{compress}) a useful and illustrative  comparison is given of 2 models;
one from the D-Brane
model (which we shall call D6) and the other from mSUGRA.
Table(\ref{compress}) actually provides some generic features over
the parameter space investigated in the D6 model. First,
the two model points live in the same 4 particle mass hierarchy with
degenerate LSP mass and light CP even Higgs mass. From Table(\ref{compress})
one observes however that the gaugino mass ratios of these models are very different.
In particular, the D6 model has a rather large wino component for a thermal relic (see Table.(\ref{dist})).
Importantly, the D6 model SUSY scale of superparticle masses are
compressed relative to the mass scale of the mSUGRA model. Thus, the LSP masses
are effectively identical, however the NLSP mass in the D6 model is about 160 GeV lighter than in
the mSUGRA case considered and we note in the D6 case the relevance of the lighter
gluino mass; indeed it is several hundred GeV lighter than the mSUGRA case.
In Table.(\ref{dist})  a direct comparison of dark matter implications for a
the mSUGRA bino-like case is shown, along with a significant mix of higgsino, while
the Bino-wino admixture seen in the D6 model point yields  different annihilation channels
allowing it satisfy the relic density constraints within a thermal paradigm.

Some general conclusions regarding the scale of the supersymmetric
particles in different model classes have recently been
emphasized~\cite{Feldman:2009jg}. The sparticle mass hierarchy
concept is extremely useful for sorting out SUSY. There are cases
however where it does not provide the full picture. In particular,
the mass hierarchy of the sparticles may be identical for the
lightest particles in the spectrum, however the scaling of gaugino
masses is a crucial ingredient and can be vastly different depending
on the pattern of softbreaking.  In conjunction with the above, it
is possible that with non-universalities, which are generic not only
to GUT models, but also to D-brane models, that the spectrum of
sparticles may be compressed. The lightness of the SUSY scale in
these models make them very appealing for collider based studies at
the LHC.

\section{M-Theory on Manifolds of $G_2$ Holonomy}

\subsection{Model description and soft terms}

The M-theory vacua we are interested in here is the fluxless
M-theory compactifications on $G_2$ manifolds where all
compactification moduli are stabilized by non-perturbative gauge
dynamics in the hidden sector~\cite{Acharya:2006ia,Acharya:2007rc}.
In addition, this strong gauge dynamics spontaneously breaks
supersymmetry and naturally generates a hierarchically small
supersymmetry breaking scale in the visible sector via dimensional
transmutation. Generically, the supersymmetric standard model
particles lives in a three-dimensional submanifold of the $G_2$
manifold which generically does not intersect the three-dimensional
submanifold where the strong gauge dynamics resides. Therefore, the
mediation of supersymmetry breaking to the MSSM sector is through
the Planck suppressed operators, and is of the ``gravity mediation"
type. This implies that the soft supersymmetry breaking terms are
expected to be of the same size as the gravitino mass.

However, gaugino masses are actually suppressed in these models
because there is no tree-level coupling between the dominant SUSY
breaking field and the gauge superfields. In the detailed analysis,
we find gaugino masses are generally one-loop suppressed compare the
scalars, and therefore the anomaly mediated contribution to gaugino
are necessary to be included. Thus, the soft supersymmetry breaking
pattern is such that there is a large mass splitting between
gauginos and scalars, and the low energy phenomenology at the weak
scale is mainly determined by the gaugino sector. Unlike
split-SUSY~\cite{ArkaniHamed:2004fb}, the Higgsinos in these vacua
are as heavy as scalars and also decoupled in the low energy. This
gives the low scale gaugino masses large threshold corrections from
the Higgs-Higgsino loop. Generically, the wino is the LSP for
$G_2$-MSSM models with light spectra, but a wino-bino mixture is
also allowed particularly for heavier spectra.

The $G_2$-MSSM models have a distinctive spectrum. One finds that at
the compactification scale ($\sim M_{\mathrm{GUT}}$), the gauginos
are light ($\lappeq 1$~TeV) and are suppressed compared to the
trilinears, scalar and  higgsino masses which are roughly equal to
the gravitino mass ($\sim 30-100$~TeV). At the electroweak scale,
the lightest top squark turns out to be significantly lighter than
the other squarks ($\sim 1-10$~TeV) because of RGE running. In
addition, there are significant finite threshold corrections to bino
and wino masses from the large Higgsino mass. Radiative electroweak
symmetry breaking is generic and $\tan \beta $ is naturally
predicted from the structure of the high scale theory to be of
$\mathcal{O}$(1). The value of $m_{Z}$ is fine-tuned, however,
implying the existence of the Little-hierarchy problem, which,
because of the larger scalar masses is worse than the usual little
hierarchy.
%
%In Table \ref{G2-benchmark}, the ``microscopic" input parameters and
%corresponding low-scale mass spectra are shown for four benchmark
%$G_2$-MSSM models.
%
These models are consistent with the precision gauge coupling
unification.

\subsection{LHC Phenomenology}

Given the fact that the only light superpartners in the $G_2$-MSSM
framework are gauginos, their productions dominate the superpartner
productions. The primary production modes for the $G_2$-MSSM models
are gluino pair production ($\tilde{g}\,\tilde{g}$),
neutralino-chargino associate production ($\tilde\chi_1^0\,\tilde
\chi^{\pm}_1$) and chargino pair production
($\tilde\chi^{+}_1\,\tilde \chi^{-}_1$). Table~\ref{XSEC} shows the
production cross sections for the four $G_2$-MSSM benchmark models.

\begin{table}[th]
%\begin{center}%
\caption{Cross sections of dominant production modes in~pb for four
$G_2$-MSSM benchmark models at the LHC. The cross sections are
calculated using {\tt PYTHIA}~\cite{PYTHIA}.} \label{XSEC}
\begin{tabular}
[c]{|lcccc|}\hline Channel & BM-1 & BM-2 & BM-3 & BM-4
\\\hline
$pp\rightarrow \tilde{g}\,\tilde{g}$ & $0.25$ & $1.9$ & $0.49$ &
$8.6$
\\\hline
$pp\rightarrow \tilde\chi_1^0\,\tilde \chi^{\pm}_1$ & $6.4$ & $8.1$
& $1.6$ & $8.4$
\\\hline
$pp\rightarrow \tilde\chi^{+}_1\,\tilde \chi^{-}_1$ & $2.2$& $2.7$&
$0.5$ & $2.8$
\\\hline
\end{tabular}
%\end{center}
\end{table}

\begin{table}[th]
%\begin{center}%
\caption{Decay channels and branching ratios of gluino for the four
$G_2$-MSSM benchmark models. The branching ratios are calculated
using SUSY-HIT~\cite{Djouadi:2006bz}.} \label{GLUINO-DECAY}
\begin{tabular}
[c]{|lcccc|}\hline Channel & BM-1 & BM-2  & BM-3 & BM-4
\\\hline
$\widetilde{g}\rightarrow\widetilde{\chi}_{1,2}^{0}t^{\mp}t^{\pm}$ &
$37\%$ & $39\%$ & $62\%$  & $36\%$  \\\hline
$\widetilde{g}\rightarrow\widetilde{\chi}_{1}^{\pm}t^{\mp}b^{\pm}$ &
$25\%$ & $21\%$ & $14\%$  & $16\%$  \\\hline
$\widetilde{g}\rightarrow\widetilde{\chi}_{1,2}^{0}b^{\mp}b^{\pm}$ &
$8\%$ & $9\%$ & $5\%$  & $10\%$  \\\hline
$\widetilde{g}\rightarrow\widetilde{\chi}_{1}^{\pm}q^{\mp}{q'}^{\pm}$
& $18\%$ & $19\%$ & $11\%$  & $21\%$  \\\hline
$\widetilde{g}\rightarrow\widetilde{\chi}_{1,2}^{0}q^{\mp}q^{\pm}$ &
$11\%$ & $12\%$ & $7\%$  & $15\%$  \\\hline
\end{tabular}
%\end{center}
\end{table}

The most interesting signals at hadron collider come from the gluino
pair production. Since $m_{\tilde{q}} > m_{\tilde{g}}$, the produced
gluinos proceed through a three-body decay into two quarks and
either a $\tilde \chi_2^0$, $\tilde \chi_1^0$, or a $\tilde
\chi_1^{\pm}$. Table~\ref{GLUINO-DECAY} shows the dominant decay
modes and the associated branching ratios for the four benchmark
models. One can see that the majority of gluino decay modes include
a pair of either top or bottom quarks, or a combination of both.
This is due to the fact that the RGE running significantly reduce
the stop mass compared to other other squarks given the small
$\tan\beta$. The top quark decays exclusively as: $t \rightarrow W +
b$, which results in at least two b-jets per decay, and four b-jets
for a $\tilde{g}\,\tilde{g}$ event for these modes. Therefore, a
typical signature for the $G_2$-MSSM models is multi-bjets plus
missing $E_T$.

There are also a fair number of leptonic events. The leptonic events
have two sources - firstly, the tops decay to $W$s which could decay
semi-leptonically. Secondly, the $\tilde \chi_2^0$ produced from
$\tilde{g} \rightarrow t\,\bar{t}\,\tilde \chi_2^0$ decays
predominantly as: $\tilde \chi_2^0\rightarrow\tilde
\chi_1^{\pm}\,W$, which could again decay semi-leptonically.
Therefore, one has an observable fraction of multi-lepton events. An
important point to notice is that since all leptons come from $W$
bosons, one expects no flavor correlation in opposite-sign dilepton
events. Finally, since gluino pair production is the dominant
mechanism leading to observable lepton events, the single lepton and
dilepton charge asymmetry is expected to be very small.

%=(1)=======================================
%\begin{comment}
\begin{figure}[t]
\caption{\label{asymm}\footnotesize{A particular slice of footprint
for the models studied. The one-lepton charge asymmetry (only
includes $e$ and $\mu$) is defined as
$A_c^{(1)}\equiv\frac{N_l^{+}-N_l^-}{N_l^{+}+N_l^-}$. The
  SSDF/1tau signature is defined as the ratio of the number of events with SSDF
  dilepton and the number of events with 1 tau lepton. All models are simulated
  with 5$fb^{-1}$ luminosity in PGS4 with L2 trigger.
  All signatures include a least two hard jets and large missing transverse energy. }}
\begin{center}
      \includegraphics[scale=0.45]{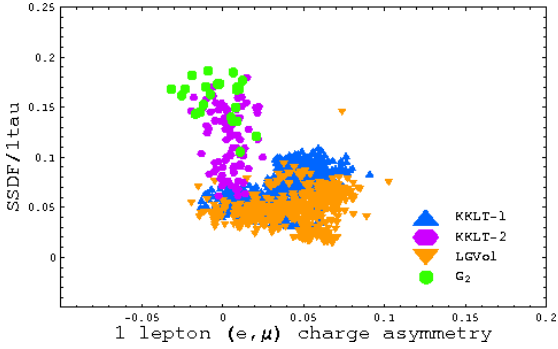}%}
\end{center}
\end{figure}
%\end{comment}
%=================================================================

%=(2)=======================================
%\begin{comment}
\begin{figure}[th]
\caption{\label{b-jet}\footnotesize{Two-dimensional slices of the
footprint of the three string-SUSY models. All models are simulated
with 5$fb^{-1}$ luminosity in PGS4 with L2 trigger. All signatures
include a least two hard jets and large missing transverse energy.}}
\begin{center}
      \includegraphics[scale=0.45]{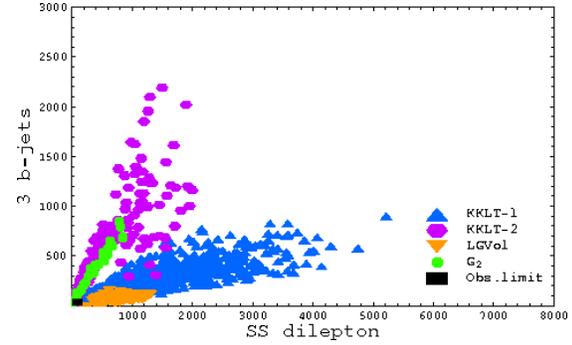}%}
\end{center}
\end{figure}
%\end{comment}
%=================================================================

Although direct production of electroweak gauginos $\tilde
\chi_1^{+}\tilde \chi_1^{-}$ and $\tilde \chi_1^0\tilde
\chi_1^{\pm}$ have large cross sections, these events are difficult
to observed. The first channel gives rise to events with two LSPs
plus some very soft particles from chargino decay, which have very
small missing $E_T$ because it is the vector sum of the $P_T$ of the
visible objects. The second channel can gives rise to additional
$W$-bosons, but again the missing $E_T$ is small because of the same
reason. Therefore, events from both channels are difficult to
trigger on since there is no hard jets or a large missing $E_T$.
%\footnote{\normalsize There could be leptons from $W$s, but it
%suffers from the huge Standard Model background.}

From the discussion above, there are three main features for the
collider signatures of $G_2$-MSSM. First,  squark-squark production
and squark-gluino associated production are negligibly small compare
to the gluino-gluino production. Therefore, there is almost no
lepton charge asymmetry in the signal events. See Fig.~\ref{asymm}
for example. Second, there is an enhancement in the faction of
events with b-jets, but no enhancement for the events with
$\tau$-lepton. See Fig.~\ref{b-jet} for the b-jet signature. Third,
the mass splitting between lightest chargino and LSP is slightly
larger than the charged pion mass. This could result in a charged
track that `kinks' when the lightest charginos decays to very soft
hadrons or leptons. Also possible is a `track-stub'; a clear,
charged track that appears to vanish when the soft decay products
are not detected. This latter scenario requires dedicated off-line
analysis to resolve.

\section{F-Theory Models}

\subsection{Review of F-theory GUTs}

In F-theory GUTs, the defining features of the GUT model are
determined by the worldvolume theory of a seven-brane which fills
our spacetime and wraps four internal directions of the six hidden
dimensions of string theory. The chiral matter of the MSSM localizes
on Riemann surfaces in the seven-brane, and interaction terms
between chiral matter localize at points in the geometry. As argued
in~\cite{Heckman:2008qt}, crude considerations based on the
existence of a limit where the effects of gravity can decouple
imposes sharp restrictions on the low energy content of the
effective field theory. In particular, because such models admit a
limit where the effects of gravity can decouple, they are
incompatible with mechanisms such as gravity mediation. Rather, in
F-theory GUTs the effects of supersymmetry breaking are communicated
to the MSSM via gauge mediation.

From the perspective of the low energy effective theory, the
defining characteristic of F-theory GUTs is that it constitutes a
deformation away from a high scale minimal gauge mediation scenario.
This is due to the fact that the theory contains an anomalous
$U(1)_{\rm PQ}$ gauge symmetry. This anomaly is canceled via the
generalized Green-Schwarz mechanism. The essential point is that
this introduces additional higher dimension operators into the
theory which have the effect of shifting by a universal amount the
soft scalar masses. For example, in a model with $N_{5}$ vector-like
pairs of messenger fields in the
$\mathbf{5}\oplus\overline{\mathbf{5}}$ of $SU(5)$, the masses of
the gauginos and scalar superpartners scale as:
\begin{eqnarray}
m_{\rm gaugino} & \sim& N_{5}\frac{\alpha}{4\pi}\Lambda \\
m_{\rm scalar} & \sim& \sqrt{N_{5}}\frac{\alpha}{4\pi}\Lambda +
e_{\Phi}\Delta_{\rm PQ}^{2}
\end{eqnarray}
where in the above, $\alpha$ is the Standard Model gauge coupling,
and $\Lambda=\frac{F_{X}}{x}$. The charge $e_{\Phi}$ is given by
$e_{\Phi}=-1$ for chiral matter and $e_{\Phi}=+2$ for the Higgs. To
leading order, the gaugino masses and trilinear couplings are
unchanged by this deformation.

The PQ deformation parameter $\Delta_{PQ}$ of F-theory GUTs lowers
the squark and slepton soft scalar masses in relation to the value
expected from a high messenger scale model of minimal gauge mediated
supersymmetry breaking. At $\Delta_{PQ}=0$, F-theory GUTs reduce to
a high messenger scale mGMSB model. In fact, the cosmology of
F-theory GUTs suggest a lower bound on $\Delta_{\rm PQ}$ on the
order of $\Delta_{\rm PQ}\gappeq 50\GeV$~\cite{Heckman:2008jy}.
There is also an upper bound to the size of $\Delta_{\rm PQ}$
because increasing $\Delta_{\rm PQ}$ decreases the soft masses of
the squarks and sleptons. Thus, for large enough values of
$\Delta_{\rm PQ}$ on the order of $500$~GeV (the precise value of
which depends on $\Lambda$ and the number of messenger fields), the
low energy spectrum will contain a tachyon. Depending on the number
of messengers as well as the size of the PQ-deformation, either a
bino-like neutralino, or a lightest stau could be the NLSP. Due to
the fact that the scale of supersymmetry breaking is so high, the
NLSP decays outside the detector, effectively behaving as a stable
particle.

In the specific context of F-theory GUTs, the $\mu$ term is roughly
given as:
\begin{equation}
\mu\sim\frac{F_{X}}{M_{X}^{KK}}\, ,
\end{equation}
where $M_{X}^{KK}\sim10^{15}$ GeV is a Kaluza-Klein mass scale of a
GUT singlet in the compactification. Thus, obtaining the correct
value of $\mu$ requires:
\begin{equation}
F_{X}\sim10^{16}-10^{18}\GeV^{2}\, .
\end{equation}
This range of values for $F_{X}$ implies that the mass of the
gravitino is $\sim 10-100$~MeV. Moreover, the fact that the scale of
supersymmetry breaking is relatively high compared to other models
of gauge mediation implies that the NLSP will decay outside the
detector due to its long lifetime.

The rough range of values for $\Lambda$ extends from $\Lambda\sim
10^{4}$ to $\Lambda\sim 10^{6}$. Beyond this range, the
mini-hierarchy problem is exacerbated. In fact, we shall typically
consider a smaller range on the order of:
\begin{equation}
10^{4}\GeV\lappeq \Lambda \lappeq 2\times10^{5}\GeV\, ,
\end{equation}
because for larger values of $\Lambda$, the masses of the gluinos
and squarks would be too heavy to be produced at the LHC. Finally,
in the context of F-theory GUTs, the $B\mu$ term and the A-terms all
vanish at the messenger scale. Thus, in this class of models, $B\mu$
and the A-terms are radiatively generated, and $\tan\beta$ is
typically in the range of $20-40$.

\subsection{LHC phenomenolgy}

%%=(3)=======================================
%%\begin{comment}
%\begin{figure}[th]
%\caption{\label{FIG:ONEMESSSPEC}\footnotesize{Plot of the mass
%spectrum of F-theory GUTs with $N_{5}=1$, $\Lambda=1.3\times10^{5}$
%GeV, and minimal (red, left part of each column) and maximal (blue,
%right part of each column) PQ-deformation.}}
%\begin{center}
%%      \includegraphics[width=4.0in ]{TOGETHER.pdf}
%\end{center}
%\end{figure}
%%\end{comment}
%%=================================================================

%\begin{figure}[ptb]
%\begin{center}
%\includegraphics[
%height=3.4272in,
%width=5.719in
%]{MGLUINO-MSQUARK-FTH1.pdf}
%\includegraphics[
%height=3.4272in,
%width=5.719in
%]{MGLUINO-MSQUARK-FTH2.pdf}
%\end{center}
%\caption{Plots of squark masses versus the gluino mass in F-theory GUTs. The top and bottom plots are for
%one and two messenger models respectively. The upper line in the plots
%corresponds to $m_{\tilde{g}}$, while the lower one corresponds to
%$m_{\tilde{g}}-m_{t}$. These figures imply that the decay of the gluino in one
%messenger models proceeds via a 3-body process, but in two messenger
%models it decays in a two-body one. The three messenger case is similar
%to the two messenger case. The gluino mass is primarily determined by $\Lambda$, and
%the variations of the squark masses in these figures are due to the change of $\Delta_{PQ}$.}
%\label{FIG:GLUINO-SQUARK-FTH}
%\end{figure}

The superpartner spectrum of the F-theory GUTs can be obtained by
solving the RG equations with the boundary condition at the
messenger scale. Compatibility with electroweak symmetry breaking
then fixes $\tan\beta$ to a large value between $20-40$, the exact
value of which depends on the specifics of the model. The dependence
of the mass spectrum on $N_{5}$ and $\Lambda$ when $\Delta_{\rm
PQ}=0$ corresponds to the case of mGMSB with a high messenger scale
$M_{\rm mess}\sim10^{12}$~GeV. In this section, we discuss the
effect of $\Delta_{\rm PQ}$ on the mass spectrum.

The mass shift due to the PQ-deformation is most prominent for
lighter sparticles. At the messenger scale, the mass shift for
squarks and sleptons is:
\begin{equation}
m=\widehat{m}\sqrt{1-\frac{\Delta_{\rm PQ}^{2}}{\widehat{m}^{2}}}\,
, \label{massshift}
\end{equation}
where $\widehat{m}$ denotes the mass at the messenger scale in the
absence of the PQ deformation. Hence, when
$\widehat{m}\gg\Delta_{\rm PQ}$, there is little change in the mass
of the sparticle, so that the squarks will shift by a comparably
small amount. On the other hand, the masses of the sleptons can
shift significantly. Since the mass spectrum is generated mainly by
gauge mediation, the absence of an $SU(2)$ gauge coupling implies
that the right-handed selectron $\widetilde{e}_{R}$, smuon
$\widetilde{\mu}_{R}$ and stau $\widetilde{\tau}_{R}$ will be
lighter, and thus more sensitive to the PQ\ deformation in
comparison with their left-handed counterparts. Depending on the
range of parameter space, the $\widetilde{e}_{R}$,
$\widetilde{\mu}_{R}$ and $\widetilde{\tau}_{R}$ mass can either be
above or below the mass of the $\widetilde{\chi}_{2}^{0}$. It is
also possible in some cases for $\widetilde{e}_{R}$,
$\widetilde{\mu}_{R}$ and $\widetilde{\tau}_{R}$ to become
comparable in mass to $\widetilde{\chi}_{1}^{0}$.

Due to the large Yukawa couplings present in the third generation,
RG flow will amplify the effects of the PQ deformation in the third
generation squarks and sleptons. The stop and sbottom can typically
become lighter than the gluino in such models, and the
$\widetilde{\tau}_{1}$ is lighter than $\widetilde{\chi}_{2}^{0}$.
Further, for large enough values of $\Delta_{PQ}$, the
$\widetilde{\tau}_{1}$ can be lighter than
$\widetilde{\chi}_{1}^{0}$.
%
%Figure \ref{FIG:ONEMESSSPEC} illustrate the mass spectrum of a
%single messenger model at minimal and maximal PQ deformation.

The phenomenology of the F-theory GUTs at the hadron collider will
highly depend on the NLSP type, i.e. whether it is the lightest stau
or Bino~\cite{Heckman:2009bi}. When the lightest stau is the NLSP,
it behaves like a charged massive particle in the detector, either
leave a highly ionizing track in the tracking chamber or
\textquotedblleft fake muons \textquotedblright \ in the muon
chamber of a detector at the LHC. The mass of the lightest stau can
be determined by the energy-loss ($dE/dt$) and time-of-flight
measurement. The other particles further up the decay chain can be
constructed as well in principle~\cite{Ellis:2006vu}. While a
completely accurate reconstruction may require about $10-30$~
fb$^{-1}$ of integrated luminosity, this can in principle be
accomplished with data from the first three years of the LHC, and
therefore provides one reliable method for determining detailed
features of the spectrum.

For the case with Bino NLSP, the collider phenomenology looks quite
similar to the typical supersymmetric model with neutralino LSP
since Bino decays outside the detector and behaves effectively like
an LSP. Therefore, naively it will be difficult to distinguish it
from mSUGRA models. However,  the relatively light $\tilde\tau_1$ in
the F-theory models results in large decay branching ratios of
$\tilde \chi_2^0$ and $\tilde \chi_1^{\pm}$ into $\tau$-leptons.
This leads to enhanced multi-$\tau$ plus missing $E_T$ signatures,
and makes F-theory GUTs distinguishable from other models without
light $\tilde \tau$, e.g. mSUGRA models with small $A$-term.
Fig.~\ref{sig-1} shows the footprints of F-theory GUTs and other
SUSY models in the LHC signature space.  One can see that F-theory
GUTs can be distinguished from mSUGRA models with small A-terms and
low scale GMSB models. Moreover, we find that at $50$~fb$^{-1}$, the
PQ deformation away from minimal gauge mediation produces observable
consequences which can also be detected to a level of order $\sim
\pm 10$ GeV. In this way, it is possible to distinguish between
models with a large and small PQ deformation.

\begin{figure*}[ptb]
\begin{center}
\includegraphics[
width=3in ]{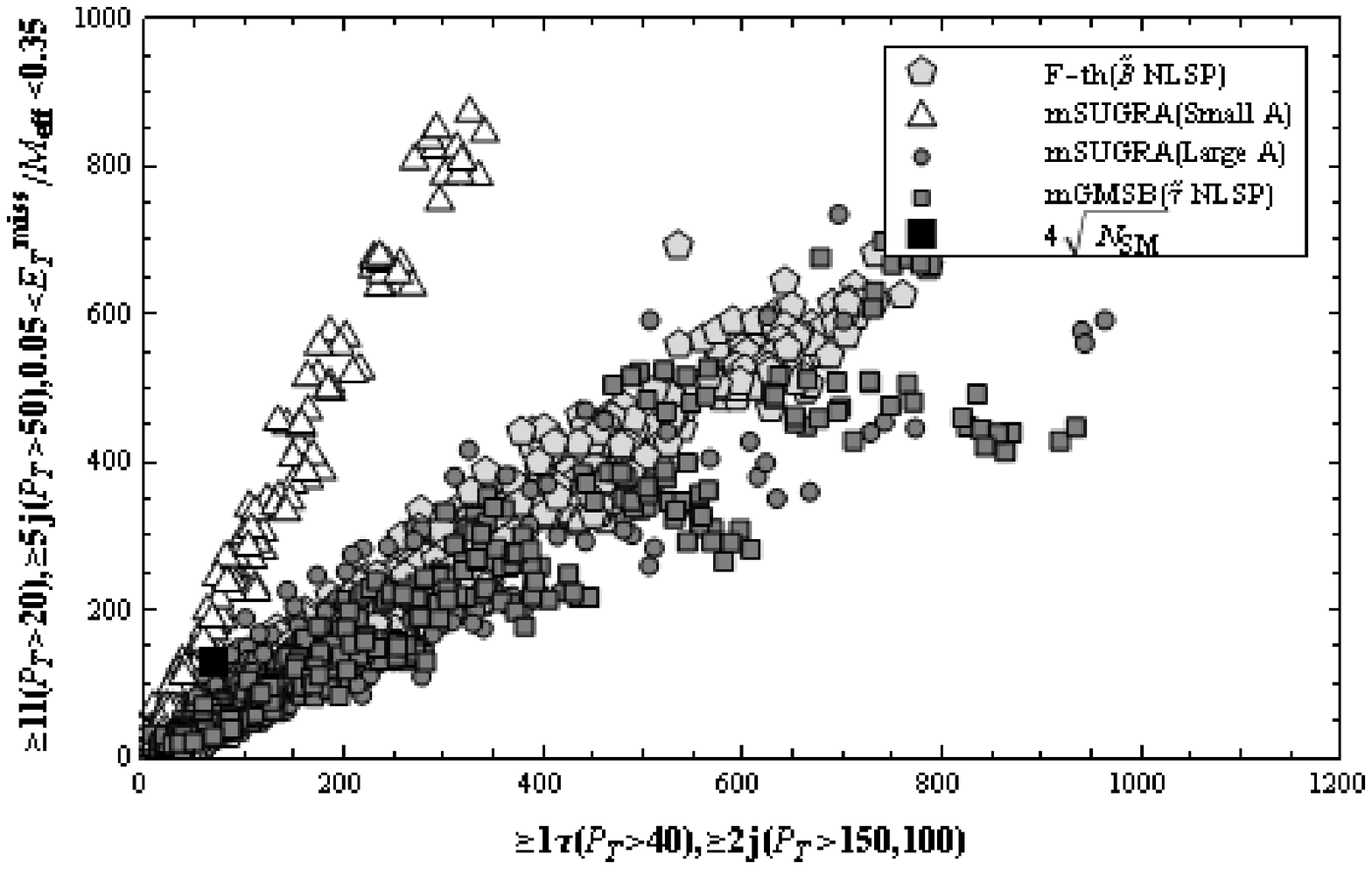}
\includegraphics[
width=3in ]{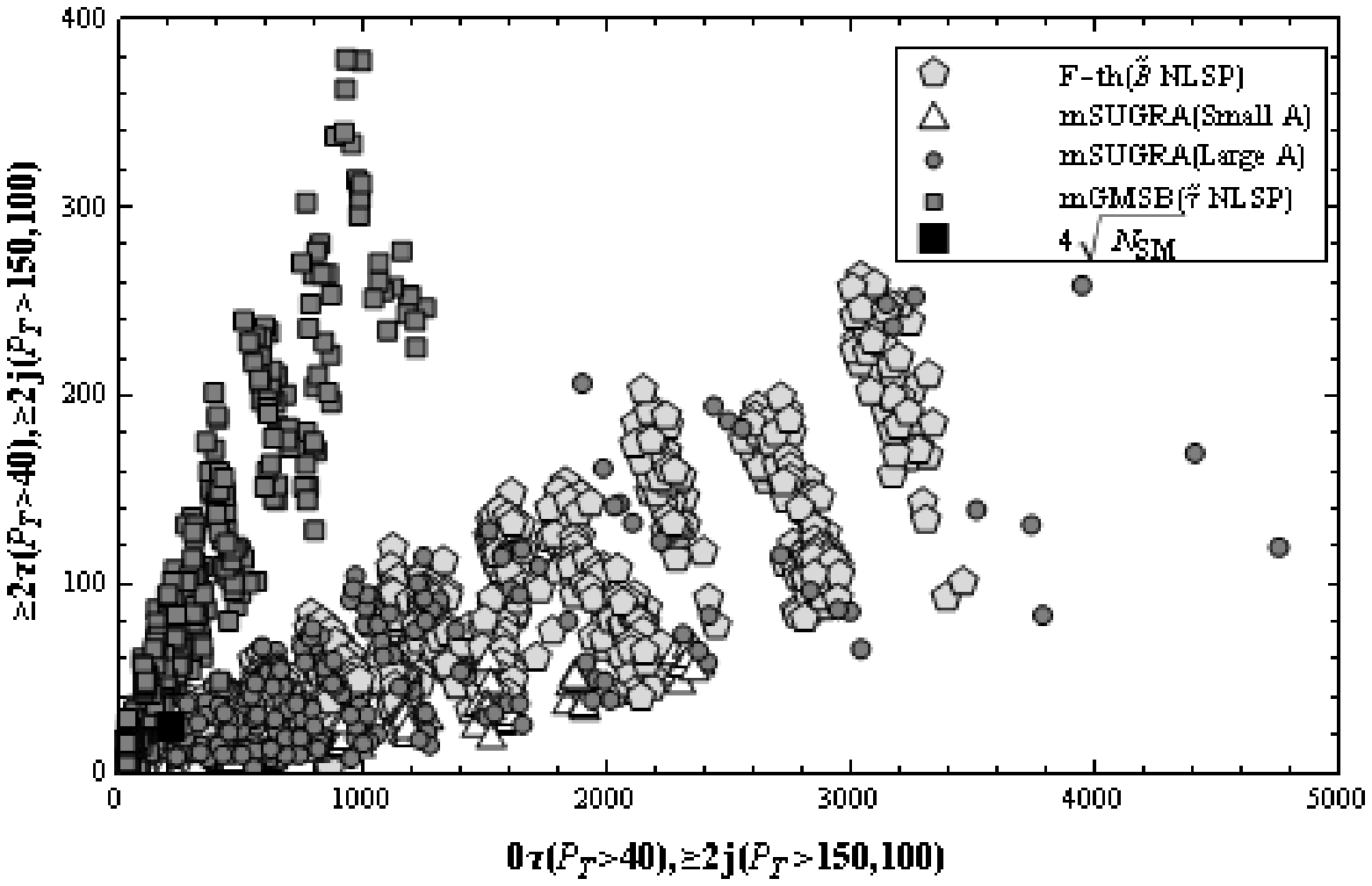}
\end{center}
\vspace{-1cm}
\caption{Footprint of LHC signatures (without SM background) for
distinguishing F-theory GUTs and small A-term mSUGRA models with
$5$~fb$^{-1}$ integrated luminosity. %The red circle denotes the
%$4\sigma$ deviation from the SM background.
}%
\label{sig-1}%
\end{figure*}

\section{Models of Supersymmetry Breaking Mediation, the LHC and Global Fits}

 We briefly describe potential  LHC signals obtained from
top-down and
bottom-up approaches to SUSY breaking. In the top-down approach, we discuss
simple models of large volume string compactification, where all moduli are
stabilised. We go on to perform global fits of the model to current indirect
data, and compare the quality of fit to other well-known models of
supersymmetry (SUSY) breaking. In the bottom-up approach, we presented in Chapter 2 global fit
results of a phenomenological parameterisation of the weak-scale minimal
supersymmetric standard model (MSSM) with 25 relevant parameters known as the
phenomenological MSSM.

\subsection{Large Volume String Scenario and LHC
Signatures}
In a top-down approach to SUSY breaking we will present a large class of string
compactifications with all moduli stabilised known as the large volume scenario
(LVS). In this scenario moduli stabilisation
with an exponentially large volume and supersymmetry breaking are achieved
via the presence of magnetic-like fluxes and controlled quantum corrections
to the
scalar potential. The standard model fields are localised either at D3 or D7
branes. Choosing which 4-cycles of the compact 6-dimensional space for the
D-branes to live in gives rise to different scenarios of soft supersymmetry
breaking.
This section is based on the following articles to which we refer for further reference:
\cite{hepth0502058,Conlon:2006tj,aqs,Conlon:2006wz,Conlon:2007xv}.

\subsubsection{The General Scenario}

We consider $N=1$ flux compactifications of IIB string theory in the presence
of D3 and D7 branes. The K\"ahler potential and superpotential for the moduli
$\Phi=S, U_a, T_i$ take the form

\bea \label{KahlerPot} \hat{K}(\Phi, \bar{\Phi}) &
= & - 2 \ln \left( {{\cal{V}}} + \frac{\hat{\xi}}{2 g_s^{3/2}} \right) - \\ \nonumber
& & \ln \left(
i \int \Omega \wedge \bar{\Omega} \right) - \ln (S + \bar{S}). \\ \hat{W}(\Phi)
& = & \int G_3 \wedge \Omega + \sum_i A_i e^{-a_i T_i}, \nonumber
\eea
respectively.
The dependence on the complex structure moduli $U$ is encoded in the Calabi-Yau
$(3,0)$ form $\Omega$. $G_3$ corresponds to the three-form fluxes and is linear
in the dilaton $S$. We have included the leading $\alpha'$ correction to the
K\"ahler potential, which depends on $\hat{\xi} = - \zeta(3) \chi(M)/(2 \pi)^3$
with $\chi(M)$ being the Euler number of the Calabi-Yau manifold $M$. Large-volume
models require $M$ to have at least two K\"ahler moduli $T_i$, one of which is
a blow-up mode, as well as a negative Euler number. These are not very stringent constraints and
are satisfied by a large class of Calabi-Yau manifolds.
The simplest model is that of ${P}^4_{[1,1,1,6,9]}$, which we use as our
working example, although our results are general. For this the volume can be written as
$ {\cal{V}} = \frac{1}{9\sqrt{2}} \left( \tau_b^{3/2} -
\tau_s^{3/2} \right). $ with  $\tau_b = \hbox{Re}(T_b)$ and $\tau_s =
\hbox{Re}(T_s)$ denote big and small cycles. The geometry is analogous to that
of a Swiss cheese: the cycle $T_b$ controls the volume (`the size of the
cheese') and $T_s$ controls a blow-up cycle (`the size of the hole').
Models with several $T_s$ fields are obviously generalised.

The ${N}=1$ scalar potential is, in a $1/{{\cal{V}}} $ expansion:
\bea
\label{pot}
V & = & \sum_{\Phi = S,U} \frac{{\cal{K}}^{\Phi \bar{\Phi}} D_\Phi W
\bar{D}_{\bar{\Phi}} \bar{W}}{{{\cal{V}}}^2}
+\\  \nonumber  && %\frac{(a_s A_s)^2
\frac{A \sqrt{\tau_s} e^{-2 a_s \tau_s}}{{\cal{V}}}  %\\ & &
  - \frac{%W_0 a_s A_s
  B \tau_s e^{-a_s \tau_s}}{{{\cal{V}}}^2}
+ \frac{\xi \vert W_0 \vert^2}{g_s^{3/2} {{\cal{V}}}^3}\nonumber
\eea
in the limit ${\cal{V}} \gg 1$. Here the constants $A,B$ are given by $A=(a_sA_s)^2$,
$B=W_0a_sA_s$.
%%%
The first terms of this scalar potential
stabilise the dilaton and complex structure moduli at $D_S W = D_U W = 0$
(up to order $1/{\cal{V}}$). The
remaining terms stabilise the K\"ahler moduli. The non-perturbative terms in
$\tau_s$ balance against the perturbative corrections in the volume, and it can
be shown that at the minimum of the scalar potential \cite{hepth0502058} $$
{\cal{V}} \sim W_0 e^{\frac{c}{g_s}}, \qquad \tau_s \sim \ln {\cal{V}}, $$ where
$W_0$ is the value of the flux superpotential at the minimum of $S$ and $U$
fields and $c \sim \xi^{2/3}$ is a numerical constant.

This simple result has
far-reaching implications since an exponentially large volume implies that the
string scale $M_s\sim M_{planck}/{{\cal{V}}}^{1/2}$ can be much smaller than the
Planck scale making string theoretical implications relevant at smaller
scales and therefore closer to be subjected to experimental scrutiny.
Notice also that the gravitino mass $m_{3/2}=e^{K/2}W\sim W_0/{{\cal{V}}}$ is
hierarchically smaller than the string scale. A combination of values for $W_0$
and the volume ${{\cal{V}}}$ give rise to several interesting physical
scenarios.
Probably the most interesting are string scales at:
\begin{enumerate}
 \item{} {\it GUT Scale}.
Here the volume is of order ${{\cal{V}}}\sim 10^{4}$ (in string units). The string
scale of order the GUT scale $10^{16}$ GeV. For the gravitino mass to be of the
TeV scale it will require a serious tuning of the flux superpotential to values
as small as $10^{-11}$. Even though this is the desired scale for unification,
it is not an ideal situation for the hierarchy problem.  Since a very small number $W_0$
has to be introduced as an input
in order to obtain the hierarchy between the weak and the GUT scales. This is technically natural
and in principle allowed by the immense number of flux compactifications, despite the fact that fluxes are quantised.
 But it is not optimal to try to explain a small number by introducing another
 small number.
\item{} {\it Intermediate
Scale}\/ For volumes of order ${\cal{V}}\sim 10^{15}$ the string scale is
intermediate $M_s\sim 10^{12}$ GeV and the gravitino mass is of order the TeV
scale even for flux superpotentials of order $W_0\sim 1$ which is the generic
case. This is appealing for the hierarchy problem since there is no fine tuning
to obtain the weak scale, although it does not naturally give rise to unification as
suggested by the LEP data for the MSSM\@.  It is worth pointing out that there are explicit
realistic models with unification precisely at this scale \cite{Aldazabal:2000sa}.

\item{} {\it TeV Scale}\/ For volumes of
order ${\cal{V}}\sim 10^{30}$ the string scale itself is the TeV scale, which
would be the most exciting scenario thinking about the prospects of string
theory physics being observable at the LHC\@. The main obstacle with this scenario
is that the volume modulus is so light in this case that would give rise to
long range interactions of the fifth force type that are not observed, although
mechanisms to ameliorate this problem may be considered.

\end{enumerate}

All of these scenarios are enriched by the freedom to have the standard model
on different types of branes. The standard model particles may live either on
D3 or D7 branes. These branes wrap different topologically non-trivial
4-cycles. There are several options. First, the size of the cycles can be
stabilised at values just larger than the string scale which
we call `small' (like $\tau_s$ above) to differentiate them with those that
are exponentially large.
 The Standard Model can only live on a small
cycle since the gauge coupling is inversely proportional to the (square root)
of the size of the cycle. In the general case, the F-term of the volume modulus
is the main source of SUSY breaking but it gives rise to no-scale soft terms
which vanish at tree level. The main source of SUSY breaking then could be the
F-term of the cycle where the standard model lives. If this is non-zero, then
the soft terms are approximately equal to the gravitino mass. Therefore the
intermediate scale scenario (scenario 2 above)
 will be the most suitable to describe the MSSM (barring the
lack of automatic unification).

If the F-term of the standard model cycle vanishes,
the main sources of supersymmetry breaking are bulk fields like the dilaton or
loop corrections of the approximately no-scale scenario driven by the volume
modulus. This gives rise to a completely different scenario that has been
recently discussed in \cite{Blumenhagen:2009gk}.
In this case the soft terms are of order $M_{soft}\sim 1/{{\cal{V}}}^2$ or
$m_{soft}\sim 1/{{\cal{V}}}^{3/2}$ and can be of order the TeV scale for
relatively small volumes,  ${{\cal{V}}} \sim 10^6$ in string units. This gives
rise to a fourth scenario.
This scenario, although at present is less under calculational control, has several
interesting features: the string scale is close to the GUT  scale $M_s\sim
10^{14}-10^{15}$ GeV. This is interesting because it has been recently realised
that the GUT scale is not actually the string scale but $M_{GUT}\sim M_s
{\cal{V}}^{1/6}$, therefore a string scale of order $10^{14}-10^{15} $ GeV would
give rise to gauge unification at $10^{16}$ GeV, where it is inferred to be at
assuming a SUSY desert from the measured values of the gauge
couplings. Furthermore, such high string
scales
can be useful in  cosmology since they are the standard inflation
scales. Moreover,
unlike the previous scenarios, in this case the lightest modulus (of mass $m\sim 1/{\cal{V}}^{3/2}$)
can be heavier
than the soft-terms and therefore free from the cosmological moduli problem.
A detailed phenomenological study of this scenario is yet to be performed.

\subsubsection{The Physical Picture}
For concreteness we will consider here scenario 2. This is following the main reason for supersymmetry as the solution to the hierarchy problem. Scenario 1 could
be considered in a similar manner by tuning $W_0$.
Scenario 3, does not need much analysis since
if it were the case, LHC would detect string states directly. Scenario 4 is
not yet sufficiently well under calculational control to be studied
systematically.
In order to study the soft terms we need two further pieces of information: the matter fields K\"ahler potential and the gauge kinetic function.

The gauge kinetic functions $f_a(\Phi)$ depend on whether the
gauge fields come from D3 or D7 branes and, in the latter case, on the
4-cycle wrapped by the D7 brane.  For D branes, $f=S$ at tree level. For D7 branes, if $T_i$ is the K\"ahler modulus
corresponding to a particular 4-cycle, reduction of the DBI action
for an unmagnetised brane wrapped on that cycle gives
$ f_i = \frac{T_i}{2 \pi}$.  We are interested
in magnetised branes wrapped on 4-cycles.
The magnetic fluxes alter this expression
 to \bea \label{FluxGaugeCouplings} f_i =
h_i(F)\, S + \frac{T_i}{2 \pi}, \eea where $h_i$ depends on the fluxes
present on that stack. The explicit form of $h_i(F)$ is not known for general compactifications.

On the chiral matter kinetic terms, again explicit expressions have not been calculated.
However, scaling arguments  allow us to find the leading order dependence on the overall volume and the
modulus determining the size of the 4-cycle that the D7 brane wraps.
\bea
\label{MatterMetric}
\tilde{K}_{\alpha \bar{\beta}} =
\frac{\tau_s^{\lambda}}{{\cal{V}}^{2/3}}k_{\alpha \bar{\beta}}(\phi),
\eea
This expression holds
in the limit of dilute fluxes and large cycle volume $\tau_s$ and
will receive corrections sub-leading in $\tau_s$. For the minimal
model in which all branes wrap the same cycle, it was shown in
\cite{Conlon:2006tj} that $\lambda = 1/3$. For
other cases $\lambda$ may take values between 0 and 1. A more precise and complete discussion
of the modular weights appearing on D7 chiral matter can be found in
\cite{Aparicio:2008wh}.

For a simple case with matter fields of the same modular weight $\lambda$, the soft terms are
\bea \label{lamb}
M_i & = & \frac{F^s}{2 \tau_s}, \nonumber \\
m_{\alpha} & = & \sqrt{\lambda} M_i, \nonumber \\
A_{\alpha \beta \gamma} & = & - 3\lambda  M_i, \nonumber \\
 B & = & - \left(\lambda +1\right)  M_i. \eea
It is worth emphasizing that the structure of soft terms in this scenario is universal to leading order. This is remarkable given
 the generic lack of universality in gravity
mediation. This is due to the fact that the source of supersymmetry breaking is the K\"ahler moduli sector which is
blind to flavour, since these moduli do not appear in the Yukawa couplings which determine
the flavour structure. It is the complex structure moduli sector that is sensitive to flavour but this sector does not participate in supersymmetry breaking.
As long as the complex structure and K\"ahler moduli have a product structure, the soft terms will be universal. The breaking
of this structure  in
higher perturbative order determines
the amount by which the  soft terms will acquire non-universal contributions, which will be suppressed with respect to the
universal contributions. A precise estimate of the size of the non universality is
not yet available.

The simplest case $\lambda=1/3$ has been studied in detail in \cite{Conlon:2007xv}.
The renormalisation group flow to low energies providing the low energy spectrum of supersymmetric particles was computed
using {\tt SOFTSUSY}~\cite{Allanach:2001kg}, event generators and detector
simulators were also used to compute observable LHC quantities. A generic
issue
of these calculations is that it is very difficult to differentiate the physical implications
of these string scenarios compared with
the standard mSUGRA scenario that has been so well studied in the literature.
The cleanest difference is the ratio of gaugino masses  that give $M_1 \, :\, M_2 \, :\, M_3\, =\, (1.5-2)\, :\, 2\, :\, 6$ which differs from the mSUGRA
relation $M_1 \, :\, M_2 \, :\, M_3\, =\, 1\, :\, 2\, :\, 6$.

\subsection{Comparison of LVS and Other Models of SUSY Breaking}
Assuming some model hypothesis $H$, Bayesian statistics helps update
some probability density function (PDF) $p(\underline m|H)$ of model
parameters $\underline m$ with data. The prior encodes our knowledge
or prejudices about the parameters. Since $p(\underline m|H)$ is a
PDF in $\underline m$, $\int p(\underline m|H) d\underline m=1$,
which defines a normalization of the prior. One talks of priors
being `flat' in some parameters, but care must be taken to refer to
the measure of such parameters. A prior that is flat between some
ranges in a parameter $m_1$ will not be flat in a parameter $x
\equiv \log m_1$, for example. The impact of the data is encoded in
the likelihood, or the PDF of obtaining data set $\underline d$ from
model point $\underline m$: $p(\underline d|\underline m,H) \equiv
{\mathcal L}(\underline m)$. The likelihood is a function of
$\chi^2$, i.e.\ a statistical measure of how well the data are fit
by the model point. The desired quantity is the PDF of the model
parameters $m$ given some observed data $\underline d$ assuming
hypothesis $H$: $p(\underline m | \underline d, H)$. Bayes' theorem
states that
\begin{equation} p(\underline m|\underline d, H) =
\frac{p(\underline d|\underline m,H)p(\underline m|H)} {p(\underline
d|H)}, \label{eq:bayes1}
\end{equation}
where $p(\underline d|H) \equiv \mathcal{Z}$ is the Bayesian evidence, the probability
density of observing data set $d$ integrated over all model parameter space.
The Bayesian evidence is given by:
\begin{equation}
\mathcal{Z} =
\int{\mathcal{L}(\underline m)p(\underline m|H)}\ d \underline m
\label{eq:3}
\end{equation}
where the integral is over $N$ dimensions of the parameter
space $\underline m$. Since the Bayesian evidence is independent of the model
parameter values $\underline m$, it is usually ignored in parameter estimation
problems
and posterior inferences are obtained by exploring the unnormalized
posterior using standard Markov Chain Monte Carlo sampling methods.

In order to select between two models $H_{0}$ and $H_{1}$ one needs to compare
their respective posterior
probabilities given the observed data set $\underline d$, as follows:
\begin{equation}
\frac{p(H_{1}|\underline d)}{p(H_{0}|\underline d)}
=\frac{p(\underline d|H_{1})p(H_{1})}{p(\underline d|
H_{0})p(H_{0})}
=\frac{\mathcal{Z}_1}{\mathcal{Z}_0}\frac{p(H_{1})}{p(H_{0})},
\label{eq:3.1}
\end{equation}
where $p(H_{1})/p(H_{0})$ is the prior probability ratio for the two models, which can often be set to unity
but occasionally requires further consideration. It can be seen from Eq.~\ref{eq:3.1} that
Bayesian model selection revolves around the evaluation of the Bayesian evidence. As the average of likelihood
over the prior, the evidence automatically implements Occam's razor.
A theory
with less parameters has a higher prior density since it integrates to
1 over the whole space.
Thus, there is an a priori preference for less parameters, unless the
data strongly require there be more.
We shall consider three different prior distributions: flat in the parameters
listed in Table~\ref{tab:ranges1}, flat in their logarithm, or flat in the MSSM
parameters $\mu$ and $B$ rather than flat in $\tan \beta$~\cite{alal}. For
robust results, we look for approximate independence to the form of the
prior. This will only happen when there is enough data to strongly constrain
the models in question.

The natural logarithm of the ratio of posterior model probabilities provides a useful guide to what constitutes a
significant difference between two models:
\begin{equation}
\Delta \log \mathcal{Z} = \log \left[ \frac{p(H_{1}|\underline
    d)}{p(H_{0}|\underline d)}\right]
=\log \left[ \frac{\mathcal{Z}_1}{\mathcal{Z}_0}\frac{p(H_{1})}{p(H_{0})}\right].
\label{eq:Jeffreys}
\end{equation}
We summarize the convention we use in this paper in Table~\ref{tab:Jeffreys}.

\begin{table}
\begin{center}
\begin{tabular}{|c|c|c|c|}
\hline
$|\Delta \log \mathcal{Z}|$ & Odds & Probability & Remark \\
\hline\hline
$<1.0$ & $< 3:1$ & $<0.750$ & Inconclusive \\
$1.0$ & $\sim 3:1$ & $0.750$ & Weak Evidence \\
$2.5$ & $\sim 12:1$ & $0.923$ & Moderate Evidence \\
$5.0$ & $\sim 150:1$ & $0.993$ & Strong Evidence \\ \hline
\end{tabular}
\end{center}
\caption{The scale we use for the interpretation of model probabilities. Here the `$\log$'
represents the natural logarithm.}
\label{tab:Jeffreys}
\end{table}

The nested sampling approach, introduced in \cite{Skilling}, is a Monte Carlo method targeted at the
efficient calculation of the evidence, but also produces posterior inferences as a by--product.
\cite{Feroz:2007kg} built on this nested sampling framework and introduced the {\sc MultiNest}
algorithm which is efficient in sampling from multi--modal posteriors that
exhibit curving degeneracies. {\sc MultiNest} produces
posterior samples and calculates the evidence and its uncertainty. This
technique has greatly reduced the
computational cost of model selection and the exploration of highly degenerate multi--modal posterior
distributions. We employ nested sampling to calculate $\Delta \log \mathcal{Z}$.

We now specify the parameter ranges over which we
sample for the different models. Firstly, we consider both signs of $\mu$ in
our analysis for  all
models. The ranges over which we vary the continuous model parameters are
shown in Table~\ref{tab:ranges1}.
\begin{table}
\begin{center}
\begin{tabular}{|c|} \hline
mSUGRA  \\ \hline
 $50\mbox{~GeV}\leq m_0 \leq 4$ TeV \\
$50\mbox{~GeV} \leq m_{1/2} \leq 2$ TeV  \\
$-4\mbox{~TeV} \leq A_0 \leq 4\mbox{~TeV}$   \\ \hline
\hline
% \end{tabular}
% \begin{tabular}{|c|c|} \hline
 LVS \\ \hline
   $50\mbox{~GeV}\leq m_0\leq 2\mbox{~TeV}$
  \\ \hline \hline
AMSB \\
 $50\mbox{~GeV}\leq m_0 \leq 4$ TeV \\
$ 20\mbox{~TeV} \leq m_{3/2} \leq 200\mbox{~TeV}$  \\ \hline
\end{tabular}
\end{center}
\caption{Ranges for the parameters.
For all models, $2\leq \tan \beta \leq 62$.}
\label{tab:ranges1}
\end{table}
$\tan \beta$ is bounded from below by 2, values lower than this are in
contravention of LEP2 Higgs searches, and from above by 62, since such large
values lead to non-perturbative Yukawa couplings below the GUT scale and
calculability is lost.
In the mSUGRA the unification scale is the standard GUT scale $m_{GUT} \approx
2\times 10^{16}$GeV, while for
the LVS the soft terms are defined at the intermediate string  scale $m_s
\approx 10^{11}$GeV as in~\cite{Allanach:2008tu}.

\begin{table*}
\begin{center}
\begin{tabular}{|c|c|c|c|c|c|c|}
\hline
& \multicolumn{3}{c|}{symmetric $\mathcal{L}_{\rm DM}$} & \multicolumn{3}{c|}{asymmetric $\mathcal{L}_{\rm DM}$} \\
\hline\hline
Model/Prior  & linear & log & natural & linear & log & natural \\
\hline
mSUGRA & $ 8.0 \pm 0.1 $ & $ 7.9 \pm 0.1 $ & $ 10.3 \pm 0.1 $ & $ 0.0 \pm 0.1 $ & $ 1.0 \pm 0.1 $ & $ 1.3 \pm 0.1 $ \\
mAMSB  & $ 0.4 \pm 0.1 $ & $ 0.6 \pm 0.1 $ & $ 0.0 \pm 0.1 $ & $ 5.1 \pm 0.1 $ & $ 6.0 \pm 0.1 $ & $ 5.0 \pm 0.1 $ \\
LVS    & $ 8.7 \pm 0.1 $ & $ 8.9 \pm 0.1 $ & $ 11.8 \pm 0.1 $ & $ 2.9 \pm 0.1 $ & $ 3.0 \pm 0.1 $ & $ 3.1 \pm 0.1 $ \\
\hline
\end{tabular}
\caption{log evidences ($\Delta \log \mathcal{Z}$) for mAMSB, LVS and the mSUGRA
  for both signs of $\mu$.
Symmetric $\mathcal{L}_{\rm DM}$ labels the assumption that the dark matter (DM) relic density is composed entirely
of the LSP and asymmetric $\mathcal{L}_{\rm DM}$ labels the assumption that
the LSP forms only a part of the
DM relic density. The log evidence of the natural prior mAMSB, $\log Z_s = 67.3$ and the log evidence of the linear prior mSUGRA, 76.7 have been subtracted
from all entries in the symmetric $\mathcal{L}_{\rm DM}$ and asymmetric $\mathcal{L}_{\rm DM}$ respectively.}
\label{tab:model-prob-odds}
\end{center}
\end{table*}

In our global fits, the following empirical data are used:
$m_W$, $\sin^2 \theta_{eff}^l$, $\delta a_\mu$, $\Omega_{DM} h^2$, $m_h$,
$\Gamma_Z^{tot}$, $R_l^0$, $R_b^0$, $R_c^0$, $A^{0,b}_{fb}$, $A^{0,c}_{fb}$,
$A^0_{LR}(SLD)$, $A_b$, $A_c$, $BR(B \rightarrow X_s \gamma)$, $BR(B_s
\rightarrow \mu^+ \mu^-)$, $BR(B \rightarrow D \tau \nu)$, $R_{\Delta M_s}$,
$\Delta_{0-}$, $R_{l23}$, $m_t$, $m_b$, $m_Z$, $\alpha_s(M_Z)$,
$\alpha(M_Z)$
as well as current sparticle search constraints. Thus, the likelihood receives
contributions from cosmological, electroweak and $b-$physics data. See
Ref.~\cite{all} for the precise numbers used and their sources.
When including the WMAP cold dark matter inferred relic density $\Omega_{DM}
h^2$, two
different assumptions are made: either the relic density comes {\em solely}
\/from a neutralino $\chi_1^0$ which is the lightest supersymmetric particle, or
alternatively,
that an additional component of cold dark matter is allowed.
The combined log likelihood is the sum of the individual log likelihoods for each
measurement,
\begin{equation}
\log\mathcal{L}^{tot} = \sum_i \log\mathcal{L}_i.
\end{equation}
To calculate the MSSM spectrum we use
\texttt{Softsusy2.0.18}~\cite{Allanach:2001kg}.
If a point survives the cuts above, it is passed via the SUSY Les
Houches Accord~\cite{hep-ph/0311123} to
\texttt{microMEGAS2.2}~\cite{Belanger:2004yn},
\texttt{SuperIso2.3}~\cite{arXiv:0808.3144}
and \texttt{SusyPOPE}~\cite{arXiv:0710.2972}.
From \texttt{microMEGAS} we obtain the
DM relic density, the rare
branching ratio $BR(B_s\to\mu^+\mu^-)$, the SUSY component $\delta a_\mu$ of
the anomalous magnetic moment of the
muon $(g-2)_{\mu}$ and DM direct detection rates.From
\texttt{SuperIso2.3} the branching ratios $BR(B\to X_s\gamma)$, $BR(B\to
D\tau\nu)$, the quantities $R_{\Delta M_s}$, $R_{l23}$, $R_{B\tau\nu}$ and the
isospin asymmetry $\Delta_{0-}$
are obtained\footnote{We note that in the process of preparing this paper and
  after our fits were performed a new version of
  \texttt{SusyBSG}\cite{arXiv:0712.3265} appeared. This more accurate
  calculation could result in a change in our $BR(B\to X_s \gamma)$
  prediction similar in size to (but smaller than) its uncertainty.}.
{\tt SusyPOPE} is used to predict the electroweak observables for every
point.

We see from the results, presented in Table~\ref{tab:model-prob-odds}, that
the model preferred by the data depends on what we assume for the DM relic
density: whether it is
made entirely of neutralinos (symmetric constraint) or whether we allow for
the presence of non-neutralino dark matter (asymmetric constraint).
An analysis of the
constraining power of the various observables showed that it
resides dominantly in the DM constraint in the case of the mSUGRA and the
LVS\@. This is not the case in mAMSB
where the relic density is uniformly too small by an order of magnitude across
parameter space, and the main constraint comes from the combined electroweak
observables.
However, for the symmetric constraint, mAMSB is strongly disfavoured (since it
predicts
essentially no neutralino dark matter) over the mSUGRA and LVS\@.
With the asymmetric constraint and using the Jeffrey's scale, we deduce that
mAMSB is at least moderately favoured over
the mSUGRA and weakly preferred to the LVS scenario. Although the log evidences
shown still show some prior dependence, it is small enough such that the
inference in terms of the Jeffrey's categorisation is robust.

Experience and
familiarity with the methods of model selection and Bayesian inference
from work such as that contained here will be
invaluable once further more constraining data become available, hopefully
from SUSY signals at colliders.

\section{TeV-Scale String Excitations}

Superstring theory provides a consistent framework to explain the underlying symmetries of nature, e.g., the unification of gravity with standard model (SM) gauge interactions and the probable existence and breaking of supersymmetry (SUSY). Earnest progress were fuelled by the realization of the vital role played by D-branes~\cite{joe} in bridging the gap between string theory and phenomenology~\cite{Blumenhagen}.  This has empower the formulation of string theories with compositeness setting in at TeV scales and large extra dimensions~\cite{Antoniadis:1998ig}.

TeV-scale superstring theory provides a brane-world description of the SM, which is localized on hyperplanes extending in $p+3$ spatial dimensions, the so-called D-branes. Gauge interactions emerge as excitations of open strings with endpoints attached on the D-branes. The basic unit of gauge invariance for D-brane constructions is a $U(1)$ field, and so one can stack up $N$ identical D-branes to generate a $U(N)$ theory with the associated $U(N)$ gauge group. Gauge bosons and associated gauginos (in a supersymmetric theory) arise from strings terminating on {\em one} stack of D-branes, whereas chiral matter fields are due to strings stretching between intersecting D-branes.  Gravitational interactions are described as closed strings propagating freely in all nine dimensions of string theory, i.e., the flat parallel dimensions extended along the $(p+3)$-branes and the transverse dimensions.  In this radically new view of spacetime gravity is not intrinsically weak, but it appears weak at the relatively ``low energies'' of common experience only because its effects are diluted by propagation in large extra dimensions. Perhaps the most remarkable consequence of TeV-scale D-brane string physics is the emergence of Regge recurrences (at parton collision energies $\sqrt{\hat s} \sim {\rm string\ scale} \equiv M_s$) that could become smoking guns at the Large Hadron Collider (LHC).

The ensuing discussion is framed within the context of a minimal model~\cite{Antoniadis:2000ena}. We consider scattering processes which take place on the (color) $U(3)_a$ stack of D-branes, which is intersected by the (weak doublet) $U(2)_b$ stack of D-branes, as well by a third (weak singlet) $U(1)_c$ stack of D-brane. These three stacks of D(3+p)-branes entirely fill the uncompactified part of space-time and wrap certain $p$-cycles $\Sigma^{(a,b,c)}$ inside the compact six-dimensional manifold $M_6$. In the bosonic sector, the open strings terminating on the $U(3)_a$ stack contain the $SU(3)_{\rm C}$ gluon octect $g$ and an additional $U(1)_a$ gauge boson $C$; on the $U(2)_b$ stacks the open strings correspond to the weak gauge bosons $W$, and again an additional $U(1)_b$ gauge field.  So the associated gauge groups for these stacks are $SU(3)_{\rm C} \times U(1)_a,$ $SU(2)_{\rm EW} \times U(1)_b$, and $U(1)_c$, respectively.  The $U(1)_Y$ boson, which gauges the usual electroweak hypercharge symmetry, is a linear combination of $C$, the $U(1)$ boson $B$ terminating on the $U(1)_c$ stack, a third additional $U(1)$ sharing a $U(2)_b$ stack which is also a terminus for the $SU(2)_L$ electroweak gauge bosons, plus in general a forth $U(1)_d$ that is not relevant for the following discussion.  The fermionic matter consists of open strings, which stretch between different stacks of D$(p+3)$-branes and are hence located at the intersection points.  Concretely, the left-handed quarks are sitting at the intersection of the $a$ and the $b$ stacks, whereas the right-handed $u$ quarks comes from the intersection of the $a$ and $c$ stacks and the right-handed $d$ quarks are situated at the intersection of the $a$ stack with the $c'$ (orientifold mirror) stack. All the scattering amplitudes between these SM particles, which we will need in the following, essentially only depend on the local intersection properties of these D-brane stacks.

Only one assumption is necessary in order to set up a solid framework: the string coupling must be small in order to rely on perturbation theory in the computations of scattering amplitudes. In this case, black hole production and other strong gravity effects occur at energies above the string scale; therefore at least a few lowest Regge recurrences are available for examination, free from interference with some complex quantum gravitational phenomena.  Starting from a small string coupling, the values of standard model coupling constants are determined by D-brane configurations and the properties of extra dimensions, hence that part of superstring theory requires intricate model-building; however, as argued in~\cite{Anchordoqui:2007da,Anchordoqui:2008ac,Anchordoqui:2008hi,Lust:2008qc,Anchordoqui:2008di,Anchordoqui:2009mm,Lust:2009pz,Anchordoqui:2009ja}, some basic properties of Regge resonances like their production rates and decay widths are completely model-independent.

The physical processes underlying dijet production at the LHC are the collisions of two partons $ij$, producing two final partons $kl$ that fragment into hadronic jets. The corresponding $2\to 2$ scattering amplitudes ${\cal M}(ij \to kl)$, computed at the leading order in string perturbation theory, are collected in~\cite{Lust:2008qc}. The amplitudes involving four gluons as well as those with two gluons plus two quarks do not depend on the compactification details of the transverse space.\footnote{The only remnant of the compactification is
  the relation between the Yang-Mills coupling and the string
  coupling. We take this relation to reduce to field theoretical
  results in the case where they exist, e.g., $gg \to gg$. Then,
  because of the require correspondence with field theory, the
  phenomenological results are independent of the compactification of
  the transverse space. However, a different phenomenology would
  result as a consequence of warping one or more parallel
  dimensions~\cite{Hassanain:2009at,Perelstein:2009qi}.}  All string
effects are encapsulated in these amplitudes in one ``form factor''
function of Mandelstam variables $\hat s,~\hat t,~\hat u$ (constrained
by $\hat s+\hat t+\hat u=0$)
\begin{eqnarray}
V(  \hat s,   \hat t,   \hat u) & = & \frac{\hat s\,\hat u}{\hat tM_s^2}B(-\hat s/M_s^2,-\hat u/M_s^2) \nonumber \\ & = & {\Gamma(1-   \hat s/M_s^2)\ \Gamma(1-   \hat u/M_s^2)\over
    \Gamma(1+   \hat t/M_s^2)}. \nonumber \\ \label{formf}
\end{eqnarray}
The physical content of the form factor becomes clear after using the
well-known expansion in terms of $s$-channel resonances~\cite{Veneziano:1968yb}:
\begin{eqnarray}
B(-\hat s/M_s^2,-\hat u/M_s^2) & = & -\sum_{n=0}^{\infty}\frac{M_s^{2-2n}}{n!}\frac{1}{\hat s-nM_s^2} \nonumber \\
& \times &
\Bigg[\prod_{J=1}^n(\hat u+M^2_sJ)\Bigg], \nonumber \\ \label{bexp}
\end{eqnarray}
which exhibits $s$-channel poles associated to the propagation of
virtual Regge excitations with masses $\sqrt{n}M_s$. Thus near the
$n$th level pole $(\hat s\to nM^2_s)$:
\begin{eqnarray}
V(  \hat s,   \hat t,   \hat u) &\approx &\frac{1}{\hat s-nM^2_s} \frac{M_s^{2-2n}}{(n-1)!} %\nonumber \\
%& \times &
\prod_{J=0}^{n-1}(\hat u+M^2_sJ)\ . \nonumber \\
\label{nthpole}
\end{eqnarray}
In specific amplitudes, the residues combine with the remaining
kinematic factors, reflecting the spin content of particles exchanged
in the $s$-channel, ranging from $J=0$ to $J=n+1$.

The amplitudes for the four-fermion processes like quark-antiquark
scattering are more complicated because the respective form factors
describe not only the exchanges of Regge states but also of heavy
Kaluza-Klein (KK) and winding states with a model-dependent spectrum
determined by the geometry of extra dimensions. Fortunately, they are
suppressed, for two reasons. First, the QCD $SU(3)$ color group
factors favor gluons over quarks in the initial state. Second, the
parton luminosities in proton-proton collisions at the LHC, at the
parton center of mass energies above~1 TeV, are significantly lower
for quark-antiquark subprocesses than for gluon-gluon and
gluon-quark~\cite{Anchordoqui:2008ac}. The collisions of valence
quarks occur at higher luminosity; however, there are no Regge
recurrences appearing in the $s$-channel of quark-quark
scattering~\cite{Lust:2008qc}.

We proceed by isolating the contribution to the partonic cross
section from the first resonant state. Note that far below the string
threshold, at partonic center of mass energies $\sqrt{\hat s}\ll M_s$,
the form factor $V(\hat s,\hat t,\hat u)\approx 1-\frac{\pi^2}{6}{\hat
  s \hat u}/M^4_s$~\cite{Lust:2008qc} and therefore the contributions
of Regge excitations are strongly suppressed. The $s$-channel pole
terms of the average square amplitudes contributing to dijet
production at the LHC can be obtained from the general formulae given
in~\cite{Lust:2008qc}, using Eq.(\ref{nthpole}). However, for
phenomenological purposes, the poles need to be softened to a
Breit-Wigner form by obtaining and utilizing the correct {\em total}
widths of the resonances~\cite{Anchordoqui:2008hi}. The contributions of the various channels to the spin and color averaged matrix elements are given elsewhere~\cite{Anchordoqui:2008di}.

The dominant $s$-channel pole terms of the average square amplitudes contributing to $pp \to \gamma$ + jet are given in~\cite{Anchordoqui:2007da,Anchordoqui:2008ac,Anchordoqui:2009ja}.
The $C-Y$ mixing coefficient ($\kappa$) is model dependent: in the $U(3) \times Sp(1) \times U(1)$ model~\cite{Berenstein:2006pk} it is quite small, around $\kappa \simeq 0.12$ for couplings evaluated at the $Z$ mass, which is modestly enhanced to $\kappa \ \simeq 0.14$ as a result of RG running of the couplings up to 2.5~TeV.

Events with a single jet plus missing energy ($\MET$) with balancing
transverse momenta (so-called ``monojets'') are incisive probes of new
physics. As in the SM, the source of this topology is $ij \to k Z$
followed by $Z \to \nu \bar \nu.$ Both in the SM and string theory
the cross section for this process is of order $g^4$. Virtual KK
graviton emission ($ij \to k G$) involves emission of closed strings,
resulting in an additional suppression of order $g^2$ compared to
$Z$ emission. A careful discussion of this suppression is given
in~\cite{Cullen:2000ef}. However, in some scenarios compensation for
this suppression can arise from the large multiplicity of graviton
emission, which is somewhat dependent on the cutoff
mechanism~\cite{Bando:1999di,Anchordoqui:2001cg,Hewett:2007st}. Ignoring
the $Z$-mass (i.e., keeping only transverse $Z$'s), the quiver
contribution to $pp \to Z + {\rm jet}$ is suppressed relative to the
$pp \to \gamma+ {\rm jet}$ by a factor of $\tan^2\theta_W = 0.29.$

\begin{figure}[htb]
%\vspace{9pt}
\includegraphics[
width=3in ]{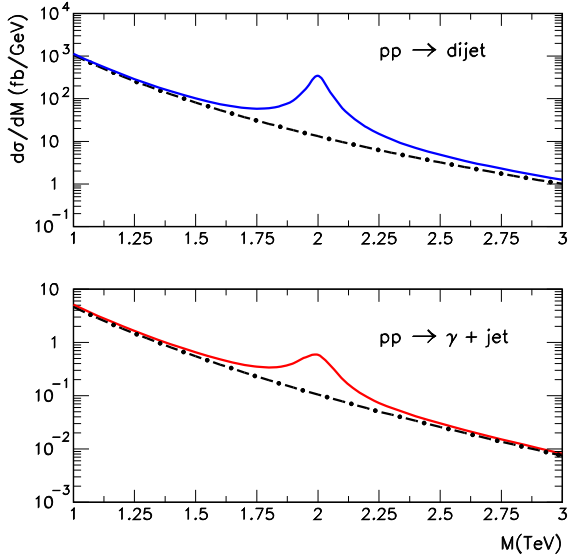}
%\postscript{whitebump.eps}{0.99}
%\framebox[55mm]{\rule[-21mm]{0mm}{43mm}}
 \caption{$d\sigma/dM$ (units of fb/GeV) {\em vs.} $M$ (TeV) is plotted for
the case of SM QCD background (dot-dashed) and (first resonance)  string signal
+ background (solid).}
\label{fig:bump}
\end{figure}

\begin{figure}[htb]
%\vspace{9pt}
\includegraphics[
width=3in ]{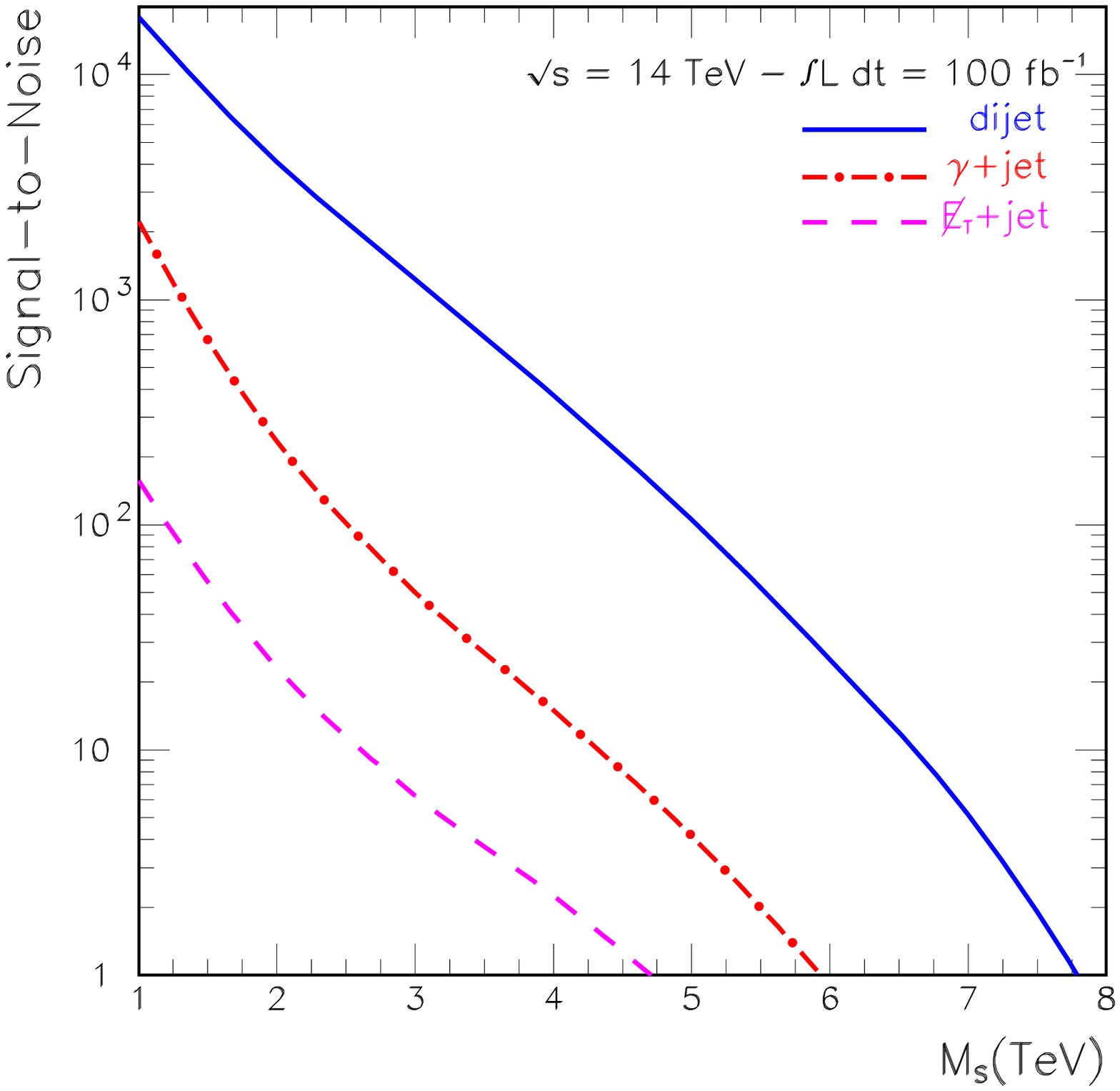}
%\postscript{whiteS2N.eps}{0.99}
%\framebox[55mm]{\rule[-21mm]{0mm}{43mm}}
\caption{Signal-to-noise ratio of $pp \to {\rm
    dijet}$, $pp \to \gamma + {\rm jet}$, and $pp \to \MET + {\rm
    jet}$, for $\sqrt{s} = 14$~TeV,  $\kappa^2
  \simeq 0.02$, and an integrated luminosity of 100~fb$^{-1}$. The approximate
  equality of the background due to misidentified $\pi^0$'s and the
  QCD background, across a range of large $p_T^\gamma$ as implemented
  in~\cite{Anchordoqui:2008ac}, is maintained as an approximate
  equality over a range of invariant $\gamma$-jet invariant masses
  with the rapidity cuts imposed. The monojet signal is obtained from
  the intermediate state $pp \to Z+$ jet multiplied by the
  corresponding branching ratio $Z \to \nu\bar \nu$.}
\label{fig:s2n}
\end{figure}

\begin{figure}[htb]
%\vspace{9pt}
\includegraphics[
width=3in ]{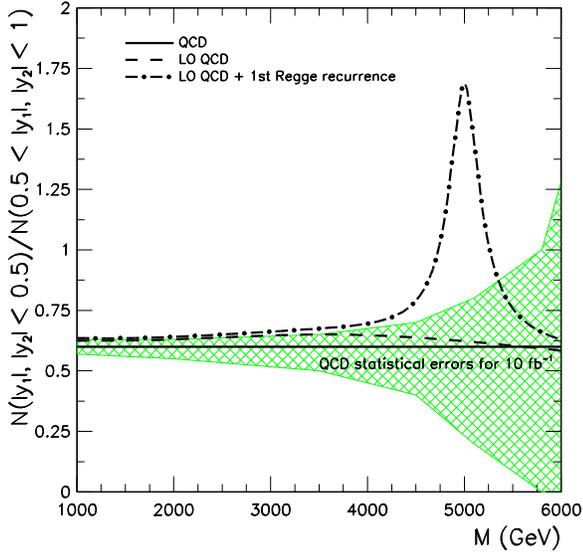}
%\postscript{white_f2.eps}{0.99}
%\framebox[55mm]{\rule[-21mm]{0mm}{43mm}}
\caption{For a luminosity of 10~fb$^{-1}$ and $\sqrt{s} = 14$~TeV,
  the expected value (solid line) and statistical error (shaded
  region) of the dijet ratio of QCD in the CMS detector is compared
  with LO QCD (dashed line) and LO QCD plus lowest massive string
  excitation (dot-dashed line), at a scale $M_s = 5$~TeV.}
\label{fig:dijet_ratio}
\end{figure}

The first Regge recurrence would be visible in data binned according
to the invariant mass $M$ of the final state, after setting cuts on
rapidities $|y_1|, \, |y_2| \le y_{\rm max}$ and transverse momenta
$p_{\rm T}^{1,2}>50$~GeV, where $y_{\rm max} = 2.4$ for photons and
$y_{\rm max} = 1$ for jets. The QCD background is calculated at the
partonic level making use of the CTEQ6D parton distribution
functions~\cite{Pumplin:2002vw}. Standard bump-hunting methods, such
as obtaining cumulative cross sections,
\begin{equation}
\sigma (M_0 ) =
\int_{M_0}^\infty \frac{d\sigma}{dM}\, dM \,,
\end{equation}
from the data and searching for regions with significant deviations from the QCD background, may reveal an interval of $M$ suspected of containing a bump (see Fig.~\ref{fig:bump}). With the establishment of such a region, one may calculate a signal-to-noise ratio, with the signal rate estimated in the invariant mass window $[M_s - 2\Gamma, M_s + 2\Gamma].$ The noise is defined as the square root of the number of background events in the same dijet mass interval for the same integrated luminosity. The LHC discovery reach (at the parton level) is encapsulated in Fig.~\ref{fig:s2n}. The solid, dot-dashed, and dashed lines show the behavior of the signal-to-noise (S$/\sqrt{\rm B}$) ratio as a function of the string scale for three different event topologies (dijet, $\gamma +$ jet, and $\MET +$ jet; respectively), at $\sqrt{s} = 14$~TeV with an integrated luminosity of 100~fb$^{-1}$. It is remarkable that with 100~fb$^{-1}$ of data collection, {\it string scales as large as 6.8~TeV are open to
  discovery at the ≥ $5\sigma$ level.} Although the discovery reach is
not as high as that for dijets, the measurement of
$pp\rightarrow\gamma~ +$ jet and $pp \to \MET +$ jet can potentially
provide an interesting corroboration for the stringy origin for new
physics manifest as a resonant structure in LHC data. Once more, we stress that
these results contain no unknown parameters. They depend only on the
D-brane construct for the SM, and {\it are independent of
  compactification details of the transverse space.}

We now turn to the study of the angular distributions. QCD
parton-parton cross sections are dominated by $t$-channel exchanges
that produce dijet angular distributions which peak at small center
of mass scattering angles. In contrast, non--standard contact
interactions or excitations of resonances result in a more isotropic
distribution. In terms of rapidity variables for standard transverse
momentum cuts, dijets resulting from QCD processes will
preferentially populate the large rapidity region, while the new
processes generate events more uniformly distributed in the entire
rapidity region. To analyze the details of the rapidity space the
D0~Collaboration~\cite{Abbott:1998wh} introduced a new parameter
$R$, the ratio of the number of events, in a given dijet mass bin,
for both rapidities $|y_1|, |y_2| < 0.5$ and both rapidities $0.5 <
|y_1|, |y_2| < 1.0$.\footnote{An illustration of the use of this
parameter in
  a heuristic model where standard model amplitudes are modified by a
  Veneziano formfactor has been presented ~\cite{Meade:2007sz}.}  In Fig.~\ref{fig:dijet_ratio} we compare the results from a full CMS detector simulation of the ratio $R$~\cite{Esen}, with predictions from LO QCD and model-independent contributions from Regge excitations~\cite{Anchordoqui:2008di}.  For an integrated luminosity of 10~fb$^{-1}$ the LO QCD contributions with $\alpha_{\rm QCD} = 0.1$ (corresponding to running scale $\mu \approx M_s$) are within statistical fluctuations of the full CMS detector simulation. (Note that the string scale is an optimal choice of the running scale which should normally minimize the role of higher loop corrections.) Since one of the purposes of utilizing NLO calculations is to fix the choice of the running coupling, we take this agreement as rationale to omit loops in QCD and in string theory. It is clear from Fig.~\ref{fig:dijet_ratio} that incorporating NLO calculation of the background and the signal would not significantly change the large deviation of the string contribution from the QCD background.

Although there are no $s$-channel resonances in $qq\rightarrow qq$ and
$qq'\rightarrow qq'$ scattering, KK modes in the $t$ and $u$ channels
generate calculable effective 4-fermion contact
terms~\cite{Lust:2008qc}. These in turn are manifest in an enhancement
in the continuum below the string scale of the $R$ ratio for dijet
events. For $M_{\rm KK}\le 3$~TeV, this contribution can be detected
at the LHC with 6$\sigma$ significance above SM
background~\cite{Anchordoqui:2009mm}.  In combination with the
simultaneous observation in dijet events of a string resonance at
$M_s> M_{\rm KK}$, this would consolidate the stringy interpretation
of these anomalies. In particular, it could serve to differentiate
between a stringy origin for the resonance as opposed to an isolated
structure such as a $Z'$, which would not modify $R$ outside the
resonant region. Moreover, because of the high multiplicity of the
angular momenta (up to $J=2$), the rapidity distribution of the decay
products of string excitations would differ significantly from those
following decay of a $Z'$ with $J=1$.  With high statistics, isolation
of lowest massive Regge excitations from KK replicas (with $J=2$) may
also be possible.

The compelling arguments for a possible discovery of Regge recurrences at the LHC discussed so far can be supplemented by the search of stringy signals in astrophysical experiments. Cosmological and astrophysical observations provide plentiful evidence that a large fraction of the universe's mass consists of non-luminous, non-baryonic material, known as dark matter~\cite{Bertone:2004pz}.  Among the plethora of dark matter candidates, weakly interacting massive particles (WIMPs) are especially well-motivated, because they combine the virtues of weak scale masses and couplings, and their stability often follows as a result of discrete symmetries that are mandatory to make electroweak theory viables (independent of cosmology)~\cite{Feng:2005uu}. Moreover, WIMPs are naturally produced with the cosmological densities required of dark matter~\cite{Griest:1989zh}.  An attractive feature of broken SUSY is that with R-parity conservation, the lightest supersymmetric particle (LSP) becomes an appealing dark matter candidate~\cite{Goldberg:1983nd,Ellis:1983ew}. Of course, to expose the identity of dark matter, it is necessary to measure its non-gravitational couplings. Efforts in this direction include direct detection experiments, which hope to observe the scattering of dark matter particles with the target material of the detector, and indirect detection experiments which are designed to search for the products of WIMP annihilation into gamma-rays, anti-matter, and neutrinos.

The galactic center (GC) has long been considered to be among the most promising targets for detection of dark matter annihilation, particularly if the halo profile of the Milky Way is cusped in its inner volume~\cite{Navarro:1996gj}.  However, a major adjustment in the prospects for indirect dark matter detection has materialized recently, following the discovery of a bright astrophysical source of TeV gamma-rays at the GC~\cite{Aharonian:2004wa,Aharonian:2006wh}. This implies that dark matter emission from the GC will not be detectable in a (quasi) background-free regime, and --unless one focus attention to other targets-- the peculiar spectral shape and angular distribution of dark matter annihilation must be used to isolate the signal from background.  The annihilation of WIMPs into photons typically proceeds via a complicated set of processes. Tree-level annihilation of WIMPs into quarks and leptons (or heavier states which decay into them) render a continuum emission of gamma-rays, with an energy cutoff at approximately the WIMP mass. For example, in the minimal supersymetric standard model (MSSM) neutralinos ($\chi^0$) dominantly annihilate to final state consisting of heavy fermions $b\bar b, t \bar t, \tau^+ \tau^-$ (i.e, bottom, top, and tau pairs, respectively), or gauge bosons. With the exception of the $\tau^+ \tau^-$ topology, these annihilation channels result in a very similar spectrum of gamma-rays (dominated by $\pi^0$ decay), which is in general rather featureless. Loop-level annihilation into a monochromatic gamma-rays can provide a stricking signal that helps discriminate against backgrounds. Unfortunately, for the MSSM, line emission typically has smaller magnitude than continuum emission and is out of the range of next-generation gamma-ray telescopes. It is therefore of interest to explore whether this can be mitigated by exploting the distinctive properties of superstring theory.

We consider the introduction of new operators, based on superstring theory, which avoids $p$-wave suppression by permitting neutralino $s$-wave annihilation into monochromatic gamma rays at an adequate rate.\footnote{It is important to stress that for a gaugino pair to annihilate into gauge bosons one needs a world-sheet with Euler characteristic $\chi= 2 -2g -h = -1$. It can be realized in two ways: a ``genus 3/2'' world-sheet ($g=1,\, h=1$)~\cite{Antoniadis:2004qn,Antoniadis:2005xa}, and a two-loop open string world-sheet ($g=0,\, h=3$)~\cite{Antoniadis:2005sd}.}  We may  choose a supersymmetric R-symmetry violating effective Lagrangian incorporating the above properties, once gauginos adquire mass through an unspecified mechanism~\cite{pincha}.  We can then constrain the free parameters of the model to acquire a neutralino relic density consistent with the measured abundance of dark matter~\cite{Komatsu:2008hk}. To generate a relic abundance consistent with the measured dark matter density of the universe ($\Omega_{\rm CDM} h^2 = 0.113 \pm 0.003$), requires a thermally averaged annihilation cross section $\langle \sigma v\rangle_{\rm eff} \approx 3\times 10^{-26}$cm$^3$/s.  With a choice of binos (hypercharge gauge bosons) as our LSP, and with the assumption of relatively small mixing with the other $U(1)$ subgroups in stacks $a$ and $b$, the bino is largely associated with the $U(1)$ stack $c$. At threshold $(s\approx 4m_{\chi^0}^2)$, the total
annihilation rate into gauge bosons must satisfy,
\begin{equation}
\left. \sigma v \right|_{WW} + \left. \sigma v \right|_{gg} +\left. \sigma v \right|_{BB}
= \langle \sigma v\rangle_{\rm eff}  \, .
\end{equation}
A property inherent to the model is that fixing the total annihilation rate yields  a 10\% branching fraction for $\chi^0\chi^0 \to \gamma \gamma$~\cite{pincha}. At this point, a comparison with the existing one-loop broken SUSY calculations of the annihilations $\chi^0\chi^0\to\gamma\gamma$~\cite{Bergstrom:1997fh,Bern:1997ng} and $\chi^0\chi^0\to\gamma Z$~\cite{Ullio:1997ke} is in order.  For all parameter space satisfying the measured dark matter abundance~\cite{Komatsu:2008hk}, the standard MSSM annihilation rates to $\gamma \gamma$ or $\gamma Z$ are always less than $\approx 10^{-28}\ {\rm cm}^3/{\rm s},$ and are typically even smaller. In contrast, the stringy model typically predicts $\sigma v_{\gamma \gamma} \sim 3 \times 10^{-27}$ cm$^3$/s. This can ease the rather severe restrictions placed on the MSSM parameter space in order to conform with WMAP data. For neutralinos with masses above a few hundred GeV, H.E.S.S.'s observations of the  GC~\cite{Aharonian:2004wa} can be used to probe the dark matter's annihilation cross section. It is this that we now turn to study.
\begin{figure}
\includegraphics[
width=3in ]{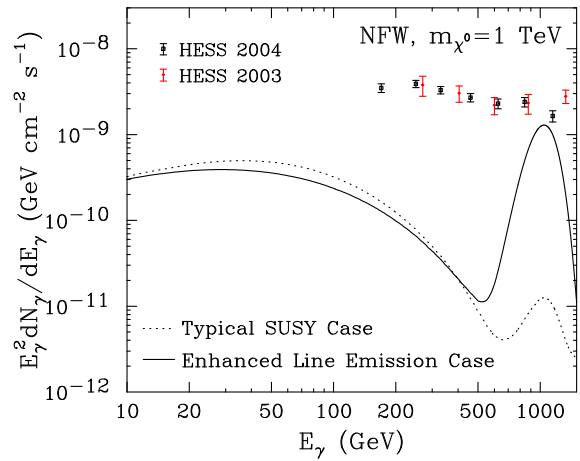}
%\postscript{spec.ps}{0.99}
%\framebox[55mm]{\rule[-21mm]{0mm}{43mm}}
\caption{Gamma ray spectrum from neutralino dark matter
  annihilating in the GC  (within a solid angle of
  $10^{-3}$~sr). The spectrum has been convolved with a gaussian of $\Delta
  E_\gamma/E_\gamma$ =15\% width, the typical energy resolution of
  H.E.S.S. and other ground based gamma ray telescopes. The solid line
  represents dark matter which annihilates 10\% of the time to
  $\gamma \gamma$. The dotted line represent dark matter
  which annihilates only 0.1\% to $\gamma \gamma$ or $\gamma Z$. In each case, we have considered $m_{\chi^0} = 1~{\rm TeV}$   and a total annihilation cross section of $3\times 10^{-26}$~cm$^3$/s. The continuum portion of the spectrum arises from the decay
  products of the W and Z bosons and QCD gluons as calculated using Pythia. Also
  shown for comparison are the measurements from
  H.E.S.S.~\cite{Aharonian:2004wa} which are generally interpreted to
  be of astrophysical origin~\cite{Aharonian:2004jr,Atoyan:2004ix}.}
\label{fig3}
\end{figure}

The differential flux of photons arising from dark matter annihilation observed in a given direction making an angle $\psi$
with the direction of the GC is given
by~\cite{Bergstrom:1997fj}
\begin{equation}
\phi^{\gamma} (\psi, E_\gamma) = \int \bar{J} \  \frac{1}{2} \ \frac{D_{\odot}}{4\pi} \ \frac{\rho^2_{\odot}}{m_\chi^2} \ \sum_f \, \langle \sigma v \rangle_f \ \frac{dN_f}{dE_{\gamma}} \, d\Omega,
\label{flux}
\end{equation}
where $\bar{J} = (1/\Delta \Omega) \int_{\Delta \Omega}
J(\psi)\,d\Omega$ denotes the average of $J$ over the solid angle
$\Delta \Omega$ (corresponding to the angular resolution of the
instrument) normalized to the local density: $J(\psi) =
(D_\odot\rho^2_{\odot})^{-1} \int_{\ell =0}^\infty \rho^2[r(\ell,
\psi)] d\ell$; the coordinate $\ell$ runs along the line of sight,
which in turn makes an angle $\psi$ with respect to the direction of
the GC ( i.e., $r^2=\ell^2+D^2_\odot-2 \ell D_\odot \cos{\psi}$); the
subindex $f$ denotes the annihilation channels with one or more
photons in the final state and $dN_f/ dE_\gamma$ is the (normalized)
photon spectrum per annihilation; $\rho(\vec x)$, $\rho_\odot =
0.3~{\rm GeV}/{\rm cm}^3$, and $D_\odot \simeq 8.5~{\rm kpc}$
respectively denote the dark matter density at a generic location
$\vec x$ with respect to the GC, its value at the solar system
location, and the distance of the Sun from the GC.  In Fig.~\ref{fig3}
we show representative gamma ray spectra from dark matter
annihilations, assuming a dark matter distribution which follows the
Navarro-Frenk-White (NFW) halo profile~\cite{Navarro:1996gj}.  The
dotted line denotes the gamma ray spectrum from a 1~TeV neutralino
with a total annihilation rate $\sigma v|_{\rm tot} = 3 \times
10^{-26}$ cm$^3$/s, but which annihilates to $\gamma \gamma$ or
$\gamma Z$ only 0.1\% of the time, which is typically for a TeV
neutralino in the MSSM. If the fraction of neutralino annihilations to
$\gamma \gamma$ were much larger, the prospects for detection would be
greatly improved. As previously noted, the stringy processes yield
much larger annihilation cross sections to this distinctive final
state. The solid line in Fig.~\ref{fig3} shows the gamma ray spectrum
predicted for a neutralino which annihilates 10\% of the time to
$\gamma \gamma$. Unlike in the case of a typical MSSM neutralino, this
leads to a very bright and potentially observable gamma ray
feature. If an experiment were to detect a strong gamma ray line flux
without a corresponding continuum signal from the cascasdes of other
annihilation products, it could indicate the presence of a low string
scale.\\

{\em In summary,} in D-brane constructions, the full-fledged string
amplitudes supplying the dominant contributions to dijet cross
sections are completely independent of the details of
compactification. If the string scale is in the TeV range, such
extensions of the standard model can be of immediate phenomenological
interest. In this section we have made use of the amplitudes evaluated
near the first resonant pole to report on the discovery potential at
the LHC for the first Regge excitations of the quark and
gluon. Remarkably, after a few years of running, the reach of LHC in
the dijet topology ($S/N = 210/42$) can be as high as 6.8~TeV. This
intersects with the range of string scales consistent with correct
weak mixing angle found in the $U(3) \times U(2) \times U(1)$ quiver
model~\cite{Antoniadis:2000ena}. For string scales as high as 5.0 TeV,
observations of resonant structures in $pp\rightarrow \gamma~ +$ jet
can provide interesting corroboration for stringy physics at the
TeV-scale. In addition, supersymmetric extensions of the D-brane
models can lead to an acceptable dark matter relic abundance of
bino-like neutralinos which annihilate a large fraction of the time
($\sim 10\%$) to $\gamma \gamma$, potentially producing a very bright
and distinctive gamma ray spectral line which could be observed by
current or next-generation gamma ray telescopes. Such a feature is
multiple orders of magnitude brighter than is typically predicted for
neutralino dark matter in the MSSM.

%\end{document}

%%%%%%%%%%%%%%%%%%%%%%%%%%%%%%%%%%%%%%%%%%%%%%%%%%%%%%%%%%%%%%%%%%%%%%%%%%%%%%%%%%%%%%%%%%%%%%
%%%%%%%%%%%%%%%%%%%%%%%%%%%%%%%%%%%%%%%%%%%%%%%%%%%%%%%%%%%%%%%%%%%%%%%%%%%%%%%%%%%%%%%%%%%%%%
\chapter{Conclusion}
As of this writing  the  Large Hadron Collider has succeeded in
producing the first collisions and has  collected a small amount of
data. In the future much more data will be forthcoming and the
energy of LHC will be ramped first to 7 TeV, then to 10 TeV and
finally to its optimum value of 14 TeV.   The LHC presents a unique
opportunity to put a variety of theoretical proposals to test.  This
report brings together diverse  views and approaches to what that
new physics is. Eighty seven active researchers working on various
theoretical aspects of new physics have contributed to this report.
Thus the report presents a very broad overview of the type of new
physics that might emerge from the LHC. It is ultimately the data
from the LHC that will determine which if any of the theoretical
models presented here will be left standing in the end. It is hoped
that the report here will be of value to the experimentalists  to
determine just that.
%%%%%%%%%%%%%%%%%%%%%%%%%%%%%%%%%%%%%%%%%%%%%%%%%%%%%%%%%%%%%%%%%%%%%%%%%%%%%%%%%%%%%%%%%%%%%%
%%%%%%%%%%%%%%%%%%%%%%%%%%%%%%%%%%%%%%%%%%%%%%%%%%%%%%%%%%%%%%%%%%%%%%%%%%%%%%%%%%%%%%%%%%%%%%
\chapter*{Acknowedgements}

Fermilab is operated by Fermi Research Alliance, LLC under Contract
No. DE-AC02-07CH11359 with the United States Department of Energy.

H.D. was supported by the US Department of Energy under Grant
Contract DE-AC02-98CH10886.

B.D. was supported in part by the DOE grant DE-FG02-95ER40917 and
would like to thank his collaborators Richard Arnowitt, Adam
Arusano, Rouzbeh Allahverdi, Alfredo Gurrola, Teruki Kamon, Nikolay
Kolev, Abram Krislock, Anupam Mazumdar, Yukihiro Mimura  and Dave
Toback for the works related to this review.

D.F. is supported in part by DOE grant DE-FG92-95ER40899.

H.G. is supported in part by NSF grant PHY-0757959

K.K. was supported by US Department of Energy contract
DE-AC02-76SF00515.

P.L. was supported by  NSF grant PHY-0503584 and by the IBM Einstein
Fellowship.

G.L. is  partially supported by the U.S.~Department of Energy under
Grant No. DE-FG02-91ER40688.

Z.L. is supported in part by NSF grant  PHY-0653342

P.N. is   supported in part by NSF grant PHY-0757959. SUSY09 and
Pre-SUSY09 were supported by NSF PHY-0834022 and DE-SC0001075.

B.D.N. was supported by National Science Foundation Grant
PHY-0653587.

E.P. was supported by the U.S.~Department of Energy under contract
DE-FG02-92ER-40699.

J.S. is funded by MICINN and projects FPA2006-05294, FQM101, FQM437
and FQM03048.

T.T. is supported in part by NSF grant PHY-0757959.

X.T. thanks the UW IceCube Group for making his visit to Wisconsin,
where this report was prepared, possible. This research was
supported in part by the United States Department of Energy.

Work of J.F.W.V. supported by the US National Science Foundation
under grant No. PHY-0652363, by European Union ITN UNILHC
(PITN-GA-2009-237920), by the Consolider Multidark project
CSD2009-00064 (MICIIN), by the FPA2008-00319/FPA grant (MICIIN), by
the PROMETEO/2009/091 grant (Generalitat Valenciana), by German
Ministry of Education and Research (BMBF) contract 05HT6WWA, and by
Colombian grant UdeA Sostenibilidad 2009-2010.

C.E.M.W.'s work at ANL is supported in part by the U.S. Department
of Energy (DOE), Div. of HEP, Contract DE-AC02-06CH11357.

%%%%%%%%%%%%%%%%%%%%%%%%%%%%%%%%%%%%%%%%%%%%%%%%%%%%%%%%%%%%%%%%%%%%%%%%%%%%%%%%%%%%%%%%%%%%%%
%%%%%%%%%%%%%%%%%%%%%%%%%%%%%%%%%%%%%%%%%%%%%%%%%%%%%%%%%%%%%%%%%%%%%%%%%%%%%%%%%%%%%%%%%%%%%%
%\twocolumn
%\chapter*{Master Reference List}
%\chapter{Master Reference List (Sample)}
%\begin{thebibliography}{999}
%\input{alpharef.tex}
%\input{ref.tex}
%
%\input{bib/1_Intro.tex}
%\input{bib/2_SUSY.tex}
%\input{bib/3_Higgs.tex}
%\input{bib/4_CP.tex}
%\input{bib/5_Dark.tex}
%\input{bib/6_Top.tex}
%\input{bib/7_ZP.tex}
%\input{bib/8_HS.tex}
%\input{bib/9_Nu.tex}
%\input{bib/10_ED.tex}
%\input{bib/11_String.tex}
%\end{thebibliography}

\end{document}